\documentclass[12pt,oneside]{report}
\pdfoutput=1
\usepackage{yalethesis}
\usepackage{graphicx}
\usepackage{amsmath}
\usepackage{amssymb}
\usepackage[footnote]{acronym}
\usepackage{makeidx}
\usepackage{fancyheadings}
\usepackage{xspace}
\usepackage{longtable}
\usepackage{textcomp}
\usepackage{epsfig}
\usepackage{cite}
\usepackage{subfig}
\usepackage{multirow}
\usepackage{epstopdf}
\usepackage[bookmarks=false]{hyperref}

\graphicspath{{theory//}{experiment//}{anintro//}{rec//}{acc//}{trigger//}{dedx//}{bre//}{res_back//}{results//}{dedx2nsigma//}{uncertainties//}{additional//}{functions//}}

\DeclareMathOperator{\MeV}{MeV\!\!}
\DeclareMathOperator{\GeV}{GeV\!\!}

 \makeindex

\begin{document}

\title{Yield and suppression of electrons from open heavy-flavor decays in heavy-ion collisions}
\author{Anders Garritt Knospe}
\awardate{December 2011} \advisor{Professor John Harris}
\copyrightyear{2011} \copyrighttrue

\begin{abstract}

Measurements by the STAR and PHENIX collaborations indicate that a quark-gluon plasma, a hot and dense state of matter in which quarks and gluons are not confined inside hadrons, is formed in heavy-ion collisions at the Relativistic Heavy Ion Collider.  Charm and bottom quarks have been predicted to interact with the medium differently than the light quarks; a study of heavy quark interactions with the medium provides an important test of theoretical models of the quark-gluon plasma.

The spectrum of non-photonic electrons and positrons $(e^{\pm})$ is dominated by $e^{\pm}$ from the semileptonic decays of $D$ and $B$ mesons.  Therefore, non-photonic $e^{\pm}$ serve as proxies for heavy quarks.  A measurement of the modification of the non-photonic $e^{\pm}$ spectrum in nucleus-nucleus collisions relative to $p+p$ collisions allows the interactions of heavy quarks with the medium to be studied.  Previous measurements indicate that high-transverse-momentum non-photonic $e^{\pm}$ are suppressed in Au + Au collisions at $\sqrt{s_{NN}}=200$ GeV relative to $p+p$ collisions at the same energy.  The magnitude of that suppression is larger than was anticipated and it has been a challenge for theoretical models to predict the in-medium energy loss of light and heavy quarks simultaneously.

This dissertation presents the first measurement of the yield of non-photonic $e^{\pm}$ from open heavy-flavor decays in Cu + Cu collisions at $\sqrt{s_{NN}}=200$ GeV and the suppression of that yield relative to $p+p$ collisions.  A comparison of this result to similar results for Au + Au collisions provides some indication that the geometry of a heavy-ion collision affects the average amount of energy loss by heavy quarks passing through the quark-gluon plasma.


\end{abstract}

\beforepreface
\prefacesection{Acknowledgements}
{
\footnotesize
``Three quarks for Muster Mark!\\
Sure he has not got much of a bark\\
And sure any he has it's all beside the mark."\\
--- James Joyce\cite{FinnegansWake}
\vspace{12pt}

``There is a theory which states that if ever anyone discovers exactly what the Universe is for and why it is here, it will instantly disappear and be replaced by something even more bizarre and inexplicable.  There is another theory which states that this has already happened."\\
--- Douglas Adams\cite{Restaurant}
\vspace{12pt}
}

I would like to thank
{\begin{itemize}
\setlength{\itemsep}{1pt}
\setlength{\parskip}{0pt}
\setlength{\parsep}{0pt}
\item John Harris, for support, insight, and being the best advisor I could possibly ask for.
\item Helen Caines, for candy and always being ready to answer a question or help with a misbehaving computer program.
\item Jaro Biel\v{c}\'{i}k, for all of his help as I worked on this project, particularly when I was starting out in this field.
\item Thomas Ullrich, for finding the $\Upsilon$ feed-down spectrum and for providing help and knowledge.
\item Keith Baker and Walter Goldberger for being so interested in serving on my committee and for asking good questions.
\item Liz Atlas for being a great assistant who is always willing to help.
\item Nikolai Smirnoff, Matt Lamont, Richard Witt, Betty Abelev, Jana Biel\v{c}\'{i}kov\'{a}, Mark Heinz, Oana Catu, J\"{o}rn Putschke, Tomas Aronsson, Ben Hicks, Rongrong Ma, and Alice Ohlson: I've really enjoyed being a part of this group.
\item Professors Iachello, Wettig, Appelquist, Alhassid, Sandweiss, Casten, Skiba, Mochrie, Gay, Moncrief, and Kharzeev and the rest of the Yale Physics department.
\item Christina Markert for helping me take the next step.
\item Alex Suaide, Gang Wang, Xin Dong, Manuel Calder\'{o}n de la Barca S\'{a}nchez, Wei Xie, Xin Li, Wenqin Xu, Priscilla Kurnadi, and Shingo Sakai for their assistance on this analysis.
\item The STAR Heavy-Flavor Working Group for their advice and assistance.
\item Jamie Dunlop, for knowing so much about so many aspects of STAR and always being willing to share that knowledge.
\item J\'{e}r\^{o}me Lauret, for his ingenuity and hard work maintaing the STAR computing system despite the efforts of many physicists to destroy it.
\item the STAR embedding team for helping me get my simulations.
\item everybody else on the STAR collaboration's author list, past and present, for making the experiment what it is.
\item the many scientists whose work is cited in the Bibliography: I have truly stood on the shoulders of giants.
\item the taxpayers, the United States Congress, George W. Bush, Barack Obama, and the Department of Energy for funding this project.
\item Christine Nattrass for computing tips, food, thesis chocolate, alcohol, and letting me dress up as Elvis.
\item Stephen Baumgart for being a friend and an office-mate who did not drive me crazy.
\item whoever invented hummus.
\item Elena Bruna for teaching me my favorite Italian word.
\item Sevil Salur, for friendship, fruit, and being a cool first shift leader.
\item Anthony Timmins for beer, advice, and being a good bloke.
\item Rashmi Raniwala: authentic, home-made Indian food made Brookhaven so much better.
\item Trey Parker, Matt Stone, Seth McFarlane, and Matt Groening for their cartoons.
\item Jon Stewart\footnote{I would thank Stephen Colbert but he doesn't believe in science or learning.}, Tina Fey, Greg Daniels, Marc Cherry, Mitch Hurwitz, Chris Carter, and the men and women of Broadway for entertainment.
\item The National Geographic Society for supporting my education.
\item The ARCS Foundation for supporting so much of my undergraduate education.
\item Richard Mawhorter, Alfred Kwok, Thomas Moore, Catalin Mitescu, David Tannenbaum, Alma Zook, Ron Hellings, Peter Saeta, and Carla Riedel for providing a great undergraduate education.
\item Dick Smith for helping me grow as a physicist.
\item Ed Adams for giving me my first research job.
\item Mrs. Rumley, Mrs. Smith, Mrs. Tyrell, Mrs. Jermunson, Mrs. Townsend, Mr. Crowe, Mrs. Miller, Mr. Gephardt, Mr. Mayer, Mr. Rein, Mr. Wilson, Mr. Watkins, Mr. Reisig, Mr. Myers, and Ms. Rogers for laying the foundations.
\item my family and friends for food, shelter, DNA, laughter, support, driving lessons, computers, birthday cakes, science lectures, teaching me how to use a table saw, airfare, education, my big TV, loans, alternate views of the universe, swimming lessons, theater tickets, companionship, and all their love.  I could not have accomplished this without you.
\end{itemize}
}
\afterpreface
\clearpage \pagenumbering{arabic}
\newpage

\chapter{Introduction}
\section{The Standard Model}
Throughout the early and mid-20th century, as the energies of particle accelerators increased and cosmic-ray measurements became more refined, a large number of particles were discovered.  Questions arose as to whether all observed particles were truly fundamental.  In 1964, Zweig and Gell-Mann independently suggested that all hadrons could be understood as being combinations of fundamental particles, which Gell-Mann called ``quarks."~\cite{GriffithsParticles}  It was hypothesized that protons, neutrons, and the other baryons are made up of three quarks, while mesons are made up of a quark and an antiquark.  Initially, the existence of three flavors of quarks - down $(d)$, up $(u)$, and strange $(s)$ - was proposed as an explanation for the observed properties of the baryons and mesons then known.  A serious theoretical objection raised against the early quark model was that the Pauli exclusion principle should preclude the existence of particles like the $\Delta^{++}$ and $\Delta^{-}$.  The quark model proposed that those spin-$\tfrac{3}{2}$ particles were composed of three identical fermions ($uuu$ for the $\Delta^{++}$ and $ddd$ for the $\Delta^{-}$) with their spins parallel, \textit{i.e.}, three identical fermions in the same quantum state, a situation not allowed by the Pauli exclusion principle.  In 1964, Greenberg proposed the existence of a new quantum number, color charge, to resolve this issue.  The existence of three color charges, called ``red," ``green," and ``blue," was proposed; each of the three $u$ quarks in the $\Delta^{++}$ would have a different color, and the Pauli exclusion principle would not be violated.  The first direct experimental support for the quark model came in the late 1960s: deep inelastic scattering experiments, studies of electron-proton scattering with large momentum transfer, indicated that the proton is not a point particle, but contains three point-like scattering centers.~\cite{HalzenMartin,GriffithsParticles}  

The quark model has been expanded to include six flavors of quarks and has become an integral part of the Standard Model of particle physics, in which the interactions among the various flavors of quarks and leptons are mediated by the exchange of vector bosons.  The photon mediates the electromagnetic interaction between electrically charged particles, the $W^{\pm}$ and $Z^{0}$ bosons mediate the weak interaction, and gluons mediate the strong interaction between quarks.  Gluons themselves carry color charge: specifically one unit of color and one unit of anticolor.  The Standard Model includes three of the four fundamental interactions; only gravity is excluded.  The only particle in the Standard Model that has not been discovered is the Higgs boson, which has been the subject of an intense search effort at the Tevatron and now at the LHC.

\section{Quantum Chromodynamics}
The interactions of quarks and gluons is described within the framework of quantum field theory by Quantum Chromodynamics (QCD).  QCD is a non-abelian gauge theory with each flavor of quark assigned to the fundamental representation of the $SU(3)$ symmetry group.  The QCD Lagrangian~\cite{PeskinSchroeder} is the Yang-Mills Lagrangian for the $SU(3)$ group:

\begin{equation}
\label{eq:theory:qcd_lagrangian}
\mathcal{L}=-\tfrac{1}{4}F_{\mu\nu}^{a}F^{\mu\nu a}+\bar{\psi}(i\gamma^{\mu}D_{\mu}-m)\psi.
\end{equation}

\noindent Here, the covariant derivative is

\begin{equation}
D_{\mu}=\partial_{\mu}-igA_{\mu}^{a}t^{a},
\end{equation}

\noindent where $g$ is the coupling constant, $A_{\mu}^{a}$ represents the gauge field (gluons), and the $t^{a}$ are the generators for $SU(3)$.  In Equation~\ref{eq:theory:qcd_lagrangian}, $F_{\mu\nu}^{a}$ is the field strength.

\begin{equation}
\label{eq:theory:qcd_field_strength}
[D_{\mu},D_{\nu}]=-igF_{\mu\nu}^{a}t^{a}\Longrightarrow F_{\mu\nu}^{a}=\partial_{\mu}A_{\nu}^{a}-\partial_{\nu}A_{\mu}^{a}+gf^{abc}A_{\mu}^{b}A_{\nu}^{c},
\end{equation}

\noindent where the $f^{abc}$ are the $SU(3)$ structure constants: $[t^{a},t^{b}]=if^{abc}t^{c}$.  When the Lagrangian in Equation~\ref{eq:theory:qcd_lagrangian} is expanded, the last term in Equation~\ref{eq:theory:qcd_field_strength} gives rise to three- and four-gluon couplings.  The coupling of the gauge field to itself, which does not occur in Quantum Electrodynamics (QED), results in the strong coupling constant being asymptotically free.

\begin{figure}[htbp]
\begin{center}
\includegraphics[height=0.7\linewidth]{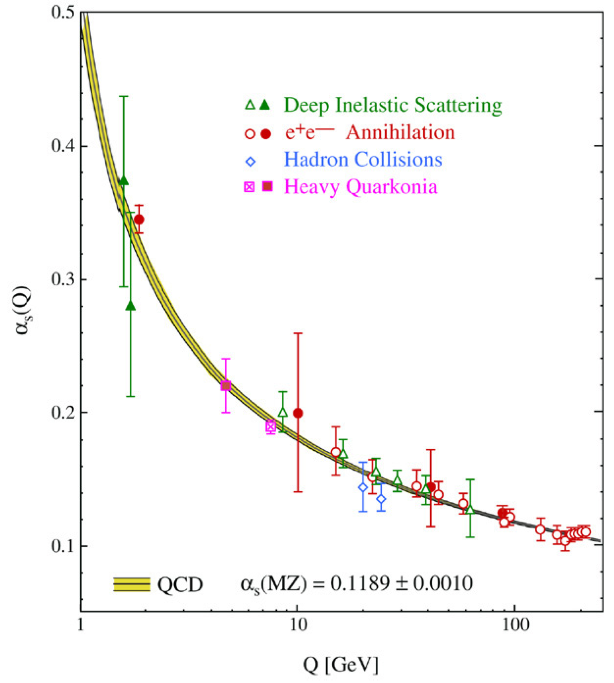}
\caption[Measured values of the strong coupling constant as a function of the momentum transfer.]{Measured values of the strong coupling constant $\alpha_{s}$ as a function of the momentum transfer $Q$.\protect\cite{Bethke2007351}  Open symbols indicate Next-to-Leading-Order (NLO) calculations, while closed symbols indicate Next-to-Next-to-Leading-Order (NNLO) calculations.}
\label{fig:intro:running_as}
\end{center}
\end{figure}

The QED vacuum contains virtual particle-antiparticle pairs which arise from the vacuum, live a short time (consistent with the Heisenberg uncertainty principle), then annihilate.  If the virtual particle and antiparticle carry electric charge, the pair constitutes an electric dipole.  In the presence of a real electric charge, the virtual dipoles are reoriented to be antiparallel to the field of the real charge, resulting in the partial screening of that charge.  The vacuum functions as a dielectric and the effective electric charge seen at large distances (or small momentum transfers) is less than the ``bare" charge seen at small distances (or large momentum transfers).~\cite{GriffithsParticles}  A similar situation arises in QCD: virtual $q\bar{q}$ pairs in the vacuum will couple to the gluon field produced by a real quark and partially screen that quark's color charge.  However, QCD also includes couplings of gluons to gluons.  Those virtual gluon loops are found to give rise to an ``antiscreening" effect, which is larger than the screening effect.  The QCD coupling constant therefore decreases with decreasing distance and increasing momentum transfer $q$.  The coupling constant $\alpha_{s}=g^{2}/4\pi$ as a function of momentum transfer has been calculated~\cite{PeskinSchroeder,Bethke2007351} to lowest order to be

\begin{equation}
\label{eq:theory:running_as}
\alpha_{s}(q^{2})=\frac{2\pi}{(\frac{11}{3}n_{c}-\frac{2}{3}n_{f})\ln(q/\Lambda)}=\frac{2\pi}{(11-\frac{2}{3}n_{f})\ln(q/\Lambda)},
\end{equation}

\noindent where $n_{c}=3$ is the number of colors, $n_{f}$ is the number of flavors, and $\Lambda\approx 200$ MeV is the QCD scale.  Figure~\ref{fig:intro:running_as} shows\footnote{Color versions of all Figures are available in the electronic version of this dissertation.} measurements of the strong coupling constant~\cite{Bethke2007351}; $\alpha_{s}$ has been observed to decrease with increasing momentum transfer in a manner consistent with theoretical predictions.  When $q$ is significantly larger than $\Lambda$, QCD can be treated perturbatively; this is the case for the production of $c$ and $b$ quarks in heavy-ion collisions.

When $q<\Lambda$, perturbation theory breaks down and other approaches must be used to perform calculations in QCD.  The most productive approach to performing non-perturbative QCD calculations has been Lattice QCD.  Continuous spacetime is replaced with a set of discrete points on a four-dimensional Euclidean lattice.  Calculations are performed numerically by evaluating path integrals between neighboring points on this lattice.~\cite{PeskinSchroeder}  Lattice QCD calculations indicate that for large distances (or small momentum transfers) the potential between a quark and an antiquark increases linearly with the distance separating them (\textit{cf.} the potential in the weak-coupling limit, which is a coulomb potential with a running coupling constant).  As the separation distance between a quark and antiquark increases, the potential energy increases without limit.  At some point it becomes energetically favorable for a second  quark-antiquark pair to be created.  If a quark-antiquark pair is separated, the result is not an isolated (anti)quark, but two mesons.  Observing an isolated quark seems to be impossible and all particles that have been directly observed are color singlets.

\begin{figure}[htbp]
\begin{center}
\includegraphics[width=0.85\linewidth]{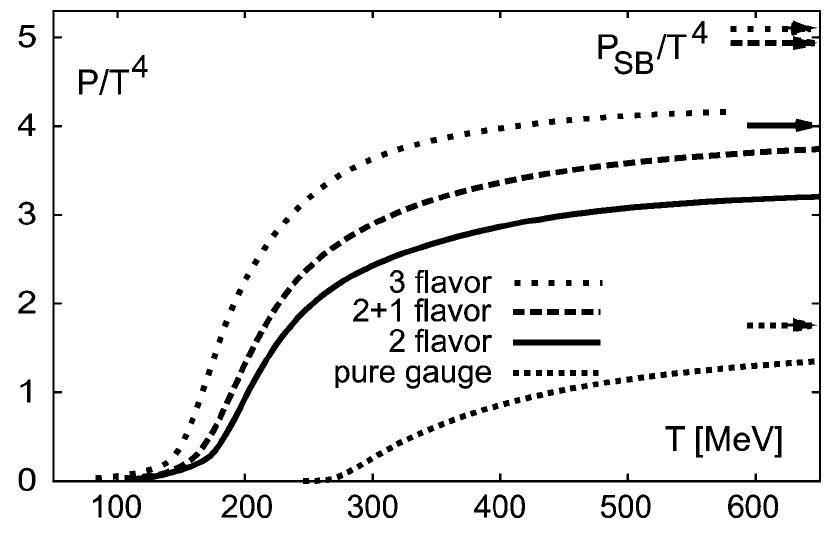}
\caption[Lattice QCD calculations of the ratio $P/T^{4}$ as a function of temperature.]{The ratio of pressure to $T^{4}$ calculated\protect\cite{KarschLattice2002} using lattice QCD as a function of temperature.  The different curves indicate the number of active quark flavors; ``2+1 flavor" indicates that the $s$ quark is heavier than the $d$ and $u$, while ``pure gauge" indicates 0 active flavors.  The arrows indicate the Stefan-Boltzmann predictions of $P/T^{4}$ for an ideal partonic gas for each set of active quark flavors.}
\label{fig:intro:lattice_phase_transition}
\end{center}
\end{figure}

While quarks are confined in the states of matter normally observed, asymptotic freedom implies that in nuclear matter at high temperature, hadrons should ``melt" and the new degrees of freedom of the system should be deconfined quarks and gluons, rather than hadrons.  Such a state of matter is called a ``Quark-Gluon Plasma" (QGP).  The transition from a hadronic state to a QGP is accompanied by an increase in the number of degrees of freedom, implying an increase in the entropy density and pressure as the temperature increases through the transition region~\cite{STAR_white_paper2005}.  For an ideal gas of quarks and gluons (in which the quarks are massless and the deconfined quarks and gluons do not interact with each other), the equation of state~\cite{KarschLattice2002} is

\begin{equation}
\label{eq:theory:ideal_partonic_gas}
\frac{P}{T^{4}}=\frac{\pi^{2}}{90}\left[2(n_{c}^{2}-1)+\frac{7}{2}n_{c}n_{f}\right]=\frac{\pi^{2}}{90}\left[16+\frac{21}{2}n_{f}\right],
\end{equation}

\noindent where $P$ is the pressure and $T$ is the temperature.  Lattice QCD calculations~\cite{KarschLattice2002} show a rapid increase in $P/T^{4}$ with increasing $T$ in the vicinity of a critical temperature $T_{c}\approx$ 160 MeV.  For temperatures $\gtrsim 2T_{c}$, $P/T^{4}$ saturates at values somewhat less than the ideal (Stefan-Boltzmann) limit given in Equation~\ref{eq:theory:ideal_partonic_gas}.  Because the QGP does not reach the limit of an ideal partonic gas at these temperatures, it is sometimes referred to as a ``strongly interacting QGP" (sQGP).  These findings are shown in Figure~\ref{fig:intro:lattice_phase_transition}.

\begin{figure}[htbp]
\begin{center}
\includegraphics[width=0.85\linewidth]{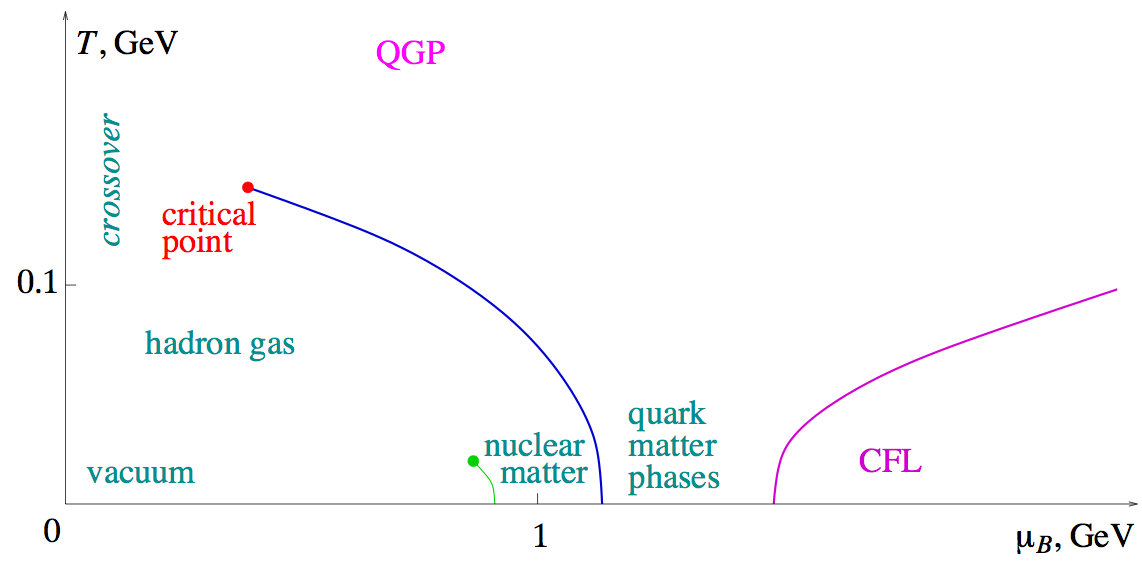}
\caption[Schematic phase diagram for strongly interacting matter.]{A schematic phase diagram for (thermally equilibrated) strongly interacting matter\protect\cite{StephanovQCDphase} with temperature on the vertical axis and baryo-chemical potential potential on the horizontal axis.  The matter produced in heavy-ion collisions is expected to lie at low baryo-chemical potentials and high temperature (in the vicinity of the ``crossover" label).}
\label{fig:intro:phase_diagram}
\end{center}
\end{figure}

Figure~\ref{fig:intro:phase_diagram} is a schematic phase diagram for strongly interacting matter~\cite{StephanovQCDphase}, with the temperature on the vertical axis and the baryo-chemical potential $(\mu_{B})$ on the horizontal axis.  Many details, including the numerical values and orders of the proposed phase transitions, are the subjects of debate and active research in the theoretical and experimental communities.  Nevertheless, the qualitative form of the phase diagram shown in Figure~\ref{fig:intro:phase_diagram} is generally accepted.  For low temperatures and baryo-chemical potentials, strongly interacting matter is in a hadronic state, either as a hadron gas or as nucleons in atomic nuclei.  For low temperatures and $\mu_{B}$ larger than nuclear values (conditions expected to be present in neutron stars) a variety of ordered quark matter phases is predicted.  As described above, at high temperatures strongly interacting matter is expected to exist as a quark-gluon plasma.  The nature of the transition from the QGP to the hadronic phase is not fully understood.  Lattice QCD calculations at $\mu_{B}=0$ tend to favor a crossover rather than a sharp phase transition, while other models for larger $\mu_{B}=0$ favor a first-order phase transition.  These results suggest the existence of a critical point in the phase diagram.  This critical point may be in the region of the phase diagram accessible to heavy-ion collision experiments.  An effort is underway at the Relativistic Heavy Ion Collider (RHIC) to find the critical point.~\cite{STAR_CriticalPoint2010}  This dissertation is concerned with the strongly interacting matter in heavy-ion collisions, which exists at high temperature and low baryo-chemical potential.  Such conditions are also predicted to have existed in the early universe, $\lesssim 10\mu\mathrm{s}$ after the Big Bang~\cite{RHIC_ScienceMag1999} and before the formation of hadrons.

\begin{figure}[htbp]
\begin{center}
\includegraphics[width=0.85\linewidth]{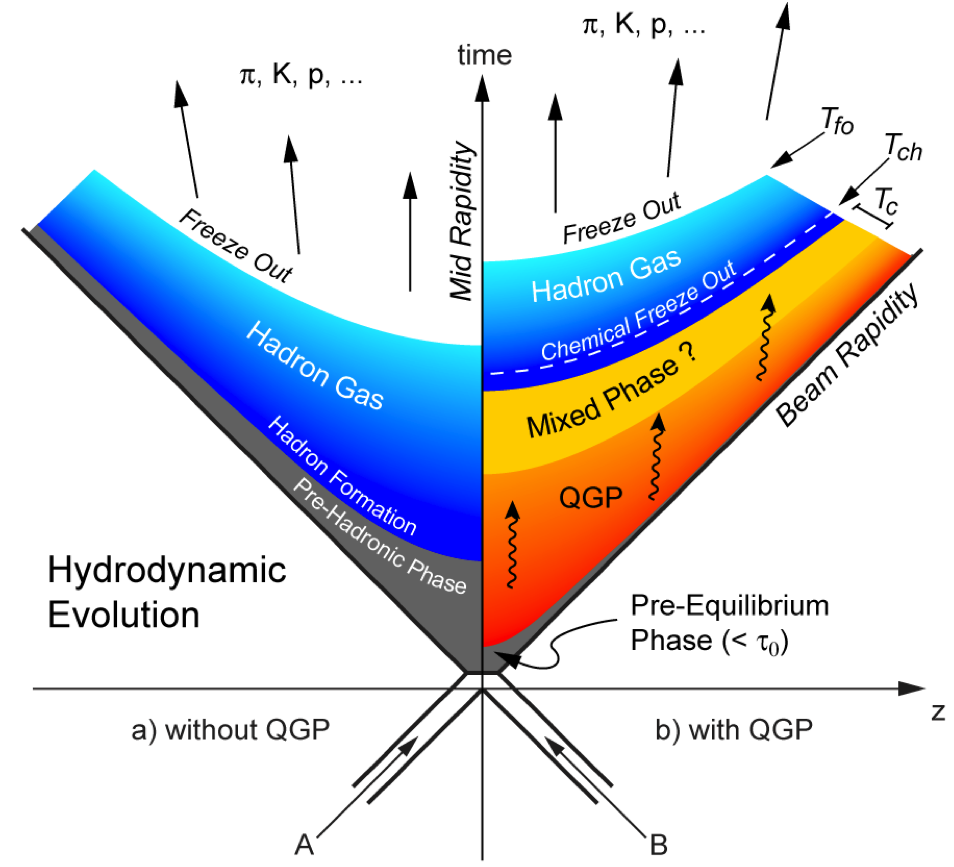}
\includegraphics[width=0.85\linewidth]{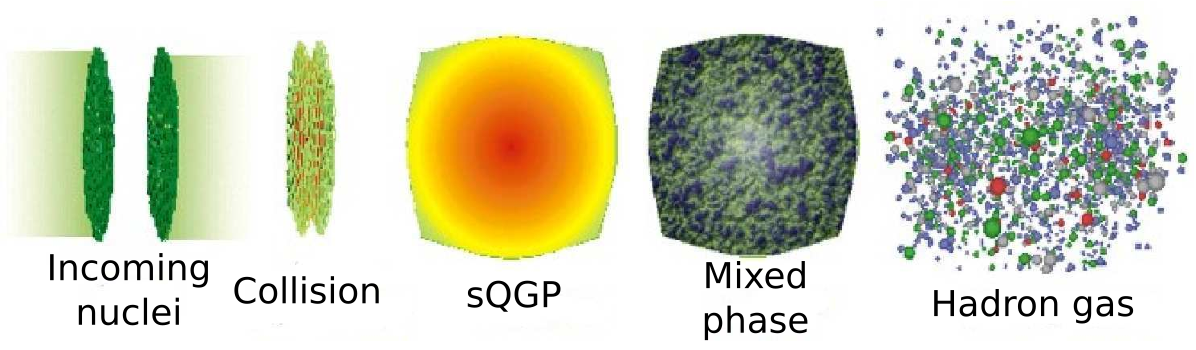}
\caption[Diagrams representing the stages of a heavy-ion collision.]{Upper plot: spacetime diagram\protect\cite{UllrichRHICOverview2007} representing the stages of a heavy-ion collision.  The right-hand side includes the formation of a quark-gluon plasma.  Lower plot: Illustration (not to scale) of some of the stages of a heavy-ion collision at various times.\protect\cite{BassQM2001}  The incoming nuclei are heavily Lorentz contracted (the Lorentz factor $\gamma\approx100$ for the highest RHIC energies).  The acronym ``sQGP" indicates ``strongly interacting QGP."}
\label{fig:intro:collision_spacetime}
\end{center}
\end{figure}

At RHIC (see Section~\ref{sec:experiment:rhic}) gold nuclei (and copper nuclei) are collided at center-of-mass energies of 200 GeV per nucleon-nucleon pair (denoted as $\sqrt{s_{NN}}$=200 GeV).  These collisions introduce a large amount of energy into a region of space the size of an atomic nucleus (and much larger than the volume of a single hadron).  At RHIC, the highest energy heavy-ion collisions are estimated~\cite{PhysRevC.70.054907} to reach energy densities of $\approx$ 4.9 GeV/fm$^{3}$.  This is seven times the critical energy density of $\approx 700\MeV/\mathrm{fm}^{3}$ (predicted by Lattice QCD calculations~\cite{KarschLattice2002}) necessary for the formation of a QGP.  Figure~\ref{fig:intro:collision_spacetime} (upper) is a schematic spacetime diagram of the expected stages of a heavy-ion collision for two scenarios, with the right-hand side including the formation of a QGP.~\cite{UllrichRHICOverview2007}  Figure~\ref{fig:intro:collision_spacetime} (lower) is a visual representation~\cite{BassQM2001} of the various stages of a heavy-ion collision.  The QGP formed in a heavy-ion collision is generally assumed to reach thermal equilibrium (at least locally), although there is debate on this point.  The QGP expands and cools, eventually reaching the critical temperature, where the system makes the transition to a hadron gas.  This transition may be a sharp phase transition, or a crossover during which QGP and hadronic gas coexist (called the ``Mixed Phase").  As described above, recent Lattice QCD calculations tend to favor a crossover for low baryo-chemical potentials (values of $\mu_{B}$ expected for RHIC collisions).  After the transition from a QGP to a hadron gas, the hadrons continue to interact with each other inelastically until the system reaches the ``chemical freezeout" temperature ($T_{ch}$ in Figure~\ref{fig:intro:collision_spacetime}).  When the system reaches chemical freezeout, the ratios of the yields of the various particle species are fixed.  The hadrons continue interacting elastically until the system reaches the ``thermal freezeout" temperature ($T_{fo}$ in Figure~\ref{fig:intro:collision_spacetime}), after which there are no hadronic interactions.  The shower of particles produced (for Au + Au collisions at $\sqrt{s_{NN}}$=200 GeV, approximately 1000 particles per unit rapidity at mid-rapidity) continues outward from the collision vertex and may be detected by experiment.

\clearpage

\clearpage

\section{Heavy-Ion Collisions: Basic Concepts}
\label{sec:theory:basic_concepts}

Some of the experimental signatures of the formation of a quark-gluon plasma will be discussed in Section~\ref{sec:theory:signatures}.  However, before that discussion the standard coordinate system and kinematic variables will be defined (Section~\ref{sec:theory:basic_concepts:kinematic}) and the geometry of a heavy-ion collision will be described (Section~\ref{sec:theory:basic_concepts:collision}).

\subsection{Coordinate System and Kinematic Variables}
\label{sec:theory:basic_concepts:kinematic}

The beam line is defined to be the $z$-axis (along which the incoming nuclei travel), with the $x$- and $y$-axes of a Cartesian coordinate system oriented horizontally and vertically, respectively.  In a spherical coordinate system, the polar angle is referred to as $\theta$, while the azimuthal angle is referred to as $\phi$.  The transverse momentum $(p_{T})$ of a particle is the component of its momentum in the direction perpendicular to the $z$-axis:

\begin{equation}
p_{T}=\sqrt{p_{x}^{2}+p_{y}^{2}}=p\sin\theta=p_{z}\tan\theta,
\end{equation}

\noindent where $p$ is the magnitude of the particle's momentum, and $p_{x}$, $p_{y}$, and $p_{z}$ are the $x$-, $y$-, and $z$-components of its momentum, respectively.  The transverse momentum is invariant under Lorentz boosts along the beam direction.  The rapidity of a particle is generally defined to be

\begin{equation}
rapidity=\mathrm{arctanh}\frac{v}{c}=\mathrm{arctanh}\frac{pc}{E}=\tfrac{1}{2}\ln\frac{E+pc}{E-pc},
\end{equation}

\noindent where $v$ is the particle's velocity, $E$ is its energy, and $c$ is the speed of light in vacuum.  However, this dissertation uses a modified definition of the rapidity: the rapidity relative to the beam axis.  This quantity, also denoted as $y$, is

\begin{equation}
y=\tfrac{1}{2}\ln\frac{E+p_{z}c}{E-p_{z}c}.
\end{equation}

\noindent For particles moving at ultrarelativistic velocities, the rapidity $y$ is well approximated by the pseudorapidity $\eta$:

\begin{equation}
y\approx\eta=\tfrac{1}{2}\ln\frac{p+p_{z}}{p-p_{z}}=-\ln\tan\frac{\theta}{2}.
\end{equation}

\noindent Unlike the rapidity, the pseudorapidity is independent of the particle's mass.  This thesis is primarily concerned with $e^{\pm}$ with momenta $>2\GeV/c$; the pseudorapidity will usually be used in place of the rapidity, and instead of the polar angle $\theta$.  For a particle with mass $m$, the ``transverse mass" is defined to be

\begin{equation}
m_{T}=\sqrt{p_{T}^{2}c^{2}+m^{2}}.
\end{equation}

For a heavy-ion collision, the energy quoted is typically the center-of-mass collision energy per nucleon-nucleon pair $\sqrt{s_{NN}}$.  Collision energies referred to in this dissertation are values of $\sqrt{s_{NN}}$ unless otherwise noted, even if the symbol ``$\sqrt{s_{NN}}$" has been omitted for the sake of brevity.\footnote{For example, ``200-GeV Cu + Cu collision" means ``Cu + Cu collision at $\sqrt{s_{NN}}$=200 GeV."} Furthermore, this dissertation frequently mentions RHIC data, many of which are for collisions at $\sqrt{s_{NN}}$=200 GeV; if no energy is mentioned in reference to a collision, it should be assumed that the energy is 200 GeV.

\subsection{Collision Geometry}
\label{sec:theory:basic_concepts:collision}

When two particles traveling along the $z$-axis collide, the plane including the $z$-axis and the centers of the two particles is called the ``reaction plane."  The distance in the transverse plane between the centers of the colliding particles is called the ``impact parameter."  For heavy-ion collisions, many quantities depend upon the impact parameter.  Head-on collisions, those with small impact parameter and a large overlap between the nuclei in the transverse plane, are said to be ``central" collisions, while collisions with large impact parameter and a small overlap are said to be ``peripheral."  In central collisions, the number of nucleons that participate in collisions (called $N_{part}$, the ``number of participants") is large, while in peripheral collisions, that number is small.  Therefore, for a given pair of colliding nuclei, the amount of QGP produced is directly related to the centrality of the collision; many of the signatures of a QGP should be stronger in central collisions than peripheral collisions.  Centrality is typically quoted as a percent of the geometric cross-section, \textit{e.g.} the ``0-20\% most central collisions."  Related to $N_{part}$ is the number of binary nucleon-nucleon collisions, which is called $N_{bin}$ in this dissertation, but is also called $N_{coll}$ in some references.

$N_{part}$ and $N_{bin}$ are estimated from simulation using a Glauber model~\cite{Glauber1959} of nucleus-nucleus interactions.  Nuclei are described~\cite{AnnualReviewGlauber} as a collection of nucleons distributed in space according to a density profile, typically (for spherical nuclei)

\begin{equation}
\rho(r)=\rho_{0}\left[1+\exp\frac{r-R}{a}\right]^{-1},
\end{equation}

\noindent where $R$ is the nuclear radius and $a$ is the ``skin depth."  It is assumed that in high-energy collisions, the nucleons in the colliding nuclei carry enough momentum that they are undeflected when the nuclei pass through each other.  The nucleons therefore travel in independent linear trajectories along the $z$-axis.  The nucleon-nucleon scattering cross-section is assumed to be the same as the inelastic $p+p$ scattering cross-section, $\sigma_{inel.}=42\pm 2$ mb at $\sqrt{s}$ = 200 GeV.~\cite{PDG_review,PhysRevLett.70.525}  Given the nuclear density profile and the nucleon-nucleon cross-section, the number of collisions experienced by a single nucleon can be calculated.  The STAR collaboration performs Glauber-model calculations using Monte Carlo simulations.  Nucleus-nucleus impact parameters are randomly generated, and nucleons are randomly distributed in space according to the nuclear density profile.  A nucleon-nucleon collision occurs if two nucleons come within a distance of $\sqrt{\sigma_{inel.}/\pi}$ of each other.  The mean values of $N_{part}$ and $N_{bin}$ ($\langle N_{part}\rangle$ and $\langle N_{bin}\rangle$) for a given range of impact parameters (centrality class) can be determined after simulations of many collisions.  The ``mean nuclear thickness function" is $\langle T_{AA}\rangle=\langle N_{bin}\rangle/\sigma_{inel.}$.  It is also possible to calculate the area and eccentricity of the nuclear overlap region.

In experiments, the impact parameters, and therefore centralities, of heavy-ion collisions cannot be controlled.  The multiplicity of charged particles produced in a collision increases with decreasing nuclear impact parameter (\textit{i.e.}, as the collisions become more central).  Collisions may be grouped into centrality classes based on the charged-particle multiplicity.  The STAR collaboration defines centrality classes based on the ``reference multiplicity," the number of charged tracks (not corrected for tracking efficiency) measured in the pseudorapidity range $|\eta|<0.5$.  Reference multiplicity values are $\approx 700$ for central Au + Au collisions and $\approx 400$ for central Cu + Cu collisions at 200 GeV.

\section{Experimental Signatures of a QGP}
\label{sec:theory:signatures}
This section will describe a few of the experimental results that indicate that a quark-gluon plasma is formed in heavy-ion collisions at RHIC.

\subsection{Jet Quenching}
\label{sec:theory:signatures:jet_quenching}

\begin{figure}[htbp]
\begin{center}
\includegraphics[width=1\linewidth]{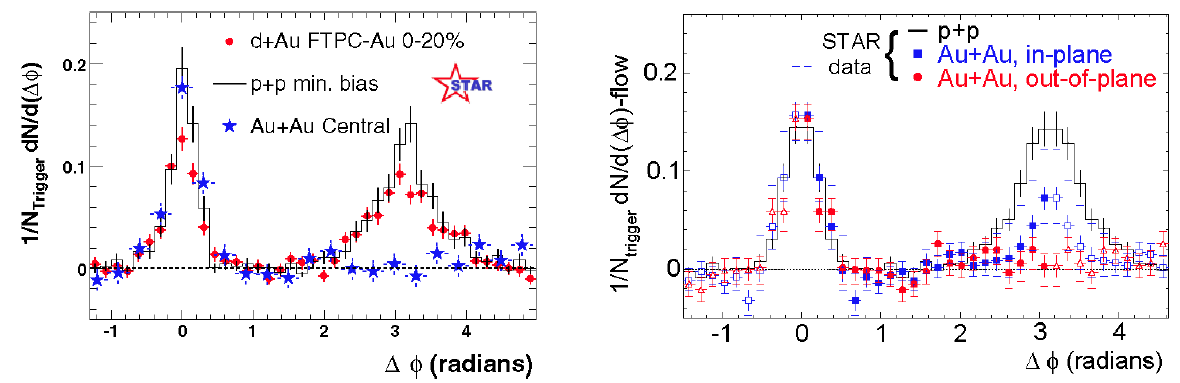}
\caption[Jet quenching in heavy-ion collisions, measured using dihadron correlations.]{High-$p_{T}$ dihadron azimuthal correlations: the difference in azimuth between a charged, high-$p_{T}$ trigger particle and other charged particles in the same event.  In the left-hand plot are the azimuthal correlation data measured in 200-GeV $p+p$ (black), central $d+$Au (red), and central Au + Au (blue) collisions.\protect\cite{MillerThesis,PhysRevLett.90.082302,PhysRevLett.91.072304}  The right-hand plot shows azimuthal correlation data for 200-GeV $p+p$ (black) collisions and Au + Au collisions (colored).\protect\cite{PhysRevLett.93.252301}  The Au + Au data are divided into two classes: in the ``in-plane" class (blue) the azimuth of the trigger particle is within $45^{\circ}$ of the reaction plane; in the ``out-of-plane" class (red) the azimuth of the trigger particle is $>45^{\circ}$ from the reaction plane.  The open symbols are reflections of the measured data (closed symbols) about $\Delta\phi=0$ and $\Delta\phi=\pi$.}
\label{fig:intro:jet_quenching}
\end{center}
\end{figure}

Figure~\ref{fig:intro:jet_quenching} shows measurements~\cite{PhysRevLett.90.082302,PhysRevLett.91.072304} by the STAR collaboration which, in the first few years of RHIC operation, were among the strongest indications that a QGP is formed in heavy-ion collisions.  The distributions are $\Delta\phi$, the difference in azimuth between a high-$p_{T}$ ``trigger" particle (with $p_{T}>4\GeV/c$) and ``associated" particles from the same event which have $p_{T}>2\GeV/c$.  When the trigger and associated particles are in the same jet, a peak near $\Delta\phi=0$ is produced, called the ``near-side" peak.  When the trigger and associated particles come from back-to-back jets, a peak near $\Delta\phi=\pi$ is produced, called the ``away-side" peak.  Figure~\ref{fig:intro:jet_quenching} shows that for 200-GeV $p+p$ and $d+$Au collisions (in which no QGP would be produced), both the near- and away-side peaks are present.  However, as shown in the left-hand panel, for central Au + Au collisions the away-side peak is not present.  This is interpreted as the absorption (or ``quenching") of one member of a back-to-back jet pair in a QGP.  Consider a $q\bar{q}$ or $gg$ pair that becomes back-to-back jets and is produced near the surface of the medium.  One parton is directed out of the medium into the vacuum and fragments into the near-side jet, largely unmodified by the presence of the medium.  The other parton passes through a large amount of QGP, which absorbs its energy, leading to a reduction in associated particles with $p_{T}>2\GeV/c$ and the observed quenching of the away-side jet.  Since the away-side peak is present in $d+$Au collisions, the quenching cannot be due merely to the presence of (cold) nuclear matter in the collision region.

Jet quenching is further illustrated in the right-hand plot of Figure~\ref{fig:intro:jet_quenching}, which shows dihadron azimuthal correlations for two subsets of Au + Au events in the 20-60\% centrality class.~\cite{PhysRevLett.93.252301}  In the ``in-plane" class, the azimuth of the trigger particle is within $45^{\circ}$ of the reaction plane, while in the ``out-of-plane" class the azimuth of the trigger particle is $>45^{\circ}$ from the reaction plane.  For a non-central heavy-ion collision, the overlap region between the nuclei is azimuthally anisotropic, as is the QGP produced (see Figure~\ref{fig:intro:overlap}).  Projected into the transverse plane, the overlap region is roughly elliptical or ``almond" shaped, with the long axis oriented perpendicular to the reaction plane.  A parton produced perpendicular to the reaction plane will therefore have a greater path length through the medium, leading to greater energy loss.  The jet quenching effect should be larger when the trigger particle is roughly perpendicular to the reaction plane than when the trigger particle is parallel to the reaction plane; as observed in the right-hand plot in Figure~\ref{fig:intro:jet_quenching}.

\subsection{Elliptic Flow}

\begin{figure}[htbp]
\begin{center}
\includegraphics[height=0.6\linewidth]{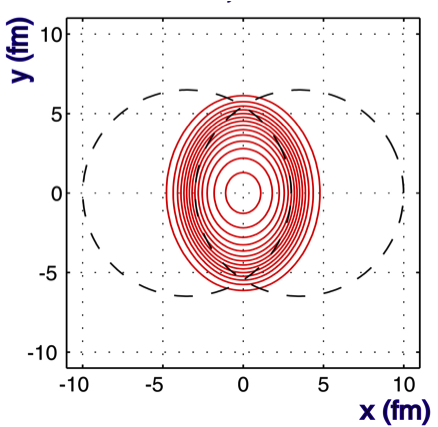}
\caption[Hydrodynamical model describing the overlap region of two colliding gold nuclei.]{A hydrodynamical model\protect\cite{KolbHeinzHydro2003} describing the overlap region of two Au nuclei (dashed circles) moving along the $z$-axis (into and out of the page) and colliding with an impact parameter of 7 fm.  The overlap region is ``almond" shaped, with the long axis oriented perpendicular to the reaction $(xz)$ plane.  The solid red curves show hydrodynamical calculations of curves of constant energy density within the QGP.}
\label{fig:intro:overlap}
\end{center}
\end{figure}

\begin{figure}[htbp]
\begin{center}
\includegraphics[height=0.6\linewidth]{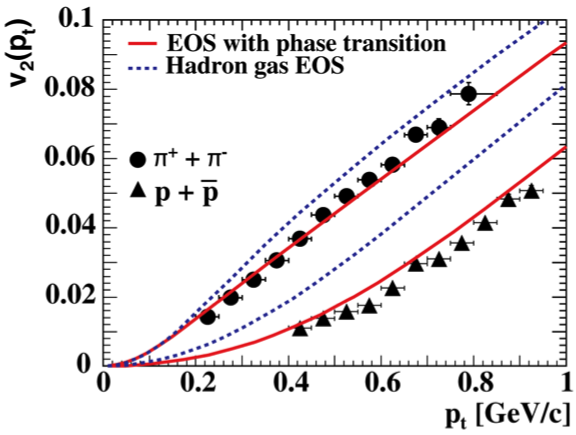}
\caption[Measurements of the elliptic-flow parameter $v_{2}$ for charged pions and (anti)protons.]{Measurements of $v_{2}$ by the STAR collaboration\protect\cite{PhysRevLett.87.182301} for charged pions and (anti)protons in 200-GeV Au + Au collisions.  The curves are from calculations assuming a hadron-gas equation of state (dotted) and calculations assuming a perfect-liquid QGP and a phase transition to a hadron gas.\protect\cite{Huovinen200158}}
\label{fig:intro:v2_hydro}
\end{center}
\end{figure}

\begin{figure}[htbp]
\begin{center}
\includegraphics[width=1\linewidth]{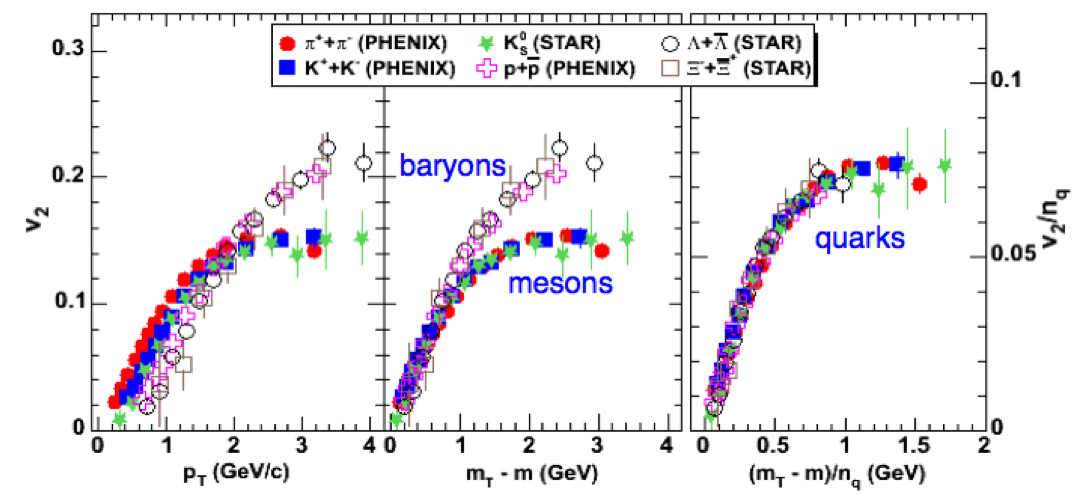}
\caption[Dependence of $v_{2}$ on $p_{T}$, particle species, and quark content.]{Measurements of $v_{2}$ by the STAR\protect\cite{PhysRevLett.92.052302,PhysRevLett.95.122301} and PHENIX\protect\cite{PhysRevLett.98.162301} collaborations for various particle species.  In the left-hand panel, $v_{2}$ is plotted as a function of transverse momentum.  In the middle panel, $v_{2}$ is plotted as a function of $m_{T}-m$ (the ``transverse kinetic energy").  In the right-hand panel, both the abscissa and the ordinate have been scaled by the number of valence quarks in each hadron.}
\label{fig:intro:v2_nq}
\end{center}
\end{figure}

Results from RHIC also indicate that the medium produced in a heavy-ion collision can be described by ultrarelativistic hydrodynamical models with low or 0 viscosity (\textit{i.e.}, a ``perfect fluid").  In a non-central nucleus-nucleus collision, azimuthal anisotropy in the shape of the overlap region results in an azimuthal anisotropy in the momentum distribution of the emitted particles.  The solid curves in Figure~\ref{fig:intro:overlap} are hydrodynamical calculations~\cite{KolbHeinzHydro2003} of curves of constant energy density within a QGP that can be described as a perfect fluid.  The pressure gradients within the QGP will impart additional transverse momentum to the particles emitted from it (a phenomenon called ``radial flow").  Since the pressure gradient is greater in the reaction plane than perpendicular to it, particles with azimuthal angle near the reaction plane receive a greater momentum boost due to flow.  This results in a momentum-space azimuthal anisotropy in the spectrum of emitted particles, referred to as ``elliptic flow."  The azimuthal distribution of particles within a given transverse-momentum bin may be written in the form of a Fourier expansion~\cite{VoloshinZhang1996}:

\begin{equation}
\label{eq:intro:v2}
E\frac{d^{3}N}{dp^{3}}=\frac{1}{2\pi}\frac{d^{2}N}{dp_{T}dy}\left\lbrace 1+\sum\limits_{n=1}\limits^{\infty} 2v_{n}\cos\left[n(\phi-\Psi_{RP})\right]\right\rbrace.
\end{equation}

\noindent Here, the $v_{n}$ are the Fourier expansion coefficients\footnote{For a system that is symmetric with respect to the reaction plane the sine terms in a Fourier expansion will be 0, hence the omission of such terms from Equation~\ref{eq:intro:v2}.} and $\Psi_{RP}$ is the azimuthal angle of the reaction plane in the lab frame.  The coefficient for the second harmonic $v_{2}$ is used as a measurement of elliptic flow.  In heavy-ion collisions at RHIC, $v_{2}$ tends to increase as collisions become more peripheral~\cite{PhysRevLett.87.182301,PhysRevLett.98.162301}.  This is consistent with the assumption that the momentum-space azimuthal anisotropy is due to the differing pressure gradients in and out of the reaction plane in a QGP: the eccentricity of the overlap region increases as collisions become more peripheral, resulting in greater difference in the pressure gradients.  Figure~\ref{fig:intro:v2_hydro} shows $v_{2}$ as a function of $p_{T}$ for identified $\pi^{\pm}$ and (anti)protons~\cite{PhysRevLett.92.052302,PhysRevLett.95.122301,PhysRevLett.87.182301} along with hydrodynamical calculations~\cite{Huovinen200158} of $v_{2}$.  For the solid curves, it is assumed that there is a phase transition between a QGP with zero viscosity and a hadron gas, while the dashed curves assume that the system is only described by a hadron-gas equation of state.  The data are better described by the hydrodynamical calculations that assume the presence of a QGP.  Figure~\ref{fig:intro:v2_nq} shows $v_{2}$ for a variety of particle species~\cite{PhysRevLett.98.162301}; in the left-hand panel at low $p_{T}$ particles with lower mass tend to have a higher $v_{2}$.  In the middle panel, $v_{2}$ is plotted as a function of $m_{T}-m$ (the kinetic energy, neglecting longitudinal momentum) rather than $p_{T}$: all baryons appear to follow one trend, while all mesons follow another, and at low $p_{T}$ all of the particle species shown appear to follow the same trend.  This baryon-meson separation can be eliminated by scaling $v_{2}$ and $m_{T}-m$ by a factor of $1/n_{q}$, where $n_{q}$ is the number of valence quarks in the hadron (right-hand panel).  This is taken to be an indication that if a fluid is created, it is the individual quarks that are flowing.

\clearpage

\subsection[High-$p_{T}$ Particle Suppression]{High-$\boldsymbol{p_{T}}$ Particle Suppression}
\label{sec:theory:high_pt_supp}

\begin{figure}[htbp]
\begin{center}
\includegraphics[width=0.85\linewidth]{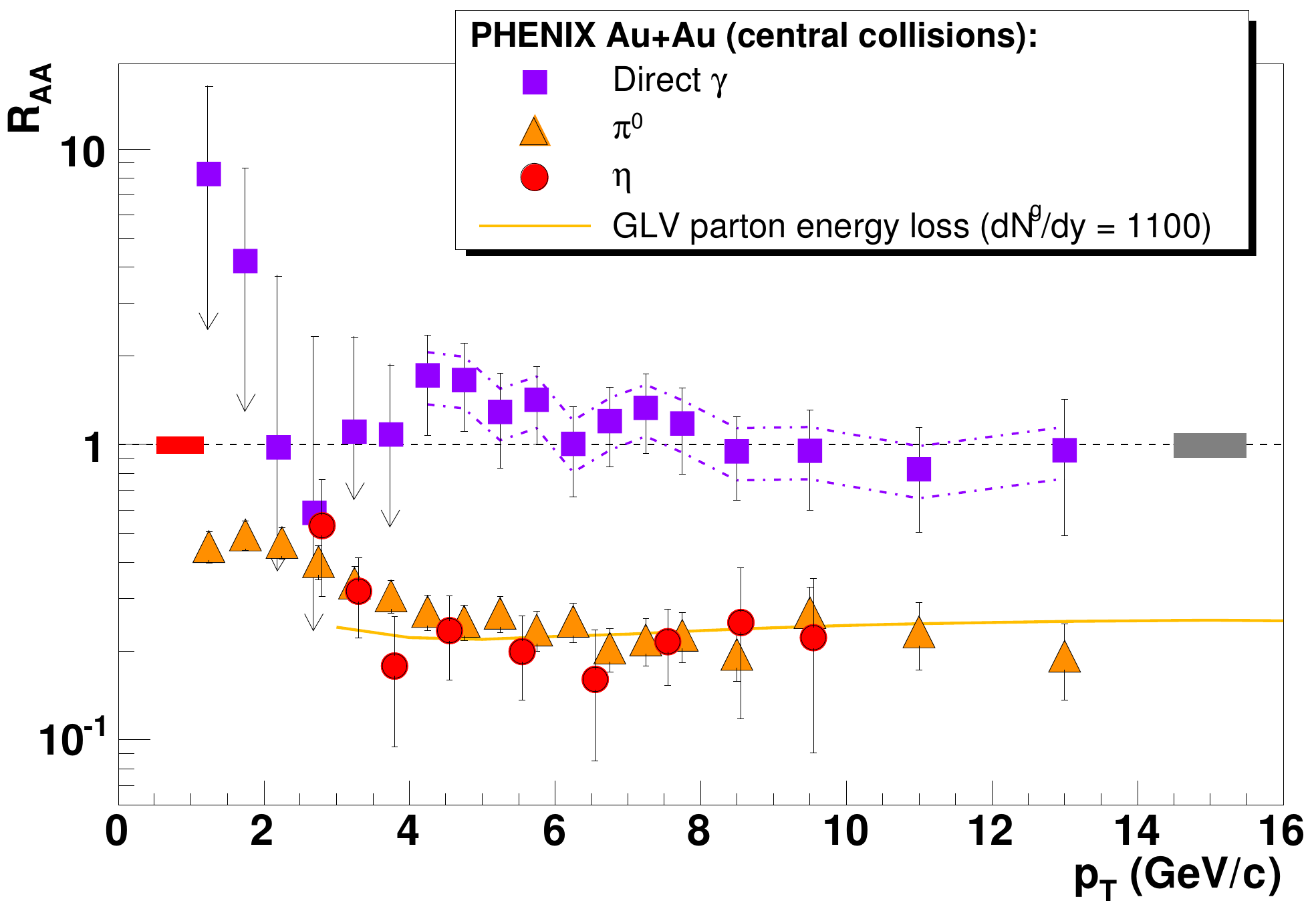}
\caption[Nuclear modification factor for neutral pions, $\eta$ mesons, and direct photons in central 200-GeV Au + Au collisions.]{The nuclear modification factor $(R_{AA})$ for neutral pions,\protect\cite{PhysRevLett.91.072301,PhysRevC.76.034904} $\eta$ mesons,\protect\cite{PhysRevC.75.024909} and direct photons\protect\cite{PhysRevLett.94.232301} in central 200-GeV Au + Au collisions.  Also shown is the suppression of light-flavor hadrons predicted by the GLV model\protect\cite{PhysRevLett.89.252301,VitevJetTomographyQM204} with gluon density $dN^{g}/dy=1100$.}
\label{fig:intro:high_pt_supp}
\end{center}
\end{figure}

As a parton passes through a quark-gluon plasma, it is expected to lose energy to the medium through gluon bremsstrahlung and collisions with other partons in the medium.~\cite{PhysRevC.72.014905,PhysRevC.74.064907}  This is expected to result in a suppression of high-$p_{T}$ particles in heavy-ion collisions (relative to $p+p$ collisions, in which no medium is present).  The ``nuclear modification factor" is used to quantify the effect of the presence of nuclear matter on particle yields.  This factor is often denoted as $R_{AA}$, although the subscripts can be used to indicate the specific collision system being considered (\textit{e.g.}, $R_{d\mathrm{Au}}$ for $d+\mathrm{Au}$ collisions, $R_{\mathrm{AuAu}}$ for Au + Au collisions, and $R_{\mathrm{CuCu}}$ for Cu + Cu collisions).  The nuclear modification factor is the ratio of the yield of a particle species in nucleus-nucleus collisions to the yield of the same particle species in proton-proton collisions, with the $p+p$ data scaled to account for the fact that there are many nucleon-nucleon collisions in an $A+A$ collision.  For a given transverse-momentum bin, the nuclear modification factor for particle species $X$ is defined as

\begin{equation}
\label{eq:theory:define_raa}
R_{AA}^{X}=\frac{\left(\dfrac{Ed^{3}N^{X}_{AA}}{dp^{3}}\right)\sigma_{inel.}}{\left(\dfrac{Ed^{3}\sigma^{X}_{pp}}{dp^{3}}\right)\langle N_{bin}\rangle},
\end{equation}

\noindent where $Ed^{3}N^{X}_{AA}/dp^{3}$ is the invariant yield of particle species $X$ in nucleus-nucleus collisions, $\langle N_{bin}\rangle$ is the mean number of binary nucleon-nucleon collisions for those nucleus-nucleus collisions, $Ed^{3}\sigma^{X}_{pp}/dp^{3}$ is the invariant cross-section for particle species $X$ in $p+p$ collisions, and collisions being studied, and $\sigma_{inel.}=$42 mb is the inelastic $p+p$ scattering cross-section~\cite{PDG_review}.  The value of $\sqrt{s_{NN}}$ is assumed to be the same for the $A+A$ and $p+p$ collisions being compared.  If the presence of nuclear matter in the collision region has no effect on the spectrum of particle species $X$, an $A+A$ collision can be viewed as a superposition of $\langle N_{bin}\rangle$ nucleon-nucleon collisions and the nuclear modification factor will be 1.  Values of $R_{AA}^{X}$ greater than 1 indicate that particle species $X$ is enhanced in $A+A$ collisions relative to $p+p$ collisions, while values of $R_{AA}^{X}<1$ indicate a suppression of particle species $X$.

Figure~\ref{fig:intro:high_pt_supp} shows measurements by the PHENIX collaboration of the nuclear modification factor for neutral pions, $\eta$ mesons, and direct photons (photons produced in the collision itself and not through the subsequent decays of hadrons) in central Au + Au collisions.  The direct photons~\cite{PhysRevLett.94.232301} are not suppressed, which is expected since photons do not interact via the strong interaction and should not be affected by the presence of a quark-gluon plasma.  The hadrons~\cite{PhysRevLett.91.072301,PhysRevC.76.034904,PhysRevC.75.024909} are observed to be suppressed by about a factor of five at high $p_{T}$.  The curve is a theoretical calculation of light-flavor-hadron suppression in one model, the GLV model.~\cite{PhysRevLett.89.252301,VitevJetTomographyQM204}  The quantity $dN_{g}/dy$ is the gluon density per unit rapidity, a parameter in the GLV model that is related to the opacity of the medium.  The observed values of $R_{AA}$ for light-flavor hadrons were used to constrain the model; the GLV model with $dN^{g}/dy\approx 1100$ best describes the data.


\subsection{Thermal Models}

\begin{figure}[htbp]
\begin{center}
\includegraphics[width=0.85\linewidth]{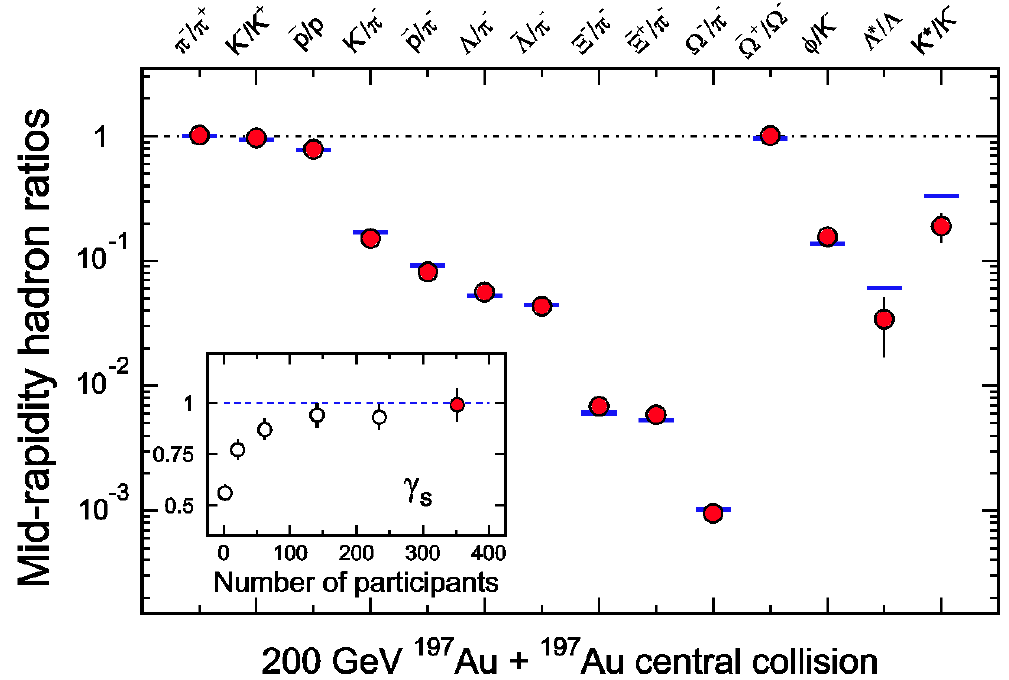}
\caption[Measured ratios of the yields of various particle species compared to thermal-model predictions.]{The ratios of the yields of various particle species measured\protect\cite{STAR_white_paper2005} by the STAR collaboration in central 200-GeV Au + Au collisions (red points).  Also shown is a thermal-model fit to those data, which is used to find the temperature and baryo-chemical potential of the system at chemical freezeout.  The inset shows how the strangeness-suppression factor $\gamma_{s}$ changes with collision centrality.}
\label{fig:intro:thermal_fit}
\end{center}
\end{figure}

Measurements of the ratios of particle yields in heavy-ion collisions indicate that the system has reached thermal equilibrium at the time it reaches chemical freezeout (which occurs after the system has passed through the phase transition from a QGP to a hadron gas).  In statistical thermal models, the system is described by a grand canonical ensemble, with the Hamiltonian usually taken to be that of a hadron resonance gas.  From a simple expression for the grand canonical partition function, the mean number of particles of species $X$ (which has mass $m_{X}$, baryon number $B_{X}$, strangeness $S_{X}$, and electric charge $Q_{X}$) is calculated~\cite{BraunMunzingerThermal2003} to be

\begin{equation}
\label{eq:theory:thermal}
\langle N_{X}\rangle=\frac{VTm_{X}^{2}g_{X}}{2\pi^{2}}\sum\limits_{j=1}\limits^{\infty}\frac{(\pm 1)^{j+1}}{j}K_{2}\left(j\frac{m_{X}}{T}\right)\exp\left[j\frac{B_{X}\mu_{B}+S_{X}\mu_{S}+Q_{X}\mu_{Q}}{T}\right].
\end{equation}

\noindent $V$ and $T$ are the volume and temperature of the system.  The quantities $\mu_{B}$, $\mu_{S}$, and $\mu_{Q}$ are the chemical potentials associated with baryon number, strangeness, and electric charge.  $K_{2}$ is the modified Bessel function.  The factor $g_{X}$ is the spin-isospin degeneracy factor~\cite{Cleymans1986217} for particle species $X$.  The upper (lower) sign is to be used for fermions (bosons) and Equation~\ref{eq:theory:thermal} is written in units where $c=\hbar=k_{B}=1$.  The chemical potentials are not independent: the initial charges of the incoming nuclei and the requirement of zero net strangeness restrict the possible configurations of the system, leaving $V$, $T$, and $\mu_{B}$ as the only independent parameters.  The volume of the system is unknown, but is removed from consideration if the ratios of particle yields are considered.  The measurement of two particle ratios is sufficient to determine the two unknown parameters; in practice, many particle ratios are measured and the set of those measurements is fit to determine the most likely values of $T$ and the baryo-chemical potential.  The ``strangeness-suppression factor" $\gamma_{s}$ is also introduced to account for the fact that strange quarks may not be in full thermal equilibrium.  Figure~\ref{fig:intro:thermal_fit} shows the results of one such fit for central 200-GeV Au + Au collisions measured by the STAR collaboration.~\cite{STAR_white_paper2005}  The thermal fit with $T=163\pm 4\MeV$, $\mu_{B}=24\pm 4\MeV$ and $\gamma_{s}=0.99\pm0.07$ describes the ratios well.  The inset shows how the value of $\gamma_{s}$ changes with collision centrality, with strangeness approaching equilibrium values as collisions become more central.  The chemical freezeout temperature of 163 MeV places a lower limit on the critical temperature $T_{c}$ at $\mu_{B}=24$ MeV.

\clearpage

\section{Heavy Flavor}

Heavy quarks, defined for the purposes of this dissertation to be the $c$ and $b$ quarks, are useful probes of the medium produced in heavy-ion collisions at RHIC.  Because their masses ($m_{c}=(1.27^{+0.07}_{-0.09})\GeV/c^{2}$ and $m_{b}=(4.19^{+0.18}_{-0.06})\GeV/c^{2}$)~\cite{PDG_review} are much larger than the RHIC temperature scale, heavy quarks are not expected to be produced thermally in significant amounts in the QGP.  At RHIC, heavy quarks are expected to be produced primarily in the hard (large momentum transfer) scattering of partons in the initial stages of the collision.  Heavy quarks are therefore expected to be present in the QGP and the subsequent hadron gas throughout their evolution.  Furthermore, because of their large masses, heavy quarks have been predicted to interact differently with the medium than light quarks.  A satisfactory model of the QGP should be able to describe gluon, light-quark, and heavy-quark interactions with the medium using the same set of parameters (see Section~\ref{sec:theory:heavy_interactions}).  This section describes some of the theoretical issues regarding heavy quarks and presents some of the major experimental heavy-flavor measurements from RHIC.  Because of their short lifetimes, hadrons containing heavy quarks do not survive long enough to leave the RHIC beam line (much less reach the tracking components of the RHIC detectors).  Heavy-flavor hadrons must therefore be studied indirectly through their decay products.  The yields of $D$ mesons have been measured by reconstructing their hadronic decays (see~\cite{PhysRevLett.94.062301,Zhang2006701,BaumgartThesis}).  The yields of the $J/\psi$ and $\Upsilon$ have been measured by reconstructing their decays to $e^{-}e^{+}$ pairs (see~\cite{PhysRevLett.98.232301,PhysRevLett.96.012304,AverbeckQM2006,PhysRevD.82.012001,PhysRevC.80.041902,PhysRevD.82.012004}).  Some measurements of heavy-flavor hadron yields have also been performed by measuring their decays to muons (see~\cite{PhysRevD.76.092002,ZhongQM2006}).

Several studies (of which this dissertation is one) have measured the suppression (relative to $p+p$ collisions) of heavy quarks at high $p_{T}$ by measuring the nuclear modification factor of non-photonic $e^{\pm}$ in various collision systems.  A non-photonic $e^{\pm}$ (sometimes abbreviated as ``\textbf{NPE}" and sometimes referred to as a ``single $e^{\pm}$") \textbf{is defined to be an} $\boldsymbol{e^{\pm}}$ \textbf{produced with an (anti)neutrino in a charged-current weak decay}.  In contrast, ``photonic $e^{\pm}$" are produced in pairs from sources including photon conversions, Dalitz decays (\textit{e.g.}, $\pi^{0}\rightarrow e^{-}e^{+}\gamma$), $J/\psi\rightarrow e^{-}e^{+}$, the Drell-Yan process, and others.  While non-photonic $e^{\pm}$ may also come from sources other than heavy-flavor decays (charged pions, muons, kaons, etc.) the spectrum of non-photonic $e^{\pm}$ is dominated by the products of heavy-flavor decays (see section~\ref{sec:res_back:lfn}).  In this dissertation, the yield and nuclear modification factor of non-photonic $e^{\pm}$ in Cu + Cu collisions at $\sqrt{s_{NN}}=200$ GeV will be measured.

\subsection{Production}

\begin{figure}[htbp]
\begin{center}
\includegraphics[width=0.85\linewidth]{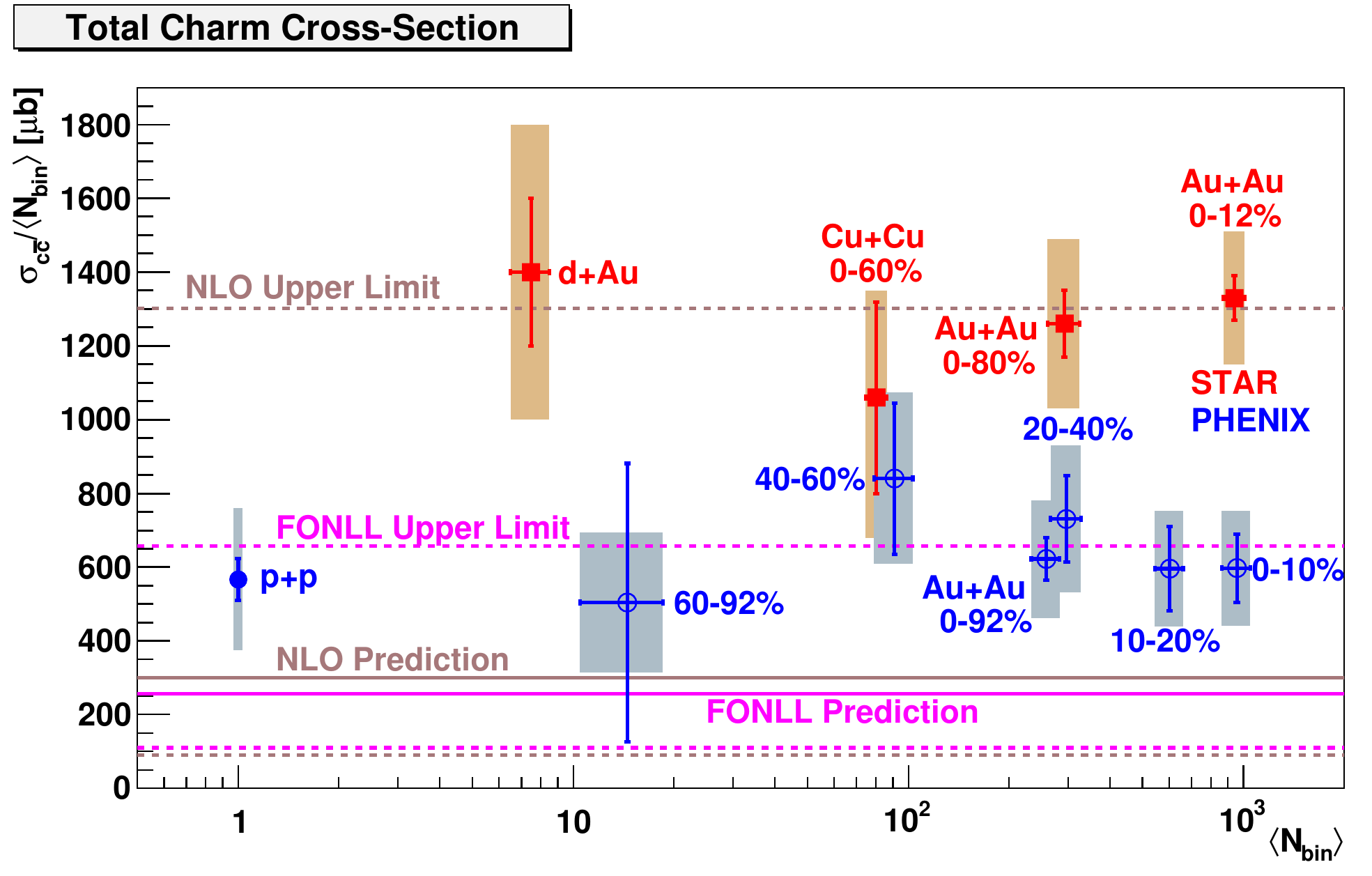}
\caption[Measurements of the total charm cross-section for various collision systems at RHIC.]{The measured total charm cross-section $(\sigma_{c\bar{c}})$ scaled by $\langle N_{bin}\rangle$ as a function of $\langle N_{bin}\rangle$ for 200-GeV $p+p$\protect\cite{PhysRevLett.97.252002}, $d+\mathrm{Au}$\protect\cite{PhysRevLett.94.062301}, Cu + Cu\protect\cite{BaumgartThesis}, and Au + Au\protect\cite{ZhongQM2006,PhysRevLett.98.172301} collisions in various centrality classes.  Also shown are NLO\protect\cite{VogtHeavyCrossSections2009} and FONLL\protect\cite{PhysRevLett.95.122001} calculations of $\sigma_{c\bar{c}}$.}
\label{fig:intro:charm_xsection}
\end{center}
\end{figure}

Since heavy quarks are primarily produced through parton hard scattering, the production cross-section may be calculated using perturbative QCD.  These perturbative calculations are typically done at the Next-to-Leading-Order (NLO) or Fixed-Order-Next-to-Leading-Log (FONLL) levels.  NLO calculations include contributions of order $\alpha_{s}^{2}$ and $\alpha_{s}^{3}$, while FONLL calculations include contributions of those orders along with logarithmic contributions of the form $\alpha_{s}^{n}\log^{n-1}(p_{T}/M_{Q})$ and $\alpha_{s}^{n}\log^{n}(p_{T}/M_{Q})$ resummed to all orders $n$ (where $M_{Q}$ is the heavy-quark mass).~\cite{PhysRevLett.95.122001,VogtHeavyCrossSections2009}  Since the heavy-quark $p_{T}$ appears in the logarithmic terms, FONLL calculations must be performed in transverse-momentum bins and then integrated to obtain the total cross-section.  The heavy quark is treated as an ``active" flavor for $p_{T}\gg M_{Q}$ and included in the calculation of the coupling constant (see Equation~\ref{eq:theory:running_as}).  Calculations using this framework give $\sigma_{c\bar{c}}^{FONLL}=0.256^{+0.400}_{-0.146}$ mb and $\sigma_{b\bar{b}}^{FONLL}=1.87^{+0.99}_{-0.67}\;\mu\mathrm{b}$ for the total charm and bottom cross-sections in nucleon-nucleon collisions,~\cite{PhysRevLett.95.122001} with the uncertainties in the cross-sections due to uncertainties in the heavy quark masses and the choice of the renormalization scale.  NLO calculations may be performed as described above for FONLL, but they may also be performed in a different framework in which the total cross-section is evaluated for all transverse momenta (\textit{i.e.}, division into $p_{T}$ bins is not necessary) and the heavy quark is treated as massive (and not counted in the evaluation of $\alpha_{s}$).  One such calculation gives $\sigma_{c\bar{c}}^{NLO}=0.301^{+1.000}_{-0.210}$ mb and $\sigma_{b\bar{b}}^{NLO}=2.06^{+1.85}_{-0.81}\;\mu\mathrm{b}$~\cite{VogtHeavyCrossSections2009} for the total charm and bottom cross-sections in nucleon-nucleon collisions.

The STAR and PHENIX collaborations have measured the total charm cross-section in 200-GeV $p+p$ and $A+A$ collisions.  The STAR collaboration measures the yield of $\tfrac{1}{2}(D^{0}+\bar{D}^{0})$ by reconstructing the $D^{0}\rightarrow\pi^{+}K^{-}(\bar{D}^{0}\rightarrow\pi^{-}K^{+})$ decays and finding the invariant masses of pion-kaon pairs.  A simultaneous fit of the $D^{0}$ measurements (mostly at $p_{T}\lesssim2\GeV/c$) and non-photonic $e^{\pm}$ measurements (at higher $p_{T}$) is performed to obtain an estimate of the total charm cross-section.  The PHENIX collaboration estimates $\sigma_{c\bar{c}}$ from non-photonic $e^{\pm}$ (which PHENIX can measure for $p_{T}>200\MeV/c$).  Figure~\ref{fig:intro:charm_xsection} shows the experimental measurements of $\sigma_{c\bar{c}}/\langle N_{bin}\rangle$ for $p+p$~\cite{PhysRevLett.97.252002}, $d+\mathrm{Au}$~\cite{PhysRevLett.94.062301}, Cu + Cu~\cite{BaumgartThesis}, and Au + Au~\cite{ZhongQM2006,PhysRevLett.98.172301} collisions at $\sqrt{s_{NN}}$=200 GeV, along with the two calculated values of $\sigma_{c\bar{c}}$ described above.  For each experiment, the measurements indicate that the total charm cross-section scales with the number of binary collisions, consistent with the assumption that charm quarks are produced predominantly in the initial hard scattering of nucleons.  This indicates that it is reasonable to use perturbative QCD (FONLL) calculations of the $c$- and $b$-quark spectra as initial spectra for theoretical studies of heavy-quark interactions with the QGP (as is done in the studies described in Section~\ref{sec:theory:heavy_interactions}).  However, there appears to be a systematic factor of $\approx 2$ difference between the values of $\sigma_{c\bar{c}}$ measured by the STAR and PHENIX collaborations; the STAR measurements are also inconsistent with the FONLL prediction by a factor of $\approx 4$.

\subsection{Interactions With the Medium}
\label{sec:theory:heavy_interactions}

\begin{figure}[htbp]
\begin{center}
\includegraphics[width=0.85\linewidth]{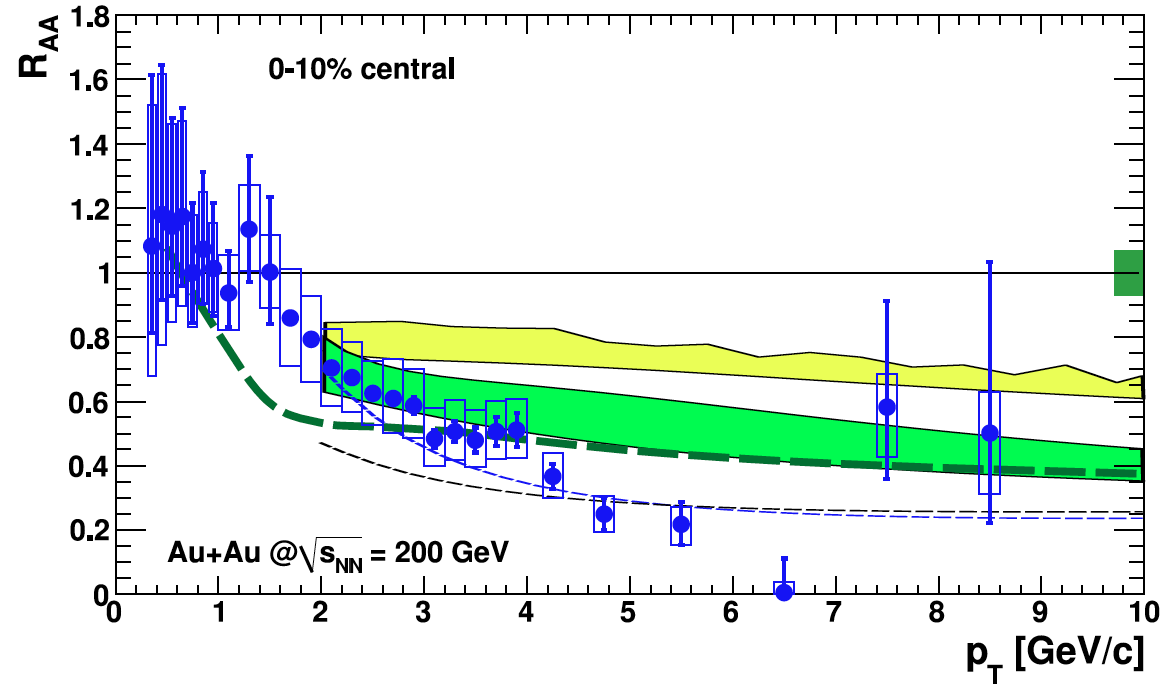}
\caption[Measured nuclear modification factor for non-photonic $e^{\pm}$ compared to predictions of the DGLV model.]{The nuclear modification factor for non-photonic $e^{\pm}$ $R_{AA}^{NPE}$ in the 0-10\% most central Au + Au collisions measured by the PHENIX collaboration\protect\cite{PHENIX_NPE2010} and predicted by the DGLV\protect\cite{PhysRevLett.94.112301,Djordjevic200681,Djordjevic2004265,Wicks2007426} and BDMPS\protect\cite{Armesto2006362,Baier1997291} models.  The green dashed line (yellow band) shows the predictions of the BDMPS (DGLV) model, including only radiative energy loss.  The green band shows the prediction of the DGLV model including both radiative and collisional energy loss.  The thin dashed curves show the prediction of the DGLV model for $e^{\pm}$ from $D$-meson decays only.}
\label{fig:intro:raa_npe_DGLV_BDMPS}
\end{center}
\end{figure}

\begin{figure}[htbp]
\begin{center}
\includegraphics[width=0.85\linewidth]{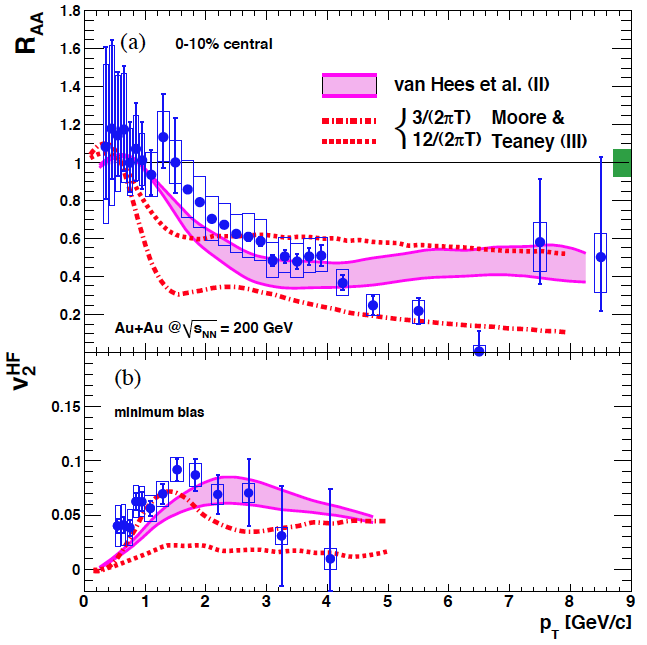}
\caption[Measured nuclear modification factor for non-photonic $e^{\pm}$ compared to predictions of the Moore/Teany and van Hees models.]{In the upper panel is the nuclear modification factor for non-photonic $e^{\pm}$ $R_{AA}^{NPE}$ in the 0-10\% most central Au + Au collisions measured by the PHENIX collaboration\protect\cite{PHENIX_NPE2010} and predicted by the Moore/Teany\protect\cite{PhysRevC.71.064904} and van Hees\protect\cite{PhysRevC.73.034913,PhysRevC.71.034907} models.  The predictions of the Moore/Teany model are shown for two different choices of the heavy-quark diffusion coefficient.  Shown in the lower panel is the elliptic flow parameter $v_{2}$ for non-photonic $e^{\pm}$ measured by the PHENIX collaboration for minimum-bias (0-92\% centrality) collisions and predicted by the two models.}
\label{fig:intro:raa_npe_vanHees_MT}
\end{center}
\end{figure}

\begin{figure}[htbp]
\begin{center}
\includegraphics[width=0.85\linewidth]{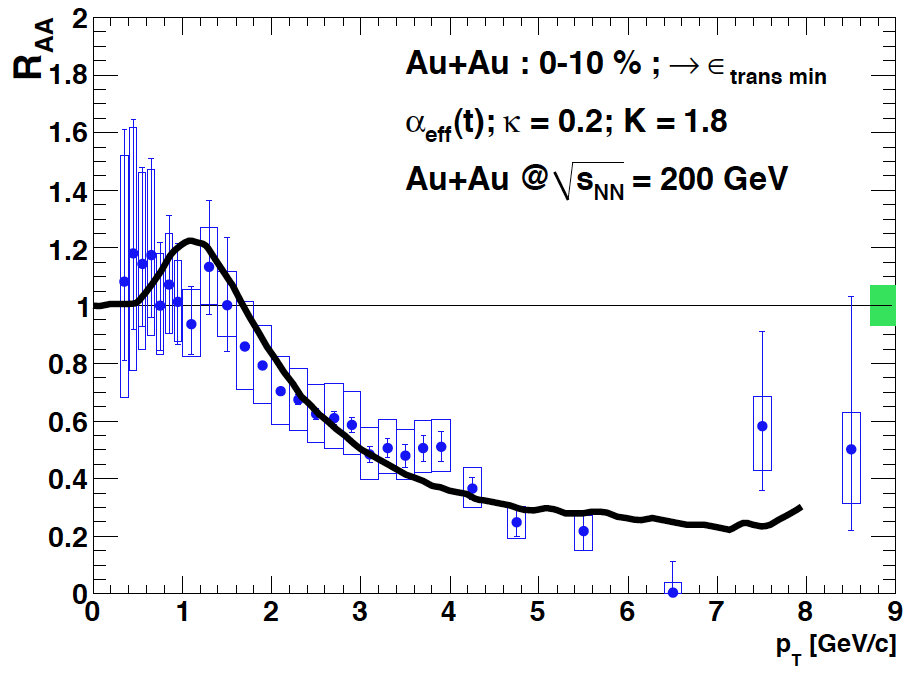}
\caption[Measured nuclear modification factor for non-photonic $e^{\pm}$ compared to predictions of the Gossiaux/Aichelin model.]{The nuclear modification factor for non-photonic $e^{\pm}$ $R_{AA}^{NPE}$ in the 0-10\% most central Au + Au collisions measured by the PHENIX collaboration\protect\cite{PHENIX_NPE2010} and predicted by the Gossiaux/Aichelin\protect\cite{PhysRevC.78.014904,PhysRevC.79.044906} model.}
\label{fig:intro:raa_npe_GA}
\end{center}
\end{figure}

\begin{figure}[htbp]
\begin{center}
\includegraphics[width=0.85\linewidth]{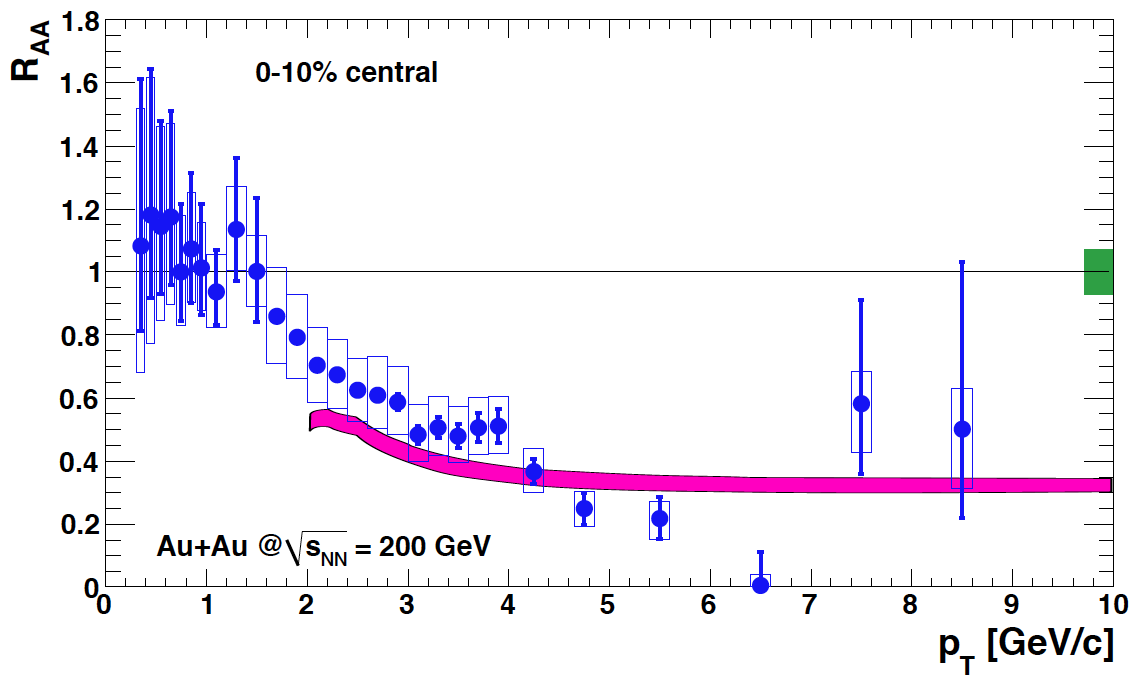}
\caption[Measured nuclear modification factor for non-photonic $e^{\pm}$ compared to predictions of the Collisional Dissociation model.]{The nuclear modification factor for non-photonic $e^{\pm}$ $R_{AA}^{NPE}$ in the 0-10\% most central Au + Au collisions measured by the PHENIX collaboration\protect\cite{PHENIX_NPE2010} and predicted by the Collisional Dissociation\protect\cite{Adil2007139,PhysRevC.80.054902} model.}
\label{fig:intro:raa_npe_CD}
\end{center}
\end{figure}

A parton passing through a strongly interacting medium is expected to lose energy to the medium through gluon bremsstrahlung.  For heavy quarks, the gluon-bremsstrahlung distribution is estimated~\cite{Dokshitzer2001199} to be reduced from the light-quark case by a factor of $[1+(M_{Q}^{2}/E_{Q}^{2})\cdot\theta^{-2}]^{-2}$, where $M_{Q}$ and $E_{Q}$ are the mass and energy of the heavy quark and $\theta$ is the angle of gluon emission (with respect to the momentum of the heavy quark).  The suppression factor is always less than 1, but is closest to zero for small angles $(\theta\lesssim M_{Q}/E_{Q})$; this suppression of small-angle gluon radiation is called the ``dead-cone effect."

Because of the dead cone effect, it was expected that heavy quarks would lose less energy to the medium than light quarks.  This would lead to less suppression of heavy quarks (and therefore of non-photonic $e^{\pm}$) at high transverse momenta.  However, measurements of the non-photonic $e^{\pm}$ nuclear modification factor~\cite{PhysRevLett.96.032301,PhysRevLett.97.252002,PhysRevLett.98.172301,PHENIX_NPE2010} at RHIC indicate that the suppression of non-photonic $e^{\pm}$ at high $p_{T}$ is similar in magnitude to the suppression of light-flavor hadrons~\cite{PhysRevLett.91.072301,PhysRevC.76.034904,PhysRevC.75.024909}.  Several models of parton energy loss in a QGP have been developed; the models describe heavy quark energy loss through different mechanisms, including (but not limited to) medium-induced gluon radiation and elastic collisions with the partons that make up the medium.  Some of these models will be described below and their predictions will be compared to the PHENIX collaboration's measurements of $R_{AA}$ and $v_{2}$ for non-photonic $e^{\pm}$ in central 200-GeV Au + Au collisions.~\cite{PHENIX_NPE2010}

In its original form, the DGLV model~\cite{PhysRevLett.94.112301,Djordjevic200681,Djordjevic2004265} describes parton energy loss through medium-induced gluon radiation, with the parton interacting with multiple scattering centers as it passes through the medium.  Perturbative QCD is used to calculate the matrix elements for gluon emission in these interactions.  The opacity of the medium is parametrized by the gluon rapidity density $dN_{g}/dy$.  The BDMPS model~\cite{Armesto2006362,Baier1997291} describes parton energy loss through medium-induced gluon radiation.  As it passes through the medium, the parton undergoes multiple independent interactions with a series of scattering centers, which have screened Coulomb potentials (the weak-coupling regime of QCD).  The magnitude of the energy loss in the medium is determined by the color charge and mass of the parton, the path length in the medium, and the time-averaged squared momentum transfer from the parton to the medium, which is represented by the ``transfer coefficient" $\hat{q}$.  The DLGV and BDMPS radiative energy-loss models with parameters $dN_{g}/dy\approx 1000$ and $4\mathrm{GeV^{2}/fm}<\hat{q}<14\mathrm{GeV^{2}/fm}$ give values of $R_{AA}$ that reproduce the suppression of light-flavor hadrons well (see Figure~\ref{fig:intro:high_pt_supp} and~\cite{PhysRevLett.91.072301,PhysRevC.76.034904,PhysRevC.75.024909}).  However, as shown in Figure~\ref{fig:intro:raa_npe_DGLV_BDMPS}, those models tend to under-predict the suppression of non-photonic $e^{\pm}$ (\textit{i.e.}, over-predict the values of $R_{AA}$) in heavy-ion collisions at RHIC.  An extension of the DGLV model~\cite{Wicks2007426} to include parton energy loss through elastic scattering in the medium is in better agreement with the data, although the model may still under-predict the suppression at high $p_{T}$.

The model of Moore and Teany~\cite{PhysRevC.71.064904} describes heavy-quark energy loss and flow through elastic scattering in a thermalized, expanding QGP, with the scattering amplitudes calculated perturbatively.  The heavy-quark diffusion coefficient, which is related to the temperature of the medium, determines the magnitude of the energy loss and elliptic flow.  To avoid infrared divergences in some of the perturbative calculations, an infrared regulator proportional to the gluon Debye mass\footnote{The Debye screening length $\lambda_{D}$ is the length scale over which charges (color or electric) are screened in a plasma~\cite{Wong1994}; the Debye mass $m_{D}$ is the inverse of this length.  With $c=\hbar=1$, $m_{D}=\lambda_{D}^{-1}=gT\sqrt{\tfrac{1}{3}n_{c}+\tfrac{1}{6}n_{f}}$ in a QCD plasma.} is introduced.  The model of van Hees \textit{et al.}~\cite{PhysRevC.73.034913,PhysRevC.71.034907} is similar to the Moore/Teany model, but heavy quarks are also allowed to scatter through $D$- and $B$-meson-like resonance states in the QGP.  For the perturbative calculations of the elastic scattering amplitudes, a non-running coupling constant in the range $0.3\leq\alpha_{s}\leq 0.5$ is used and the gluon Debye mass is used as an infrared regulator.  Calculations of the non-photonic $e^{\pm}$ nuclear modification factor and flow parameter $v_{2}$ using these two models are shown in Figure~\ref{fig:intro:raa_npe_vanHees_MT}.  The Moore/Teany model is shown for two different values of the diffusion coefficient; this model does not appear to describe both $R_{AA}$ and $v_{2}$ for non-photonic $e^{\pm}$ simultaneously.  The van Hees model describes $R_{AA}$ and $v_{2}$ reasonably well, although there are some indications that it too may under-predict the suppression of non-photonic $e^{\pm}$ at high $p_{T}$.  The uncertainty in the calculation is due in part to uncertainty in the choice of the strong coupling constant and the masses and widths of the $D$ and $B$ meson resonance states.

Like the Moore/Teany and van Hees models, the model of Gossiaux and Aichelin~\cite{PhysRevC.78.014904,PhysRevC.79.044906} describes heavy-quark energy loss through elastic scattering in a QGP.  However, this model uses a physical running coupling constant and a different infrared regulator\footnote{The infrared regulator in the Gossiaux/Aichelin model is $\approx 0.2m_{D}$, a choice guided by a QED calculation~\cite{PhysRevD.77.014015} of muon energy loss in a plasma.} in the perturbative QCD calculations of the scattering matrix elements.  This model appears to describe the observed suppression of non-photonic $e^{\pm}$ well for all but the most peripheral centrality classes (shown in Figure~\ref{fig:intro:raa_npe_GA} for the 0-10\% centrality class; see~\cite{PHENIX_NPE2010} for the other centrality classes).


In the Collisional Dissociation model~\cite{Adil2007139,PhysRevC.80.054902}, heavy quarks lose energy through fragmentation to $D$ and $B$ mesons and the subsequent dissociation of those mesons through collisions with partons in the medium.  The formation times for those mesons are expected to be less than the time required to travel through the medium and repeated cycles of fragmentation and dissociation lead to an effective energy loss.  A calculation of the non-photonic $e^{\pm}$ nuclear modification factor using this model is shown in Figure~\ref{fig:intro:raa_npe_CD}.  The Collisional Dissociation model appears to describe the high-$p_{T}$ suppression of non-photonic $e^{\pm}$.

\clearpage

\subsection{Relative Abundances of Charm and Bottom}

\begin{figure}[htbp]
\begin{center}
\includegraphics[width=0.85\linewidth]{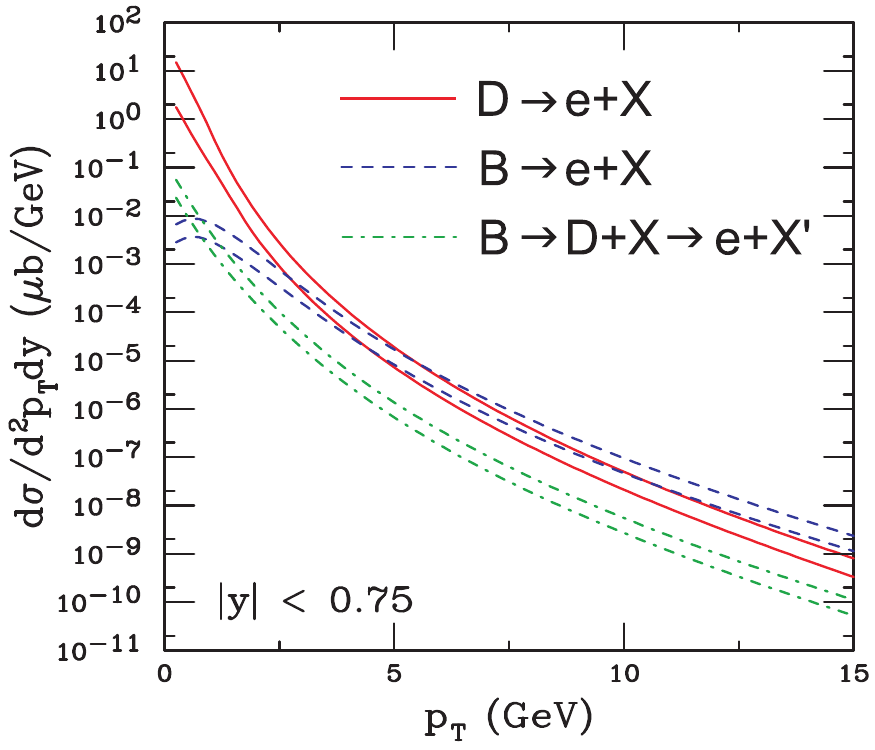}
\caption[Perturbative QCD calculations of the contributions of $D$- and $B$-meson decays to the non-photonic $e^{\pm}$ spectrum in 200-GeV $p+p$ collisions.]{Perturbative QCD (FONLL) calculations\protect\cite{PhysRevLett.95.122001} of the contributions of $D$- and $B$-meson decays to the non-photonic $e^{\pm}$ spectrum in 200-GeV $p+p$ collisions.}
\label{fig:intro:FONLL_c_b}
\end{center}
\end{figure}

\begin{figure}[htbp]
\begin{center}
\includegraphics[width=0.85\linewidth]{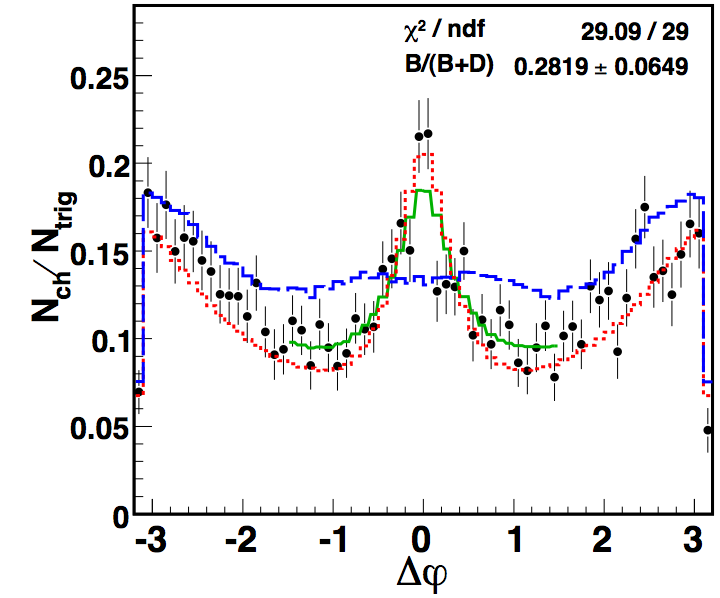}
\caption[Illustration of the fitting method used to find the ratio of non-photonic $e^{\pm}$ from $B$ decays to the total non-photonic $e^{\pm}$ yield.]{An illustration\protect\cite{XioyanLinThesis} of the fitting method used to find $r_{B}$, the ratio of non-photonic $e^{\pm}$ from $B$ decays to the total non-photonic $e^{\pm}$ yield (from both $D$ and $B$ decays) using measurements of the azimuthal correlation between $e^{\pm}$ and hadrons.  In red (blue) is the $D$ ($B$) contribution to the $\Delta\phi$ distribution, shown as black circles.  The near-side peak in the $\Delta\phi$ distribution (black circles) is fit with a linear combination (green) of the $D$-decay (red) and $B$-decay (blue) contributions calculated from simulations.  For this $p_{T}$ bin ($2.5\GeV/c<p_{T}<3.5\GeV/c$), $r_{B}=0.28\pm 0.06$.}
\label{fig:intro:DvB_fit}
\end{center}
\end{figure}

\begin{figure}[htbp]
\begin{center}
\includegraphics[width=0.85\linewidth]{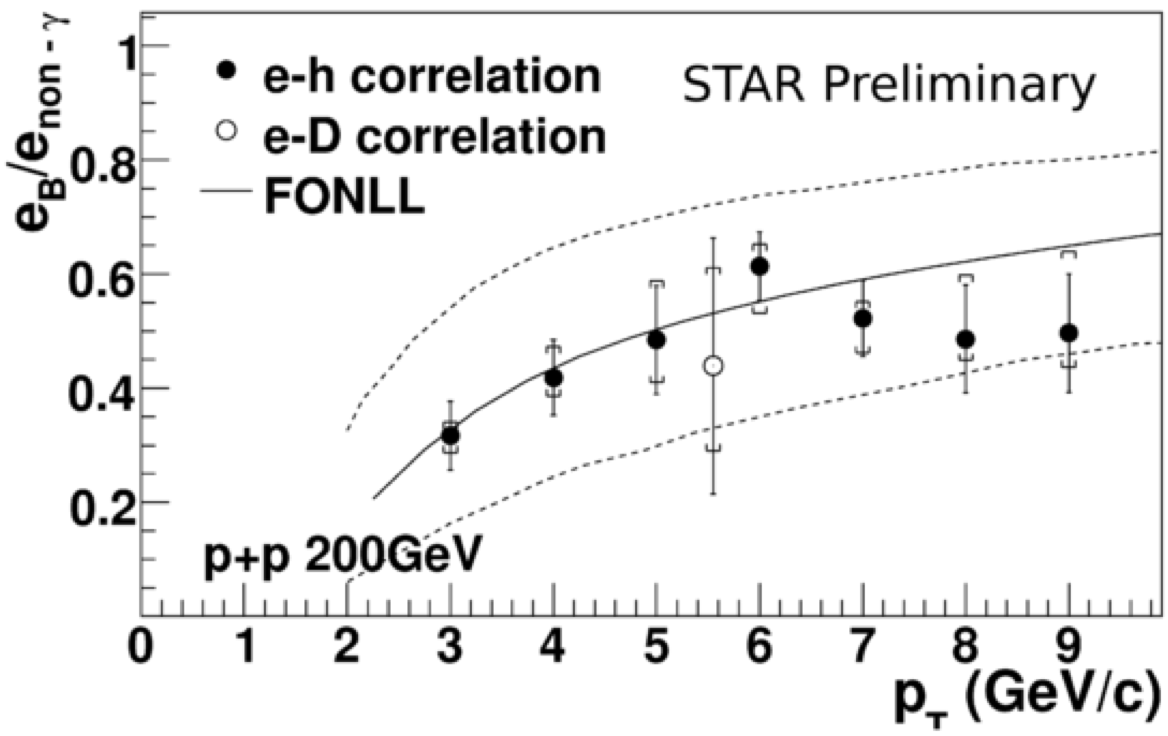}
\caption[Measurement from $e^{\pm}$-hadron correlation studies of the fraction of non-photonic $e^{\pm}$ that come from $B$-meson decays.]{The fraction $(r_{B})$ of non-photonic $e^{\pm}$ that come from $B$-meson decays as a function of $p_{T}$ predicted\protect\cite{PhysRevLett.95.122001} by FONLL calculations and measured\protect\cite{Biritz2009849c} by the STAR collaboration through $e^{\pm}$-hadron correlation studies.  The open circle indicates the measurement of $r_{B}$ from an $e^{\pm}$-$D^{0}$ correlation measurement\protect\cite{Mischke2009361}.}
\label{fig:intro:Bfraction}
\end{center}
\end{figure}

FONLL calculations~\cite{PhysRevLett.95.122001} indicate that the bottom-decay contribution to the non-photonic $e^{\pm}$ spectrum in nucleon-nucleon collisions dominates over the charm contribution at high transverse momenta (see Figure~\ref{fig:intro:FONLL_c_b}).  The calculations indicate that the crossover occurs near $p_{T}\approx 5\GeV/c$, though this estimate has large uncertainties.  The relative contribution of $c$- and $b$-quark decays to the spectrum of non-photonic $e^{\pm}$ has been measured using the azimuthal correlations of non-photonic $e^{\pm}$ with hadrons.  The method~\cite{XioyanLinThesis,Biritz2009849c} is illustrated in Figure~\ref{fig:intro:DvB_fit}.  Similar to dihadron correlation measurements (see Section~\ref{sec:theory:signatures:jet_quenching}), the difference in azimuth $\Delta\phi$ between a ``trigger" non-photonic $e^{\pm}$ and ``associated" particles in the same event is calculated.  Due to decay kinematics, the width of the near-side peak is expected to be wider when the $e^{\pm}$ originates from a $B$-meson decay than when it originates from a $D$-meson decay.  If a $D$ and a $B$ meson are traveling with the same momentum, the $D$ will be traveling with a larger Lorentz boost and its decay products will tend to be separated by smaller angles in the reference frame of the laboratory.  The $\Delta\phi$ distributions for trigger $e^{\pm}$ from $D$ and $B$ decays are simulated using PYTHIA.~\cite{PYTHIA6.4}  The near-side peak in the measured distribution is fit with a linear combination of the simulated $D$- and $B$-decay distributions to extract $r_{B}$, the fraction of non-photonic $e^{\pm}$ that come from $B$ decays in 200-GeV $p+p$ collisions.  The results of this measurement are shown in Figure~\ref{fig:intro:Bfraction}; the measured ratio is consistent with the FONLL prediction.

Measurements of the non-photonic $e^{\pm}$ nuclear modification factor $R_{AA}^{NPE}$ and $r_{B}$ will allow constraints to be placed on the values of $R_{AA}^{e_{D}}$ and $R_{AA}^{e_{B}}$, the nuclear modification factors for non-photonic $e^{\pm}$ from $D$ and $B$ decays, respectively.  The non-photonic $e^{\pm}$ nuclear modification factor can be written as

\begin{equation}
R_{AA}^{NPE}=(1-r_{B})R_{AA}^{e_{D}}+r_{B}R_{AA}^{e_{B}}.
\end{equation}

\begin{figure}[htbp]
\begin{center}
\includegraphics[width=0.75\linewidth]{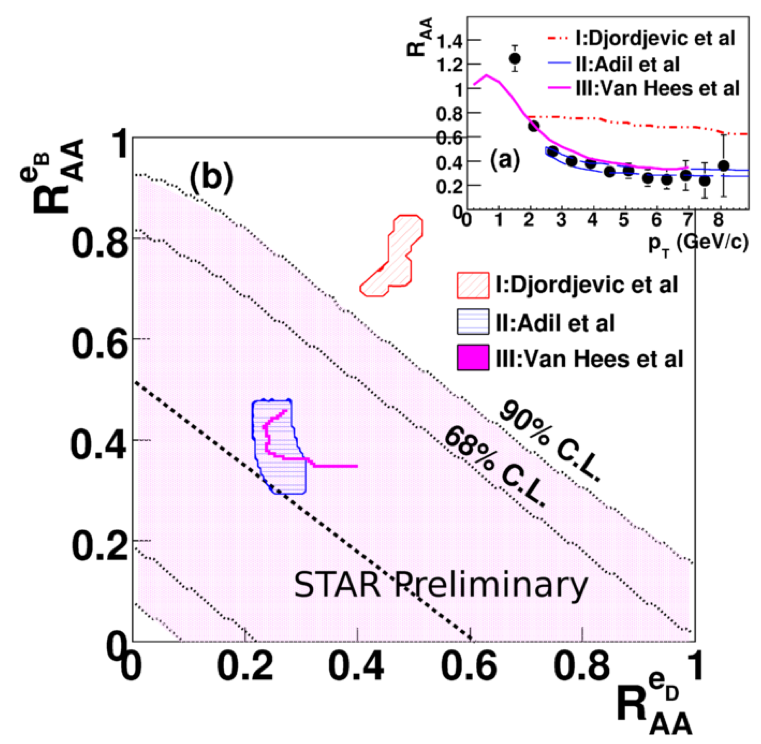}
\caption[The possible values of the nuclear modification factor for for non-photonic $e^{\pm}$ from $D$ and $B$ decays.]{An illustration\protect\cite{Dunlop2009419c} of the method which can be used to constrain the possible values of the nuclear modification factor for non-photonic $e^{\pm}$ from $D$ and $B$ decays ($R_{AA}^{e_{D}}$ and $R_{AA}^{e_{B}}$) based on the measurement of the $R_{AA}^{NPE}$ and $r_{B}$.  The predictions of some of the models discussed in Section~\ref{sec:theory:heavy_interactions} are given (``Djodjevic \textit{et al.}'' refers to DGLV, ``Adil \textit{et al.}'' refers to Collisional Dissociation).  The inset shows the total non-photonic $e^{\pm}$ $R_{AA}$ as a function of $p_{T}$.\protect\cite{PhysRevLett.98.192301,PhysRevLett.98.192301Erratum}}
\label{fig:intro:raa_DvB}
\end{center}
\end{figure}

Figure~\ref{fig:intro:raa_DvB} shows the allowed values~\cite{Dunlop2009419c} of $R_{AA}^{e_{D}}$ and $R_{AA}^{e_{B}}$ based on the STAR collaboration's measurements of $R_{AA}^{NPE}$ for $p_{T}>5\GeV/c$ in Au + Au collisions (the 0-10\% centrality class) and $r_{B}$ in $p+p$ collisions.  This measurement would exclude the predictions of the DGLV model (with only radiative energy loss) at the 90\% confidence level.  Measurements of $R_{AA}^{NPE}$ and $r_{B}$ with smaller uncertainties will allow more of the $R_{AA}^{e_{D}}$ vs. $R_{AA}^{e_{B}}$ parameter space to be excluded.  However, it should be noted that the value of $R_{AA}^{NPE}$ used in the calculation shown in Figure~\ref{fig:intro:raa_DvB} is suspect, as the spectra of non-photonic $e^{\pm}$ in $p+p$ and Au + Au~\cite{PhysRevLett.98.192301} collisions used to calculate it are subject to an erratum.\footnote{This erratum~\cite{PhysRevLett.98.192301Erratum} indicates that both the $p+p$ and Au + Au spectra decreased from the published values by a factor of $\approx 2$.  Therefore, the corrected values of $R_{AA}$ are the same as the original values within uncertainties, although the uncertainties changed.  The calculations used to produce Figure~\ref{fig:intro:raa_DvB} have not yet been redone.}  The figure is included only as an illustration of the method.

\clearpage

\section[Remarks on Non-Photonic $e^{\pm}$ in 200-GeV Cu+Cu Collisions]{Remarks on Non-Photonic $\boldsymbol{e^{\pm}}$ \\ in 200-GeV Cu+Cu Collisions}

In this dissertation, the yields of non-photonic $e^{\pm}$ in Cu + Cu collisions at $\sqrt{s_{NN}}=200$ GeV for the 0-20\% and 20-60\% centrality classes (as well as the overlap 0-60\% centrality class) are measured.  The non-photonic $e^{\pm}$ nuclear modification factor in Cu + Cu collisions $R_{\mathrm{CuCu}}^{NPE}$ are calculated by comparing those yields to measurements of the non-photonic $e^{\pm}$ spectrum in 200-GeV $p+p$ collisions by the STAR~\cite{STAR_ppNPE2011} and PHENIX~\cite{PhysRevLett.97.252002} collaborations.\footnote{The STAR collaboration's measurement is a new measurement not subject to the erratum mentioned in the previous section.}  The Cu + Cu centrality classes studied in this analysis are expected to have values of $\langle N_{part}\rangle$ similar to the 40-60\% centrality class of Au + Au collisions.  A measurement of $R_{AA}^{NPE}$ in Cu + Cu and Au + Au (centrality 40-60\%) collisions allows the suppression to be studied in two collision systems with similar QGP volumes but different collision geometries: the overlap region in a mid-central Au + Au collision should have a larger eccentricity than the overlap region in a central Cu + Cu collision.  If a difference were to be observed between $R_{\mathrm{CuCu}}^{NPE}$ and $R_{\mathrm{AuAu}}^{NPE}$, it could be an indication that the physical shape of the QGP plays a role in determining the suppression of non-photonic $e^{\pm}$.

\clearpage

\chapter{Experiment}
This chapter describes the experimental facilities used to collide heavy ions and to record data on the products of those collisions.  The Relativistic Heavy Ion Collider is described in Section~\ref{sec:experiment:rhic}, followed by an overview of the STAR detector in Section~\ref{sec:experiment:star}.  Sections~\ref{sec:experiment:tpc} and~\ref{sec:experiment:bemc} describe the STAR Time Projection Chamber and the Barrel Electromagnetic Calorimeter, the two components of the STAR detector that are used most directly in the analysis presented in the following chapters.

\section{RHIC}
\label{sec:experiment:rhic}

\begin{figure}[htbp]
\begin{center}
\includegraphics[width=0.85\linewidth]{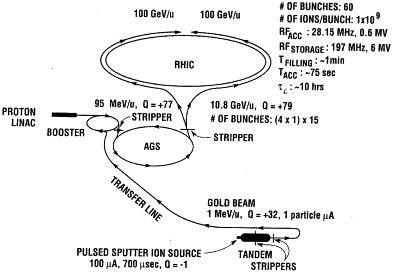}
\caption[The Relativistic Heavy Ion Collider.]{The Relativistic Heavy Ion Collider.\protect\cite{Harrison2003235}}
\label{fig:experiment:RHIC_layout}
\end{center}
\end{figure}

\subsection{Overview of RHIC}

The Relativistic Heavy Ion Collider (RHIC)~\cite{Harrison2003235,Hahn2003245} is a collider/accelerator facility at Brookhaven National Laboratory in Upton, NY.  It was conceived in 1983 to advance the study of nuclear physics at high energies.  RHIC became operational in 2000 and until recently was the highest energy heavy-ion collider in the world.  The facility (see Figure~\ref{fig:experiment:RHIC_layout}) consists of two rings that carry beams of ions ranging in mass from protons to gold nuclei.  The maximum possible energy to which ions can be accelerated decreases as their charge-to-mass ratio increases: for example, the maximum energy is 250 GeV for protons, 114.9 GeV per nucleon for Cu ions, and 100 GeV per nucleon for Au ions.~\cite{RHIC_Design}  RHIC is capable of accelerating polarized beams of protons, allowing the spin structure of the proton to be studied; RHIC is the highest energy polarized-proton accelerator in the world.

The main part of the collider consists of two rings 3.8 km in circumference, which intersect in six regions equally spaced around the ring.  Each ring consists of six curved sections of length $\sim 356$ m and six straight sections (through the interaction regions) of length $\sim 277$ m.  In each of the curved sections, each ion beam passes through a series of 32 superconducting dipole magnets with field strengths of 3.458 T at the maximum beam energy, which bend the beams in the horizontal plane. In the curved sections, the beams are horizontally separated by 90 cm, and each beam passes through a separate set of magnets.  It is therefore possible for each ring to carry a different species of ion; this capability has been used to collide beams of deuterons with Au ions at a center-of-mass energy per nucleon-nucleon pair of $\sqrt{s_{NN}}=200$ GeV.  Each curved section also contains quadrupole and sextupole magnets (and magnets with octupole and decupole components)~\cite{Anerella2003280} to focus the beams and correct for imperfections in the fields of the other magnets.  In the areas where the straight and curved sections meet, a set of dipole magnets directs the two beams into a single beam line to achieve head-on collisions.

The journey of heavy ions (Au ions in this example) begins at a pulsed sputter ion source, which produces negative ions with charge $Q=-1$.  Those negative ions are accelerated through a Tandem Van de Graaff accelerator, reaching a potential of 14 MV at the center of the Tandem.  There, the ions pass through a foil which strips electrons from them, leaving Au ions with a charge of $Q=+12$.  Those positive ions are accelerated through a potential of 14 MV, leaving the Tandem Van de Graaff accelerator with an energy of 1 MeV/u.  A second stripping foil at the exit of the Tandem leaves Au ions with a charge state $Q=+32$ upon entering the transfer line for injection into the Booster synchrotron.  There are two independent Tandem Van de Graaff accelerators, allowing for the production of beams of different species of ions.  A separate linear accelerator is used to produce beams of protons, which are injected directly into the Booster synchrotron.  In the Booster, RF cavities are used to group the ions into bunches and to accelerate them to energies of 95 MeV/u.  Upon exiting the Booster, the Au ions pass through another stripping foil, which removes all but the two most tightly bound ($K$-shell) electrons, leaving the ions with a charge state of $Q=+77$ upon injection into the Alternating Gradient Synchrotron (AGS).  There, the ions are accelerated to energies of 8.86 GeV/u, then stripped of their final two electrons (leaving them in a charge state of $Q=+79$).  Bunches of ions are then injected into one of the two RHIC rings; the ion beams in these rings circulate in opposite directions.  Upon entering the RHIC rings there are $\sim 10^{9}$ ions per bunch ($\sim 10^{10}$ protons per bunch when proton beams are being collided).  In the RHIC rings, the acceleration RF system uses two RF cavities (per ring) operating at a frequency of 28.15 MHz and per cavity voltages of 20-300 kV to accelerate ions to the desired energy.  A second ``storage" RF system operating at a frequency of 197 MHz and a voltage of 6 MV is used to prevent the bunches of ions from spreading out in the longitudinal direction.  The RF systems maintain 360 ``buckets'' around each ring, of which 120 buckets are typically filled with ion bunches.  When RHIC is filled, it typically contains $\sim 1.2\times10^{11}$ ions in each ring.

\subsection{The RHIC Experiments}

Ions may collide in the six interaction regions around the RHIC ring.  Four of those interaction regions have hosted experimental detectors.  The Solenoidal Tracker At RHIC (STAR) detector, which was used to collect the data for this dissertation, will be described in the following section.  The PHOBOS detector~\cite{Back2003603} was designed to measure charged-particle multiplicities over a large solid angle using a set of silicon detectors around the beam line, covering the full azimuth and pseudorapidity $|\eta|<5.4$.  The PHOBOS detector also included two ``arms" opposite each other in azimuth and spanning the pseudorapidity range $0\leq\eta\leq2$, with a width of 0.2 rad in azimuth.  These arms used several layers of silicon detectors and time-of-flight detectors to identify and measure the trajectories of particles with low momenta (transverse momenta down to 30 MeV/$c$ could be accurately measured).  The BRAHMS detector~\cite{Adamczyk2003437} consisted of two moveable magnetic spectrometer arms covering small solid angles.  The ``Forward Spectrometer," which covered 0.8 msr and could be rotated through the polar angle range $2.3^{\circ}<\theta<30^{\circ}$, used a combination of dipole magnets, time projection chambers, drift chambers, time-of-flight hodoscopes, and Cherenkov detectors to track and identify particles produced with small angles relative to the beam line.  The ``Mid-Rapidity Spectrometer," which covered 6.5 msr and could be rotated through the polar angle range $30^{\circ}<\theta<95^{\circ}$, used a dipole magnet, time projection chambers and a time-of-flight hodoscope to track and identify low-momentum particles produced at mid-rapidity.  BRAHMS also included a barrel of silicon-strip and scintillator-tile detectors to measure charged particle multiplicities and collision vertex position around the nominal interaction point, covering the full azimuth and the pseudorapidity range $|\eta|<2.2$.  Both PHOBOS and BRAHMS have been decommissioned.

The detector for the Pioneering High Energy Nuclear Interaction eXperiment (PHENIX)~\cite{Adcox2003469} is a large, multi-component detector with four arms, two for particle tracking and identification at mid-rapidity, and two used to track muons at large rapidities.  The central arms~\cite{Adcox2003489,Aizawa2003508,Aphecetche2003521}, which span the pseudorapidity range $|\eta|<0.35$ and two back-to-back sectors of width $90^{\circ}$ in azimuth, consist of a set of multi-wire chambers, time-of-flight detectors, Cherenkov detectors, and calorimeters that are used to track and identify particles in an axial magnetic field.~\cite{Aronson2003480}  The muon arms~\cite{Akikawa2003537}, which cover the full azimuth and the pseudorapidity ranges $-1.15\geq\eta\geq -2.25$ and $1.15\leq\eta\leq 2.44$, consist of three multi-plane drift chambers that track charged particles in a radial magnetic field~\cite{Aronson2003480} and a multi-layer muon identifier (alternating layers of steel absorbers and streamer tubes) to distinguish muons from other charged particles.  PHENIX also includes a two-layer silicon strip detector~\cite{Allen2003549} close to the beam, covering the full azimuth and the pseudorapidity range $|\eta|<2.6$ to measure charged-particle multiplicities, collision vertex position, and the orientation of the event reaction plane.

The Zero-Degree Calorimeters (ZDCs)~\cite{Adler2003433,Bieser2003766} are a set of eight identical hadronic calorimeters, with one placed near the beam on each side of each of the four RHIC experiments.  The ZDCs act as minimum-bias triggers and luminosity monitors for heavy-ion collisions, and measure event characteristics in ways that are the same for each of the four experiments.  The ZDCs are described in Section~\ref{sec:experiment:star_overview}.

\section{STAR}
\label{sec:experiment:star}

\begin{figure}[htbp]
\begin{center}
\includegraphics[width=1\linewidth]{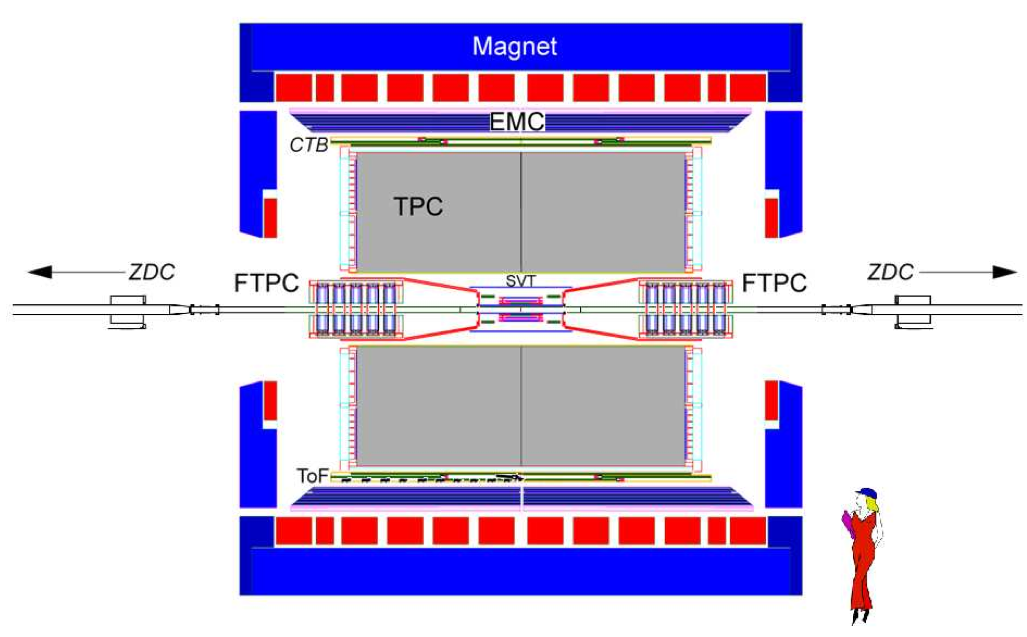}
\caption[Cross-sectional view of the STAR detector.]{A cross-sectional view (in the $yz$-plane) of the STAR detector.\protect\cite{Ackermann2003624}  Note that the BEMC is labeled ``EMC" here, the EEMC is not shown, and the SSD is not labeled.}
\label{fig:experiment:STAR_layout}
\end{center}
\end{figure}

The STAR detector~\cite{Ackermann2003624} is a large, multi-component detector primarily designed to measure hadron production over a large solid angle while operating in an environment with a high multiplicity of charged particles ($\sim 1000$ charged particles per unit pseudorapidity for central Au + Au collisions).  The major components of the STAR detector (see Figure~\ref{fig:experiment:STAR_layout}) will be briefly described in this section.  The analysis described in this dissertation is most directly concerned with the Time Projection Chamber (TPC) and the Barrel Electromagnetic Calorimeter (BEMC); those detectors will be described in greater detail in subsequent sections.  The primary tracking component of the STAR detector is the Time Projection Chamber (TPC), a large gas detector that sits inside a room-temperature solenoidal magnet that produces a uniform magnetic field of 0.5 T.  Silicon detectors provide particle tracking near the nominal interaction point, while electromagnetic calorimeters sit outside of the TPC (but within the magnet) to provide energy measurements and assist in particle identification.

\subsection{Overview of STAR}
\label{sec:experiment:star_overview}

The TPC~\cite{Anderson2003659} is a large-volume, cylindrical gas detector with multi-wire proportional chambers as endcaps.  It extends from cylindrical radius 50 cm to radius 200 cm in the range $|z|<$ 210 cm, covering the full azimuth.  At its outer (inner) radius it covers the pseudorapidity range $|\eta|<0.9$ ($|\eta|<1.8$).  The TPC will be described in greater detail in Section~\ref{sec:experiment:tpc}.

Outside the TPC is the Barrel Electromagnetic Calorimeter (BEMC)~\cite{Beddo2003725}, a large lead-scintillator sampling calorimeter that extends from cylindrical radius 223 cm to 263 cm, covering the pseudorapidity range $|\eta|<1$.  The BEMC is $\approx 20$ radiation lengths deep, allowing for nearly full containment of electromagnetic showers.  A sub-detector of the BEMC, the Barrel Shower Maximum Detector, sits 5-6 radiation lengths from the inner face of the BEMC and provides measurements of the shapes and sizes of electromagnetic showers.  The BEMC will be described in greater detail in Section~\ref{sec:experiment:bemc}.

The TPC, BEMC, and most other components of the detector sit inside a magnetic field~\cite{Bergsma2003633} produced by a solenoidal magnet of inner diameter 5.3 m and length 6.85 m.  This magnet produces a uniform magnetic field parallel to the beam line.  The field strength may be adjusted within the range 0.25 T $\leq B\leq$ 0.5 T; the polarity may be changed so that the field points along either the $+z$ or $-z$ directions.  The magnet is typically operated with the maximum field strength, as was the case when the data analyzed in this dissertation were recorded.  The field is produced by 10 main coils, which are connected in series and draw a current of 4500 A when producing the maximum field.  The magnet also has four additional coils, called ``trim coils," two at each end of the solenoid, that are used to improve field uniformity.  The STAR magnetic field has been mapped using an array of Hall probes; the azimuthal component of the magnetic field is observed to have a magnitude of $<3$ gauss, while the radial component is observed to have a magnitude $<50$ gauss at the outer radius of the TPC (with even smaller deviations from 0 closer to the center of the detector).

Between the beam pipe and the inner boundary of the TPC were two silicon detectors (the SVT and the SSD), which were present in STAR at the time the data analyzed in this dissertation were recorded, but were removed a few years ago.  Closest to the beam pipe was the Silicon Vertex Tracker (SVT)~\cite{Bellwied2003640}, which consisted of 216 silicon drift detectors arranged in three concentric cylindrical barrels with radii of 6.9 cm, 10.8 cm, and 14.5 cm.  The SVT was segmented into 13 million pixels, providing 20-$\mu$m spatial resolution and energy-loss resolution of $\approx 7\%$.  The SVT contributed about 0.06 radiation lengths to the total material budget inside the TPC.  The SVT was intended to provide precise tracking in the inner part of the STAR detector, allowing the position of the primary collision vertex to be reconstructed accurately and allowing the decay vertices of short-lived particles (such as strange and multi-strange baryons) to be reconstructed and distinguished from the collision vertex.


Between the SVT and the inner boundary of the TPC, at a radius of 23 cm, was a fourth layer of silicon, the Silicon Strip Detector (SSD).~\cite{Arnold2003652}  The SSD covered the pseudorapidity range $|\eta|<1.2$ and consisted of $\approx$ 500,000 silicon strip readout channels, providing spatial resolution of 740 $\mu$m in the $z$-direction and 20 $\mu$m in the transverse direction.  By providing a fourth layer of position and energy-loss measurements, the SSD was intended to improve tracking in the inner portion of the STAR detector and improve the extrapolation of TPC tracks to hits in the SVT.

The Forward Time Projection Chambers (FTPCs)~\cite{Ackermann2003713} are two cylindrical gas detectors which surround the beam pipe and span the pseudorapidity ranges $2.5\leq|\eta|\leq 4$, ranges not covered by the TPC.  While the electric field in the TPC points in the $\pm z$-direction, the FTPCs have a radial electric field.  Signals are read out in five ring-shaped multi-wire chambers on the outer surface of each cylinder.

A single Endcap Electromagnetic Calorimeter (EEMC)~\cite{Allgower2003740} sits beyond the west end of the TPC.  This annular lead-scintillator sampling calorimeter covers the full azimuth and the pseudorapidity range $1<\eta<2$, a range not covered by the BEMC.  Similar to the BEMC, the EEMC is approximately 20 radiation lengths deep and has a shower maximum detector at a depth of about 5 radiation lengths.  The Endcap Shower Maximum Detector (ESMD) is composed of scintillator strips (\textit{cf.} the Barrel Shower Maximum Detector, which is a set of wire chambers, see section~\ref{sec:experiment:bemc:bsmd}).

The Zero-Degree Calorimeters (ZDCs)~\cite{Adler2003433,Bieser2003766} are hadronic calorimeters which sit along the $z$-axis $\pm 18$ m from the center of the STAR detector and subtend an angle of 2.5 mrad.  The ZDCs are located beyond the RHIC dipole magnets which bend the beams to merge them into a single beam line through the interaction region.  The ion beams and charged collision fragments are bent away from the ZDCs by the dipole magnets, while neutral collision fragments (mainly neutrons) strike the ZDCs, producing hadronic showers.  Each ZDC consists of alternating layers of tungsten absorbers and wavelength-shifting fibers, longitudinally segmented into three modules.  The fibers collect the Cherenkov radiation produced as the shower particles pass through the calorimeter and direct that light to photomultiplier tubes for conversion into an amplified electronic signal.  The STAR collaboration uses a coincidence in the ZDCs as a minimum-bias trigger for heavy-ion collisions, with the signal in each ZDC required to be greater than 40\% of a single neutron signal.  A precise measurement of the time difference between ZDC signals after a collision allows for a measurement of the $z$-position of the collision vertex\footnote{In STAR, this can also be measured by the tracking detectors.}.  Identical ZDCs sit in similar positions near each of the other RHIC experiments.

\section{The TPC}
\label{sec:experiment:tpc}

\begin{figure}[htbp]
\begin{center}
\includegraphics[width=0.85\linewidth]{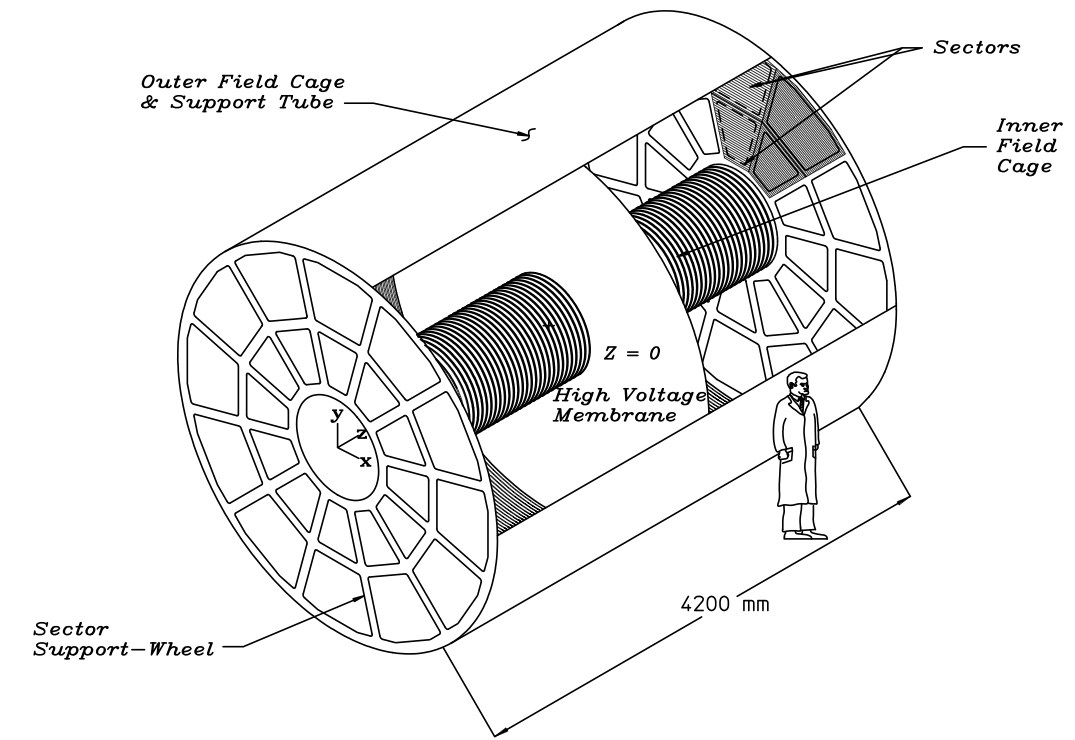}
\caption[Cutaway diagram of the STAR TPC.]{A cutaway diagram of the STAR Time Projection Chamber.\protect\cite{Anderson2003659}}
\label{fig:experiment:TPC_full}
\end{center}
\end{figure}

\subsection{Overview of the TPC}

The STAR Time Projection Chamber (TPC)~\cite{Anderson2003659} is the primary tracking component of the STAR detector.  It allows the trajectory of a charged particle moving through the STAR magnetic field to be recorded in three dimensions, permitting the momentum of the track to be reconstructed.  The TPC is also used to measure the ionization energy loss (typically denoted $dE/dx$) of the particle, information which is useful in determining the identity of the particle.

The TPC, shown in Figure~\ref{fig:experiment:TPC_full}, is a large volume of P10 gas (a mixture of 10\% methane and 90\% argon) contained between two concentric cylindrical ``field cages" (50 cm and 200 cm from the beamline) and two endcaps (at $z=\pm 210$ cm).  Within this volume, a uniform 140-kV/cm electric field is maintained.  A thin, conductive central membrane (cathode) sits at $z=0$ at a potential of 28 kV, while planes of anode wires at the endcaps are held at ground.  Field uniformity between these planes is maintained by the inner and outer field cages,\footnote{The field cages serve two purposes: containing the P10 drift gas and maintaining field uniformity.} the structures of which include a series of closely spaced equipotential rings.  The rings are held at potentials that decrease linearly from the central-membrane potential to ground.

The endcaps of the TPC are multi-wire proportional chambers with pad readout.  The endcaps consist of three planes of wires (the anode grid, the shielding grid, and the gating grid) and a plane of readout pads.  When a charged particle passes through the TPC, it ionizes the gas.  The ionization electrons drift parallel to the electric field towards the endcaps with a drift velocity of $\sim 5\mathrm{cm}/\mu\mathrm{s}$.  The transverse (longitudial) diffusion of the drifting cloud of electrons is $230\mu\mathrm{m}/\sqrt{\mathrm{cm}}$ $(360\mu\mathrm{m}/\sqrt{\mathrm{cm}})$.  The fact that the electric and magnetic fields in the TPC are parallel helps to reduce the transverse diffusion.  The electron cloud drifts to the anode-wire plane, where the large electric field near each wire accelerates the electrons enough to induce secondary ionization, resulting in an avalanche of charge and amplification by a factor of 1000-3000.  The positive ions produced in the avalanche induce an image charge on the nearby readout pads; the signal on each pad is then measured~\cite{Anderson2003679} by a preamplifier/shaper/waveform digitizer system.  The ionization track produced by the passage of a charged particle is thus broken into a set of discrete ``clusters."  The $x$- and $y$-positions of each cluster are determined by the position(s) of the affected readout pad(s).  The $z$-position is reconstructed by combining measurements of the drift velocity and the time at which the cluster signal was recorded.

Each endcap is divided azimuthally into twelve sectors; each sector is divided into inner and outer sub-sectors.  There is a 3-mm gap between sectors.  Within each sector, the wire planes and rows of readout pads run roughly in the azimuthal direction (\textit{i.e.}, perpendicular to the radial direction at the center of each sector).  There are 45 padrows in each sector.  The pad width along the wire direction is 2.85 mm (6.20 mm) for the inner (outer) sub-sectors.  With these pad widths, the image charge from a single avalanche is spread out over several adjacent pads; measurements of the relative signal amplitude in each pad allow the centroid of a cluster to be determined with a spatial resolution less than the width of a single pad.

The function of the shielding grid, which sits 2 mm into the gas volume from the anode grid, is to shape the electric field as it makes the transition from the avalanche region to the uniform field of the drift volume.  The gating grid, which sits 6 mm into the gas volume from the shielding grid, acts as a ``shutter" to control the entry of electrons from the drift volume and to prevent positive ions from the avalanche region from entering the drift region (where they could distort the electric field).  In its ``open" configuration, the wires of the gating grid are held at 110 V.  In the ``closed" configuration, alternate wires are held at potentials of 35 V and 185 V; the 150-V potential difference between adjacent wires creates an electric field roughly perpendicular to the $z$ (drift) direction.  The gating grid is opened when a trigger~\cite{Bieser2003766} signal is received; the 2.1-$\mu$s combined response time of the trigger and gating grid reduce the effective length of the TPC's tracking volume by 12 cm.

The STAR tracking algorithm~\cite{STAR_Note190} reconstructs charged-particle tracks starting with clusters in the outermost few padrows of the TPC (where the track density is lowest).  The algorithm works inward, finding nearby clusters and then sets of three clusters that lie along a straight line.  As the track length increases, the algorithm begins fitting the track with a helix and adding clusters that fall near that helix.  The full track is reconstructed once the algorithm reaches the innermost padrows.  If it is assumed that the particle that produced the track had charge $\pm e$, the transverse momentum can be determined~\cite{GriffithsParticles} from the magnetic field strength $B$ and the radius of curvature $r$ of the helix: $p_{T}=|eB|r$.  The sign of the charge can be determined from the directions of the track's curvature and the magnetic field.  The momentum of the track is determined from the measured transverse momentum and polar angle.

\begin{figure}[htbp]
\begin{center}
\includegraphics[width=0.85\linewidth]{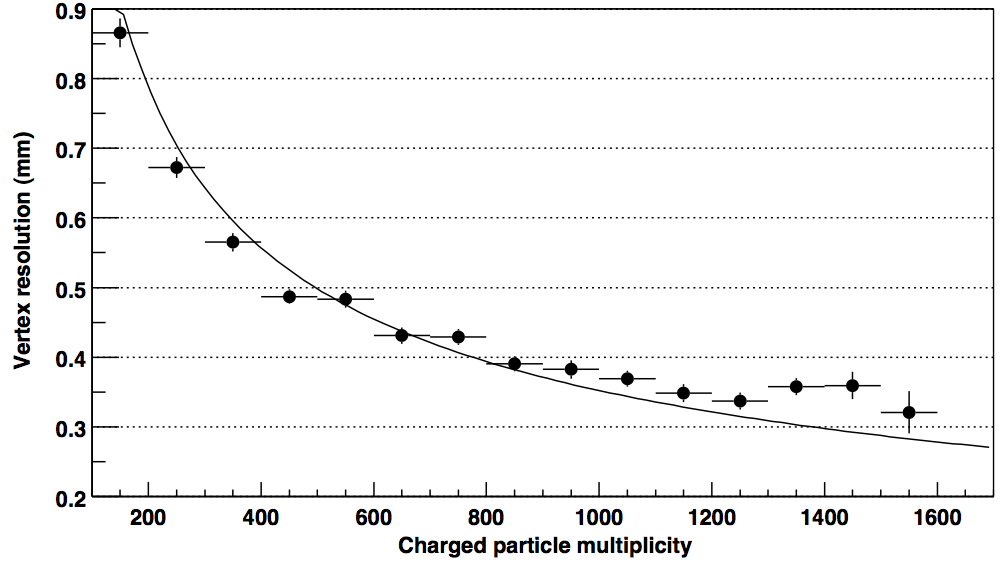}
\caption[Resolution in the TPC of the position of the primary vertex in the $xy$-plane.]{The resolution of the position in the $xy$-plane of the primary vertex; the resolution improves with increasing charged-particle multiplicity.\protect\cite{Anderson2003659}}
\label{fig:experiment:TPC_vxy_res}
\end{center}
\end{figure}

\begin{figure}[htbp]
\begin{center}
\includegraphics[width=0.85\linewidth]{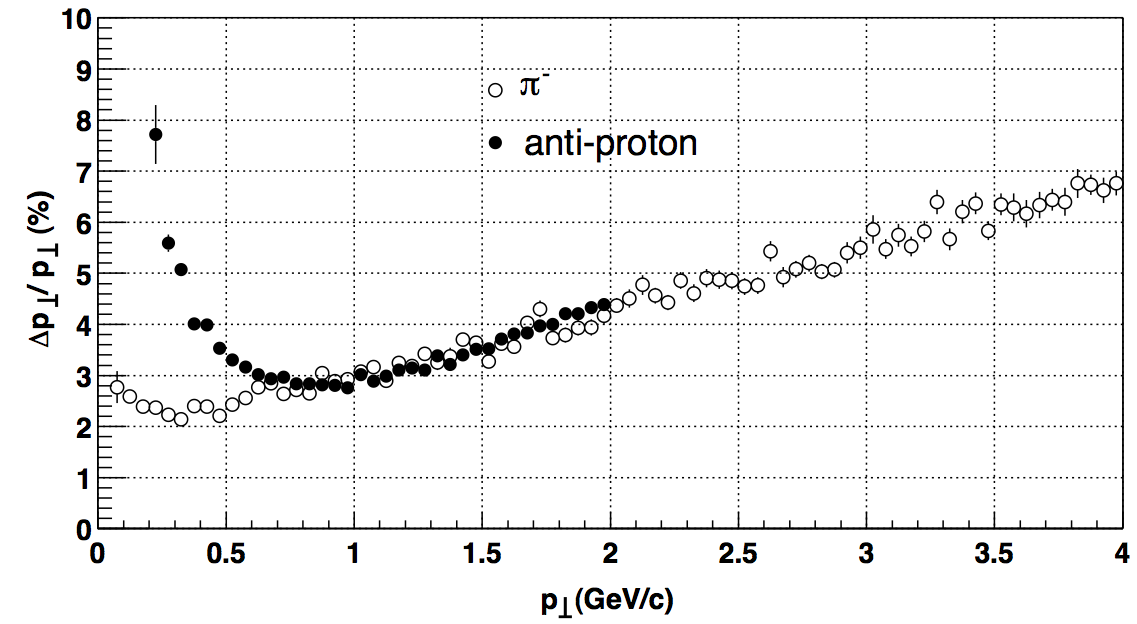}
\caption[Transverse-momentum resolution of the TPC.]{Relative $p_{T}$ resolution in the TPC for a 0.25-T magnetic field.\protect\cite{Anderson2003659}}
\label{fig:experiment:TPC_momentum_res}
\end{center}
\end{figure}

The location of the collision (the ``primary vertex") is found by extrapolating all tracks towards the center of the detector and finding the point $V$ for which $\chi^{2}=\sum_{j}(d_{j}/\sigma_{j})^2$ is at a minimum, where $d_{j}$ is the distance between the point $V$ and the $j$th track helix and $\sigma_{j}$ is the uncertainty of that distance.~\cite{STAR_Note89}  This procedure is repeated several times with large, outlying values of $d_{j}$ removed in each iteration.  Figure~\ref{fig:experiment:TPC_vxy_res} shows the resolution of the position of the primary vertex in the $xy$-plane.  The primary vertex resolution is less than 1 mm in the $xy$-plane and in the $z$-direction.

The position resolution for an individual cluster in the $xy$-plane and in the $z$-direction is $<$ 4 mm.  The transverse momentum resolution, shown in Figure~\ref{fig:experiment:TPC_momentum_res} is typically $\lesssim 10\%$.  The degradation in momentum resolution at low $p_{T}$ is due to multiple Coulomb scattering,\footnote{The difference between pions and antiprotons is due to the fact that the RMS scattering angle is proportional to $\beta^{-1}$~\cite{PDG_review} and for a given momentum, $\beta$ is larger for pions than for protons.}  while the degradation at high $p_{T}$ is due to the difficulty in determining the radius of curvature.

\subsection{Energy Loss}
\label{sec:experiment:tpc:dedx}

Moderately relativistic charged particles (with $0.1\lesssim \beta\gamma\lesssim 100$) passing through matter typically lose energy through ionization.  The mean energy-loss rate is given~\cite{PDG_review} by the Bethe-Bloch equation.  For a particle with mass $M$ and charge $z$ traveling with speed $\beta c$ and Lorentz factor $\gamma$,

\begin{equation}
\label{eq:experiment:bethe_bloch}
-\left\langle\frac{dE}{dx}\right\rangle=\frac{e^{4}}{4\pi\varepsilon_{0}^{2}m_{e}c^{2}}\cdot\frac{\rho N_{A}Z}{A}\cdot\frac{z^{2}}{\beta^{2}}\left(\frac{1}{2}\ln\frac{2m_{e}c^{2}\beta^{2}\gamma^{2}T_{max}}{I^{2}}-\beta^{2}-\frac{\delta(\beta\gamma)}{2}\right),
\end{equation}

\noindent where $m_{e}$ is the electron mass and $N_{A}$ is Avogadro's number.  $Z$, $A$, and $\rho$ are the atomic number, atomic mass, and density of the matter being traversed.  $T_{max}$ is the maximum kinetic energy that can be imparted to a free electron in a single collision:

\begin{equation}
T_{max}=\frac{2m_{e}c^{2}\beta^{2}\gamma^{2}}{1+2\gamma m_{e}/M+(m_{e}/M)^{2}}.
\end{equation}

\noindent The mean ionization potential per electron is $I\approx Z\cdot$(10 eV).  The quantity $\delta$ accounts for the screening of the projectile's electric field in dense materials.~\cite{Fernow}  The only dependence on the projectile particle's mass enters Equation~\ref{eq:experiment:bethe_bloch} through the factor of $T_{max}$.  The values of energy loss are distributed about this mean in a non-normal distribution, with a ``tail" at high values of $dE/dx$.

In the STAR TPC, energy loss is measured through the amount of charge collected for each TPC cluster.  Ideally, the mean energy loss could be measured by calculating the average energy loss for all of the TPC clusters on a track.  Since the energy loss is measured on a cluster-by-cluster basis, a track is effectively divided into segments of length $\sim 2$ cm (the width of a readout pad in the radial direction) for the purposes of energy-loss measurements.  That segment length is too short to average out the large fluctuations in ionization energy loss (due to the ``tail" in the $dE/dx$ distribution).  It was found that the energy loss could not be measured accurately by simply finding the mean of the cluster-by-cluster measurements.  Instead, the STAR collaboration finds the \textit{most probable} energy loss\footnote{The most probable value is not the same as the mean due to the skewed shape of the energy-loss distribution.} by excluding the fraction (typically 30\%) of clusters having the largest signals.  This ``truncated mean" value is found~\cite{Bichsel2006154} through simulations to be a good estimate of the most probable value of the energy loss.  The energy-loss resolution of the TPC is $\lesssim 8\%$ for tracks that cross 40 padrows.

Instead of using the truncated mean energy loss directly, the STAR collaboration typically uses the normalized energy loss.~\cite{Xu201028}  The measured energy-loss value (the truncated mean, which will be denoted as $\langle dE/dx\rangle$) is compared to the expected value for a given particle species.  The difference between the measured and expected values is divided by the energy-loss resolution.

\begin{equation}
n\sigma_{X}=\ln\left( \frac{\langle dE/dx\rangle}{B_{X}} \right)\frac{1}{\sigma_{X}},
\end{equation}

\noindent where $B_{X}$ is the most probable value of $\langle dE/dx\rangle$ for particle species $X$ and $\sigma_{X}$ is the expected width of that distribution.  For pions, the distribution of $n\sigma_{\pi}$ should be a Gaussian with mean 0 and width 1 for all momenta (and the same should hold true for other particle species).  A single value of $\langle dE/dx\rangle$ corresponds to multiple values of $n\sigma$, one for each particle species.  For example, a value $\langle dE/dx\rangle=2.5$ keV/cm might correspond to $n\sigma_{e}=-5$ and $n\sigma_{\pi}=0$. The values of $B_{X}$ and $\sigma_{X}$ are found through simulations.~\cite{Bichsel2006154}  For particle species $X$, the value of $B_{X}$ as a function of momentum or velocity is called the ``Bichsel function."  The Bichsel functions have the same general shape as the Bethe-Bloch function and depend upon the number of track segments used in the calculation of $\langle dE/dx\rangle$ and the lengths of those segments.

\section{The BEMC}
\label{sec:experiment:bemc}

\begin{figure}[htbp]
\begin{center}
\includegraphics[width=0.85\linewidth]{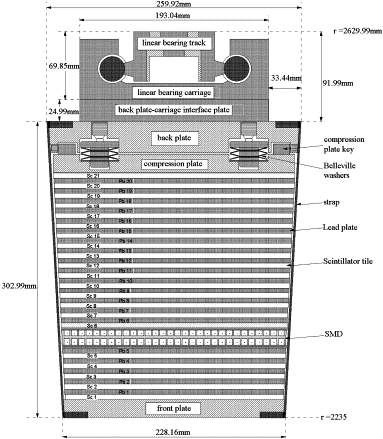}
\caption[Cross-sectional view of two BEMC towers.]{A cross-sectional view (in the $xy$-plane) of a BEMC module (two towers wide).\protect\cite{Beddo2003725}}
\label{fig:experiment:BEMC_module}
\end{center}
\end{figure}

\subsection{BEMC Towers}

The STAR Barrel Electromagnetic Calorimeter (BEMC)~\cite{Beddo2003725} is a sampling calorimeter that sits outside of the TPC, covering the full azimuth and the pseudorapidity range $|\eta|<1$.  The BEMC is segmented into 4800 towers, 120 towers in $\phi$ by 40 towers in $\eta$; for each tower, $\Delta\eta\times\Delta\phi\approx 0.05\times 0.05$.  A cross-section of a two-tower BEMC module is shown in Figure~\ref{fig:experiment:BEMC_module}.  Each tower consists of 21 plastic scintillator tiles, most of which are 5 mm thick,\footnote{The first two scintillator layers are 6 mm thick to accomodate additional wavelength-shifting fibers for the Preshower Detector.} separated by 5-mm-thick layers of lead absorber.  Photons or $e^{\pm}$ striking the inner face of the BEMC will initiate electromagnetic showers in the lead absorber layers.  Hadrons striking the calorimeter will initiate hadronic showers, which develop over longer distance scales than electromagnetic showers; hadronic showers may have an electromagnetic component due to the production of neutral pions, which decay predominantly to photons.  The charged particles in the shower interact with the scintillator, producing ultraviolet light.  This light is collected by wavelength-shifting fibers which run in a $\sigma$-shaped groove around the outside of each scintillator tile.  These fibers guide the scintillation light out of the calorimeter and beyond the STAR magnet to a set of photomultiplier tubes (PMTs).  The light from all 21 scintillator layers in a single tower is directed to a single PMT, which converts the light pulse to an electronic signal and amplifies it by a factor of $\sim 10^{5}$.  The wavelength-shifting fibers increase the wavelength of the scintillation light into a range to which the PMTs are sensitive.  The signal from each PMT is integrated by a charge integrator and converted to a digital value by an Analog-to-Digital Converter (ADC).  The ADC value is then converted to an energy using a linear conversion function.  The relative energy resolution of the BEMC is 12\% for an $e^{\pm}$ with energy 1.5 GeV; the energy resolution scales with $1/\sqrt{E}$ and is degraded by a few percent for events with high charged-particle multiplicity.

The Preshower Detector is a sub-detector of the BEMC which is designed to measure energy deposition near the beginning of a shower and thus improve $e^{\pm}$-hadron discrimination.  The first two scintillator layers each contain two wavelength-shifting fibers.  The ``extra" fibers direct scintillation light to illuminate one pixel of a multi-anode PMT, providing a separate measurement of the energy deposition in the first two scintillation layers.

Electromagnetic showers typically~\cite{PDG_review} reach maximum energy deposition at a depth of

\begin{equation}
t_{max}\approx[\ln(E/E_{c})\pm 0.5]X_{0},
\end{equation}

\noindent where $E$ is the energy of the particle that initiated the shower, $E_{c}$ is the critical energy (7.43 MeV for lead), and $X_{0}$ is the radiation length ($637\mathrm{g/cm^{2}}=0.5612$ cm for lead).  The upper (lower) sign is used for showers initiated by photons $(e^{\pm})$.  The average depth needed for containment of 98\% of a shower's energy is $L_{C}\approx(t_{max}+13.6)X_{0}$.~\cite{FabjanCalorimetry}  For electrons with $E=5\GeV/c$, $t_{max}=6X_{0}$ and $L_{C}=19.6X_{0}$.  The BEMC is approximately 20 radiation lengths (11 cm of lead) deep at $\eta\approx 0$, approximately the depth needed for near-full containment of an electromagnetic shower in the energy range of interest for many STAR measurements $(1\GeV\lesssim E\lesssim 10\GeV\;)$.  In contrast, one parametrization~\cite{FabjanCalorimetry} of the longitudinal development of hadronic showers estimates the shower maximum to occur at a depth of $\lambda_{max}\approx[0.2\ln(E/1\GeV\;)+0.7]\lambda_{I}$, where $\lambda_{I}$ is the nuclear interaction length ($199.6 \mathrm{g/cm^{2}}=17.59$ cm for lead).  The depth for 95\% containment of a hadronic shower is $\lambda_{C}\approx\lambda_{max}+2.5\lambda_{I}(E/1\GeV\;)^{0.13}$.  Since the depth of the calorimeter is less than one interaction length, hadronic showers are not well contained within the BEMC.

\subsection{Shower Maximum Detector}
\label{sec:experiment:bemc:bsmd}

\begin{figure}[htbp]
\begin{center}
\includegraphics[width=0.85\linewidth]{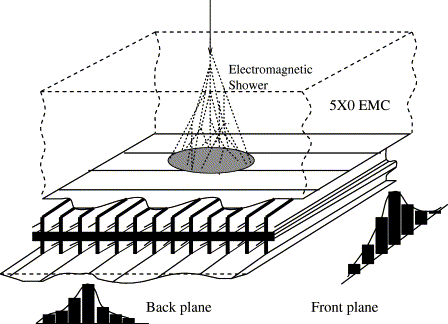}
\caption[Schematic diagram of the BSMD.]{A schematic diagram of a portion of the Barrel Shower Maximum Detector (BSMD).\protect\cite{Beddo2003725}}
\label{fig:experiment:BSMD}
\end{center}
\end{figure}

The Barrel Shower Maximum Detector (BSMD) is a sub-detector of the BEMC which sits 5-6 radiation lengths from the inner face of the calorimeter, approximately the location of the maximum of an electromagnetic shower.  The BSMD allows the shape and size of a shower to be measured with greater resolution than the BEMC towers.  Figure~\ref{fig:experiment:BSMD} shows a schematic diagram of the BSMD.  The BSMD is two layers of gas wire proportional counters with strip readout.  Two layers of wires separated by an aluminum extrusion run in the $z$-direction, with 15 wires running over the width of a tower.  In an electromagnetic shower, $e^{\pm}$ will cause avalanches (with a gas amplification factor of $\approx 3000$) in the high electric field around the anode wires, this induces an image charge on the cathode (readout) strips.  There are 36,000 cathode strips in two planes.  The (inner) BSMD-$\eta$ plane is divided into 300 strips in the $z$-direction (and 60 strips azimuthally), allowing a measurement of the pseudorapidity (or, equivalently, $z$ or $\theta$).  The (outer) BSMD-$\phi$ plane is divided into 900 strips azimuthally (and 20 strips in the $z$-direction) allowing for a measurement of the azimuth.  The strip widths and pitches are all between 1 and 2 cm (see~\cite{Beddo2003725} for the exact values), a small fraction of the dimensions of a tower (which is approximately 12 cm $\times$ 12 cm).  In each plane, adjacent strips that record a signal are grouped into ``clusters."  The centroid of a cluster can be determined with a resolution smaller than the width of a strip.  The position resolution of the BSMD, measured using test beams in the AGS, is $2.4\,\mathrm{mm}+5.6\,\mathrm{mm}/\sqrt{E/1\GeV\;}$ $(3.2\,\mathrm{mm}+5.8\,\mathrm{mm}/\sqrt{E/1\GeV\;})$ for the $\eta$ $(\phi)$ plane.

\clearpage

\chapter{Analysis of Non-Photonic $\boldsymbol{e^{\pm}}$}
\label{sec:anintro}

This chapter describes the method used to extract the yield of non-photonic $e^{\pm}$ from Cu + Cu collisions at $\sqrt{s_{NN}}=200$ GeV.  Section~\ref{sec:anintro:event_selection} describes the criteria used to select the events from which the $e^{\pm}$ yields will be extracted.  Section~\ref{sec:anintro:ele_id} describes the selection cuts used to identify $e^{\pm}$ in those events and section~\ref{sec:anintro:back_sub} describes the method used to extract the yield of photonic $e^{\pm}$.  Section~\ref{sec:anintro:correction} summarizes the correction factors that are applied to the measured data to find the fully corrected non-photonic $e^{\pm}$ spectra; those correction factors will be discussed in detail in subsequent chapters.

Table~\ref{table:anintro:full} summarizes the selection criteria and cuts used in this analysis.  The symbols will be defined in the following sections.  Note that the high-tower trigger condition is not always applied.

\begin{table}
\caption[Summary of cuts used in this analysis.]{Summary of cuts used in this analysis.  See the following sections for more details.}
\label{table:anintro:full}
\begin{tabular}{| r | c |}
\hline
System/Measurement & Condition\\\hline\hline
\multicolumn{2}{| c |}{Run Selection Criteria}\\\hline
run number & 6031103 $\leq$ run number $\leq$ 6065058\\\hline
STAR detector & TPC and BEMC included in run\\\hline
BEMC geometrical acceptance & $A_{BEMC}\geq 0.45$\\\hline\hline
\multicolumn{2}{| c |}{Event Selection Criteria}\\\hline
ZDC & coincidence in both STAR ZDCs\\\hline
trigger ID & trigger id = 66007 (MB) and/or 66203 (HT)\\\hline
high-tower trigger & $E_{Tower}\gtrsim 3.75\GeV$\\\hline
$z$ position of primary vertex & $|v_{z}|<$ 20 cm\\\hline
reference multiplicity & $refmult\geq 19$\\\hline\hline
\multicolumn{2}{| c |}{Track Quality Cuts}\\\hline
radius of first TPC point & $R_{1TPC}<$102 cm\\\hline
number of TPC fit points & $N_{TPCfit}>20$\\\hline
TPC fit points ratio & $R_{FP}>0.52$\\\hline\hline
\multicolumn{2}{| c |}{Non-Photonic $e^{\pm}$ Selection Cuts}\\\hline
DCA to primary vertex & $GDCA<1.5$ cm\\\hline
track pseudorapidity & $-0.1<\eta<0.7$\\\hline\hline
\multicolumn{2}{| c |}{$e^{\pm}$ Selection Cuts}\\\hline
transverse momentum & $p_{T}>2\GeV/c$\\\hline
TPC energy loss & $-1.5<n\sigma_{e}<3.5$\\\hline
energy in EMC & $p/E_{Tower}<2/c$\\\hline
BSMD cluster displacement & $\Delta_{SMD}<0.02$\\\hline
BSMD cluster sizes & $N_{SMD\eta}\geq 2$ and $N_{SMD\phi}\geq 2$\\\hline
BEMC status & ``good" for tower, BSMD-$\eta$, and BSMD-$\phi$\\\hline\hline
\multicolumn{2}{| c |}{Partner Track Selection Cuts}\\\hline
TPC fit points ratio & $R_{FP}(partner)>0.52$\\\hline
partner track $p_{T}$ & $p_{T}(partner)>0.3\GeV/c$\\\hline
TPC energy loss & $\langle dE/dx\rangle (partner)>2.8$ keV/cm\\\hline\hline
\multicolumn{2}{| c |}{Pair Selection Cuts}\\\hline
DCA & $DCA(e^{-}e^{+})<1.5$ cm\\\hline
invariant mass & $M_{inv.}(e^{-}e^{+})<150\MeV/c^{2}$\\\hline
\end{tabular}
\end{table}

\clearpage

\section{Event Selection}
\label{sec:anintro:event_selection}

This section describes the criteria used to select the events to be analyzed.  The data analyzed in this dissertation were recorded from 31 January to 6 March 2005, when RHIC was colliding beams of Cu ions at $\sqrt{s_{NN}}=200\GeV$.  The STAR TPC and BEMC are both used to identify $e^{\pm}$ in this analysis, so only runs during which both of those detectors were operational were analyzed.  A run is a period of data recording usually lasting from a few minutes to a half hour, during which thousands of collisions are recorded.  200-GeV Cu + Cu data were recorded before 31 January, but the STAR BEMC was not properly calibrated during those runs.  A small number of runs exhibited corruption in the BEMC data and were excluded from the analysis.  For some runs, an abnormally low BEMC geometrical acceptance $(A_{BEMC})$ was observed; only runs with $A_{BEMC}\geq 0.45$ are included in this analysis.  See Chapter~\ref{sec:acc:bemc} for a more detailed discussion of the BEMC geometrical acceptance.

For the Cu + Cu collisions analyzed in this dissertation, event readout is triggered by a coincidence in the STAR Zero-Degree Calorimeters during a bunch crossing.  This is called the ``minimum-bias" trigger (abbreviated as ``MB").  During some runs, an additional trigger condition was applied: the Barrel Electromagnetic Calorimeter is required to have at least one tower with energy above a threshold value, $E_{Tower}\gtrsim 3.75$ GeV.  This is called the ``high-tower" trigger (abbreviated as ``HT").  The actual trigger threshold is not in units of energy, but in ADC counts: the amplified signal produced by the photomultiplier tube attached to a tower is integrated over time by a charge integrator and then converted to a digital value by an Analog-to-Digital Converter (ADC).~\cite{Beddo2003725}  Since not all towers have the exact same calibration, a single ADC threshold corresponds to a range of energy values.  Typically, only one tower per event satisfies the HT trigger condition.  The HT trigger results in an enhancement of particle yields at high $p_{T}$.  In this dissertation, the spectra of $e^{\pm}$ from the MB and HT data sets will be combined, with MB data used for $p_{T}<4\GeV/c$ and HT data used for $p_{T}>4\GeV/c$.


\begin{figure}[htbp]
\begin{center}
\includegraphics[width=0.85\linewidth]{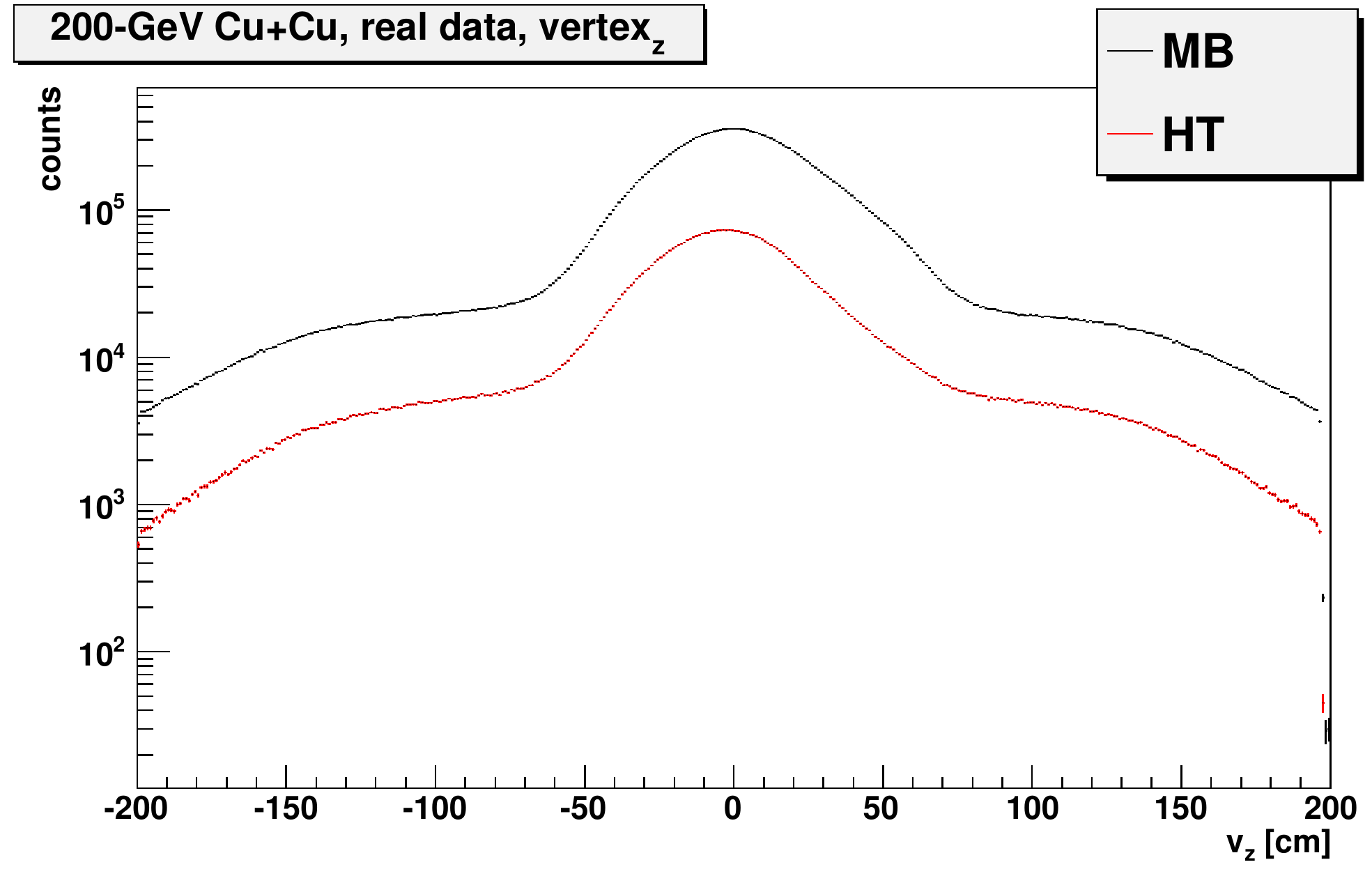}
\caption[Primary collision vertex position along the $z$-axis.]{The $z$-position $(v_{z})$ of the primary collision vertex for minimum-bias (black) and high-tower-triggered (red) events.}
\label{fig:anintro:vz}
\includegraphics[width=0.85\linewidth]{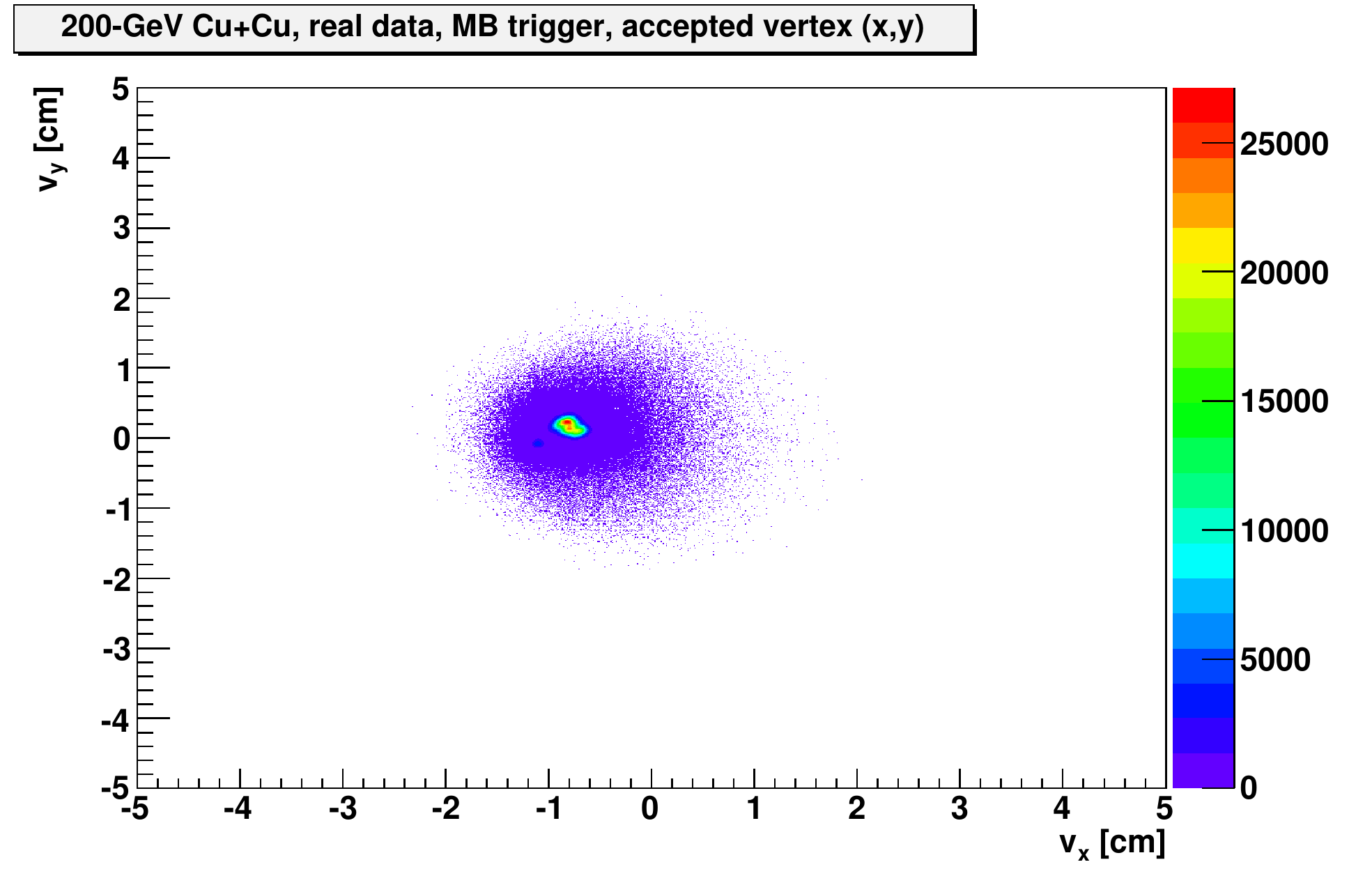}
\caption[Primary collision vertex position in the $xy$-plane.]{The position of the primary collision vertex in the $xy$-plane for MB events that pass the event selection cuts.}
\label{fig:anintro:vxy}
\end{center}
\end{figure}

\begin{figure}[htbp]
\begin{center}
\includegraphics[width=0.85\linewidth]{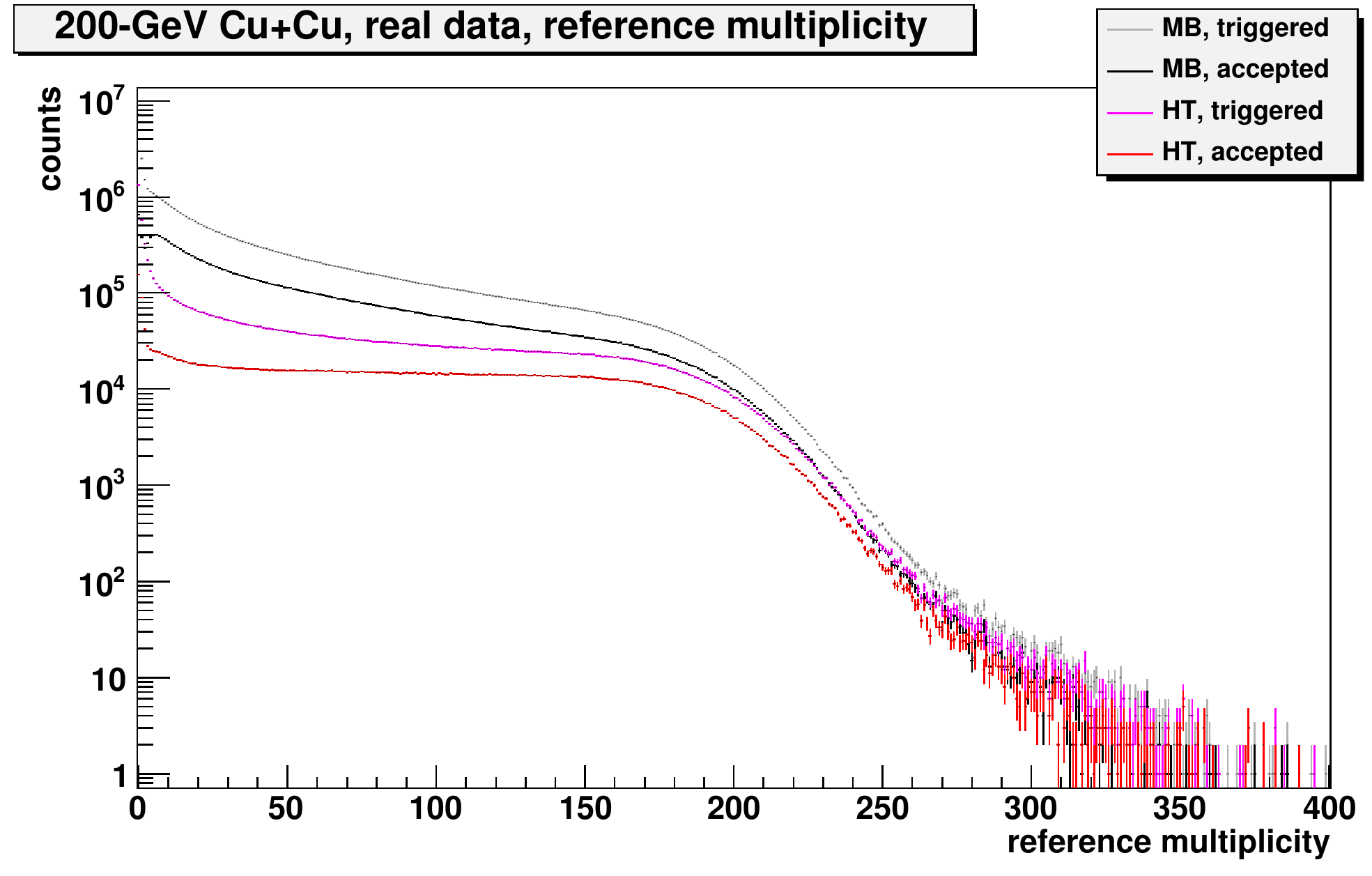}
\caption[Reference-multiplicity distributions for Cu + Cu collisions.]{The distribution of reference multiplicity for MB and HT triggered events.  The distributions for all triggered events (gray and magenta) and for the subset of those events that were accepted by the event selection cuts (black and red) are shown.}
\label{fig:anintro:refmult}
\end{center}
\end{figure}

For each event, it is required that a primary vertex was found and that the $z$-position of that vertex is within the range $|v_{z}|<20$ cm.  This cut is applied in order to reduce the photonic $e^{\pm}$ background.  The support structures for the Silicon Vertex Tracker begin at $z\approx\pm 30$ cm and extend outward in the direction of the beam line.  Photons produced in Cu + Cu collisions with $|v_{z}|\gtrsim 30$ cm therefore have a higher probability of converting to photonic $e^{-}e^{+}$ pairs than photons produced closer to the center of the STAR detector.  Figure~\ref{fig:anintro:vz} shows the distribution of $v_{z}$ for the two trigger types.  Figure~\ref{fig:anintro:vxy} shows the position of the primary vertex in the $xy$-plane.  No cut is applied to $v_{x}$ or $v_{y}$; this figure is included only for reference.

In addition, the reference multiplicity of each event is used to group events into centrality classes.  Table~\ref{table:anintro:centrality} gives the reference-multiplicity ranges that correspond to each of the three centrality classes used in this analysis~\cite{PhysRevC.81.054907}.  Also given for each centrality class are the values of $\langle N_{bin}\rangle$ and $\langle N_{part}\rangle$ (the mean number of binary nucleon-nucleon collisions and the mean number of nucleons participating in collisions, respectively) for each of these centrality classes.  These values were found from simulations based on the Glauber model of nucleus-nucleus collisions, see Section~\ref{sec:theory:basic_concepts:collision}.  Figure~\ref{fig:anintro:refmult} shows the reference-multiplicity distributions for MB (black) and HT (red) triggered events.  Note that the high-tower-triggered data set is biased towards more central (higher reference multiplicity) events.

\begin{table}
\caption{Reference-multiplicity ranges, collision characteristics, and event totals for the centrality classes of 200-GeV Cu + Cu collisions used in this analysis.}
\label{table:anintro:centrality}
\begin{tabular}{| r | c | c | c |}
\hline
Centrality & 0-20\% & 20-60\% & 0-60\%\\\hline\hline
Ref. Mult. & $98\leq refmult$ & $19\leq refmult\leq 97$ & $19\leq refmult$\\\hline
$\langle N_{bin}\rangle$ & $156.35^{+12.14}_{-11.01}$ & $42.68^{+2.80}_{-2.43}$ & $80.41^{+5.90}_{-5.61}$\\\hline
$\langle N_{part}\rangle$ & $86.84^{+1.23}_{-1.18}$ & $33.61^{+0.67}_{-0.48}$ & $51.30^{+0.78}_{-0.87}$\\\hline
$N_{evt}^{MB}$ & 3,708,138 & 8,960,915 & 12,669,053\\\hline
$N_{evt}^{HT}(accepted)$ & 1,323,950 & 1,240,857 & 2,564,807\\\hline
$N_{evt}^{HT}(true)$ & 1,139,315 & 1,076,621 & 2,215,936\\\hline
$N_{evt}^{HT}(normalized)$ & 331,754,960 & 868,834,688 & 1,200,589,648\\\hline
\end{tabular}
\end{table}

Table~\ref{table:anintro:centrality} also gives the total number of events analyzed for each trigger type and centrality class.  $N_{evt}^{MB}$ indicates the number of minimum-bias events that passed the event selection cuts.  For high-tower triggered events, three types of event totals are given.  $N_{evt}^{HT}(accepted)$ indicates the number of events that passed the selection cuts and were labeled as having satisfied the HT trigger condition.  However, in about 14\% of those events, the tower that recorded the signal that fired the HT trigger is recorded as having 0 energy.  This is due to the fact that whether or not the HT trigger fires is determined by the raw data (ADC counts) read out from each tower.  In contrast, the recorded energy values are determined using each tower's calibration equation.  In the event of a bad calibration, or if a tower is later determined to have bad status,\footnote{see Section~\ref{sec:acc:bemc}} a tower which recorded an ADC count above the threshold may subsequently be assigned 0 energy.  $N_{evt}^{HT}(true)$ is the number of high-tower triggered events for which the tower energy was nonzero.

During runs in which the HT trigger was active, the MB trigger was prescaled: only a small fraction (usually $<1\%$) of events that satisfy the MB trigger conditions were actually recorded.  This was done to conserve data storage space.  For each run, one out of every $F_{prescale}$ events was recorded.  The value of $F_{prescale}$, the prescale factor, was not constant, but rather decreased as the RHIC luminosity (collision rate) decreased.  For this analysis, the values of the prescale factor are within the range $23\leq F_{prescale}\leq 595$, though for most runs $100\leq F_{prescale}\leq 400$.  When the HT non-photonic $e^{\pm}$ spectra are normalized, the correct number of events by which the particle yield must be divided is not $N_{evt}^{HT}(accepted)$ or $N_{evt}^{HT}(true)$.  Rather, the correct number of events for normalization is the number of MB events that were observed (whether recorded or not) during the time that the HT events were recorded.  This quantity is the sum over all runs (during which the HT trigger was active) of $F_{prescale}$ times the number of recorded MB events that passed the appropriate event- and centrality-selection cuts.

\begin{equation}
\label{eq:anintro:prescale}
N_{evt}^{HT}(normalized)=\sum_{runs}[F_{prescale}(run)\times N_{evt}^{MB}(run)]
\end{equation}

\noindent The values of $N_{evt}^{HT}(normalized)$ are also given in Table~\ref{table:anintro:centrality}.

\section[$e^{\pm}$ Identification]{$\boldsymbol{e^{\pm}}$ Identification}
\label{sec:anintro:ele_id}

This section describes the cuts used to identify $e^{\pm}$.

\begin{figure}[htbp]
\begin{center}
\includegraphics[width=0.85\linewidth]{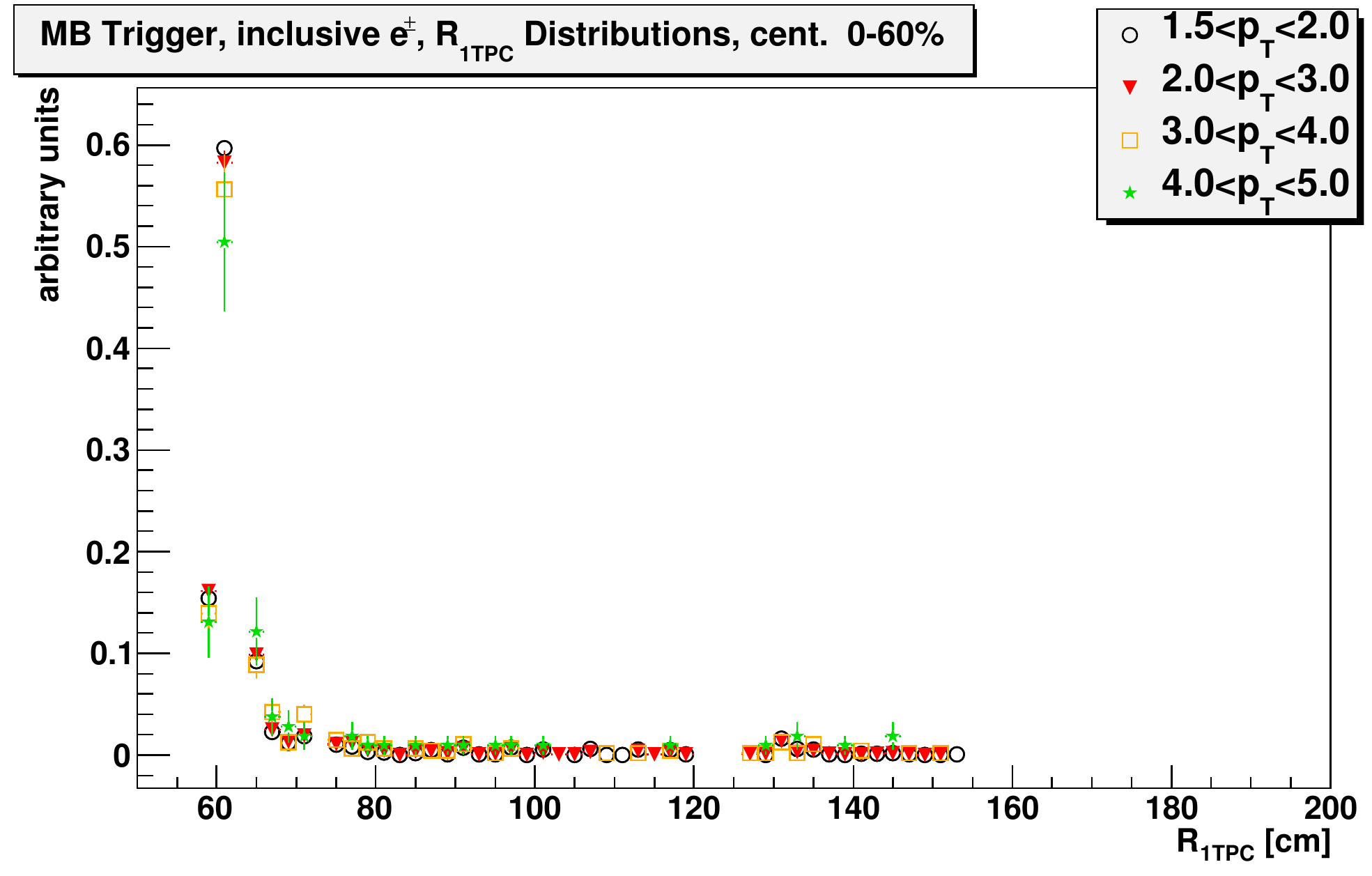}
\caption[Distribution of the radial position of the first TPC hit of a track.]{The radial position (in cylindrical coordinates) of the first TPC hit for tracks that pass the other $e^{\pm}$ identification cuts.  Most of these tracks have the first hit in the first few TPC padrows and satisfy the $R_{1TPC}<102$ cm cut.  The distributions shown are for MB triggered events in several $p_{T}$ bins; the distributions are similar for HT triggered events.  The gap near $R_{1TPC}=120$ cm is due to a masked-out padrow in the TPC.}
\label{fig:anintro:r1tpc}
\end{center}
\end{figure}

Several cuts are used to select good-quality TPC tracks.  Longer tracks with more TPC hits will tend to have their momenta reconstructed and their energy loss measured more accurately than shorter tracks.  It is required that the radial position (in STAR's cylindrical global coordinate system) of a track's first recorded TPC hit $(R_{1TPC})$ be less than 102 cm (\textit{i.e.}, the first TPC hit is in the innermost 9 pad rows of the TPC).  Figure~\ref{fig:anintro:r1tpc} shows the distribution of $R_{1TPC}$ for tracks that have passed all other $e^{\pm}$ identification cuts.  For most of the tracks, the first TPC hit is in the first few pad rows.

\begin{figure}[htbp]
\begin{center}
\includegraphics[width=0.85\linewidth]{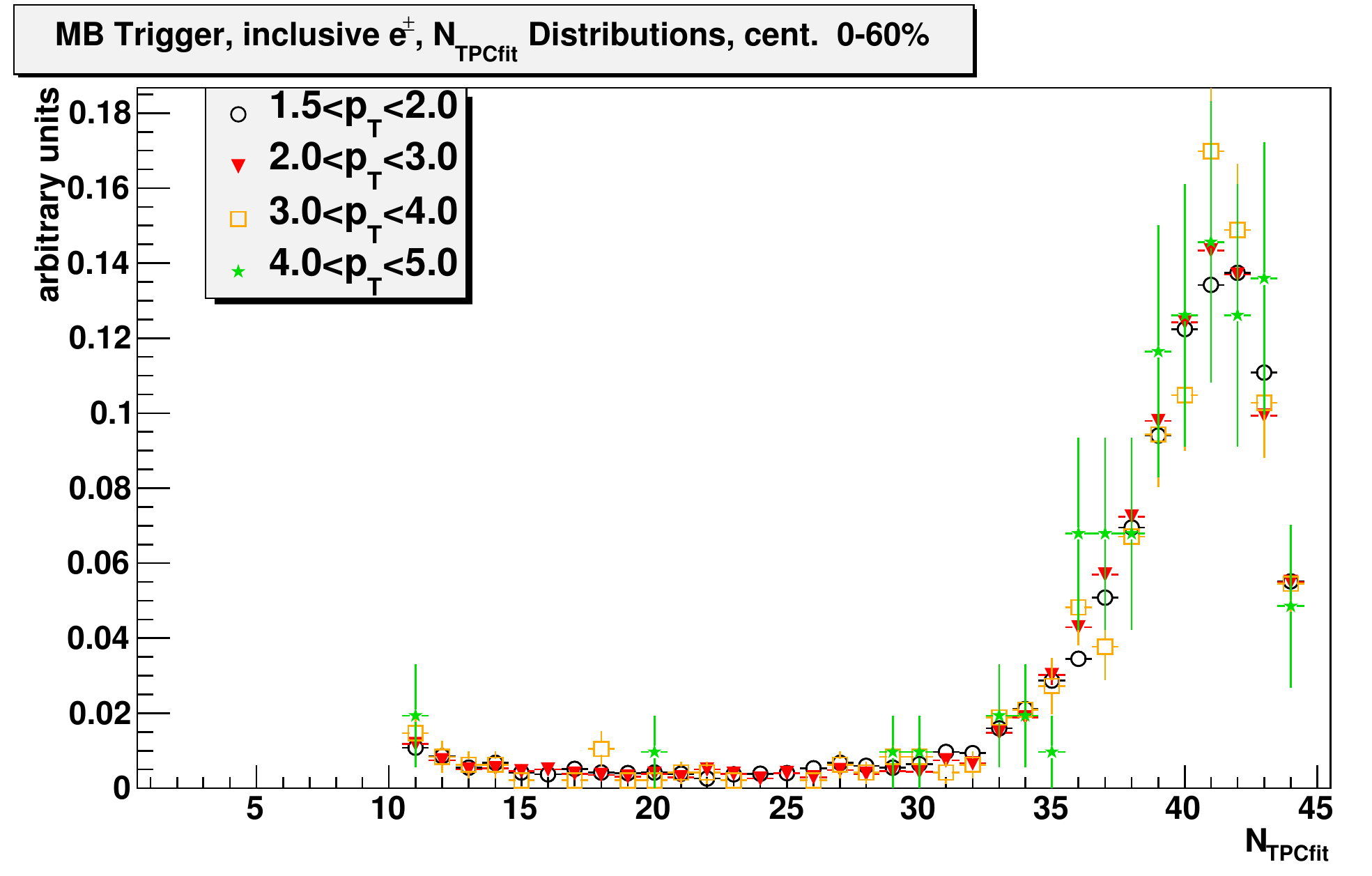}
\caption[Distribution of the number of TPC fit points.]{The number of TPC fit points for tracks that pass the other $e^{\pm}$ identification cuts.  Most of these tracks satisfy the $N_{TPCfit}>20$ cut.  The distributions shown are for MB triggered events in several $p_{T}$ bins; the distributions are similar for HT triggered events.}
\label{fig:anintro:tfp}
\includegraphics[width=0.85\linewidth]{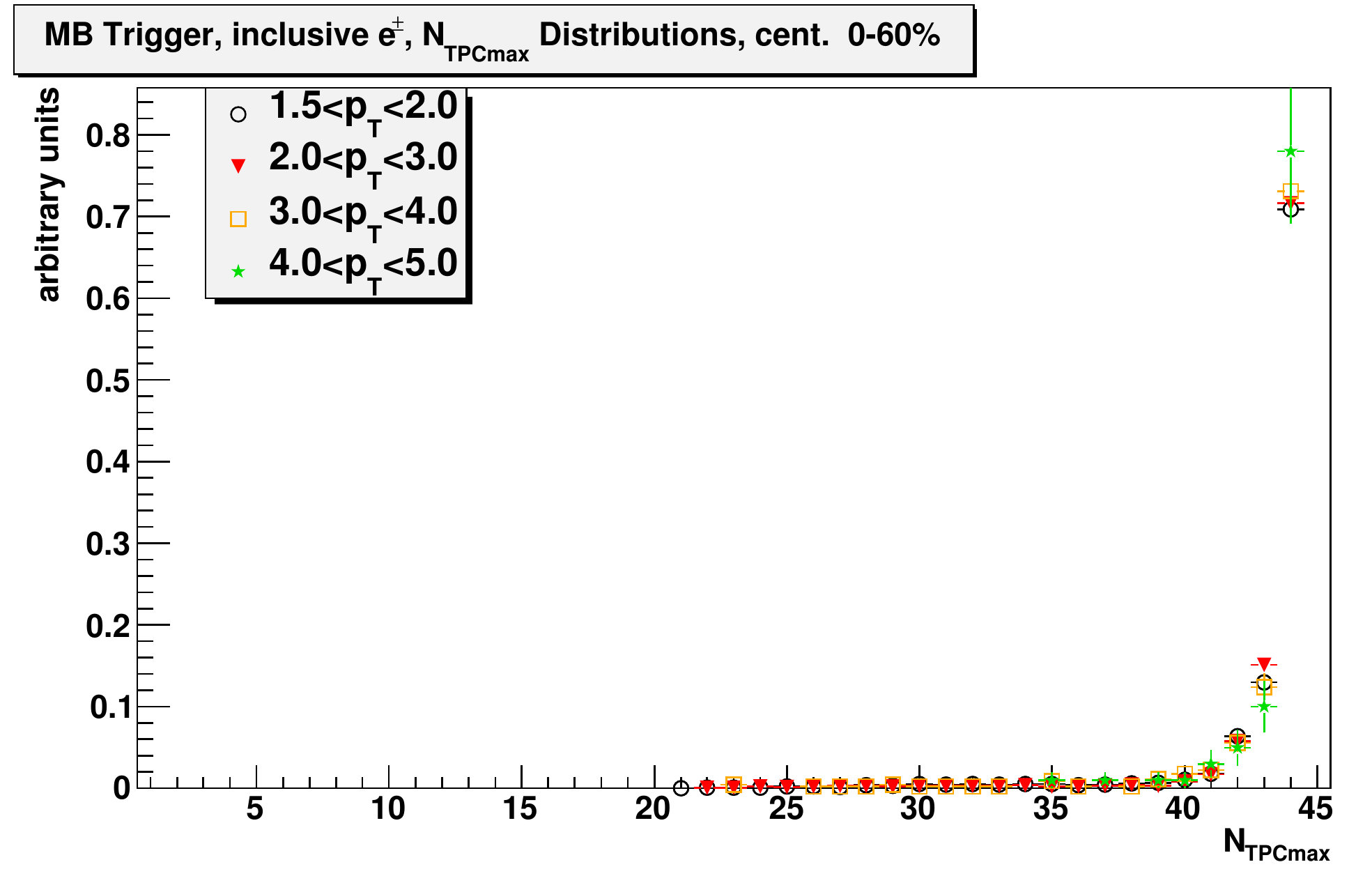}
\caption[Distribution of the maximum possible number of TPC fit points.]{The maximum possible number of TPC fit points for tracks that pass the other $e^{\pm}$ identification cuts.  Most tracks have $N_{TPCmax}$ near 44.}
\label{fig:anintro:mtp}
\end{center}
\end{figure}

\begin{figure}[htbp]
\begin{center}
\includegraphics[width=0.85\linewidth]{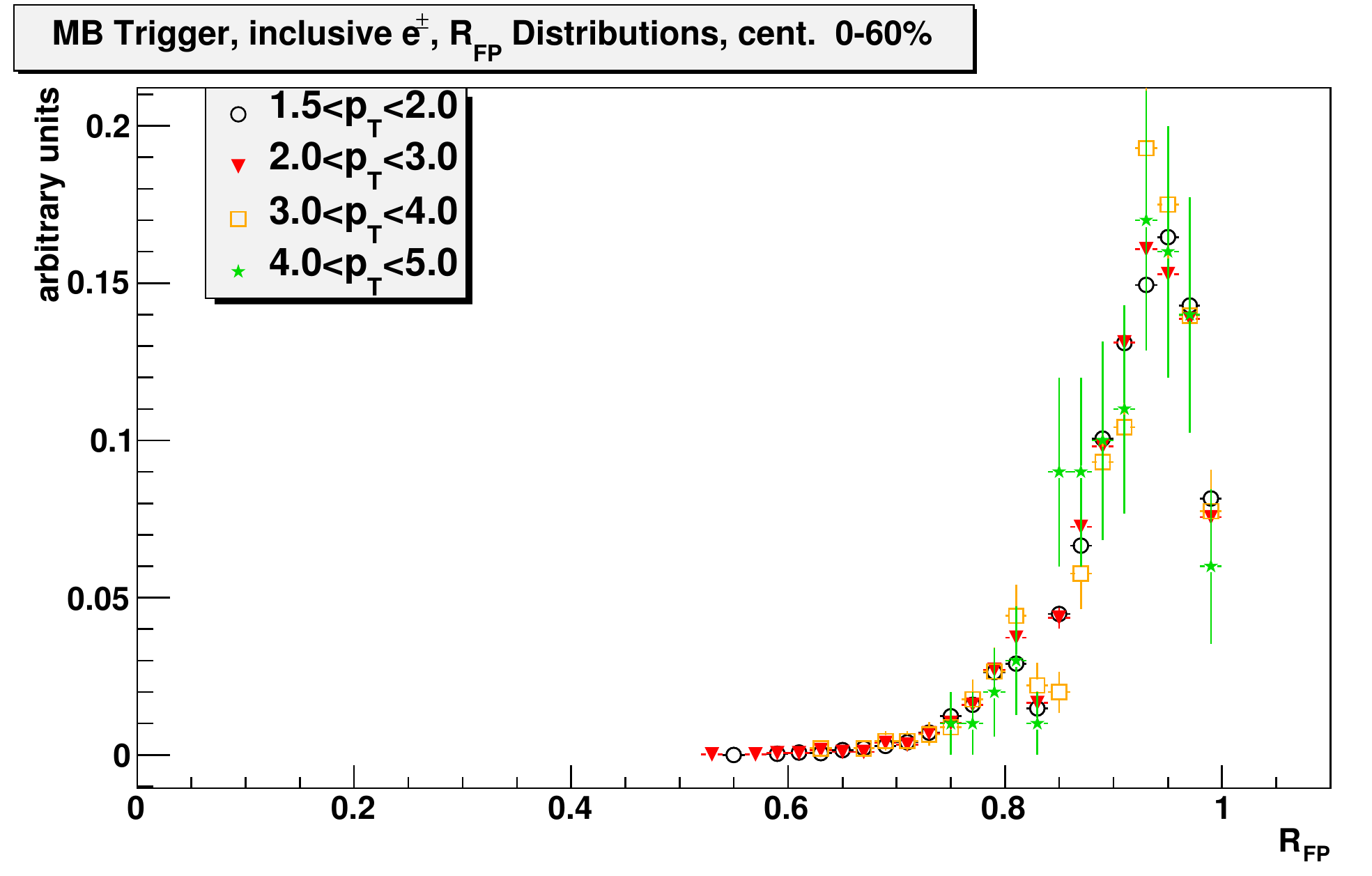}
\caption[Distribution of the fit-points ratio $R_{FP}=N_{TPCfit}/N_{TPCmax}$.]{The ratio $R_{FP}=N_{TPCfit}/N_{TPCmax}$ for tracks that pass the other $e^{\pm}$ identification cuts.  The tracks shown here all satisfy the $R_{FP}>0.52$ cut.  The distributions shown are for MB triggered events in several $p_{T}$ bins; the distributions are similar for HT triggered events.}
\label{fig:anintro:fpr}
\includegraphics[width=0.8\linewidth]{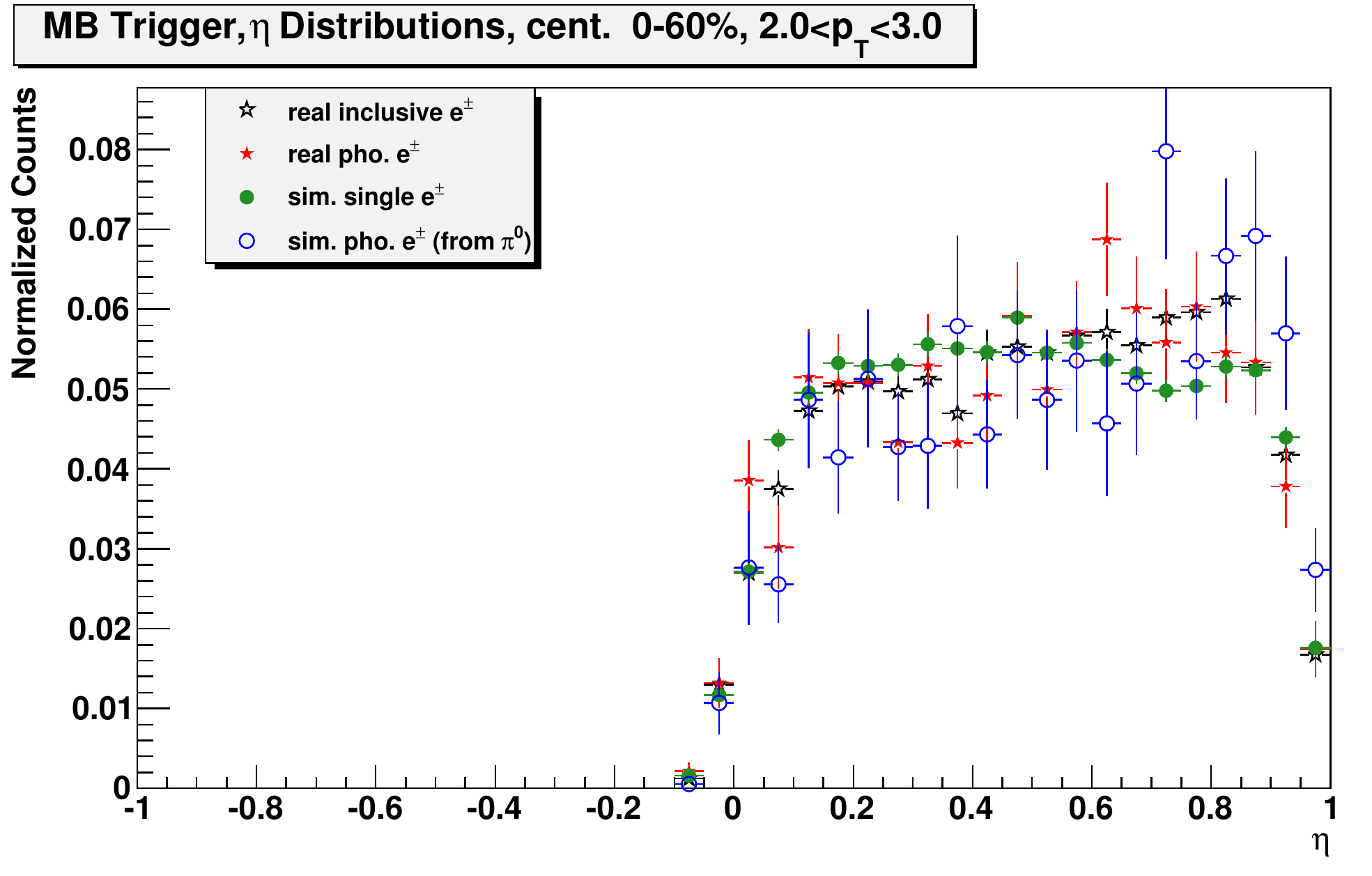}
\caption[Pseudorapidity distribution for real and simulated $e^{\pm}$.]{The pseudorapidity distributions for inclusive $e^{\pm}$ (in the range $2\GeV/c<p_{T}<3\GeV/c$) in real events (black), photonic $e^{\pm}$ in real events (red), simulated non-photonic $e^{\pm}$ (green), and simulated photonic $e^{\pm}$ from photon conversions (blue).  The yield of photonic $e^{\pm}$ is slightly larger for larger values of pseudorapidity.  The distributions for real data are from MB triggered events.  The method used to extract the photonic $e^{\pm}$ distribution from the real data is described in Section~\ref{sec:anintro:back_sub}.  The simulations will be discussed in Chapters~\ref{sec:rec} and~\ref{sec:bre}.}
\label{fig:anintro:eta}
\end{center}
\end{figure}

Cuts are also applied to the number of TPC fit points $(N_{TPCfit}>20)$ and the fit points ratio $(R_{FP}>0.52)$.  The fit points ratio $(R_{FP})$ is the ratio $N_{TPCfit}/N_{TPCmax}$, where $N_{TPCmax}$ is the maximum possible number of TPC fit points.  The value of $N_{TPCmax}$ can vary from track to track, but is usually equal to 44, the number\footnote{The TPC has 45 pad rows~\cite{Anderson2003659}, but pad row 13, the innermost pad row of the outer sub-sectors, was masked out.~\cite{Caines_Private_2011}} of functioning pad rows in the TPC.  Lower values of $N_{TPCmax}$ are possible, such as when a track lies at the edges of a TPC sector.  Figures~\ref{fig:anintro:tfp},~\ref{fig:anintro:mtp}, and~\ref{fig:anintro:fpr} show distributions of $N_{TPCfit}$, $N_{TPCmax}$, and $R_{FP}$ for 200-GeV Cu + Cu collisions; in all cases all other $e^{\pm}$ identification cuts have been applied.

Cuts on track pseudorapidity and the distance of closest approach of a track with the primary collision vertex (this quantity is referred to as the ``global DCA" or $GDCA$) are intended to increase the fraction of non-photonic $e^{\pm}$ in the $e^{\pm}$ sample.  The radiation length of the STAR detector increases with the absolute value of the pseudorapidity for photons originating at the center of the detector.  This is because photons with larger $|\eta|$ will pass through the layers of the Silicon Vertex Tracker (SVT) and other inner structures at larger angles.  In addition, support structures for the SVT begin at $z\approx\pm 30$ cm and extend outward in the direction of the beam line.  Photons with higher $|\eta|$ will have a greater chance of striking these structures and converting to produce background $e^{-}e^{+}$ pairs.  The combination of the cut on the $z$-position of the collision vertex ($|v_{z}|<20$ cm, discussed above) and a cut on track pseudorapidity $(|\eta|<0.7)$ reduces the number of $e^{\pm}$ produced by photon conversions that are included in the sample.  Due to the fact that only one half of the Barrel Electromagnetic Calorimeter (BEMC) was functioning when the data analyzed in this dissertation were recorded, an asymmetric pseudorapidity cut $(-0.1<\eta<0.7)$ is used.  The functioning half of the BEMC covers the half of the STAR detector with $z>0$ in STAR's global coordinate system.  A track with a slightly negative value of pseudorapidity originating from a primary vertex near $v_{z}=+20$ cm can strike the BEMC at a positive $z$ coordinate; this is the reason for the use of -0.1 as the lower bound on $\eta$.  Figure~\ref{fig:anintro:eta} shows the pseudorapidity distributions of $e^{\pm}$ in real MB events and simulated data.  An increase in the photonic $e^{\pm}$ yield for larger values of $|\eta|$ is observed.

A cut on the distance of closest approach between a track and the primary collision vertex ($GDCA<1.5$ cm) is applied in order to preferentially select non-photonic $e^{\pm}$ from the decays of heavy-flavor hadrons.  A $D^{\pm}$ meson with momentum 2 GeV/$c$ $(20\GeV/c)$ produced at the collision vertex has a probability of only $3\times 10^{-20}$ $(1.1\%)$ of decaying farther than 1.5 cm from the vertex (based on mass and lifetime measurements in~\cite{PDG_review}); the other $D$ and $B$ mesons have even lower probabilities of decaying far from the vertex.  In contrast, a $K^{\pm}$ with momentum 2 GeV/$c$ $(20\GeV/c)$ produced at the collision vertex has a probability of $99.9\%$ $(1-10^{-4})$ of decaying farther than 1.5 cm from the vertex.  Photon-conversion $e^{\pm}$ will not be produced in any significant amount within the radius of the beampipe (4 cm).  However, photonic $e^{-}e^{+}$ pairs from $\pi^{0}$ Dalitz decays and the decays of other neutral mesons will all originate from near the collision vertex and will not be excluded by the $GDCA<1.5$ cm cut.  It should be noted that even if an $e^{\pm}$ originates far from the vertex, the helix of its track may still have a DCA of less than 1.5 cm with the vertex, a situation which becomes more common as transverse momentum increases and the track radius of curvature becomes larger.  Figure~\ref{fig:rec:par_eff_gdca} (page~\pageref{fig:rec:par_eff_gdca}) shows a $GDCA$ distribution for simulated non-photonic $e^{\pm}$ and for photonic $e^{\pm}$ from both real and simulation data.  The distribution for non-photonic $e^{\pm}$ is narrower than the distributions for photonic $e^{\pm}$, although all the distributions lie predominantly below $GDCA=1.5$ cm.  The simulations used to create these distributions will be discussed in Chapters~\ref{sec:rec} and~\ref{sec:bre}.

\clearpage

\begin{figure}[htbp]
\begin{center}
\includegraphics[width=0.85\linewidth]{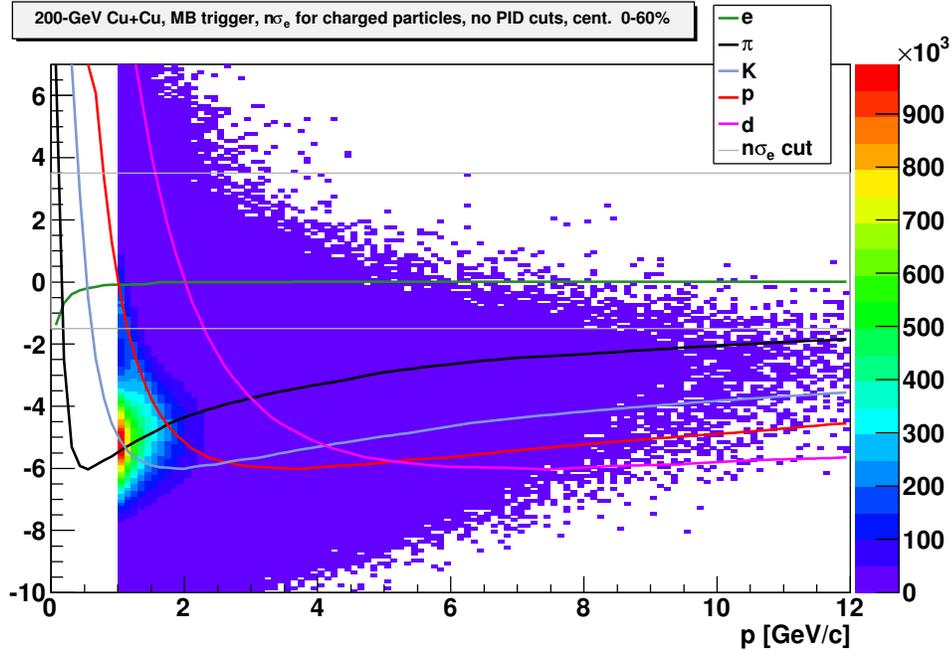}
\caption[Normalized TPC energy loss vs. particle momentum.]{Normalized TPC energy loss $(n\sigma_{e})$ vs. particle momentum for tracks in MB events that have passed the track quality cuts; the $e^{\pm}$ identification cuts using data from the BEMC have not been applied.  The curves indicate the estimated most probable value of $n\sigma_{e}$ for each of five particle species.  The horizontal lines show the range of the $n\sigma_{e}$ cut used in this analysis.}
\label{fig:anintro:nsigma_p_nocuts}
\end{center}
\end{figure}

\begin{figure}[htbp]
\begin{center}
\includegraphics[width=0.8\linewidth]{nsigma_ee_mb_c03_0}
\caption[Normalized TPC energy loss $(n\sigma_{e})$ vs. particle momentum for tracks in MB triggered events that have passed all other $e^{\pm}$ identification cuts.]{Normalized TPC energy loss $(n\sigma_{e})$ vs. particle momentum for tracks in MB triggered events that have passed all other $e^{\pm}$ identification cuts.  The curves indicate the estimated most probable value of $n\sigma_{e}$ for each of five particle species.  The horizontal lines show the range of the $n\sigma_{e}$ cut used in this analysis.}
\label{fig:anintro:nsigma_p_e_mb}
\includegraphics[width=0.8\linewidth]{nsigma_ee_ht_c03_0}
\caption[Normalized TPC energy loss $(n\sigma_{e})$ vs. particle momentum for tracks in HT triggered events that have passed all other $e^{\pm}$ identification cuts.]{Normalized TPC energy loss $(n\sigma_{e})$ vs. particle momentum for tracks in HT triggered events that have passed all other $e^{\pm}$ identification cuts.}
\label{fig:anintro:nsigma_p_e_ht}
\end{center}
\end{figure}

Several quantities measured using the TPC and BEMC are used to distinguish $e^{\pm}$ from hadrons.  The single most important cut used to identify $e^{\pm}$ is a cut on particle energy loss in the TPC.  Figure~\ref{fig:anintro:nsigma_p_nocuts} shows $n\sigma_{e}$, the normalized TPC energy loss, versus momentum for tracks that pass the track quality cuts; the $e^{\pm}$ identification cuts using data from the BEMC have not been applied.  Figures~\ref{fig:anintro:nsigma_p_e_mb} and~\ref{fig:anintro:nsigma_p_e_ht} show $n\sigma_{e}$ versus momentum for tracks that pass all other $e^{\pm}$ identification cuts for MB and HT triggered events.  See Section~\ref{sec:experiment:tpc:dedx} for a more detailed discussion of $n\sigma_{e}$.  The curves in Figures~\ref{fig:anintro:nsigma_p_nocuts},~\ref{fig:anintro:nsigma_p_e_mb}, and~\ref{fig:anintro:nsigma_p_e_ht} show the estimated\footnote{See Appendix~\ref{sec:dedx2nsigma} for a description of the calculation of those curves.} most probable values of $n\sigma_{e}$ for five different particle species.  For $e^{\pm}$, the most probable values of $n\sigma_{e}$ are around 0.  For this analysis, a cut on the normalized TPC energy loss of $-1.5<n\sigma_{e}<3.5$ is applied to exclude particle species other than $e^{\pm}$.  This cut on $n\sigma_{e}$ approximately corresponds to a cut of 3.35 keV/cm $<\langle dE/dx\rangle <$ 5 keV/cm, where $\langle dE/dx\rangle$ is the $70\%$ truncated mean value of the TPC energy loss (see Appendix~\ref{sec:dedx2nsigma}).  Figures~\ref{fig:anintro:nsigma_cuts_mb} and~\ref{fig:anintro:nsigma_cuts_ht} show $n\sigma_{e}$ distributions in single transverse-momentum bins for MB and HT triggered events with various combinations of $e^{\pm}$ selection cuts applied (the track-quality cuts have been applied in all cases).  The BSMD cuts are observed to be more effective at removing hadron contamination than the $p/E_{Tower}$ cut.  When the HT trigger condition is applied, the $p/E_{Tower}$ is largely redundant.

For momenta less than 1.5 GeV/$c$, it becomes difficult for the TPC energy-loss cut to distinguish between $e^{\pm}$ and charged hadrons.  For this reason, and due to problems in the simulation data for photonic $e^{\pm}$ (data which are used in calculating the background rejection efficiency $\varepsilon_{B}$), this analysis is restricted to particles with $p_{T}>2\GeV/c$.


\begin{figure}[htbp]
\begin{center}
\includegraphics[width=0.75\linewidth]{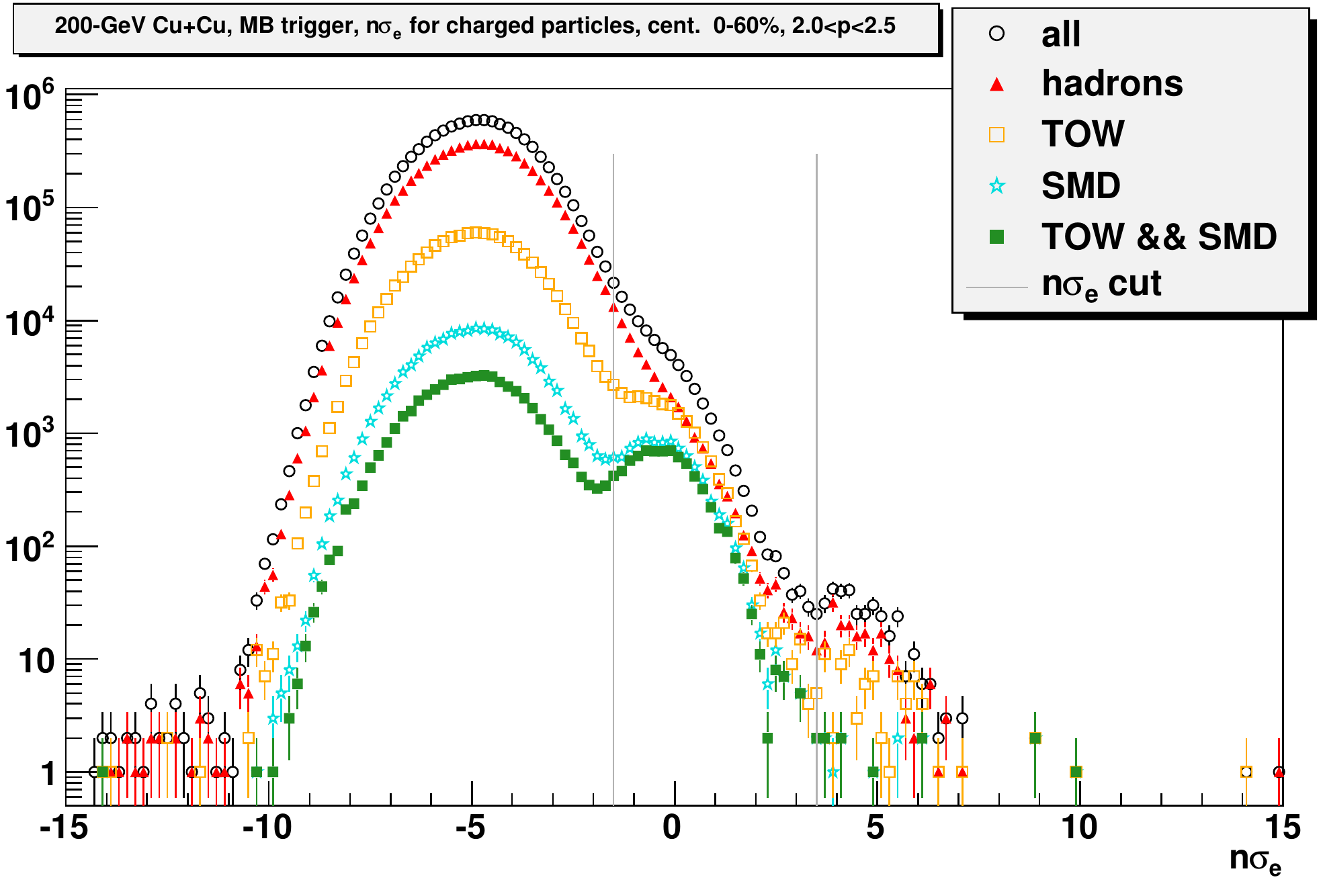}
\caption[Normalized TPC energy loss distributions in MB triggered events with various sets of cuts applied.]{Normalized TPC energy loss $(n\sigma_{e})$ for particles with momenta in the range $2\GeV/c<p<2.5\GeV/c$ in MB triggered events.  The effects of various sets of cuts on the $n\sigma_{e}$ distribution are shown (the track-quality cuts have been applied in all cases).  The black histogram shows the $n\sigma_{e}$ distribution for all particles (no $e^{\pm}$ selection cuts), while the red histogram shows the $n\sigma_{e}$ distribution for hadrons (identified by requiring that $p/E_{Tower}>2.5$ and that $N_{SMD\eta}=1$ or $N_{SMD\phi}=1$).  Also shown are the $n\sigma_{e}$ distributions for particles that pass the $p/E_{Tower}<2$ cut (orange), particles identified as $e^{\pm}$ using the BSMD (cyan), and particles identified as $e^{\pm}$ using all cuts (green).  These BEMC-related cuts are described in the text.  The BSMD cuts are observed to be more effective at removing hadron contamination than the $p/E_{Tower}$ cut.}
\label{fig:anintro:nsigma_cuts_mb}
\end{center}
\end{figure}

\begin{figure}[htbp]
\begin{center}
\includegraphics[width=0.75\linewidth]{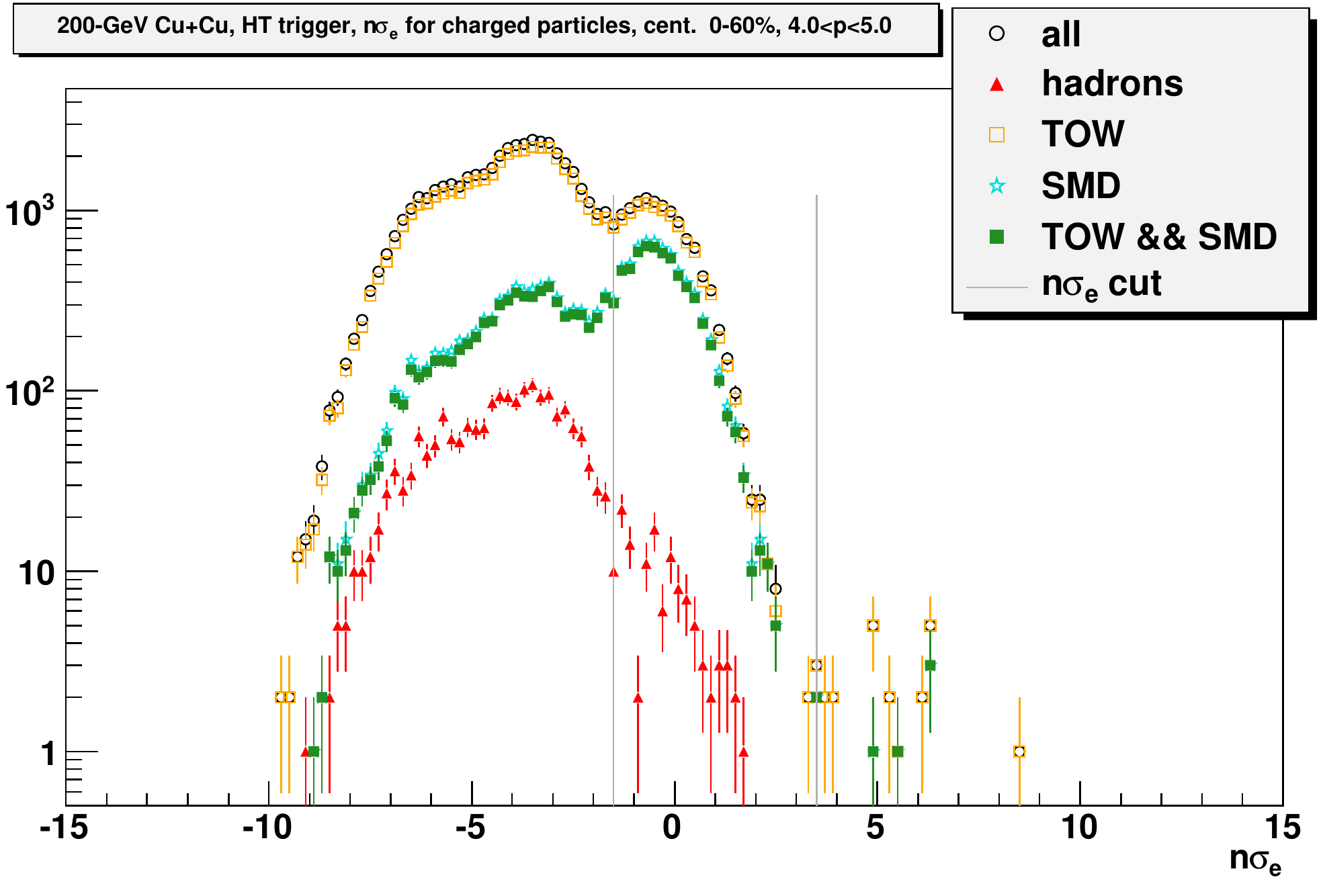}
\caption[Normalized TPC energy loss distributions in HT triggered events with various sets of cuts applied.]{Normalized TPC energy loss $(n\sigma_{e})$ for particles with momenta in the range $4\GeV/c<p<5\GeV/c$ in HT triggered events.  See Figure~\ref{fig:anintro:nsigma_cuts_mb} (caption) for a description of each histogram.  All tracks are required to point to a tower that satisfies the HT trigger condition, which renders the $p/E_{Tower}$ cut redundant.}
\label{fig:anintro:nsigma_cuts_ht}
\end{center}
\end{figure}

\clearpage


\begin{figure}[htbp]
\begin{center}
\includegraphics[width=0.8\linewidth]{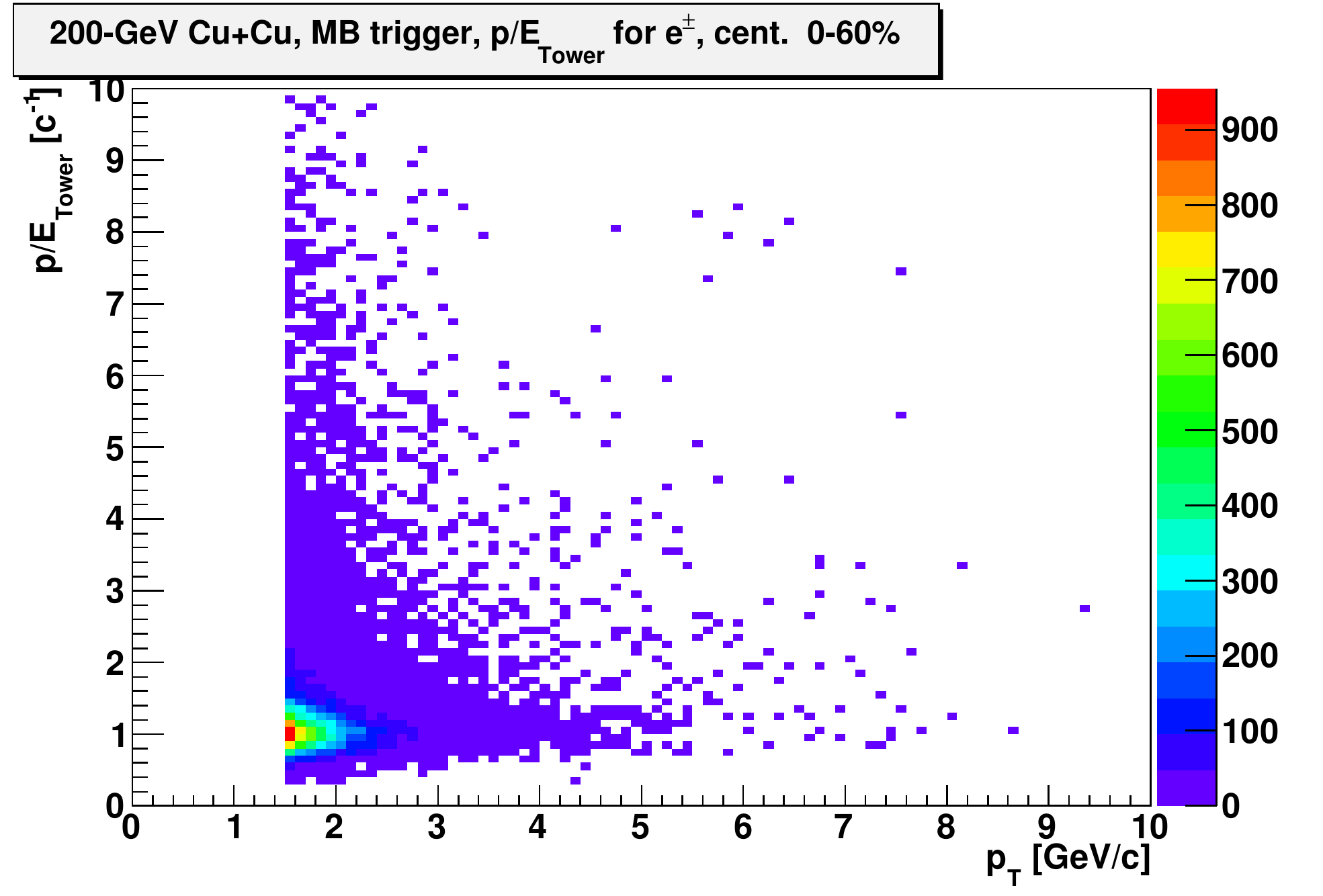}
\caption[The ratio $p/E_{Tower}$ vs. $p_{T}$ for $e^{\pm}$.]{The ratio $p/E_{Tower}$ vs. $p_{T}$ for tracks in MB triggered events that pass all other $e^{\pm}$ identification cuts.  Most $e^{\pm}$ have $p/E_{Tower}\approx 1/c$.}
\label{fig:anintro:poe_e_mb}
\includegraphics[width=0.8\linewidth]{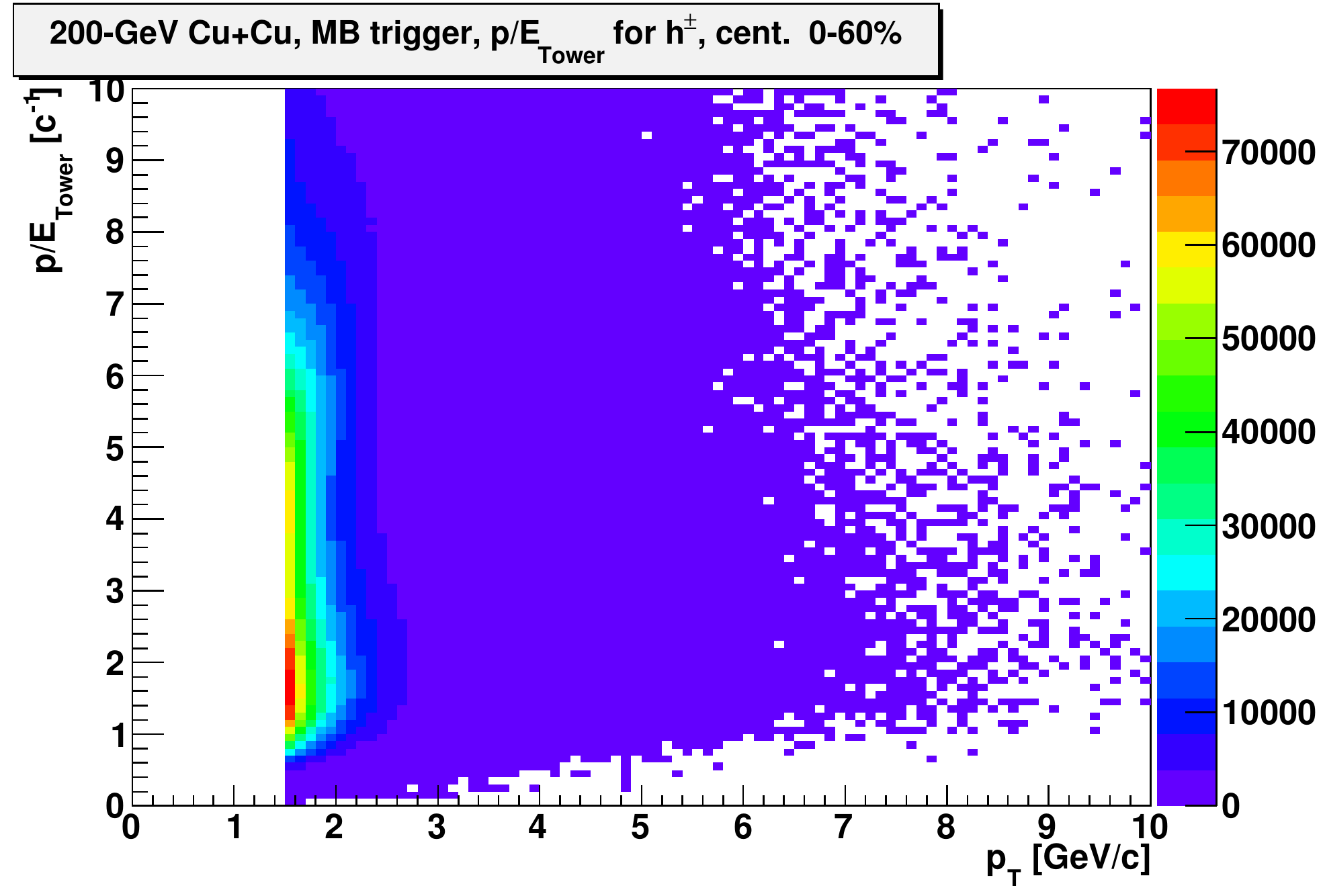}
\caption[The ratio $p/E_{Tower}$ vs. $p_{T}$ for charged hadrons.]{The ratio $p/E_{Tower}$ vs. $p_{T}$ for tracks in MB triggered events that are identified as hadrons (requiring that $n\sigma_{e}<-1.5$ and without BSMD cuts).  Hadrons have a much broader $p/E_{Tower}$ distribution than do $e^{\pm}$.}
\label{fig:anintro:poe_h_mb}
\end{center}
\end{figure}

\begin{figure}[htbp]
\begin{center}
\includegraphics[width=0.85\linewidth]{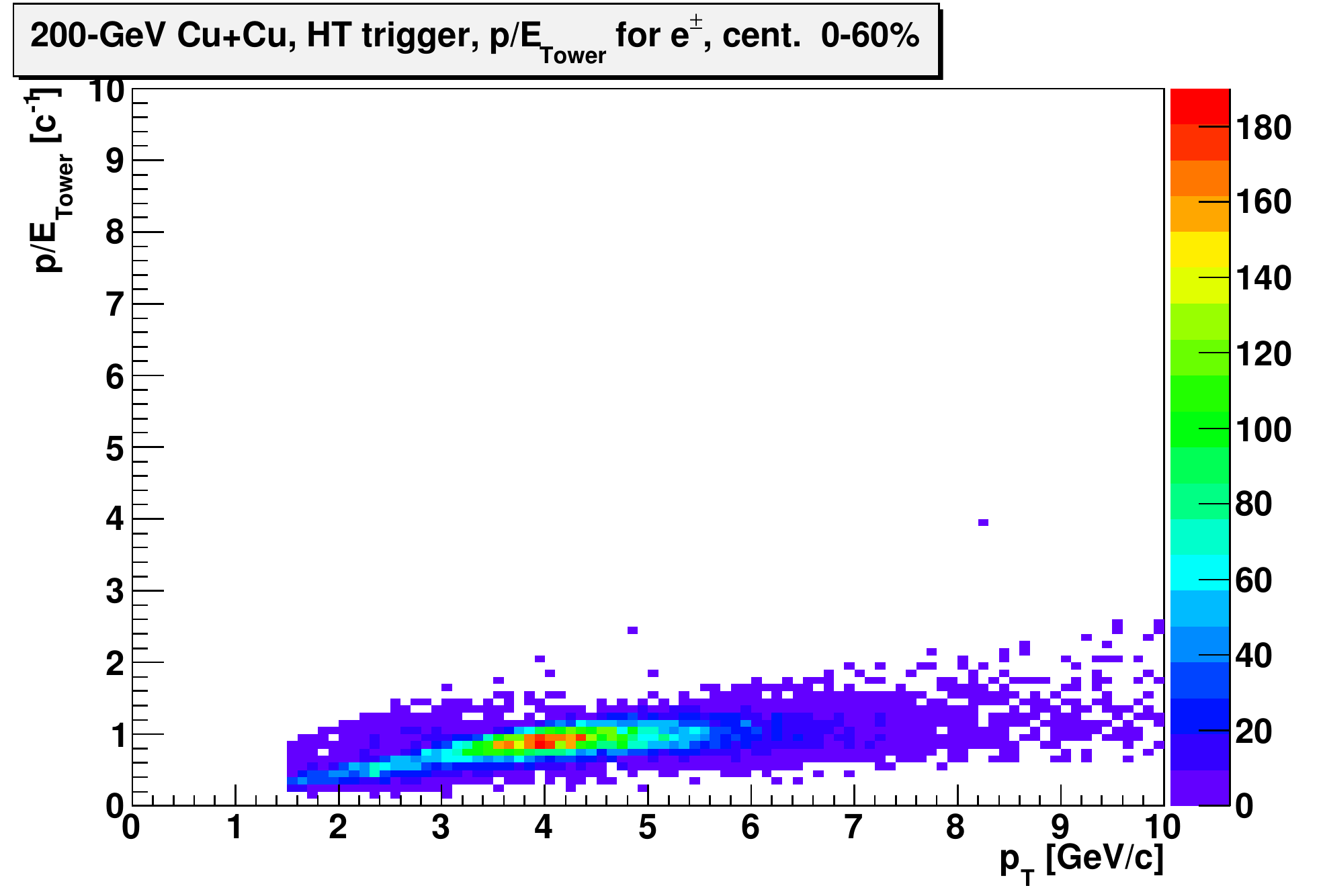}
\caption[Distributions of $p/E_{Tower}$ vs. $p_{T}$ for $e^{\pm}$ in HT triggered events.]{The ratio $p/E_{Tower}$ vs. $p_{T}$ for tracks in HT triggered events that pass all other $e^{\pm}$ identification cuts.  Since all tracks were required to point to a tower that satisfied the HT trigger condition, the $p/E_{Tower}$ distribution is heavily biased and the $p/E_{Tower}<2/c$ cut is largely redundant.}
\label{fig:anintro:poe_e_ht}
\end{center}
\end{figure}

\begin{figure}[htbp]
\begin{center}
\includegraphics[width=0.85\linewidth]{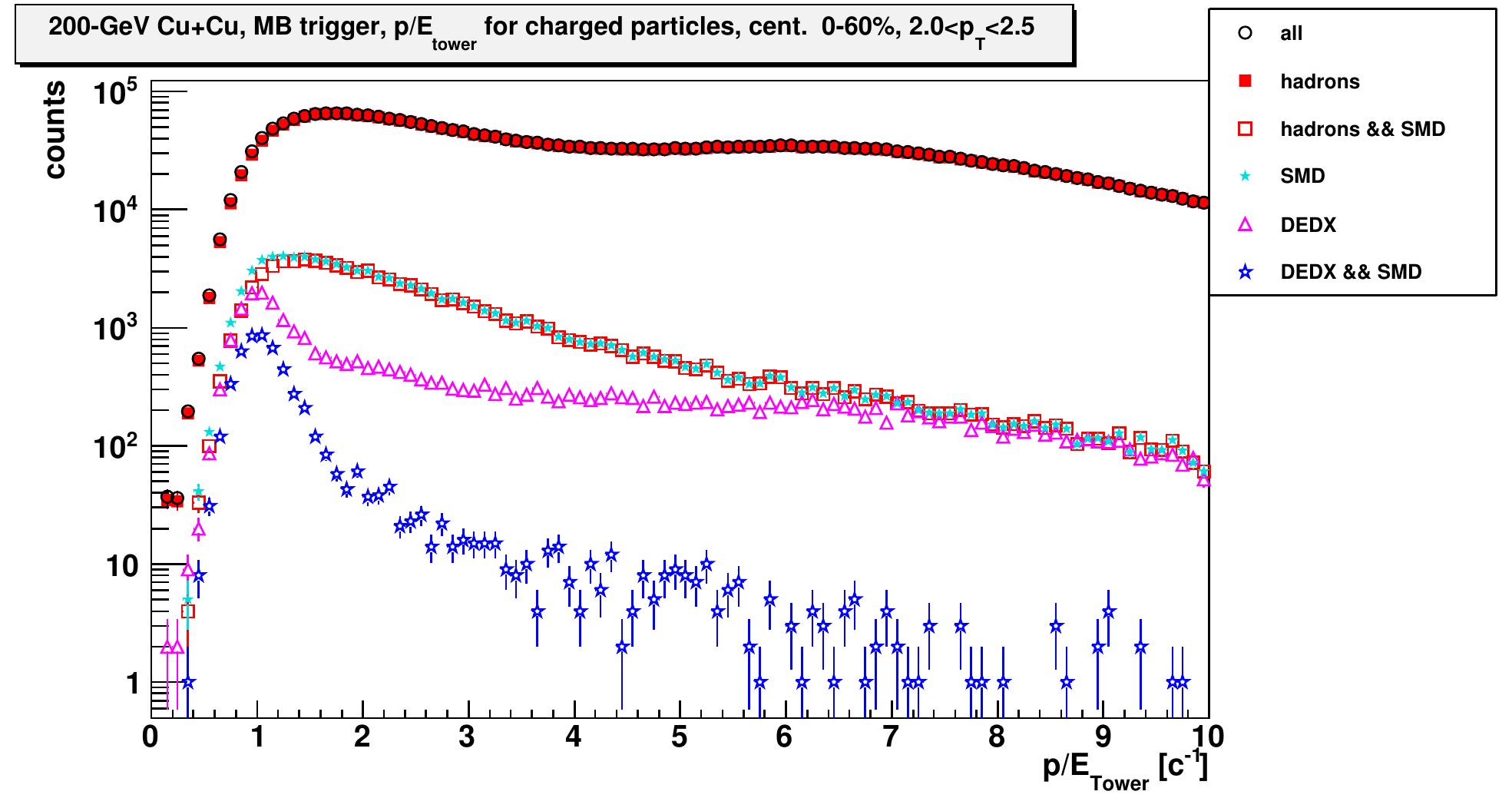}
\caption[Distributions of $p/E_{Tower}$ with various sets of cuts applied.]{Distributions of $p/E_{Tower}$ for particles with transverse momenta in the range $2\GeV/c<p_{T}<2.5\GeV/c$ in MB triggered events.  The effects of various sets of cuts on the $p/E_{Tower}$ distribution are shown (the track-quality cuts have been applied in all cases).  The black histogram shows the $p/E_{Tower}$ distribution for all particles (no $e^{\pm}$ selection cuts).  The solid red histogram shows the $p/E_{Tower}$ distribution for hadrons identified by requiring that $n\sigma_{e}<-1.5$ (and without BSMD cuts), while the open red histogram shows the $p/E_{Tower}$ distribution for hadrons that have passed the $e^{\pm}$ identification cuts in the BSMD.  Also shown are the $p/E_{Tower}$ distributions for particles that pass the TPC energy-loss cut (magenta), particles identified as $e^{\pm}$ using the BSMD (cyan), and particles identified as $e^{\pm}$ using all cuts (blue).  The BSMD cuts are described in the text.  The TPC energy-loss cut is observed to be more effective at removing hadron contamination than the BSMD cuts.}
\label{fig:anintro:poe_cuts}
\end{center}
\end{figure}

The STAR Barrel Electromagnetic Calorimeter is also used to distinguish $e^{\pm}$ from hadrons.  The BEMC is approximately 20 radiation lengths deep, the depth required to contain approximately 95\% of the energy of an electromagnetic shower with energy $<20$ GeV (see Section~\ref{sec:experiment:bemc}).  In contrast, the BEMC is $\approx 0.6$ nuclear interaction lengths deep, much less than the depth (several interaction lengths) needed to contain a hadronic shower with energy $<20$ GeV.  Therefore, $e^{\pm}$ will deposit, on average, more energy in the calorimeter than hadrons with the same momentum.

In this analysis, a cut of $p/E_{Tower}<2/c$ is used, where $p$ is the momentum of a track in the TPC and $E_{Tower}$ is the energy measured in the BEMC tower to which that track projects.  An additional requirement that $E_{Tower}>0$ is implied whenever the $p/E_{Tower}$ cut is mentioned.  It is also required that the tower to which the track projects has ``good" status: \textit{i.e.}, it has been determined that the tower was operating properly (see Section~\ref{sec:acc:bemc}).

Figure~\ref{fig:anintro:poe_e_mb} shows the ratio $p/E_{Tower}$ versus transverse momentum for $e^{\pm}$ candidates; these particles are mostly $e^{\pm}$, with some (a few percent) hadron contamination.  A hadron that appears as contamination in this distribution must have had both large TPC energy loss and also large electromagnetic component early in the development of its shower, leading to an increased likelihood that it would pass the BSMD cuts (see below).  Figure~\ref{fig:anintro:poe_h_mb} shows the ratio $p/E_{Tower}$ versus $p_{T}$ for particles identified as hadrons by requiring that $n\sigma_{e}<-1.5$.

An $e^{\pm}$ may also fail the $p/E_{Tower}$ cut if it struck the calorimeter near the edge or corner of a tower and its energy was deposited in multiple towers.  This would be expected to skew the $p/E_{Tower}$ distribution, producing a ``tail" at large values of $p/E_{Tower}$.  Transverse to an electromagnetic shower's direction of propagation, 95\% of the shower's energy is expected to be contained within a cylinder of diameter $4\rho_{M}$,~\cite{Fernow} where $\rho_{M}$ is the Moli\`ere radius (1.602 cm for lead)~\cite{PDG_review}.  The towers in the BEMC are approximately 12 cm $\times$ 12 cm (at $\eta=0$) while the expected shower diameter is about half of those dimensions, 6.4 cm.  Most electromagnetic showers will therefore be contained within a single tower.  Any losses in the efficiency of the $E_{Tower}/p$ cut due to the energy of a shower being shared between multiple towers are accounted for in the $e^{\pm}$ reconstruction efficiency ($\varepsilon_{R}$, see Chapter~\ref{sec:rec}).

Figure~\ref{fig:anintro:poe_h_mb} indicates that many hadrons pass the $p/E_{Tower}<2/c$ cut.  Those hadrons likely produced neutral pions in collisions near the beginning of their showers, leading to large electromagnetic components and more of their energy being deposited in the calorimeter.~\cite{Fernow}  In the high-tower triggered data set (but not the minimum-bias data set), the EMC tower to which a track projects is required to have satisfied the high-tower trigger condition.  Figure~\ref{fig:anintro:poe_e_ht} shows the distribution of $p/E_{Tower}$ versus transverse momentum for $e^{\pm}$ in HT triggered events.  The distribution is biased towards low values of $p/E_{Tower}$ and the cut on that ratio has little effect.  Figure~\ref{fig:anintro:poe_cuts} shows $p/E_{Tower}$ distributions in a single transverse-momentum bin for MB triggered events with various combinations of $e^{\pm}$ selection cuts applied (the track-quality cuts have been applied in all cases).

\clearpage

\begin{figure}[htbp]
\begin{center}
\includegraphics[width=0.85\linewidth]{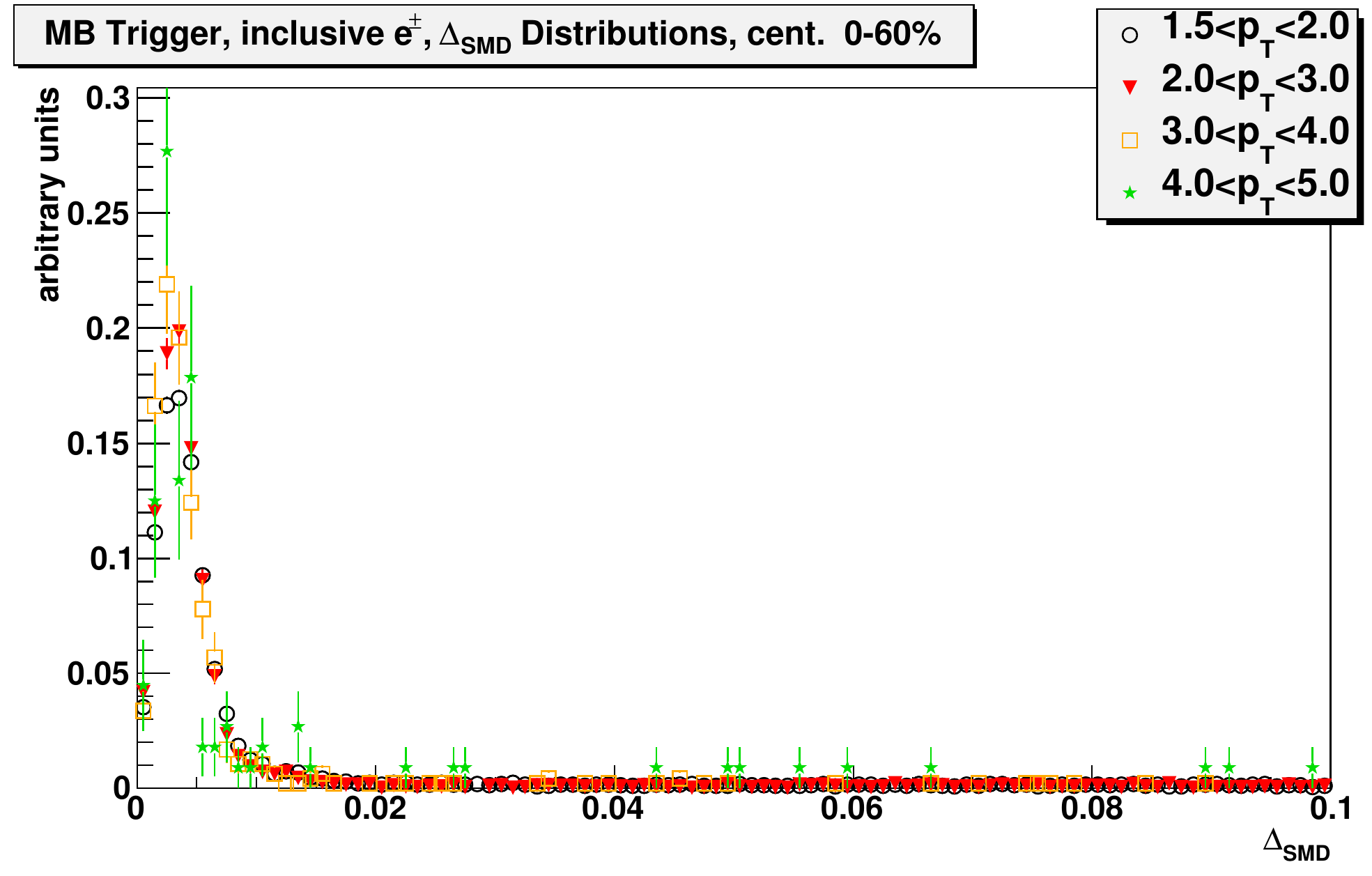}
\caption[Distributions of the BSMD cluster displacement $(\Delta_{SMD})$.]{The BSMD cluster displacement $\Delta_{SMD}$ for tracks that pass the other $e^{\pm}$ identification cuts.  The peaks of the distributions are below $\Delta_{SMD}=0.02$.  The distributions shown are for MB triggered events in several $p_{T}$ bins; the distributions are similar for HT triggered events.}
\label{fig:anintro:dsmd}
\end{center}
\end{figure}

\begin{figure}[htbp]
\begin{center}
\includegraphics[width=0.47\linewidth]{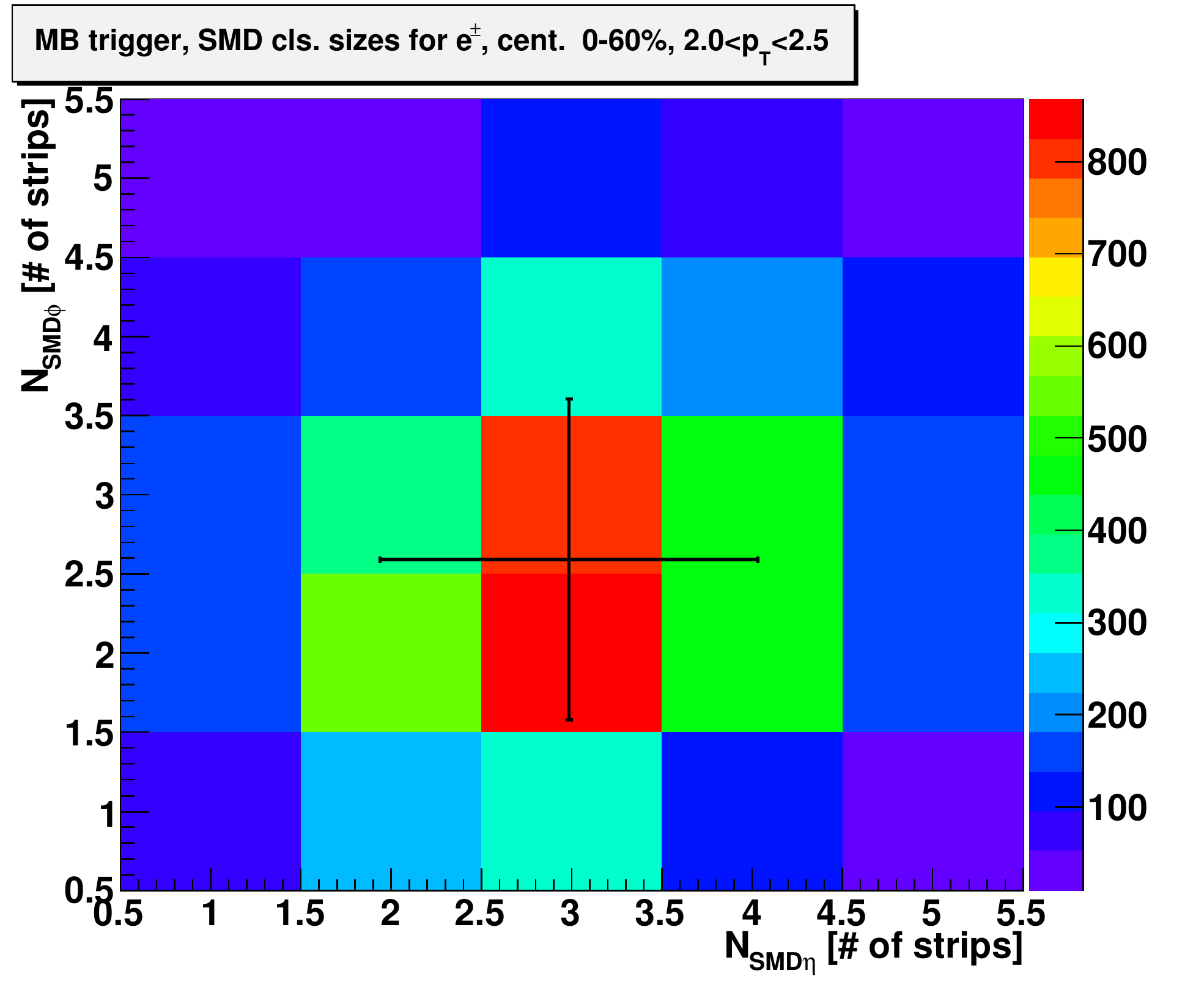}
\includegraphics[width=0.47\linewidth]{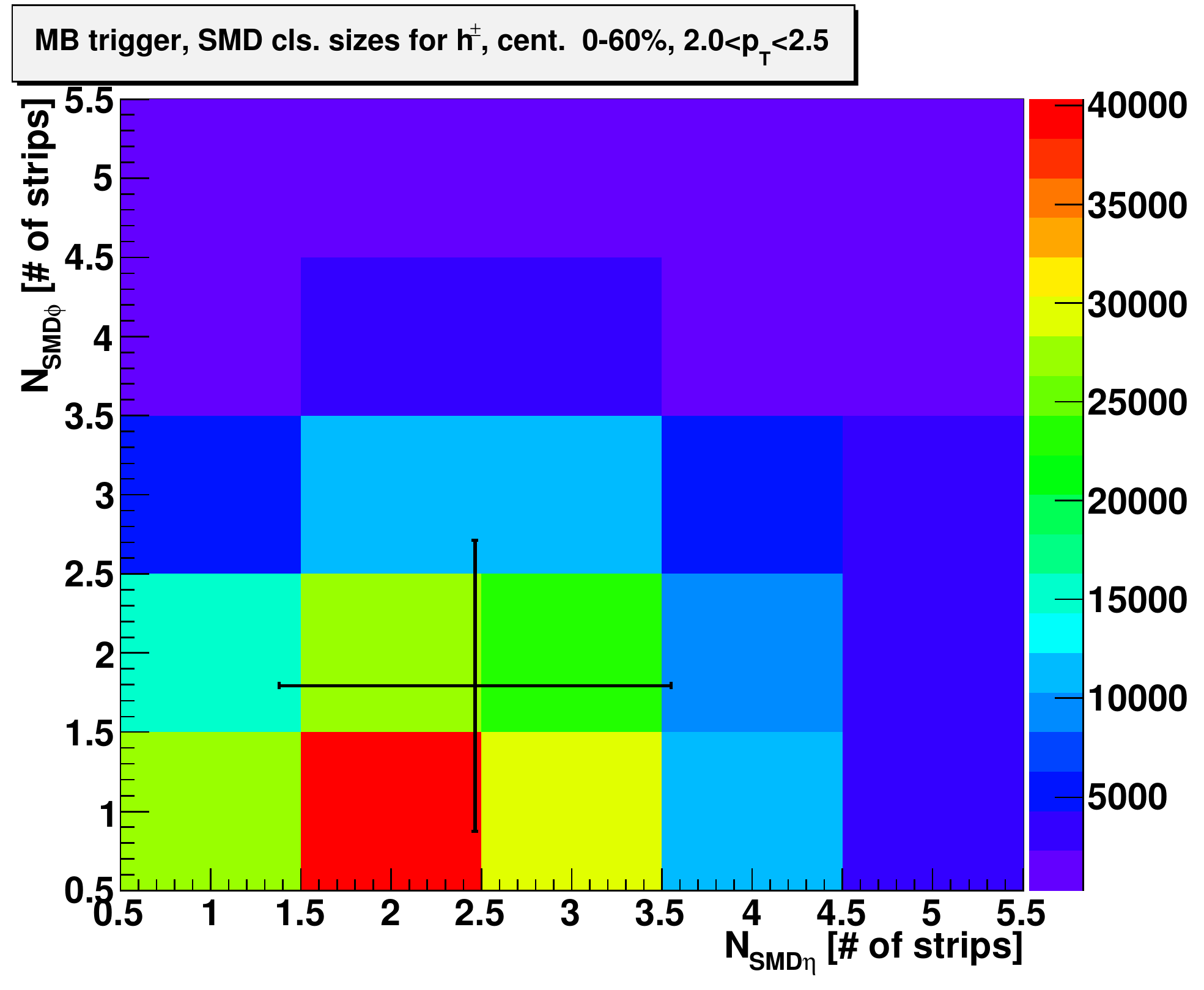}
\caption[Distributions of cluster sizes in the two planes of the BSMD.]{Cluster sizes in the $\eta$ and $\phi$ planes of the BSMD for tracks with transverse momenta in the range $2\GeV/c<p_{T}<2.5\GeV/c$ in MB events.  The center of the each black cross indicates the mean BSMD cluster sizes in each plane, while the bars indicate the standard deviation in the BSMD cluster sizes.  The left plot shows $N_{SMD\phi}$ vs. $N_{SMD\eta}$ for tracks that satisfy all other $e^{\pm}$ identification cuts.  The right plot shows $N_{SMD\phi}$ vs. $N_{SMD\eta}$ for tracks identified as hadrons by requiring that $n\sigma_{e}<-1.5$.}
\label{fig:anintro:smd_sizes_X}
\end{center}
\end{figure}



\begin{figure}[htbp]
\begin{center}
\includegraphics[width=0.85\linewidth]{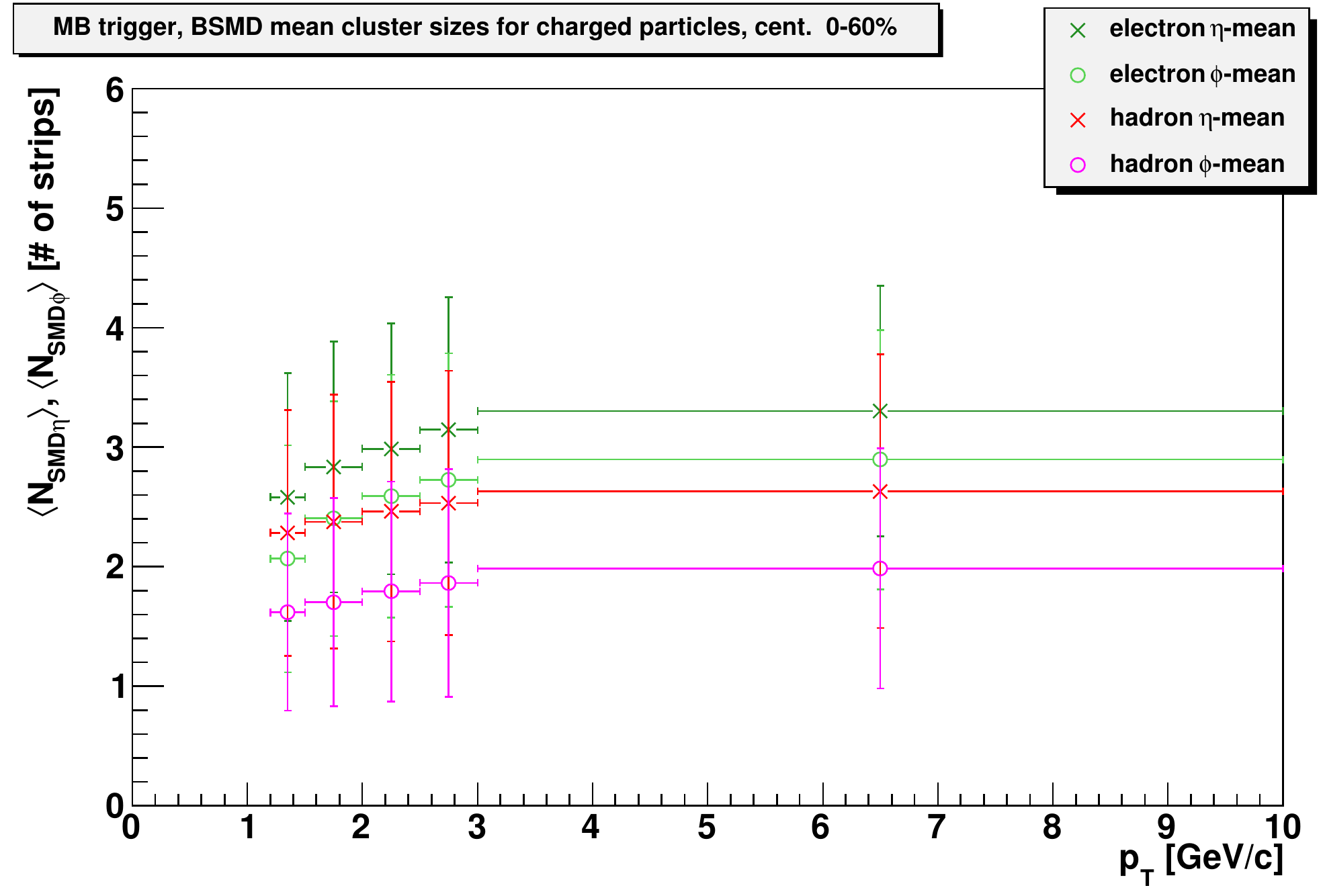}
\caption[Mean cluster sizes in the $\eta$ and $\phi$ planes of the BSMD versus $p_{T}$.]{Mean cluster sizes in the $\eta$ and $\phi$ planes of the BSMD versus transverse momentum for tracks in MB events.  The dark (light) green histogram indicates $\langle N_{SMD\eta}\rangle$ $(\langle N_{SMD\phi}\rangle)$ for tracks that pass all other $e^{\pm}$ identification cuts.  The red (magenta) histogram indicates $\langle N_{SMD\eta}\rangle$ $(\langle N_{SMD\phi}\rangle)$ for tracks identified as hadrons by requiring that $n\sigma_{e}<-1.5$.  The vertical bars indicate the standard deviations of $N_{SMD\eta}$ and $N_{SMD\phi}$.  It is observed that $e^{\pm}$ have higher mean cluster sizes than hadrons and that $\langle N_{SMD\eta}\rangle$ is consistently larger than $\langle N_{SMD\phi}\rangle$ for a given particle type.}
\label{fig:anintro:mean_smd_sizes}
\end{center}
\end{figure}

The STAR Barrel Shower Maximum Detector (BSMD, see Section~\ref{sec:experiment:bemc:bsmd}) is used to measure the shape and size of electromagnetic showers with finer resolution than the sizes of the towers.  The BSMD sits 5-6 radiation lengths inside the calorimeter, near the maximum density for electromagnetic shower with energies greater than about 1-2 GeV~\cite{Beddo2003725}; for a 2-GeV shower one parametrization~\cite{PDG_review} predicts a maximum at $[\ln(E/7.43\MeV\;)-0.5]X_{0}=5.1X_{0}$ or 2.9 cm of lead.  Hadronic showers are expected to develop more slowly on average, reaching a maximum around one nuclear interaction length inside a calorimeter; for a 2-GeV shower, one parametrization~\cite{FabjanCalorimetry} predicts a maximum at $[0.2\ln (E/1\GeV\;)+0.7]\lambda_{I}=0.83\lambda_{I}=26X_{0}$ in lead.  At the depth of the BSMD, many hadronic showers will not be fully developed.

The STAR collaboration uses an algorithm to find groups of adjacent strips (called clusters) for which signals were recorded.  Each track in this analysis is projected to the location of the BSMD and the nearest clusters in the $\eta$ and $\phi$ planes of the BSMD are found.  A cut is applied on the distance between the track projection and the two clusters.  The BSMD cluster displacement $\Delta_{SMD}$ is defined as

\begin{equation}
\Delta_{SMD}=\sqrt{\left(\Delta\eta_{SMD}\right)^{2}+\left(\Delta\phi_{SMD}\right)^{2}},
\end{equation}

\noindent where $\Delta\eta_{SMD}$ is the difference in pseudorapidity between the track projection and the centroid of the nearest cluster in the BSMD-$\eta$ plane, and $\Delta\phi_{SMD}$ is the difference (in radians) of the azimuth between the track projection and the centroid of the nearest cluster in the BSMD-$\phi$ plane.  It is required that $\Delta_{SMD}<0.02$, equivalent to a distance of $\approx 4.6$ cm.  Figure~\ref{fig:anintro:dsmd} shows distributions of $\Delta_{SMD}$ for particles in minimum-bias events that have passed all other $e^{\pm}$ identification cuts.  Most of these distributions lie well below $\Delta_{SMD}=0.02$.  It is also required that the centroid of each cluster lies in a strip that had ``good" status (was determined to have been working properly).

A cut on the sizes of the nearest cluster in each plane of the BSMD is applied.  The size of a cluster is measured by the number of strips in which a signal was recorded.  It is required that $N_{SMD\eta}\geq 2$ and $N_{SMD\phi}\geq 2$, where $N_{SMD\eta}$ $(N_{SMD\phi})$ is the size of the nearest cluster in the $\eta$ $(\phi)$ plane of the BSMD.  Figure~\ref{fig:anintro:smd_sizes_X} shows $N_{SMD\phi}$ vs. $N_{SMD\eta}$ for particles in the transverse-momentum range $2\GeV/c<p_{T}<2.5\GeV/c$ for $e^{\pm}$ (particles that have passed the other $e^{\pm}$ identification cuts) and for hadrons (identified by requiring that $n\sigma_{e}<-1.5$).  Figure~\ref{fig:anintro:mean_smd_sizes} shows the mean values of $N_{SMD\eta}$ and $N_{SMD\phi}$ for $e^{\pm}$ and hadrons as functions of $p_{T}$.  The $e^{\pm}$ are observed to have larger values of $\langle N_{SMD\eta}\rangle$ and $\langle N_{SMD\phi}\rangle$.  A difference in the performance of the two planes of the BSMD is also observed: $\langle N_{SMD\eta}\rangle$ is observed to be larger (0.3-0.4 strips) than $\langle N_{SMD\phi}\rangle$.  This is due to the loss of the low-energy component of showers in the inner $(\eta)$ plane of the BSMD, leading to reduced spatial resolution and lower energy deposition in the outer $(\phi)$ plane.~\cite{Akimenko199592}

\section[Photonic $e^{\pm}$ Background Subtraction]{Photonic $\boldsymbol{e^{\pm}}$ Background Subtraction}
\label{sec:anintro:back_sub}

Once $e^{\pm}$ have been identified, it is necessary to identify the large photonic $e^{\pm}$ background, which will be subtracted from the inclusive $e^{\pm}$ yield to give the non-photonic $e^{\pm}$ signal.  Photonic $e^{\pm}$ are identified by reconstructing low-invariant-mass $e^{-}e^{+}$ pairs.  Photon conversions tend to produce $e^{-}e^{+}$ pairs with invariant mass near 0, and the pairs produced in neutral-pion Dalitz decays $(\pi^{0}\rightarrow e^{-}e^{+}\gamma)$ have invariant masses less than the $\pi^{0}$ mass.  In this analysis, a cut on the pair invariant mass of $M_{inv.}(e^{-}e^{+})<150\MeV/c^{2}$, slightly above the $\pi^{0}$ mass, is applied.  Other sources (including $\eta$ and $\eta^{\prime}$ meson Dalitz decays) may also produce $e^{-}e^{+}$ pairs with $M_{inv.}(e^{-}e^{+})<150\MeV/c^{2}$.  Photonic $e^{-}e^{+}$ pairs with invariant mass greater than the $150\MeV/c^{2}$ cut are not identified using this method; the yield of photonic $e^{\pm}$ from those sources (``residual photonic $e^{\pm}$") will be estimated in Chapter~\ref{sec:res_back} and subtracted from the efficiency-corrected non-photonic $e^{\pm}$ yield.

\begin{figure}[htbp]
\begin{center}
\includegraphics[width=0.85\linewidth]{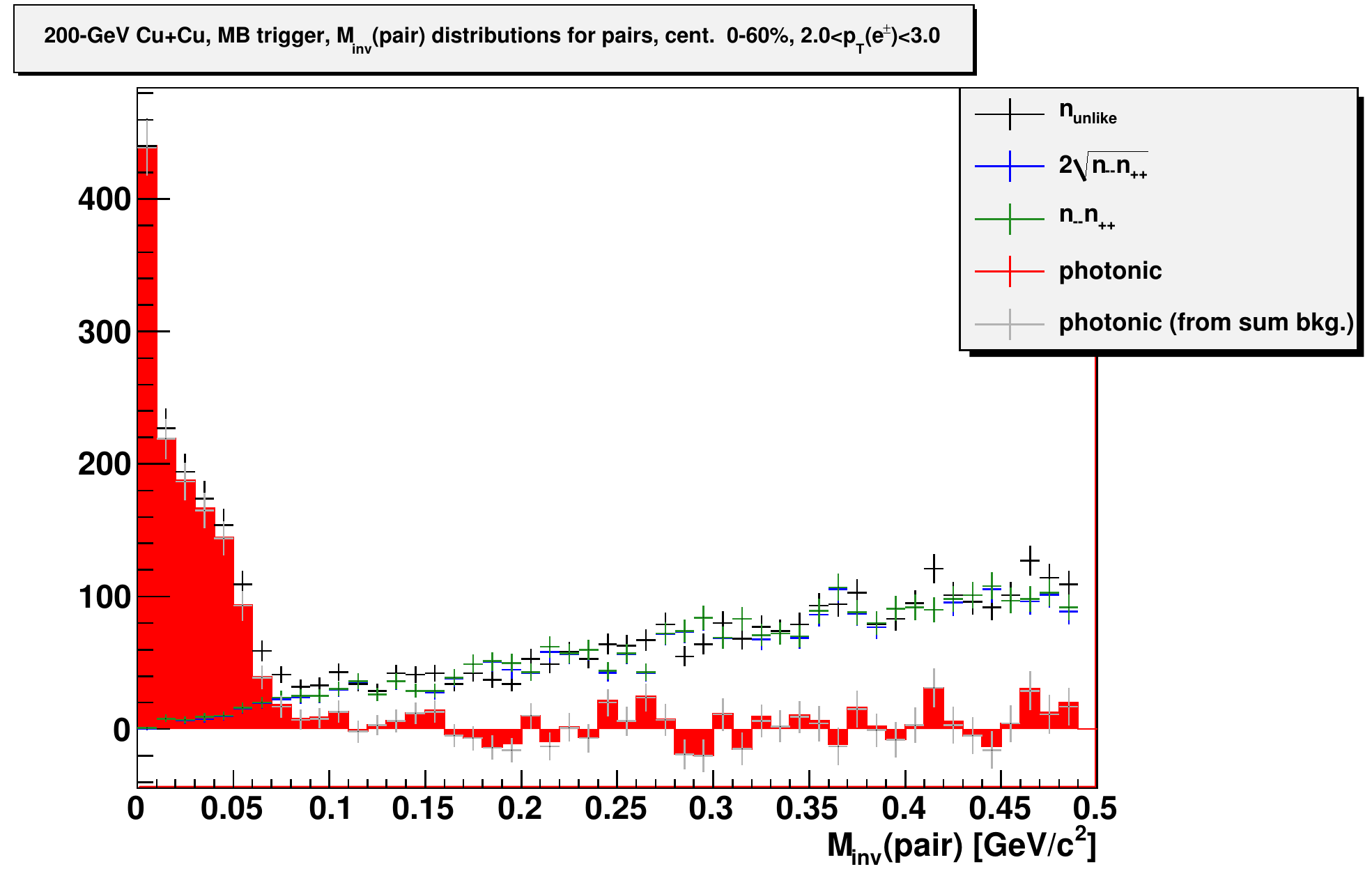}
\caption[Distributions of invariant mass for pairs of $e^{\pm}$ candidates and partners in MB triggered events.]{Distributions of invariant mass for pairs of $e^{\pm}$ candidates and partners in MB triggered events.  The transverse momenta of the $e^{\pm}$ candidates are in the range $2\GeV/c<p_{T}<3\GeV/c$; all other $e^{\pm}$ identification, pair, and partner selection cuts have been applied.  In black is the distribution for unlike-charge pairs $(n_{unlike})$.  The combinatorial background calculated as $2\sqrt{n_{--}n_{++}}$ (see Equation~\ref{eq:anintro:back_sub}) is shown in blue, while an alternate calculation of the combinatorial background $(n_{--}+n_{++}$) is shown in green.  The red and gray histograms are the invariant-mass distributions for photonic $e^{-}e^{+}$ pairs: the combinatorial background subtracted from the number of unlike-charge pairs.  The invariant mass distribution for photonic $e^{-}e^{+}$ pairs is peaked near 0.  The alternate method of calculating the combinatorial background gives essentially the same result as Equation~\ref{eq:anintro:back_sub}.}
\label{fig:anintro:invmass}
\end{center}
\end{figure}

\begin{figure}[htbp]
\begin{center}
\includegraphics[width=0.85\linewidth]{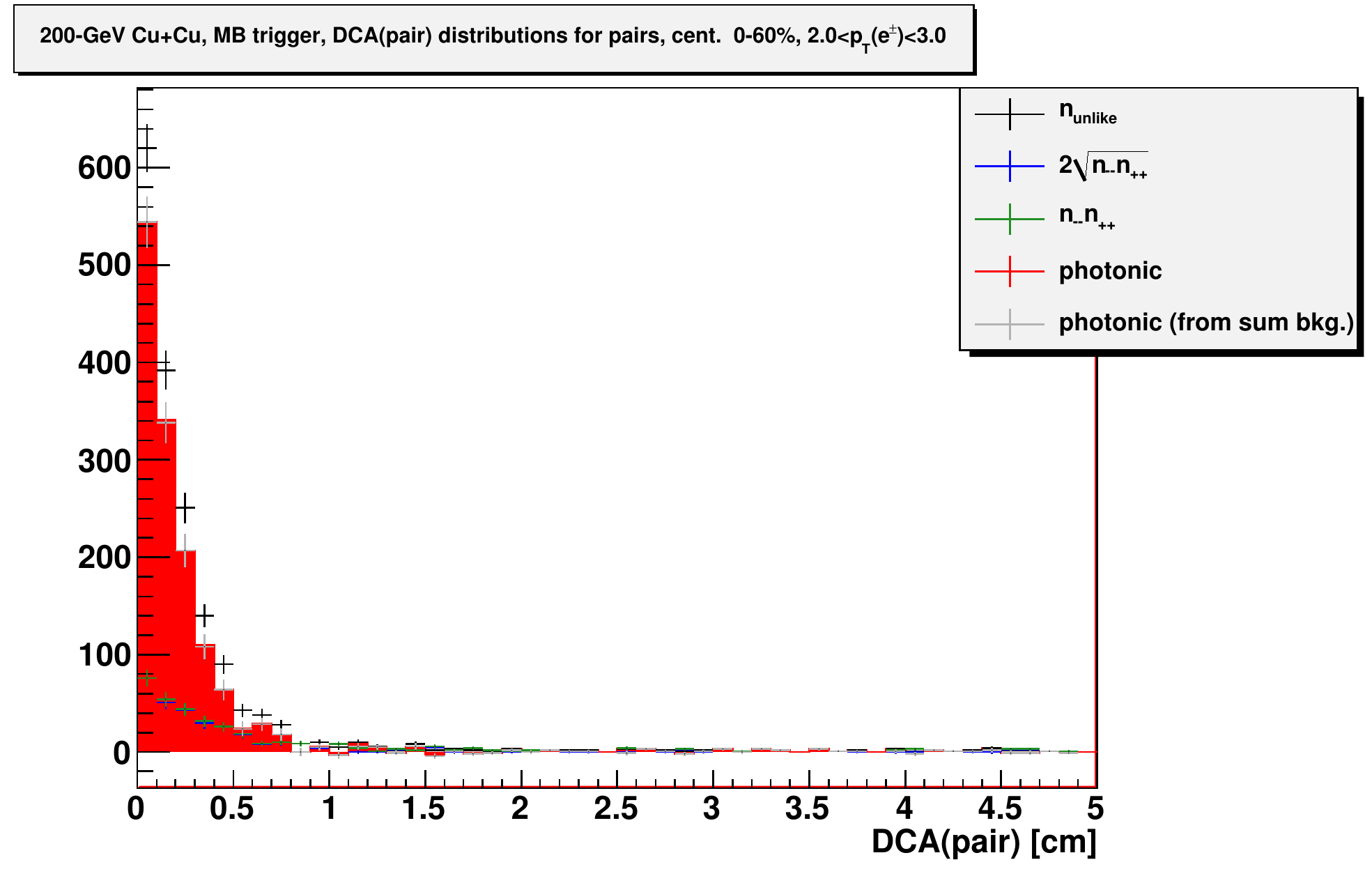}
\caption[Distributions of the distance of closest approach for pairs of $e^{\pm}$ candidates and partners in MB triggered events.]{Distributions of the distance of closest approach for pairs of $e^{\pm}$ candidates and partners in MB triggered events.  The transverse momenta of the $e^{\pm}$ candidates are in the range $2\GeV/c<p_{T}<3\GeV/c$; all other $e^{\pm}$ identification, pair, and partner selection cuts have been applied.  See the caption of Figure~\ref{fig:anintro:invmass} for a definition of each histogram.  Photonic $e^{-}e^{+}$ pairs have pair DCA distributions peaked near 0, although most random pairs also have DCA near 0.}
\label{fig:anintro:pair_dca}
\end{center}
\end{figure}

The number of photonic $e^{-}e^{+}$ pairs is found as follows.  Each particle that has passed the $e^{\pm}$ identification cuts described in the previous section (called an ``$e^{\pm}$ candidate") is paired with ``partner" tracks from the same event.  The number of pairs of oppositely charged tracks for which $M_{inv.}(pair)<150\MeV/c^{2}$ and $DCA(pair)<1.5$ cm is recorded.  However, it is possible that a random combination of an $e^{\pm}$ candidate with a partner track could satisfy these conditions.  The combinatorial background is constructed from pairs in which both tracks have the same charge.  The combinatorial background is then subtracted from the number of oppositely charged pairs to give the number of photonic $e^{-}e^{+}$ pairs.  Mathematically, this is expressed as

\begin{equation}
\label{eq:anintro:back_sub}
P=n_{unlike}-2\sqrt{n_{--}n_{++}},
\end{equation}

\noindent where $P$ is the number of true (non-random) photonic $e^{-}e^{+}$ pairs and $n_{unlike}$ is the number of oppositely charged pairs that pass the invariant-mass and $DCA$ cuts.  The quantity $n_{--}$ $(n_{++})$ is the number of pairs of electrons (positrons) with negatively (positively) charged partners that pass the invariant-mass and $DCA$ cuts.  It should be noted that when both members of a pair pass the $e^{\pm}$ identification cuts (\textit{i.e.}, both tracks qualify as $e^{\pm}$ candidates) that pair is counted twice.  Given this, and the fact that photonic $e^{\pm}$ can have only one true conversion partner, the quantity $P$ in Equation~\ref{eq:anintro:back_sub} is the same as the number of photonic $e^{\pm}$ (modulo efficiency corrections).  Figure~\ref{fig:anintro:invmass} illustrates the combinatorial background subtraction procedure using the invariant mass distribution.  It is possible to calculate $n_{unlike}$, the combinatorial background, and the photonic $e^{\pm}$ yield as functions of other quantities.  Figure~\ref{fig:anintro:pair_dca} illustrates the calculation of the DCA distribution for photonic $e^{-}e^{+}$ pairs. Figure~\ref{fig:dedx:nsigma_pho_pairs} (page~\pageref{fig:dedx:nsigma_pho_pairs}) shows those three quantities as functions of the $e^{\pm}$ candidate's value\footnote{When both members of a pair qualify as $e^{\pm}$ candidates, the pair is counted twice: once at the value of $n\sigma_{e}$ of the first candidate and once at the value of $n\sigma_{e}$ of the second candidate.} of $n\sigma_{e}$.

A few selection cuts are applied to the partner tracks.  Simulations by the STAR collaboration~\cite{Anderson2003659} indicate that for transverse momenta less than a few hundred MeV, the $p_{T}$ resolution of the STAR detector is degraded due to multiple Coulomb scattering in the inner structures of the detector.  The fractional difference between the simulated $p_{T}$ and the transverse momentum reconstructed by the STAR detector is observed to increase sharply with decreasing transverse momentum for $p_{T}\lesssim 300\MeV/c$ for low-mass particles (charged pions in the simulations).  Therefore, in this analysis partner tracks are required to have $p_{T}(partner)>300\MeV/c$.  A track-quality cut is also applied to partner tracks: it is required that $R_{FP}(partner)>0.52$ (the same fit-points-ratio cut used for $e^{\pm}$ candidates).  Finally, a TPC energy-loss cut is used to exclude some charged hadrons from the set of partner tracks and reduce the number of random pairs in the sample.  It is required that $\langle dE/dx\rangle<2.8$ keV/cm, which generally corresponds to a normalized TPC energy loss of $n\sigma_{e}<4$, far enough from the $e^{\pm}$ peak that the number of $e^{\pm}$ excluded by the cut is expected to be negligible.

\section{Correction of Spectra}
\label{sec:anintro:correction}

The uncorrected non-photonic $e^{\pm}$ yield is simply the difference between the number of $e^{\pm}$ identified (the inclusive $e^{\pm}$ yield) and the number of photonic $e^{\pm}$ identified.  A number of factors have been computed to correct for inefficiencies in the analysis method described above.  Subsequent chapters describe the calculation of each correction factor in detail; see Table~\ref{table:anintro:corrections} for a short summary of the correction factors and the chapters in which their calculations are discussed.  Simulations of non-photonic $e^{\pm}$ and their interactions with the STAR detector are used to find the $e^{\pm}$ reconstruction efficiency $(\varepsilon_{R})$, the efficiency with which $e^{\pm}$ tracks are reconstructed in the STAR TPC and then identified as $e^{\pm}$ using the selection cuts described in Section~\ref{sec:anintro:ele_id}.  The mean BEMC geometrical acceptance $(\langle A_{BEMC}\rangle)$ is used to correct for temporary losses in BEMC geometrical acceptance.  The $e^{\pm}$ spectra in the high-tower triggered data set are corrected by the HT trigger efficiency $\varepsilon_{T}$.  The simulations used to calculate $\varepsilon_{R}$ do not properly reproduce TPC energy loss; therefore, the efficiency of the TPC energy-loss cut $(\varepsilon_{dE/dx})$ is calculated separately using the distributions of $n\sigma_{e}$ in real data.  To correct for the hadron contamination that remains after the application of the $e^{\pm}$ identification cuts, the purity of the inclusive $e^{\pm}$ sample $(K_{inc.})$ is calculated, also using the the distributions of $n\sigma_{e}$ in real data.  The background rejection efficiency $(\varepsilon_{B})$ is the efficiency with which photonic $e^{\pm}$ are identified using the method described in Section~\ref{sec:anintro:back_sub}; this efficiency is calculated using simulations of photonic $e^{\pm}$ and their interactions with the STAR detector.

The efficiency-corrected non-photonic $e^{\pm}$ yield is

\begin{equation}
\label{eq:anintro:effcorr}
N_{EC}=\frac{1}{\langle A_{BEMC}\rangle\varepsilon_{R}\cdot\varepsilon_{T}\cdot\varepsilon_{dE/dx}}\left(K_{inc.}I-\frac{1}{\varepsilon_{B}}P\right),
\end{equation}

\noindent where $I$ is the uncorrected inclusive $e^{\pm}$ yield and $P$ is the uncorrected photonic $e^{\pm}$ yield given in Equation~\ref{eq:anintro:back_sub}.

The residual background is the sum of the estimated yields of $e^{\pm}$ from non-open heavy-flavor sources, which are subtracted from the efficiency-corrected non-photonic $e^{\pm}$ yields.  The components of the residual background are photonic $e^{\pm}$ from light-flavor sources with $M_{inv}.(e^{-}e^{+})>150\MeV/c^{2}$ (denoted $B_{LFP}$), non-photonic $e^{\pm}$ from light-flavor sources (mainly the semileptonic decays of kaons, denoted $B_{LFN}$), photonic $e^{\pm}$ from $J/\psi$ and $\Upsilon$ decays ($B_{J/\psi}$ and $B_{\Upsilon}$, respectively), and photonic $e^{\pm}$ from the Drell-Yan process $(B_{D-Y})$.  Most of the components of the residual background are estimated using simulations of particles decaying into $e^{\pm}$; the Drell-Yan contribution is calculated~\cite{Vogelsang_Private_2010} using perturbative QCD.  The yield of non-photonic $e^{\pm}$ from open-heavy-flavor sources (mostly $D$ and $B$ meson decays) is

\begin{equation}
\label{eq:anintro:ydb}
Y_{D,B}=N_{EC}-B_{LFP}-B_{LFN}-B_{J/\psi}-B_{\Upsilon}-B_{D-Y}.
\end{equation}

\noindent This correction procedure will be discussed further in Chapter~\ref{sec:results}.

\begin{table}
\caption{A summary of the correction factors used in this analysis and the chapters or sections in which their calculations are described.}
\label{table:anintro:corrections}
\begin{tabular}{| r | c | c | c |}
\hline
 & & & Chapter/\\
Name & Symbol & Corrects for & Section\\\hline\hline
$e^{\pm}$ reconstruction eff.  & $\varepsilon_{R}$ & track reconstruction &~\ref{sec:rec}\\
 & & $e^{\pm}$ ID cuts & \\\hline
 mean BEMC & $\langle A_{BEMC}\rangle$ & losses in BEMC &~\ref{sec:acc:bemc}\\
 geometrical acceptance & & geometrical acceptance & \\\hline
HT trigger eff. & $\varepsilon_{T}$ & HT trigger ``turn-on" &~\ref{sec:trigger}\\\hline
 energy-loss cut eff. & $\varepsilon_{dE/dx}$ & TPC energy-loss cut &~\ref{sec:dedx}\\\hline
 inclusive $e^{\pm}$ purity & $K_{inc.}$ & hadron contamination &~\ref{sec:dedx}\\\hline
 background rejection eff. & $\varepsilon_{B}$ & photonic $e^{\pm}$ ID &~\ref{sec:bre}\\\hline\hline
 \multicolumn{4}{| c |}{Residual Background Components}\\\hline
 residual photonic $e^{\pm}$ & $B_{LFP}$ & photonic $e^{\pm}$ from &~\ref{sec:res_back:lfp}\\
 from light-flavor sources & & $\eta$, $\rho^{0}$, $\omega$, $\eta^{\prime}$, $\phi$, and $K^{0}_{S}$ & \\\hline
 residual non-photonic $e^{\pm}$ & $B_{LFN}$ & non-photonic $e^{\pm}$ from &~\ref{sec:res_back:lfn}\\
  & & non-heavy-flavor sources & \\\hline
 $e^{\pm}$ from $J/\psi$ & $B_{J/\psi}$ & photonic $e^{\pm}$ from $J/\psi$ decays &~\ref{sec:res_back:jpsi}\\\hline
 $e^{\pm}$ from $\Upsilon$ & $B_{\Upsilon}$ & photonic $e^{\pm}$ from $\Upsilon$ decays &~\ref{sec:res_back:upsilon}\\\hline
 Drell-Yan $e^{\pm}$ & $B_{D-Y}$ & photonic $e^{\pm}$ from Drell-Yan process &~\ref{sec:res_back:dy}\\\hline
\end{tabular}
\end{table}

\clearpage

\chapter{$\boldsymbol{e^{\pm}}$ Reconstruction Efficiency}
\label{sec:rec}

\section{Introduction}
\label{sec:rec:intro}

This section describes the calculation of the $e^{\pm}$ reconstruction efficiency $(\varepsilon_{R})$, the efficiency with which an $e^{\pm}$ track is identified as an $e^{\pm}$ using all identification cuts except the TPC energy-loss cut.  The efficiency is calculated with the use of simulated non-photonic $e^{\pm}$ embedded into real 200-GeV Cu + Cu events.

\begin{table}
\caption[The combined efficiency of these $e^{\pm}$ identification cuts is found using the method described in this chapter.]{The combined efficiency of these $e^{\pm}$ identification cuts is found using the method described in this chapter.  These cuts are discussed in Section~\ref{sec:anintro:ele_id}.  For reasons discussed in the text, the TPC energy-loss cut is not included here.}
\label{table:rec:eid}
\begin{tabular}{| r | c |}
\hline
radius of first TPC point & $R_{1TPC}<$102 cm\\\hline
number of TPC fit points & $N_{TPCfit}>20$\\\hline
TPC fit points ratio & $R_{FP}>0.52$\\\hline
DCA to primary vertex & $GDCA<1.5$ cm\\\hline
track pseudorapidity & $-0.1<\eta<0.7$\\\hline
energy in BEMC & $p/E_{Tower}<2/c$\\\hline
BSMD cluster displacement & $\Delta_{SMD}<0.02$\\\hline
BSMD cluster sizes & $N_{SMD\eta}\geq 2$ and $N_{SMD\phi}\geq 2$\\\hline
BEMC status & ``good" for tower, BSMD-$\eta$, and BSMD-$\phi$\\\hline
\end{tabular}
\end{table}

The efficiency $\varepsilon_{R}$ is the $p_{T}$-dependent ratio of $N_{rec.}$ (the number of \textit{reconstructed} $e^{\pm}$) to $N_{sim.}$ (the number of \textit{simulated} $e^{\pm}$).  For a given transverse-momentum bin with $p_{T}^{min.}<p_{T}<p_{T}^{max.}$, $N_{rec.}$ is the number of reconstructed tracks that satisfy the following conditions:

\begin{itemize}
\item The reconstructed track shares at least one half of its TPC points with a simulated $e^{\pm}$ track.
\item The track passes all $e^{\pm}$ identification cuts (see Table~\ref{table:rec:eid}) except the TPC energy-loss cut.
\item The reconstructed transverse momentum and pseudorapidity of the track are within the ranges $p_{T}^{min.}<p_{T}(rec.)<p_{T}^{max.}$ and $-0.1<\eta(rec.)<0.7$.
\end{itemize}

For a given $p_{T}$ bin, $N_{sim.}$ is the number of simulated $e^{\pm}$ tracks for which the simulated transverse momentum and pseudorapidity are within the ranges $p_{T}^{min.}<p_{T}(MC)<p_{T}^{max.}$ and $-0.1<\eta(MC)<0.7$.  Note that the tracks in $N_{rec.}$ are not strictly a subset of the tracks in $N_{sim.}$.  Imperfect momentum resolution can cause a reconstructed track to be in a different $p_{T}$ bin than the corresponding simulated track.  The factor $\varepsilon_{R}$ therefore allows the non-photonic $e^{\pm}$ spectrum to be corrected for the STAR detector's momentum (and pseudorapidity) resolution.

The calculation of $\varepsilon_{R}$ presented in this chapter uses embedding data.  In embedding, simulated particle tracks are generated and their interactions with the STAR detector are simulated using GEANT~\cite{Agostinelli2003250}, producing hits in the TPC, BEMC, and other detector components.  These simulated hits are then mixed with real data and the combination of the two is run through STAR's event reconstruction (tracking, vertex-finding, etc.) algorithms.  This allows the efficiency of the STAR detector to be studied in a realistic (high-particle-multiplicity) background.

The BEMC simulator uses the calibration and status of the BEMC for the real event into which the simulated data are embedded.  Therefore, any losses in BEMC acceptance in the real data will also appear in the embedding data.  For this reason, the $e^{\pm}$ reconstruction efficiency is corrected by the mean BEMC acceptance (the calculation of this factor is described in Chapter~\ref{sec:acc:bemc}).

\begin{equation}
\label{eq:rec:definition}
\varepsilon_{R}(p_{T})=\frac{1}{\langle A_{BEMC}(sim.)\rangle}\cdot\frac{N_{rej.}(p_{T})}{N_{sim.}(p_{T})},
\end{equation}

\noindent where $\langle A_{BEMC}(sim.)\rangle$ is the BEMC acceptance for the set of events used in embedding, which is a subset of all real Cu + Cu events.  Since the BEMC acceptance changes with time, the value of $\langle A_{BEMC}\rangle$ may be different for the embedding subset.

Three embedding data sets are used in this chapter.

\begin{itemize}
\item Data Set $N_{1}$: This is the primary embedding data set used to calculate $\varepsilon_{R}$.  Simulated single $e^{\pm}$ are embedded into $\approx$ 670,000 real 200-GeV Cu + Cu events.  The $e^{\pm}$ were generated uniformly in the transverse-momentum range $1\GeV/c<p_{T}(MC)<15\GeV/c$ and the pseudorapidity range $-0.3<\eta(MC)<1.3$, covering the full azimuth.  The $e^{\pm}$ were embedded at a rate of $\approx 0.05$ times the reference multiplicity of the underlying event.  For this data set, $\langle A_{BEMC}(N_{1})\rangle=0.606$.  This data set was created using an incorrect function in the vertex-finding algorithm; as described below, this error did not have an effect on the calculated efficiency.
\item Data Set $N_{2}$: This data set is used to study the effect of a flaw in data set $N_{1}$.  This data set is very similar to data set $N_{1}$, but was generated using the correct function in the vertex-finding algorithm.  This smaller data set included only $\approx$ 150,000 events.  For this data set, $\langle A_{BEMC}(N_{2})\rangle=0.613$.
\item Data Set $P_{1}$: This data set is not directly used in the calculation of $\varepsilon_{R}$, but some distributions of measured quantities from this data set are shown in Section~\ref{sec:rec:compare2real}.  Simulated photonic $e^{\pm}$ (from $\pi^{0}\rightarrow\gamma\gamma$ decays and the subsequent conversion of those photons to $e^{-}e^{+}$ pairs in the inner structures of the STAR detector) are embedded into $~\approx$ 800,000 real Cu + Cu events.  This data set is used to calculate the background rejection efficiency (see Chapter~\ref{sec:bre}, which contains a more detailed description of this data set).
\end{itemize}

Note that the embedding data sets used for this calculation did not include a satisfactory simulation of TPC energy loss; the efficiency of the TPC energy-loss cut is calculated separately in Chapter~\ref{sec:dedx}.  Unless otherwise noted, the statistical uncertainties in $\varepsilon_{R}$ shown in Figures in this chapter are estimated uncertainties calculated using the method described in Section~\ref{sec:uncertainties:weff}.

\clearpage

\section{Weighting}
\label{sec:rec:wpt}

In embedding data sets $N_{1}$ and $N_{2}$, the simulated $e^{\pm}$ are distributed uniformly in transverse momentum over the range $1\GeV/c<p_{T}<15\GeV/c$.  If the correction factor $\varepsilon_{R}$ is to properly account for the STAR detector's momentum resolution, the $e^{\pm}$ spectrum must be weighted to have a realistic dependence on $p_{T}$.  For example, consider two $e^{\pm}$ tracks, one with simulated transverse momentum $p_{T}(MC)=3\GeV/c$, the other with $p_{T}(MC)=5\GeV/c$, with both tracks having the same reconstructed transverse momentum $p_{T}(rec.)=4\GeV/c$.  In a realistic $e^{\pm}$ spectrum, the particle yield decreases with increasing $p_{T}$.  Thus, there should be fewer electrons with $p_{T}(MC)=5\GeV/c$ than with $p_{T}(MC)=3\GeV/c$.  If $p_{T}$ weights are not applied, the two tracks will make equal contributions to the calculation of $\varepsilon_{R}$ and the value of the efficiency at $p_{T}=4\GeV/c$ will be too large (given that $\varepsilon_{R}$ increases with $p_{T}$).

\begin{figure}[htbp]
\begin{center}
\includegraphics[width=0.85\linewidth]{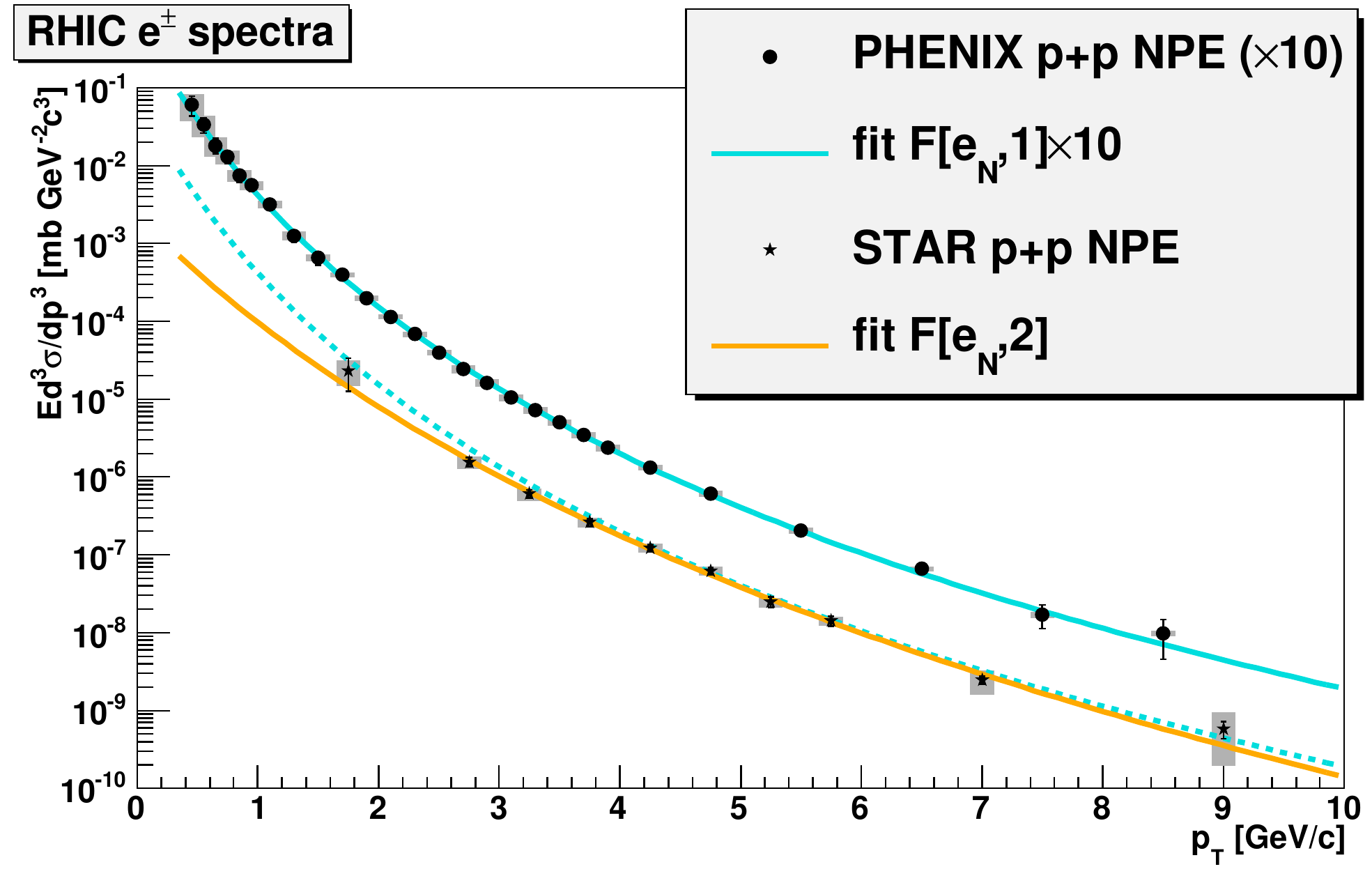}
\caption[Non-photonic $e^{\pm}$ cross-sections in $p+p$ collisions used as initial weighting functions in calculating $\varepsilon_{R}$.]{Non-photonic $e^{\pm}$ cross-sections (vs. $p_{T}$) for 200-GeV $p+p$ collisions measured by the PHENIX\protect\cite{PhysRevLett.97.252002} and STAR\protect\cite{STAR_ppNPE2011} collaborations.  Statistical uncertainties are indicated by error bars; systematic uncertainties are indicated by the gray shaded boxes.  Also shown are power-law fits to the data (for the fits, the statistical and systematic uncertainties of each data point were added in quadrature).  The fit to the PHENIX data, $F[e_{N},1]$ is used as the initial weighting function in the calculation of $\varepsilon_{R}$.  The dashed cyan curve is $F[e_{N},1]$ without the scaling factor of 10.  The parameters for the fit to the STAR data $(F[e_{N},2])$ are given in Section~\ref{sec:functions:npe}.}
\label{fig:rec:rhic_ele}
\end{center}
\end{figure}

The weighting function should have the same shape as the spectrum of non-photonic $e^{\pm}$ in 200-GeV Cu + Cu collisions.\footnote{In data set $P_{1}$, each photonic $e^{\pm}$ was weighted according to the transverse momentum of the pion that produced it.  The weighting function, $F[\pi^{0},2]$, is described in Section~\ref{sec:bre:weighting}.} However, this spectrum is not known \textit{a priori}, so the following iterative weighting procedure is used.  An initial weighting function is chosen as the expected shape of the non-photonic $e^{\pm}$ spectrum.  The $e^{\pm}$ reconstruction efficiency is calculated with each $e^{\pm}$ weighted according to its simulated transverse momentum.  The values of $\varepsilon_{R}$ calculated with this initial weighting function are used to find the fully corrected non-photonic $e^{\pm}$ spectrum for 200-GeV Cu + Cu collisions.  A fit to that spectrum is then used as a weighting function in a new calculation of $\varepsilon_{R}$.  This procedure is repeated until the values of the $e^{\pm}$ reconstruction efficiency stabilize.

Figure~\ref{fig:rec:rhic_ele} shows the initial weighting function used in the iterative momentum weighting procedure: the non-photonic $e^{\pm}$ cross-section found by the PHENIX collaboration~\cite{PhysRevLett.97.252002} in 200-GeV $p+p$ collisions.  That spectrum is fit with a power-law function $F[e_{N},1](p_{T})=A(1+p_{T}/p_{0})^{n}$.  The parameters of this fit function, as well as others discussed in this dissertation are given in Appendix~\ref{sec:functions}.  The $e^{\pm}$ spectra used to calculate $\varepsilon_{R}$ are weighted by factors of $p_{T}(MC)\times F[e_{N},1]$.

\begin{figure}[htbp]
\begin{center}
\includegraphics[width=0.85\linewidth]{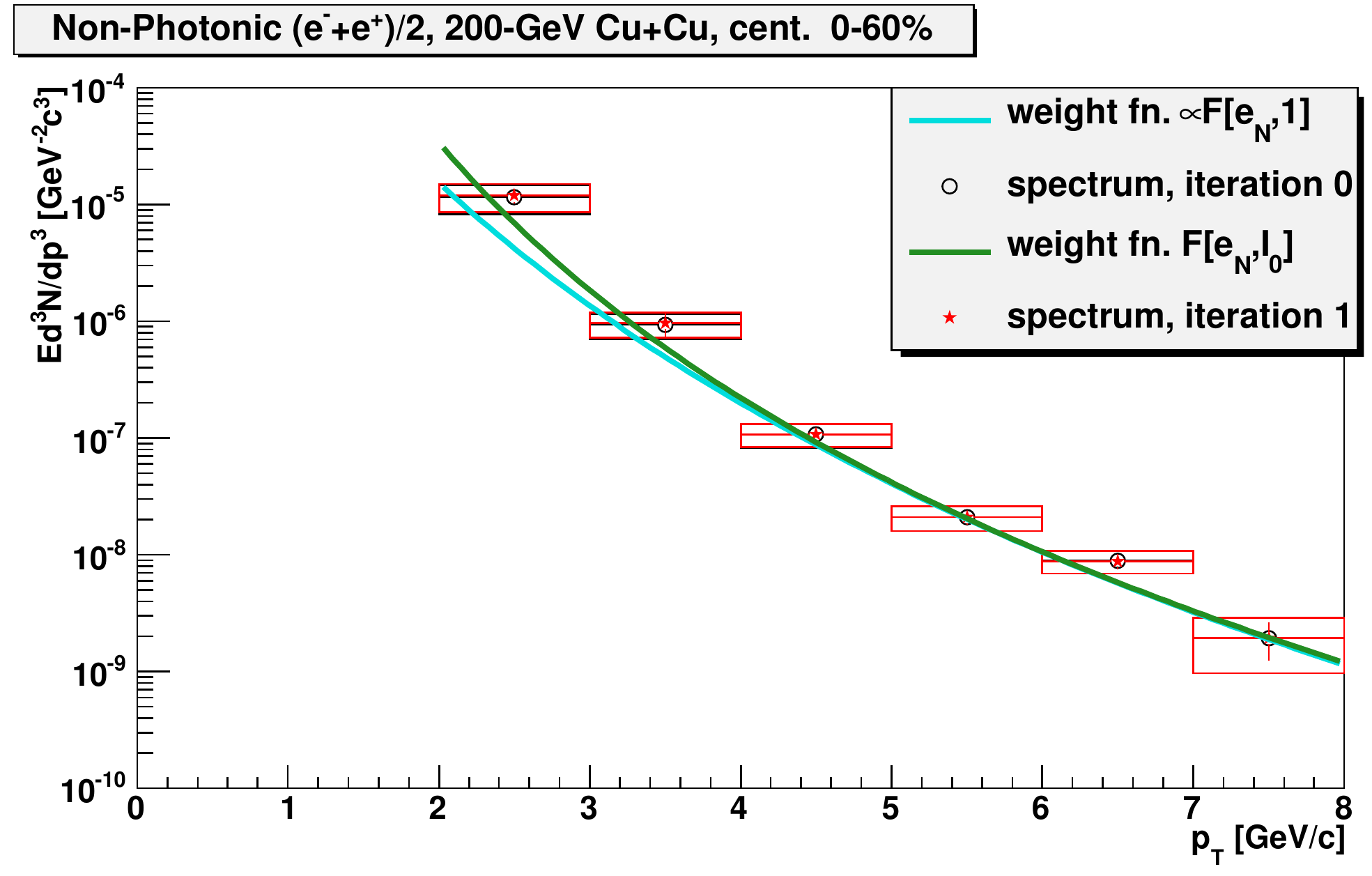}
\caption[Two iterations of the corrected non-photonic $e^{\pm}$ spectrum in 200-GeV Cu + Cu collisions.]{Two iterations of the corrected non-photonic $e^{\pm}$ spectrum (vs. $p_{T}$) in 200-GeV Cu + Cu collisions.  Statistical uncertainties are indicated by error bars; systematic uncertainties are indicated by the boxes around each data point.  Also shown are two of the weight functions used in the calculation of $\varepsilon_{R}$.  The initial weighting function is $F[e_{N},1]$, the fit to the PHENIX collaboration's non-photonic $e^{\pm}$ spectrum in $p+p$ collisions.  It was used to generate iteration 0 of $\varepsilon_{R}$, which was used to generate iteration 0 of the non-photonic $e^{\pm}$ spectrum for Cu + Cu collisions (black).  That spectrum was fit with a power-law function $F[e_{N},I_{0}]$, which was used as a weighting function in the calculation of iteration 1 of $\varepsilon_{R}$.  This was used to generate iteration 1 of the non-photonic $e^{\pm}$ spectrum for Cu + Cu collisions (red).  Note that $F[e_{N},I_{0}]$ was fit to the data using the function's integral over each $p_{T}$ bin, not its value at the bin center.}
\label{fig:rec:spectra_iter}
\end{center}
\end{figure}

\begin{figure}[htbp]
\begin{center}
\includegraphics[width=0.75\linewidth]{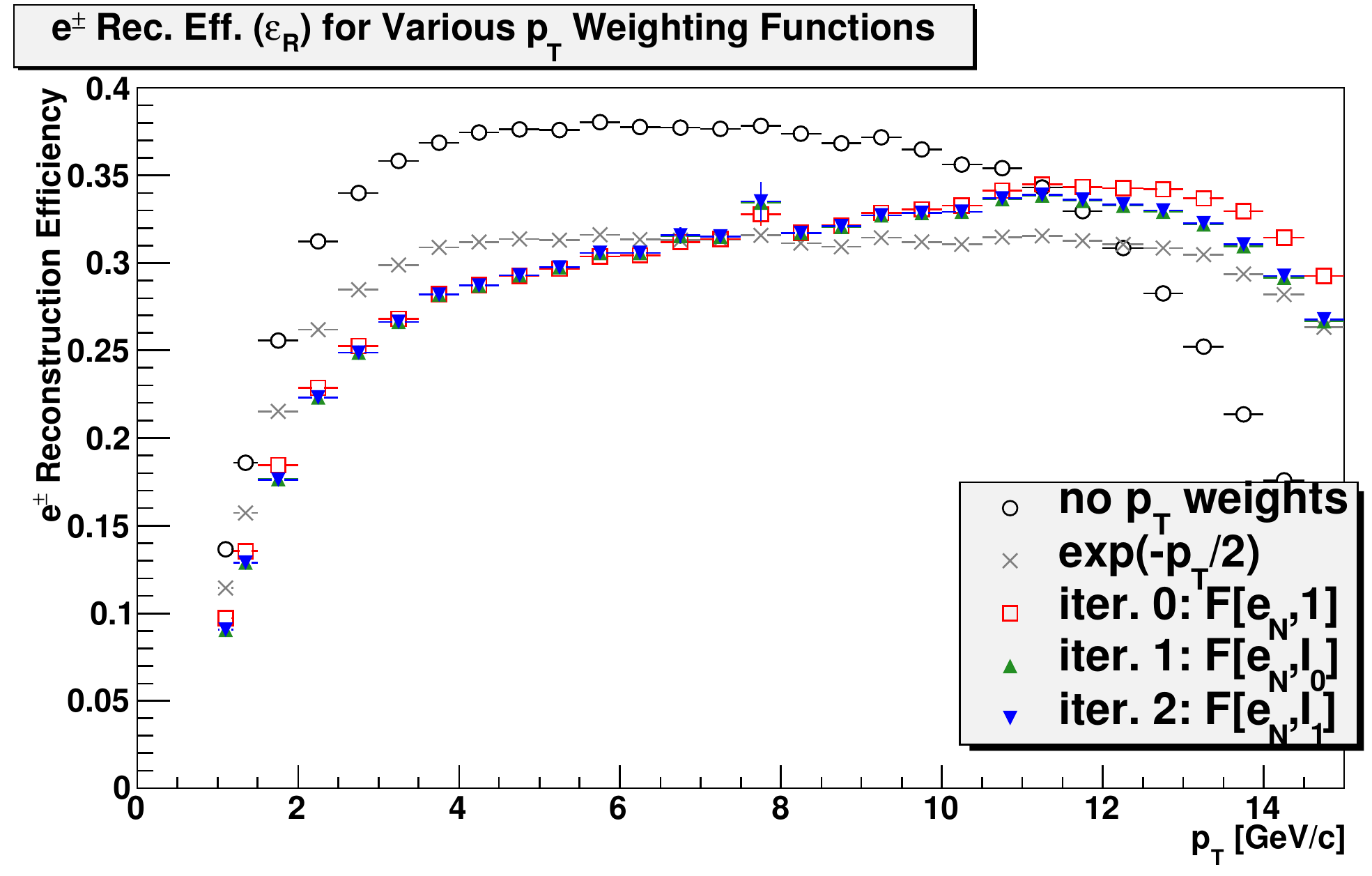}
\caption[The $e^{\pm}$ reconstruction efficiency calculated using various $p_{T}$ weighting functions.]{The $e^{\pm}$ reconstruction efficiency as a function of $p_{T}$, calculated using various $p_{T}$ weighting functions.  The differences between iterations 1 and 2 are negligible.  The decreasing efficiencies at high $p_{T}$ are due to momentum resolution and the fact that the simulated transverse momentum does not exceed 15 GeV/$c$.}
\label{fig:rec:eff_wpt}
\includegraphics[width=0.75\linewidth]{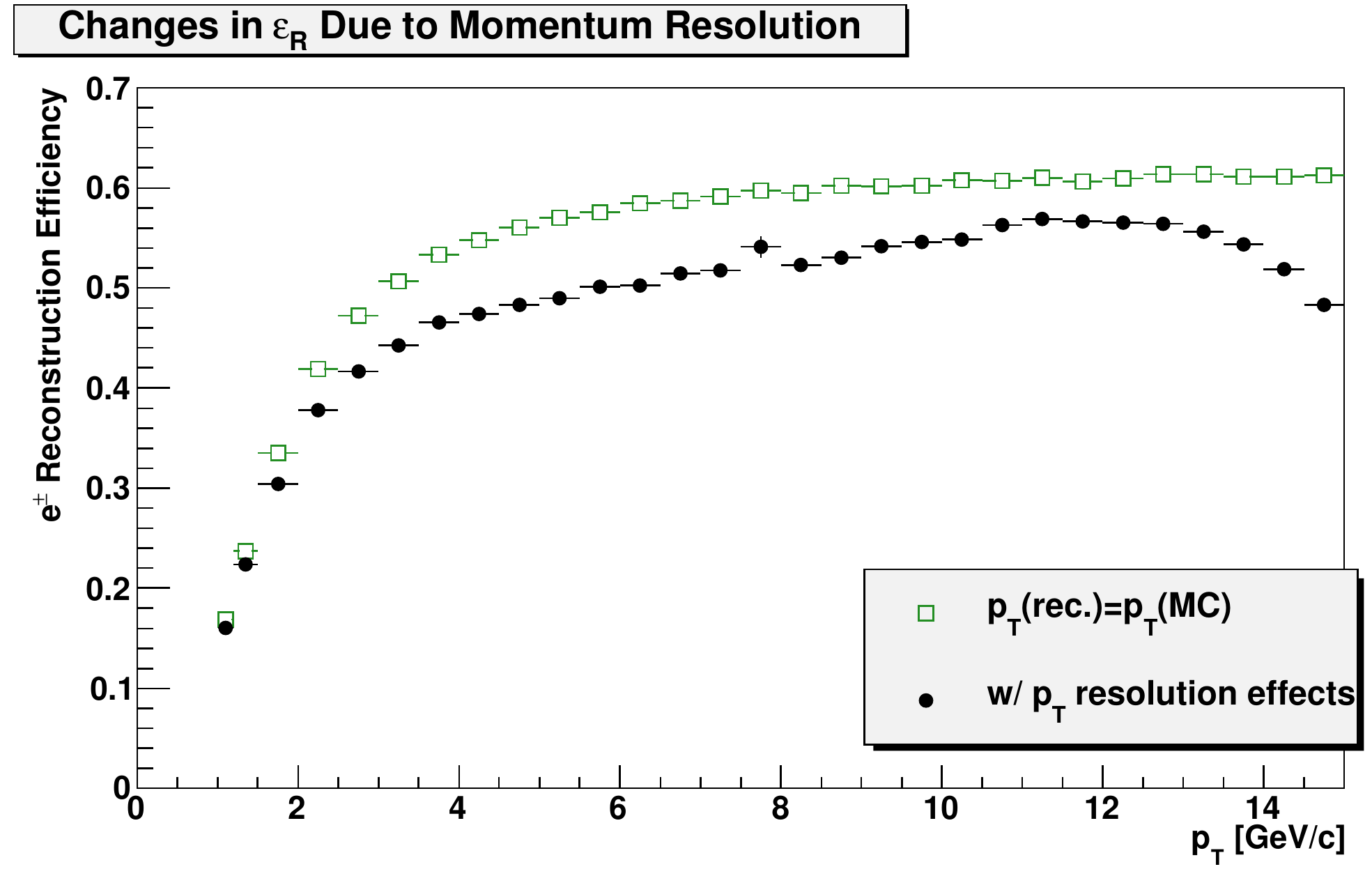}
\caption[Changes in $\varepsilon_{R}$ as a function of $p_{T}$ when the STAR detector's momentum resolution is taken into account.]{The changes in $\varepsilon_{R}$ as a function of $p_{T}$ when the STAR detector's momentum resolution is taken into account.  The black histogram (which includes momentum resolution effects) is calculated as defined in Section~\ref{sec:rec:intro}.  To generate the green histogram, the numerator of the calculation was taken to be a function of the simulated transverse momentum $(p_{T}(MC))$ rather than the reconstructed transverse momentum $(p_{T}(rec.))$.}
\label{fig:rec:eff_pres}
\end{center}
\end{figure}

Figure~\ref{fig:rec:spectra_iter} shows the corrected non-photonic $e^{\pm}$ spectrum for the $0-60\%$ most central 200-GeV Cu + Cu collisions after the first two iterations of the momentum weighting procedure.  See Chapter~\ref{sec:results} for a detailed description of the correction procedure.  The spectrum from iteration 0 is used as the input weighting function to find $\varepsilon_{R}$ for iteration 1.  Figure~\ref{fig:rec:eff_wpt} shows $\varepsilon_{R}$ calculated for various weighting functions, including three iterations of the weighting procedure.  The efficiencies for iteration 1 and iteration 2 are virtually identical, so the iteration procedure was stopped and the non-photonic $e^{\pm}$ spectra from iteration 1 were used in subsequent calculations of the nuclear modification factor $R_{AA}$.

Figure~\ref{fig:rec:eff_pres} shows the effect of the STAR detector's momentum resolution on $\varepsilon_{R}$.  The black histogram shows $\varepsilon_{R}$ calculated according to its usual definition given in Section~\ref{sec:rec:intro}.  The green histogram shows $\varepsilon_{R}$ calculated assuming perfect momentum resolution (\textit{i.e.}, the spectrum in the numerator of the calculation is a function of the simulated transverse momentum $p_{T}(MC)$, instead of $p_{T}(rec.)$); the weighting function is $F[e_{N},1]$.

\clearpage

\section{Calculated Efficiencies from Embedding}
\label{sec:rec:eff_from_emb}

In this section, various calculations of the $e^{\pm}$ reconstruction efficiency are shown.  Unless otherwise stated, the efficiencies are extracted from embedding data set $N_{1}$, with weighting function $F[e_{N},1]$.

\begin{figure}[htbp]
\begin{center}
\includegraphics[width=0.85\linewidth]{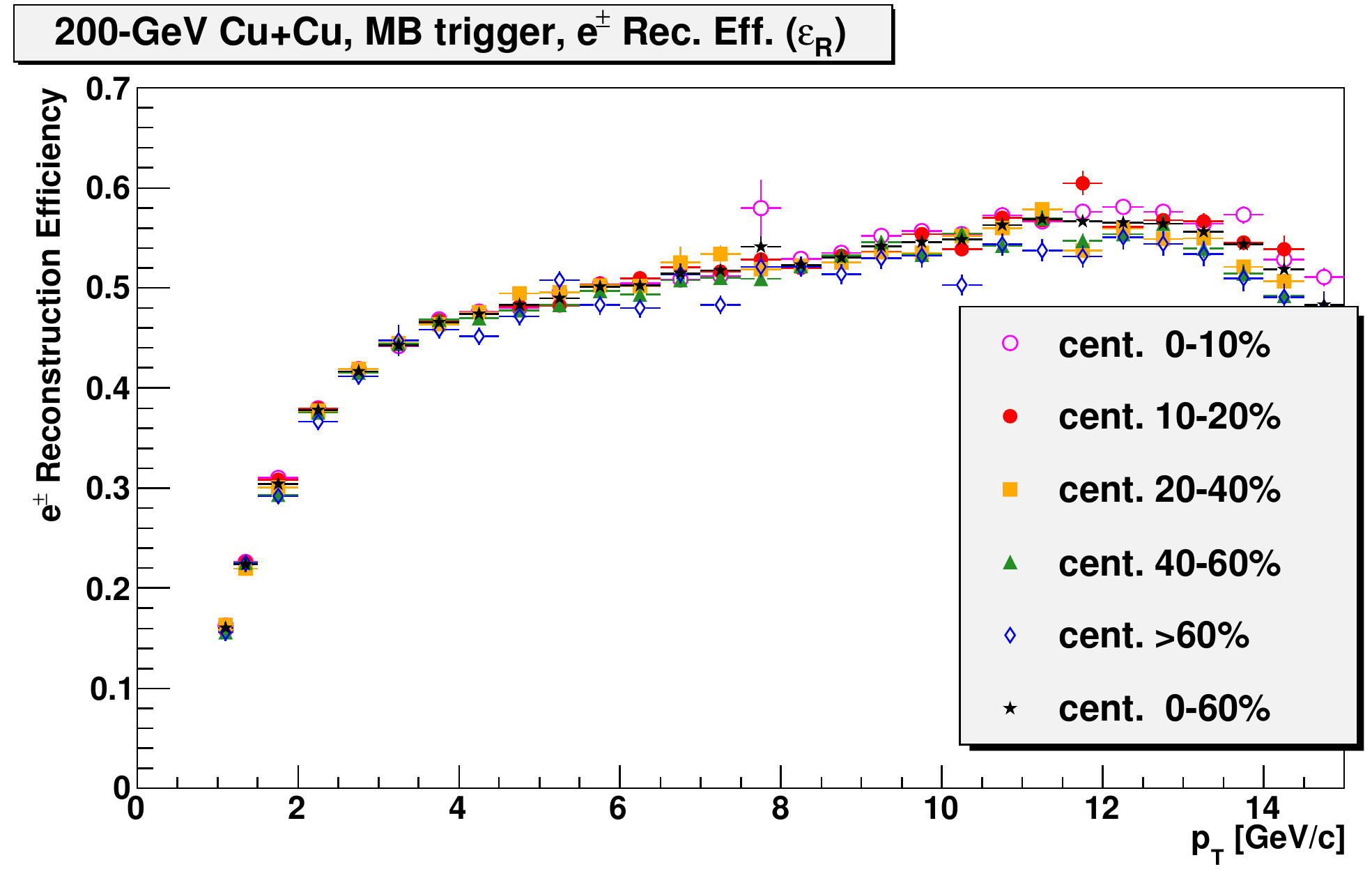}
\caption[The $e^{\pm}$ reconstruction efficiency as a function of $p_{T}$ for various collision centrality classes.]{The $e^{\pm}$ reconstruction efficiency as a function of $p_{T}$ for various collision centrality classes; $\varepsilon_{R}$ is not observed to change with the collision centrality of the underlying event.}
\label{fig:rec:eff_cent}
\includegraphics[width=0.85\linewidth]{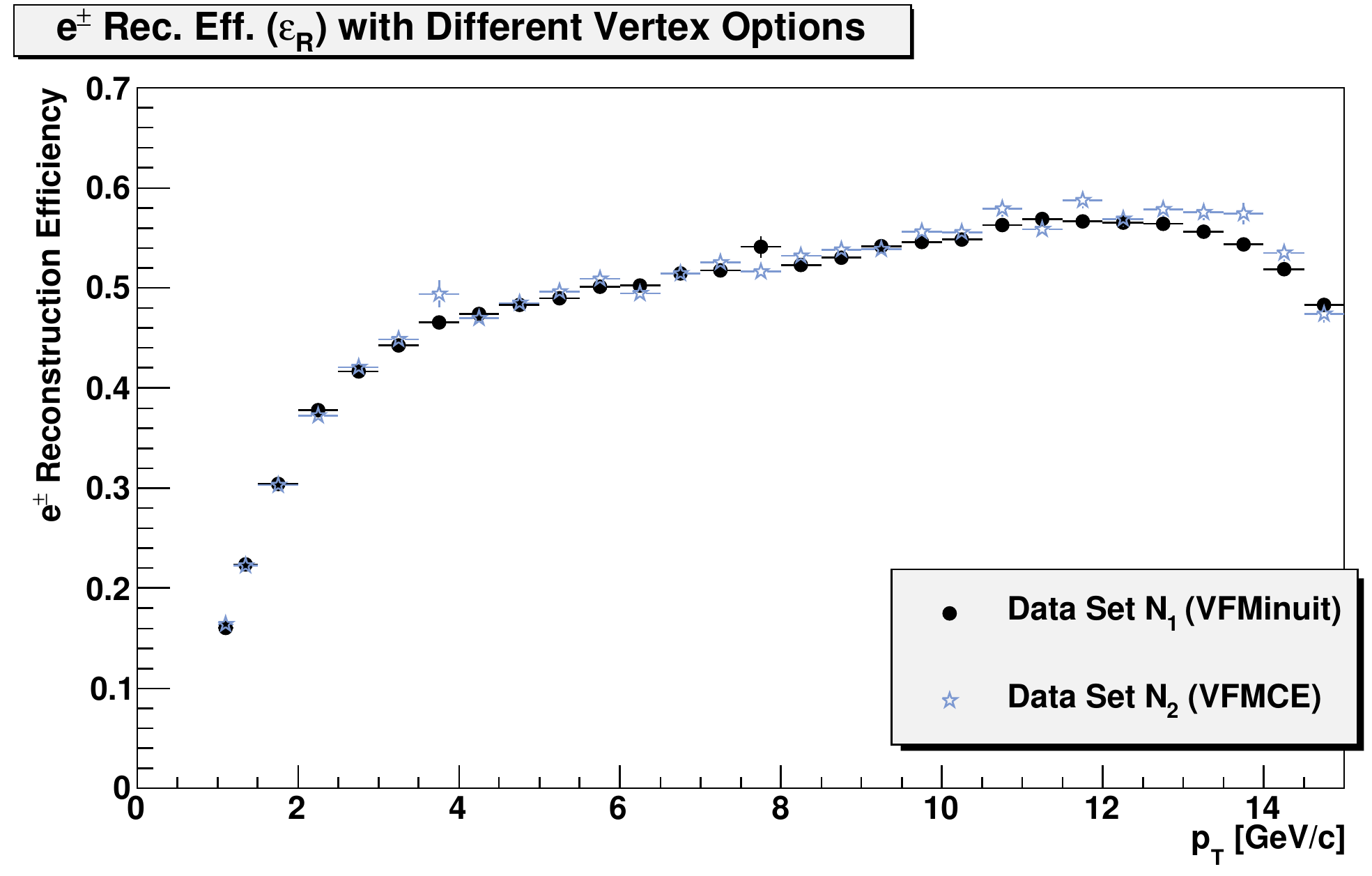}
\caption[The $e^{\pm}$ reconstruction efficiency as a function of $p_{T}$, calculated using two different vertex finders.]{The $e^{\pm}$ reconstruction efficiency as a function of $p_{T}$, calculated using two different vertex finders (see the text for an explanation).  Using the incorrect vertex finder (data set $N_{1}$) does not affect the value of $\varepsilon_{R}$.}
\label{fig:rec:eff_vf}
\end{center}
\end{figure}

Figure~\ref{fig:rec:eff_cent} shows the $e^{\pm}$ reconstruction efficiency for various collision centrality classes.  No correlation between $\varepsilon_{R}$ and collision centrality is observed.

Figure~\ref{fig:rec:eff_vf} shows the values of $\varepsilon_{R}$ extracted from embedding data sets $N_{1}$ and $N_{2}$.  As was mentioned earlier, data set $N_{2}$ was created due to the existence of a potential flaw in data set $N_{1}$.  In both of those embedding data sets, the simulated non-photonic $e^{\pm}$ tracks originate from the primary vertex of the underlying event.  In data set $N_{1}$ the STAR vertex-finding algorithm was applied to all reconstructed tracks (including tracks in the underlying event and tracks associated with simulated $e^{\pm}$) and a new primary vertex was found.  The global DCA for each track was calculated relative to this new vertex, which would not be at the same position as the initial vertex of the embedded particles.  If the two vertices were separated by more than the global DCA cut (1.5 cm), a simulated non-photonic $e^{\pm}$ would not be counted in the numerator of $\varepsilon_{R}$ even if its track were reconstructed perfectly.  In data set $N_{2}$, the global DCA of each reconstructed track was calculated relative to the original vertex of the collision (the vertex from which the simulated tracks also originated).\footnote{For a reader familiar with the STAR collaboration's nomenclature: data set $N_{1}$ was generated using the \texttt{VFMinuit} vertex-finding option, while data set $N_{2}$ was generated using the \texttt{VFMCE} option.} No difference in $\varepsilon_{R}$ based on the choice of vertex finding method is observed in Figure~\ref{fig:rec:eff_vf}.  Since data set $N_{1}$ contains a greater number of events, that data set is used for subsequent calculations of $\varepsilon_{R}$.

\begin{figure}[htbp]
\begin{center}
\includegraphics[width=0.85\linewidth]{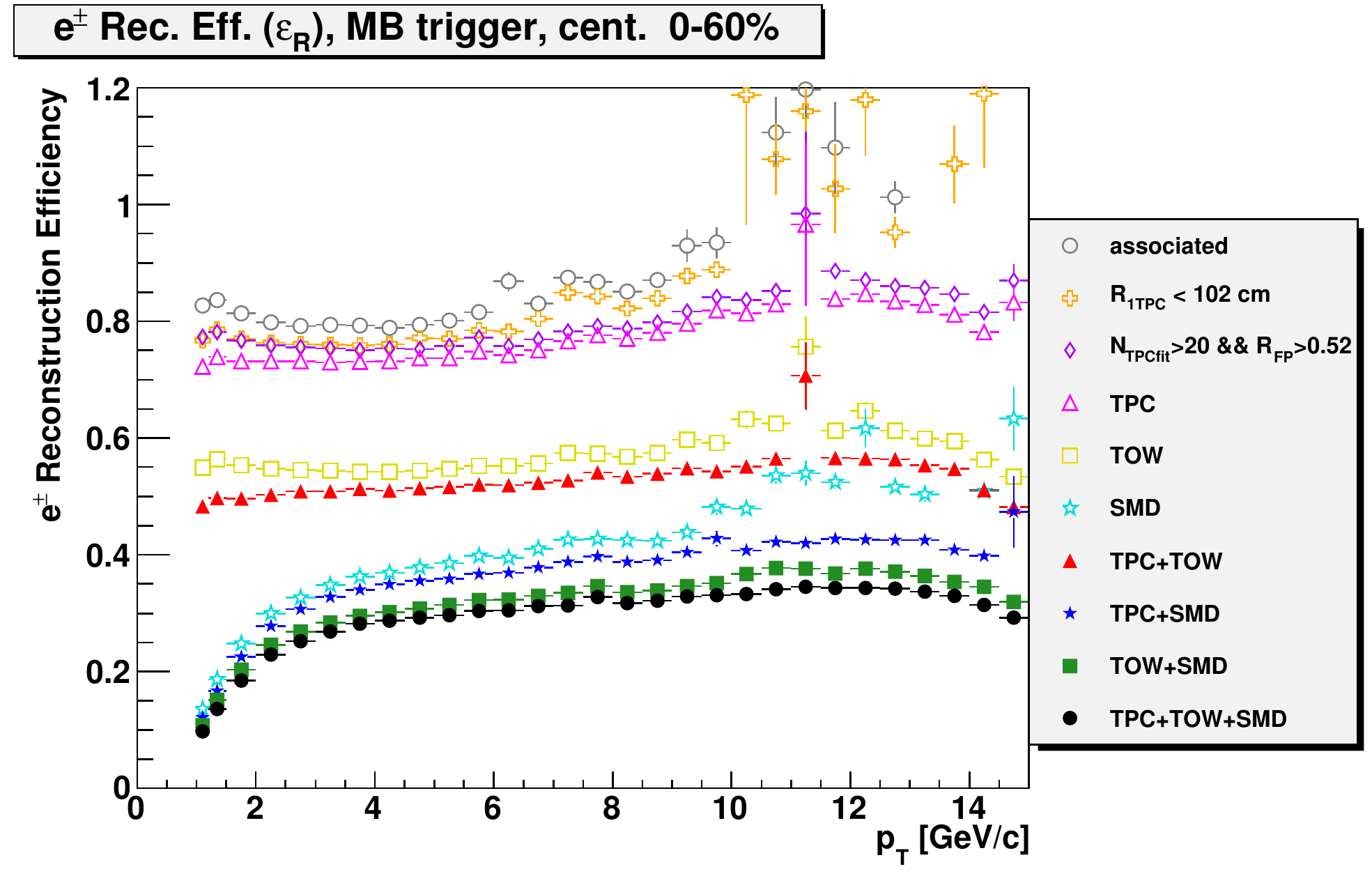}
\caption[The $e^{\pm}$ reconstruction efficiency as a function of $p_{T}$, calculated when various sets of $e^{\pm}$ identification cuts are applied.]{The $e^{\pm}$ reconstruction efficiency as a function of $p_{T}$, calculated when various sets of $e^{\pm}$ identification cuts are applied.  The $GDCA$ cut is always applied.  For the ``associated" (gray) histogram, no other cut is applied.  ``TPC" indicates the combination of the cuts on $R_{1TPC}$, $N_{TPCfit}$, and $R_{FP}$ (but not TPC energy loss);Ê``TOW" indicates the cuts relating to BEMC towers; ``SMD" indicates the cuts relating to the BSMD.  Note that the correction factor of $\langle A_{BEMC}(N_{1})\rangle$ has not been applied to these calculations of $\varepsilon_{R}$.}
\label{fig:rec:eff_cuts}
\end{center}
\end{figure}

\begin{figure}[htbp]
\begin{center}
\includegraphics[width=0.85\linewidth]{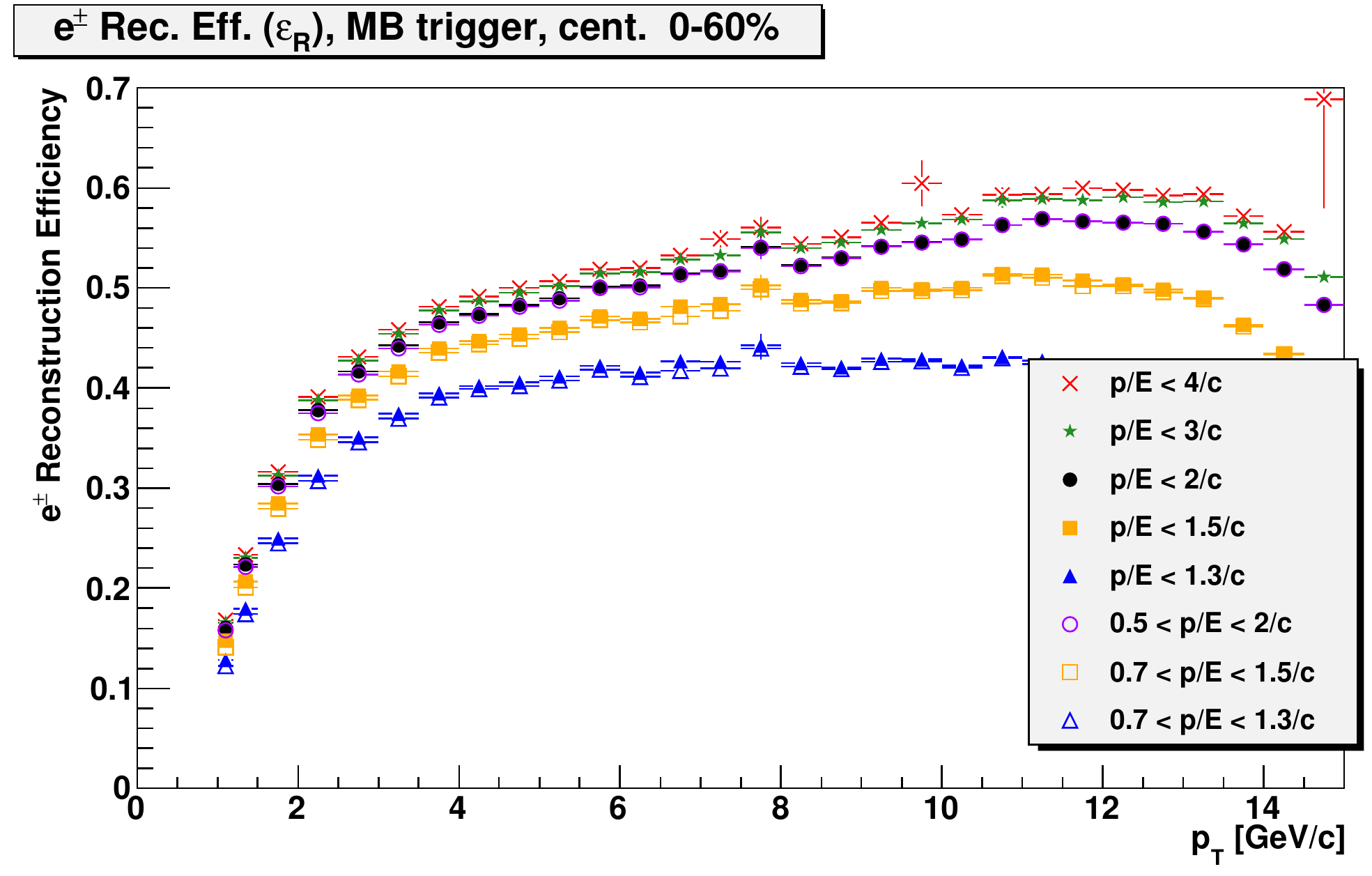}
\caption{The $e^{\pm}$ reconstruction efficiency as a function of $p_{T}$ with various upper and lower cuts on $p(rec.)/E_{Tower}$.}
\label{fig:rec:eff_poe}
\includegraphics[width=0.85\linewidth]{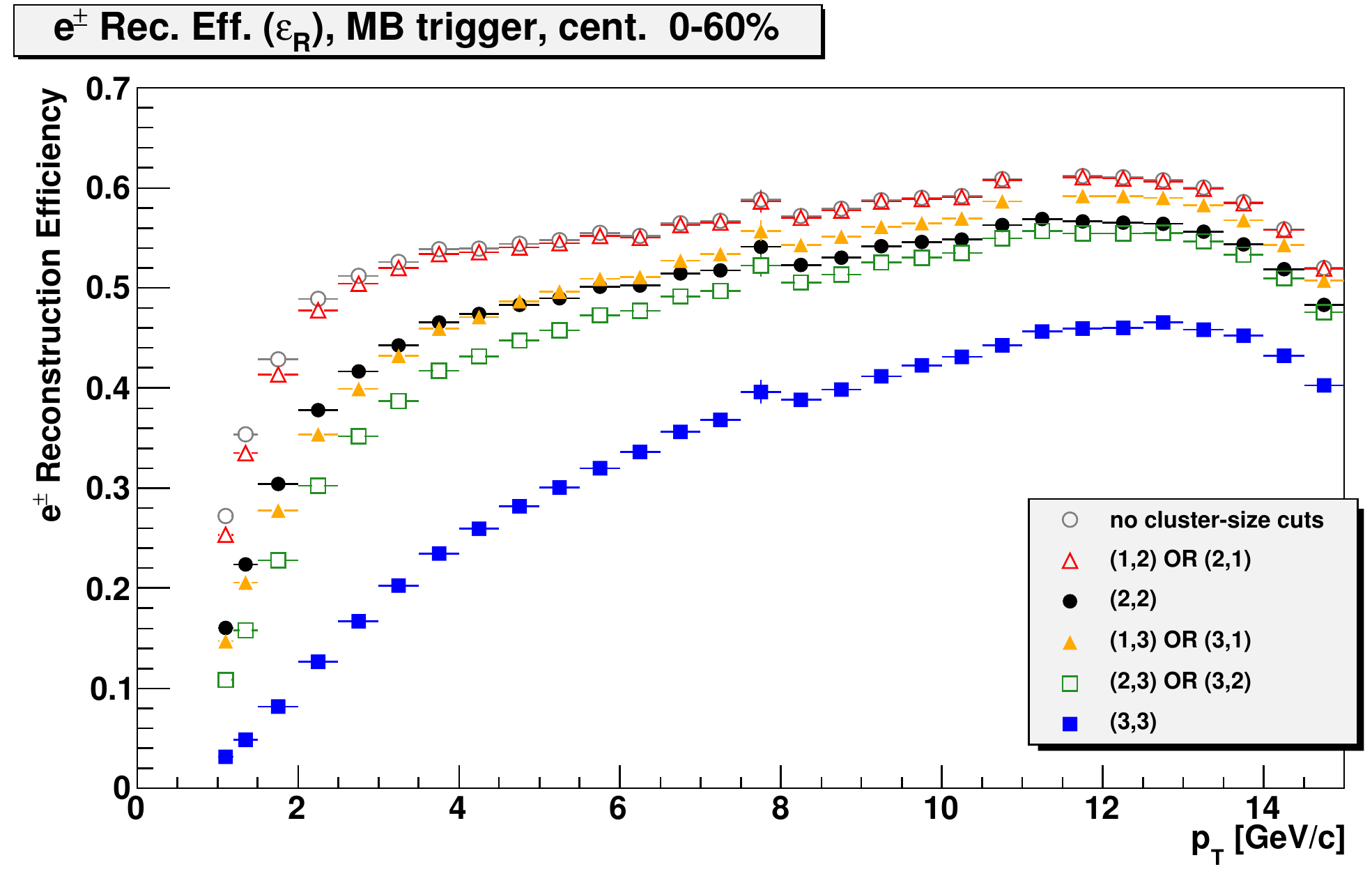}
\caption[The $e^{\pm}$ reconstruction efficiency as a function of $p_{T}$ for various cuts on the minimum BSMD shower sizes $N_{SMD\eta}$ and $N_{SMD\phi}$.]{The $e^{\pm}$ reconstruction efficiency as a function of $p_{T}$ for various cuts on the minimum shower sizes $N_{SMD\eta}$ and $N_{SMD\phi}$.  In the legend $(X,Y)$ stands for $(N_{SMD\eta}\geq X\;\mathrm{AND}\;N_{SMD\phi}\geq Y)$.}
\label{fig:rec:eff_smd_size}
\end{center}
\end{figure}

Calculations of $\varepsilon_{R}$, with the values of various cuts changed (see Table~\ref{table:rec:eid}), are shown in Figures~\ref{fig:rec:eff_cuts} through~\ref{fig:rec:eff_smd_size}, with some additional Figures presented in Section~\ref{sec:add:rec:embedding}.  In these Figures, the black circles indicate the calculation of $\varepsilon_{R}$ with all cuts having their standard values.

Figure~\ref{fig:rec:eff_cuts} shows $\varepsilon_{R}$ with various sets of cuts applied.  Note that with no BSMD cuts applied, the efficiency has little dependence on transverse momentum.

Figures~\ref{fig:rec:eff_poe} and~\ref{fig:rec:eff_smd_size} show $\varepsilon_{R}$ calculated with various values for the cuts on $p(rec.)/E_{Tower}$ and the minimum shower sizes in the BSMD-$\eta$ and BSMD-$\phi$ planes, respectively.

\begin{figure}[htbp]
\begin{center}
\includegraphics[width=0.85\linewidth]{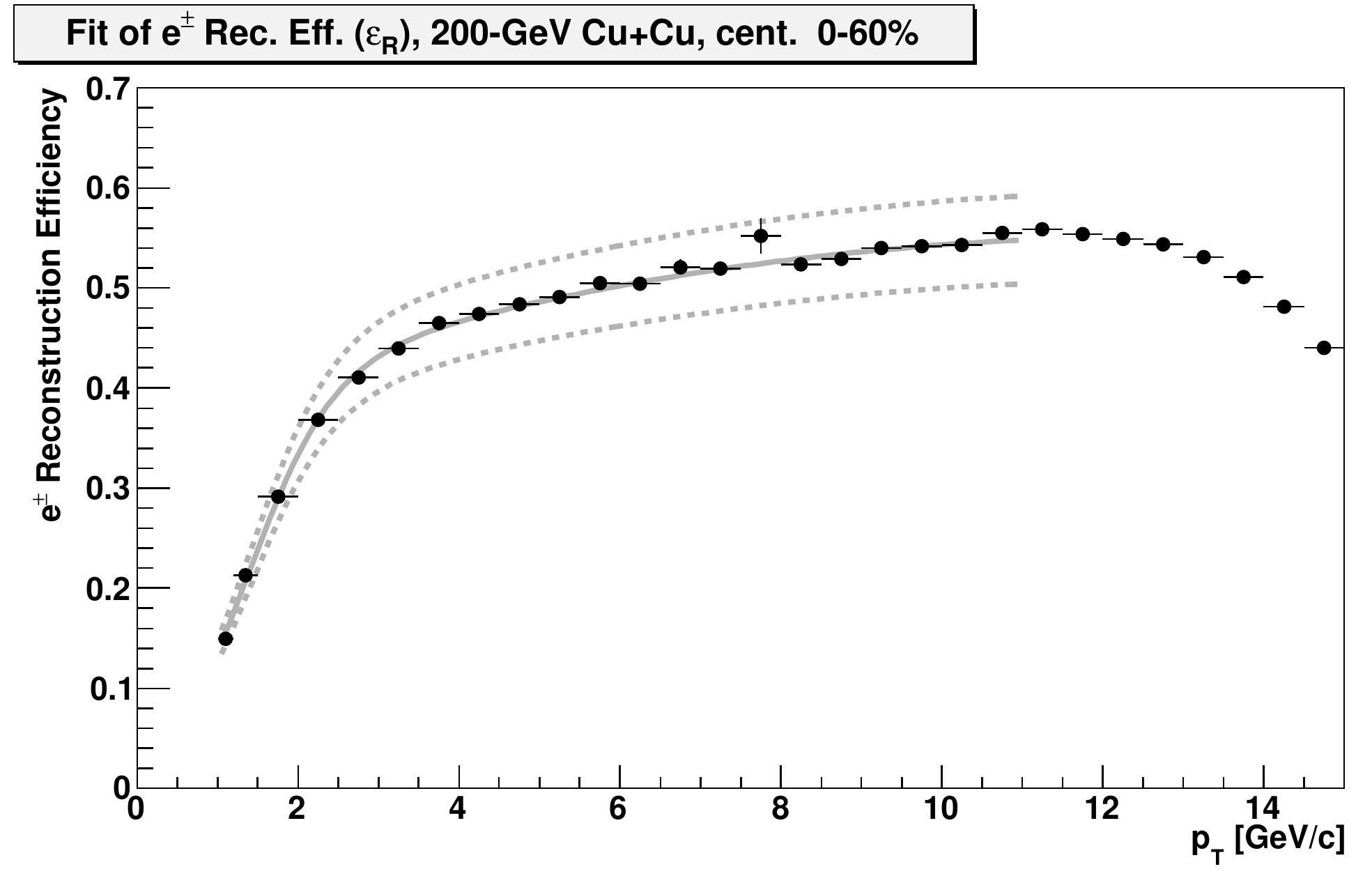}
\caption[Fit of $\varepsilon_{R}$ as a function of $p_{T}$.]{Fit of $\varepsilon_{R}$ as a function of $p_{T}$ with the form of equation~\ref{eq:rec:eff_fit_fun}, a polynomial multiplied by a sigmoid function.  The dashed curves indicate $\pm8\%$ relative systematic uncertainties, see Section~\ref{sec:rec:compare2real}.  See Table~\ref{table:rec:fit_pars} (page~\pageref{table:rec:fit_pars}) for the values of the fit parameters.}
\label{fig:rec:eff_fits}
\end{center}
\end{figure}

Figure~\ref{fig:rec:eff_fits} shows the $e^{\pm}$ reconstruction efficiency fit with a function of the form

\begin{equation}
\label{eq:rec:eff_fit_fun}
\tfrac{1}{2}\lbrace\tanh\left[a_{0}\left(p_{T}-a_{1}\right)\right]+1\rbrace\left(b_{0}+b_{1}p_{T}+b_{2}p_{T}^{2}\right)
\end{equation}

\begin{table}
\caption{Parameters of fits of the $e^{\pm}$ reconstruction efficiency.}
\label{table:rec:fit_pars}
\begin{center}
\begin{tabular}{| r | c |}
\hline
Parameter & Value\\\hline\hline
$a_{0}$ & $0.97539\pm0.02464$\\\hline
$a_{1}$ & $1.3218\pm0.0142$\\\hline
$b_{0}$ & $0.37873\pm0.00811$\\\hline
$b_{1}$ & $(2.6216\pm0.2498)\!\cdot\! 10^{-2}$\\\hline
$b_{2}$ & $(-9.5544\pm1.7578)\!\cdot\! 10^{-4}$\\\hline
$\chi^{2}/NDF$ & 57.2/16\\\hline
\end{tabular}
\end{center}
\end{table}


\noindent The fit parameters are given in Table~\ref{table:rec:fit_pars}.  The dashed curves in Figure~\ref{fig:rec:eff_fits} show the $\pm8\%$ relative systematic uncertainties assumed for $\varepsilon_{R}$ (see Section~\ref{sec:rec:compare2real}).

\clearpage

\clearpage


\section{Comparison of Embedding Data to Real Data and Systematic Uncertainties}
\label{sec:rec:compare2real}

In this section, the $e^{\pm}$ in the embedding data are compared to the $e^{\pm}$ found in the real data; discrepancies between the two data sets are quantified and incorporated into the systematic uncertainties of $\varepsilon_{R}$.  The effect of each $e^{\pm}$ identification cut on $\varepsilon_{R}$ (when all other cuts are applied) is defined to be the the partial efficiency of that cut.  For example, the partial efficiency of the $N_{TPCfit}$ cut is a ratio: the numerator is the number of reconstructed $e^{\pm}$ tracks that pass all $e^{\pm}$ identification cuts; the denominator is the number of reconstructed $e^{\pm}$ tracks that pass all $e^{\pm}$ identification cuts except the $N_{TPCfit}$ cut.  This is equivalent to the ratio of $\varepsilon_{R}$ (with the $N_{TPCfit}$ cut applied) to the value of $\varepsilon_{R}$ calculated without the $N_{TPCfit}$ cut.

\begin{figure}[htbp]
\begin{center}
\includegraphics[width=0.85\linewidth]{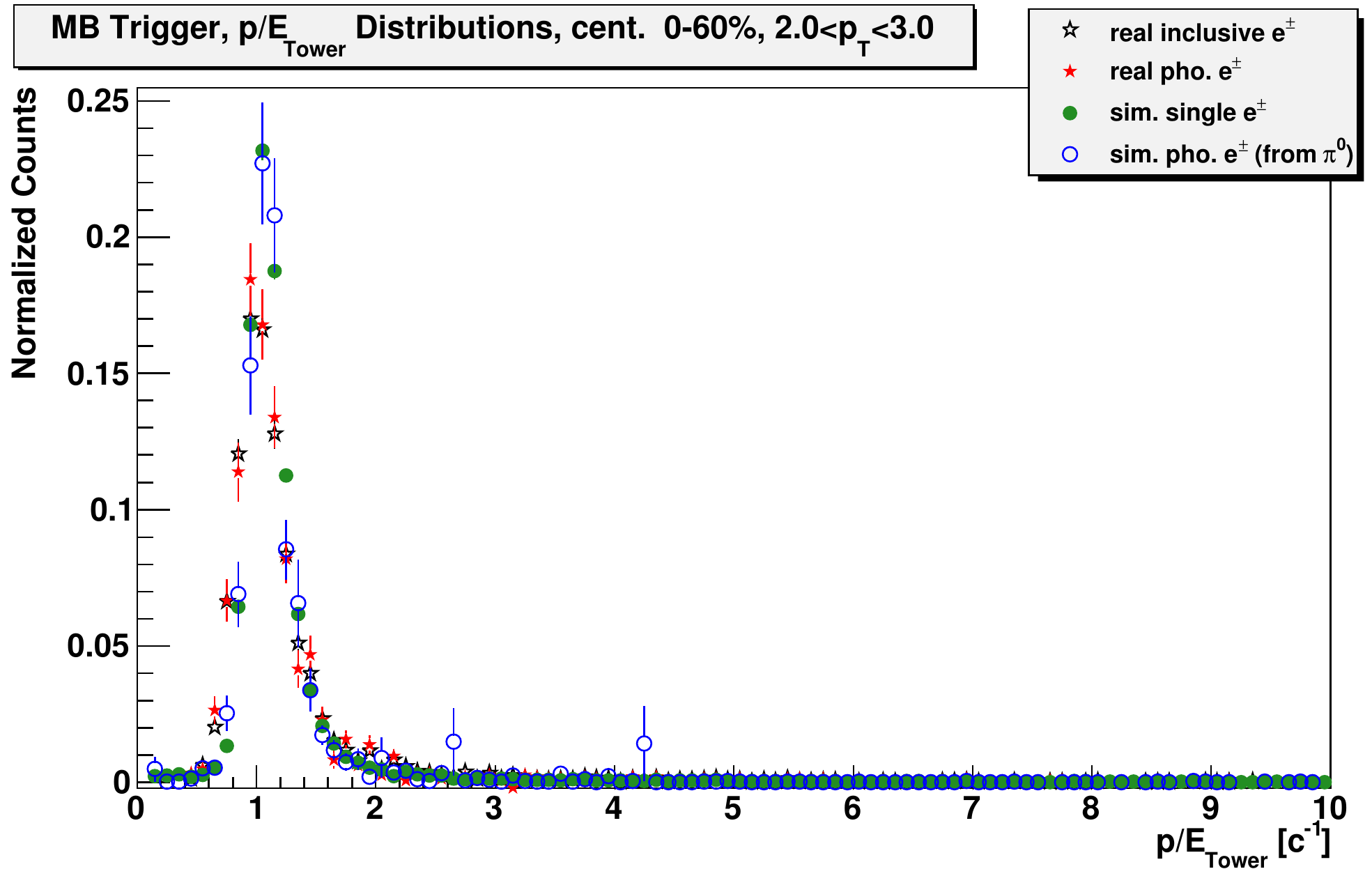}
\caption{Distributions of $p/E_{Tower}$ for real and simulated $e^{\pm}$.}
\label{fig:rec:dist_poe}
\includegraphics[width=0.85\linewidth]{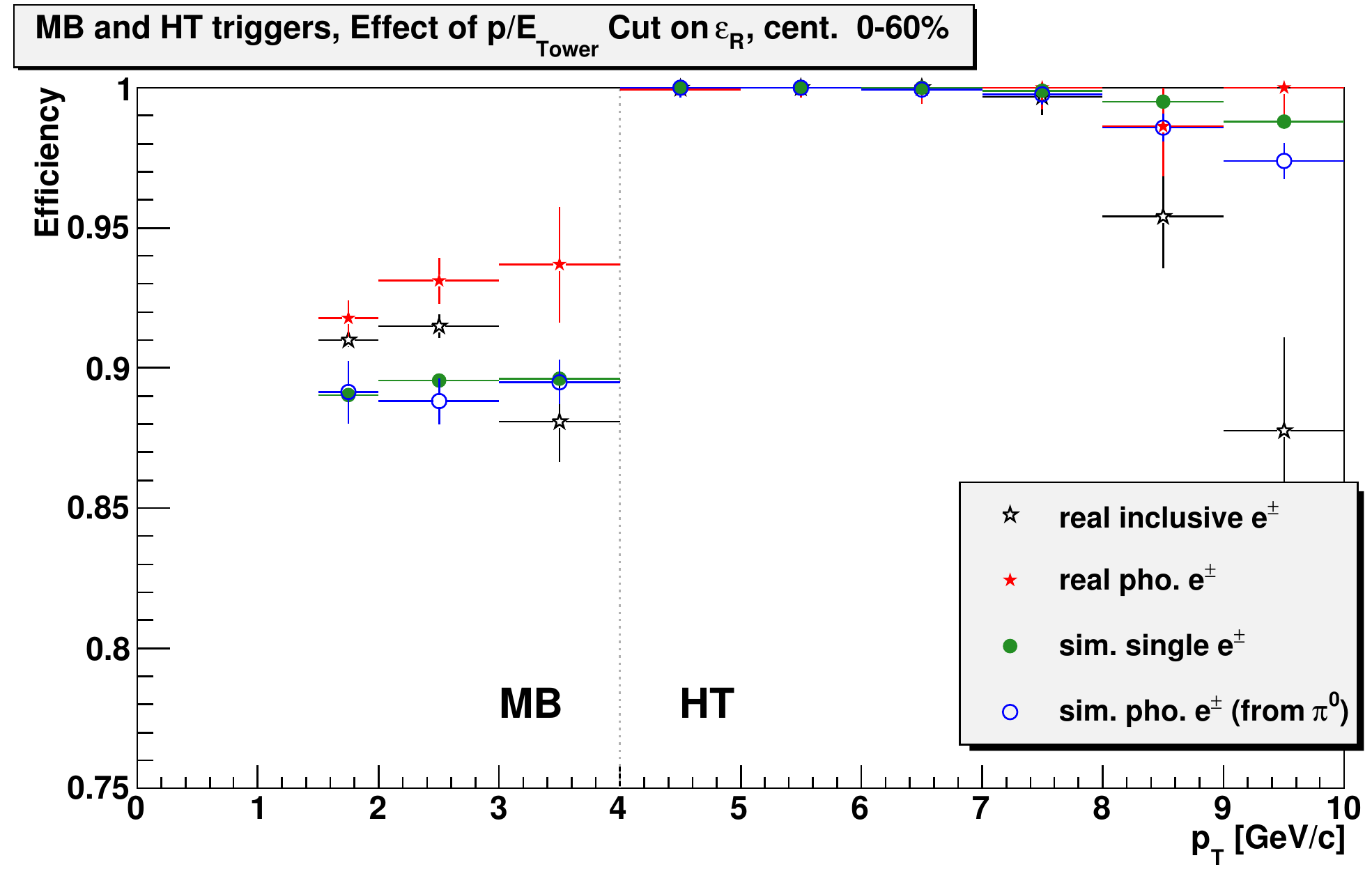}
\caption{Partial efficiencies of the cut $p/E_{Tower}<2/c$.}
\label{fig:rec:par_eff_poe}
\end{center}
\end{figure}

Figure~\ref{fig:rec:dist_poe} shows four distributions of $p/E_{Tower}$: for reconstructed non-photonic $e^{\pm}$ from embedding data set $N_{1}$ (green), reconstructed photonic $e^{\pm}$ from embedding data set $P_{1}$ (blue), inclusive $e^{\pm}$ in real data (black), and photonic $e^{\pm}$ in real data (red, found by subtracting the combinatorial background from the number of unlike-charge pairs).  Note that the distributions have been normalized so that their integrals are all unity.  To produce these distributions, all other $e^{\pm}$ identification cuts were applied (the TPC energy-loss cut was applied to find $e^{\pm}$ in the real data only).  Note that the distribution for inclusive $e^{\pm}$ in the real data does include some hadron contamination (although the purity is high for the $p_{T}$ bin shown); the distribution for photonic $e^{\pm}$ in the real data should not be contaminated by hadrons. 

As shown in Figure~\ref{fig:rec:dist_poe}, the $p/E_{Tower}$ distributions for simulated $e^{\pm}$ are slightly narrower and have maxima at higher values of $p/E_{Tower}$ than the distributions for real $e^{\pm}$.\footnote{These observations apply only for minimum-bias triggered $e^{\pm}$.}  For each transverse-momentum bin, distributions like those shown in Figure~\ref{fig:rec:dist_poe} can be used to calculate the partial efficiency of the $p/E_{Tower}<2/c$ cut; the results of these calculations are shown in Figure~\ref{fig:rec:par_eff_poe}.  If the embedding data sets precisely reproduce the interactions of $e^{\pm}$ with the STAR detector, there should be no difference between the partial efficiencies calculated for real data and embedding.  For the purposes of this section, it is most relevant to compare the partial efficiency for photonic $e^{\pm}$ in the real data (red), which are very pure, to the partial efficiency for $e^{\pm}$ in data set $N_{1}$ (green).  The differences between the real and simulated $p/E_{Tower}$ distributions (Figure~\ref{fig:rec:dist_poe}) for minimum-bias events result in noticeable (though not large) differences in the partial efficiencies.\footnote{For high-tower triggered events, the partial efficiencies are essentially 1 until $p_{T}$ becomes greater than approximately twice the trigger threshold (the value of the threshold is approximately 3.75 GeV).}  These discrepancies will be incorporated into the systematic uncertainty of $\varepsilon_{R}$ (explained below).

\clearpage

\begin{figure}[htbp]
\begin{center}
\includegraphics[width=0.85\linewidth]{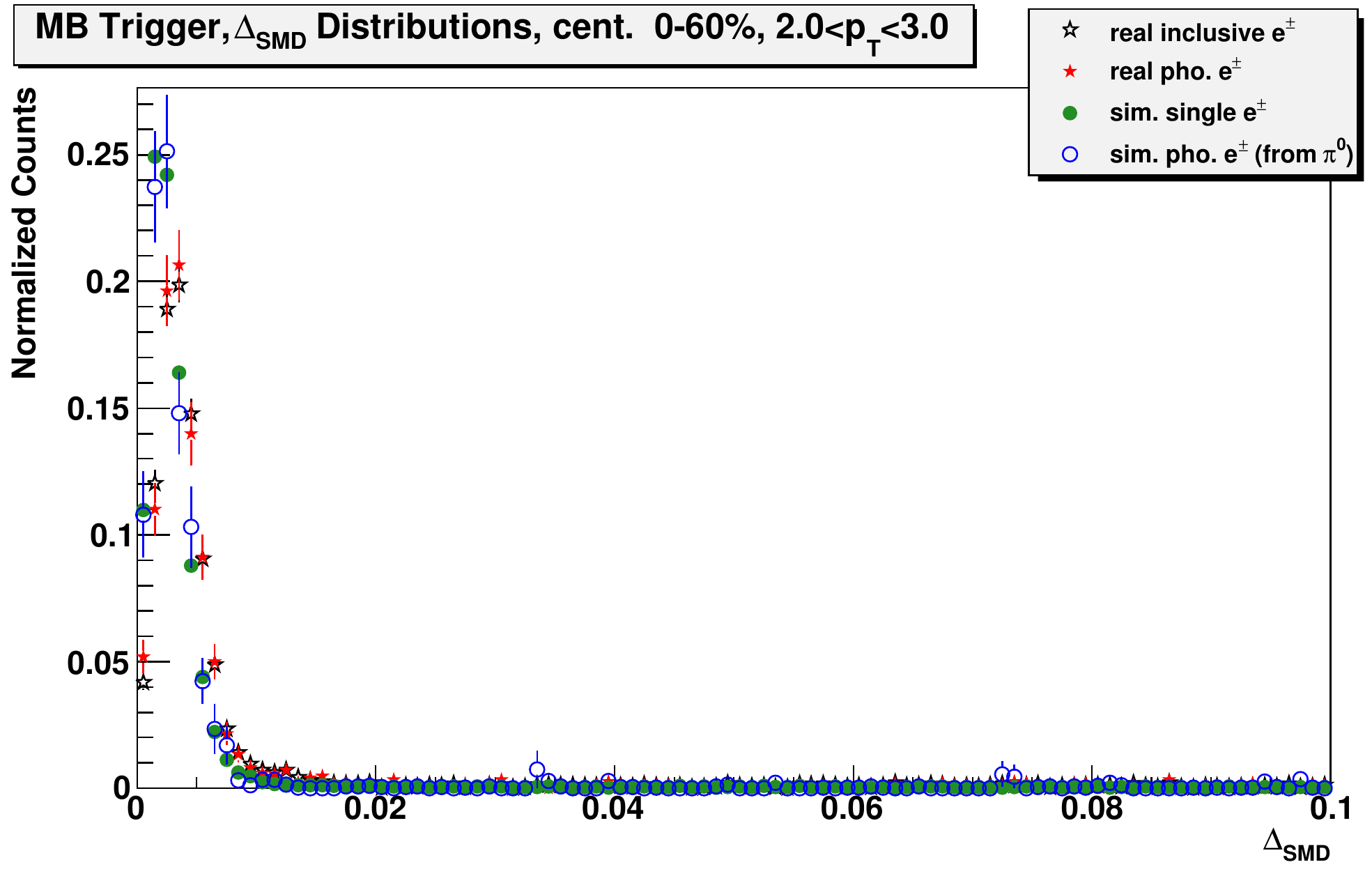}
\caption{Distributions of $\Delta_{SMD}$ for real and simulated $e^{\pm}$.}
\label{fig:rec:dist_dsmd}
\includegraphics[width=0.85\linewidth]{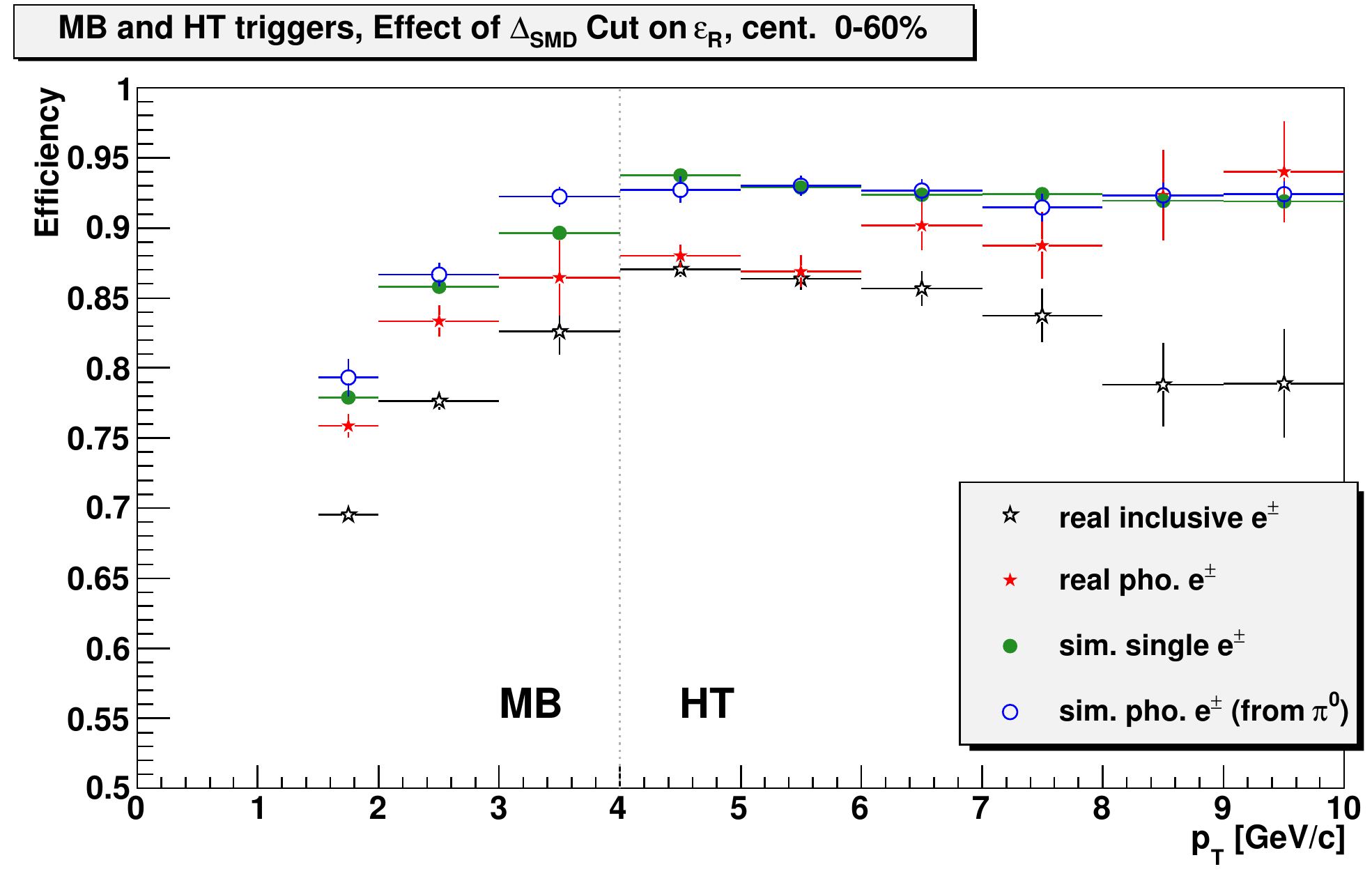}
\caption{Partial efficiencies of the cut $\Delta_{SMD}<0.02$.}
\label{fig:rec:par_eff_dsmd}
\end{center}
\end{figure}

\begin{figure}[htbp]
\begin{center}
\includegraphics[width=0.85\linewidth]{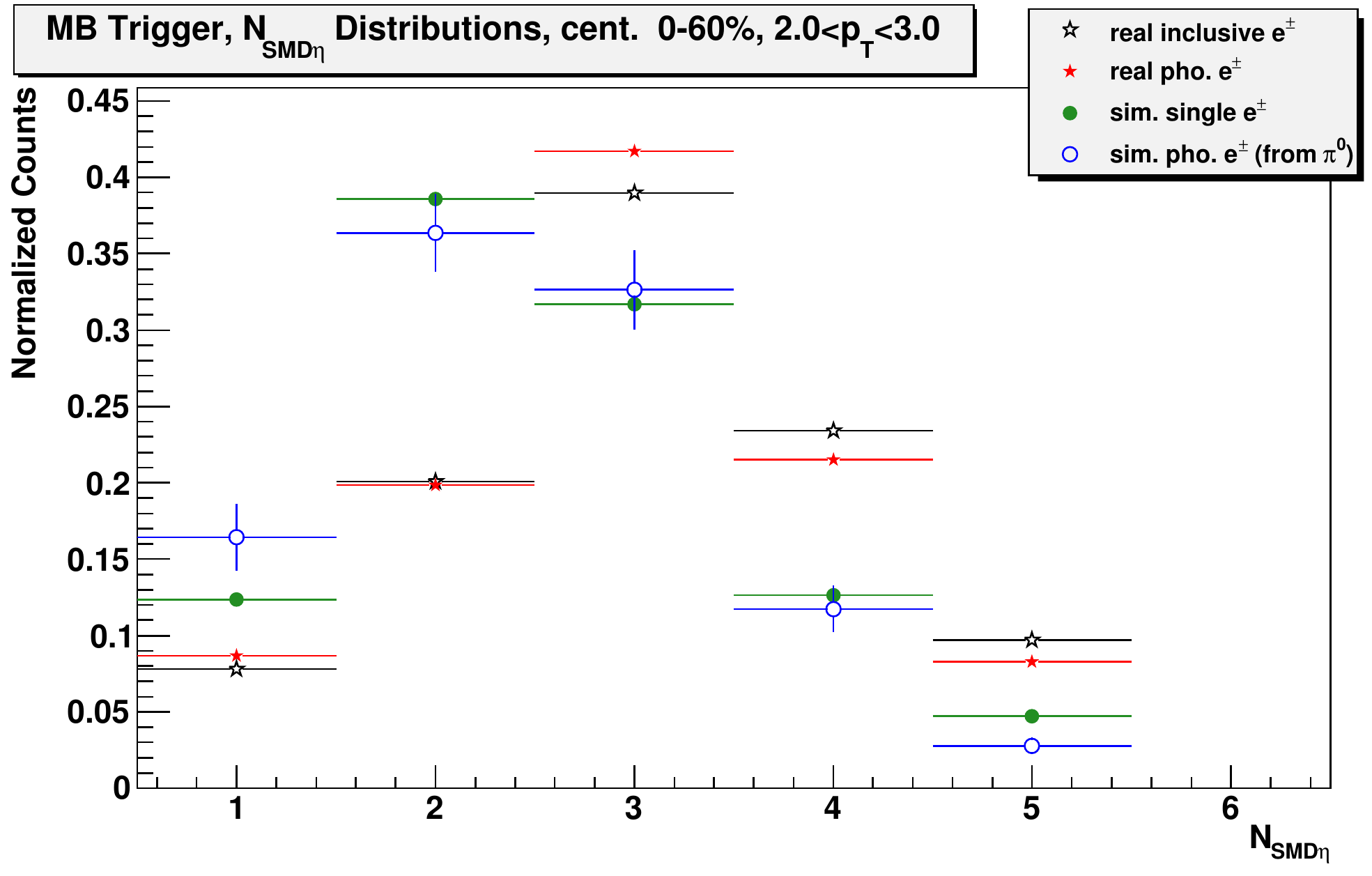}
\caption{Distributions of $N_{SMD\eta}$ for real and simulated $e^{\pm}$.}
\label{fig:rec:dist_smde_size}
\includegraphics[width=0.85\linewidth]{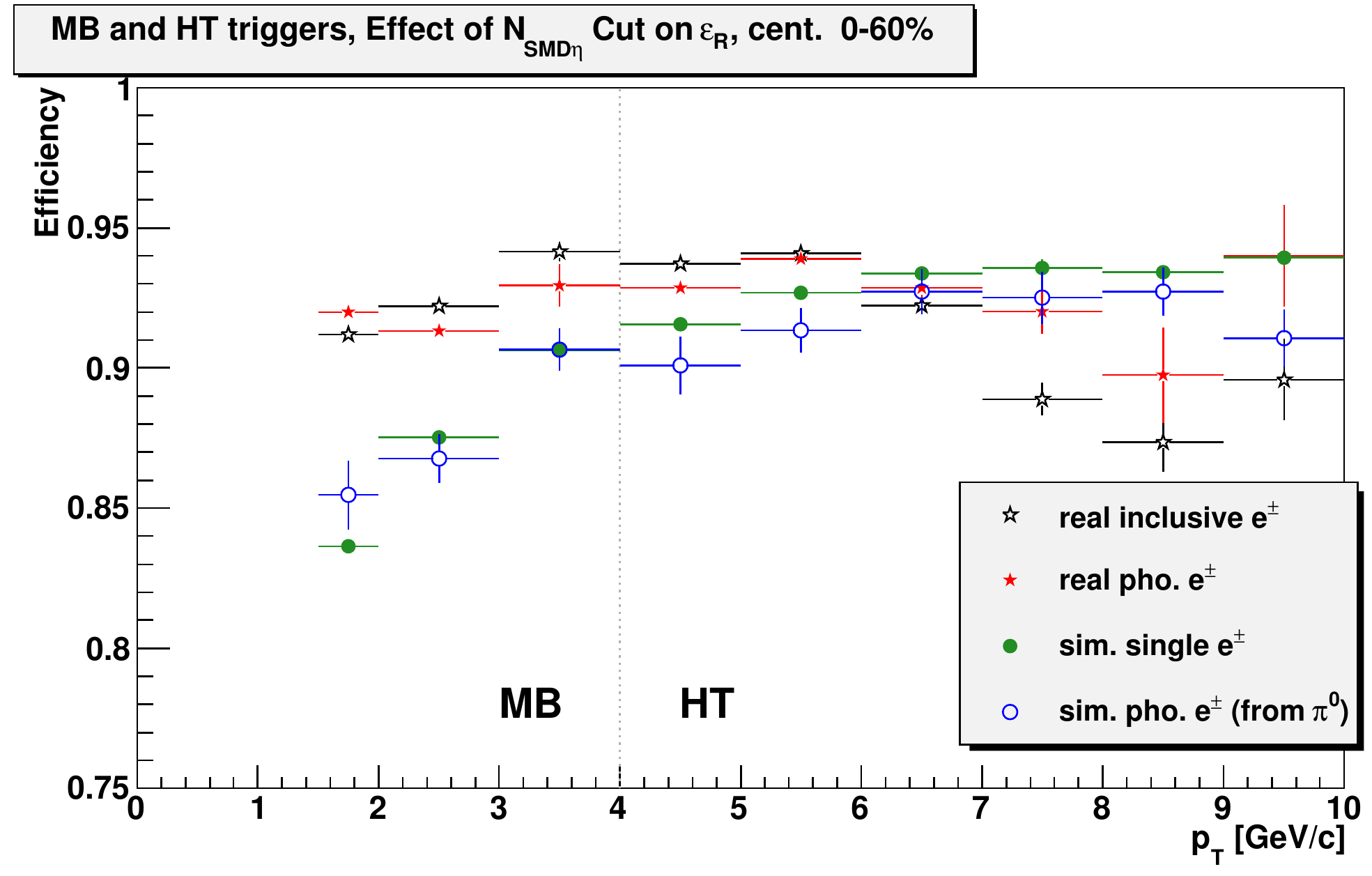}
\caption{Partial efficiencies of the cut $N_{SMD\eta}\geq 2$.}
\label{fig:rec:par_eff_smde_size}
\end{center}
\end{figure}

\begin{figure}[htbp]
\begin{center}
\includegraphics[width=0.85\linewidth]{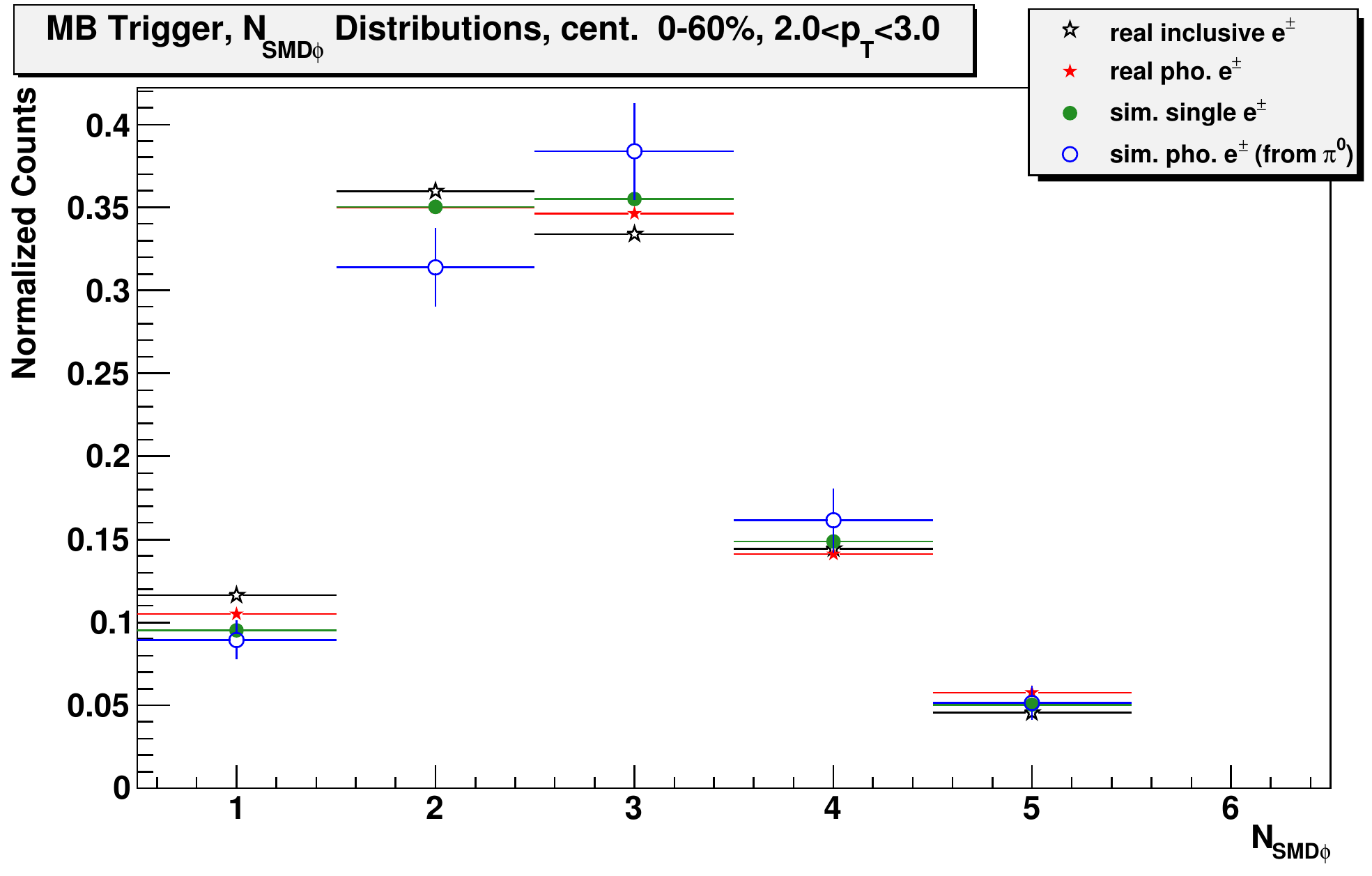}
\caption{Distributions of $N_{SMD\phi}$ for real and simulated $e^{\pm}$.}
\label{fig:rec:dist_smdp_size}
\includegraphics[width=0.85\linewidth]{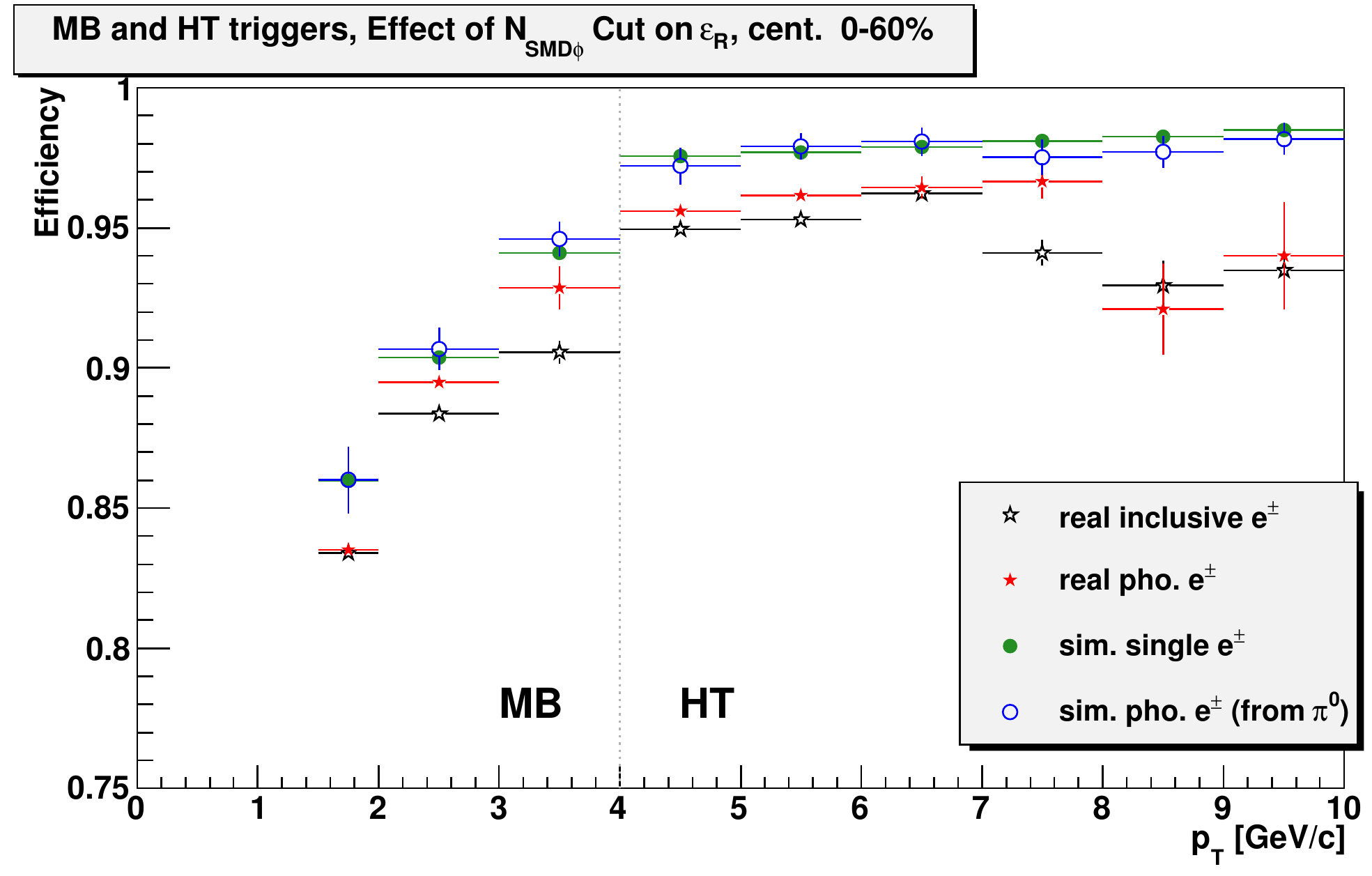}
\caption{Partial efficiencies of the cut $N_{SMD\phi}\geq 2$.}
\label{fig:rec:par_eff_smdp_size}
\end{center}
\end{figure}

\begin{figure}[htbp]
\begin{center}
\includegraphics[width=0.85\linewidth]{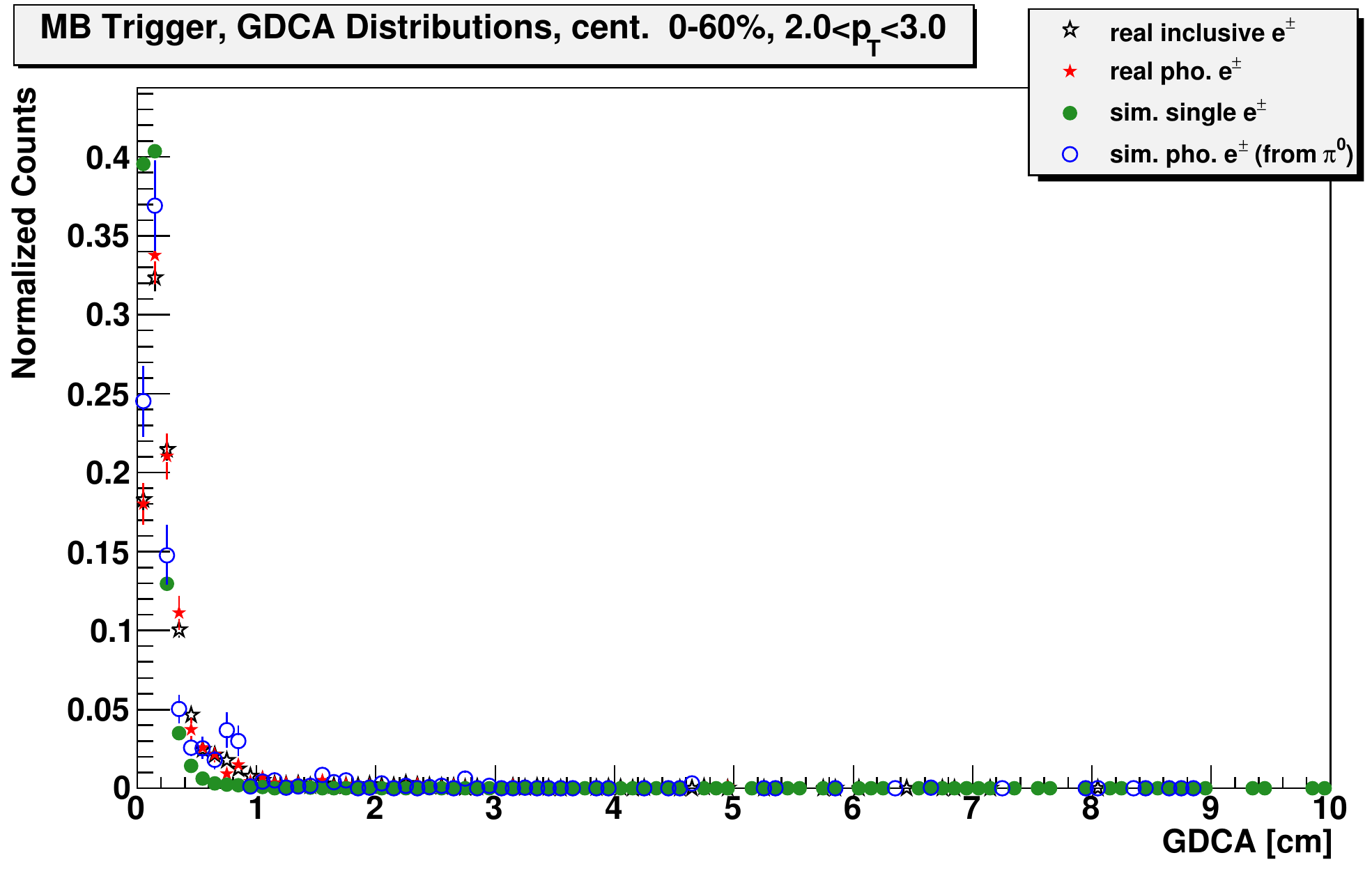}
\caption{Distributions of the global DCA for real and simulated $e^{\pm}$.}
\label{fig:rec:dist_gdca}
\includegraphics[width=0.85\linewidth]{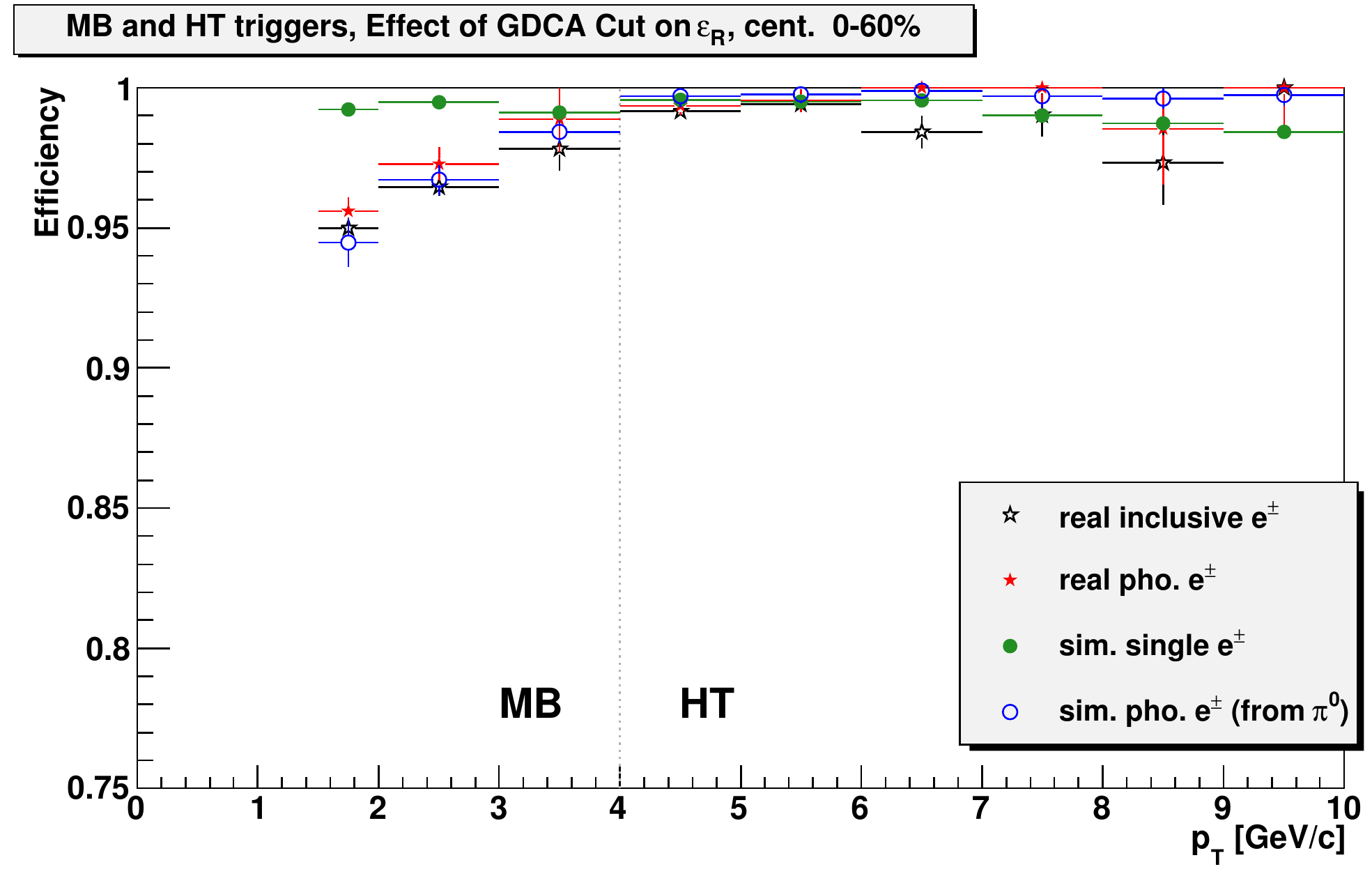}
\caption{Partial efficiencies of the cut $GDCA<1.5$ cm.}
\label{fig:rec:par_eff_gdca}
\end{center}
\end{figure}

The partial efficiencies have been calculated for each of the $e^{\pm}$ identification cuts.  Some of these calculations are presented in this section, while additional material is presented in Section~\ref{sec:add:rec:partial}.  Figures~\ref{fig:rec:dist_poe} through~\ref{fig:rec:par_eff_gdca} show example distributions for each quantity (in one $p_{T}$ bin) and the calculated partial efficiencies for each cut.  Partial efficiencies for minimum-bias $e^{\pm}$ are shown for $p_{T}<4\GeV/c$, while partial efficiencies for high-tower triggered $e^{\pm}$ are shown for $p_{T}>4\GeV/c$.

Figure~\ref{fig:rec:dist_dsmd} shows the distributions of $\Delta_{SMD}=\sqrt{\left(\Delta\eta_{SMD}\right)^{2}+\left(\Delta\phi_{SMD}\right)^{2}}$; the distributions for simulated $e^{\pm}$ have maxima at lower values of $\Delta_{SMD}$ than the distributions for real $e^{\pm}$.  Figure~\ref{fig:rec:par_eff_dsmd} shows the partial efficiencies of the $\Delta_{SMD}<0.02$ cut; the noticeable differences between the real and simulated $\Delta_{SMD}$ distributions do not result in large differences in the partial efficiencies.

Figure~\ref{fig:rec:dist_smde_size} shows the distributions of $N_{SMD\eta}$; the distributions for simulated $e^{\pm}$ have maxima at lower values of $N_{SMD\eta}$ than the distributions for real $e^{\pm}$.  These differences are present for all transverse-momentum bins.  Figure~\ref{fig:rec:par_eff_smde_size} shows the partial efficiencies of the $N_{SMD\eta}\geq 2$ cut; the obvious differences in the $N_{SMD\eta}$ distributions do not lead to large differences in the partial efficiencies (this would not be the case for a more restrictive cut, \textit{e.g.} $N_{SMD\eta}\geq 3$).

Figure~\ref{fig:rec:dist_smdp_size} shows the distributions of $N_{SMD\phi}$; the differences among the $N_{SMD\phi}$ distributions are less pronounced than the differences among the $N_{SMD\eta}$ distributions.  Figure~\ref{fig:rec:par_eff_smdp_size} shows the partial efficiencies of the $N_{SMD\phi}\geq 2$ cut.

Figure~\ref{fig:rec:dist_gdca} shows the distributions of $GDCA$; there are visible differences between the $GDCA$ distributions for real and simulated photonic $e^{\pm}$.  Figure~\ref{fig:rec:par_eff_gdca} shows the partial efficiencies of the $GDCA<1.5$ cm cut.  At low transverse momenta, the differences in the $GDCA$ distribution between non-photonic and photonic $e^{\pm}$ produce a small difference in the partial efficiency of the $GDCA$ cut.  Note that photonic $e^{\pm}$ would be expected to have a wider $GDCA$ distribution than non-photonic $e^{\pm}$.

\begin{figure}[htbp]
\begin{center}
\includegraphics[width=0.85\linewidth]{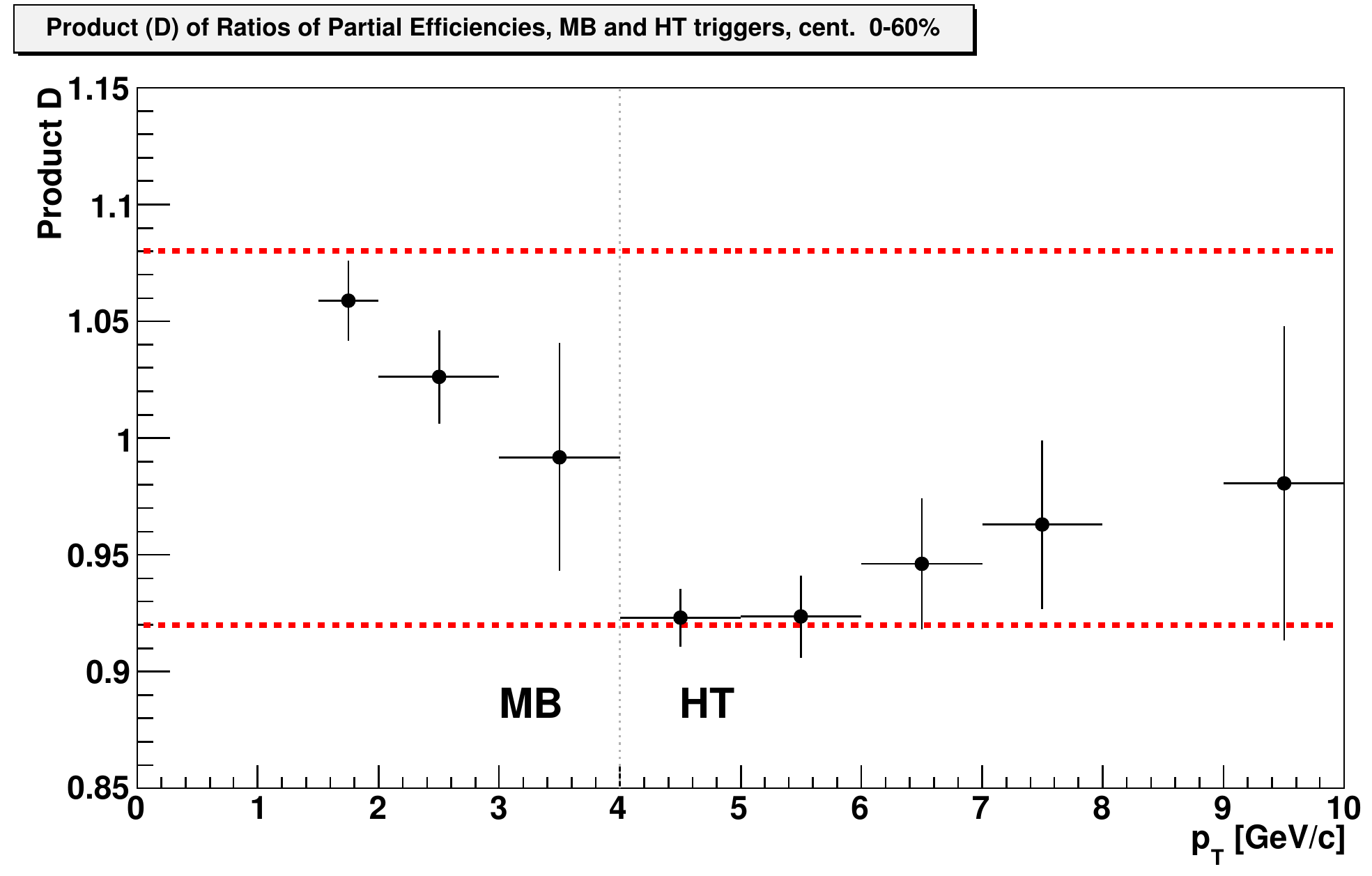}
\caption[The product $(D)$ of the ratios of partial efficiencies for all $e^{\pm}$ identification cuts, used in calculating the systematic uncertainty of $\varepsilon_{R}$.]{The product $(D)$ of the ratios of partial efficiencies for all cuts ($D$ is defined in equation~\ref{eq:rec:true_rec_eff}).  For reasons discussed in the text, $|D-1|$ is treated only as an approximate upper limit of deviations of the calculated value of $\varepsilon_{R}$ from its true value.  With the exception of one $p_{T}$ bin, $D$ lies within the range $1\pm0.08$, therefore $\pm8\%$ relative systematic uncertainties are assumed for $\varepsilon_{R}$.}
\label{fig:rec:partial_ratio_prod}
\end{center}
\end{figure}

For each cut, the discrepancy between the real and embedding data can be quantified by finding the ratio of partial efficiencies $r_{X}(p_{T})/e_{X}(p_{T})$, where $r_{X}$ is the partial efficiency for cut $X$ from the real data (photonic $e^{\pm}$) and $e_{X}$ is the partial efficiency for cut $X$ from embedding data set $N_{1}$.  If the $e^{\pm}$ identification cuts were all independent of each other (\textit{i.e.}, applying or removing cut $X$ has no effect on the efficiency of cut $Y$), the true value of the $e^{\pm}$ reconstruction efficiency could be found by correcting the value of $\varepsilon_{R}$ found from embedding:

\begin{equation}
\varepsilon_{R,true}(p_{T})=\varepsilon_{R}(p_{T})\prod\limits_{X}\frac{r_{X}(p_{T})}{e_{X}(p_{T})}=\varepsilon_{R}(p_{T})D(p_{T}),
\label{eq:rec:true_rec_eff}
\end{equation}

\noindent where $D$ denotes the product for all cuts of the ratios of the partial efficiencies.  In practice, the cuts are probably not all independent of each other and the partial efficiencies for the real data exhibit large statistical fluctuations, so Equation~\ref{eq:rec:true_rec_eff} would not be expected to give a reliable value for the true $e^{\pm}$ reconstruction efficiency.  Also, in some cases (such as the $GDCA$ distribution) differences between the photonic and non-photonic $e^{\pm}$ would be expected, and the use of photonic $e^{\pm}$ to represent all real $e^{\pm}$ would be misleading.  Nevertheless, the product $D$ can be used as an estimated upper limit on the deviation of the true $e^{\pm}$ reconstruction efficiency from the calculated $\varepsilon_{R}$.  Figure~\ref{fig:rec:partial_ratio_prod} shows the product $D$, which is generally within a few percent of unity.  A relative systematic uncertainty of $\pm8\%$ is assumed for the calculated values of $\varepsilon_{R}$ (see also Figure~\ref{fig:rec:eff_fits}).

\clearpage

\clearpage

\clearpage

\clearpage

\clearpage

\clearpage

\chapter[BEMC Geometrical Acceptance and \\ High-Tower Trigger Efficiency]{BEMC Geometrical Acceptance and High-Tower Trigger Efficiency}
\section{BEMC Geometrical Acceptance}
\label{sec:acc:bemc}

This section describes the calculation of correction factors to account for time-dependent changes in the geometrical acceptance of the STAR Barrel Electromagnetic Calorimeter (BEMC).  The data presented in this dissertation were recorded over the course of 33 days.  During that time the geometrical acceptance frequently changed as problems were encountered and the various towers and BSMD strips were turned on and off.  It is necessary to correct the $e^{\pm}$ yields to account for these changes in acceptance.

The BEMC geometrical acceptance $A_{BEMC}$ is calculated for each run (a data-recording period usually lasting from a few minutes to one half hour, during which thousands of collisions are recorded).  For each run, the status of every tower and every BSMD strip is recorded.  The following method is used to determine the status of BSMD strips (a similar method was used to determine the status of BEMC towers).  For every event, the signal in each strip is read out by an Analog-to-Digital Converter (ADC).  The rate at which each strip records a signal greater than five standard deviations above the pedestal\footnote{The pedestal value is the mean ADC signal when no shower is present in the detector.} value is recorded and the mean rate rate for all strips is calculated.  Strips that deviate from this mean rate by less than a factor of two are assigned ``good" status, while strips that deviate above or below the mean rate by more than a factor of two are assigned ``bad" status.

\begin{figure}[htbp]
\begin{center}
\includegraphics[width=1\linewidth]{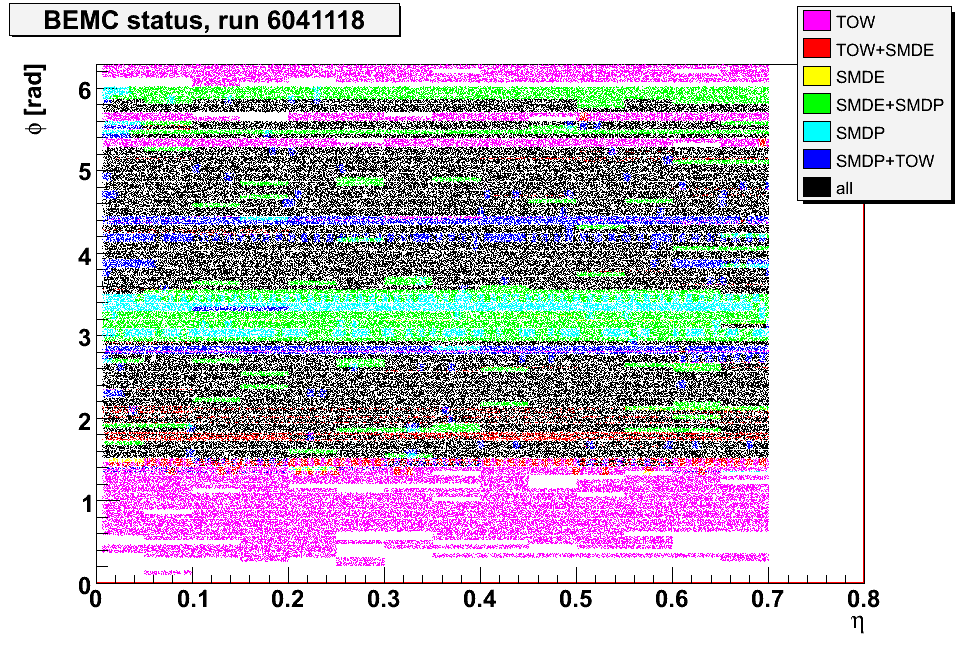}
\caption[The status of the components of the BEMC for one example run.]{The status of the BEMC for one run.  ``TOW" indicates that the tower at the given location has good status; ``SMDE" (``SMDP") indicates that the BSMD-$\eta$ (BSMD-$\phi$) strip at the given location has good status.}
\label{fig:dedx:bemc_acc_example}
\end{center}
\end{figure}

To calculate $A_{BEMC}$, $10^{6}$ points are generated uniformly distributed in pseudorapidity and azimuth in the ranges $0<\eta<0.7$ and $0<\phi<2\pi$.  The BEMC geometrical acceptance is the fraction of those points for which the corresponding tower and both BSMD strips all have good status.  An example of this procedure is shown in figure ~\ref{fig:dedx:bemc_acc_example}.  The mean BEMC geometrical acceptance $\langle A_{BEMC}\rangle$ is calculated by finding the weighted average of $A_{BEMC}$ for all runs, with the weight for each run being the number of events in that run that passed the event selection cuts used in this analysis.  For minimum-bias (high-tower) triggered runs, the mean BEMC geometrical acceptance is 0.634 (0.642).

To avoid possible calibration problems, towers/strips with bad status were excluded from the analysis of the embedding data.  To produce the embedding data sets, simulated events were embedded into a subset of sixteen runs of real events.  The mean acceptance $\langle A_{BEMC}\rangle$ was calculated for two embedding data sets (see Chapter~\ref{sec:rec}) in the manner described above, with the weight for each run being the number of simulated events that passed the event selection cuts.  For data sets $N_{1}$ and $N_{2}$ the values of the mean BEMC geometrical acceptance are $\langle A_{BEMC}(N_{1})\rangle=0.606$ and $\langle A_{BEMC}(N_{2})\rangle=0.613$.  These values will be used to correct the calculated values of the $e^{\pm}$ reconstruction efficiency $(\varepsilon_{R})$ for the BEMC geometrical acceptance in the embedding data.

\clearpage

\section{High-Tower Trigger Efficiency}
\label{sec:trigger}

As described in Section~\ref{sec:anintro:event_selection}, the high-tower (HT) trigger is used to select events in which at least one BEMC tower has energy $\gtrsim 3.75\GeV$.  The HT triggered data set therefore contains an enhancement of high-momentum $e^{\pm}$.  The trigger enhancement is corrected by dividing the $e^{\pm}$ yield by $N_{evt}^{HT}(normalized)$, the number of minimum-bias (MB) triggered events that were observed during the time the HT triggered events were recorded (see Section~\ref{sec:anintro:event_selection}).  However, it is still necessary to correct for the HT trigger efficiency $(\varepsilon_{T})$, which is momentum-dependent.

\begin{figure}[htbp]
\begin{center}
\includegraphics[width=0.85\linewidth]{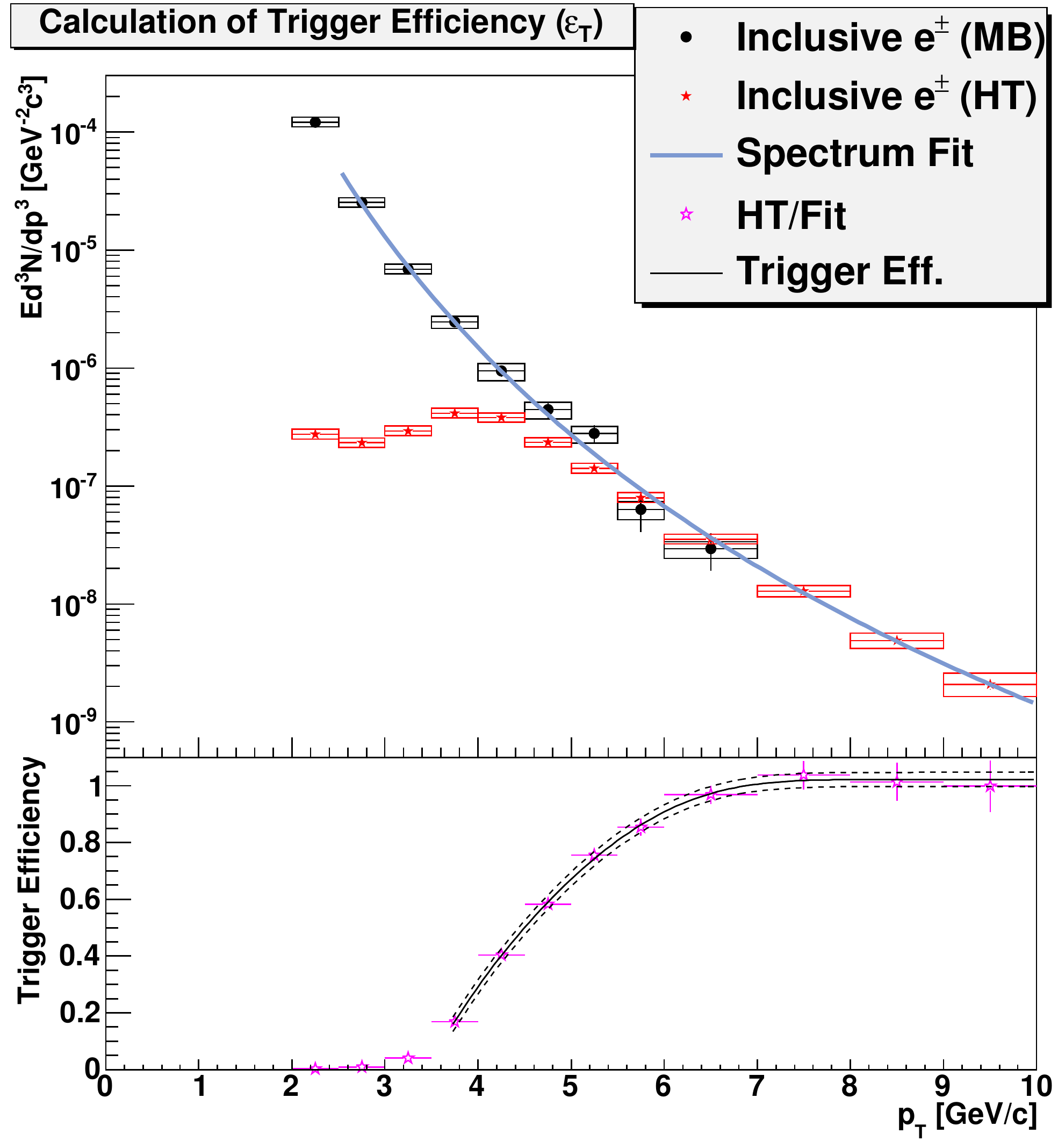}
\caption[Illustration of the calculation of the high-tower trigger efficiency.]{Illustration of the calculation of $\varepsilon_{T}$, the high-tower trigger efficiency.  The upper plot shows the inclusive $e^{\pm}$ spectra for the minimum-bias (black circles) and high-tower triggered (red stars) data sets.  All correction factors have been applied to the HT spectrum except $\varepsilon_{T}$.  A combined spectrum (MB for $p_{T}<5.5\GeV/c$ and HT for $p_{T}\geq5.5\GeV/c$) is fit with a fit function (blue curve).  The lower plot shows the ratio of the HT spectrum to the spectrum fit.  The trigger efficiency is fit with a function of the form of Equation~\ref{eq:trigger:trig_fit}; the systematic uncertainties assigned to the trigger efficiency are shown as dashed lines.  See the text for further explanation.}
\label{fig:trigger:trig_eff}
\end{center}
\end{figure}

Figure~\ref{fig:trigger:trig_eff} illustrates the calculation of $\varepsilon_{T}$.  The efficiency-corrected inclusive $e^{\pm}$ yields for the MB (black) and HT triggered (red) data sets are shown (for the 0-60\% centrality class).  For the HT triggered spectrum, all correction factors except $\varepsilon_{T}$ have been applied.  A combined spectrum, consisting of the MB spectrum for $p_{T}<5.5\GeV/c$ and the HT spectrum otherwise, is fit with a function of the form $A(1+p_{T}/B)^{n}+\exp[-(p_{T}-C)/D]$.  The HT spectrum is divided by this fit (lower panel) and fit with a function of the form\footnote{This function is not a sigmoid and does not describe the scaled HT spectrum at low $p_{T}$.  It does, however, describe the scaled spectrum for $p_{T}>4\GeV/c$.}


\begin{equation}
\label{eq:trigger:trig_fit}
\varepsilon_{T}=A-\exp\sinh\frac{p_{T}-B}{C}.
\end{equation}

\noindent The HT trigger efficiency is assigned an absolute $\pm 2.5\%$ systematic uncertainty, which covers the statistical scatter of the scaled HT histogram.

\begin{figure}[htbp]
\begin{center}
\includegraphics[width=0.85\linewidth]{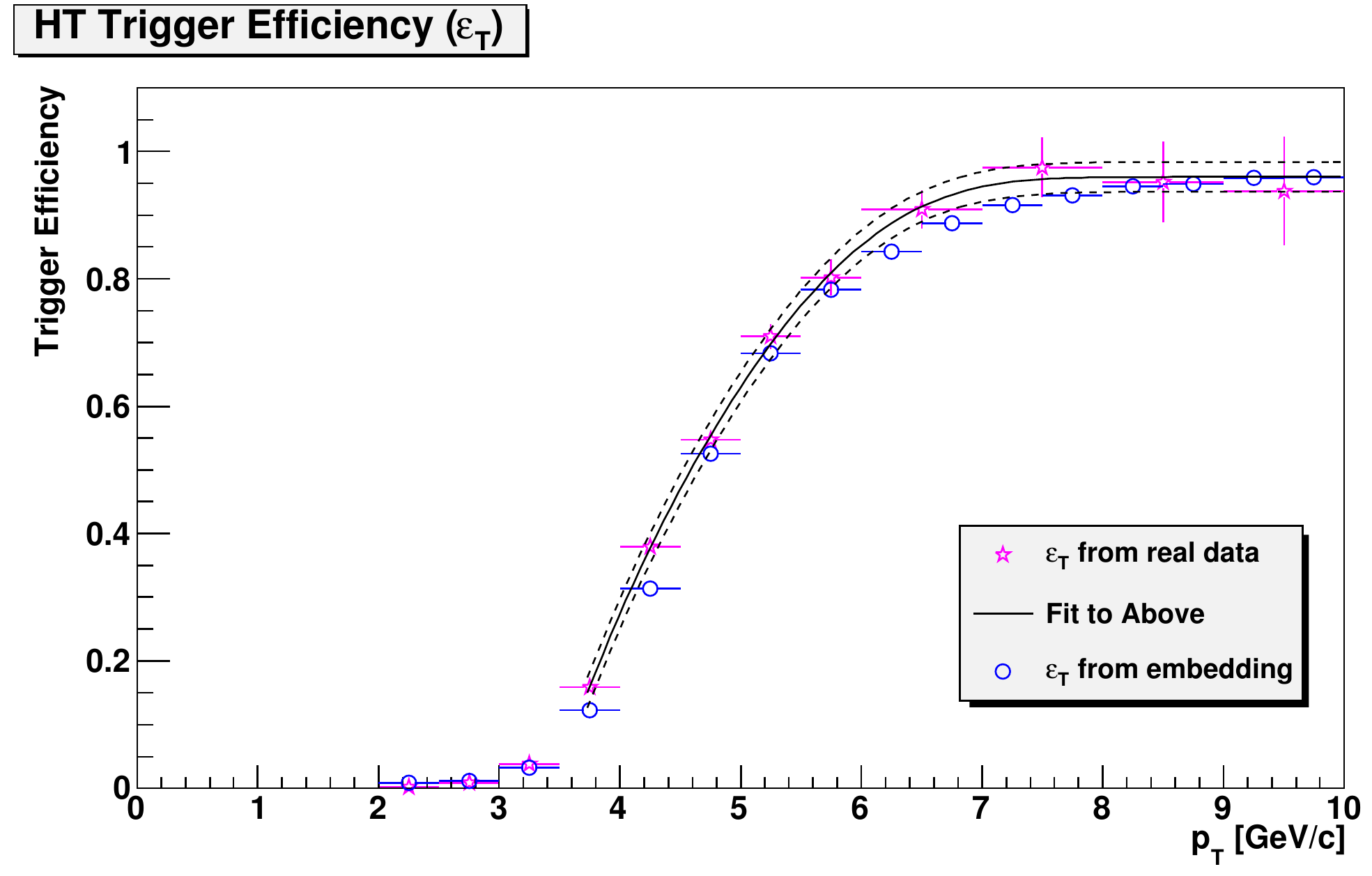}
\caption[HT trigger efficiency calculated from real data (scaled) and embedding.]{The HT trigger efficiency $(\varepsilon_{T})$ vs. transverse momentum calculated using real data (stars) and embedding data (circles).  The efficiency from real data (and the fit thereof) is scaled to match the embedding data at high $p_{T}$.}
\label{fig:trigger:trig_eff_scaled}
\end{center}
\end{figure}

The method described above permits the shape of the trigger efficiency to be determined, but does not determine the overall magnitude of $\varepsilon_{T}$.  The high-$p_{T}$ value of $\varepsilon_{T}$ is estimated using embedding (data set $N_{1}$, see Chapter~\ref{sec:rec}).  The simulated $\varepsilon_{T}$ is the ($p_{T}$-dependent) ratio $N_{rec.,trig.}/N_{rec.}$.  Here, $N_{rec.}$ is the number of simulated $e^{\pm}$ tracks that are reconstructed and identified as $e^{\pm}$.  $N_{rec.,trig.}$ is the subset of those $e^{\pm}$ that fulfill the HT trigger condition.  The simulated $\varepsilon_{T}$ is shown (circles) in Figure~\ref{fig:trigger:trig_eff_scaled}.  The HT trigger efficiency calculated using real data (stars in Figure~\ref{fig:trigger:trig_eff_scaled}) is scaled so that it matches the simulated efficiency at high $p_{T}$.

\clearpage

\chapter{Energy-Loss Cut Efficiency and Purity}
\label{sec:dedx}

\section{Introduction}
\label{sec:dedx:intro}

The energy-loss cut efficiency $(\varepsilon_{dE/dx})$ is calculated by fitting the normalized energy-loss $(n\sigma_{e})$ distribution for photonic $e^{\pm}$ with a Gaussian.  The inclusiveÊ$\;e^{\pm}$ purity $(K_{inc.})$ is calculated using the normalized energy-loss distribution of particles that pass all other $e^{\pm}$ identification cuts: that distribution is fit with multiple Gaussians to account for the contributions of $e^{\pm}$ and various hadron species.  These calculations will be described in greater detail below.

As described in Section~\ref{sec:experiment:tpc:dedx}, $n\sigma_{e}$, the normalized energy loss (relative to the expected value for $e^{\pm}$) is defined as

\begin{equation}
n\sigma_{e}=\ln\left( \frac{\langle dE/dx\rangle}{B_{e}} \right)\frac{1}{\sigma_{e}}.
\end{equation}

\noindent Here, $\langle dE/dx\rangle$ is the measured 70\% truncated mean value of the TPC energy loss, $B_{e}$ is the most probable value of the 70\% truncated mean, and $\sigma_{e}$ is the expected width of the $\ln\left(\langle dE/dx\rangle/B_{e}\right)$ distribution. The factors $B_{e}$ and $\sigma_{e}$ are both derived from Bichsel's simulations~\cite{Bichsel2006154} of particle energy loss in the TPC gas.  Figures~\ref{fig:dedx:nsigma_p_mb} and~\ref{fig:dedx:nsigma_p_ht} show $n\sigma_{e}$ versus momentum for minimum-bias and high-tower triggered triggered events, respectively.  The curves indicate the approximate locations of the most probable value of $n\sigma_{e}$ as a function of momentum for five particle species (see Appendix~\ref{sec:dedx2nsigma} for more information on the calculation of these curves).  Also shown (horizontal lines) are the values of the cut $(-1.5<n\sigma_{e}<3.5)$ used to identify $e^{\pm}$.



\begin{figure}[htbp]
\begin{center}
\includegraphics[width=0.75\linewidth]{nsigma_ee_mb_c03_0}
\caption[Normalized TPC energy loss vs. particle momentum for MB triggered events.]{Normalized TPC energy loss $(n\sigma_{e})$ vs. particle momentum for MB triggered events.  The tracks shown here have passed all $e^{\pm}$ identification cuts except the energy-loss cut.  The curves show the estimated most probable value of $n\sigma_{e}$ for each of five particle species.  The horizontal lines show the range of the $n\sigma_{e}$ cut used in this analysis.}
\label{fig:dedx:nsigma_p_mb}
\includegraphics[width=0.75\linewidth]{nsigma_ee_ht_c03_0}
\caption[Normalized TPC energy loss vs. particle momentum for HT triggered events.]{Normalized TPC energy loss $(n\sigma_{e})$ vs. particle momentum for HT triggered events.  The tracks shown here have passed all $e^{\pm}$ identification cuts except the energy-loss cut.}
\label{fig:dedx:nsigma_p_ht}
\end{center}
\end{figure}

\clearpage

\section{Energy-Loss Cut Efficiency}
\label{sec:dedx:eff}

\begin{figure}[htbp]
\begin{center}
\includegraphics[width=0.85\linewidth]{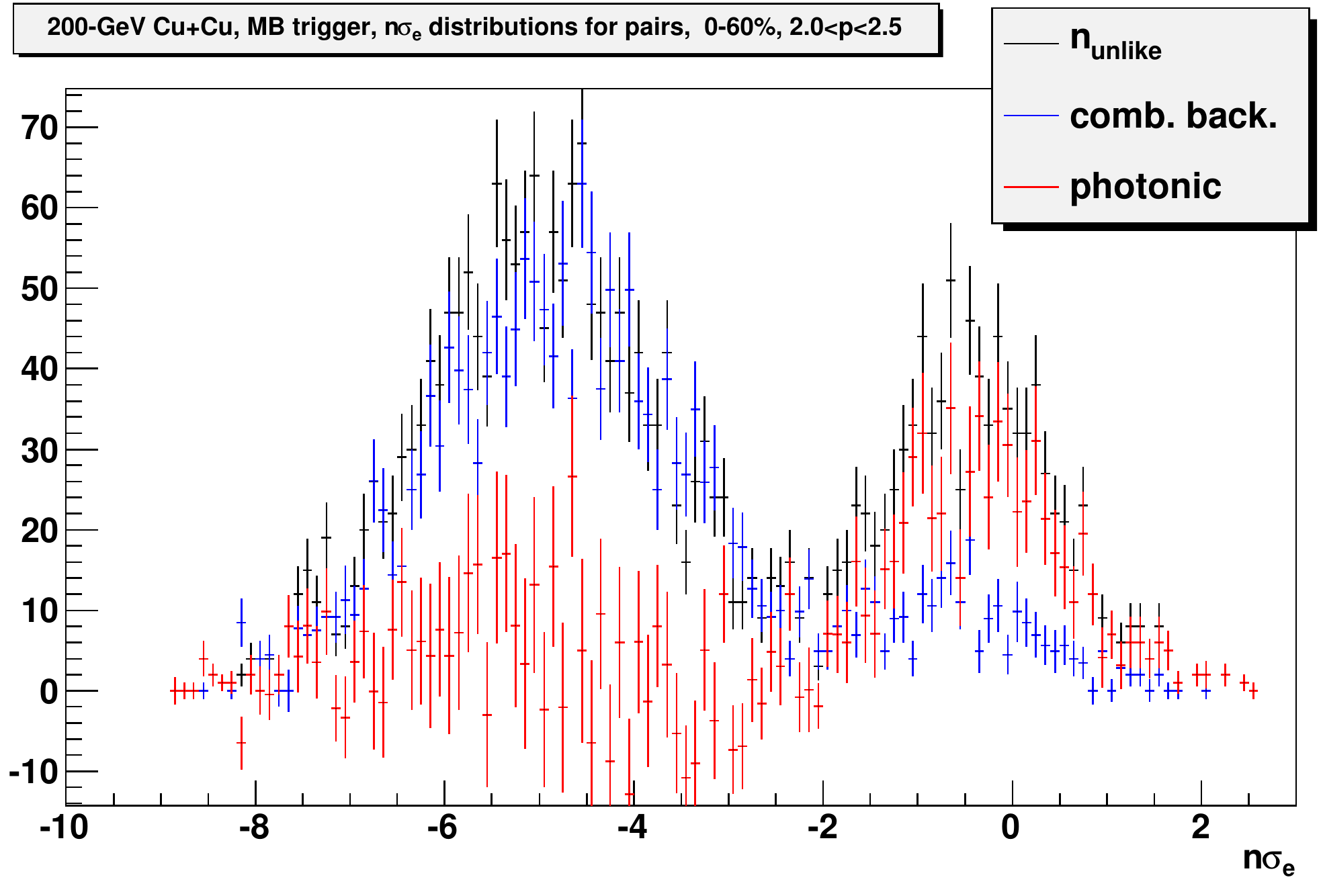}
\caption[Calculation of the $n\sigma_{e}$ distribution for photonic $e^{\pm}$.]{The calculation of the $n\sigma_{e}$ distribution for photonic $e^{\pm}$.  The abscissa is the value of $n\sigma_{e}$ for the $e^{\pm}$ candidate (\textit{i.e.}, the member of the pair that has passed all other $e^{\pm}$ identification cuts).  Shown are the distribution of $n\sigma_{e}$ for unlike-charge pairs (black), the combinatorial background computed from the $n\sigma_{e}$ distributions for like-charge pairs (see Section~\ref{sec:anintro:back_sub}), and the $n\sigma_{e}$ distribution for photonic $e^{\pm}$ (the difference between $n_{unlike}$ and the combinatorial background).}
\label{fig:dedx:nsigma_pho_pairs}
\end{center}
\end{figure}

The efficiency of the energy-loss cut is calculated by fitting the $n\sigma_{e}$ distribution for photonic $e^{\pm}$ with a Gaussian.  Figure~\ref{fig:dedx:nsigma_pho_pairs} illustrates the calculation of the $n\sigma_{e}$ distribution for photonic $e^{\pm}$.  In black is the $n\sigma_{e}$ distribution for pairs of oppositely charged tracks; in blue is the combinatorial background calculated from the $n\sigma_{e}$ distributions for like-charge tracks.  The red histogram shows the difference between the other two: the distribution of $n\sigma_{e}$ for photonic $e^{\pm}$.\footnote{For more information on the background-subtraction method, see Section~\ref{sec:anintro:back_sub}.}  Note that the contribution from hadrons (tracks with negative $n\sigma_{e}$) is largely cancelled out.  This is the reason for using photonic $e^{\pm}$ to determine $\varepsilon_{dE/dx}$: a fit of the $e^{\pm}$ peak is less likely to be biased in the negative direction due to hadron contamination.

\begin{figure}[htbp]
\begin{center}
\includegraphics[width=0.85\linewidth]{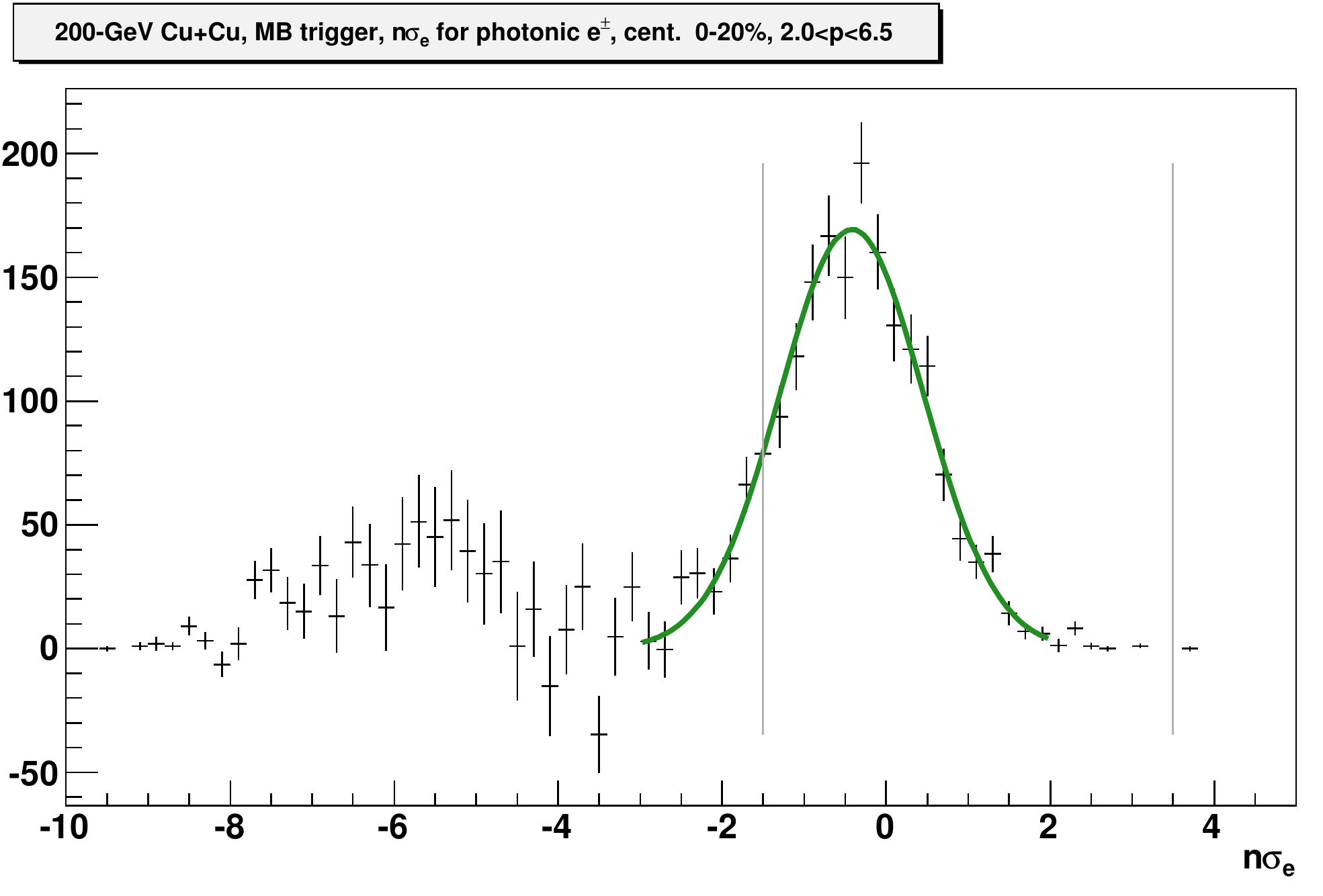}
\caption[Calculation of the energy-loss cut efficiency.]{The $n\sigma_{e}$ distribution for photonic $e^{\pm}$ is fit with a Gaussian.  The cuts on $n\sigma_{e}$ used for this analysis are shown by the vertical lines at $n\sigma_{e}=-1.5$ and $n\sigma_{e}=3.5$; the area of the Gaussian fit within those limits is the energy-loss cut efficiency $\varepsilon_{dE/dx}$.}
\label{fig:dedx:nsigma_pho_pairs_fit}
\end{center}
\end{figure}

\begin{figure}[htbp]
\begin{center}
\includegraphics[width=0.75\linewidth]{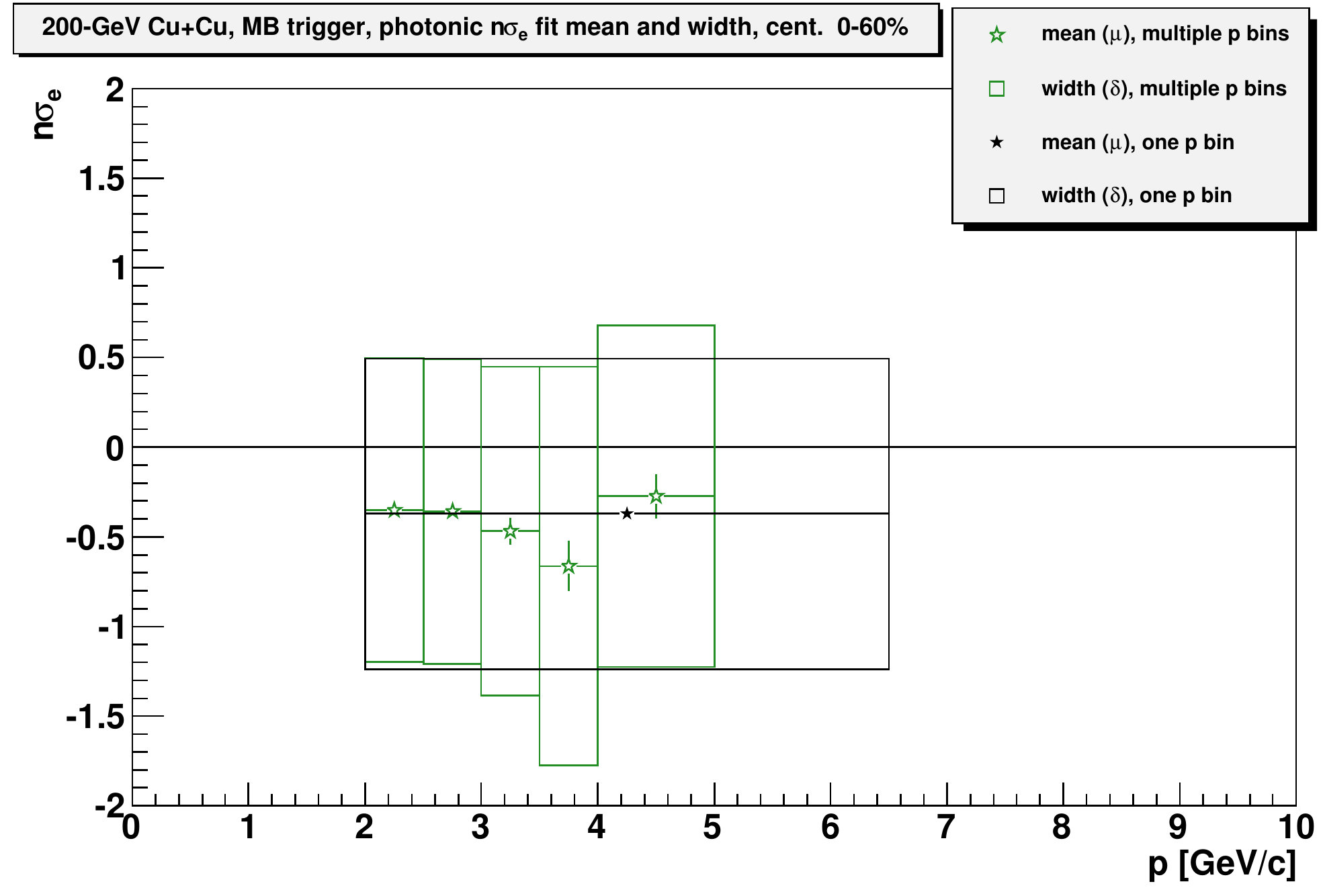}
\caption[Parameters of fits to the $n\sigma_{e}$ distribution for photonic $e^{\pm}$ in MB triggered events.]{Means $(\mu)$ and widths $(\delta)$ of fits to the $n\sigma_{e}$ distribution for photonic $e^{\pm}$ as a function of momentum for minimum-bias triggered events.  The vertical error bars indicate the uncertainties in the fit means.  The boxes indicate the widths of the fits $(\mu\pm\delta)$.}
\label{fig:dedx:pho_fit_pars_mb}
\includegraphics[width=0.75\linewidth]{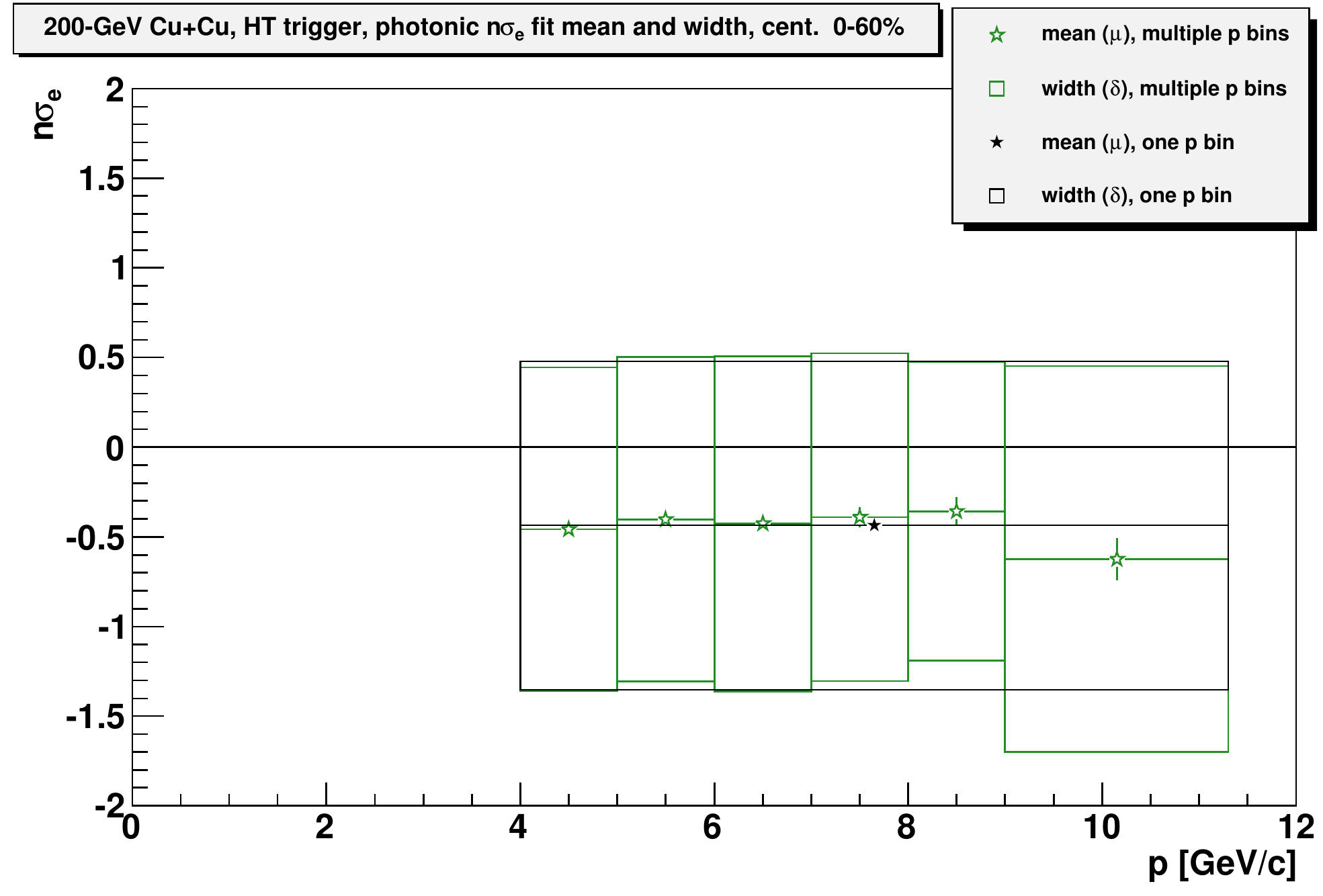}
\caption{Parameters of fits to the $n\sigma_{e}$ distribution for photonic $e^{\pm}$ in HT triggered events.}
\label{fig:dedx:pho_fit_pars_ht}
\end{center}
\end{figure}

Figure~\ref{fig:dedx:nsigma_pho_pairs_fit} shows the $n\sigma_{e}$ distribution for photonic $e^{\pm}$ in minimum-bias collisions (0-60\% centrality class).  For each centrality class and trigger type, the $e^{\pm}$ peak is fit with a Gaussian\footnote{of the form $A\exp\lbrace-\tfrac{1}{2}[(x-\mu)/\delta]^{2}\rbrace$, where $A$ is the amplitude} with mean $\mu$ and width $\delta$.  These fit parameters, which are given in Table~\ref{table:dedx:dedx_eff}, deviate significantly from their expected values ($\mu=0$ and $\delta=1$, respectively), indicating differences in the TPC calibration between the real data and the simulations used to determine the normalization factors $B_{e}$ and $\sigma_{e}$.  The fit parameters for different momentum bins are shown in Figures~\ref{fig:dedx:pho_fit_pars_mb} (for minimum-bias triggered events) and~\ref{fig:dedx:pho_fit_pars_ht} (for high-tower triggered events) for the 0-60\% centrality class.  No momentum dependence is observed.\footnote{For minimum-bias triggered events, the $n\sigma_{e}$ distributions for $3.5\GeV/c<p<4\GeV/c$ have low numbers of entries, leading to fluctuations in the fit parameters.}

The energy-loss cut efficiency is the ratio of the area of the Gaussian fit within the range $-1.5<n\sigma_{e}<3.5$ to the total area of the Gaussian.  For each trigger type, the fit for a single large momentum bin ($2\GeV/c<p<6.5\GeV/c$ for minimum bias, $4\GeV/c<p<11.3\GeV/c$ for high tower) is used to calculate $\varepsilon_{dE/dx}$.  The uncertainties in the efficiency are calculated by changing the fit means and widths from their central values until the fit $\chi^{2}$ has increased by one and again calculating the efficiency.  The uncertainties in the efficiency due to changing the fit mean and changing the fit width are added in quadrature to give the quoted uncertainty.

\begin{table}
\caption{Fit parameters for the photonic $e^{\pm}$ peak and energy-loss cut efficiencies for various centrality classes and event triggers.}
\label{table:dedx:dedx_eff}
\begin{tabular}{| r || c | c | c |}
\hline
& Mean $(\mu)$ & Width $(\delta)$ & $\varepsilon_{dE/dx}$\\\hline\hline
MB, 0-20\% & $-0.41932\pm 0.03007$ & $0.87788\pm 0.02519$ & $0.89084^{+0.02011}_{-0.01992}$\\\hline
MB, 20-60\% & $-0.33308\pm 0.02638$ & $0.86441\pm 0.02268$ & $0.91148^{+0.01899}_{-0.01878}$\\\hline
MB, 0-60\% & $-0.37130\pm 0.02070$ & $0.86616\pm 0.01769$ & $0.90373^{+0.01456}_{-0.01444}$\\\hline\hline
HT, 0-20\% & $-0.46847\pm 0.01945$ & $0.92948\pm 0.01587$ & $0.86645^{+0.01164}_{-0.01160}$\\\hline
HT, 20-60\% & $-0.41376\pm 0.02059$ & $0.91234\pm 0.01689$ & $0.88309^{+0.01284}_{-0.01277}$\\\hline
HT, 0-60\% & $-0.43650\pm 0.01435$ & $0.91620\pm 0.01174$ & $0.87712^{+0.00882}_{-0.00879}$\\\hline

\end{tabular}
\end{table}

\clearpage

\section{Purity}
\label{sec:dedx:purity}

\begin{figure}[htbp]
\begin{center}
\includegraphics[width=0.85\linewidth]{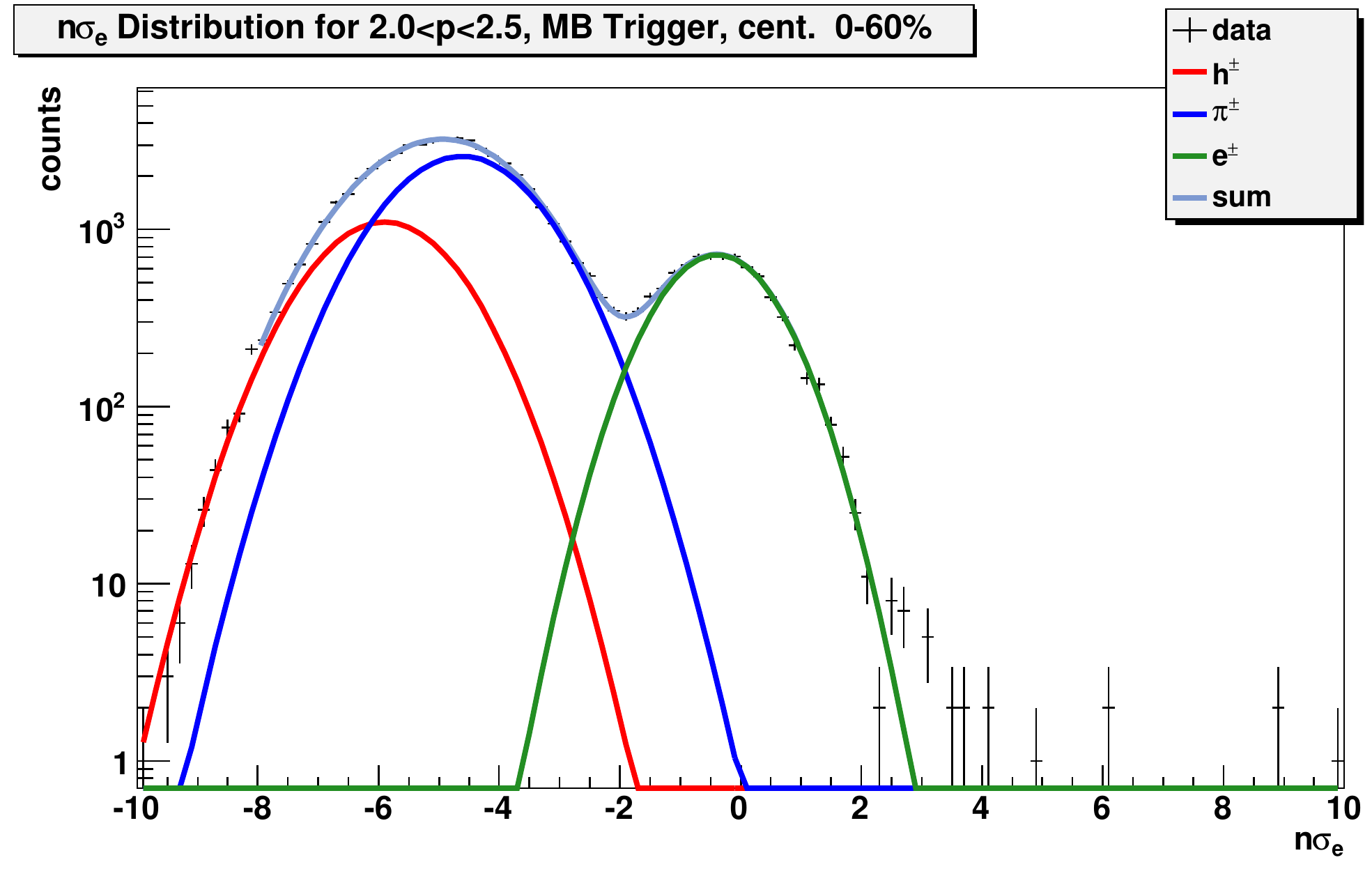}
\caption[Illustration of the multi-Gaussian fitting procedure used to determine the purity of the $e^{\pm}$ sample.]{The $n\sigma_{e}$ distribution for tracks that have passed all other $e^{\pm}$ identification cuts is fit with three Gaussians, to account for the contributions of $e^{\pm}$ (green peak), $\pi^{\pm}$ (blue), and other hadron species (red).}
\label{fig:dedx:3g_fit}
\end{center}
\end{figure}

\begin{figure}[htbp]
\begin{center}
\includegraphics[width=0.85\linewidth]{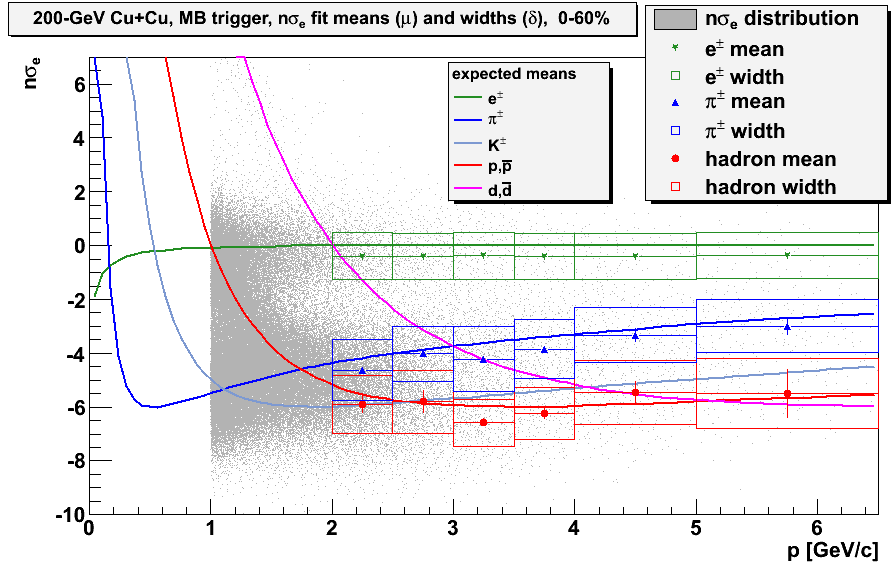}
\caption[Multi-Gaussian fit parameters for MB triggered events.]{Means $(\mu)$ and widths $(\delta)$ of the three Gaussians used to fit the $n\sigma_{e}$ distributions for particles that pass the other $e^{\pm}$ identification cuts for minimum-bias triggered events.  The vertical error bars indicate the uncertainties in the fit means.  The boxes indicate the widths of the fits $(\mu\pm\delta)$.}
\label{fig:dedx:inc_fit_pars_mb}
\includegraphics[width=0.85\linewidth]{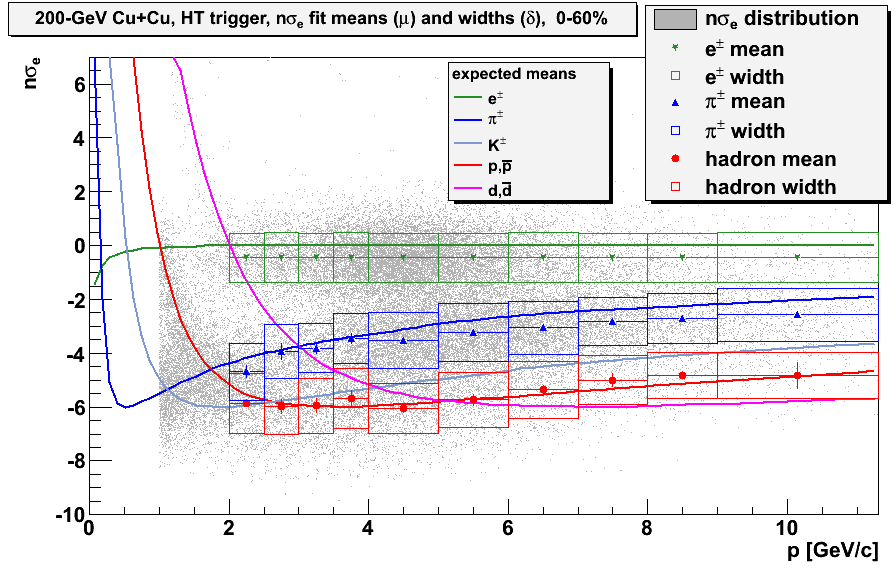}
\caption[Multi-Gaussian fit parameters for HT triggered events.]{Means and widths of the three Gaussians used to fit the $n\sigma_{e}$ distributions for particles that pass the other $e^{\pm}$ identification cuts for high-tower triggered events.}
\label{fig:dedx:inc_fit_pars_ht}
\end{center}
\end{figure}

The inclusive $e^{\pm}$ purity $(K_{inc.})$ is calculated by fitting with three Gaussians the $n\sigma_{e}$ distribution for all tracks that pass the other $e^{\pm}$ identification cuts, as illustrated in Figure~\ref{fig:dedx:3g_fit}.  The Gaussian centered near $n\sigma_{e}=0$ accounts for the contribution from $e^{\pm}$ and its mean and width are constrained to the values found from the (very pure) photonic $e^{\pm}$ $n\sigma_{e}$ distributions described in Section~\ref{sec:dedx:eff}.  The other two Gaussians account for the contributions from hadrons, expected to be predominantly $\pi^{\pm}$, $K^{\pm}$, $p$, and $\bar{p}$.  The middle (blue) peak is intended to account primarily for charged pions (although it is not possible to guarantee that this peak accounts for all pions and only pions).  Such three-Gaussian fits are performed for multiple momentum bins.  When possible, the hadron peak means and widths are allowed to vary freely.  However, it is occasionally necessary to constrain hadron fit parameters to avoid nonsensical results or deviations from the expected behavior of the fit (\textit{e.g.}, the mean of the pion peak suddenly decreasing as $p$ increases, or a sudden change in a peak width).  Figures~\ref{fig:dedx:inc_fit_pars_mb} and~\ref{fig:dedx:inc_fit_pars_ht} show the parameters of these fits for all momentum bins for the minimum-bias and high-tower-triggered data sets, along with approximate expected peak locations for five particle species.  The uncertainties of the hadron fit parameters are found by varying the value of each parameter until the fit $\chi^{2}$ value increases by one.

\begin{figure}[htbp]
\begin{center}
\includegraphics[width=0.85\linewidth]{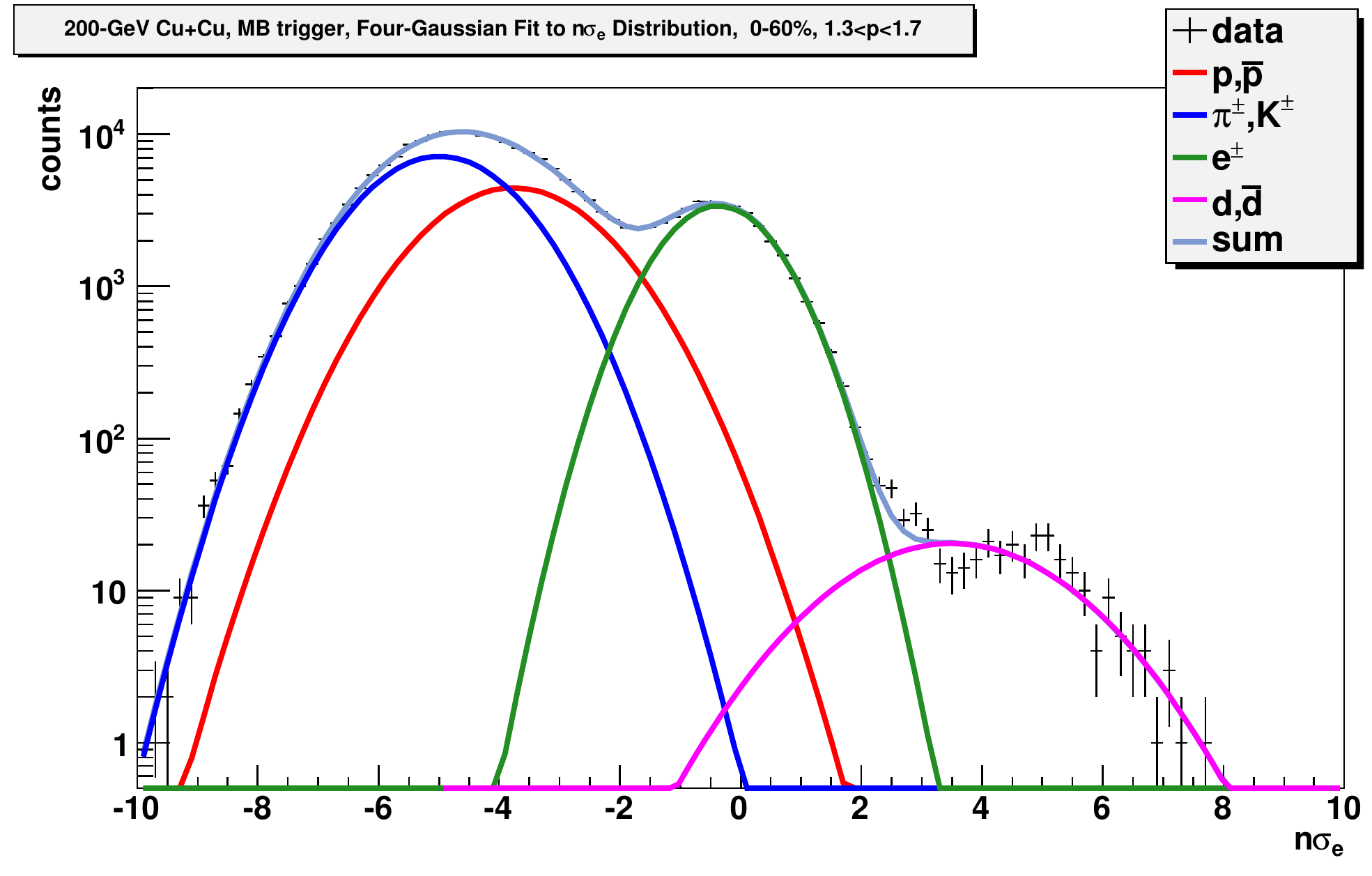}
\caption[Four-Gaussian $n\sigma_{e}$ fit used in estimation of (anti)deuteron yield.]{In a procedure similar to that illustrated in Figure~\ref{fig:dedx:3g_fit}, the $n\sigma_{e}$ distribution in the momentum bin $1.3\GeV/c<p<1.7\GeV/c$ is fit with four Gaussians, with the fourth (magenta) peak used to account for the contributions of (anti)deuterons.  The ratio of the areas of the $e^{\pm}$ and $\bar{d}(d)$ peaks is used to estimate the $\bar{d}(d)$ contamination in the $e^{\pm}$ sample for the momentum bin $2\GeV/c<p<2.5\GeV/c$.}
\label{fig:dedx:4g_fit}
\end{center}
\end{figure}

Figure~\ref{fig:dedx:nsigma_p_mb} indicates that the (anti)deuteron band crosses through the range $-1.5<n\sigma_{e}<3.5$ above $p=2\GeV/c$, resulting in contamination of the $e^{\pm}$ that is not accounted for in the three-Gaussian fit.  Figure~\ref{fig:dedx:4g_fit} shows a four-Gaussian fit to the $n\sigma_{e}$ distribution for the momentum range $1.3\GeV/c<p<1.7\GeV/c$, where there is still a decent separation between the $e^{\pm}$ and (anti)deuteron peaks.  The $\bar{d}(d)$ peak is allowed to have a large width to account for the fact that the mean value of $n\sigma_{e}$ for deuterons is not constant and changes significantly across that momentum bin.  The ratio of the areas of those two peaks is 0.011.  This ratio is used to estimate the (anti)deuteron yield in the momentum range $2\GeV/c<p<2.5\GeV/c$.

\clearpage

For a given momentum range, the $n\sigma_{e}$ distributions for the 0-20\% and 20-60\% centrality classes are not fit independently of each other.  Instead, the fit for the full 0-60\% centrality class is used to constrain the fits for the 0-20\% and 20-60\% centrality classes.  As an example, consider a single momentum bin for the minimum-bias triggered data set.  The $n\sigma_{e}$ distribution in the full 0-60\% centrality class is fit with the function $f$, which consists of three Gaussians with amplitudes $A_{j}$ (where $j=1,2,3$), means $\mu_{j}$, and widths $\delta_{j}$.  The $n\sigma_{e}$ distribution in the 0-20\% centrality class is fit with the function $f^{\prime}$, with parameters $A^{\prime}_{j}$, $\mu^{\prime}_{j}$, and $\delta^{\prime}_{j}$.  The values of some of the parameters of $f^{\prime}$ are constrained by the values of the parameters of $f$.  Each amplitude $A^{\prime}_{j}$ is allowed to vary freely.  Each width $\delta^{\prime}_{j}$ is constrained to lie within the range $\delta_{j}-\Delta(\delta_{j})\leq\delta^{\prime}_{j}\leq\delta_{j}+\Delta(\delta_{j})$, where $\Delta(\delta_{j})$ indicates the uncertainty of $\delta_{j}$.  Each mean $\mu^{\prime}_{j}$ is constrained to be within the range

\begin{equation}
\mu_{j}-\Delta(\mu_{j})-|s|\leq\mu^{\prime}_{j}\leq\mu_{j}+\Delta(\mu_{j}),\;\;\;\;\;(s<0)
\end{equation}

\noindent or

\begin{equation}
\mu_{j}-\Delta(\mu_{j})\leq\mu^{\prime}_{j}\leq\mu_{j}+\Delta(\mu_{j})+s,\;\;\;\;\;(s>0).
\end{equation}

\noindent The variable $s$ is the shift in the electron peak mean between the 0-60\% centrality class and the 0-20\% centrality class, which is determined by fitting the $n\sigma_{e}$ distribution for photonic $e^{\pm}$ (see Section~\ref{sec:dedx:eff}).  The same procedure is also used to constrain the fit parameters for the 20-60\% centrality class.

\begin{figure}[htbp]
\begin{center}
\includegraphics[width=0.85\linewidth]{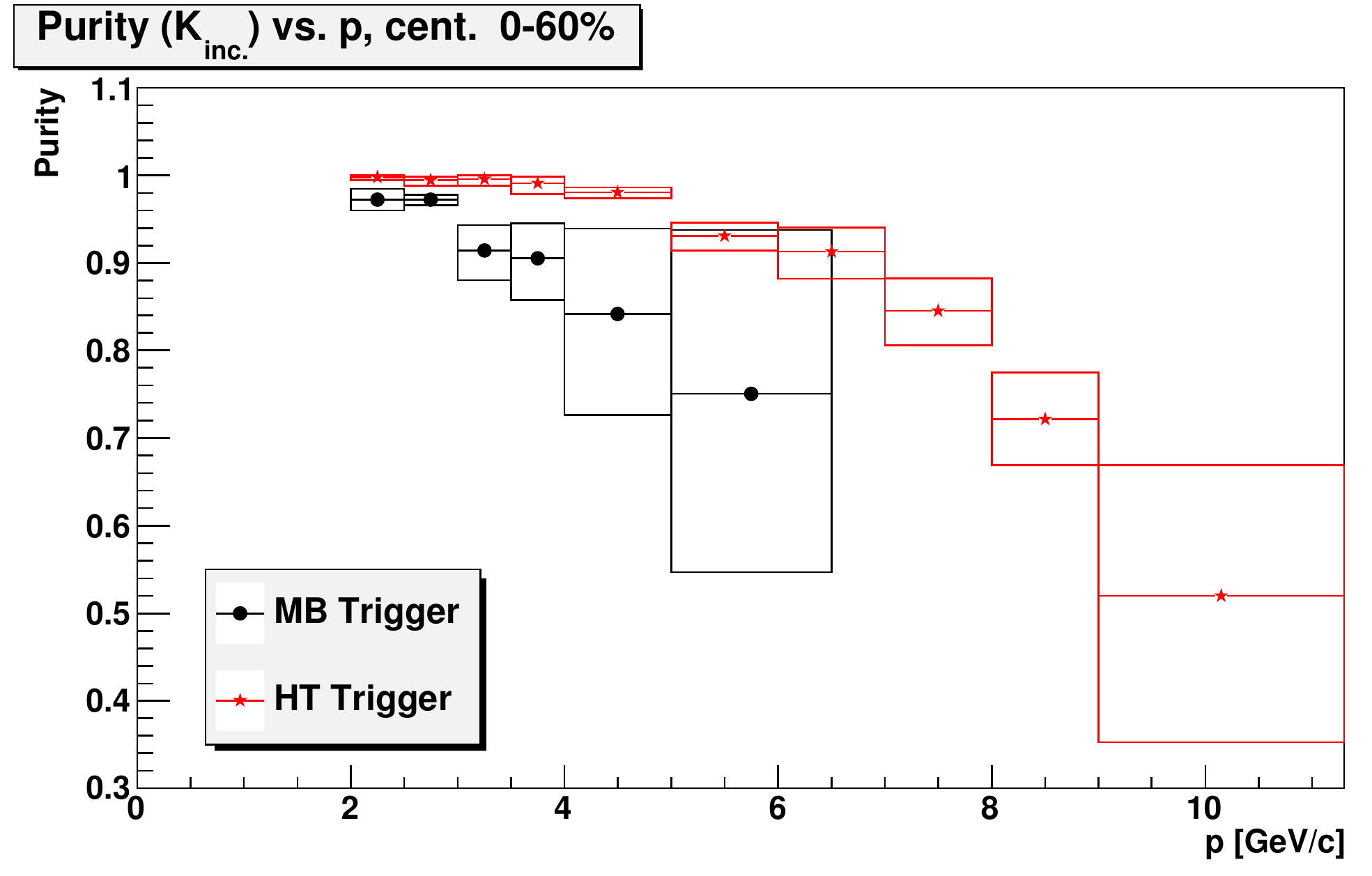}
\caption[Inclusive $e^{\pm}$ purity as a function of momentum.]{The inclusive $e^{\pm}$ purity $(K_{inc.})$ as a function of momentum for minimum-bias (black circles) and high-tower triggered (red stars) events.}
\label{fig:dedx:purity_p}
\includegraphics[width=0.85\linewidth]{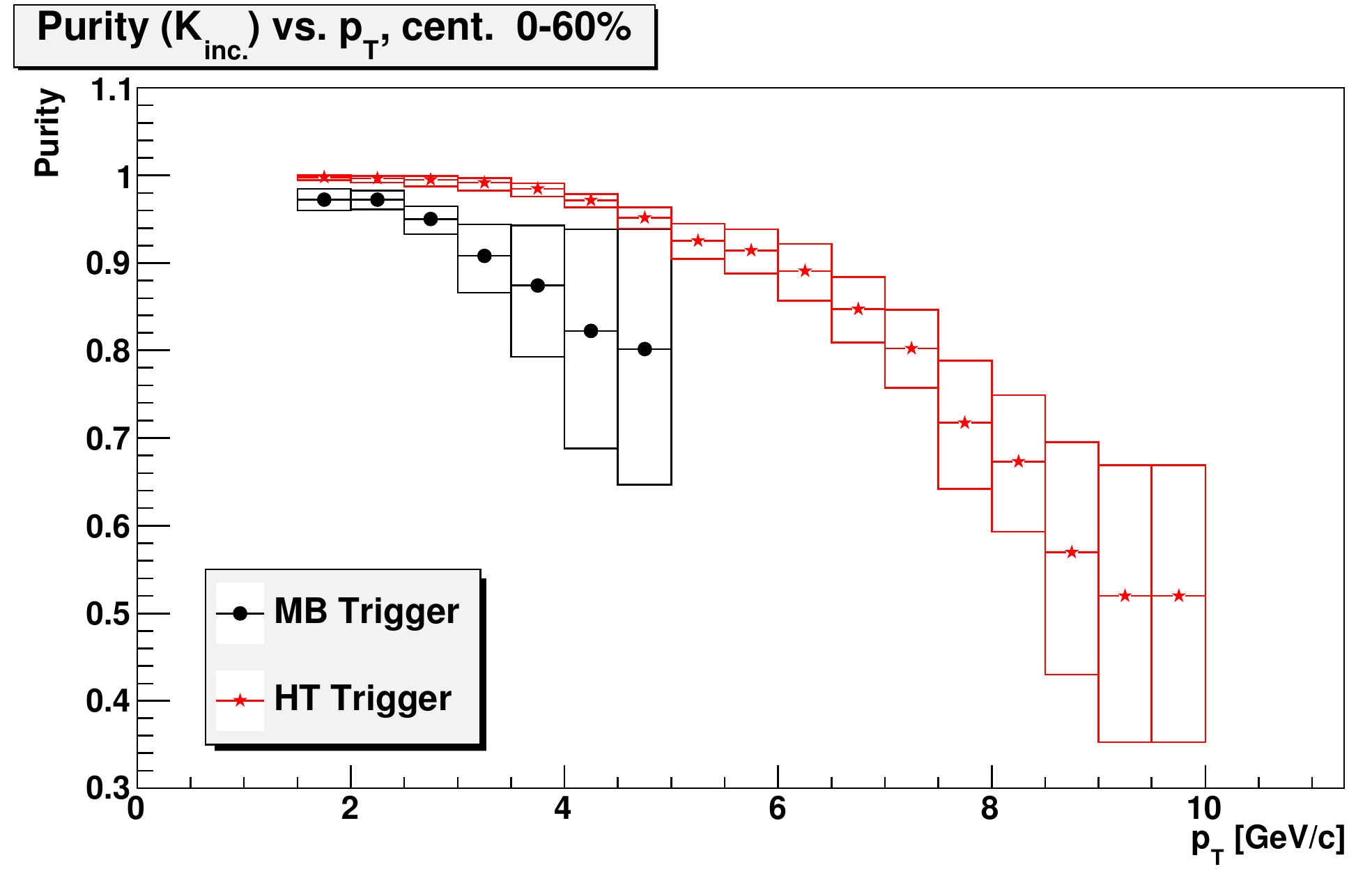}
\caption[Inclusive $e^{\pm}$ purity as a function of $p_{T}$.]{The inclusive $e^{\pm}$ purity $(K_{inc.})$ as a function of $p_{T}$ for minimum-bias (black circles) and high-tower triggered (red stars) events.}
\label{fig:dedx:purity_pt}
\end{center}
\end{figure}

The inclusive $e^{\pm}$ purity, shown in Figure~\ref{fig:dedx:purity_p}, is the ratio of the area of the $e^{\pm}$ peak within the range $-1.5<n\sigma_{e}<3.5$ to the total area of the three peaks within that range (the purity in the range $2\GeV/c<p<2.5\GeV/c$ is reduced to account for the presence of (anti)deuterons).  To find the lower uncertainty in $K_{inc.}$, each fit parameter is changed from its central value by its uncertainty in a manner that decreases the purity (the hadron peak means, widths, and amplitudes and the $e^{\pm}$ peak width are increased by their own uncertainties, while the $e^{\pm}$ peak amplitude and mean are decreased).  The changes in $K_{inc.}$ caused by changing each parameter are then added in quadrature to obtain the total uncertainty.  The upper uncertainty in $K_{inc.}$ is calculated in a similar fashion by changing the values of the fit parameters in the opposite directions.  In the momentum range $2\GeV/c<p<2.5\GeV/c$, the lower uncertainty due to the (anti)deuteron contribution is found by calculating the purity with the $d(\bar{d})$ yield doubled; the upper uncertainty is found by removing the $d(\bar{d})$ contribution.  These $d(\bar{d})$ uncertainties are added in quadrature with the other uncertainties.

The calculations of the purity described above have been divided into bins in momentum, but the non-photonic $e^{\pm}$ spectrum is presented in transverse momentum bins.  It is useful (but not necessary) to convert the momentum-dependent purity calculation described above to a $p_{T}$-dependent measurement (see Figure~\ref{fig:dedx:purity_pt}).  This is accomplished by taking a weighted average.  For each transverse momentum bin,

\begin{equation}
K_{inc.}(p_{T})=\frac{\sum\limits_{p}K_{inc.}(p)I(p,p_{T})}{\sum\limits_{p}I(p,p_{T})},
\end{equation}

\noindent where $I(p,p_{T})$ is the measured (uncorrected) inclusive $e^{\pm}$ yield in each bin in momentum and transverse momentum.

Additional material regarding the calculation of $K_{inc.}$ can be found in Section~\ref{sec:add:purity}.  That section includes plots similar to Figure~\ref{fig:dedx:3g_fit} showing the results of multi-Gaussian fits to the $n\sigma_{e}$ distributions for various momentum bins.  That section also includes tables with the values of $K_{inc.}(p_{T})$ for all centrality classes and trigger types.

\clearpage

\clearpage

\clearpage

\clearpage

\chapter{Background Rejection Efficiency}
\label{sec:bre}

\section{Introduction}
\label{sec:bre:intro}

The background rejection efficiency $(\varepsilon_{B})$ is the efficiency with which photonic $e^{\pm}$ are identified, using the method described in Section~\ref{sec:anintro:back_sub}.  This efficiency is calculated from simulations of photon conversions and Dalitz decays of neutral pions $(\pi^{0}\rightarrow\gamma e^{-}e^{+})$.  The efficiency is found by considering the set of all reconstructed tracks that are identified as $e^{\pm}$ and that are associated with simulated photonic $e^{\pm}$ tracks.  The fraction of those $e^{\pm}$ that is identified as photonic $e^{\pm}$ is $\varepsilon_{B}$.

More precisely, \textit{reconstructed} $e^{\pm}$ are defined to be those $e^{\pm}$ that pass the following cuts.

\begin{itemize}
\item The reconstructed track shares at least one half of its TPC points with a simulated photonic $e^{\pm}$ track.
\item $GDCA(rec.)<1.5$ cm.
\item $-0.1<\eta(rec.)<0.7$
\item $R_{1TPC}<102$ cm
\item $N_{TPCfit}>20$
\item $R_{FP}>0.52$
\item $p(rec.)/E_{Tower}<2c^{-1}$
\item $\Delta_{SMD}<0.02$
\item $N_{SMD\eta},N_{SMD\phi}\geq 2$
\item The BEMC has good status at the point to which the reconstructed track projects.
\end{itemize}

\noindent \textit{Rejected} $e^{\pm}$ are defined to be the subset of reconstructed $e^{\pm}$ for which the following conditions are also satisfied.
\begin{itemize}
\item The true conversion (or Dalitz-decay) partner of the simulated $e^{\pm}$ is also reconstructed (\textit{i.e.}, a reconstructed track shares at least one half of its TPC points with the simulated photonic $e^{\pm}$ partner).
\item The reconstructed true partner track has $R_{FP}(partner)>0.52$.
\item The reconstructed true partner track has $p_{T}(partner)>300$ MeV/$c$.
\item The reconstructed tracks of the photonic $e^{-}e^{+}$ pair have opposite charges.
\item The reconstructed tracks of the photonic $e^{-}e^{+}$ pair have a distance of closest approach of $DCA(e^{-}e^{+})<1.5$ cm. 
\item The reconstructed tracks of the photonic $e^{-}e^{+}$ pair have an invariant mass of $M_{inv.}(e^{-}e^{+})<150$ MeV/$c^{2}$ at their point of closest approach.
\end{itemize}

The efficiency is the ratio of the number of rejected $e^{\pm}$ to the number of reconstructed $e^{\pm}$.  In the analysis of real data, a cut on the normalized TPC energy loss of $-1.5<n\sigma_{e}<3.5$ is used to identify $e^{\pm}$, while a cut of $\langle dE/dx\rangle>$ 2.8 keV/cm is used to increase the fraction of $e^{\pm}$ in the set of partner tracks.  Due to difficulties in simulating TPC energy loss, these cuts are not applied in the analysis of simulation data.  The energy-loss cut efficiency $(\varepsilon_{dE/dx})$ is used to correct for the first cut; the second cut is assumed to be 100\% efficient (See Chapter~\ref{sec:dedx}).

Four separate simulation data sets are used in the calculation of the background rejection efficiency.
\begin{itemize}
\item Data Set $P_{1}$: This data set is used to calculate $\varepsilon_{B}$ for $e^{\pm}$ from photon conversions, with the photons originating from $\pi^{0}$ decays.  Simulated $\pi^{0}\rightarrow\gamma\gamma$ decays and the subsequent conversion of those photons to $e^{-}e^{+}$ pairs in the inner structures of the STAR detector (simulated using GEANT~\cite{Agostinelli2003250}) were embedded into $\approx$ 800,000 real Cu + Cu events.  The simulated $\pi^{0}$ were distributed uniformly in transverse momentum from 0.5 GeV/$c$ $<p_{T}(\pi^{0})<$15 GeV/$c$ and uniformly in pseudorapidity from $-0.5<\eta(\pi^{0})<1.5$.  The pions were embedded at a rate of $\approx0.05$ times the reference multiplicity of the underlying event.  Neutral-pion Dalitz decays were also embedded into this data set at the natural branching ratio (1.174\%)~\cite{PDG_review}, but these decays were simulated incorrectly and their products are not considered in this analysis.



\item Data Set $P_{2}$: This data set is used to calculate $\varepsilon_{B}$ for $e^{\pm}$ from $\pi^{0}$ Dalitz decays $(\pi^{0}\rightarrow e^{-}e^{+}\gamma)$.  Neutral-pion Dalitz decays (with the correct~\cite{PhysRev.98.1355} invariant-mass distribution and form factor) were simulated in $\approx 10^6$ $p+p$ events generated using PYTHIA.~\cite{PYTHIA6.4}  The interactions of the decay products with the STAR detector were simulated using GEANT.  The simulated pions were distributed uniformly in transverse momentum for $p_{T}(\pi^{0})<20$ GeV/$c$ and uniformly in pseudorapidity for $|\eta|<1.5$.  Calculations of $\varepsilon_{B}$ were done for the $e^{-}e^{+}$ pairs produced directly in these Dalitz decays as well as for the $e^{-}e^{+}$ produced by the conversion of the Dalitz-decay photons (the former contribution is larger).  Two-photon decays of neutral pions were also simulated in this data set, but are not included in the final calculation of the background rejection efficiency.

\item Data Set $P_{3}$: This data set is used to calculate $\varepsilon_{B}$ for $e^{\pm}$ from photon conversions, with the photons from sources other than $\pi^{0}$ decays.  Photons were simulated in $\approx 10^6$ $p+p$ events generated using PYTHIA for the same $p_{T}$ and $\eta$ ranges used for the simulated $\pi^{0}$ in data set $P_{2}$.  The interactions of the decay products with the STAR detector were simulated using GEANT.  The conversions of these photons were used to calculate the effect on $\varepsilon_{B}$ of photons that are not $\pi^{0}$ decay products (specifically, direct photons and photons from $\eta\rightarrow\gamma\gamma$ decays).

\item Data Set $P_{4}$: This data set was used to determine the expected shape of the spectrum of photons from $\eta$-meson decays.  Approximately $5\times 10^5$ two-photon decays of $\eta$ mesons were simulated using PYTHIA.  Unlike data sets $P_{1}$, $P_{2}$, and $P_{3}$, the particles in this data set were not embedded into a real or simulated event.  The interactions of the decay photons with the STAR detector were not simulated, so photonic $e^{\pm}$ and the background rejection efficiency could not be studied directly for this data set.
\end{itemize}

\clearpage

\section{Weighting}
\label{sec:bre:weighting}

\begin{figure}[htbp]
\begin{center}
\includegraphics[width=0.85\linewidth]{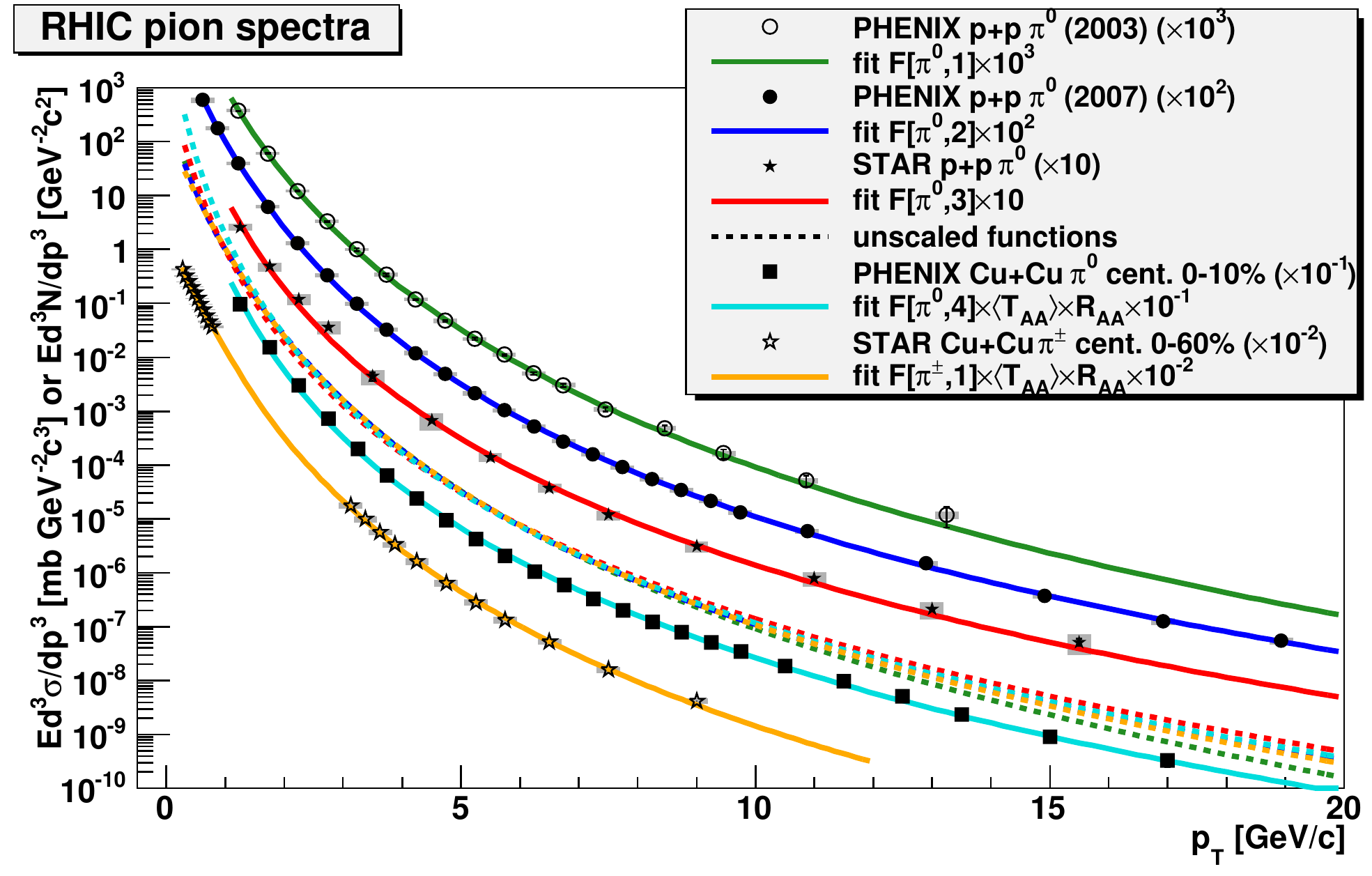}
\caption[RHIC pion spectra in $p+p$ and Cu + Cu collisions.]{STAR\protect\cite{PhysRevC.81.064904,STAR_CuCu_freezeout,PhysRevC.81.054907} and PHENIX measurements\protect\cite{PhysRevLett.91.241803,PhysRevD.76.051106,PhysRevLett.101.162301} of neutral and charged pion spectra:  Plotted on the vertical axis is $Ed^3\sigma/dp^3$ for $p+p$ spectra and $Ed^{3}N/dp^{3}$ for the Cu + Cu spectra.  Fits to the spectra are also shown; the dashed curves are the unscaled functions.}
\label{fig:bre:rhic_pion}
\end{center}
\end{figure}

In the three simulation data sets described above, the photonic $e^{\pm}$ are produced from simulated particles ($\pi^{0}$ or $\gamma$) that are uniformly distributed in transverse momentum.  This is, of course, not a realistic distribution for pions and photons; the photonic $e^{\pm}$ spectra used to calculate $\varepsilon_{B}$ must be weighted to correct for this.  The weighting procedure is described below.

In the calculation of the background rejection efficiency, each photonic $e^{\pm}$ is assigned a weight according to the $p_{T}$ of the $\pi^{0}$ (for data sets $P_{1}$ and $P_{2}$) or photon (for data set $P_{3}$) from which it originates.  The choice of weighting function can have a large effect on the calculated value of the background rejection efficiency.  Consider two different photonic electrons, each with $p_{T}(e^{-})=2$ GeV/$c$.  One of these came from a photon with $p_{T}(\gamma)=2.2$ GeV/$c$ and the other came from a photon with $p_{T}(\gamma)=10$ GeV/$c$.  The conversion partner of the first electron must have low transverse momentum, $p_{T}(partner)\approx 0.2$ GeV/$c$, and will likely fail to be reconstructed or fail the 0.3-GeV/$c$ cut on $p_{T}(partner)$.  The first electron is unlikely to be identified as photonic.  The conversion partner of the second electron will have high transverse momentum, $p_{T}(partner)\approx 8$ GeV/$c$; the second electron has a greater chance of being identified as photonic. A photon spectrum that decreases with $p_{T}(\gamma)$ will give the first electron greater weight than the second, resulting in a lower efficiency than the un-weighted case.

For data sets $P_{1}$ and $P_{2}$, each $e^{\pm}$ is assigned a weighting factor equal to the expected $\pi^{0}$ cross-section multiplied by $p_{T}(\pi^{0})$.  Figure~\ref{fig:bre:rhic_pion} shows one STAR~\cite{PhysRevC.81.064904} and two PHENIX~\cite{PhysRevLett.91.241803,PhysRevD.76.051106} measurements of the $\pi^{0}$ cross-section in $p+p$ collisions.  Also shown are the PHENIX collaboration's measurement~\cite{PhysRevLett.101.162301} of the $\pi^{0}$ yield in Cu + Cu collisions and low- and high-$p_{T}$ STAR measurements~\cite{STAR_CuCu_freezeout,PhysRevC.81.054907} of the charged-pion ($(\pi^{-}+\pi^{+})/2$) yield in Cu + Cu collisions.  Five fit functions are used to describe these cross-section measurements.  The fit parameters of this function, as well as all other weighting functions discussed in this section, are given in Appendix~\ref{sec:functions}.

\begin{itemize}
\item Function $F[\pi^{0},1]$: The PHENIX $p+p\rightarrow\pi^{0}$ cross-section published in 2003 was fit~\cite{PhysRevLett.91.241803} using a power-law function with the form $A(1+p_{T}/p_{0})^n$.
\item Function $F[\pi^{0},2]$: The PHENIX $p+p\rightarrow\pi^{0}$ cross-section published in 2007~\cite{PhysRevD.76.051106} was fit using a function with the form $A(1+p_{T}^{2}/B)^n+C\exp(-p_{T}/D)$. 
\item Function $F[\pi^{0},3]$: The STAR $p+p\rightarrow\pi^{0}$ cross-section was fit~\cite{PhysRevC.81.064904} using a power-law function with the form $A(1+p_{T}/p_{0})^n$.
\item Function $F[\pi^{0},4]$: The PHENIX Cu + Cu $\rightarrow\pi^{0}$ yield (0-10\% centrality class)~\cite{PhysRevLett.101.162301} was fit using a power-law function with the form $A(1+p_{T}/p_{0})^n$.  The data were scaled by $1/(\langle T_{AA}\rangle R_{AA})=42\,\mathrm{mb}/(182.7\times 0.515)$ before fitting.
\item Function $F[\pi^{\pm},1]$: The combined STAR Cu + Cu $\rightarrow\pi^{\pm}$ yield (0-60\% centrality class)~\cite{STAR_CuCu_freezeout,PhysRevC.81.054907} was fit using a function with the form $A(1+p_{T}^{2}/B)^{n}+Ce^{-p_{T}/D}$.  The data were scaled by $1/(\langle T_{AA}\rangle R_{AA})=42\,\mathrm{mb}/(80.41\times 0.7)$ before fitting.
\end{itemize}

The $p+p$ neutral-pion weighting functions appear to match the shape of the Cu + Cu pion spectra.  Function $F[\pi^{0},2]$ is used as the primary weighting function for pions because the measurement from which it is derived has the smallest uncertainties and spans the largest transverse-momentum range of the measurements shown in Figure~\ref{fig:bre:rhic_pion}.  The effect of the different pion weighting functions upon $\varepsilon_{B}$ is explored in Sections~\ref{sec:bre:p1} and~\ref{sec:bre:combined}.

\begin{figure}[htbp]
\begin{center}
\includegraphics[width=0.8\linewidth]{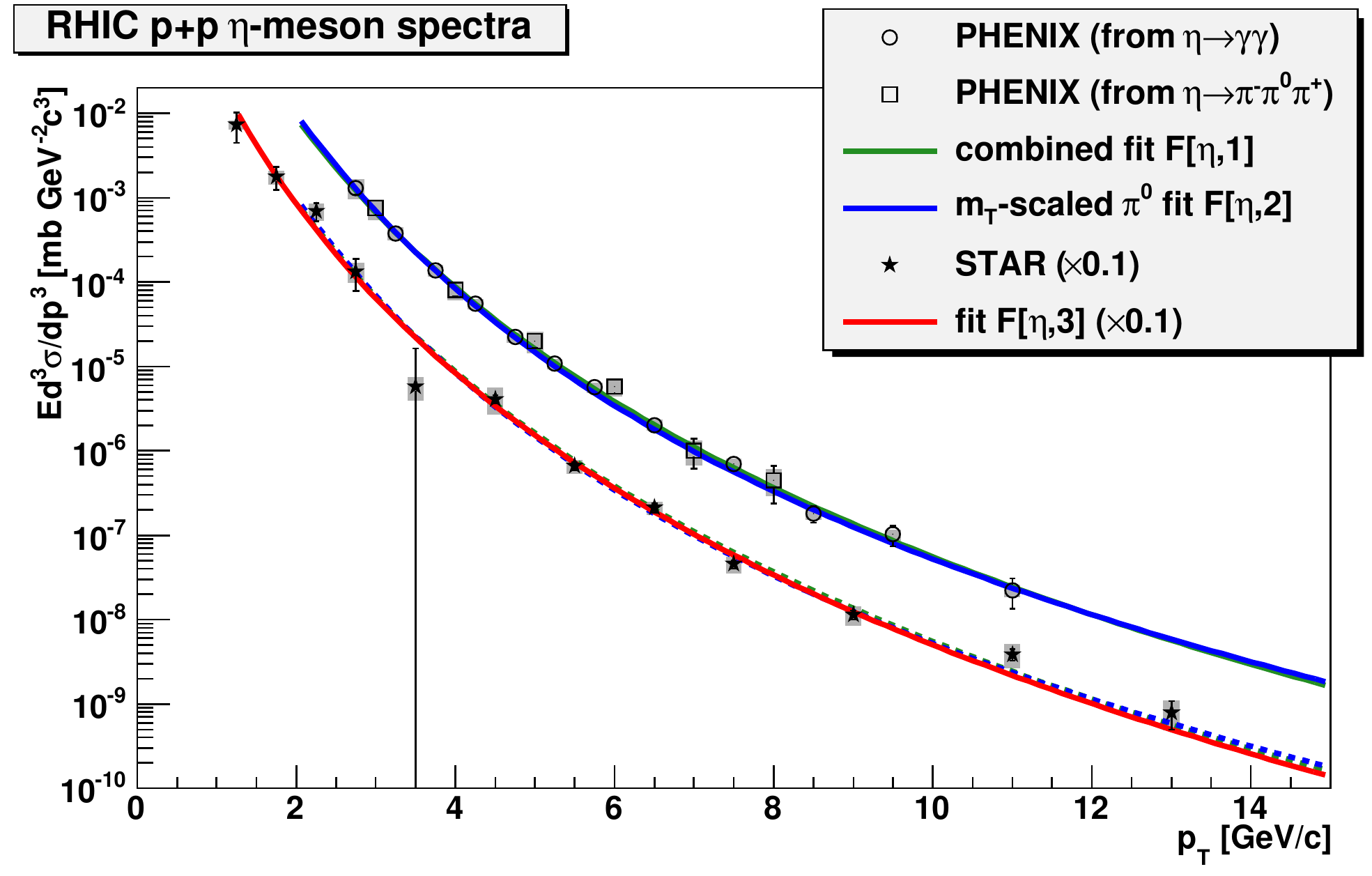}
\caption[RHIC $\eta$-meson spectra in $p+p$ collisions.]{STAR\protect\cite{PhysRevC.81.064904} and PHENIX\protect\cite{PhysRevC.75.024909} measurements of $\eta$-meson spectra in $p+p$ collisions.  Fits to the spectra are also shown.}
\label{fig:bre:rhic_eta}
\includegraphics[width=0.8\linewidth]{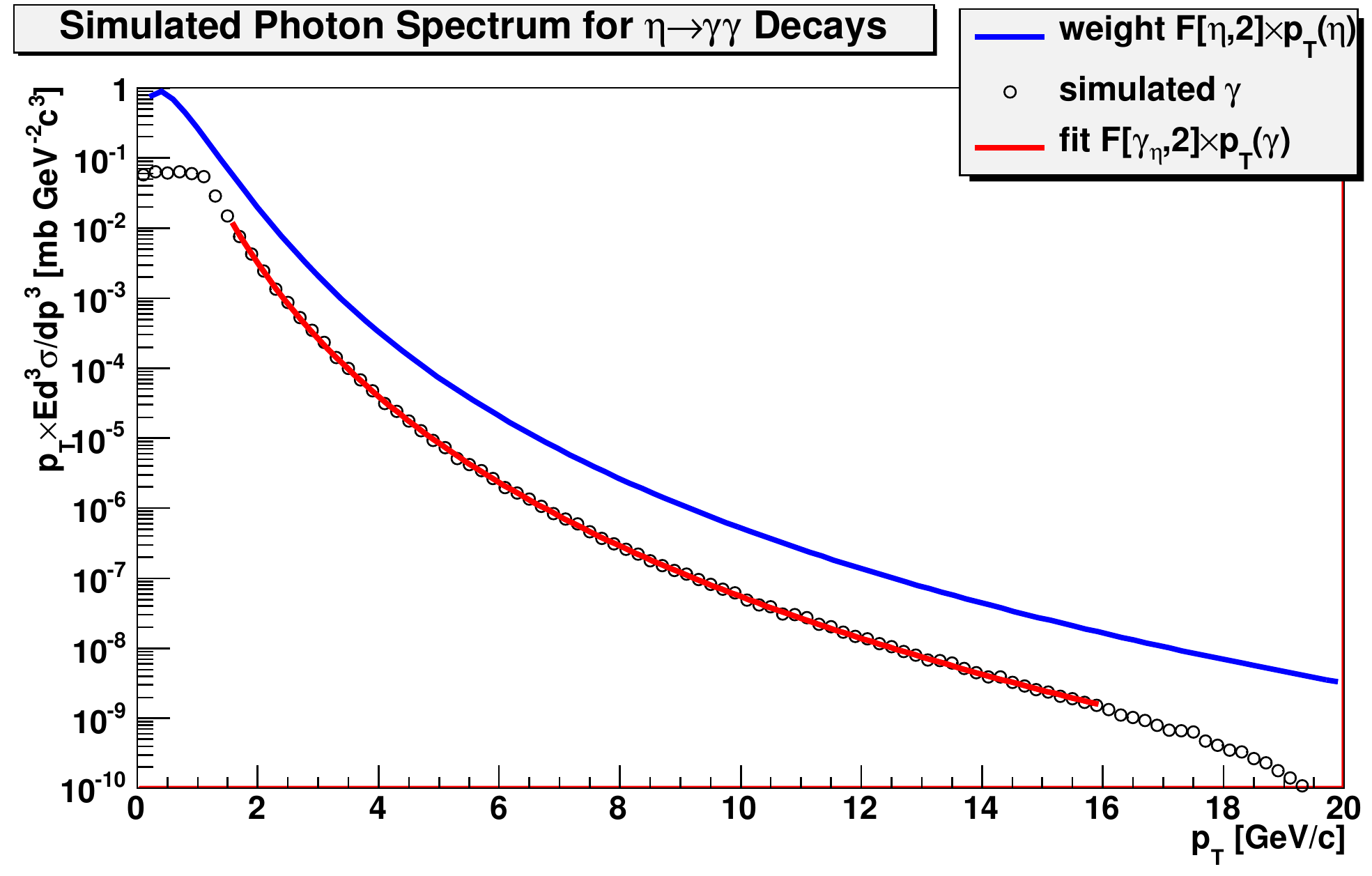}
\caption[Generation of a simulated photon spectrum from an $\eta$-meson spectrum.]{Generation of a simulated photon spectrum from fit $F[\eta,2]$: The simulated photon spectrum is scaled by the 39.31\% branching ratio for the $\eta\rightarrow\gamma\gamma$ decay.  The simulated $\eta$ mesons are distributed uniformly in the transverse-momentum range $1\GeV/c<p_{T}(\eta)<20\GeV/c$,  which causes the photon spectrum to behave non-physically below $p_{T}(\gamma)=1\GeV/c$ and near $p_{T}(\gamma)=20\GeV/c$.}
\label{fig:bre:pho_F_eta_2}
\end{center}
\end{figure}

For data set $P_{3}$, several different weighting functions are added together to describe the shape of the spectrum of $\eta$-meson-decay photons and direct photons.  Figure~\ref{fig:bre:rhic_eta} shows STAR~\cite{PhysRevC.81.064904} and PHENIX~\cite{PhysRevC.75.024909} measurements of the $\eta$-meson cross-section in $p+p$ collisions (PHENIX measured this cross section using two different $\eta$-meson decay modes).
Three different functions are used to describe the shape of the $\eta$-meson spectrum.

\begin{itemize}
\item Function $F[\eta,1]$:  The PHENIX cross-sections (from both measurements) are fit with a function of the form $A(1+p_{T}/p_{0})^{n}$.
\item Function $F[\eta,2]$: The shape of the $\eta$-meson spectrum is estimated from the PHENIX $p+p\rightarrow\pi^{0}$ spectrum by making the substitution\\ $p_{T}(\pi^{0})\rightarrow \sqrt{p_{T}(\eta)^{2}+m(\eta)^{2}c^{2}-m(\pi^{0})^{2}c^{2}}$ in function $F[\pi^{0},2]$.  The resulting function is multiplied by the measured~\cite{PhysRevC.75.024909} $\eta/\pi^{0}$ ratio of 0.48.
\item Function $F[\eta,3]$: The STAR cross section is fit~\cite{PhysRevC.81.064904} with a function of the form $A(1+p_{T}/p_{0})^{n}$.
\end{itemize}

Function $F[\eta,2]$ is used as the primary weighting function for $\eta$-mesons.  That function is derived from the PHENIX $\pi^{0}$ cross-section measurement in $p+p$ collisions, which has smaller uncertainties and covers a larger transverse-momentum range than either of the $\eta$-meson cross-section measurements shown.  The spectrum of photons expected from the decays of $\eta$ mesons is estimated by applying all three of these weighting functions to the simulated photons in data set $P_{4}$, a set of simulated $\eta\rightarrow\gamma\gamma$ decays.  The resulting photon spectra were fit with functions of the form $A(1+p_{T}/p_{0})^{n}\times p_{T}$.  This is illustrated by Figure~\ref{fig:bre:pho_F_eta_2}, in which $F[\eta,2]$ is used as the $\eta$-meson weighting function.  The simulated photon spectrum is scaled by the 39.31\% branching ratio~\cite{PDG_review} for the $\eta\rightarrow\gamma\gamma$ decay.

\begin{itemize}
\item Function $F[\gamma_{\eta},1]$ is the fit to the photon spectrum generated using $F[\eta,1]$ as a weighting function.
\item Function $F[\gamma_{\eta},2]$ is the fit to the photon spectrum generated using $F[\eta,2]$ as a weighting function.  This function is used as the primary weighting function for $\eta$-meson-decay photons because it is derived from weighting function $F[\pi^{0},2]$, the primary weighting function used for neutral pions.
\item Function $F[\gamma_{\eta},3]$ is the fit to the photon spectrum generated using $F[\eta,3]$ as a weighting function.
\end{itemize}

\begin{figure}[htbp]
\begin{center}
\includegraphics[width=0.8\linewidth]{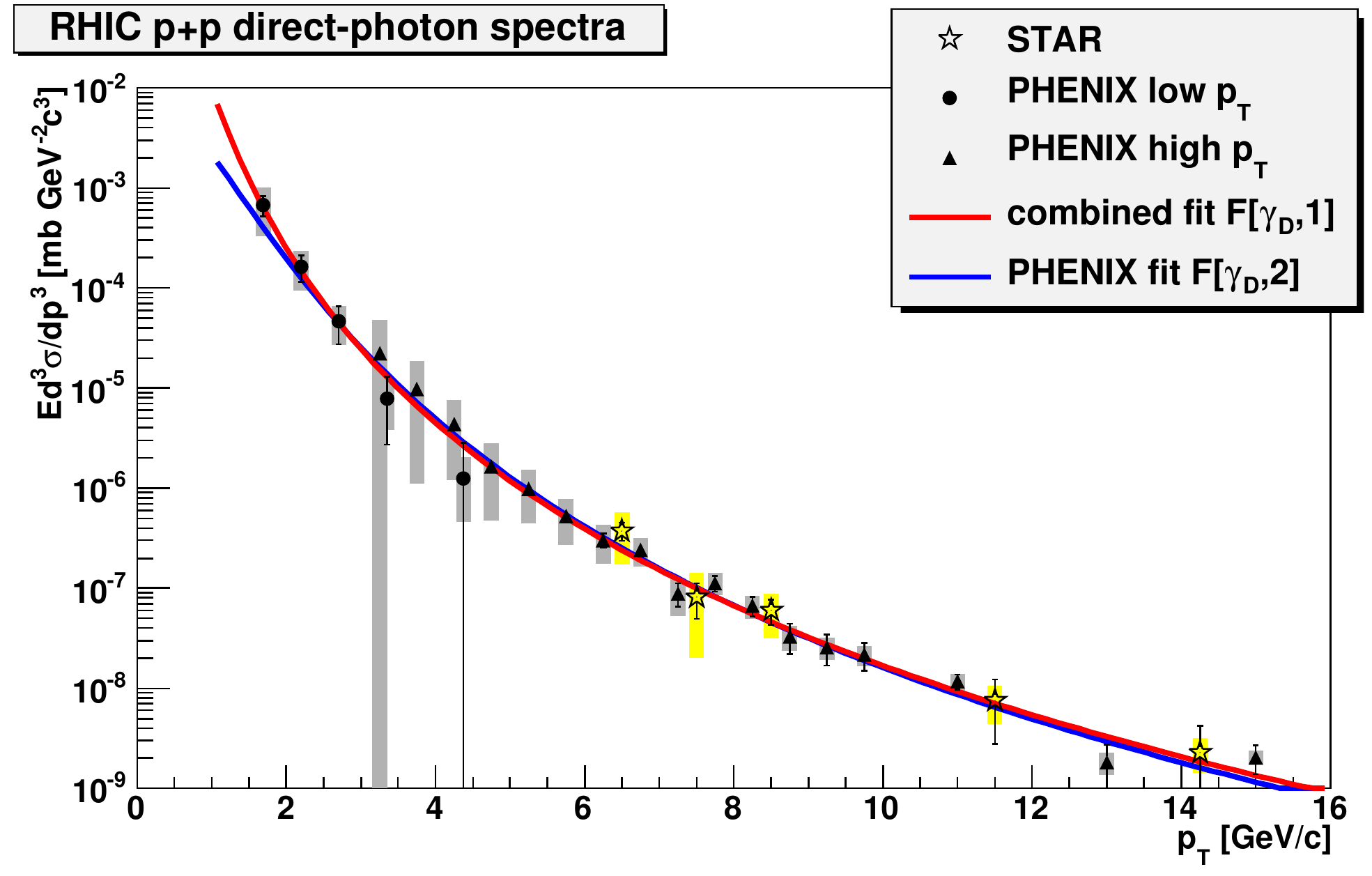}
\caption[RHIC direct-photon spectra in $p+p$ collisions.]{STAR\protect\cite{PhysRevC.81.064904} and PHENIX\protect\cite{PhysRevLett.104.132301,PhysRevLett.98.012002} measurements of direct-photon spectra in $p+p$ collisions:  Fits to the combined PHENIX spectra are also shown.}
\label{fig:bre:rhic_pp_pho}
\includegraphics[width=0.8\linewidth]{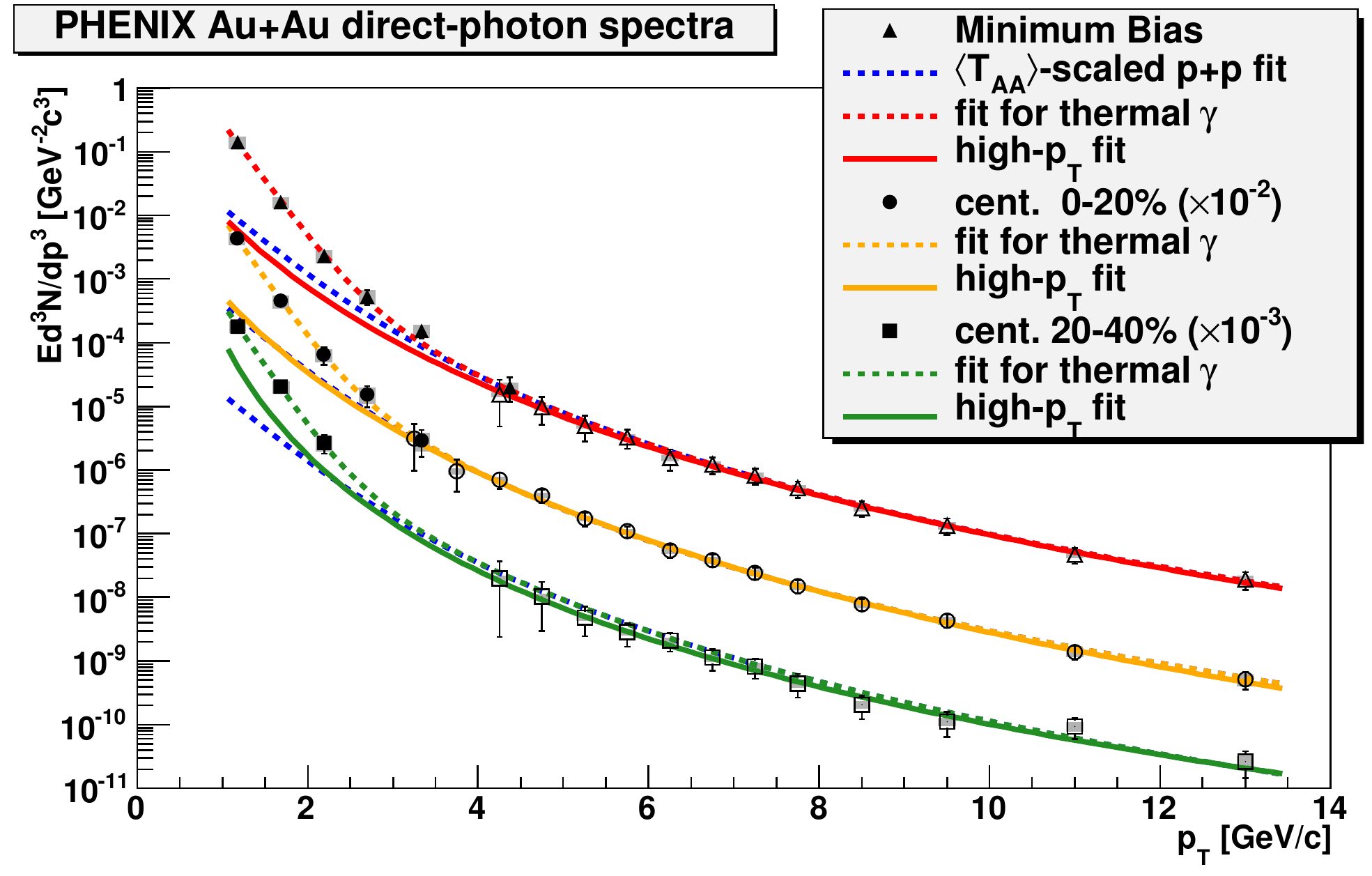}
\caption[PHENIX direct-photon spectra in Au + Au collisions.]{PHENIX measurements\protect\cite{PhysRevLett.104.132301,PhysRevLett.94.232301} of direct-photon spectra in Au + Au collisions for three centrality classes:  The solid curves are fits to the spectra above $p_{T}=4$ GeV/$c$.  The dashed blue curves are the function $F[\gamma_{D},2]$, a fit to the PHENIX $p+p$ direct-photon spectrum, scaled by the value of $\langle T_{AA}\rangle$ for each centrality class.  The other dashed fit curves have the form of decaying exponentials, which describe the thermal-photon component, added to the scaled $p+p$ fit.}
\label{fig:bre:phenix_auau_pho}
\end{center}
\end{figure}

The direct-photon components of the photon weighting function were determined from the PHENIX measurements of the direct-photon spectra for $p+p$ and Au + Au collisions.  The low-$p_{T}$~\cite{PhysRevLett.104.132301} and high-$p_{T}$~\cite{PhysRevLett.98.012002} cross-section measurements for $p+p$ collisions are shown in Figure~\ref{fig:bre:rhic_pp_pho}.  Two fits to the $p+p$ cross-section were studied.

\begin{itemize}
\item Function $F[\gamma_{D},1]$ is a fit to the cross-section which was generated by the author of this dissertation.
\item Function $F[\gamma_{D},2]$ is a fit to the cross-section generated by the PHENIX collaboration~\cite{PhysRevLett.104.132301}.  This function is used as the primary weighting function for direct photons for this analysis.
\end{itemize}

Additional direct-photon weighting functions were generated from PHENIX direct photon cross-section measurements in Au + Au collisions.  Figure~\ref{fig:bre:phenix_auau_pho} shows low-$p_{T}$~\cite{PhysRevLett.104.132301} and high-$p_{T}$~\cite{PhysRevLett.94.232301} measurements of the direct-photon yields in Au + Au collisions in three centrality classes: 0-92\% (minimum bias), 0-20\%, and 20-40\%.

These spectra were fit for $p_{T}>4$ GeV/$c$ to determine three additional direct-photon weighting functions.

\begin{itemize}
\item Function $F[\gamma_{D},3]$ is a fit to the minimum-bias spectrum.
\item Function $F[\gamma_{D},4]$ is a fit to the centrality 0-20\% spectrum.
\item Function $F[\gamma_{D},5]$ is a fit to the centrality 20-40\% spectrum.
\end{itemize}

In Au + Au collisions, an additional component of the direct-photon yield has been observed~\cite{PhysRevLett.104.132301}.  This component, which is most important at low $p_{T}$, is believed to be thermal photons produced in a quark-gluon plasma.  The three cross-section measurements shown in Figure~\ref{fig:bre:phenix_auau_pho} were also used to generate three weighting functions describing the thermal-photon component. To describe the shapes of the spectra, the fit of the $p+p$ direct-photon spectrum ($F[\gamma_{D},2]$) is scaled by the nuclear thickness function $\langle T_{AA}\rangle$.\footnote{$\langle T_{AA}\rangle=6.14\,\mathrm{mb}^{-1}$ for the 0-92\% centrality class, 18.6 mb$^{-1}$ for the 0-20\% centrality class, and 7.07 mb$^{-1}$ for the 20-40\% centrality class~\cite{PhysRevLett.91.072301}.}  A decaying exponential is added to account for the thermal-photon component; only this component was allowed to vary during fitting.  Three fits to the low-$p_{T}$ Au + Au direct-photon spectra were generated.

\begin{itemize}
\item Function $F[\gamma_{T},1]$ has the form $C\exp(-p_{T}/T))$.  The minimum-bias spectrum was fit with a function of the form $\langle T_{AA}\rangle\times(F[\gamma_{D},2]+F[\gamma_{T},1])$, with only the parameters of $F[\gamma_{T},1]$ allowed to vary.  This function is used as the primary weighting function for thermal photons for this analysis.
\item Function $F[\gamma_{T},2]$ was generated in the same fashion by fitting the spectrum in the 0-20\% centrality class.
\item Function $F[\gamma_{T},3]$ was generated in the same fashion by fitting the spectrum in the 20-40\% centrality class.
\end{itemize}

\begin{figure}[htbp]
\begin{center}
\includegraphics[width=0.85\linewidth]{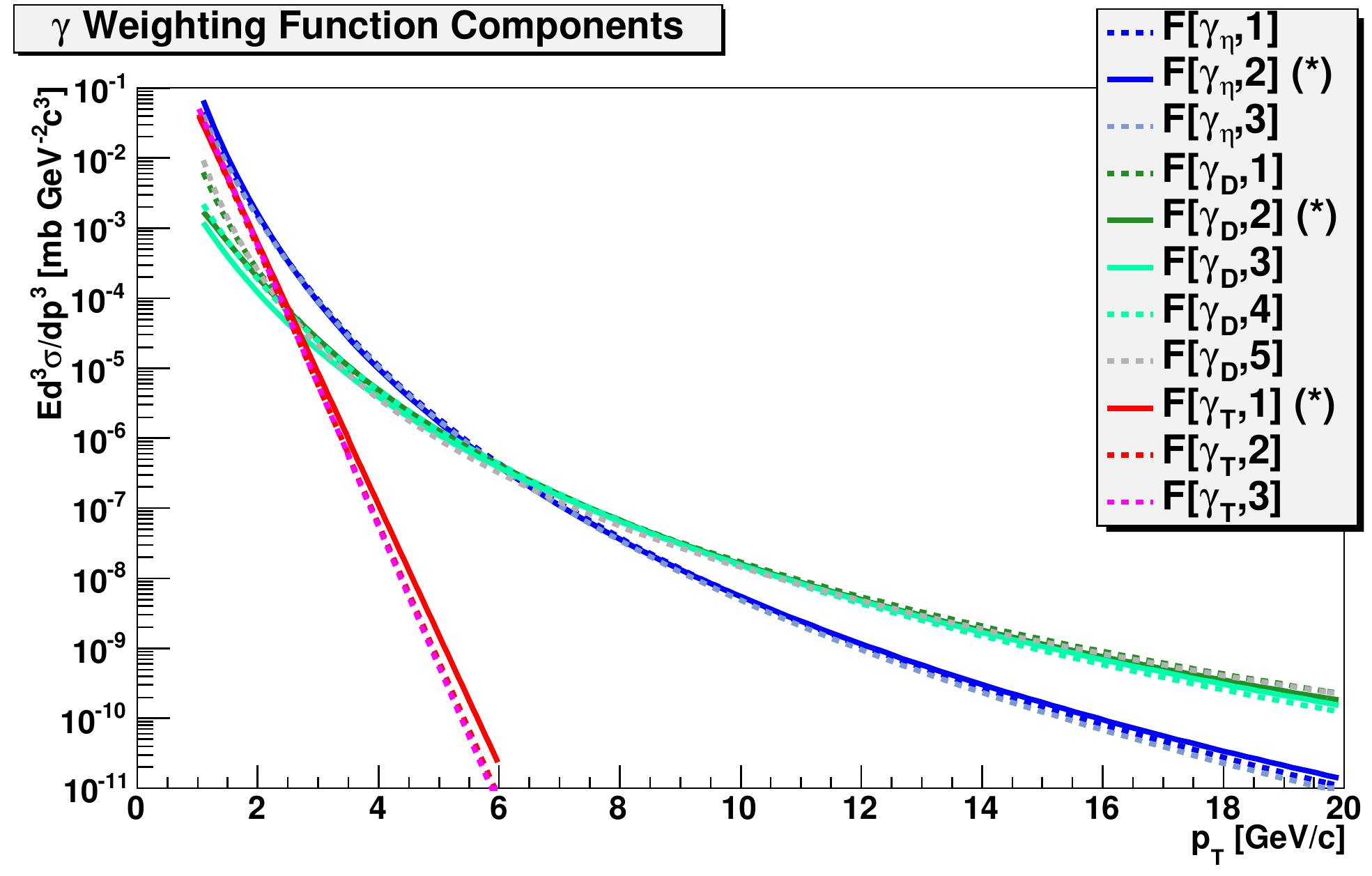}
\caption[Components of the photon weighting function used in calculating the background rejection efficiency.]{Components of the photon weighting function used in this analysis: The components used to generate the primary weighting function are marked with asterisks.}
\label{fig:bre:photon_weight_components}
\end{center}
\end{figure}

The contributions of photons from $\eta$-meson decays, direct photons, and thermal direct photons are added together and used as a weighting function for data set $P_{3}$.  Figure~\ref{fig:bre:photon_weight_components} shows the eleven different weighting-function components used for data set $P_{3}$: three to describe photons from $\eta$-meson decays, five to describe non-thermal direct photons, and three to describe thermal photons.

\clearpage

\section[Background Rejection Efficiency for Data Set $P_{1}$: $\pi^{0}\rightarrow\gamma\gamma$ Decays]{Background Rejection Efficiency for \\ Data Set $\boldsymbol{P_{1}}$: $\boldsymbol{\pi^{0}\rightarrow\gamma\gamma}$ Decays}
\label{sec:bre:p1}

\begin{figure}[htbp]
\begin{center}
\includegraphics[width=0.85\linewidth]{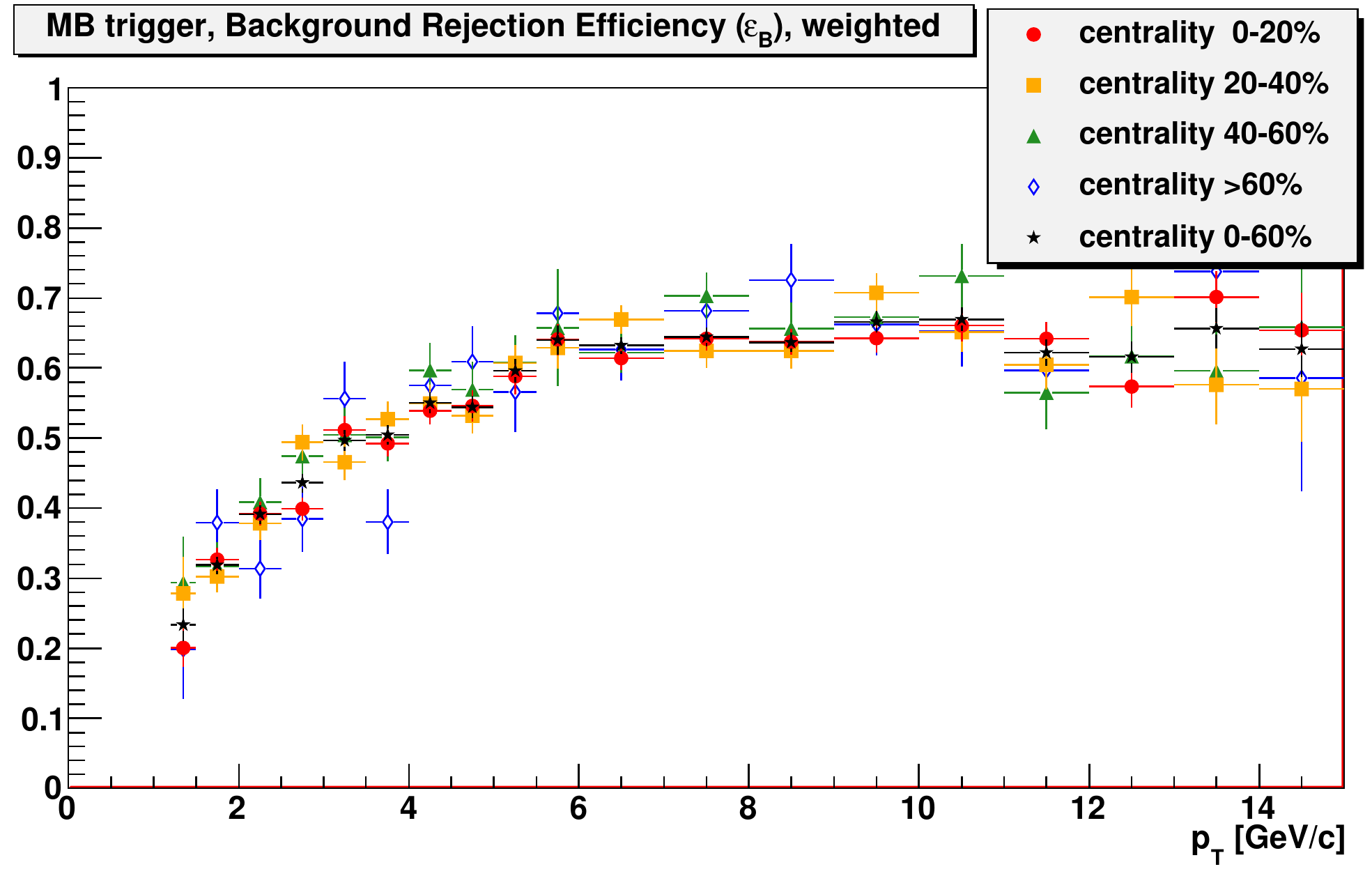}
\caption[Background rejection efficiency as a function of $p_{T}$ for five centrality classes.]{Background rejection efficiency as a function of $p_{T}$ for five centrality classes: The efficiency does not appear to depend strongly (if at all) upon the collision centrality of the underlying event.}
\label{fig:bre:cent_dep}
\end{center}
\end{figure}

\begin{figure}[htbp]
\begin{center}
\includegraphics[width=0.85\linewidth]{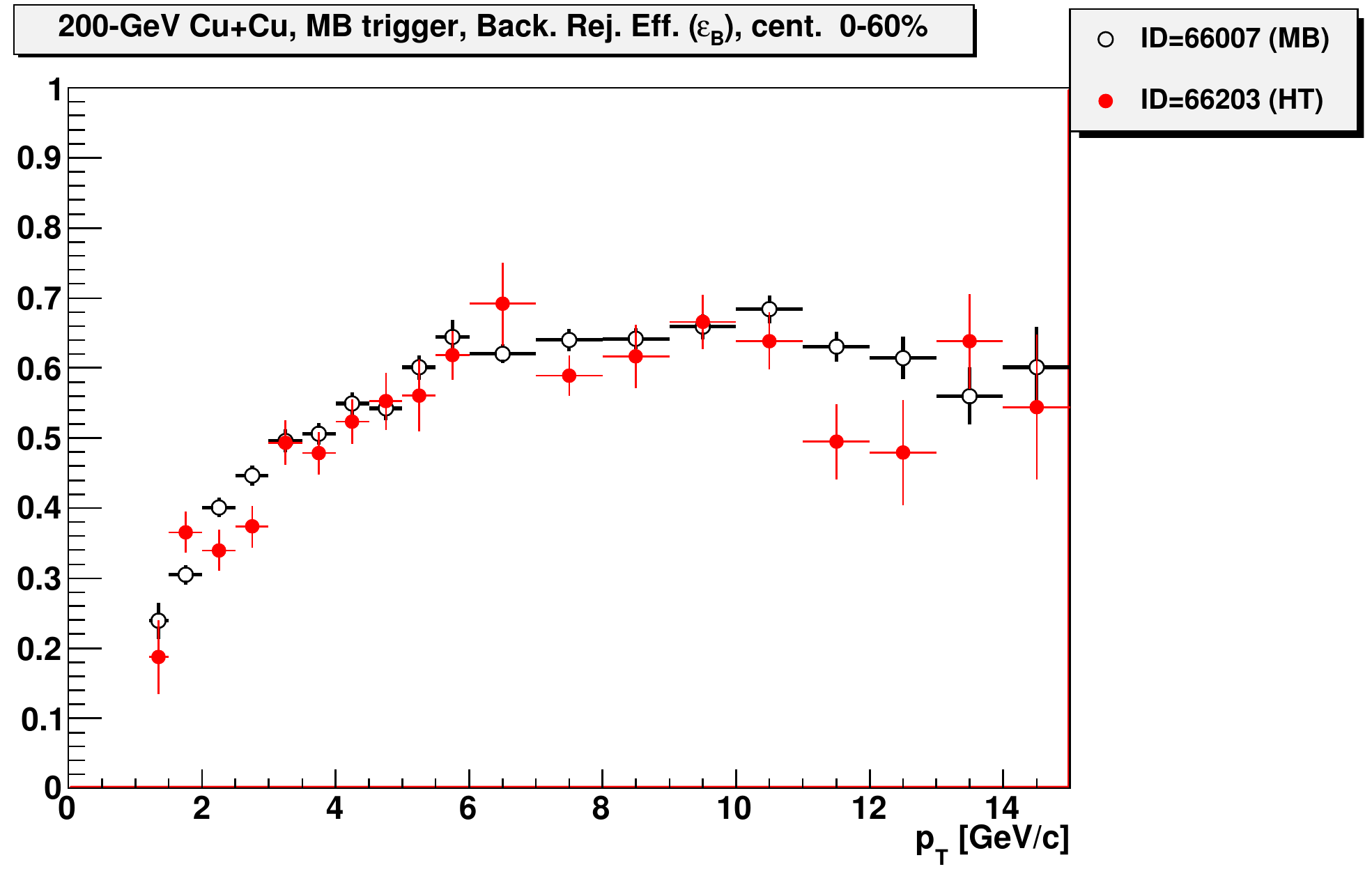}
\caption[Background rejection efficiency as a function of $p_{T}$ for two different STAR trigger IDs.]{Background rejection efficiency as a function of $p_{T}$ for the two different STAR trigger IDs considered in this analysis: The efficiency does not appear to depend upon the trigger ID of the underlying event.}
\label{fig:bre:trig_dep}
\includegraphics[width=0.85\linewidth]{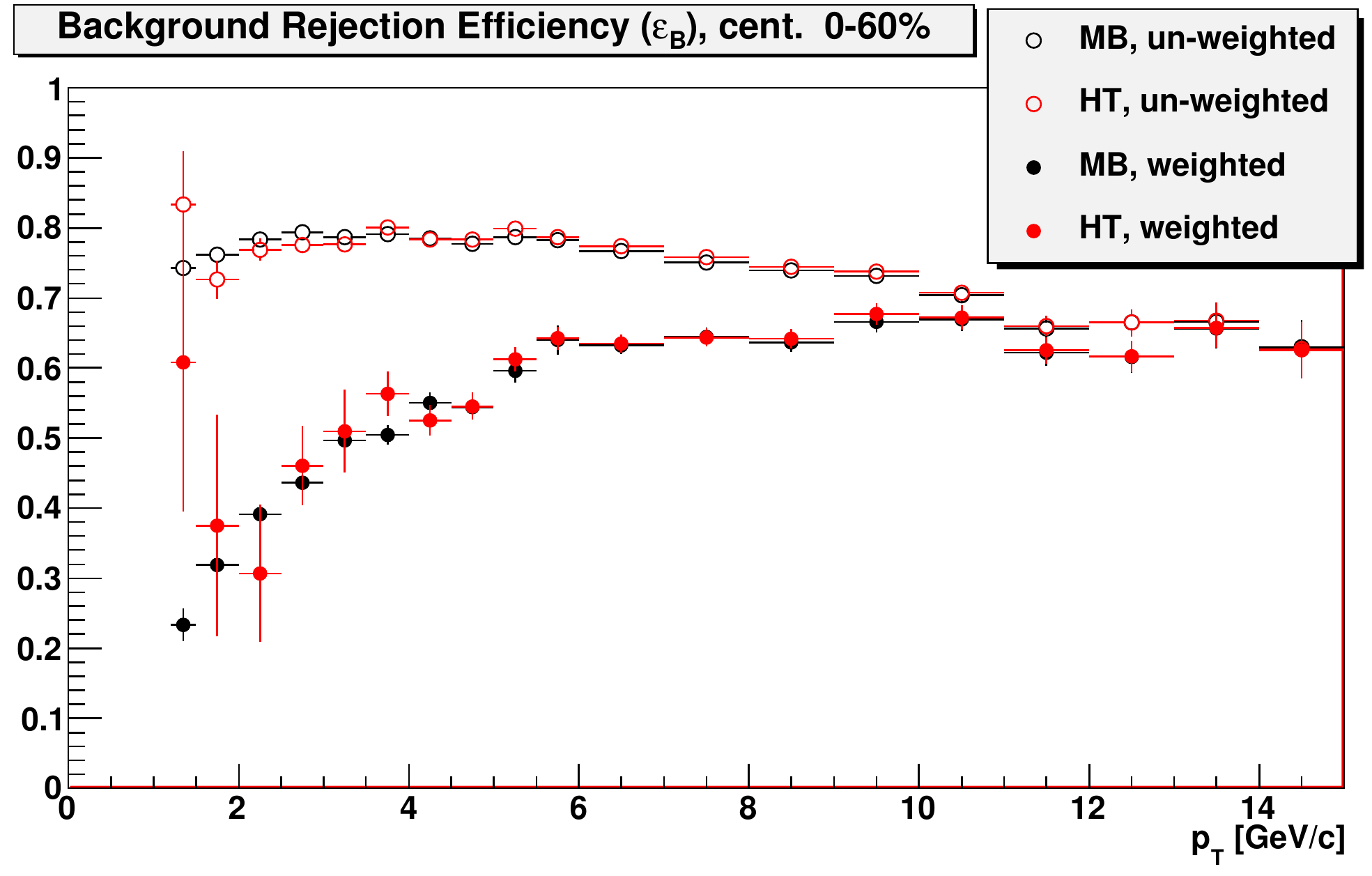}
\caption[Background rejection efficiency as a function of $p_{T}$ with and without the HT trigger condition.]{Background rejection efficiency as a function of $p_{T}$: The efficiency is calculated for tracks that point to towers which satisfied the high-tower trigger condition (red).  Also shown are efficiency calculations when that condition is not required.}
\label{fig:bre:track_trig_dep}
\end{center}
\end{figure}

In this section, various calculations\footnote{See the note on statistical uncertainties in Section~\ref{sec:uncertainties:weff}.} of the background rejection efficiency for data set $P_{1}$ are presented.  Unless otherwise stated, the weighting function was $F[\pi^{0},2]$.  Figure~\ref{fig:bre:cent_dep} shows the efficiency for five different centrality classes.  No dependence of the background rejection efficiency upon collision centrality is observed.  Figure~\ref{fig:bre:trig_dep} shows $\varepsilon_{B}$ for two different types of underlying events: events with trigger id 66007 (the minimum-bias trigger) and events with trigger id 66203 (high-tower triggered events).  No strong dependence of $\varepsilon_{B}$ on the trigger id of the underlying event is observed.  Taken together, these results support the assumption that the background rejection efficiency does not depend on the characteristics of the underlying event in which the photonic $e^{\pm}$ is observed.

Figure~\ref{fig:bre:track_trig_dep} shows $\varepsilon_{B}$ for $e^{\pm}$ that hit a tower which satisfied the high-tower trigger condition.  Above 3 GeV/$c$ transverse momentum, $\varepsilon_{B}$ does not appear to depend on whether or not the trigger condition was satisfied.  The high-tower trigger condition removes almost all particles with transverse momentum less than 3 GeV/$c$, so there are not enough particles in the calculation to determine whether or not $\varepsilon_{B}$ is affected by applying the trigger condition.  Low-$p_{T}$ $e^{\pm}$ from the real high-tower triggered data set are excluded from the calculation of the non-photonic $e^{\pm}$ yield and nuclear modification factor.

\begin{figure}[htbp]
\begin{center}
\includegraphics[width=0.85\linewidth]{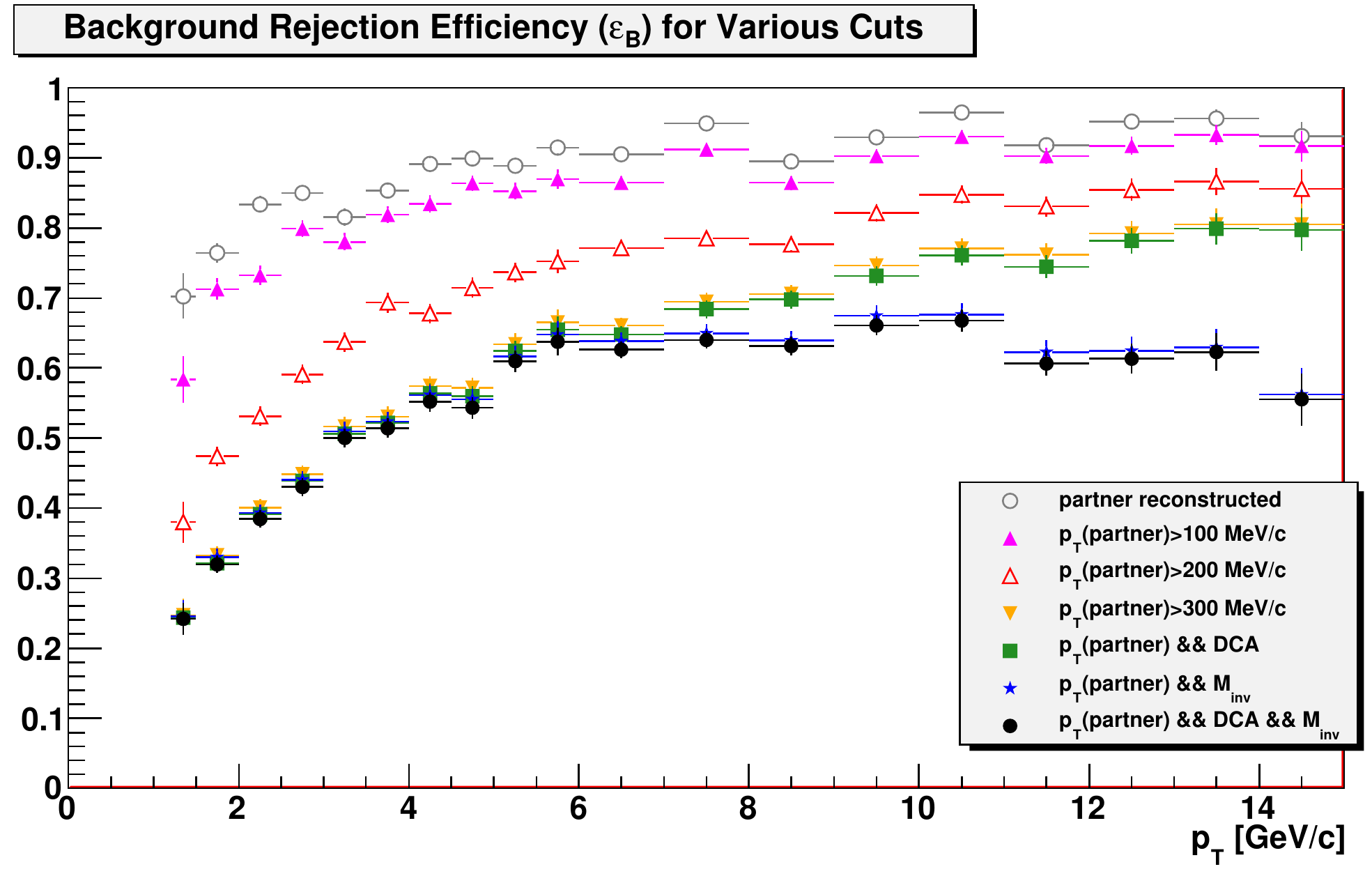}
\caption[Background rejection efficiency as a function of $p_{T}$ for various combinations of partner and pair selection cuts.]{Background rejection efficiency as a function of $p_{T}$ for various combinations of partner and pair selection cuts:  For \textit{partner reconstructed}, the true conversion partner shares at least one half of its TPC points with a reconstructed track, but no other cuts are applied.}
\label{fig:bre:partner_cuts}
\includegraphics[width=0.85\linewidth]{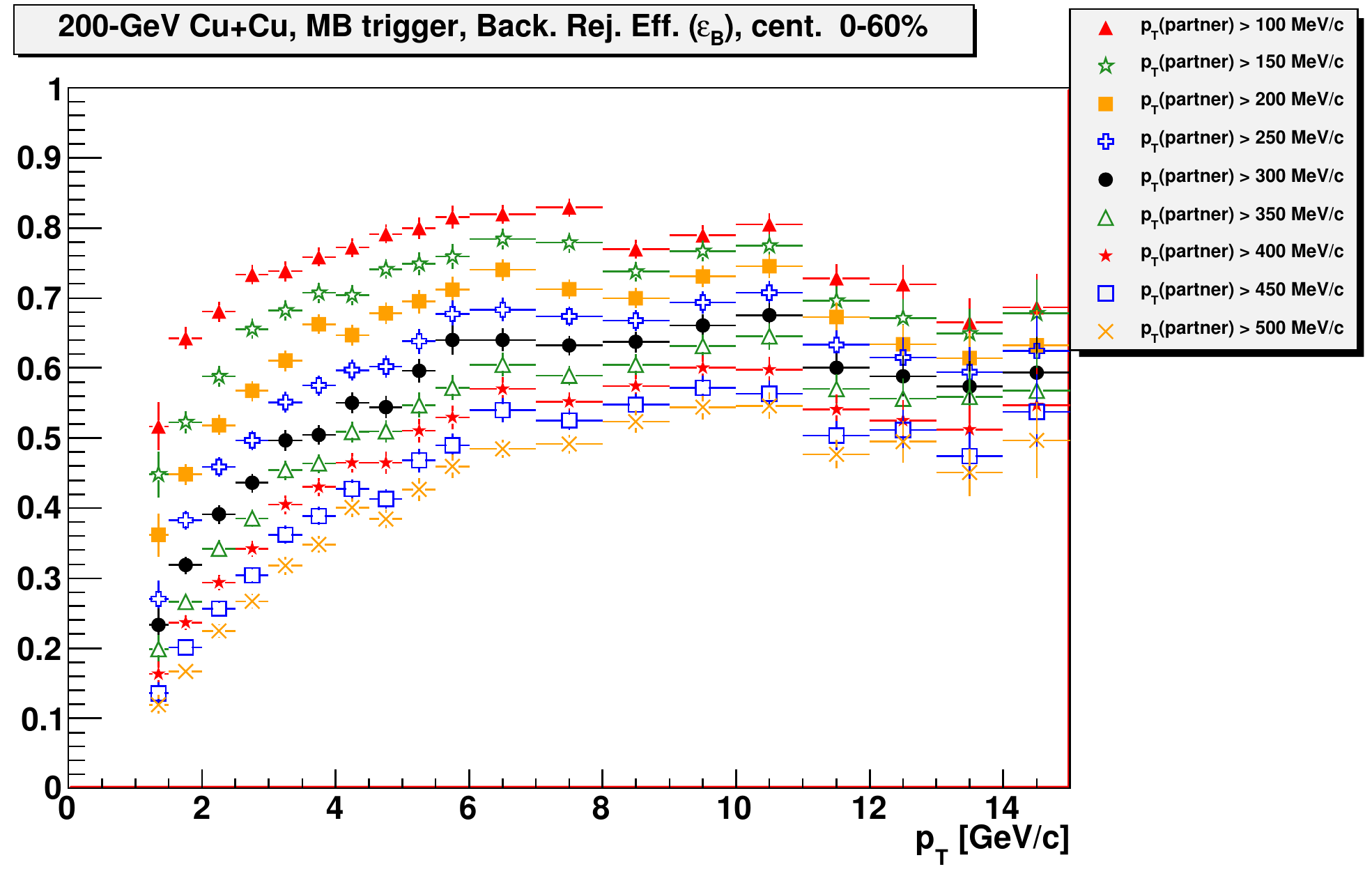}
\caption[Background rejection efficiency as a function of $p_{T}$ for various values of the $p_{T}(partner)$ cut.]{Background rejection efficiency as a function of $p_{T}$ for various values of the $p_{T}(partner)$ cut:  All other cuts have been applied.}
\label{fig:bre:pt2_dep}
\end{center}
\end{figure}

Figure~\ref{fig:bre:partner_cuts} shows $\varepsilon_{B}$ for various combinations of partner and pair selection cuts.

Figure~\ref{fig:bre:pt2_dep} shows the effect upon the background rejection efficiency of various values of the $p_{T}(partner)$ cut.

\begin{figure}[htbp]
\begin{center}
\includegraphics[width=0.85\linewidth]{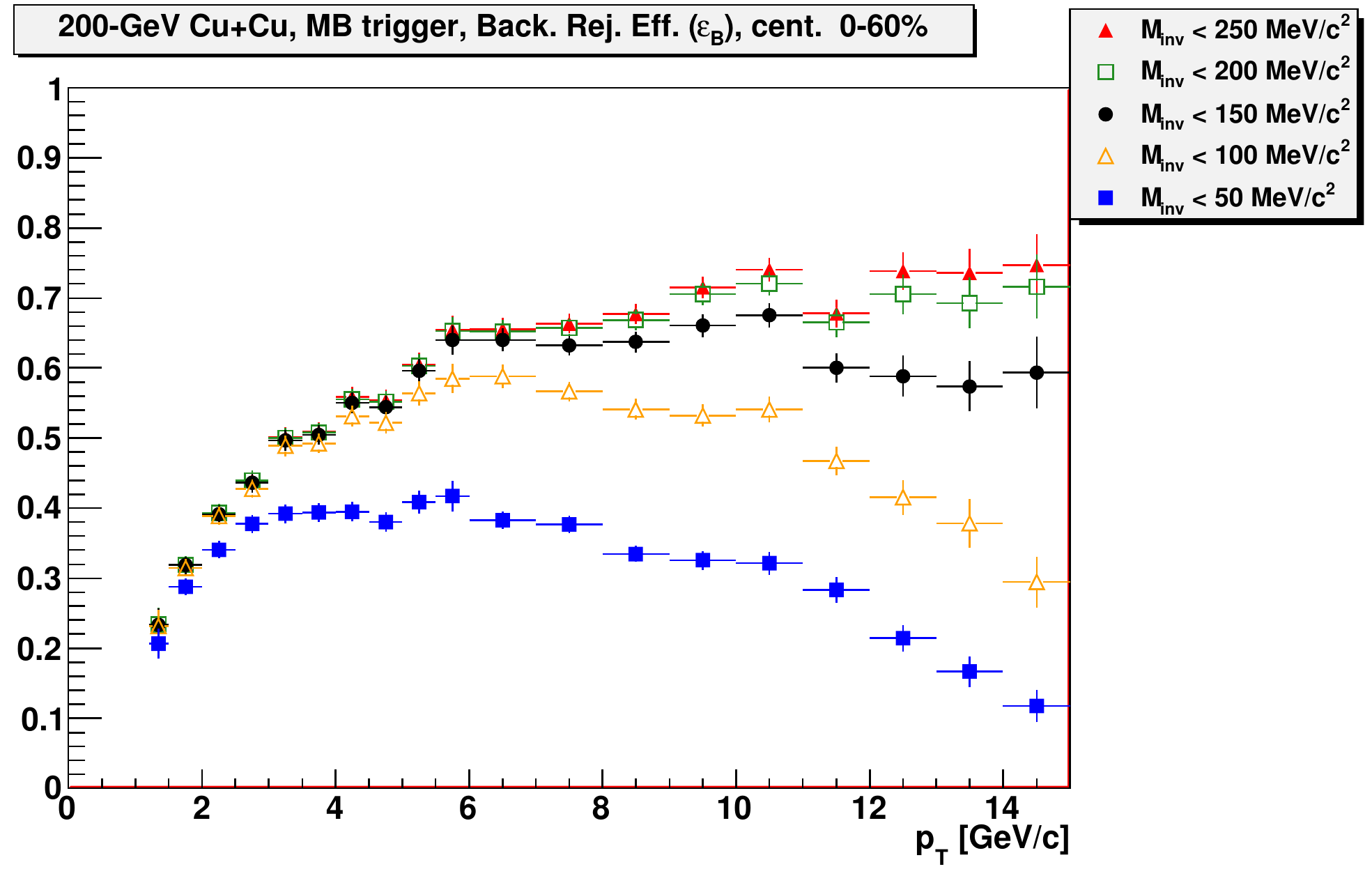}
\caption[Background rejection efficiency as a function of $p_{T}$ for various values of the invariant-mass cut.]{Background rejection efficiency as a function of $p_{T}$ for various cuts on the invariant mass of the reconstructed photonic $e^{-}e^{+}$ pair:  All other cuts have been applied.}
\label{fig:bre:mass_dep}
\includegraphics[width=0.85\linewidth]{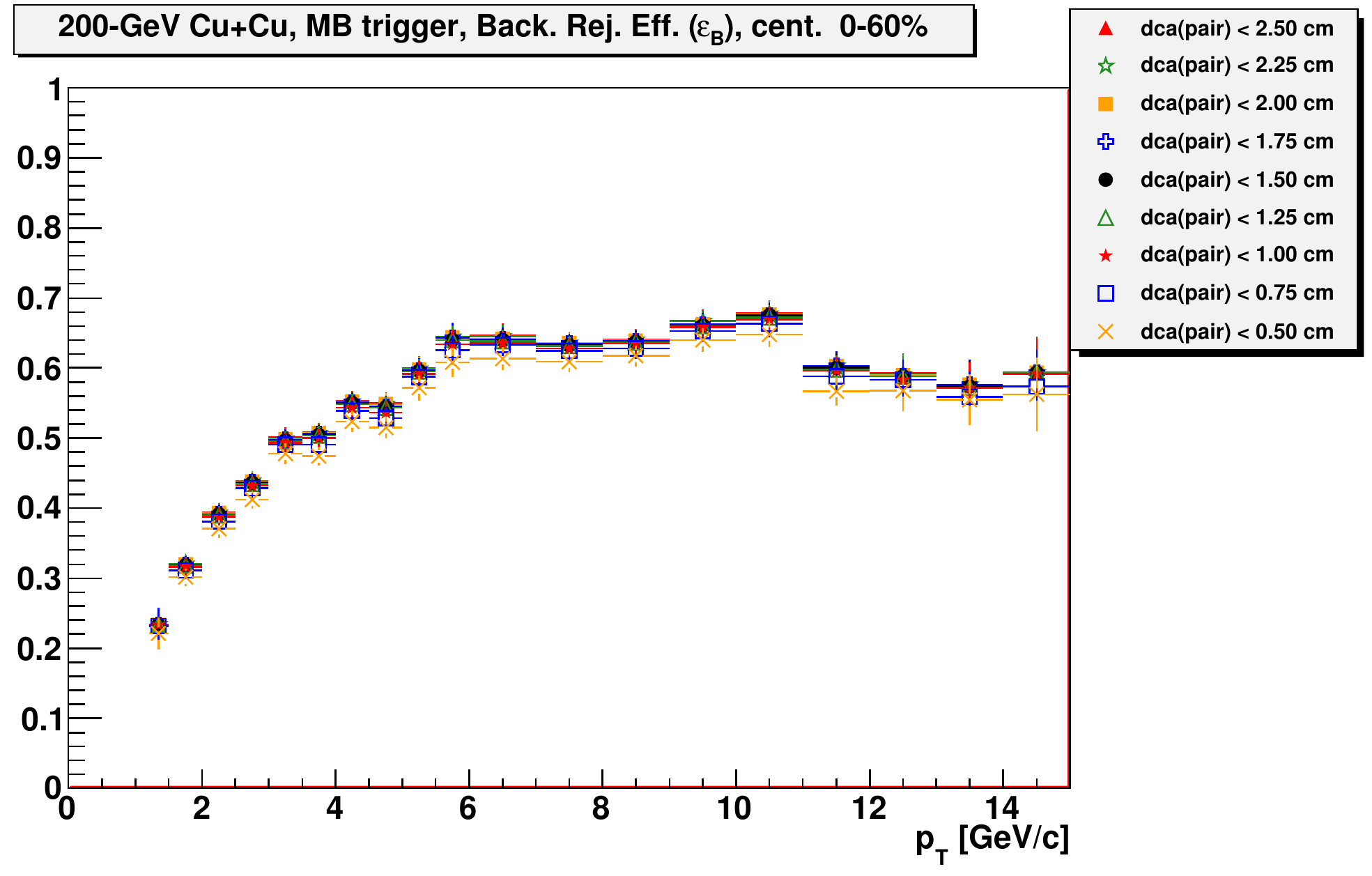}
\caption[Background rejection efficiency as a function of $p_{T}$ for various cuts on the pair DCA.]{Background rejection efficiency as a function of $p_{T}$ for various cuts on the distance of closest approach between the two reconstructed photonic $e^{\pm}$ tracks:  All other cuts have been applied.}
\label{fig:bre:dca_dep}
\end{center}
\end{figure}

Figure~\ref{fig:bre:mass_dep} shows the effect upon the background rejection efficiency of various values of the invariant-mass cut.

Figure~\ref{fig:bre:dca_dep} shows the effect upon $\varepsilon_{B}$ of various values of the pair DCA cut.

\begin{figure}[htbp]
\begin{center}
\includegraphics[width=0.85\linewidth]{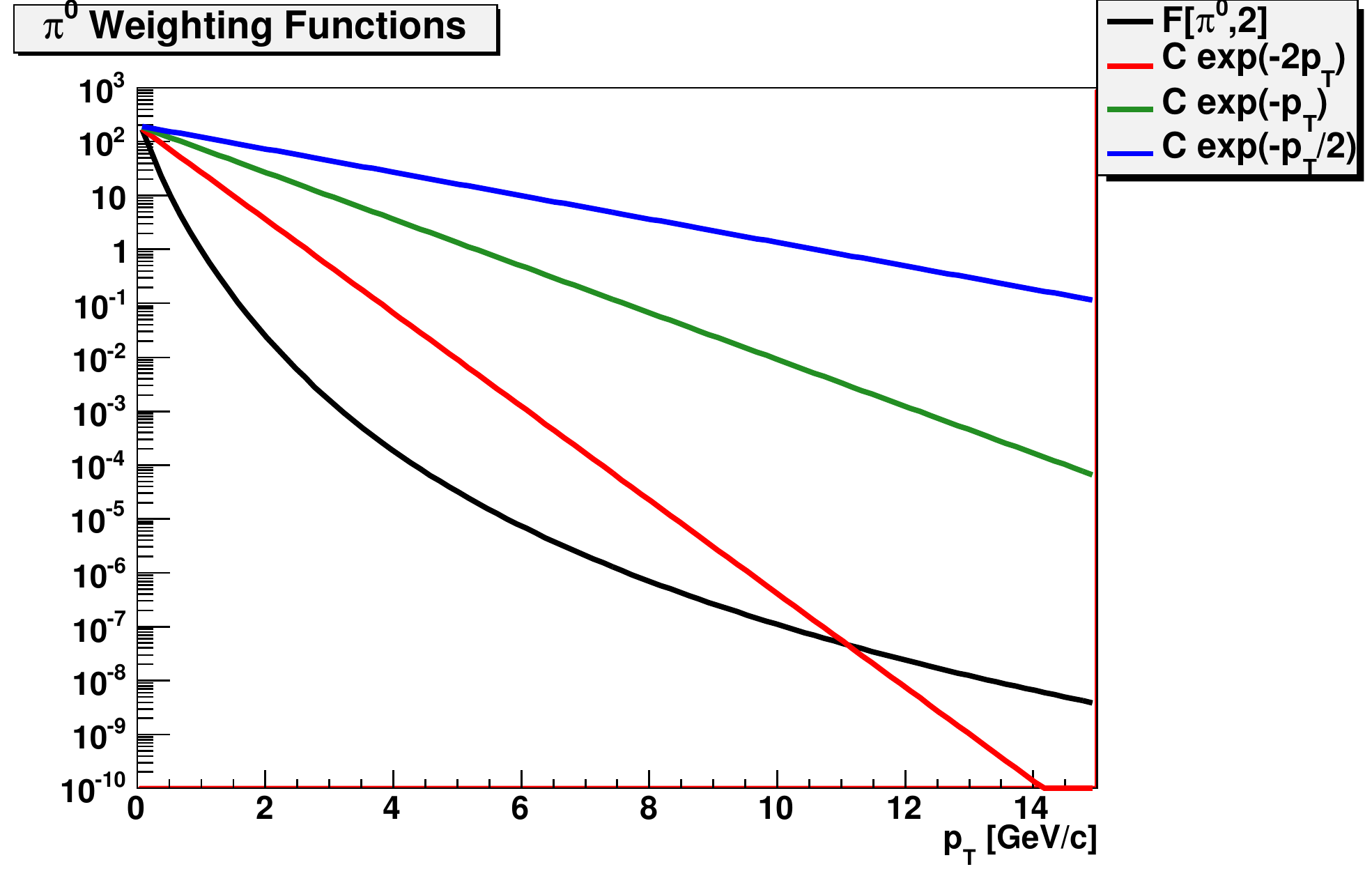}
\caption[Different $\pi^{0}$ weighting functions used in the calculations of $\varepsilon_{B}$.]{Different $\pi^{0}$ weighting functions used in the calculations of $\varepsilon_{B}$ in Figure~\ref{fig:bre:eff_diff_weights}.}
\label{fig:bre:diff_weights}
\includegraphics[width=0.85\linewidth]{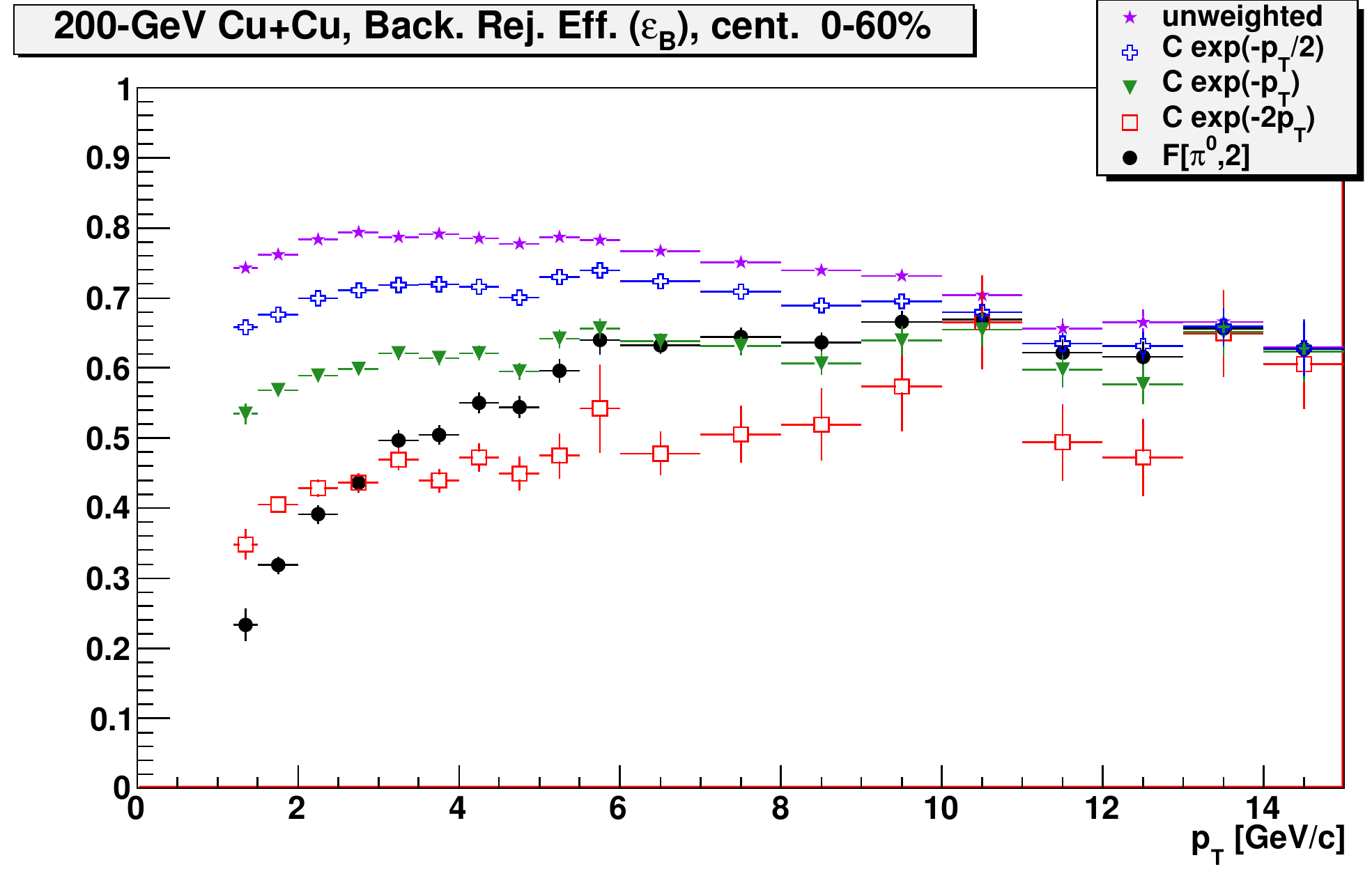}
\caption{Background rejection efficiency as a function of $p_{T}$ for different weighting functions.}
\label{fig:bre:eff_diff_weights}
\end{center}
\end{figure}

\begin{figure}[htbp]
\begin{center}
\includegraphics[width=0.85\linewidth]{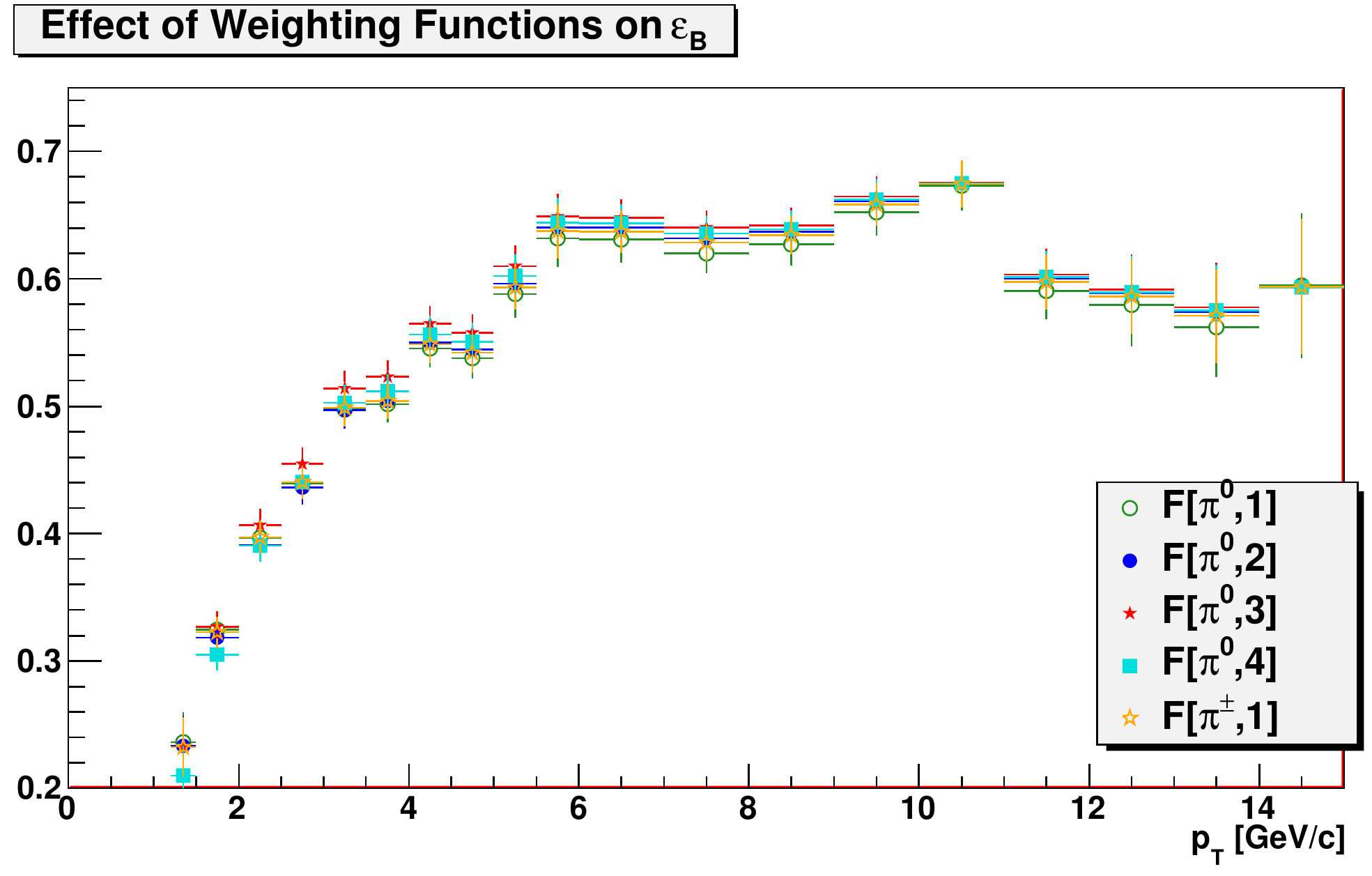}
\caption[Background rejection efficiency as a function of $p_{T}$, with five different $\pi^{0}$ weighting functions.]{Background rejection efficiency as a function of $p_{T}$, with five different fits to measured $\pi^{0}$ spectra (see Figure~\ref{fig:bre:rhic_pion}) used as weighting functions.}
\label{fig:bre:eff_diff_pi0_weights}
\end{center}
\end{figure}

Figures~\ref{fig:bre:diff_weights},~\ref{fig:bre:eff_diff_weights}, and~\ref{fig:bre:eff_diff_pi0_weights} provide a demonstration of the effect of various weighting functions on the background rejection efficiency.  Figure~\ref{fig:bre:diff_weights} shows a set of example weighting functions used in efficiency calculations.  In addition to weighting function $F[\pi^{0},2]$, several exponential weighting functions are shown.  Figure~\ref{fig:bre:eff_diff_weights} shows $\varepsilon_{B}$ for each of these functions, as well as the un-weighted efficiency.  Figure~\ref{fig:bre:eff_diff_weights} shows the efficiency calculated for each of the five pion weighting functions defined in Section~\ref{sec:bre:weighting}.  The systematic uncertainties in the background rejection efficiency due to different weighting functions will be explored in Section~\ref{sec:bre:combined}.

\clearpage

\clearpage

\section[Background Rejection Efficiency for Data Set $P_{2}$: $\pi^{0}$ Dalitz Decays]{Background Rejection Efficiency for \\ Data Set $\boldsymbol{P_{2}}$: $\boldsymbol{\pi^{0}}$ Dalitz Decays}
\label{sec:bre:dalitz}

\begin{figure}[htbp]
\begin{center}
\includegraphics[width=0.85\linewidth]{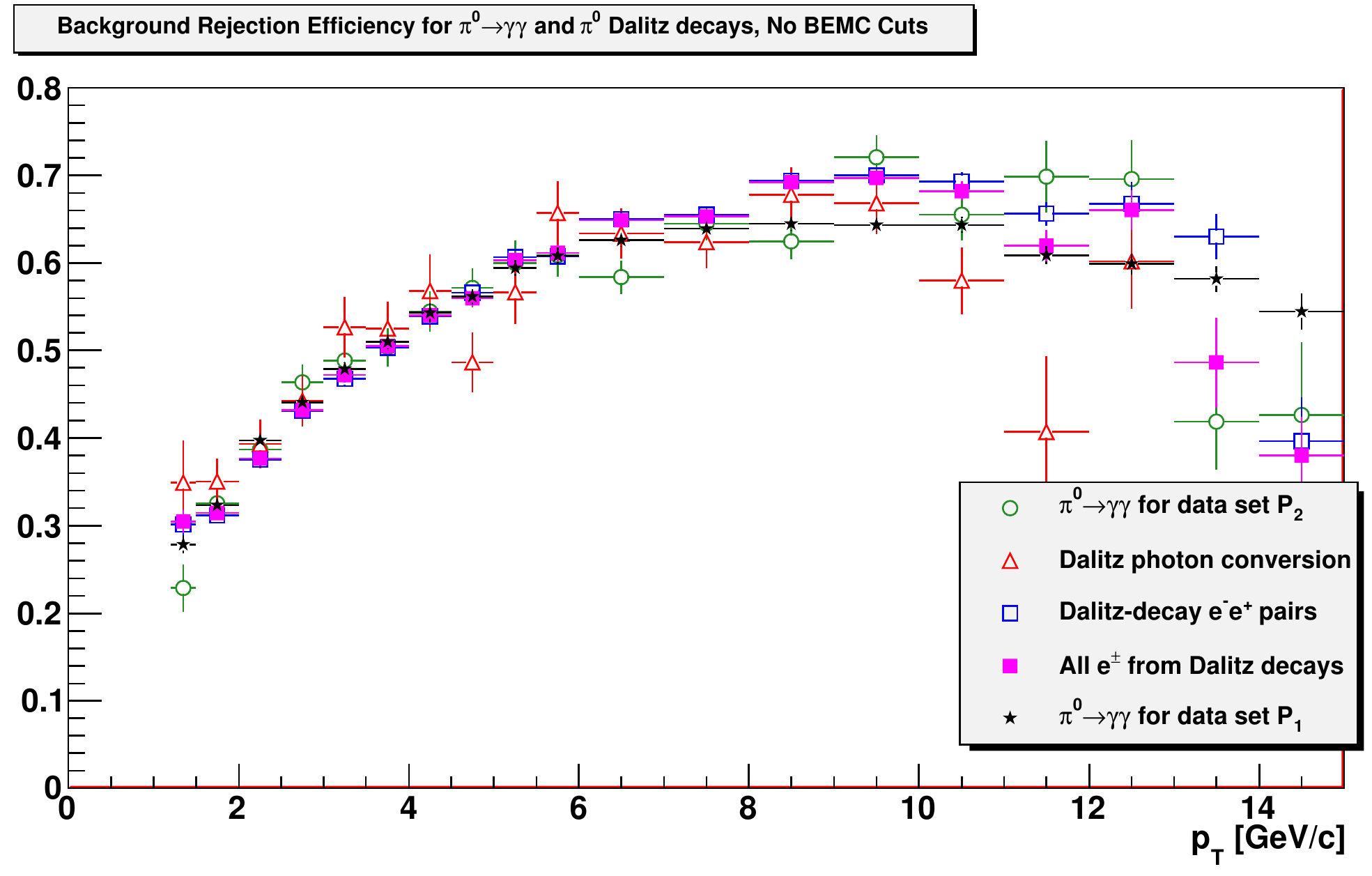}
\caption[Background rejection efficiency as a function of $p_{T}$ for various sources of $e^{\pm}$ (without BEMC cuts).]{Background rejection efficiency as a function of $p_{T}$ for various sources of $e^{\pm}$:  To increase the numbers of tracks in the calculation, no cuts using the BEMC have been applied.}
\label{fig:bre:bre_dalitz_noemc}
\includegraphics[width=0.85\linewidth]{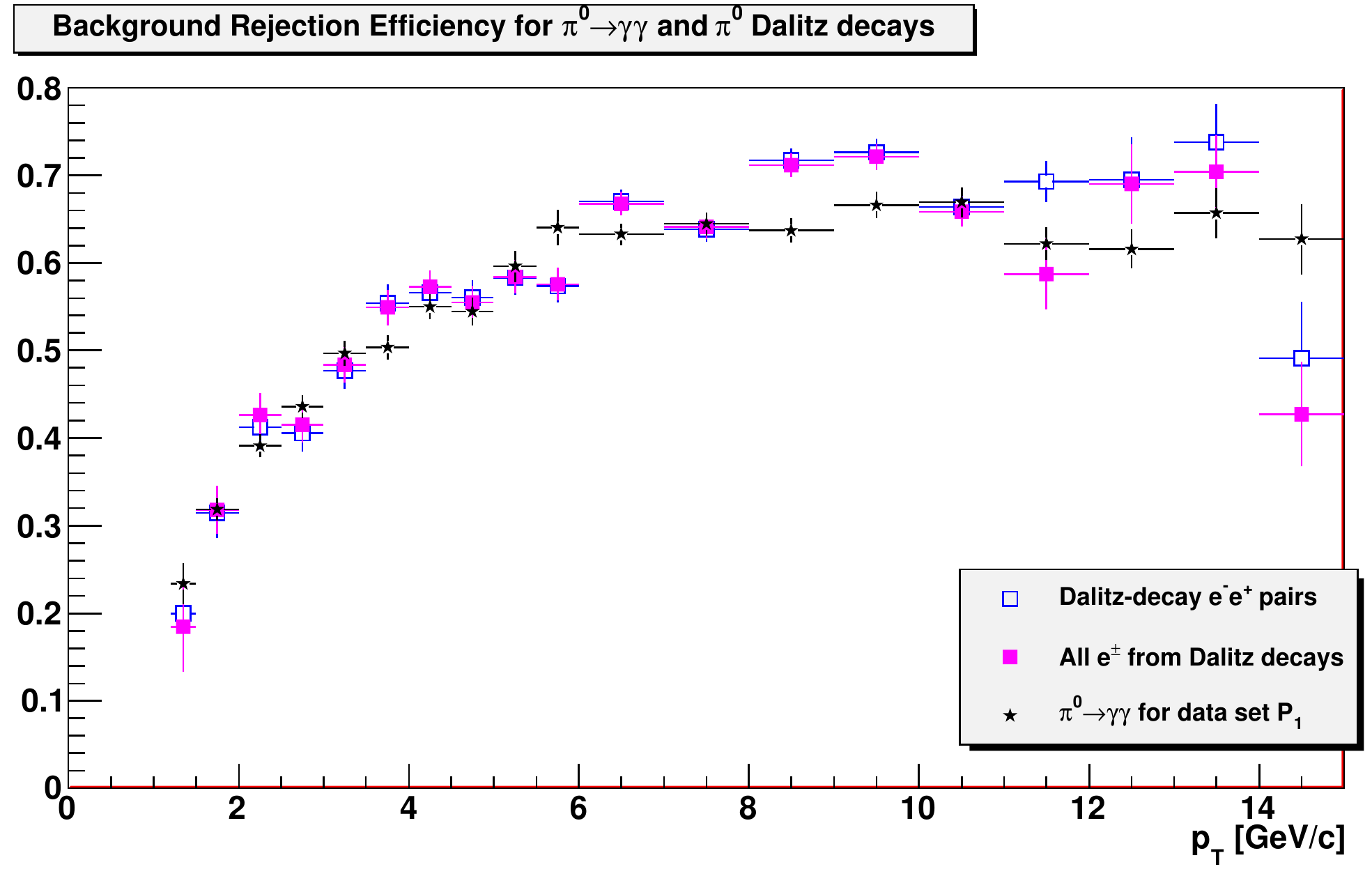}
\caption[Background rejection efficiency as a function of $p_{T}$ for various sources of $e^{\pm}$.]{Background rejection efficiency as a function of $p_{T}$ for various sources of $e^{\pm}$; the standard set of cuts has been applied.}
\label{fig:bre:bre_dalitz_emc}
\end{center}
\end{figure}

In this section, calculations of the background rejection efficiency for data set $P_{2}$ are presented.  Figure~\ref{fig:bre:bre_dalitz_noemc} shows the background rejection efficiency calculated for data set $P_{2}$ for photonic $e^{\pm}$ from three sources.

\begin{itemize}
\item conversions of photons from the $\pi^{0}\rightarrow\gamma\gamma$ decay
\item conversions of the real photons produced in $\pi^{0}$ Dalitz decays
\item $e^{\pm}$ produced directly in $\pi^{0}$ Dalitz decays
\end{itemize}

In order to increase the number of $e^{\pm}$ in the calculations, the efficiencies shown in Figure~\ref{fig:bre:bre_dalitz_noemc} were calculated without BEMC cuts and with a different pseudorapidity cut $(|\eta|<0.7)$.  For comparison, Figure~\ref{fig:bre:bre_dalitz_noemc} also shows the background rejection efficiency for data set $P_{1}$ (calculated with the same set of cuts).  The calculations of $\varepsilon_{B}$ for the two-photon decay mode appear to agree for the two simulation data sets $P_{1}$ and $P_{2}$.  At high $p_{T}$, $\varepsilon_{B}$ for Dalitz-decay $e^{\pm}$ appears to be higher than the efficiency for $e^{\pm}$ from photon conversions.  At low $p_{T}$, the opposite appears to be true.  Figure~\ref{fig:bre:bre_dalitz_emc} shows calculations of $\varepsilon_{B}$ (with the standard set of cuts) for electrons from $\pi^{0}$ Dalitz decays along with the efficiency calculation for data set $P_{1}$ for comparison.

\clearpage

\section[Background Rejection Efficiency for Data Set $P_{3}$: Other Sources]{Background Rejection Efficiency for \\ Data Set $\boldsymbol{P_{3}}$: Other Sources}
\label{sec:bre:photon}

\begin{figure}[htbp]
\begin{center}
\includegraphics[width=0.85\linewidth]{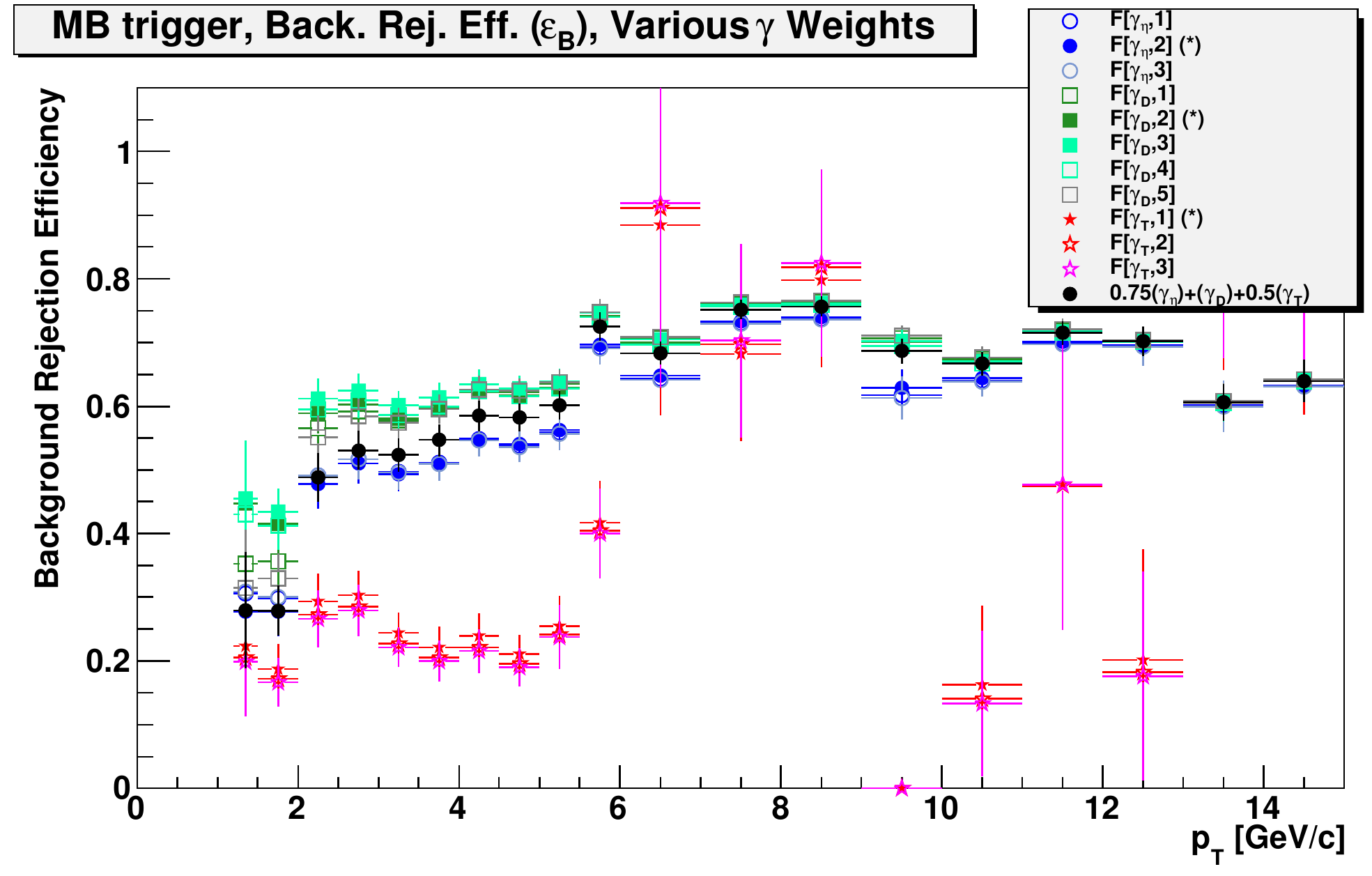}
\caption[Background rejection efficiency for different photon weighting-function components.]{Background rejection efficiency for the photon weighting components shown in Figure~\ref{fig:bre:photon_weight_components}:  Also shown is the efficiency calculated using the primary mixture of these components. }
\label{fig:bre:diff_pho_components}
\includegraphics[width=0.85\linewidth]{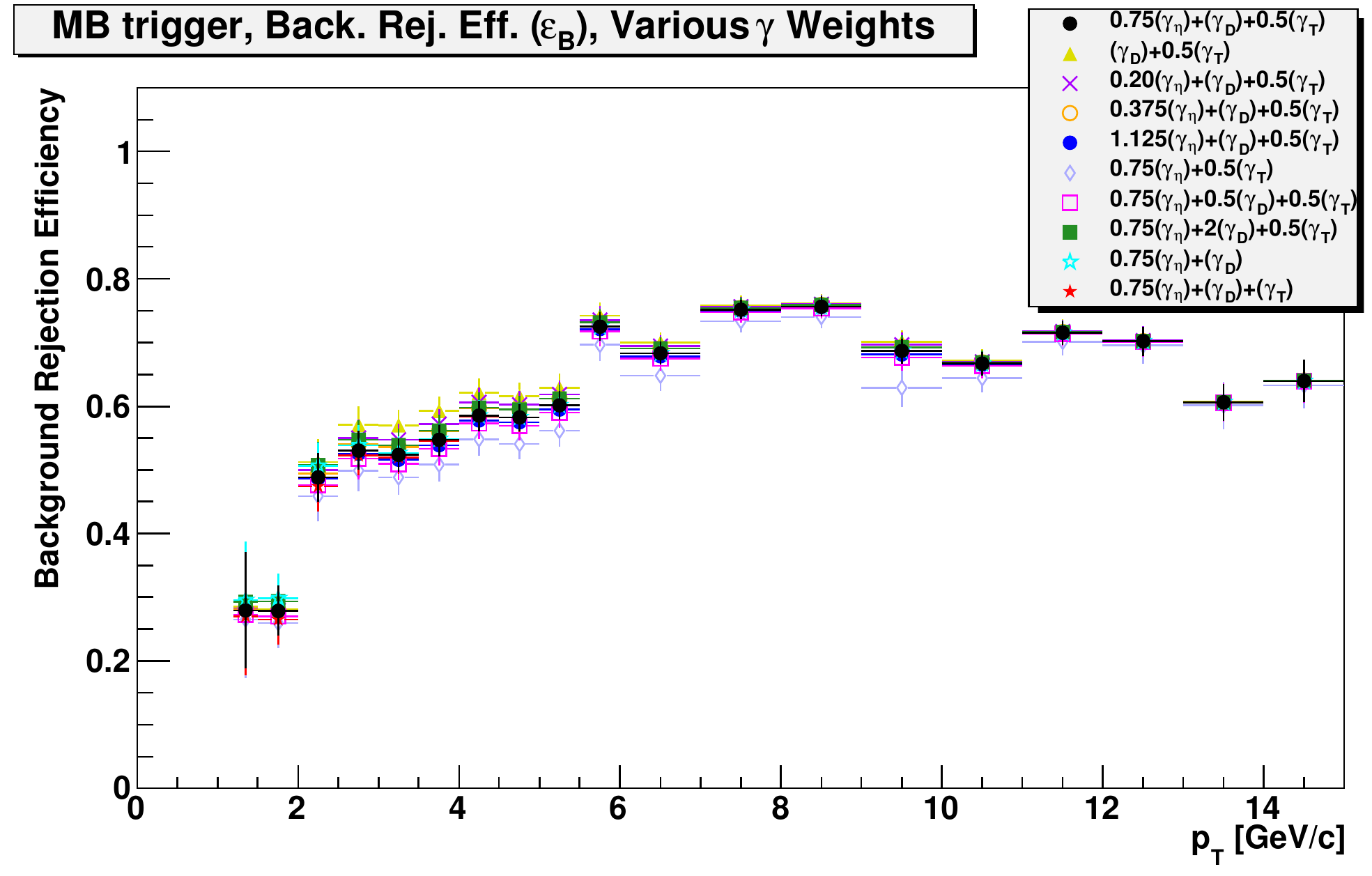}
\caption[Background rejection efficiency for different mixtures of photon weighting-function components.]{Background rejection efficiency for different mixtures of photon weighting function components, accounting for photons from $\eta$-meson decays, direct photons, and thermal direct photons.}
\label{fig:bre:eff_diff_pho_w}
\end{center}
\end{figure}

Data set $P_{3}$ is used to determine the effect on the efficiency of photonic $e^{\pm}$ sources other than pions.  The photon weighting function has three components:

\begin{itemize}
\item $(\gamma_{\eta})$: photons from $\eta\rightarrow\gamma\gamma$ decays
\item $(\gamma_{D})$: direct photons (non-thermal)
\item $(\gamma_{T})$: thermal direct photons
\end{itemize}

Unless otherwise noted, these components are described by functions $F[\gamma_{\eta},2]$, $F[\gamma_{D},2]$, and $F[\gamma_{T},1]$, respectively.  The relative strengths of these components should be different for Cu + Cu collisions than for $p+p$ or Au + Au collisions.  For this analysis, the combination $R_{\mathrm{CuCu}}^{\eta}(\gamma_{\eta})+(\gamma_{D})+0.5(\gamma_{T})$ was primarily used.  The values of $R_{\mathrm{CuCu}}^{\eta}$ are assumed to be the same as the pion nuclear modification factor at high $p_{T}$,~\cite{PhysRevC.81.054907} which varies with the collision centrality: $R_{\mathrm{CuCu,0-20\%}}^{\eta}=0.64$, $R_{\mathrm{CuCu,20-60\%}}^{\eta}=0.94$, and $R_{\mathrm{CuCu,0-60\%}}^{\eta}=0.75$.  These values of the nuclear modification factor will be allowed to vary by $\pm 50\%$.  The PHENIX collaboration has shown~\cite{PhysRevLett.94.232301} that non-thermal direct photons are not suppressed in Au + Au collisions with respect to $p+p$ collisions.  This analysis assumes that the direct photon yield is also not suppressed in Cu + Cu collisions.  A factor of 0.5 is used as a rough estimate of the suppression of thermal photons in Cu + Cu collisions relative to Au + Au collisions.  However, the thermal-photon contribution is only important at low $p_{T}$ and does not affect the background rejection efficiency for the transverse-momentum range of interest for this dissertation.

Figure~\ref{fig:bre:diff_pho_components} shows the background rejection efficiency calculated separately for each of the eleven alternate photon weighting function components (shown in Figure~\ref{fig:bre:photon_weight_components}).

Figure~\ref{fig:bre:eff_diff_pho_w} shows the background rejection efficiency for mixtures of photon weighting-function components, with the relative strengths of the three components varied.

\clearpage

\section{Combined Background Rejection Efficiency}
\label{sec:bre:combined}

\begin{figure}[htbp]
\begin{center}
\includegraphics[width=0.85\linewidth]{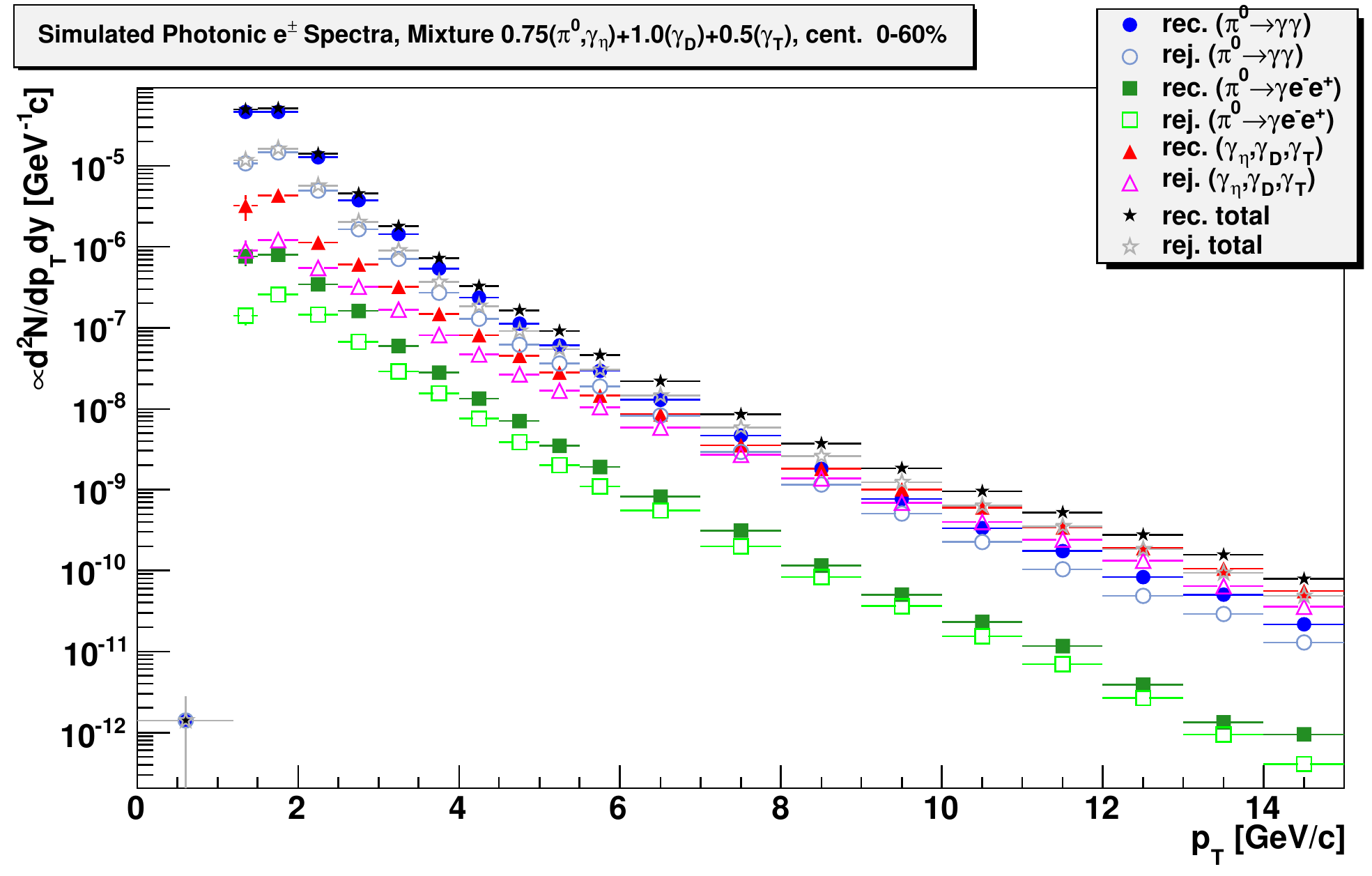}
\caption{Spectra of reconstructed and rejected photonic $e^{\pm}$ from all three simulation data sets.}
\label{fig:bre:combined_spectra}
\includegraphics[width=0.85\linewidth]{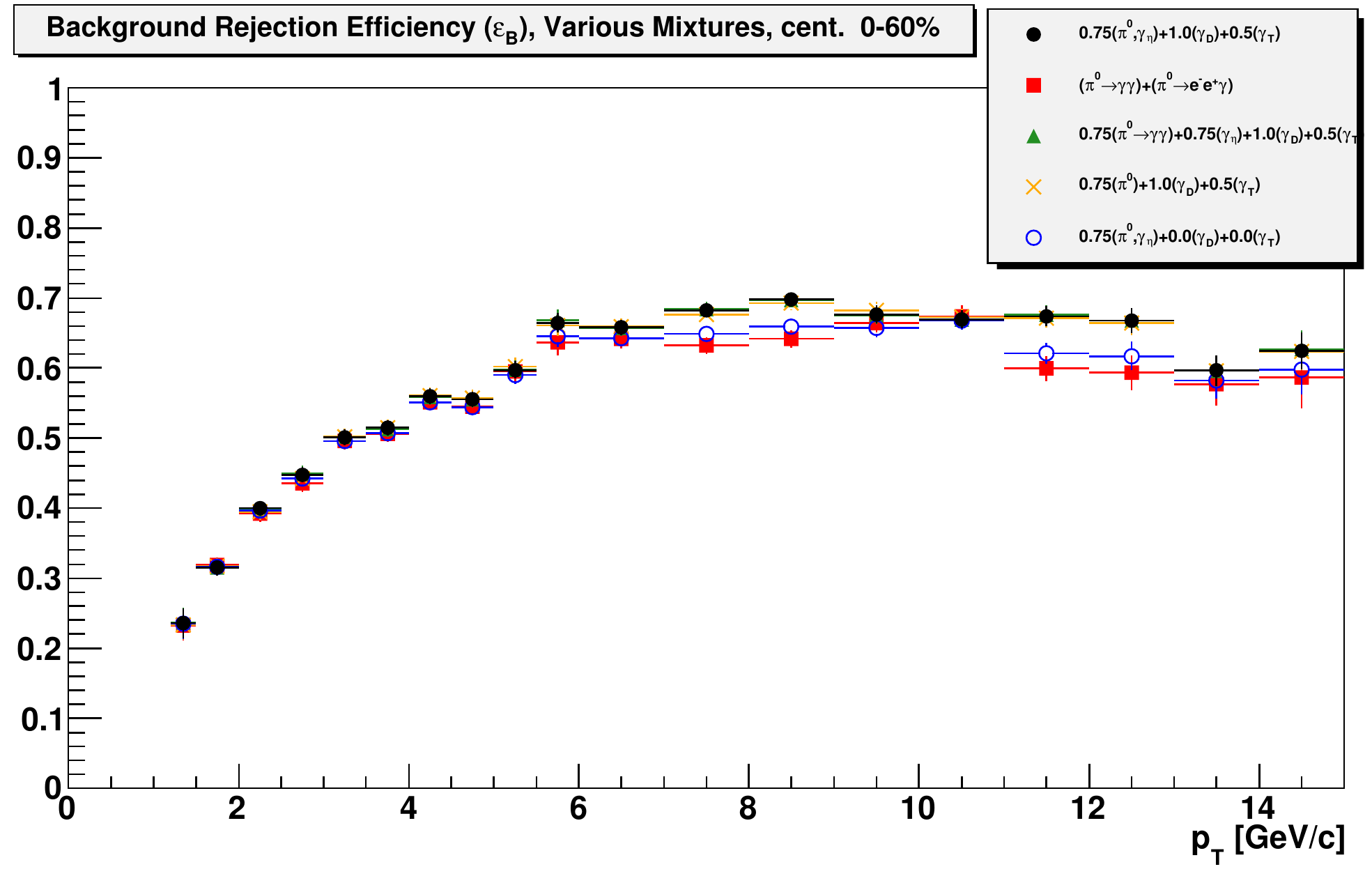}
\caption[Background rejection efficiency for various mixtures of simulated photonic $e^{\pm}$.]{Background rejection efficiency for various mixtures of simulated photonic $e^{\pm}$: Completely removing the Dalitz-decay or $\eta$-meson contributions has little effect on the efficiency.}
\label{fig:bre:combined_eff_1}
\end{center}
\end{figure}

The background rejection efficiency for all sources is calculated by adding together the reconstructed and rejected $e^{\pm}$ yields from $\pi^{0}\rightarrow\gamma\gamma$ decays (data set $P_{1}$), $\pi^{0}$ Dalitz decays (data set $P_{2}$), and other photon sources (data set $P_{3}$), as shown in Figure~\ref{fig:bre:combined_spectra}.  For each data set, the simulated $e^{\pm}$ spectra are scaled by the number of simulated $\pi^{0}$ or photons with 1 GeV/$c$ $<p_{T}(\pi^{0},\gamma)<$ 15 GeV/$c$.  Photonic $e^{\pm}$ from $\pi^{0}\rightarrow\gamma\gamma$ decays (data set $P_{1}$) dominate the combined spectrum at low $p_{T}$, while the contribution from direct photons (in data set $P_{3}$) is larger at high $p_{T}$.

\begin{figure}[htbp]
\begin{center}
\includegraphics[width=0.85\linewidth]{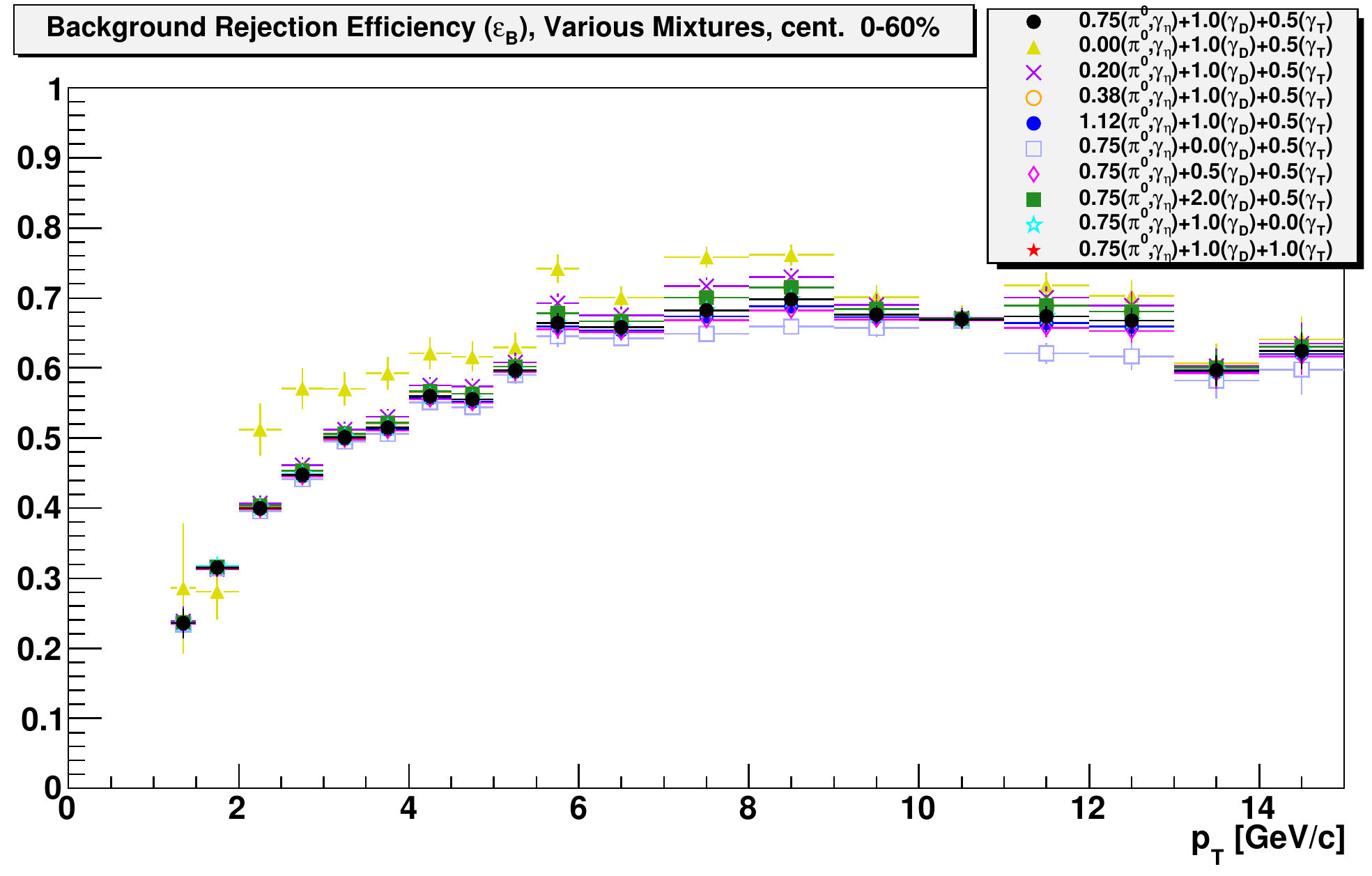}
\caption{Background rejection efficiency for various mixtures of simulated photonic $e^{\pm}$.}
\label{fig:bre:combined_eff_2}
\includegraphics[width=0.85\linewidth]{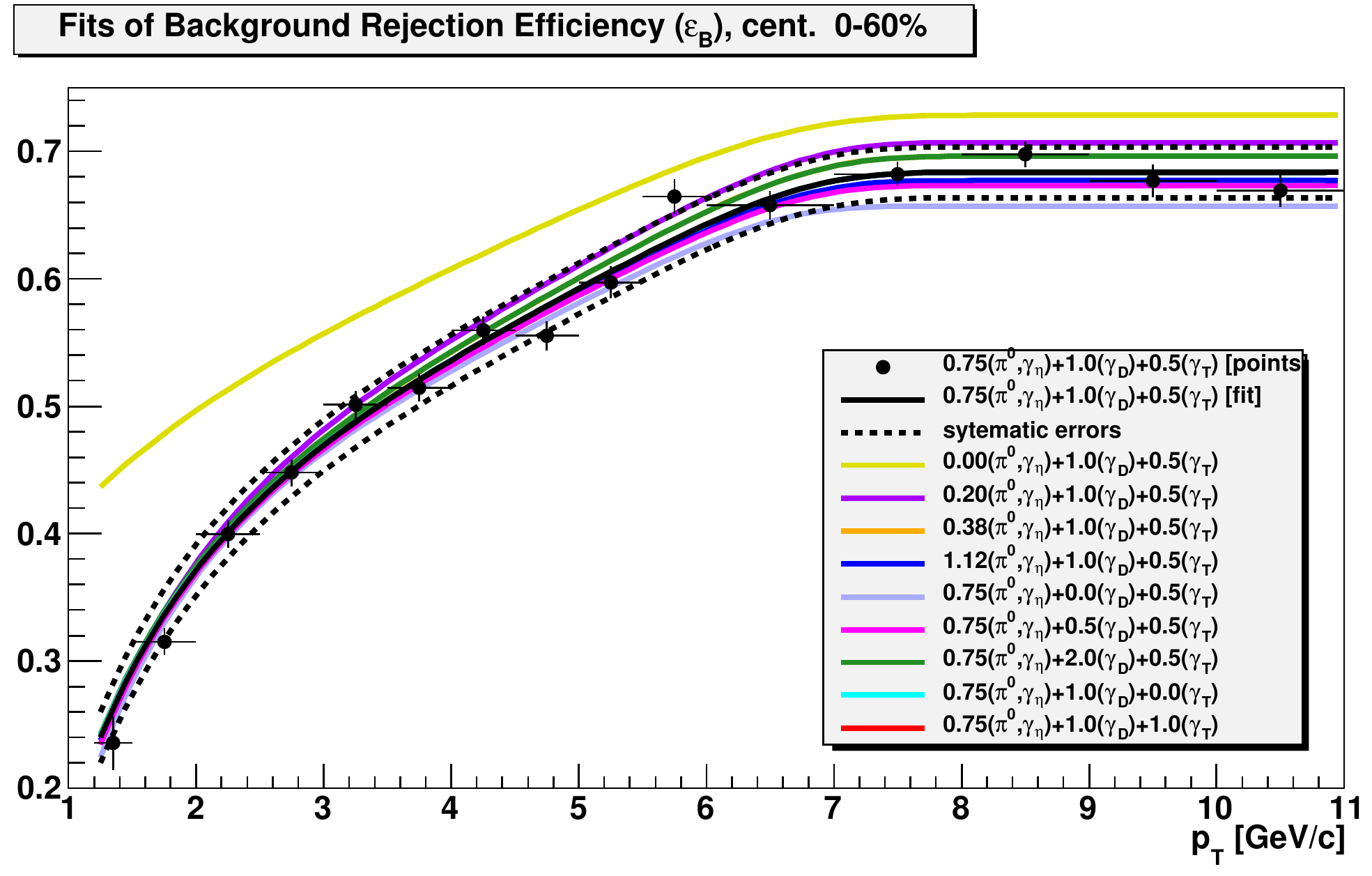}
\caption[Fits to the background rejection efficiency for various mixtures of simulated photonic $e^{\pm}$.]{Fits to the background rejection efficiency calculations shown in Figure~\ref{fig:bre:combined_eff_2}: Also shown are the primary efficiency measurement (points) and the $\pm 2\%$ estimate for the systematic uncertainty of the fits.}
\label{fig:bre:combined_eff_2_fits}
\end{center}
\end{figure}

Figures~\ref{fig:bre:combined_eff_1} and~\ref{fig:bre:combined_eff_2} show the background rejection efficiency calculations (for the 0-60\% centrality class) for various mixtures of photonic-$e^{\pm}$ sources.  Calculations with a greater direct-photon contribution tend to have a higher efficiency, due to the fact that the (non-thermal) direct-photon spectrum has a less steeply falling slope than the neutral-pion and $\eta$-meson spectra.  The effects on $\varepsilon_{B}$ of the $\pi^{0}$-Dalitz-decay and thermal-photon components are negligible.  Figure~\ref{fig:bre:combined_eff_2_fits} shows fits to the efficiency calculations shown in Figure~\ref{fig:bre:combined_eff_2} for the transverse-momentum range 2 GeV/$c<p_{T}<11$ GeV/$c$.  The fits have the form

\begin{equation}
C+A\exp\sinh\sinh\left(b(p_{T}-p_{0})\right).
\end{equation}

The fit parameters for the primary efficiency calculation are given in Table~\ref{table:bre:fits} for three centrality classes.  The systematic uncertainty curves are obtained by adding or subtracting 2\% from the primary mixture (black curve).  These uncertainties include the fits for almost all of the alternate mixtures shown.  The exceptions are the mixture with no direct photon contribution and the mixtures with meson contributions much smaller than expected.

\begin{figure}[htbp]
\begin{center}
\includegraphics[width=0.85\linewidth]{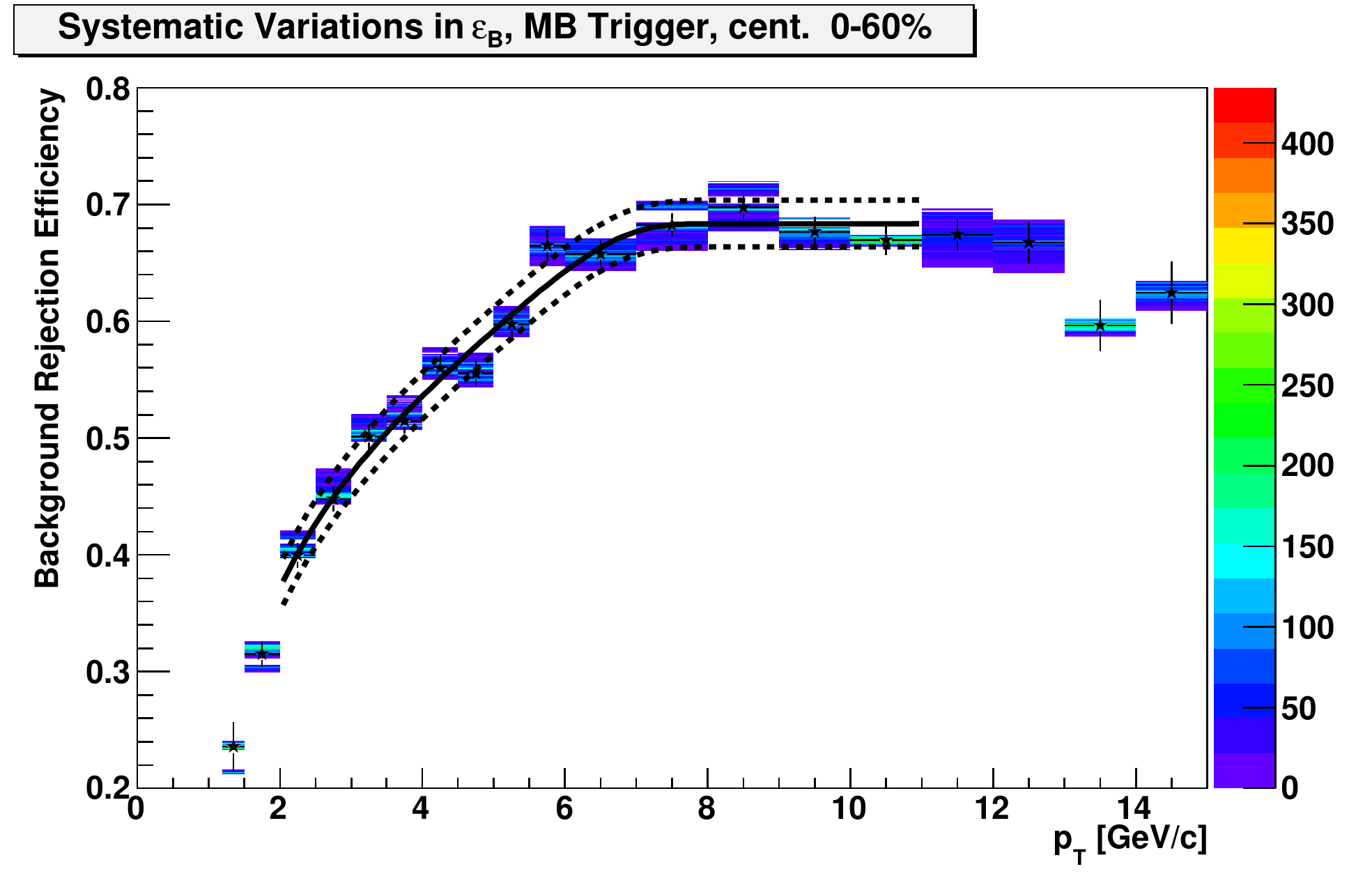}
\caption[Background rejection efficiency calculated for various permutations and mixtures of weighting functions.]{Background rejection efficiency (for the 0-60\% centrality class) calculated for various permutations of weighting functions and various mixtures of functions:  The primary calculation is shown in black; the dashed curves are the estimated systematic uncertainties to the primary fit.}
\label{fig:bre:systematic}
\includegraphics[width=0.85\linewidth]{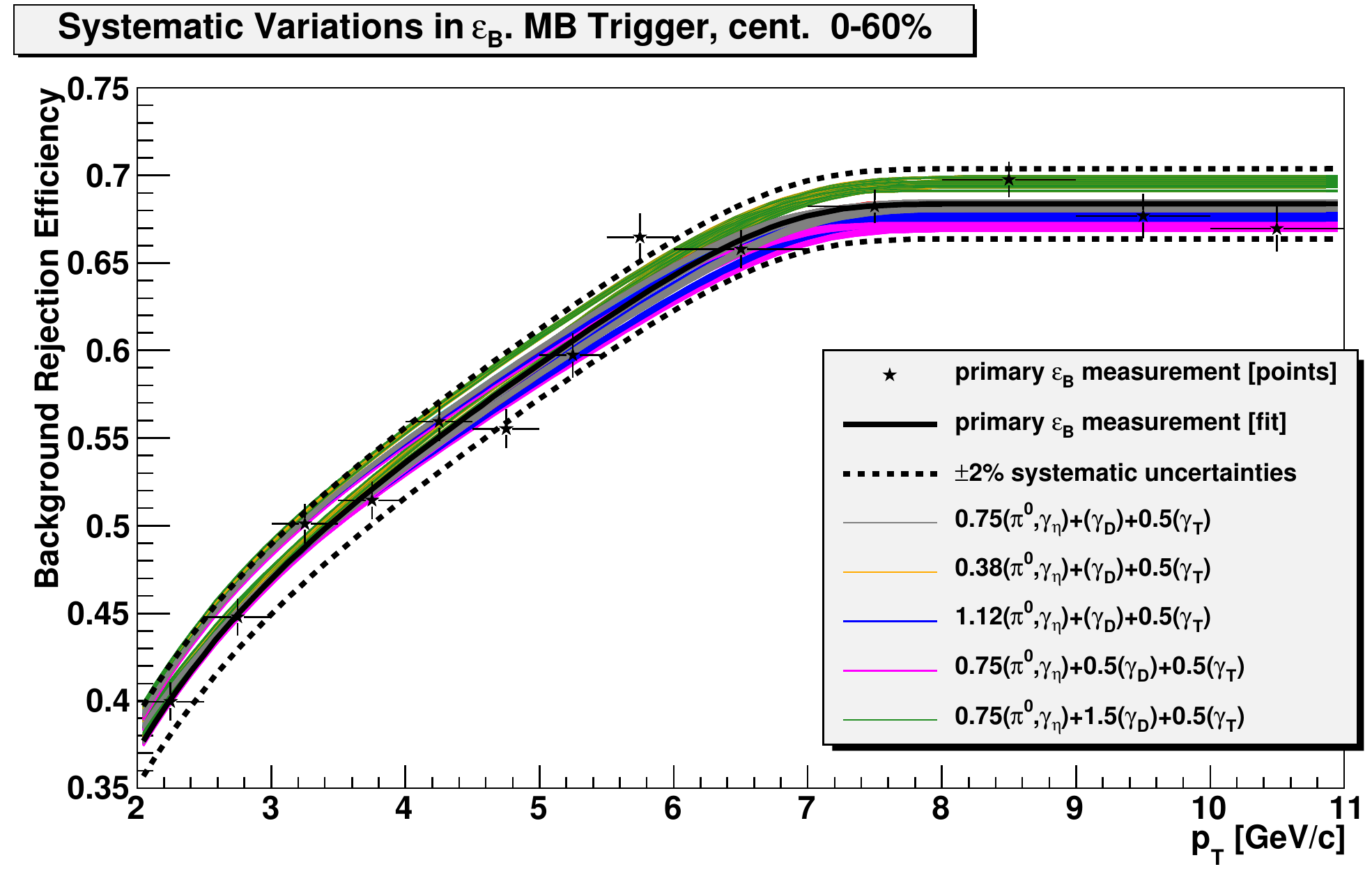}
\caption[Fits to the background rejection efficiency calculated for various permutations and mixtures of weighting functions.]{Fits to the measurements of $\varepsilon_{B}$ (for the 0-60\% centrality class) shown in Figure~\ref{fig:bre:systematic}:  The $\pm 2\%$ systematic uncertainty on the primary fit includes nearly all of the fits shown here.}
\label{fig:bre:fit_systematic}
\end{center}
\end{figure}

Figure \ref{fig:bre:systematic} shows the systematic variation in $\varepsilon_{B}$ for different mixtures of all of the alternate weighting function components.  The efficiency is calculated for each of the following seven mixtures of components, with the relative strengths of the meson $(\pi^{0},\gamma_{\eta})$, direct photon $(\gamma_{D})$, and thermal direct photon $(\gamma_{T})$ contributions varying within reasonable limits.

\begin{itemize}
\item $R_{\mathrm{CuCu}}^{\pi}(\pi^{0},\gamma_{\eta})+(\gamma_{D})+0.5(\gamma_{T})$
\item $0.5R_{\mathrm{CuCu}}^{\pi}(\pi^{0},\gamma_{\eta})+(\gamma_{D})+0.5(\gamma_{T})$
\item $1.5R_{\mathrm{CuCu}}^{\pi}(\pi^{0},\gamma_{\eta})+(\gamma_{D})+0.5(\gamma_{T})$
\item $R_{\mathrm{CuCu}}^{\pi}(\pi^{0},\gamma_{\eta})+0.5(\gamma_{D})+0.5(\gamma_{T})$
\item $R_{\mathrm{CuCu}}^{\pi}(\pi^{0},\gamma_{\eta})+2(\gamma_{D})+0.5(\gamma_{T})$
\item $R_{\mathrm{CuCu}}^{\pi}(\pi^{0},\gamma_{\eta})+(\gamma_{D})+0.25(\gamma_{T})$
\item $R_{\mathrm{CuCu}}^{\pi}(\pi^{0},\gamma_{\eta})+(\gamma_{D})+(\gamma_{T})$
\end{itemize}

For each of these mixtures, the efficiency is calculated with all permutations of the alternate weighting function components described in Section~\ref{sec:bre:weighting}.

\begin{itemize}
\item five possible neutral-pion weighting functions
\item three possible weighting functions describing $\eta$-meson-decay photons
\item five possible weighting functions describing non-thermal direct photons
\item three possible weighting functions describing thermal direct photons
\end{itemize}

\begin{figure}[htbp]
\begin{center}
\includegraphics[width=0.85\linewidth]{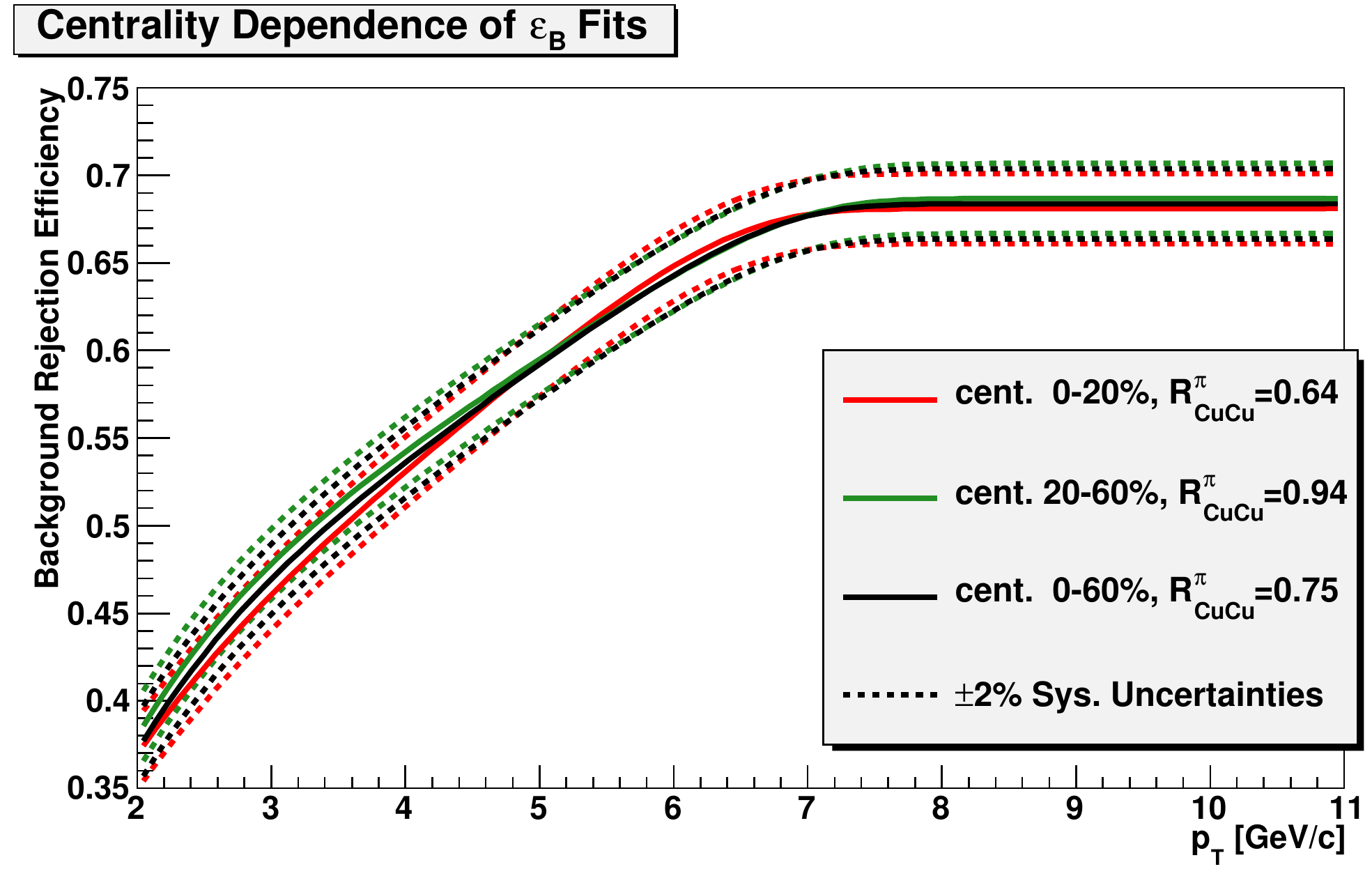}
\caption[Fits to the measurements of $\varepsilon_{B}$ for three centrality classes.]{Fits to the measurements of $\varepsilon_{B}$ for three centrality classes.  The dashed curves indicate the $\pm 2\%$ systematic uncertainty assigned to these fits.}
\label{fig:bre:fit_cent}
\end{center}
\end{figure}

Figure \ref{fig:bre:systematic} shows the results of all 1575 of these efficiency calculations; fits to each result are shown in Figure~\ref{fig:bre:fit_systematic}.  Figure~\ref{fig:bre:fit_cent} shows the fits of $\varepsilon_{B}$ for three centrality classes; these fits are used to find the efficiency-corrected non-photonic $e^{\pm}$ yield (see Chapter~\ref{sec:results}).

\clearpage

\clearpage

\begin{table}
\caption[Fit parameters for the background rejection efficiency (primary calculation) for three centrality classes.]{Fit parameters for the background rejection efficiency (primary calculation) for three centrality classes.  The parameter $b$ has units of $(\GeV/c)^{-1}$; $p_{0}$ has units of $\GeV/c$.}
\label{table:bre:fits}
\begin{tabular}{| c | r | r | r |}
\hline
Parameter & 0-20\% & 20-60\% & 0-60\%\\\hline\hline
$C$ & $0.68025\pm 0.00644$ & $0.68601\pm 0.00679$ & $0.68215\pm 0.00579$\\\hline
$A$ & $-0.25726\pm 0.09650$ & $-0.19389\pm 0.04691$ & $-0.21004\pm 0.04182$\\\hline
$b$ & $-0.34214\pm 0.07763$ & $-0.35903\pm 0.04248$ & $-0.35923\pm 0.03442$\\\hline
$p_{0}$ & $0.87181\pm 0.57172$ & $1.1478\pm 0.3312$ & $1.0928\pm 0.2731$\\\hline\hline
$\chi^{2}/NDF$ & 21.9/9 & 13.3/9 & 15.9/9\\\hline
\end {tabular}
\end{table}

\clearpage

\chapter{Residual Background}
\label{sec:res_back}

This chapter describes the calculation of the residual $e^{\pm}$ background.  Residual photonic $e^{\pm}$ come from the decays of mesons more massive than pions, which may produce $e^{-}e^{+}$ pairs with $M_{inv.}(e^{-}e^{+})>150\MeV/c^{2}$.  The yield of residual photonic $e^{\pm}$ from light-flavor sources (denoted $B_{LFP}$) is estimated in Section~\ref{sec:res_back:lfp}.  The yield $(B_{LFN})$ of residual non-photonic $e^{\pm}$, which come predominantly from the decays of kaons, is estimated in Section~\ref{sec:res_back:lfn}.  Decays of $J/\psi$ and $\Upsilon$ and the Drell-Yan process also contribute to the residual photonic $e^{\pm}$ background; estimates of the $e^{\pm}$ spectra from those sources (denoted $B_{J/\psi}$, $B_{\Upsilon}$, and $B_{D-Y}$, respectively) are described in Section~\ref{sec:res_back:other}.

\section[Residual Photonic $e^{\pm}$ from Light-Flavor Sources]{Residual Photonic $\boldsymbol{e^{\pm}}$ \\ from Light-Flavor Sources}
\label{sec:res_back:lfp}

The analysis method described in Chapter~\ref{sec:anintro} identifies photonic $e^{-}e^{+}$ pairs with invariant mass less than $150\MeV/c^{2}$, allowing the removal of photonic $e^{\pm}$ from photon conversions and $\pi^{0}$ Dalitz decays from the $e^{\pm}$ sample.  However, other sources produce photonic $e^{-}e^{+}$ pairs with $M_{inv.}(e^{-}e^{+})>150\MeV/c^{2}$.  The $\rho^{0}$, $\omega$, or $\phi$ mesons may undergo two-body decays to $e^{-}e^{+}$ pairs.  Those vector mesons, as well as the $\eta$, $\eta^{\prime}$, and $K^{0}_{S}$ mesons, may undergo three- and four-body decays which include $e^{-}e^{+}$ pairs among the products; some of those pairs may have $M_{inv.}(e^{-}e^{+})>150\MeV/c^{2}$.  In this section, the yield of residual photonic $e^{\pm}$ from light-flavor sources $(B_{LFP})$ is estimated using PYTHIA~\cite{PYTHIA6.4} simulations.

In these simulations, $10^{7}$ of each particle ($\pi^{0}$, $K^{0}_{S}$, $\eta$, $\rho^{0}$, $\omega$, $\eta^{\prime}$, and $\phi$) are generated.  These simulated particles are distributed uniformly in transverse momentum $(0\GeV/c<p_{T}<30\GeV/c)$, rapidity $(|y|<1.5)$, and azimuth $(0<\phi<2\pi)$.  These particles are allowed to decay via modes that produce $e^{-}e^{+}$ pairs.  Also simulated are the decays of $\eta^{\prime}$ and $\phi$ mesons to $\eta$, $\rho^{0}$, and $\omega$ mesons, with the subsequent decays of those particles to $e^{-}e^{+}$ pairs.  Table~\ref{table:res_back:pho_decay} summarizes the decays simulated, along with the physical branching ratios~\cite{PDG_review}.

\begin{table}
\caption[Simulated decays studied to estimate the residual photonic $e^{\pm}$ background, with their natural branching ratios.]{Simulated decays studied to estimate the residual photonic $e^{\pm}$ background.  The values of the branching ratios are from\protect\cite{PDG_review}.  Photonic $e^{-}e^{+}$ pairs are removed from the $e^{\pm}$ by the invariant mass cut; simulations of $\pi^{0}$ are included in this chapter only for reference.  There are two decays for which only an upper limit is given; the value of the branching ratio is assumed to be that upper limit, with a fractional uncertainty of $\pm 100\%$.  Regarding the $\eta^{\prime}\rightarrow\gamma\rho^{0}$ and $\phi\rightarrow\pi^{0}\rho^{0}$ decays: the branching ratios given in\protect\cite{PDG_review} include non-resonant modes with $\pi^{-}\pi^{+}$ instead of $\rho^{0}$.  Those values are used here, although they are likely overestimates of the branching ratios for decays to $\rho^{0}$.}
\label{table:res_back:pho_decay}
\begin{center}
\begin{tabular}{| l | l |}
\hline
Decay & Branching Ratio\\
\hline\hline
$\pi^{0}\rightarrow e^{-}e^{+}\gamma$ & $(1.174\pm0.035)\%$\\\hline
$K^{0}_{S}\rightarrow e^{-}e^{+}\pi^{-}\pi^{+}$ & $(4.69\pm0.30)\times 10^{-5}$\\\hline
$\eta\rightarrow e^{-}e^{+}\gamma$ & $(7.0\pm0.7)\times 10^{-3}$\\\hline
$\eta\rightarrow e^{-}e^{+}\pi^{-}\pi^{+}$ & $(2.68\pm0.11)\times 10^{-4}$\\\hline
$\rho^{0}\rightarrow e^{-}e^{+}$ & $(4.72\pm0.05)\times 10 ^{-5}$\\\hline
$\rho^{0}\rightarrow e^{-}e^{+}\pi^{0}$ & $<1.2\times 10^{-5}$ (CL=90\%)\\\hline
$\omega\rightarrow e^{-}e^{+}$ & $(7.28\pm0.14)\times 10^{-5}$\\\hline
$\omega\rightarrow e^{-}e^{+}\pi^{0}$ & $(7.7\pm0.6)\times 10^{-4}$\\\hline
$\eta\prime\rightarrow e^{-}e^{+}\gamma$ & $<9\times 10^{-4}$ (CL=90\%)\\\hline
$\eta\prime\rightarrow e^{-}e^{+}\pi^{-}\pi^{+}$ & $(2.4^{+1.3}_{-1.0})\times 10^{-4}$\\\hline
$\phi\rightarrow e^{-}e^{+}$ & $(2.954\pm0.030)\times 10^{-4}$\\\hline
$\phi\rightarrow e^{-}e^{+}\pi^{0}$ & $(1.12\pm0.28)\times 10^{-5}$\\\hline
$\phi\rightarrow e^{-}e^{+}\eta$ & $(1.15\pm0.10)\times 10^{-4}$\\\hline\hline
$\eta\prime\rightarrow \pi^{0}\pi^{0}\eta$ & $(21.7\pm0.8)\%$\\\hline
$\eta\prime\rightarrow\pi^{-}\pi^{+}\eta$ & $(43.2\pm0.7)\%$\\\hline
$\eta\prime\rightarrow\gamma\rho^{0}$ & $(29.3\pm0.5)\%$\\\hline
$\eta\prime\rightarrow\gamma\omega$ & $(2.75\pm0.22)\%$\\\hline
$\phi\rightarrow\gamma\eta$ & $(1.309\pm0.024)\%$\\\hline
$\phi\rightarrow\pi^{0}\rho^{0}$ & $(15.32\pm0.32)\%$\\\hline
\end{tabular}
\end{center}
\end{table}

As was the case in the calculation of the background rejection efficiency (see Chapter~\ref{sec:bre}), the spectra produced in these simulations must be weighted to correct for the fact that the decaying particles are generated uniformly in transverse momentum.  The $\pi^{0}$ and $\eta$-meson weighting functions that were used in Chapter~\ref{sec:bre} (functions $F[\pi^{0},2]$ and $F[\eta,2]$, respectively) are used as weighting functions here.  For the other particle species, the shapes of the weighting functions are estimated from the $\pi^{0}$ weighting function by making the substitution $p_{T}(\pi^{0})\rightarrow\sqrt{p_{T}(X)^{2}+m(X)^{2}c^{2}-m(\pi^{0})^2c^{2}}$ ($m_{T}$ scaling), where $X$ denotes the various particle species.  In addition, the weighting function for each particle species is scaled by constant factors according to Equation~\ref{eq:res_back:wf_scale} to give the invariant yield $(Ed^{3}N/dp^{3})$ of particle species $X$ in each of the three Cu + Cu centrality classes.

\begin{equation}
\label{eq:res_back:wf_scale}
F[X,\mathrm{Cu+Cu}]=E\frac{d^{3}N_{\mathrm{Cu+Cu}}}{dp^{3}}=R_{\mathrm{CuCu}}^{\pi}\times\frac{\langle N_{bin}\rangle}{\sigma_{inel.}}\times R(X/\pi^{0})\times F[X].
\end{equation}

\noindent Here, $F[X]$ denotes the $m_{T}$-scaled $\pi^{0}$ weighting function and $R(X/\pi^{0})$ denotes the ratio of the measured yield of particle species $X$ to the $\pi^{0}$ yield.  The values of $R(X/\pi^{0})$ are given in Table~\ref{table:res_back:pho_rpi}; the PHENIX and STAR collaborations use the same ratios in their calculations of photonic $e^{\pm}$ spectra in 200-GeV $p+p$ and Au + Au collisions.  (Those analyses do not consider the $K^{0}_{S}$ contribution; the choice of $0.6\pm0.3$ as the kaon-to-pion ratio is explained in the next section.)  In Equation~\ref{eq:res_back:wf_scale}, $R_{\mathrm{CuCu}}^{\pi}$ indicates the high-$p_{T}$ charged-pion nuclear modification factor measured by the STAR collaboration in 200-GeV Cu + Cu collisions~\cite{PhysRevC.81.054907}.  It is assumed that neutral pions will have the same suppression as the charged pions.  The values of $R_{\mathrm{CuCu}}^{\pi}$ and $\langle N_{bin}\rangle$ (the number of binary nucleon-nucleon collisions) for the three Cu + Cu centrality classes are given in Table~\ref{table:res_back:pho_scale}.  The inelastic $p+p$ cross-section, $\sigma_{inel.}$, is 42 mb~\cite{PDG_review}.

\begin{table}
\caption[Values of $R(X/\pi^{0})$ used to scale the meson weighting functions.]{The values of $R(X/\pi^{0})$ used to scale the meson weighting functions.  The $K^{0}_{S}$ ratio is estimated from\protect\cite{PhysRevC.69.034909}.  The other ratios are the same as those used in\protect\cite{PHENIX_NPE2010} and\protect\cite{STAR_ppNPE2011}.}
\label{table:res_back:pho_rpi}
\begin{center}
\begin{tabular}{| l | l |}
\hline
Ratio & Value\\
\hline\hline
$K^{0}_{S}/\pi^{0}$ & $0.6\pm0.3$\\\hline
$\eta/\pi^{0}$ & $0.48\pm0.10$\\\hline
$\rho^{0}/\pi^{0}$ & $1.0\pm0.3$\\\hline
$\omega/\pi^{0}$ & $0.90\pm0.27$\\\hline
$\eta\prime/\pi^{0}$ & $0.250\pm0.075$\\\hline
$\phi/\pi^{0}$ & $0.4\pm0.12$\\\hline
\end{tabular}
\end{center}
\end{table}

\begin{table}
\caption[High-$p_{T}$ charged-pion nuclear modification factors and values of $\langle N_{bin}\rangle$ for various 200-GeV Cu + Cu centrality classes.]{High-$p_{T}$ charged-pion nuclear modification factors and values of $\langle N_{bin}\rangle$\protect\cite{PhysRevC.81.054907} for various 200-GeV Cu + Cu centrality classes.}
\label{table:res_back:pho_scale}
\begin{center}
\begin{tabular}{| r | c | c |}
\hline
Centrality & $R_{\mathrm{CuCu}}^{\pi}$ & $\langle N_{bin}\rangle$\\
\hline\hline
$0-20\%$ & $0.66\pm0.10$ & $86.84^{+1.23}_{-1.18}$\\\hline
$20-60\%$ & $0.99\pm0.14$ & $33.61^{+0.67}_{-0.48}$\\\hline
$0-60\%$ & $0.82\pm0.12$ & $51.30^{+0.78}_{-0.87}$\\\hline
\end{tabular}
\end{center}
\end{table}

Each of the simulated photonic $e^{\pm}$ is weighted by the branching ratio for the process that produced it.  Each $e^{\pm}$ is also weighted according to the transverse momentum of its parent particle by a factor of $p_{T}(X)\times F[X,\mathrm{Cu+Cu}]$.  For each particle species $X$, the factor $p_{T}(X)$ is needed because

\begin{equation}
F[X,\mathrm{Cu+Cu}]=E\frac{d^{3}N}{dp^{3}}\propto\frac{1}{p_{T}}\times\frac{d^{2}N}{dp_{T}dy}.
\end{equation}

\begin{figure}[htbp]
\begin{center}
\includegraphics[width=0.85\linewidth]{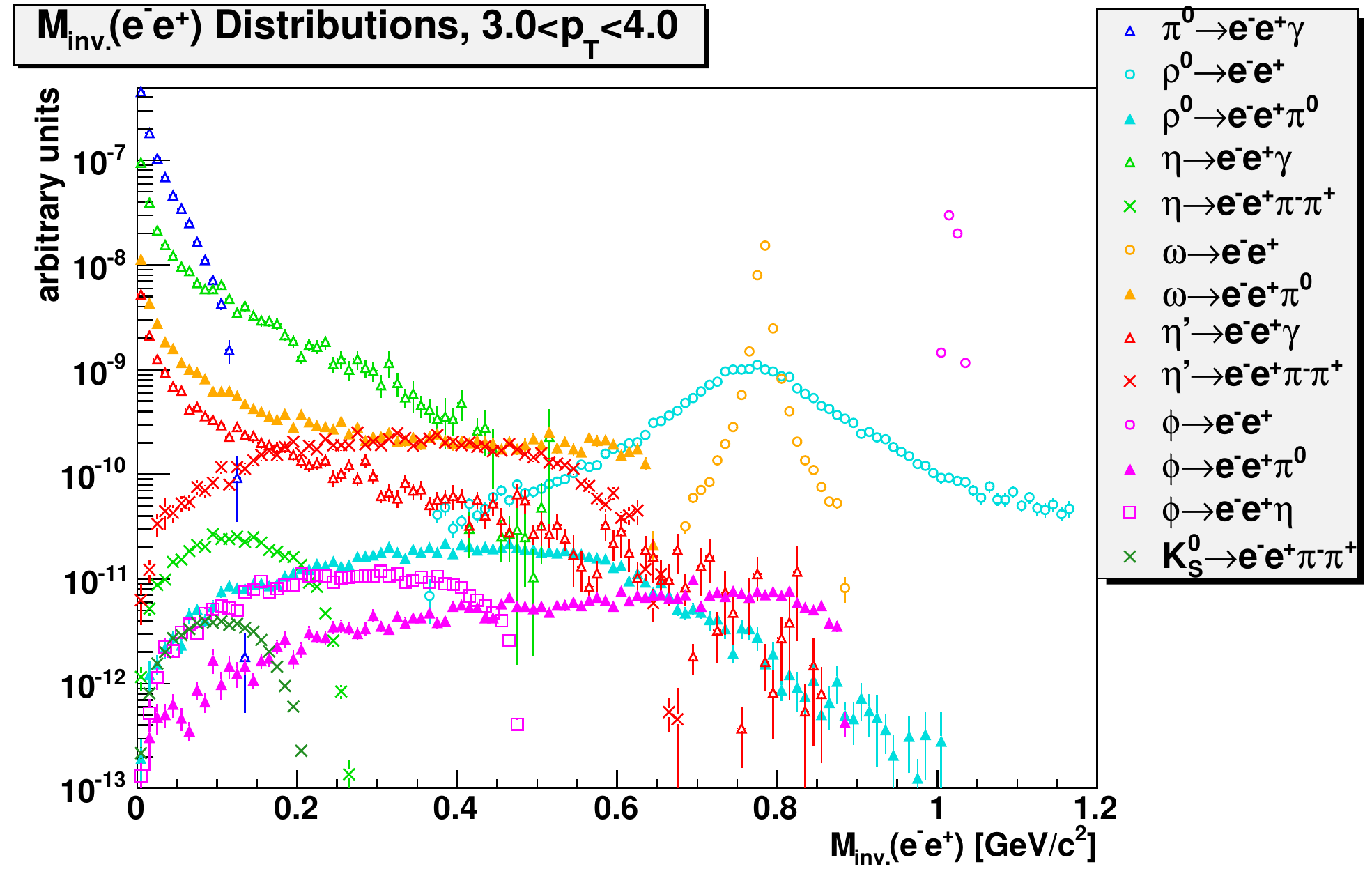}
\caption{Invariant mass of photonic $e^{-}e^{+}$ pairs from various meson-decay processes $(3\GeV/c<p_{T}<4\GeV/c)$.}
\label{fig:res_back:mass}
\end{center}
\end{figure}



\begin{figure}[htbp]
\begin{center}
\includegraphics[width=0.85\linewidth]{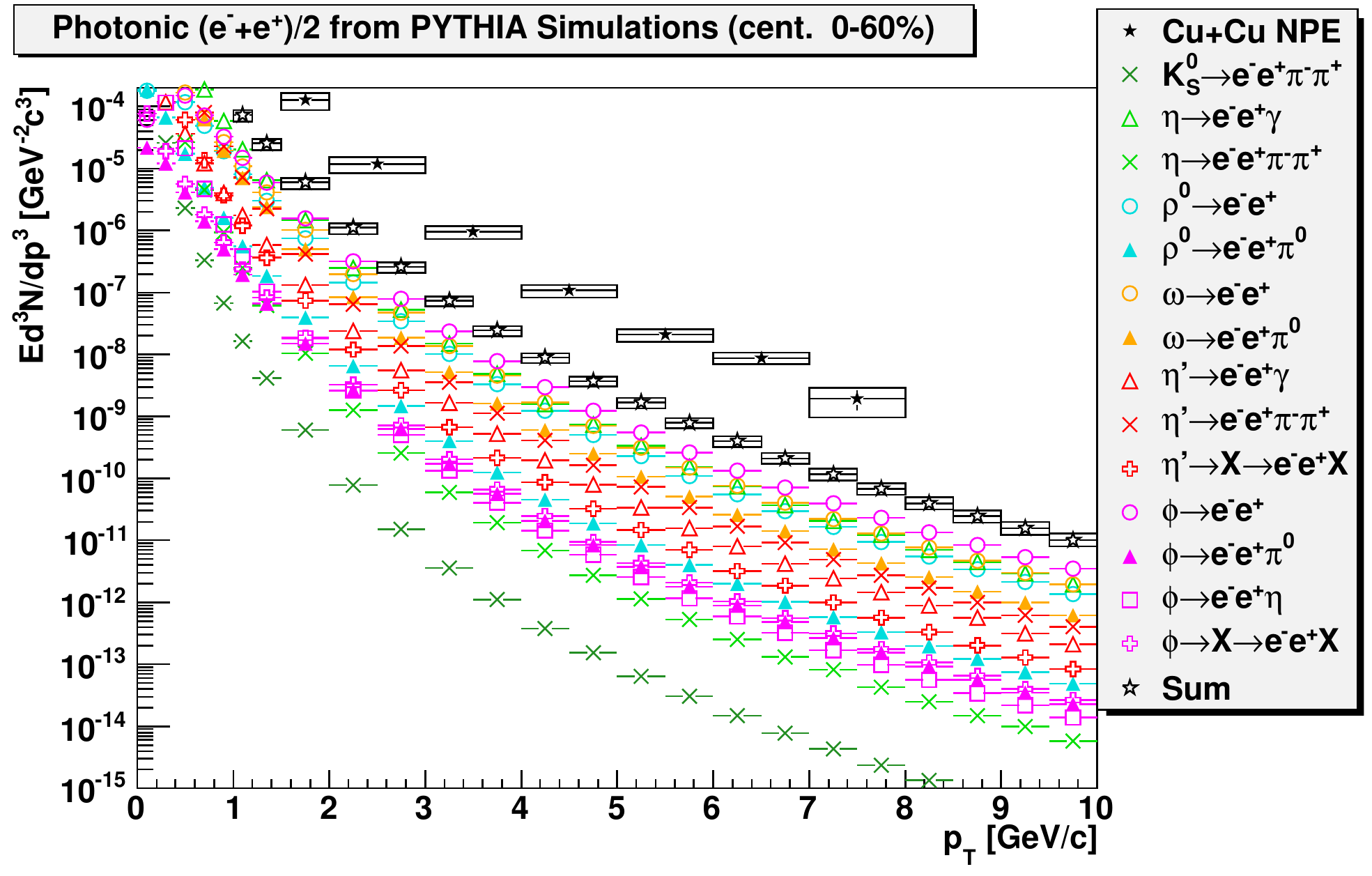}
\caption{Spectrum of photonic $e^{\pm}$ from pairs that fail the $M_{inv.}(e^{-}e^{+})<150\MeV/c^{2}$ cut, scaled for the 0-60\% centrality class of 200-GeV Cu + Cu collisions.}
\label{fig:res_back:pho_spectrum_c03}
\end{center}
\end{figure}

\begin{figure}[htbp]
\begin{center}
\includegraphics[width=0.85\linewidth]{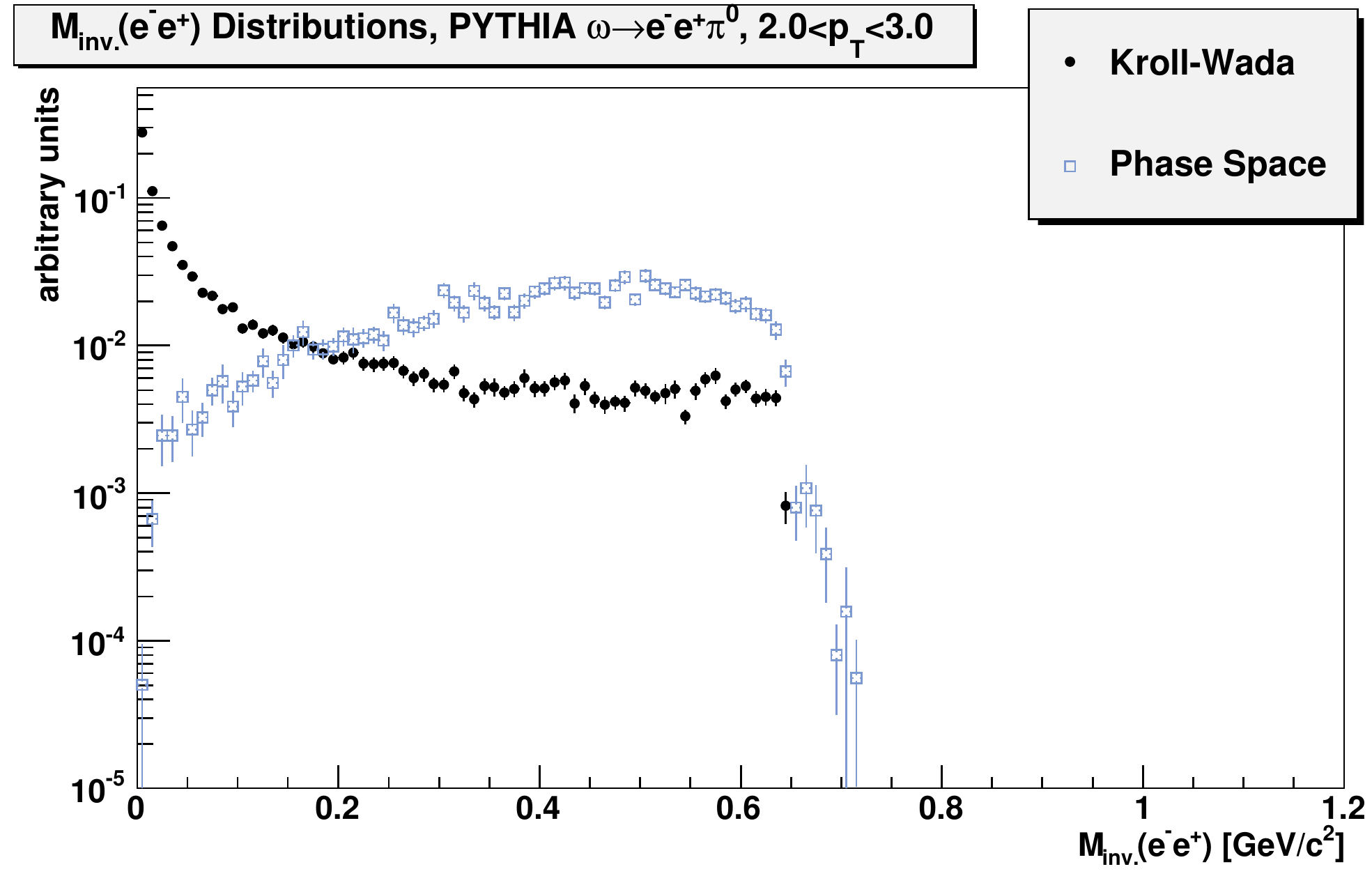}
\caption[The effect on the $e^{-}e^{+}$ pair invariant mass distribution of using different momentum distributions in the simulation of the decay $\omega\rightarrow e^{-}e^{+}\pi^{0}$.]{The effect on the $e^{-}e^{+}$ pair invariant mass distribution of using different momentum distributions in the simulation of the decay $\omega\rightarrow e^{-}e^{+}\pi^{0}$.  The blue distribution was generated by allowing the momenta of the decay products to be uniformly distributed in the available phase space.  The more realistic black distribution was generated according to\protect\cite{PhysRev.98.1355} (Kroll-Wada).}
\label{fig:res_back:form_factor}
\end{center}
\end{figure}

Figure~\ref{fig:res_back:mass} shows invariant-mass distributions (for $3\GeV/c<p_{T}<4\GeV/c$) for the $e^{-}e^{+}$ pairs produced in the simulated decays.  Peaks in the distribution due to the two-body decays of the vector mesons are visible, as are wide distributions with peaks near 0 due to Dalitz decays.  Figure~\ref{fig:res_back:pho_spectrum_c03} shows the simulated invariant yield of photonic $\tfrac{1}{2}(e^{-}+e^{+})$ for the 0-60\% centrality class in 200-GeV Cu + Cu collisions.  For the spectra shown in this figure, the cut on pair invariant mass has been applied.  Also shown is the efficiency-corrected invariant yield for non-photonic $e^{\pm}$ found in this analysis (prior to the subtraction of the residual background).

The systematic uncertainties in $B_{LFP}$, the total residual light-flavor photonic $e^{\pm}$ yield, are calculated by adding in quadrature the uncertainties in the total spectrum due to the uncertainties in the branching ratios, the $R(X/\pi^{0})$ values, $R_{\mathrm{CuCu}}^{\pi}$, and the uncertainty due to the choice of pion weighting function.  As described above, function $F[\pi^{0},2]$ was chosen as the primary $\pi^{0}$ weighting function and the weighting functions for the other particle species were derived from $F[\pi^{0},2]$ by $m_{T}$ scaling.  Four other sets of weighting functions were also derived, each set based on a different $\pi^{0}$ weighting function (see Chapter~\ref{sec:bre}).  $B_{LFP}$ was calculated using each set of weighting functions.  In each $p_{T}$ bin, the difference between the primary calculation and the most extreme calculations above (below) was used as an upper (lower) systematic uncertainty.  These weighting-function systematic uncertainties were then added in quadrature to the uncertainties due to the scaling factors.

Note that the three-body decays of $\rho^{0}$ and $\phi$ may not have been simulated using the correct invariant-mass distributions: the products of those decays have momenta distributed uniformly in the available phase space.  The $\omega\rightarrow e^{-}e^{+}\pi^{0}$ decay was simulated using the Kroll-Wada distribution~\cite{PhysRev.98.1355}, as well as the ``phase-space" distribution; the $M_{inv.}(e^{-}e^{+})$ distributions for these cases are shown in Figure~\ref{fig:res_back:form_factor}.  For this decay, the incorrect phase-space distribution has a greater fraction of pairs with $M_{inv.}(e^{-}e^{+})>150\MeV/c^{2}$; using the phase-space distribution has given an overestimate of the residual background.  A similar effect is observed when the phase-space distributions are used for the pseudoscalar-meson Dalitz decays.  It is assumed that the same will hold true for the three-body $\rho^{0}$ and $\phi$ decays and that the spectra of $e^{\pm}$ from those decays shown in Figure~\ref{fig:res_back:pho_spectrum_c03} are overestimates.  Note that those decays make negligible contributions to the residual background.

Similarly, the four-body decays of $K^{0}_{S}$, $\eta$, and $\eta^{\prime}$ (none of which are studied in the other RHIC non-photonic $e^{\pm}$ analyses~\cite{PHENIX_NPE2010,STAR_ppNPE2011}) were also simulated by distributing their products uniformly in the available phase space.  However, these decays make only small contributions to the residual background.  Even the contribution from the four-body $\eta^{\prime}$ decay is only about 5\% of $B_{LFP}$, a contribution that is smaller than the uncertainties in the background due to the uncertainty in $R_{\mathrm{CuCu}}^{\pi}$ and the uncertainty due to the choice of weighting function.  Note also that for the calculations in this section (though not in the next section) the $K^{0}_{S}$ were assumed to have lifetimes similar to the other mesons and to decay at the collision vertex.  If this has any effect, the (negligible) $K^{0}_{S}$ contribution to the residual photonic background is an overestimate (some of the decay $e^{\pm}$ tracks may have failed the $GDCA$ cut if the decays occurred far from the vertex).

\clearpage

\section[Residual Non-Photonic $e^{\pm}$]{Residual Non-Photonic $\boldsymbol{e^{\pm}}$}
\label{sec:res_back:lfn}

The non-photonic $e^{\pm}$ sample includes contributions from sources other than heavy-flavor decays.  The largest such sources are the decays of kaons; there are also contributions from charged pions, direct muons, and baryons.  The total yield of non-photonic $e^{\pm}$ from these sources (denoted $B_{LFN}$) is estimated in this section using PYTHIA~\cite{PYTHIA6.4} simulations.  In these simulations, $10^{7}$ of each particle ($\mu^{-}$, $\pi^{+}$, $K^{+}$, $K^{0}_{S}$, $K^{0}_{L}$, $\Lambda^{0}$, $\Sigma^{+}$, $\Sigma^{-}$, $\Xi^{0}$, $\Xi^{-}$, and $\Omega^{-}$) are generated.  These simulated particles are generated uniformly in transverse momentum $(0\GeV/c<p_{T}<30\GeV/c)$, rapidity $(|y|<1.5)$, and azimuth $(0<\phi<2\pi)$.  These particles are allowed to decay via modes that produce $e^{\pm}$.  Table~\ref{table:res_back:npe_decay} summarizes the decays studied, along with the physical branching ratios~\cite{PDG_review}.

\begin{table}
\caption[Simulated decays studied to estimate the residual non-photonic $e^{\pm}$ background.]{Simulated decays studied to estimate the residual non-photonic $e^{\pm}$ background.  The values of the branching ratios come from\protect\cite{PDG_review}.}
\label{table:res_back:npe_decay}
\begin{center}
\begin{tabular}{| l | l |}
\hline
Decay & Branching Ratio\\
\hline\hline
$\mu^{-}\rightarrow e^{-}\bar{\nu}_{e}\nu_{\mu}$ & 1\\\hline
$\pi^{+}\rightarrow e^{+}\nu_{e}$ & $(1.230\pm0.004)\times 10^{-4}$\\\hline
$K^{+}\rightarrow e^{+}\nu_{e}$ & $(1.584\pm0.020)\times 10^{-5}$\\\hline
$K^{+}\rightarrow e^{+}\nu_{e}\pi^{0}$ & $(5.07\pm0.04)\%$\\\hline
$K^{0}_{S}\rightarrow e^{-}\bar{\nu}_{e}\pi^{+}(e^{+}\nu_{e}\pi^{-})$ & $(7.04\pm0.08)\times 10^{-4}$\\\hline
$K^{0}_{L}\rightarrow e^{-}\bar{\nu}_{e}\pi^{+}(e^{+}\nu_{e}\pi^{-})$ & $(40.55\pm0.12)\%$\\\hline\hline
$\Lambda^{0}\rightarrow e^{-}\bar{\nu}_{e}p$ & $(8.32\pm0.14)\times 10^{-4}$\\\hline
$\Sigma^{+}\rightarrow e^{+}\nu_{e}\Lambda^{0}$ & $(2.0\pm0.5)\times 10^{-5}$\\\hline
$\Sigma^{-}\rightarrow e^{-}\bar{\nu}_{e}n$ & $(1.017\pm0.034)\times 10^{-3}$\\\hline
$\Sigma^{-}\rightarrow e^{-}\bar{\nu}_{e}\Lambda^{0}$ & $(5.73\pm0.27)\times 10^{-5}$\\\hline
$\Xi^{0}\rightarrow e^{-}\bar{\nu}_{e}\Sigma^{+}$ & $(2.53\pm0.08)\times 10^{-4}$\\\hline
$\Xi^{-}\rightarrow e^{-}\bar{\nu}_{e}\Lambda^{0}$ & $(5.63\pm0.31)\times 10^{-4}$\\\hline
$\Xi^{-}\rightarrow e^{-}\bar{\nu}_{e}\Sigma^{0}$ & $(8.7\pm1.7)\times 10^{-5}$\\\hline
$\Omega^{-}\rightarrow e^{-}\bar{\nu}_{e}\Xi^{0}$ & $(5.6\pm2.8)\times 10^{-3}$\\\hline
\end{tabular}
\end{center}
\end{table}

As was the case in the previous section, the spectra produced in these simulations must be weighted to correct for the fact that the decaying particles are uniformly distributed in transverse momentum.  The same weighting function used for the $\pi^{0}$ in the previous section $(F[\pi^{0},2])$ is used to describe the charged-pion spectrum here.  The shapes of the kaon weighting functions were estimated from the pion weighting function using $m_{T}$ scaling.  An $m_{T}$-scaled pion spectrum was also used as a rough guess of the muon spectrum.  However, for reasons described below, the muon-decay contribution to the residual background is believed to be negligible, so the use of a questionable muon weighting function will have no effect on the background.

\begin{figure}[htbp]
\begin{center}
\includegraphics[width=0.85\linewidth]{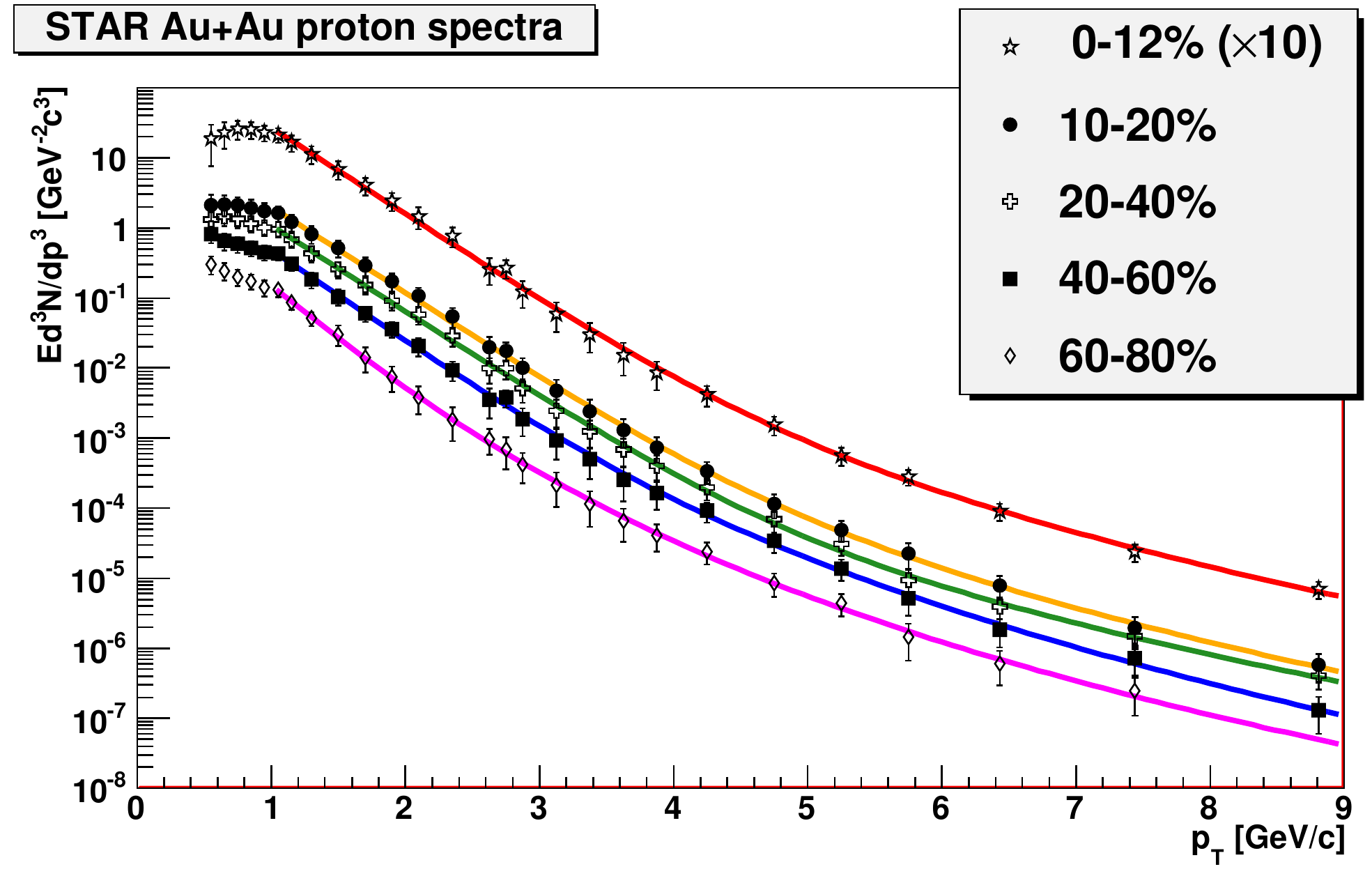}
\caption[STAR measurements of $\tfrac{1}{2}(p+\bar{p})$ spectra in 200-GeV Au + Au collisions.]{The STAR Collaboration's measurements of $\tfrac{1}{2}(p+\bar{p})$ spectra in 200-GeV Au + Au collisions\protect\cite{PhysRevLett.97.152301}.  The spectra are corrected for hyperon ($\Lambda$ and $\Sigma^{+}$) feed-down.  The error bars indicate combined statistical and systematic uncertainties.  The fit to the spectrum for the 40-60\% centrality class (blue) is used as the basis for the baryon weighting functions in this section.}
\label{fig:res_back:star_proton}
\end{center}
\end{figure}

The STAR collaboration has measured~\cite{PhysRevLett.97.152301} the $\tfrac{1}{2}(p+\bar{p})$ yield in various centrality classes in 200-GeV Au + Au collisions (see Figure~\ref{fig:res_back:star_proton}).  The 40-60\% centrality class (for which $\langle N_{part}\rangle$ and $\langle N_{bin}\rangle$ have similar values to the Cu + Cu 0-60\% centrality class) is fit with function $F[p,4]$ of the form $A(1+p_{T}^{2}/B)^{n}+C\exp(-p{T}/D)$ (see Section~\ref{sec:functions:proton} for the fit parameters).  The shapes of the baryon weighting functions are estimated from this proton fit using $m_{T}$ scaling.

The weighting function for each particle species is scaled by constant factors according to Equation~\ref{eq:res_back:wf_meson_scale} for muons and mesons and according to Equation~\ref{eq:res_back:wf_baryon_scale} for baryons.  This gives the invariant yield $(Ed^{3}N/dp^{3})$ of particle species $X$ in each of the three Cu + Cu centrality classes.

\begin{equation}
\label{eq:res_back:wf_meson_scale}
F[X,\mathrm{Cu+Cu}]=R_{\mathrm{CuCu}}^{\pi}\times\frac{\langle N_{bin}\rangle}{\sigma_{inel.}}\times R(X/\pi^{+})\times (1+R_{anti})\times F[X].
\end{equation}

\begin{equation}
\label{eq:res_back:wf_baryon_scale}
F[X,\mathrm{Cu+Cu}]=\frac{\langle N_{bin}\rangle}{\langle N_{bin}^{\mathrm{Au+Au}}\rangle}\times R(X/p)\times (1+R_{anti})\times F[X].
\end{equation}

\noindent $R_{anti}$ is the antiparticle-to-particle ratio for species $X$.  For $K^{0}_{S}$ and $K^{0}_{L}$, which are their own antiparticles, $R_{anti}$ is set to 0.  For $\Omega$, only the combined (particle plus antiparticle) yield is given in~\cite{PhysRevLett.98.062301}; $R_{anti}$ is set to 0 and the $\bar{\Omega}^{+}$ contribution is accounted for in the $\Omega$-to-proton ratio.  $\langle N_{bin}^{\mathrm{Au+Au}}\rangle=90$ is the number of binary nucleon-nucleon collisions~\cite{PhysRevC.79.034909} for the 40-60\% centrality class of 200-GeV Au + Au collisions (the baryon weighting functions are based on a fit to the $\tfrac{1}{2}(p+\bar{p})$ spectrum for this centrality class).

The values used for $R(\pi^{-}/\pi^{+})$ and $R(K^{-}/K^{+})$ are based on the PHENIX collaboration's measurements of those quantities in 200-GeV Au + Au collisions~\cite{PhysRevC.69.034909}.  The values quoted in Table~\ref{table:res_back:npe_ratios} are the means of the measurements for the 40-50\% and 50-60\% centrality classes.  These measurements are for $p_{T}<2\GeV/c$, but no transverse-momentum dependence is observed and it is assumed that the ratios will retain those values at high $p_{T}$.  The value of $R(K^{+}/\pi^{+})$ is a rough estimate from Figure 17 in~\cite{PhysRevC.69.034909}.  The desired value is the $K^{+}/\pi^{+}$ ratio at high-$p_{T}$, for which no data are available.  The value $R(K^{+}/\pi^{+})=0.6$ is an estimate based on the apparent trend of the ratio as $p_{T}$ increases; large uncertainties are therefore assigned.  It is assumed that $R(\mu^{+}/\mu^{-})=1$ and that the muon-to-pion ratio is also 1.  The latter assumption is not intended to be a realistic estimate of the muon yield, but rather a ``worst-case" scenario.  The resulting spectrum of $e^{\pm}$ from muon decays will be discussed below.

The baryon ratios are the ratios of the STAR collaboration's measurement of the integrated rapidity densities $(dN/dy)$ for each particle species for the 40-60\% centrality class in 200-GeV Au + Au collisions.  The value of $dN/dy$ for protons is taken from~\cite{PhysRevC.79.034909}, while the values of $dN/dy$ for $\Lambda$, $\Xi^{-}$, $\Omega^{-}$, and their anti-particles are taken from~\cite{PhysRevLett.98.062301}.  For the $\Sigma$ baryons, it is assumed that $R(\Sigma^{+}/p)=R(\Sigma^{-}/p)=R(\Lambda/p)$ and $R(\bar{\Sigma}^{-}/\Sigma^{+})=R(\bar{\Sigma}^{+}/\Sigma^{-})=R(\bar{\Lambda}/\Lambda)$.  For $\Xi^{0}$ it is assumed that $R(\Xi^{0}/p)=R(\Xi^{-}/p)$ and $R(\bar{\Xi}^{0}/\Xi^{0})=R(\bar{\Xi}^{+}/\Xi^{-})$.

\begin{table}
\caption[Particle ratios used to scale the weighting functions for the calculation of the residual non-photonic $e^{\pm}$ background.]{Particle ratios used to scale the weighting functions for the calculation of the residual non-photonic $e^{\pm}$ background.  Meson ratios are from\protect\cite{PhysRevC.69.034909}, proton yields are from\protect\cite{PhysRevC.79.034909}, and other baryon yields are from\protect\cite{PhysRevLett.98.062301}.  Values in parentheses are assumptions and not based on independent measurements.  See the text for further explanation.}
\label{table:res_back:npe_ratios}
\begin{center}
\begin{tabular}{| r | l || r | l |}
\hline
$R(X/\pi^{+})$& & & \\
or $R(X/p)$ & Value & $R_{anti}$ & Value\\
\hline\hline
 & & $\pi^{-}/\pi^{+}$ & $0.97\pm0.06$\\\hline
$K^{+}/\pi^{+}$ & $0.6\pm0.3$ & $K^{-}/K^{+}$ & $0.92\pm0.05$\\\hline\hline
$\Lambda/p$ & $0.206$ & $\bar{\Lambda}/\Lambda$ & $0.792$\\\hline
$\Sigma^{+}/p$ & $(0.206)$ & $\bar{\Sigma}^{-}/\Sigma^{+}$ & $(0.792)$\\\hline
$\Sigma^{-}/p$ & $(0.206)$ & $\bar{\Sigma}^{+}/\Sigma^{-}$ & $(0.792)$\\\hline
$\Xi^{0}/p$ & $(0.259)$ & $\bar{\Xi}^{0}/\Xi^{0}$ & $(0.885)$\\\hline
$\Xi^{-}/p$ & $0.259$ & $\bar{\Xi}^{+}/\Xi^{-}$ & $0.885$\\\hline
$(\Omega^{-}+\bar{\Omega}^{+})/p$ & $0.00627$ & & \\\hline
\end{tabular}
\end{center}
\end{table}

As in the previous section, each of the simulated non-photonic $e^{\pm}$ is weighted by the branching ratio for the process that produced it and by a factor related to the transverse momentum of its parent particle, $p_{T}(X)\times F[X,\mathrm{Cu+Cu}]$.  Unlike the previous section, the parent particles here may have long enough lifetimes that a significant number of decays may occur farther than 1.5 cm from the primary vertex.  It is possible that an $e^{\pm}$ produced in such a decay will fail the $GDCA$ cut and not be included in the $e^{\pm}$ sample.  For each parent particle $X$, a decay point along its track is randomly generated.  Using that decay point a helical track is constructed for the produced $e^{\pm}$, the DCA to the primary vertex is calculated, and the cut of $GDCA<1.5$ cm is applied.

To generate a random decay point, a decay length $L$ is generated with a uniform distribution in the range $0<L<\Delta_{L}$, with $\Delta_{L}$=3 m for mesons and muons and $\Delta_{L}$=4.7 m for baryons.  Those values were chosen so that a $K^{+}\;(\Omega^{-})$ with $p_{T}=1\GeV/c$ traveling the full distance $\Delta_{L}$ along a helical path in the STAR magnetic field (0.497952 T) would always decay a radial distance $>102$ cm from the $z$-axis, even when the particle had the extremal value of rapidity, $y=1.5$.  However, the decays are still constrained to be within a few meters of the STAR detector, rather than the large distances possible if the physically correct exponential distribution of decay lengths was used.  For each decay, where the decaying particle had mass $m$, lifetime $\tau$, and momentum $p$, a decay probability $D$ was calculated and used to scale the uniform decay-length distribution to match the physical exponential distribution.

\begin{equation}
\label{eq:res_back:decay_prob}
D(L)=\frac{mc^{2}}{c\tau(pc)}\Delta_{L}\exp\left(-\frac{mc^{2}L}{c\tau(pc)}\right).
\end{equation}

\begin{figure}[htbp]
\begin{center}
\includegraphics[width=0.85\linewidth]{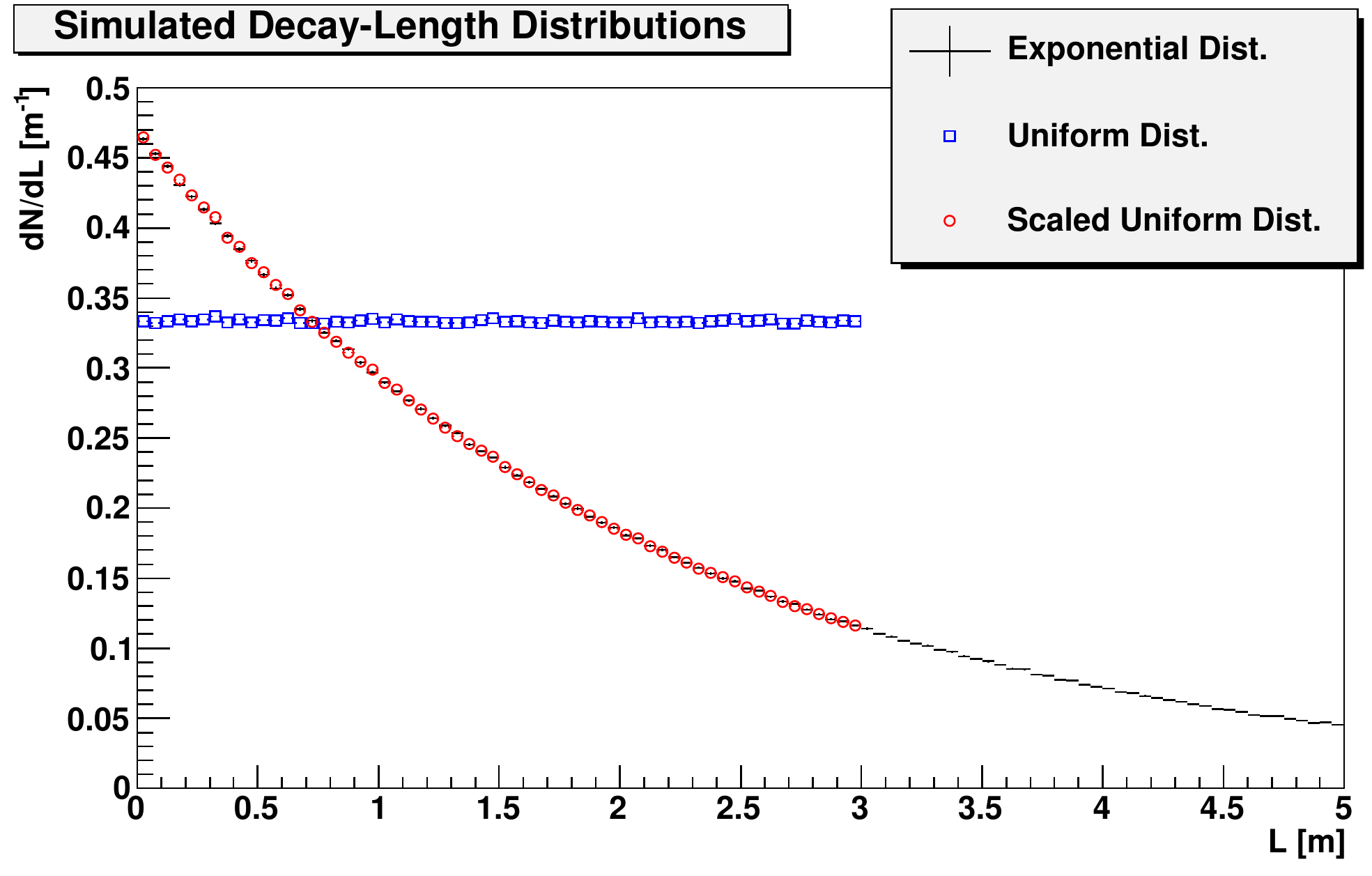}
\caption[Demonstration of how a uniform distribution is scaled to match an exponential distribution.]{Demonstration of how a uniform distribution is scaled to match an exponential distribution using Equation~\ref{eq:res_back:decay_prob}.  See the text for explanation.}
\label{fig:res_back:decay_prob}
\end{center}
\end{figure}

This is demonstrated in Figure~\ref{fig:res_back:decay_prob}.  In black is an exponential distribution of decay lengths (with decay constant $m/\tau p$ =0.47 m$^{-1}$), while in blue is a uniform distribution of decay lengths (with $\Delta_{L}=3$m).  Both distributions are normalized by the number of entries $(5\times 10^{6})$ and the bin width.  The red distribution is the blue uniform distribution scaled by the decay probability $D$ calculated using Equation~\ref{eq:res_back:decay_prob}.  The uniform distribution scaled by $D$ matches the exponential distribution for $L<\Delta_{L}$.  Each simulated $e^{\pm}$ is weighted by the decay probability $D$ for the decay that produced it.



\begin{figure}[htbp]
\begin{center}
\includegraphics[width=0.85\linewidth]{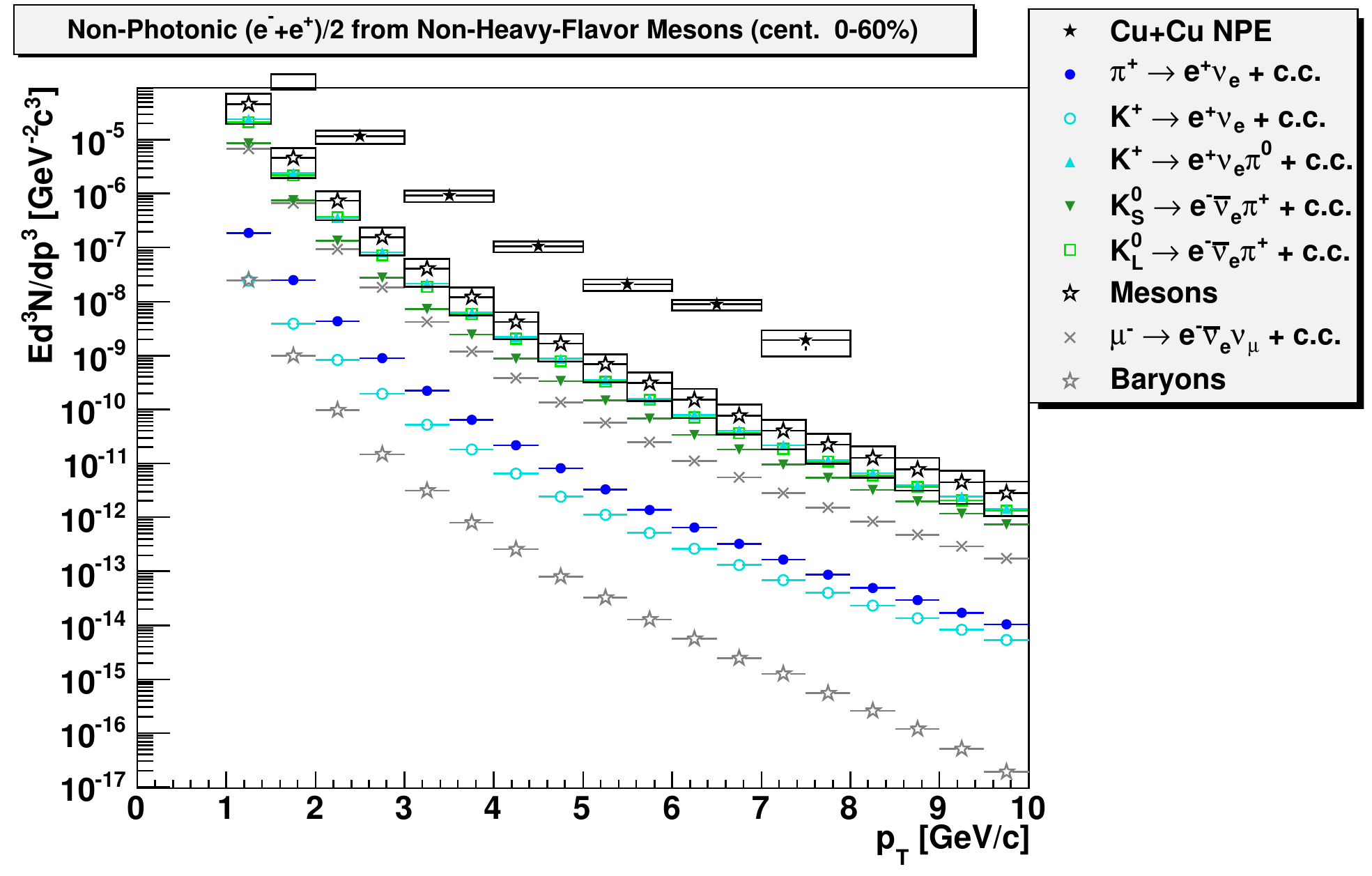}
\caption{Spectrum of non-photonic $e^{\pm}$ from decays of mesons and muons, scaled for the 0-60\% centrality class of 200-GeV Cu + Cu collisions.}
\label{fig:res_back:meson_spectrum_c03}
\end{center}
\end{figure}

\begin{figure}[htbp]
\begin{center}
\includegraphics[width=0.85\linewidth]{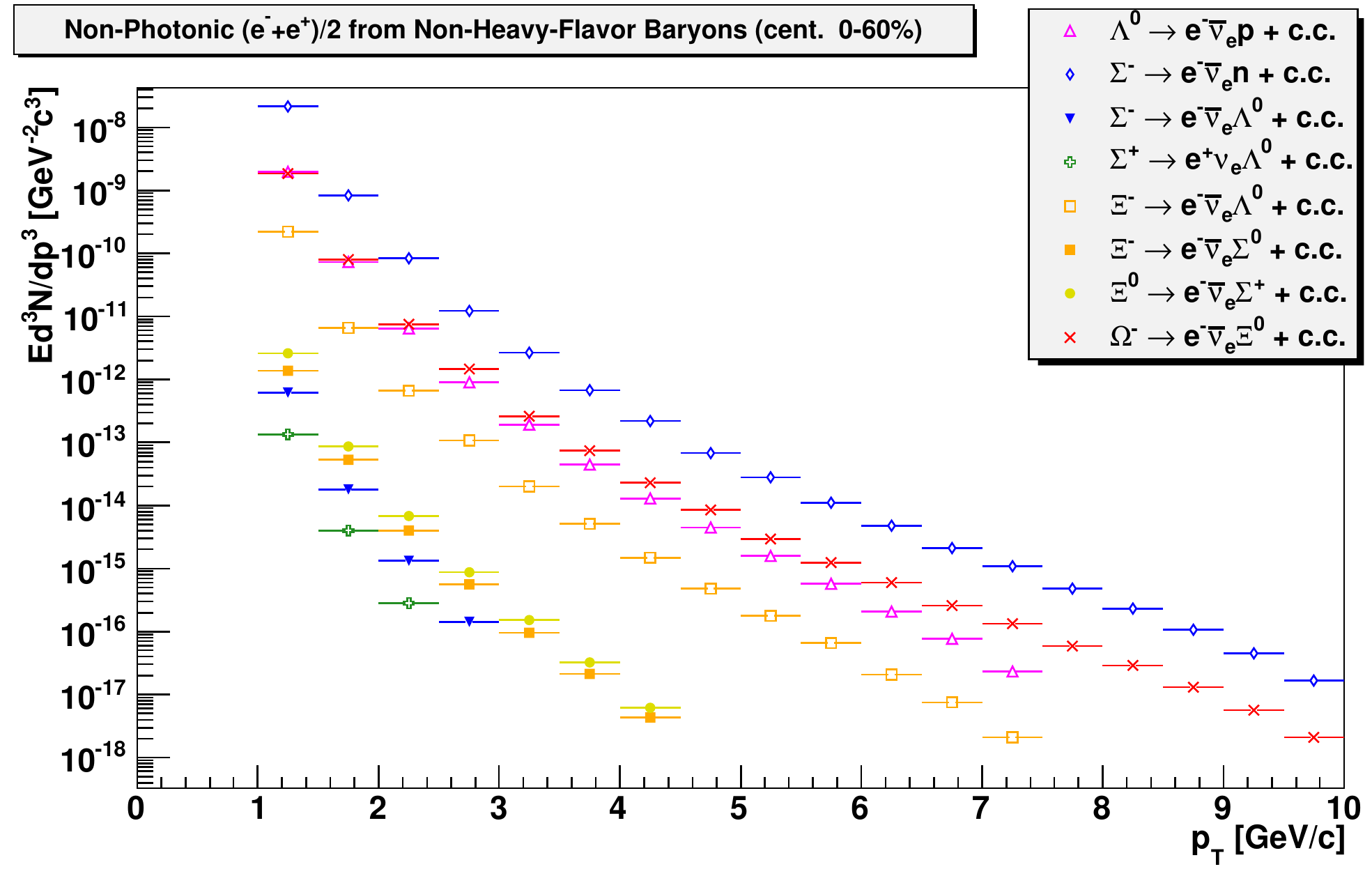}
\caption{Spectrum of non-photonic $e^{\pm}$ from decays of baryons, scaled for the 0-60\% centrality class of 200-GeV Cu + Cu collisions.}
\label{fig:res_back:baryon_spectrum_c03}
\end{center}
\end{figure}

The residual non-photonic $e^{\pm}$ background $(B_{LFN})$ calculated using the method described above is shown in Figures~\ref{fig:res_back:meson_spectrum_c03} and~\ref{fig:res_back:baryon_spectrum_c03}.  Figure~\ref{fig:res_back:meson_spectrum_c03} shows the background spectra for $e^{\pm}$ from meson and muon decays in the three Cu + Cu centrality classes.  This spectrum is dominated by the $K^{+}\rightarrow e^{+}\nu_{e}\pi^{0}$ decay (called $K^{+}_{e3}$) and the $K^{0}_{L}\rightarrow e^{+}\nu_{e}\pi^{-}$ decay (called $K^{0}_{e3}$) and their charge conjugates.

The systematic uncertainty in the total background non-photonic $e^{\pm}$ spectrum from meson decays is the sum in quadrature of the uncertainty in each $p_{T}$ bin due to the uncertainty in $R(K^{+}/\pi^{+})$ and the uncertainty due to the choice of weighting function shape.  As in the previous section, four additional sets of weighting functions were generated, each set based on a different $\pi^{0}$ weighting function.  Residual non-photonic $e^{\pm}$ spectra were calculated using each set of weighting functions.  In each $p_{T}$ bin, the difference between the primary calculation and the most extreme calculations above (below) was used as an upper (lower) systematic uncertainty.  These weighting-function systematic uncertainties were then added in quadrature to the uncertainties due to the $R(K^{+}/\pi^{+})$.

The PHENIX collaboration has measured~\cite{PhysRevD.76.092002} the low-$p_{T}$ direct-muon yield at forward rapidity $(1.5\leq\eta\leq 1.8)$ in 200-GeV $p+p$ collisions; those values are $<1\%$ of the mid-rapidity pion yields~\cite{PhysRevD.76.051106}.  The main sources of feed-down muons should be $\pi^{\pm}$, $K^{\pm}$, and $K^{0}_{L}$ decays but those particles have long enough lifetimes that (with $p>1\GeV/c$) most of them will decay outside of the TPC.  Even in the extreme scenario in which the spectrum of direct muons is the same shape and magnitude as the charged-pion spectrum, the yield of $e^{\pm}$ from the decays of those muons would amount to less than 10\% of the total yield of $e^{\pm}$ from pion and kaon decays and less than 1\% of the efficiency-corrected non-photonic $e^{\pm}$ yield.  The simulated spectrum of $e^{\pm}$ from muon decays is therefore not included in the estimates of $B_{LFN}$.

Figure~\ref{fig:res_back:baryon_spectrum_c03} shows the background spectra for $e^{\pm}$ from baryon decays for the 0-60\% centrality class in 200-GeV Cu + Cu collisions.  The baryon contribution is many orders of magnitude below the contribution from the $K_{e3}$ decays and the efficiency-corrected non-photonic $e^{\pm}$ signal.  The baryon contribution is therefore disregarded.

\clearpage

\section{Other Sources}
\label{sec:res_back:other}

This section describes the calculation of the residual photonic $e^{\pm}$ background due to $J/\psi$ and $\Upsilon$ decays and the Drell-Yan process.  This analysis estimates the yields of $e^{\pm}$ from those sources (denoted $B_{J/\psi}$, $B_{\Upsilon}$, and $B_{D-Y}$) using the same estimated spectra that the STAR collaboration uses in its measurement~\cite{STAR_ppNPE2011} of non-photonic $e^{\pm}$ in $p+p$ collisions.  Those spectra are scaled by a nuclear modification factor $R_{\mathrm{CuCu}}$ (for the quarkonia) and by $\langle N_{bin}\rangle/\sigma_{inel.}$ to convert them to an expected yield in Cu + Cu collisions.  The spectra of photonic $e^{\pm}$ from $J/\psi$ and $\Upsilon$ decays and the Drell-Yan process will be shown in Figure~\ref{fig:results:bsub_c01} (page~\pageref{fig:results:bsub_c01}); the calculations that produced those spectra are described below.

\subsection[Photonic $e^{\pm}$ from $J/\psi$ Decays]{Photonic $\boldsymbol{e^{\pm}}$ from $\boldsymbol{J/\psi}$ Decays}
\label{sec:res_back:jpsi}

\begin{figure}[htbp]
\begin{center}
\includegraphics[width=0.85\linewidth]{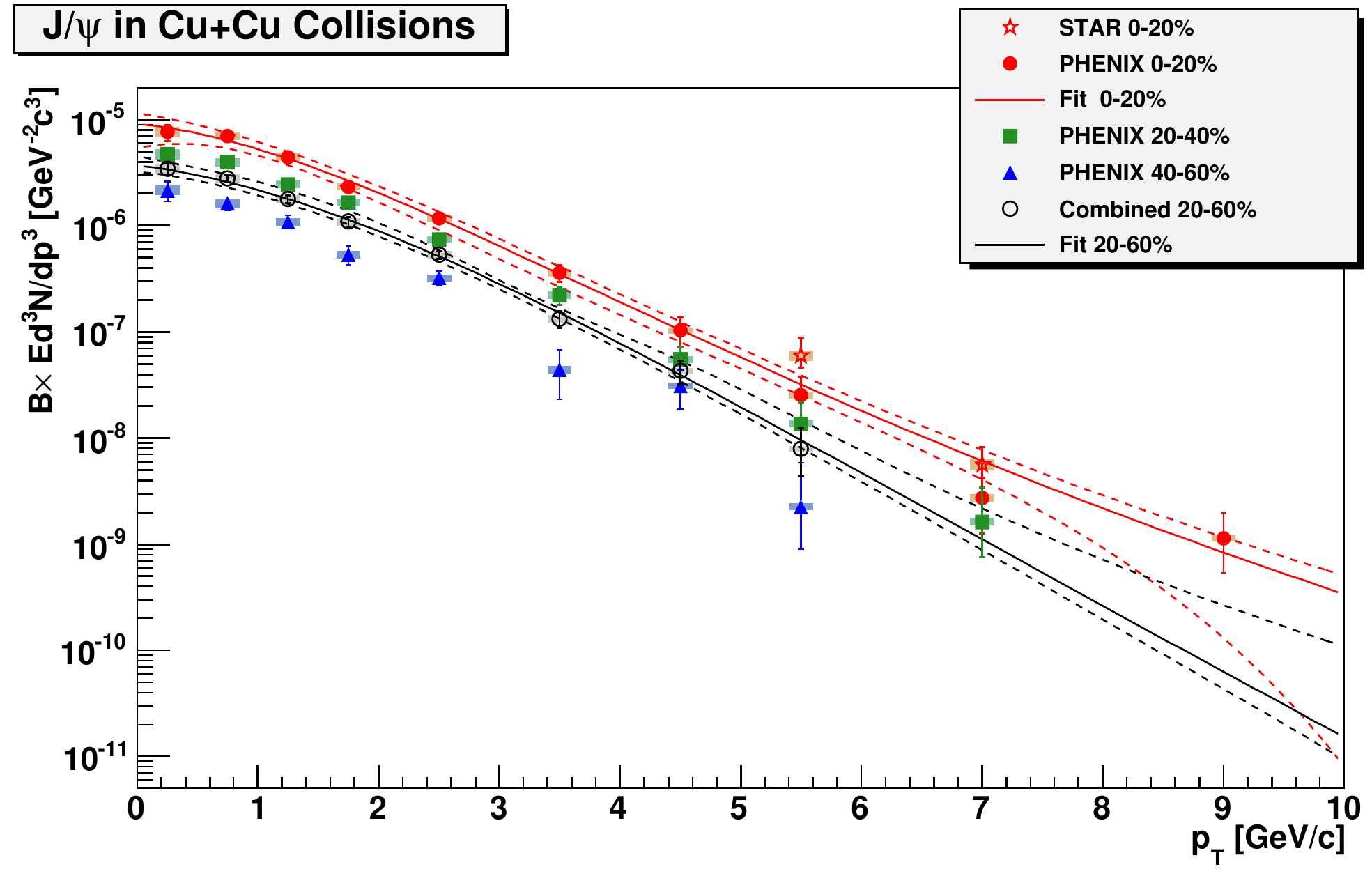}
\caption[Spectra of $J/\psi$ in 200-GeV Cu + Cu collisions.]{Spectra of $J/\psi$ in various centrality classes of 200-GeV Cu + Cu collisions measured by the STAR\protect\cite{PhysRevC.80.041902} and PHENIX\protect\cite{PhysRevLett.101.122301} collaborations.  Also shown are fits to these spectra.}
\label{fig:res_back:jpsi_spectra}
\end{center}
\end{figure}

\begin{figure}[htbp]
\begin{center}
\includegraphics[width=0.85\linewidth]{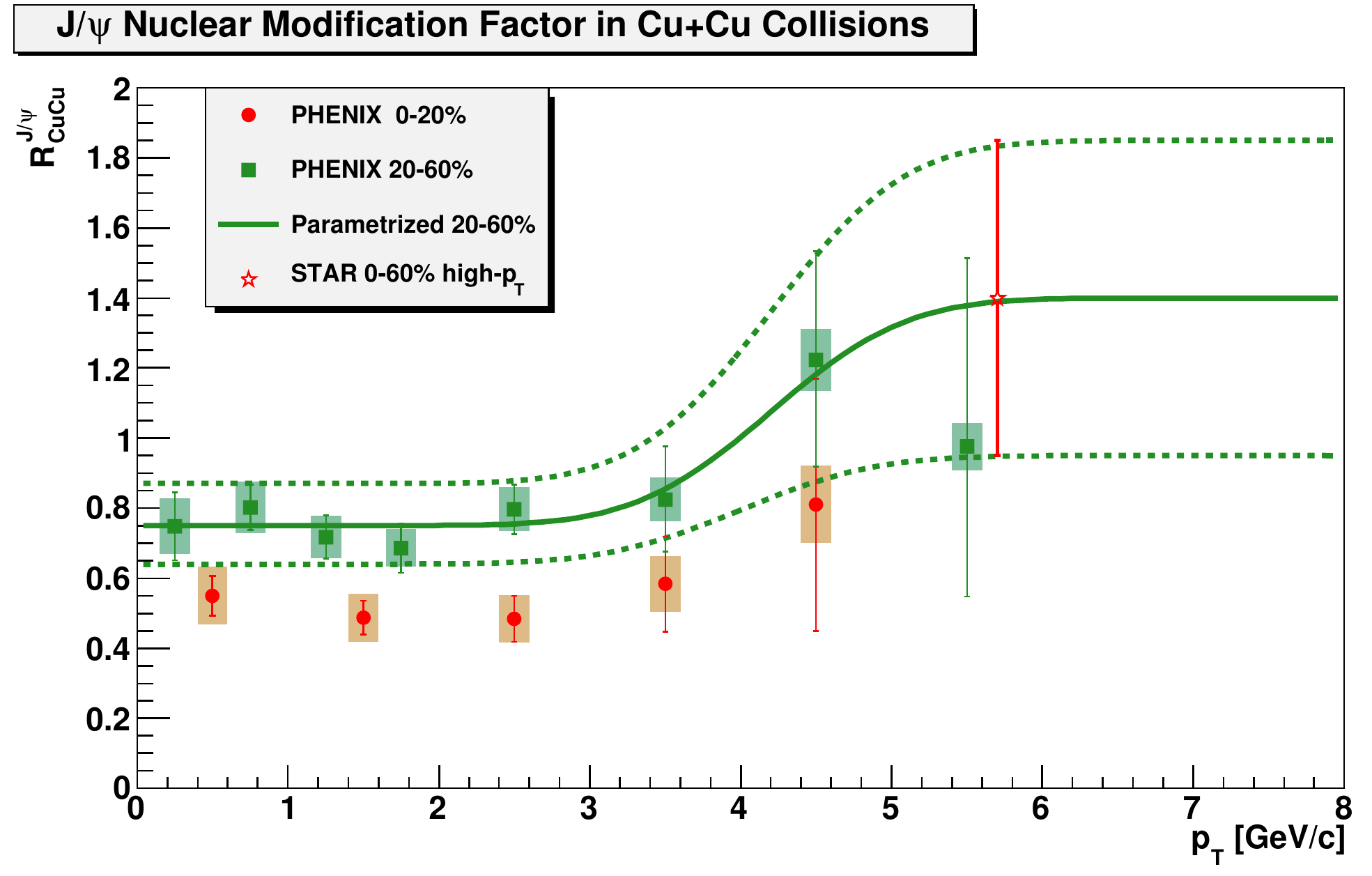}
\caption[The nuclear modification factor for $J/\psi$ in Cu + Cu collisions.]{The nuclear modification factor measured\protect\cite{PhysRevLett.101.122301} by the PHENIX collaboration for $J/\psi$ in Cu + Cu collisions for the 0-20\% (circles) and 20-60\% (squares) centrality classes.  Also shown in the STAR collaboration's measurement\protect\cite{PhysRevC.80.041902} of the nuclear modification factor for $p_{T}>5\GeV/c$ in the 0-60\% centrality class.  The curves indicate the parametrized nuclear modification factor for the 20-60\% centrality class; the upper dashed curve is used to estimate the upper limit of the $J/\psi$ yield in that centrality class.}
\label{fig:res_back:rcucu_jpsi_par}
\end{center}
\end{figure}

\begin{figure}[htbp]
\begin{center}
\includegraphics[width=0.85\linewidth]{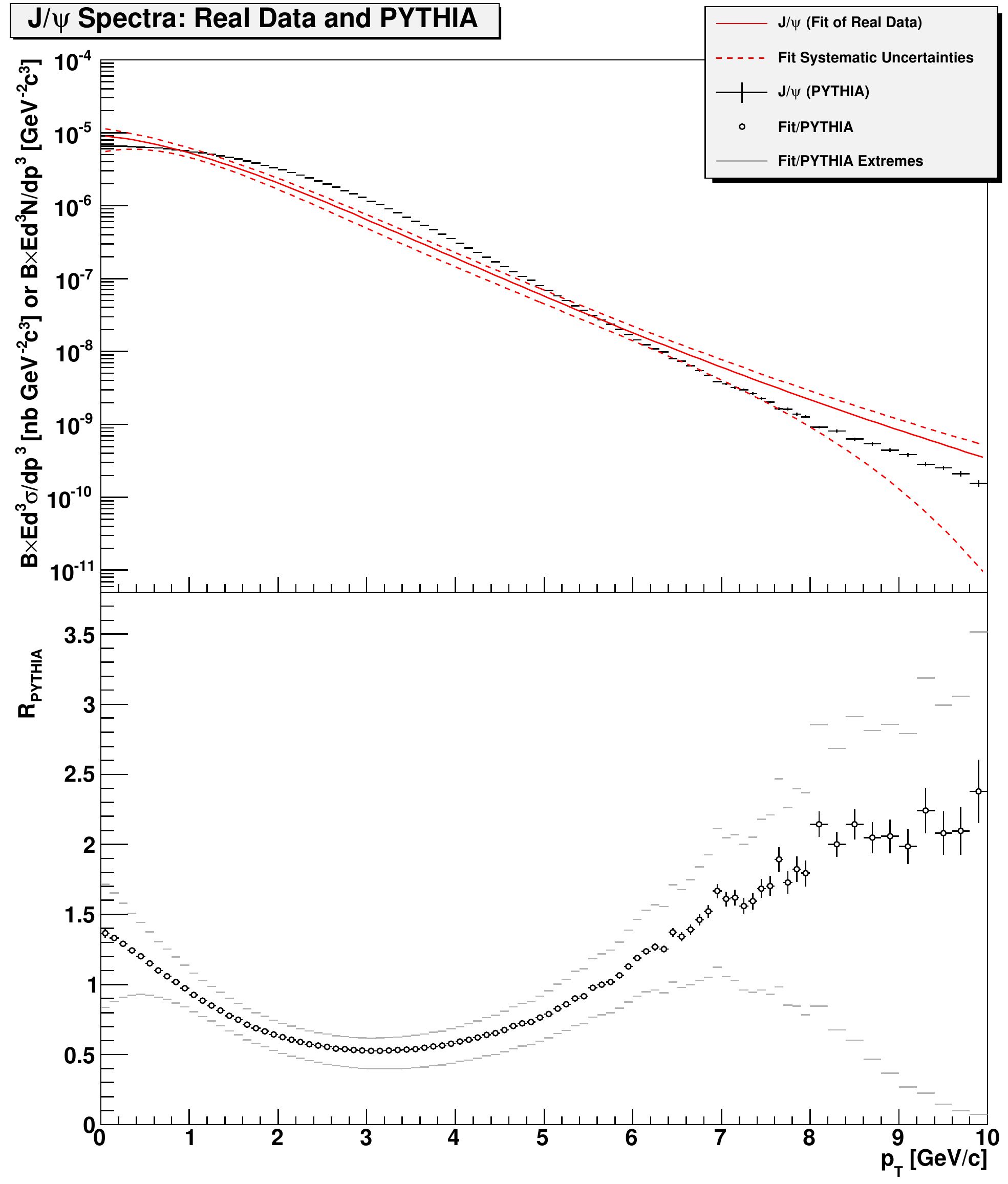}
\caption[Illustration of the calculation of $R_{PYTHIA}$.]{Upper panel: The curves represent a fit (with systematic uncertainties) to the spectrum of $J/\psi$ in the 0-20\% centrality class of Cu + Cu collisions.  The black histogram represents the cross-section of $J/\psi$ in $p+p$ collisions simulated using PYTHIA.  Both quantities have been multiplied by the branching ratio for the $J/\psi\rightarrow e^{-}e^{+}$ decay.  Lower panel: The ratio $(R_{PYTHIA})$ of the fit to the simulated PYTHIA spectrum; the gray lines indicate the ratio of the dashed curves to the simulated spectrum.}
\label{fig:res_back:r_pythia}
\end{center}
\end{figure}

The decay $J/\psi\rightarrow e^{-}e^{+}$, which has a branching ratio of $(5.94\pm 0.06)\%$,~\cite{PDG_review} produces $e^{-}e^{+}$ pairs that fail the invariant-mass cut.  While these $e^{\pm}$ do come from a heavy-flavor source, they are photonic and must be subtracted from the efficiency-corrected non-photonic $e^{\pm}$ spectrum.  The yield of $e^{\pm}$ from $J/\psi$ decays is estimated using PYTHIA simulations of $J/\psi$ production in 140 million $p+p$ collisions, with the subsequent decays of the $J/\psi$ to $e^{-}e^{+}$ pairs.  The shape of the simulated $J/\psi$ spectrum as a function of transverse momentum is not the same as the shape of the spectrum of $J/\psi$ in real Cu + Cu collisions; the simulated $J/\psi$ are weighted to account for this.  Figure~\ref{fig:res_back:jpsi_spectra} shows measurements~\cite{PhysRevC.80.041902,PhysRevLett.101.122301} of the $J/\psi$ yields for Cu + Cu collisions in several centrality classes.  For the 0-20\% centrality class, the combined STAR and PHENIX data are fit with a function $F[J/\psi,0-20\%]$, which has the form

\begin{equation}
\label{eq:res_back:jpsi:form}
F(p_{T})=A[\exp(-ap_{T}-bp_{T}^{2})+p_{T}/p_{0}]^{-n}
\end{equation}

\noindent The fit parameters are given in Section~\ref{sec:functions:jpsi}.  The dashed curves, which indicate systematic uncertainties of the fit, are derived from the calculated covariance matrix.\footnote{The calculated covariance matrix is used to evaluate the upper and lower limits of the 68\% confidence interval at many values of $p_{T}$; the results are then fit to obtain the curves describing the lower and upper limits.}

The $J/\psi$ spectrum for the 20-60\% centrality class is parametrized by a function $F[J/\psi,20-60\%]$, which has the form given by Equation ~\ref{eq:res_back:jpsi:form}.  The parameters of $F[J/\psi,20-60\%]$ are found as follows.  The $J/\psi$ spectrum for the 20-40\% centrality class is fit with a function $F[J/\psi,20-40\%]$ (which also has the form given by Equation ~\ref{eq:res_back:jpsi:form}).  The values of the parameters $a$, $b$, $p_{0}$, and $n$ in $F[J/\psi,20-60\%]$ are set equal to the values those parameters have in $F[J/\psi,20-40\%]$ (\textit{i.e.}, $F[J/\psi,20-60\%]$ is constrained to have the same shape as $F[J/\psi,20-40\%]$); the value of the scale parameter $A$ for $F[J/\psi,20-60\%]$ is found by $\chi^{2}$ minimization.  The curve describing the lower limit of $F[J/\psi,20-60\%]$ is found using the method described previously.  A different method is used to find the curve describing the upper limit of $F[J/\psi,20-60\%]$: the $J/\psi$ nuclear modification factor is parametrized and then multiplied by a fit of the $J/\psi$ yield in $p+p$ collisions.  Figure~\ref{fig:res_back:rcucu_jpsi_par} shows the measured values of the $R_{\mathrm{CuCu}}^{J/\psi}$, the nuclear modification factor for $J/\psi$ in Cu + Cu collisions.  Sigmoid functions of the form

\begin{equation}
R_{\mathrm{CuCu},20-60\%}^{J/\psi}(p_{T})=A+\frac{B-A}{2}\left[\mathrm{erf}\frac{p_{T}-a}{b}+1\right]
\end{equation}

\noindent are used to parametrize the central value and the lower and upper limits of $R_{\mathrm{CuCu}}^{J/\psi}$ over the range of transverse momentum shown.  At low $p_{T}$, the functions are constrained to the values of $R_{\mathrm{CuCu},20-60\%}^{J/\psi}$ measured by the PHENIX collaboration, while at high $p_{T}$, the functions are constrained to $R_{\mathrm{CuCu},0-60\%}^{J/\psi}=1.40\pm0.45$, measured\footnote{A separate measurement for the 20-60\% centrality class is not available.} by the STAR collaboration for $p_{T}>5\GeV/c$.  The parameters of the function $f_{up}$ used to describe the upper limit of $R_{\mathrm{CuCu},20-60\%}^{J/\psi}$ are $A=0.87$, $B=1.85$, $a=4.2\GeV/c$ and $b=1\GeV/c$.  The upper limit of the $J/\psi$ yield in the 20-60\% centrality class of Cu + Cu collisions is taken to be

\begin{equation}
F[J/\psi,20-60\%+](p_{T})=f_{up}(p_{T})\cdot\langle T_{AA}\rangle\cdot F[J/\psi,pp](p_{T}),
\end{equation}

\noindent where the function $F[J/\psi,pp]$ is a fit to the combined STAR~\cite{PhysRevC.80.041902} and PHENIX~\cite{PhysRevD.82.012001} measurements of the $J/\psi$ yield in $p+p$ collisions (see Section~\ref{sec:functions:jpsi} for the fit parameters).

For each centrality class, the ratio of the $J/\psi$ fit functions for Cu + Cu collisions to the simulated PYTHIA $J/\psi$ spectrum is calculated, as illustrated in Figure~\ref{fig:res_back:r_pythia}.  The upper panel shows the $J/\psi$ cross-section simulated using PYTHIA, as well as the fit function $F[J/\psi,0-20\%]$ and its systematic uncertainties.  The lower panel shows $R_{PYTHIA}$, the fit divided by the PYTHIA spectrum, with the gray lines indicating the the lower and upper limits of the fit divided by the simulated spectrum.  For each centrality class, the spectrum of $e^{\pm}$ from $J/\psi$ decays was calculated by weighting each $e^{\pm}$ by the value of $R_{PYTHIA}$ (evaluated at the transverse momentum of the parent $J/\psi$).  The results of these calculations will be shown in the next chapter.

\subsection[Photonic $e^{\pm}$ from $\Upsilon$ Decays]{Photonic $\boldsymbol{e^{\pm}}$ from $\boldsymbol{\Upsilon}$ Decays}
\label{sec:res_back:upsilon}

\begin{figure}[htbp]
\begin{center}
\includegraphics[width=0.85\linewidth]{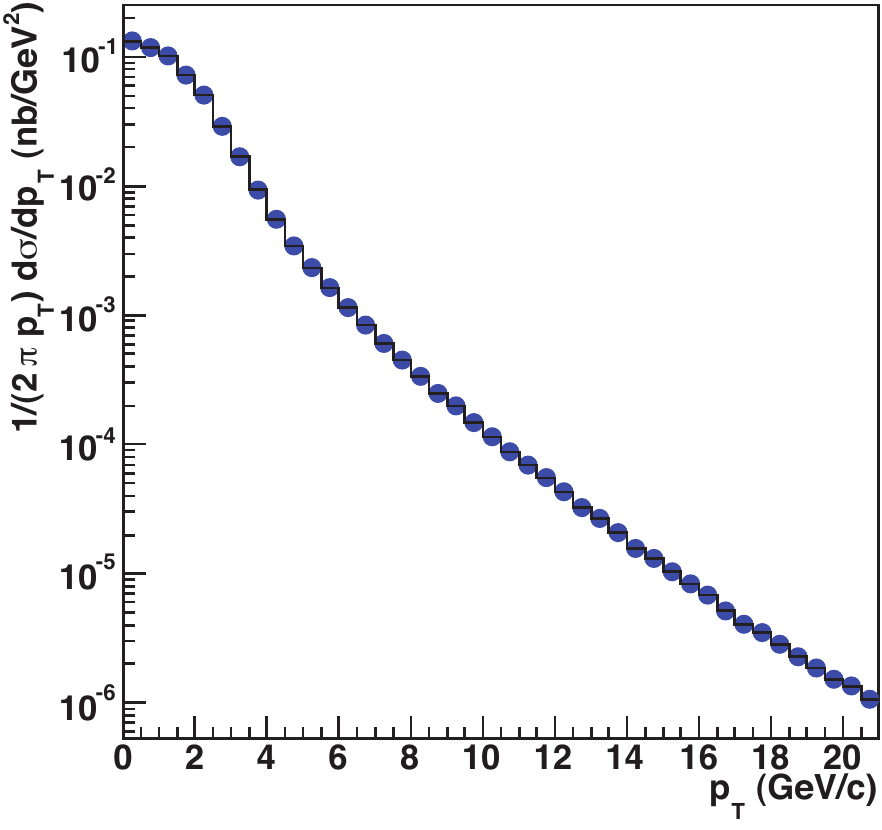}
\caption[Cross-section of $\Upsilon$ in 200-GeV $p+p$ collisions simulated using the Color Evaporation Model.]{Cross-section of $\Upsilon$ in 200-GeV $p+p$ collisions simulated using the Color Evaporation Model (CEM).\protect\cite{Frawley2008125,Barger1980253,Vogt_RHICII2005}}
\label{fig:res_back:upsilon}
\end{center}
\end{figure}

There are currently no RHIC measurements of the $\Upsilon$ cross-section as a function of transverse momentum at $\sqrt{s_{NN}}=200$ GeV.  Instead, the shape of the $\Upsilon$ spectrum is estimated using the Color Evaporation Model (CEM).  In this model,~\cite{Frawley2008125,Barger1980253,Vogt_RHICII2005} the $b\bar{b}$ production cross-section is calculated perturbatively at the Next-to-Leading Order (``NLO", $\mathcal{O}(\alpha_{s}^{3})$) level for values of $\sqrt{\hat{s}}$, the parton-parton CM collision energy, in the range $2\times($mass of $b$ quark$)\leq\sqrt{\hat{s}}\leq 2\times($mass of $B^{\pm}$ meson).  It is assumed that any net color charge in the $b\bar{b}$ pair will be neutralized through non-perturbative interactions with the color field in the collision region (color evaporation).  The $b$ quark and $\bar{b}$ antiquark may then combine with each other to form a $b\bar{b}$ bound state, or may combine with light quarks to form $B$ mesons.  However, since $\sqrt{\hat{s}}$ is less than the threshold for the production of two $B$ mesons, any energy needed to form these mesons is assumed to come from non-perturbative interactions with the color field in the collision region.  The $\Upsilon$ cross-section\footnote{The CEM can also be used to estimate the production cross-sections for other heavy quarkonium states, including $J/\psi$.} will be a fraction $F_{\Upsilon}$ of the calculated $b\bar{b}$ cross-section; $F_{\Upsilon}$ has been found~\cite{Gavai_IJMPA1995} to be independent of $\sqrt{s}$, rapidity, and $p_{T}$.  Therefore, $F_{\Upsilon}$ need not be calculated at 200 GeV, but can be determined from measurements at lower collision energies.  Figure~\ref{fig:res_back:upsilon} shows a CEM calculation~\cite{Frawley2008125,Vogt_RHICII2005} of the $\Upsilon$ cross-section (including the $\Upsilon(1S)$, $\Upsilon(2S)$, and $\Upsilon(3S)$ states) at mid-rapidity as a function of transverse momentum.  The $p_{T}$-integrated $\Upsilon$ cross-section estimated using the CEM is found to be consistent with the STAR collaboration's measurement~\cite{PhysRevD.82.012004} of that quantity: $\sigma_{\Upsilon(1S+2S+3S)}=[114\pm 38(\mathrm{stat.})^{+23}_{-24}(\mathrm{sys.})]\mathrm{pb}$.  Adding the statistical and systematic uncertainties in quadrature gives a 39\% uncertainty in this measurement.

PYTHIA was used to simulate proton-proton collisions, producing 4.8 million $\Upsilon$, which were allowed to decay to $e^{-}e^{+}$ pairs.  These PYTHIA calculations do not reproduce the shape of the $\Upsilon$ spectrum given by the CEM, so the PYTHIA data are weighted by the ratio of the CEM spectrum to the PYTHIA spectrum.  This ratio is found to vary between 4 and 0.5, generally decreasing with $p_{T}$.  For each Cu + Cu centrality class, the spectrum of $e^{\pm}$ from $\Upsilon$ decays is assumed to be the spectrum calculated for $p+p$ collisions, scaled by $\langle T_{AA}\rangle R_{\mathrm{CuCu}}^{\Upsilon}$.  Since no measurement of $R_{\mathrm{CuCu}}^{\Upsilon}$ is available, it is assumed that the nuclear modification factor is 1, with large uncertainties: $R_{\mathrm{CuCu}}^{\Upsilon}=1\pm0.7$.  The results of this calculation are shown in the next chapter; the contribution of $\Upsilon$ decays to the $e^{\pm}$ spectrum is small enough that the large uncertainties in the nuclear modification factor do not have a large effect on the final corrected spectrum.

\subsection[Photonic $e^{\pm}$ from the Drell-Yan Process]{Photonic $\boldsymbol{e^{\pm}}$ from the Drell-Yan Process}
\label{sec:res_back:dy}

A Leading-Order calculation~\cite{Vogelsang_Private_2010} of $e^{-}e^{+}$ pair production via the Drell-Yan process $(q\bar{q}\rightarrow\gamma^{*}\rightarrow e^{-}e^{+})$ is used by the PHENIX collaboration as a correction to their measurement~\cite{PHENIX_NPE2010} of non-photonic $e^{\pm}$ in 200-GeV $p+p$ collisions.  This same spectrum, scaled by $\langle N_{bin}\rangle/\sigma_{inel.}$, is used in this dissertation.

\chapter{Results}
\label{sec:results}

\section{Correction of Spectra}
The normalized efficiency-corrected yields of inclusive\footnote{all $e^{\pm}$, the sum of the photonic and non-photonic contributions}, photonic, and non-photonic $e^{\pm}$ ($I_{EC}$, $P_{EC}$, and $N_{EC}$ respectively) are found as follows.

\begin{equation}
\label{eq:results:iec}
I_{EC}=\nu\frac{K_{inc.}I}{\langle A_{BEMC}\rangle\varepsilon_{R}\cdot\varepsilon_{T}\cdot\varepsilon_{dE/dx}}
\end{equation}

\begin{equation}
\label{eq:results:pec}
P_{EC}=\nu\frac{P}{\langle A_{BEMC}\rangle\varepsilon_{R}\cdot\varepsilon_{T}\cdot\varepsilon_{dE/dx}\cdot\varepsilon_{B}}
\end{equation}

\begin{equation}
\label{eq:results:nec}
N_{EC}=I_{EC}-P_{EC}=\nu\frac{1}{\langle A_{BEMC}\rangle\varepsilon_{R}\cdot\varepsilon_{T}\cdot\varepsilon_{dE/dx}}\left(K_{inc.}I-\frac{1}{\varepsilon_{B}}P\right).
\end{equation}

\begin{figure}[htbp]
\begin{center}
\includegraphics[width=0.85\linewidth]{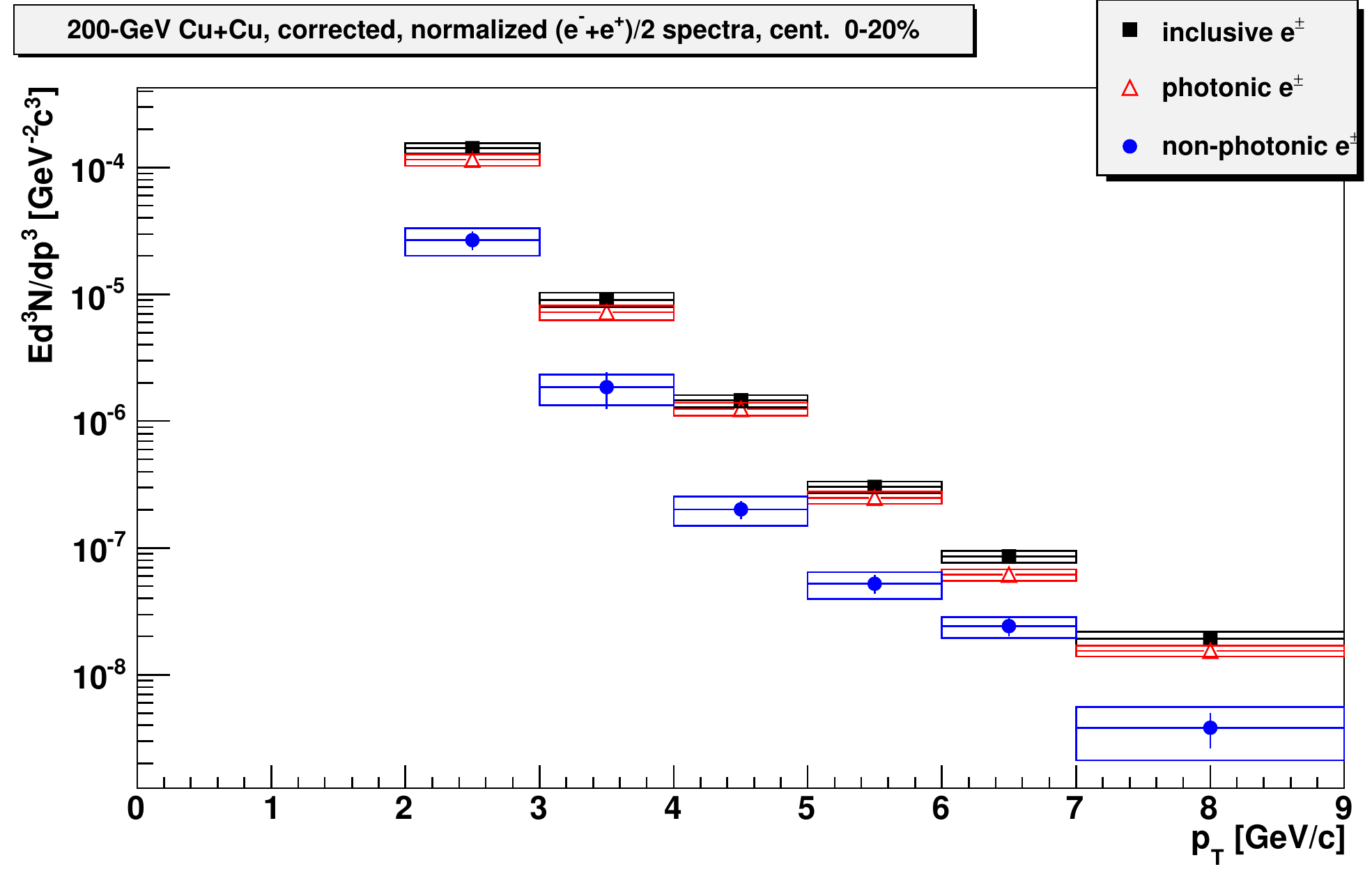}
\caption[The efficiency-corrected $e^{\pm}$ yields in the 0-20\% most central Cu + Cu collisions.]{The efficiency-corrected $e^{\pm}$ yields in the 0-20\% most central Cu + Cu collisions: inclusive (black), photonic (red), non-photonic (blue).}
\label{fig:results:eff_corr_yields_c01}
\end{center}
\end{figure}

\noindent Here, $I$ is the uncorrected yield of inclusive $e^{\pm}$ and $P=n_{unlike}-2\sqrt{n_{--}n_{++}}$ is the uncorrected yield of photonic $e^{\pm}$.  The correction factors\footnote{The spectra of $e^{\pm}$ in the minimum-bias data set are not corrected by the high-tower trigger efficiency $\varepsilon_{T}$.} are summarized in Table~\ref{table:anintro:corrections} (page~\pageref{table:anintro:corrections}).  The normalization factor $\nu$ is

\begin{equation}
\label{eq:results:norm}
\nu=\frac{\langle p_{T}^{-1}\rangle c}{2\cdot\Delta\phi\cdot\Delta y\cdot\Delta p_{T}\cdot N_{evt}},
\end{equation}

\noindent where $\Delta\phi=2\pi$ is the azimuthal range of the measurement, $\Delta y=0.8$ is the rapidity range (for ultra-relativistic $e^{\pm}$, $\Delta y$ is very well approximated by the pseudorapidity range), and $\Delta p_{T}$ is the width of each transverse momentum bin (usually 1 GeV/$c$).  The initial factor of 2 accounts for the fact that two particle types, electrons and positrons, are included in the measurement.  (For the sake of brevity in the following discussion, a normalized $\tfrac{1}{2}(e^{-}+e^{+})$ yield will continue to be referred to as an $e^{\pm}$ yield.)  For each centrality class, the number of events ($N_{evt}$) is given by Table~\ref{table:anintro:centrality} (page~\pageref{table:anintro:centrality}), with $N_{evt}^{MB}$ used for the minimum-bias data set and $N_{evt}^{HT}(normalized)$ used for the high-tower-triggered data set.  The weighting factor $\langle p_{T}^{-1}\rangle$ is the mean value of $1/p_{T}$ for the inclusive $e^{\pm}$ spectrum in each $p_{T}$ bin.  The factor of $\langle p_{T}^{-1}\rangle c$ is necessary for the conversion to an invariant yield of the form $Ed^{3}N/dp^{3}$.  Figure~\ref{fig:results:eff_corr_yields_c01} shows the efficiency-corrected $e^{\pm}$ yields as functions of transverse momentum for the 0-20\% most central 200-GeV Cu + Cu collisions.  These spectra are combinations of the $e^{\pm}$ spectra measured in minimum-bias triggered events (for $p_{T}<4\GeV/c$) and high-tower triggered events (for $p_{T}>4\GeV/c$).

\begin{figure}[htbp]
\begin{center}
\includegraphics[width=1\linewidth]{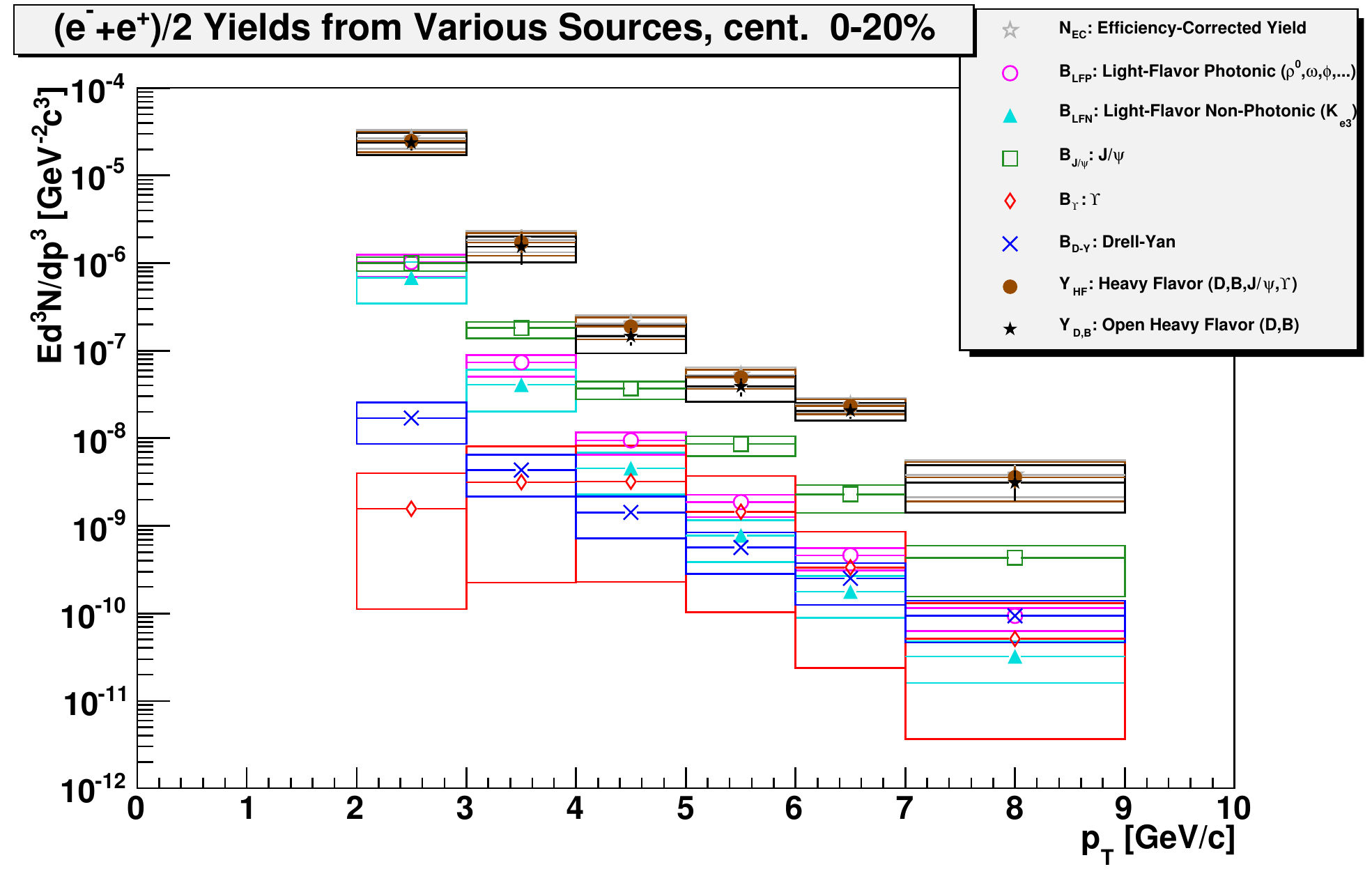}
\caption[Components of the residual background that are subtracted from the efficiency-corrected non-photonic $e^{\pm}$ yield.]{The efficiency-corrected non-photonic $e^{\pm}$ yield $(N_{EC})$, various components of the residual background (see the text in this section and Chapter~\ref{sec:res_back}), and the yields of $e^{\pm}$ from heavy-flavor $(Y_{HF})$ and open-heavy-flavor sources $(Y_{D,B})$.  The spectra shown are for the 0-20\% centrality class of 200-GeV Cu + Cu collisions.}
\label{fig:results:bsub_c01}
\end{center}
\end{figure}

The efficiency-corrected non-photonic $e^{\pm}$ yield, $N_{EC}$, includes some contamination.  Chapter~\ref{sec:res_back} describes the estimation of the various components of the residual background, which is subtracted from $N_{EC}$.  The yield $Y_{HF}$ of $e^{\pm}$ from the decays of hadrons containing heavy quarks is

\begin{equation}
\label{eq:results:yhf}
Y_{HF}=N_{EC}-B_{LFP}-B_{LFN}-B_{D-Y},
\end{equation}

\noindent where $B_{LFP}$ is the residual yield of photonic $e^{\pm}$ from light-flavor sources ($\eta$, $\rho^{0}$, $\omega$, $\eta^{\prime}$, $\phi$, and $K^{0}_{S}$; see Section~\ref{sec:res_back:lfp}), $B_{LFN}$ is the yield of non-photonic $e^{\pm}$ from light-flavor sources (predominantly the semileptonic decays of kaons, see Section~\ref{sec:res_back:lfn}), and $B_{D-Y}$ is the yield of $e^{\pm}$ produced via the Drell-Yan process.  The yield $Y_{HF}$ includes photonic $e^{\pm}$ from the decays of mesons made up of $c\bar{c}$ and $b\bar{b}$ pairs, predominantly $J/\psi$ and $\Upsilon$.  The yields of $e^{\pm}$ from those sources are subtracted from $Y_{HF}$ to give $Y_{D,B}$ the yield of non-photonic $e^{\pm}$ from the decays of $D$ and $B$ mesons, which are said to have ``open heavy flavor" because they have non-zero net charm or bottomness.  The yield of non-photonic $e^{\pm}$ from open-heavy-flavor sources is

\begin{equation}
\label{eq:results:ydb}
Y_{D,B}=Y_{HF}-B_{J/\psi}-B_{\Upsilon}=N_{EC}-B_{LFP}-B_{LFN}-B_{D-Y}-B_{J/\psi}-B_{\Upsilon},
\end{equation}

\noindent where $B_{J/\psi}$ $(B_{\Upsilon})$ is the yield of photonic $e^{\pm}$ from the decays of $J/\psi$ $(\Upsilon)$.  Figure~\ref{fig:results:bsub_c01} shows the corrected $e^{\pm}$ yields $N_{EC}$, $Y_{HF}$, and $Y_{D,B}$ for the 0-20\% most central Cu + Cu collisions, as well as the various components of the residual background.  The non-photonic $e^{\pm}$ yields $Y_{HF}$ and $Y_{D,B}$ are presented for three centrality classes in Section~\ref{sec:results:binshift}.

\section{Uncertainties}
For each transverse-momentum bin, the statistical uncertainty of the uncorrected photonic $e^{\pm}$ yield, $P$, is calculated based on the assumptions that $n_{--}$ and $n_{++}$ are statistically independent, and that the statistical uncertainty of the combinatorial background $2\sqrt{n_{--}n_{++}}$ is uncorrelated with the uncertainty in $n_{unlike}$.  For  $n_{unlike}$, $n_{--}$, and $n_{++}$, the uncertainty of each quantity is assumed to be the square root of that quantity.  The statistical uncertainty of $P$ is

\begin{equation}
\label{eq:results:stat_err_p}
\sigma P=\sqrt{n_{unlike}+n_{--}+n_{++}}.
\end{equation}

\noindent The statistical uncertainty of $I_{EC}$ is

\begin{equation}
\label{eq:results:stat_err_iec}
\sigma I_{EC}=\frac{\sigma I}{I}\cdot I_{EC}=\frac{I_{EC}}{\sqrt{I}}
\end{equation}

\noindent and the statistical uncertainty of $P_{EC}$ is

\begin{equation}
\label{eq:results:stat_err_pec}
\sigma P_{EC}=\frac{\sigma P}{P}\cdot P_{EC}.
\end{equation}

\noindent As shown in Section~\ref{sec:uncertainties:spectra}, the statistical uncertainty of $N_{EC}$ is calculated based on the assumption that the inclusive $e^{\pm}$ yield is statistically independent of the ratio $r=P/K_{inc.}I\varepsilon_{B}$, the fraction of $e^{\pm}$ that are photonic.  The statistical uncertainty of $N_{EC}$ is found to be

\begin{equation}
\label{eq:results:stat_err_nec}
\sigma N_{EC}=N_{EC}\sqrt{\frac{1}{I}+\frac{r^{2}}{(1-r)^{2}}\left[\left(\frac{\sigma P}{P}\right)^{2}-\frac{1}{I}\right]}.
\end{equation}

\noindent The absolute statistical uncertainties of $Y_{HF}$ and $Y_{D,B}$ are also assumed to be $\sigma N_{EC}$; the uncertainties in the components of the residual background are accounted for in the systematic uncertainties.

The systematic uncertainties in $I_{EC}$, $P_{EC}$, and $N_{EC}$ are calculated by assuming that the uncertainties in each of the correction factors are uncorrelated.  See Section~\ref{sec:uncertainties:spectra_sys} for the equations used.  The systematic uncertainties of $Y_{HF}$ and $Y_{D,B}$ are the quadrature sums of the systematic uncertainty of $N_{EC}$ and the systematic uncertainties of the components of the residual background that were subtracted.


\section{Bin-Shift Correction}
\label{sec:results:binshift}

In the next section, the non-photonic $e^{\pm}$ yield $(Y_{D,B})$ in Cu + Cu collisions will be compared to the non-photonic $e^{\pm}$ yield in $p+p$ collisions measured by the STAR~\cite{STAR_ppNPE2011} and PHENIX~\cite{PhysRevLett.97.252002} collaborations, and the nuclear modification factor will be calculated.  In Reference~\cite{STAR_ppNPE2011} one more correction is applied to $Y_{D,B}$: the bin-shift correction factor $C_{bs}$.  The calculation of $C_{bs}$ for non-photonic $e^{\pm}$ in Cu + Cu collisions is described in this section.

In Reference~\cite{STAR_ppNPE2011} the non-photonic $e^{\pm}$ cross-section in $p+p$ collisions is presented as the value of $Ed^{3}\sigma/dp^{3}$ at the \textit{center} of each bin.  In contrast, the non-photonic $e^{\pm}$ yield in Cu + Cu collisions has been presented in this dissertation as the \textit{average value} of the invariant yield $Ed^{3}N/dp^{3}$ for each bin (\textit{i.e.}, the integral of $Ed^{3}N/dp^{3}$ over the bin, divided by the bin width).  Unless $Ed^{3}N/dp^{3}$ is constant or varies linearly across the bin, the average value will not be the same as the value of $Ed^{3}N/dp^{3}$ at the bin center.  The bin-shift correction factor $C_{bs}$ accounts for this difference.

\begin{figure}[htbp]
\begin{center}
\includegraphics[width=0.85\linewidth]{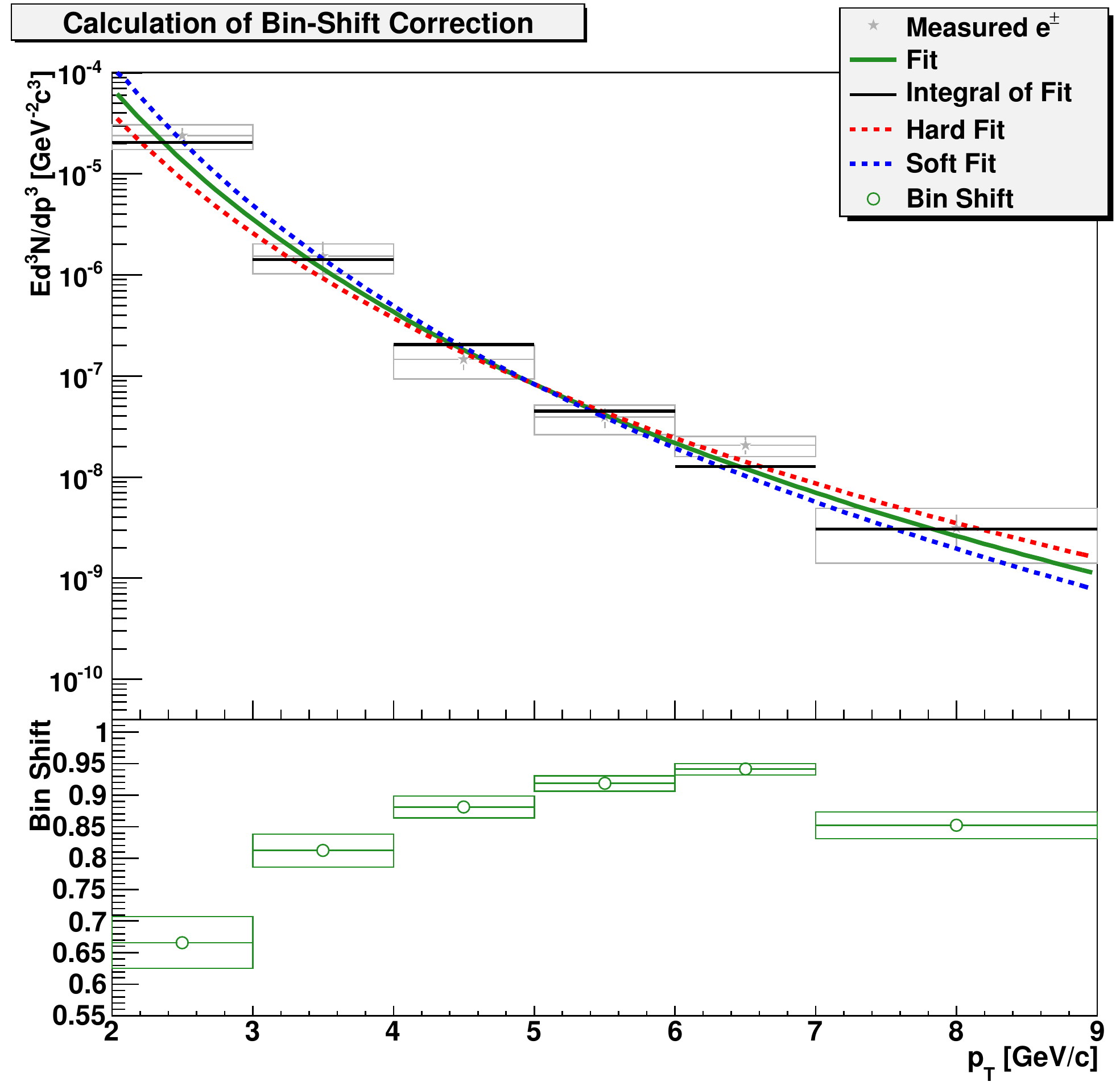}
\caption[Calculation of the bin-shift correction.]{Upper panel: illustration of the calculation of the bin-shift correction $C_{bs}$ for the 0-20\% centrality class of Cu + Cu collisions.  See the text for a description of the quantities plotted.  Lower panel: the bin-shift correction.}
\label{fig:results:bin_shift}
\end{center}
\end{figure}

\begin{figure}[htbp]
\begin{center}
\includegraphics[width=0.85\linewidth]{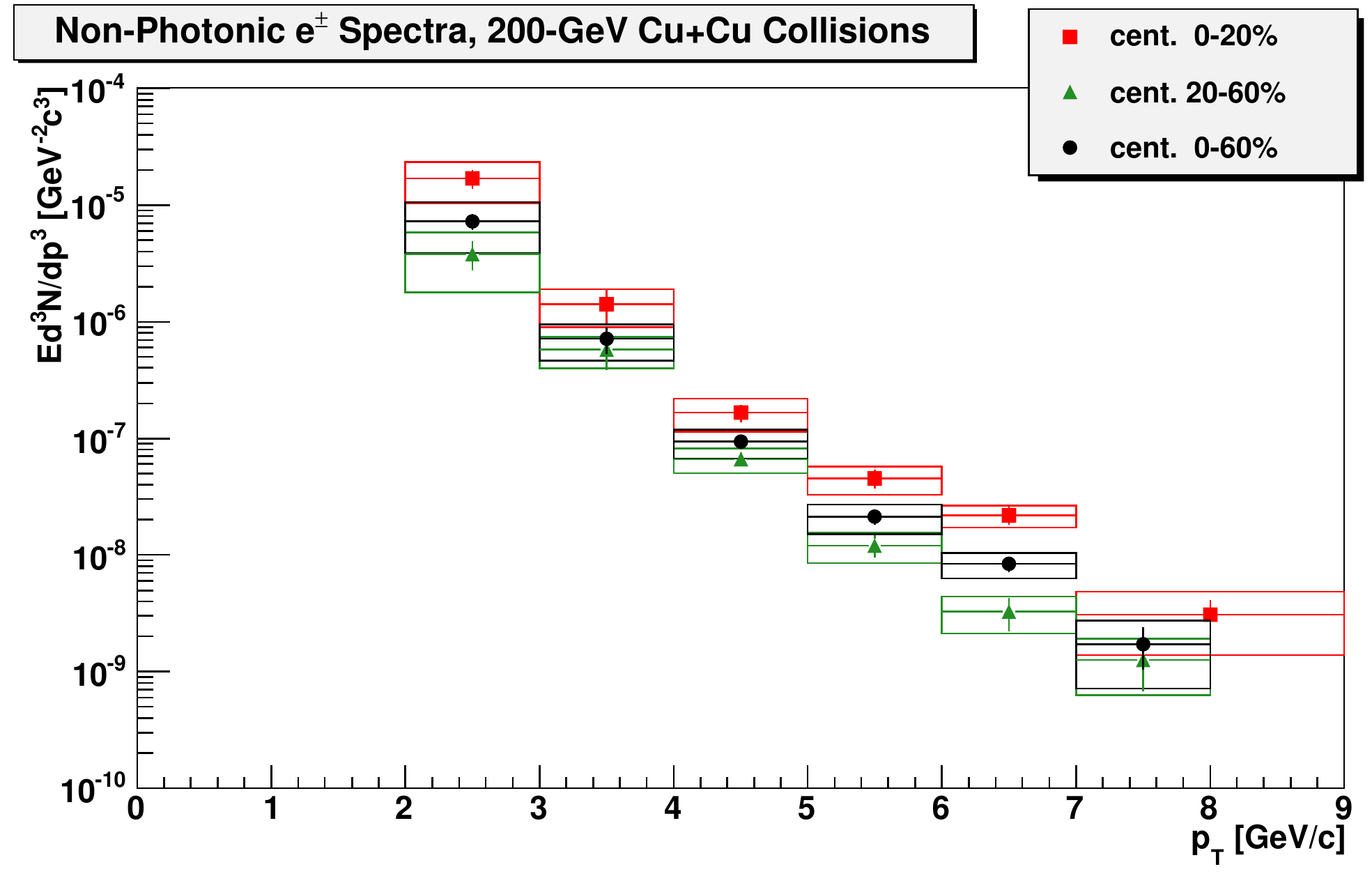}
\caption{The yields $(Y_{HF})$ of $e^{\pm}$ from heavy-flavor decays as functions of transverse momentum for three centrality classes of 200-GeV Cu + Cu collisions.}
\label{fig:results:yhf_cX}
\includegraphics[width=0.85\linewidth]{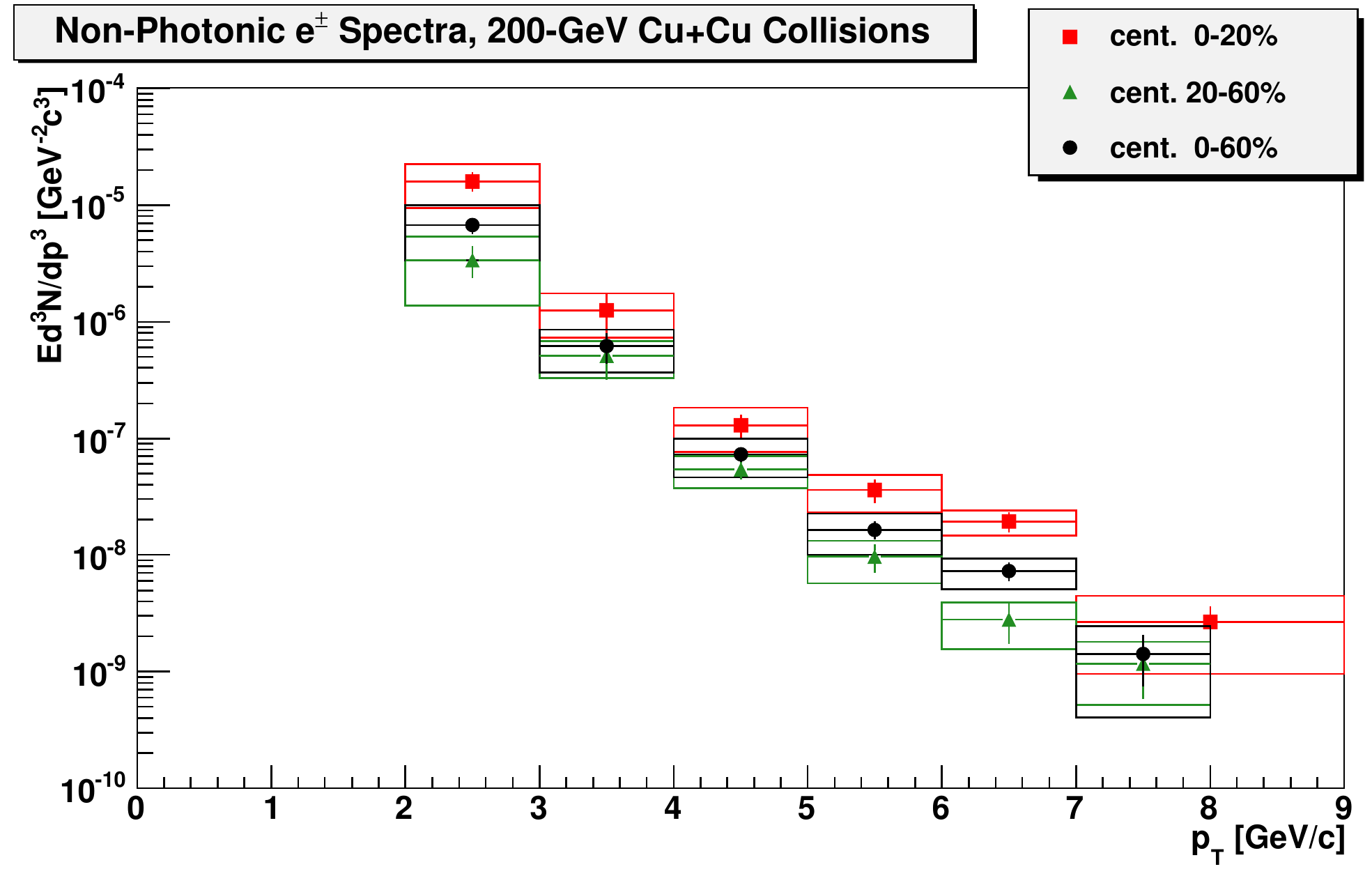}
\caption{The yields $(Y_{D,B})$ of $e^{\pm}$ from open-heavy-flavor decays ($D$ and $B$ meson decays) as functions of transverse momentum for three centrality classes of 200-GeV Cu + Cu collisions.}
\label{fig:results:ydb_cX}
\end{center}
\end{figure}

For each Cu + Cu centrality class, the non-photonic $e^{\pm}$ yield $Y_{D,B}$ is fit with a function of the form $f=A(1+p_{T}/B)^{n}$.  In the calculation of $\chi^{2}$, the fit \textit{intergal} in each bin is compared to the measured value of the spectrum.  For each transverse-momentum bin, $C_{bs}$ is the ratio of the value of $f$ at the bin center to the average value of $f$.  This method is illustrated in Figure~\ref{fig:results:bin_shift} for the 0-20\% centrality class.  In the upper panel, the gray points show the values of $Y_{D,B}$.  The green curve is a fit to that spectrum (the fit parameters are given in Section~\ref{sec:functions:npe}) and the black lines represent the fit integral in each bin.  For each bin, the bin-shift correction is the ratio of the green curve evaluated at the bin center to the fit integral.  The values of $C_{bs}$ are shown in the lower panel.  The dashed curves are used to calculate the systematic uncertainties of $C_{bs}$.  The ``soft fit" (``hard fit") curve is generated from the fit by decreasing (increasing) the parameter $n$ by its uncertainty.\footnote{The dashed curves are also multiplied by scale factors, but this has no effect on the calculation of the bin-shift correction.}  The bin-shift correction is calculated using these two extremal curves; the uncertainties in $C_{bs}$ are the difference between the extreme lower or upper calculations and the central value.  For each centrality class, the non-photonic $e^{\pm}$ yield $Y_{D,B}$ is multiplied by $C_{bs}$, giving the final corrected spectrum.\footnote{An alternative approach, described in~\cite{Lafferty1995541}, is to shift the abscissa of each data point away from the bin center to the point where the fit function is equal to the fit integral (leaving the ordinate unchanged).}  These spectra are shown for three centrality classes in Figures~\ref{fig:results:yhf_cX} (for $Y_{HF}$) and ~\ref{fig:results:ydb_cX} (for $Y_{D,B}$).

\clearpage

\section{Nuclear Modification Factor}
\label{sec:results:raa}

The nuclear modification factor, which was defined in Section~\ref{sec:theory:high_pt_supp} (page~\pageref{sec:theory:high_pt_supp}), is used to quantify the differences in particle yields between nucleus-nucleus collisions and proton-proton collisions.  This factor is the ratio of the yield of a particle species in nucleus-nucleus collisions to the yield of the same particle species in $p+p$ collisions.  The ratio is scaled to account for the fact that a nucleus-nucleus collision consists of many nucleon-nucleon collisions.   For a given $p_{T}$ bin, the nuclear modification factor for particle species $X$ is

\begin{equation}
\label{eq:results:define_raa}
R_{AA}^{X}=\frac{\left(\dfrac{Ed^{3}N^{X}_{AA}}{dp^{3}}\right)\sigma_{inel.}}{\left(\dfrac{Ed^{3}\sigma^{X}_{pp}}{dp^{3}}\right)\langle N_{bin}\rangle}.
\end{equation}

\noindent Here, $Ed^{3}N^{X}_{AA}/dp^{3}$ is the invariant yield of particle species $X$ in nucleus-nucleus collisions and $Ed^{3}\sigma^{X}_{pp}/dp^{3}$ is the cross-section for particle species $X$ in proton-proton collisions.  The inelastic nucleon-nucleon cross-section is $\sigma_{inel.}=42\,\mathrm{mb}$, and $\langle N_{bin}\rangle$ is the mean number of binary nucleon-nucleon collisions for the $A+A$ collision system.  A value of $R_{AA}^{X}=1$ indicates that the yield of particle species $X$ is not suppressed in nucleus-nucleus collisions relative to proton-proton collisions.  The symbol ``$R_{AA}$" is generic; the subscripts may be used to indicate a specific collision system, which will be the practice in this chapter.

\begin{figure}[htbp]
\begin{center}
\includegraphics[width=0.85\linewidth]{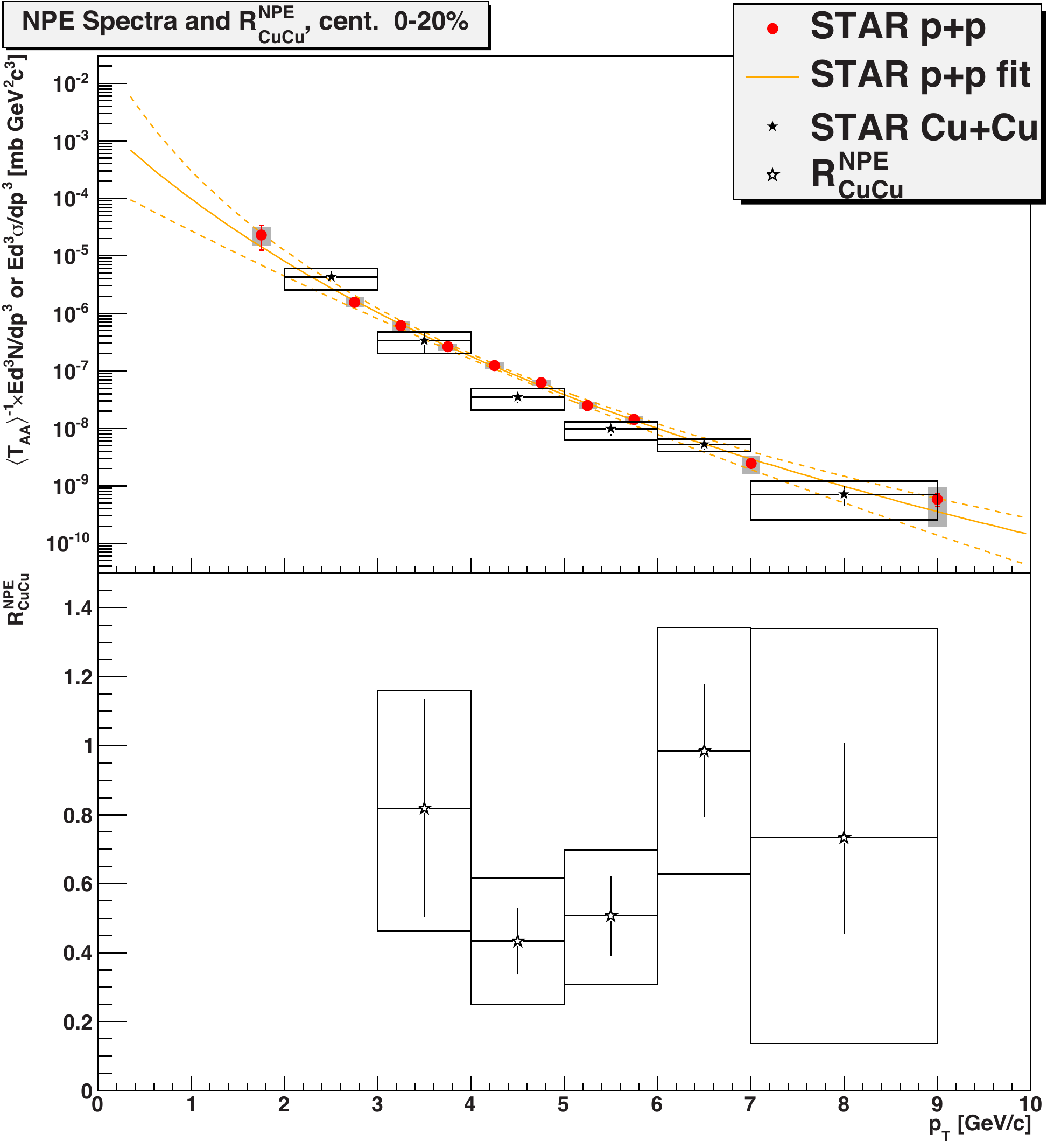}
\caption[Nuclear modification factor for non-photonic $e^{\pm}$ in the 0-20\% centrality class of 200-GeV Cu + Cu collisions, computed using a STAR measurement of non-photonic $e^{\pm}$ in $p+p$ collisions.]{Illustration of the calculation of $R_{\mathrm{CuCu}}^{NPE}$, the nuclear modification factor for non-photonic $e^{\pm}$ in the 0-20\% centrality class of 200-GeV Cu + Cu collisions.  The upper panel shows the non-photonic $e^{\pm}$ yield $(Y_{D,B})$ found in this analysis as a function of $p_{T}$ along with the STAR collaboration's measurement\protect\cite{STAR_ppNPE2011} of the same quantity in $p+p$ collisions (and a fit to those data).  The lower panel shows the nuclear modification factor, the scaled ratio of the Cu + Cu measurement to the $p+p$ measurement, also as a function of $p_{T}$.}
\label{fig:results:raa_STAR_c01}
\end{center}
\end{figure}

\begin{figure}[htbp]
\begin{center}
\includegraphics[width=0.85\linewidth]{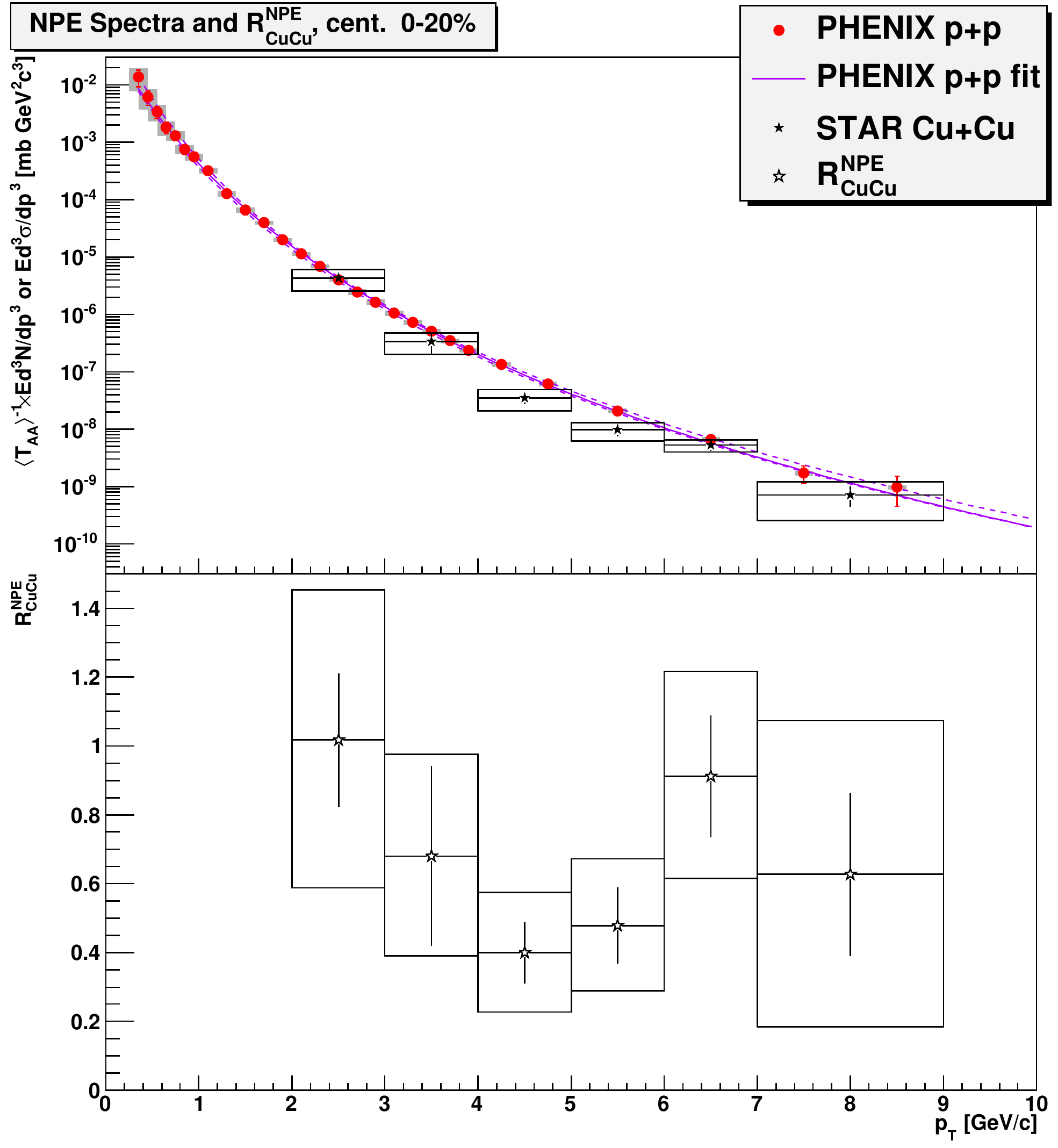}
\caption[Nuclear modification factor for non-photonic $e^{\pm}$ in the 0-20\% centrality class of 200-GeV Cu + Cu collisions, computed using a PHENIX measurement of non-photonic $e^{\pm}$ in $p+p$ collisions.]{Illustration of the calculation of $R_{\mathrm{CuCu}}^{NPE}$, the nuclear modification factor for non-photonic $e^{\pm}$ in the 0-20\% centrality class of 200-GeV Cu + Cu collisions (see also Figure~\ref{fig:results:raa_STAR_c01}).  The upper panel shows the non-photonic $e^{\pm}$ yield $(Y_{D,B})$ found in this analysis as a function of $p_{T}$ along with the PHENIX collaboration's measurement\protect\cite{PhysRevLett.97.252002} of the same quantity in $p+p$ collisions (and a fit to those data).  The lower panel shows the nuclear modification factor, the scaled ratio of the Cu + Cu measurement to the $p+p$ fit, also as a function of $p_{T}$.}
\label{fig:results:raa_PHENIX_c01}
\end{center}
\end{figure}

Equation~\ref{eq:results:define_raa} is used to calculate $R_{\mathrm{CuCu}}^{NPE}$, the nuclear modification factor for non-photonic $e^{\pm}$ in 200-GeV Cu + Cu collisions (in comparison to 200-GeV $p+p$ collisions).  The STAR~\cite{STAR_ppNPE2011} and PHENIX~\cite{PhysRevLett.97.252002} collaborations have measured the cross-sections of non-photonic $e^{\pm}$ in 200-GeV $p+p$ collisions.  In both of those measurements, the contributions from $J/\psi$ and $\Upsilon$ decays have been removed; those spectra are therefore measurements in $p+p$ collisions of the quantity called $Y_{D,B}$ in this dissertation.  The upper panel of Figure~\ref{fig:results:raa_STAR_c01} shows the STAR measurement of $Y_{D,B}$ in $p+p$ and the 0-20\% most-central Cu + Cu collisions; the Cu + Cu data have been scaled by $1/\langle T_{AA}\rangle =\sigma_{inel.}/\langle N_{bin}\rangle$.  The lower panel of Figure~\ref{fig:results:raa_STAR_c01} shows the nuclear modification factor $R_{\mathrm{CuCu}}^{NPE}$ as a function of transverse momentum.  The $p+p$ spectrum is fit with a function of the form $A(1+p_{T}/B)^{n}$.  The parameters of this fit function, called $F[e_{N},2]$, are given in Section~\ref{sec:functions:npe}.  The lower (upper) systematic uncertainty in the fit is estimated by moving all $p+p$ data points down (up) by the quadrature sum of their statistical and systematic uncertainties, then fitting the resulting spectrum with another function of the same form.  The statistical uncertainties of $R_{\mathrm{CuCu}}^{NPE}$ are due to the statistical uncertainties of the Cu + Cu spectrum only; the systematic uncertainties of the Cu + Cu spectrum and the $p+p$ fit are assumed to be uncorrelated.  Figure~\ref{fig:results:raa_PHENIX_c01} is similar to Figure~\ref{fig:results:raa_STAR_c01}, but the PHENIX $p+p$ measurement is used in place of the STAR $p+p$ measurement.  The nuclear modification factor $R_{\mathrm{CuCu}}^{NPE}$ is calculated by dividing the scaled Cu + Cu spectrum by a fit (called $F[e_{N},1]$, of the same functional form as $F[e_{N},2]$) to the PHENIX $p+p$ measurement of $Y_{D,B}$.

\clearpage

\section{Discussion}
\label{sec:results:discussion}

\begin{figure}[htbp]
\begin{center}
\includegraphics[width=0.85\linewidth]{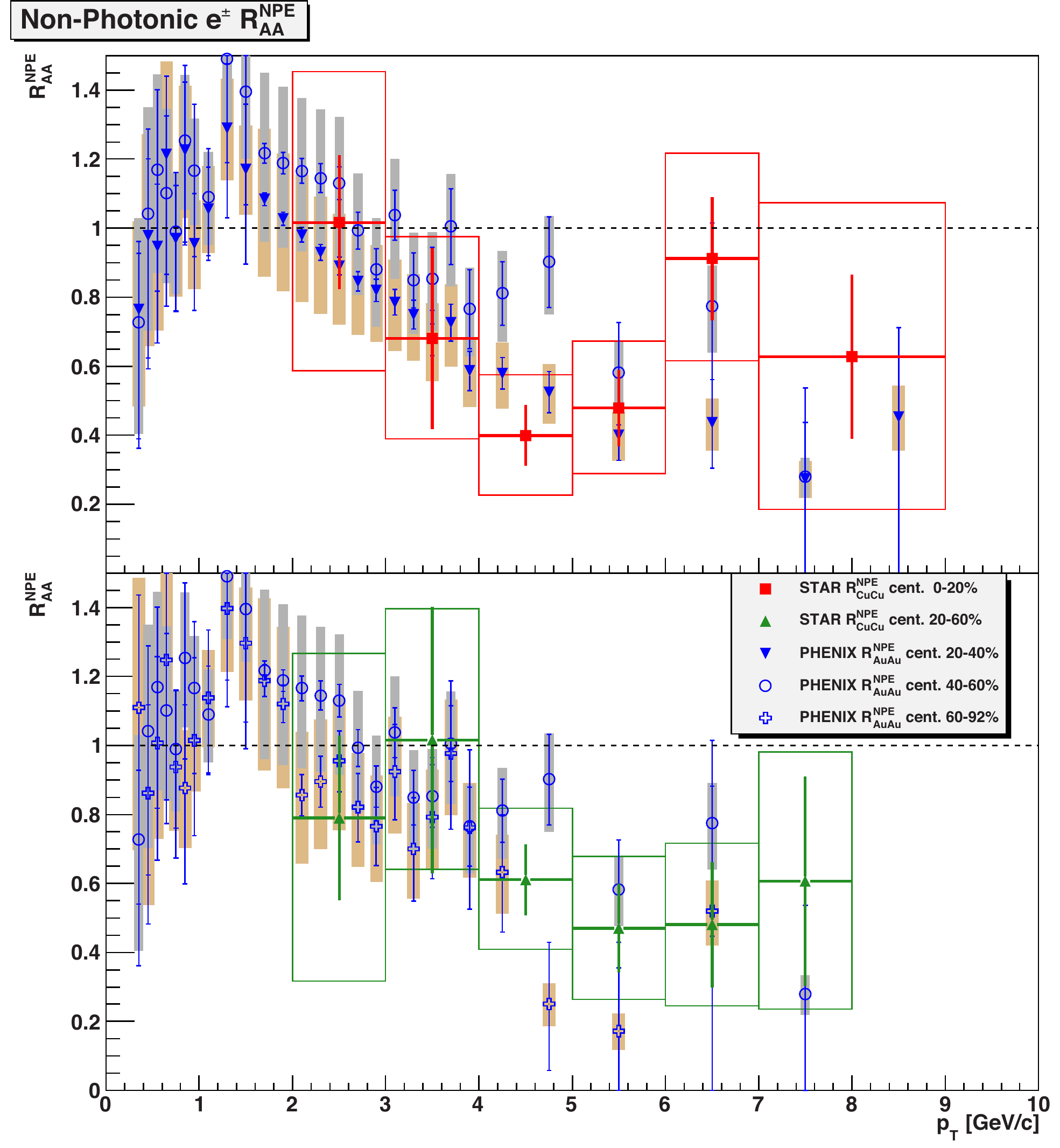}
\caption[Nuclear modification factors of non-photonic $e^{\pm}$ as functions of $p_{T}$ in collision systems with similar values of $\langle N_{part}\rangle$.]{Nuclear modification factors of non-photonic $e^{\pm}$ as functions of $p_{T}$ in Cu + Cu and Au + Au collisions for centrality classes with similar values of $\langle N_{part}\rangle$.  In the upper panel, $R_{\mathrm{CuCu}}^{NPE}$ for the 0-20\% centrality class is compared to the PHENIX collaboration's measurements\protect\cite{PhysRevLett.98.172301} of $R_{\mathrm{AuAu}}^{NPE}$ for the 20-40\% and 40-60\% centrality classes.  In the lower panel, $R_{\mathrm{CuCu}}^{NPE}$ for the 20-60\% centrality class is compared to the PHENIX collaboration's measurements of $R_{\mathrm{AuAu}}^{NPE}$ for the 40-60\% and 60-92\% centrality classes.  The gray shaded boxes represent the systematic uncertainties of $R_{\mathrm{AuAu}}^{NPE}$ for the 40-60\% centrality class, while the brown shaded boxes represent the systematic uncertainties of $R_{\mathrm{AuAu}}^{NPE}$ for the other two centrality classes.}
\label{fig:results:npe_raa_pt}
\end{center}
\end{figure}

In this section, $R_{\mathrm{CuCu}}^{NPE}$, the nuclear modification factor for non-photonic $e^{\pm}$ in 200-GeV Cu + Cu collisions, will be compared with related measurements of the nuclear modification factor.  Figure~\ref{fig:results:npe_raa_pt} shows the nuclear modification factors for non-photonic $e^{\pm}$ in 200-GeV Cu + Cu and Au + Au collisions~\cite{PhysRevLett.98.172301} with similar values of $\langle N_{part}\rangle$.  The upper panel shows $R_{\mathrm{CuCu},0-20\%}^{NPE}$, the nuclear modification factor for non-photonic $e^{\pm}$ in the 0-20\% most central Cu + Cu collisions (for which $\langle N_{part}\rangle=86$); the lower panel shows $R_{\mathrm{CuCu},20-60\%}^{NPE}$, the nuclear modification factor for non-photonic $e^{\pm}$ in the 20-60\% centrality class of Cu + Cu collisions ($\langle N_{part}\rangle=33$).\footnote{$\langle N_{part}\rangle=51$ for the 0-60\% most central Cu + Cu collisions.}  Measurements of the nuclear modification factor for non-photonic $e^{\pm}$ in three centrality classes of Au + Au collisions are also shown: $R_{\mathrm{AuAu},20-40\%}^{NPE}$ ($\langle N_{part}\rangle=140$) is shown in the upper panel, $R_{\mathrm{AuAu},40-60\%}^{NPE}$ ($\langle N_{part}\rangle=60$) is shown in both panels, and $R_{\mathrm{AuAu},60-92\%}^{NPE}$ ($\langle N_{part}\rangle=14$) is shown in the lower panel.  In the upper panel, the values of $R_{\mathrm{AuAu},20-40\%}^{NPE}$ tend to be lower than the values of $R_{\mathrm{AuAu},40-60\%}^{NPE}$: non-photonic $e^{\pm}$ appear to be suppressed more when $\langle N_{part}\rangle$ (and therefore the volume of the QGP) is greater.  This behavior is typical and is observed in several collision systems for various particle species.  However, the measured values of $R_{\mathrm{AuAu},60-92\%}^{NPE}$ deviate from this trend: the values of $R_{\mathrm{AuAu},60-92\%}^{NPE}$ appear to be below the measured values of $R_{\mathrm{AuAu},40-60\%}^{NPE}$.  For both Cu + Cu centrality classes, the measured values of $R_{\mathrm{CuCu}}^{NPE}$ are generally consistent with the measured values of $R_{\mathrm{AuAu}}^{NPE}$ for collision systems with similar values of $\langle N_{part}\rangle$.

\begin{figure}[htbp]
\begin{center}
\includegraphics[width=0.85\linewidth]{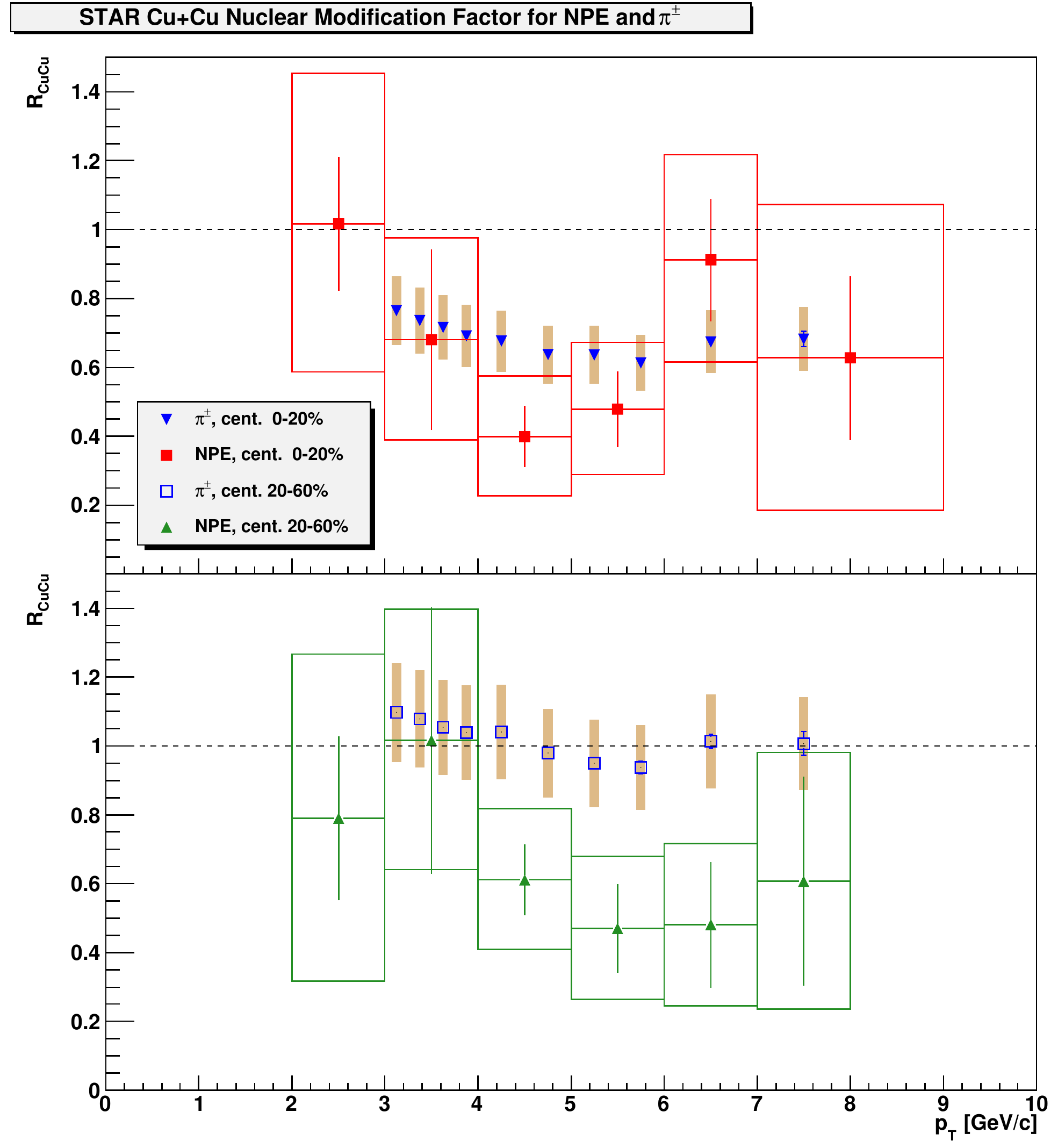}
\caption[Nuclear modification factors of non-photonic $e^{\pm}$ and $\pi^{\pm}$ in Cu + Cu collisions.]{Nuclear modification factors of non-photonic $e^{\pm}$ and $\pi^{\pm}$\protect\cite{PhysRevC.81.054907} in Cu + Cu collisions as functions of $p_{T}$.}
\label{fig:results:raa_npe_pipm}
\end{center}
\end{figure}

Measurements of the nuclear modification factor in Au + Au collisions indicate that, at high $p_{T}$, the suppression of non-photonic $e^{\pm}$~\cite{PhysRevLett.96.032301,PhysRevLett.97.252002,PhysRevLett.98.172301,PHENIX_NPE2010} is similar in magnitude to the suppression of pions~\cite{PhysRevLett.91.072301,PhysRevLett.97.152301}.\footnote{See also the discussion of Figure~\ref{fig:results:raa_npart} below.}  Figure~\ref{fig:results:raa_npe_pipm} shows measurements of $R_{\mathrm{CuCu}}$ for non-photonic $e^{\pm}$ and charged pions~\cite{PhysRevC.81.054907} in two centrality classes.\footnote{Note that $R_{\mathrm{CuCu}}^{\pi}$ increases as collisions become more peripheral.  This trend is expected, and differs from the behavior of $R_{\mathrm{AuAu}}^{NPE}$ seen in the lower panel of Figure~\ref{fig:results:npe_raa_pt}.}  The upper panel shows $R_{\mathrm{CuCu},0-20\%}^{NPE}$ plotted with $R_{\mathrm{CuCu},0-20\%}^{\pi}$, while the lower panel shows $R_{\mathrm{CuCu},20-60\%}^{NPE}$ plotted with $R_{\mathrm{CuCu},20-60\%}^{\pi}$.  At high transverse momenta, the measured values of $R_{\mathrm{CuCu},0-20\%}^{NPE}$ and $R_{\mathrm{CuCu},0-20\%}^{\pi}$ are consistent with each other.  However, for the 20-60\% centrality class, non-photonic $e^{\pm}$ appear to be more suppressed at high $p_{T}$ than pions.

\begin{figure}[htbp]
\begin{center}
\includegraphics[width=0.85\linewidth]{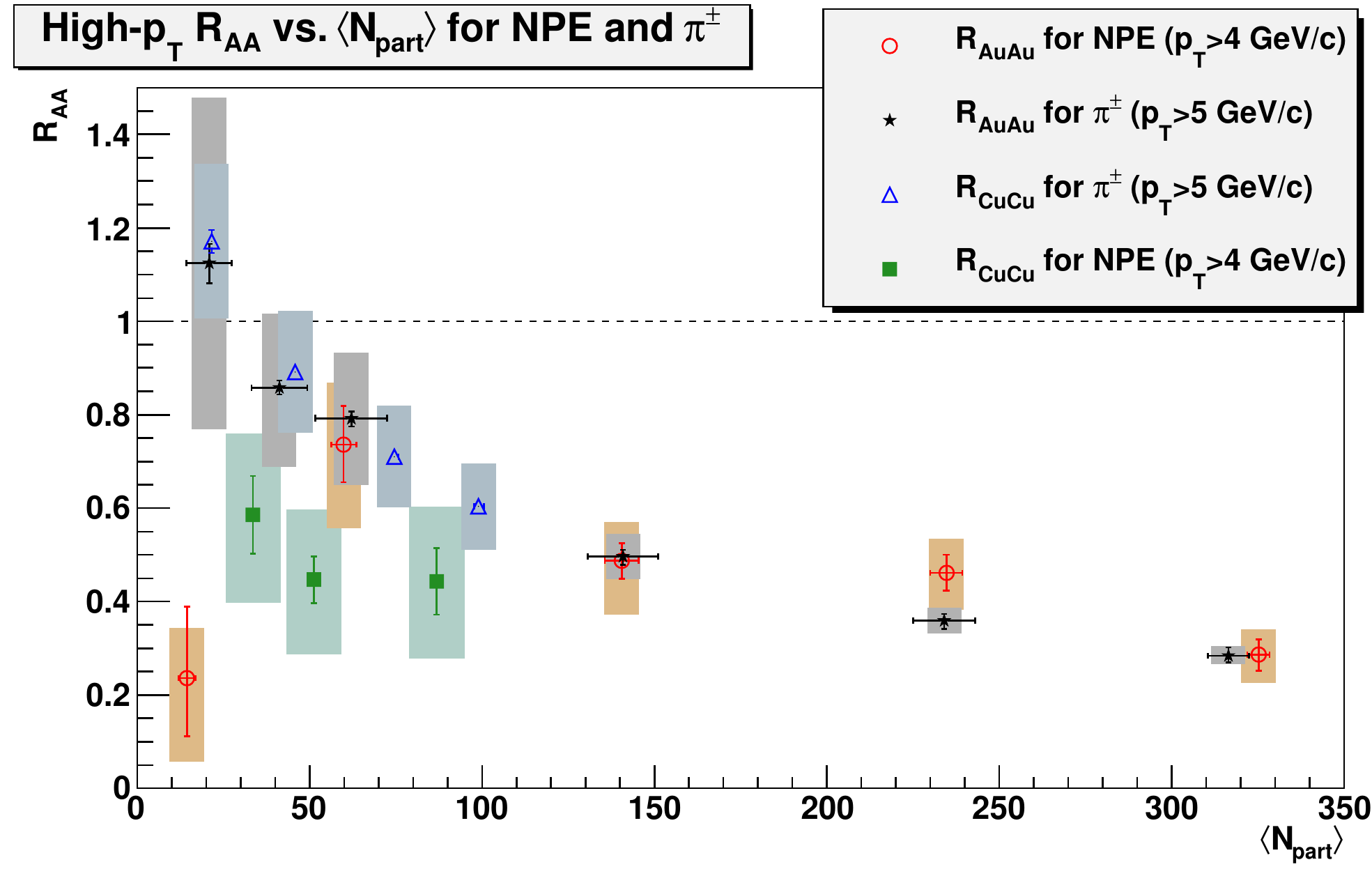}
\caption[High-$p_{T}$ nuclear modification factor of non-photonic $e^{\pm}$ and $\pi^{\pm}$ as a function of $\langle N_{part}\rangle$.]{High-$p_{T}$ nuclear modification factor of non-photonic $e^{\pm}$ in Cu + Cu collisions (green) and Au + Au collisions (red)\protect\cite{PhysRevLett.98.172301} as a function of $\langle N_{part}\rangle$ compared with the nuclear modification factors of $\pi^{\pm}$ in Cu + Cu collisions (blue)\protect\cite{PhysRevC.81.054907} and Au + Au collisions (black)\protect\cite{PhysRevLett.97.152301}.}
\label{fig:results:raa_npart}
\end{center}
\end{figure}

Figure~\ref{fig:results:raa_npart} shows measurements (as functions of $\langle N_{part}\rangle$) of the nuclear modification factor at high transverse momenta for non-photonic $e^{\pm}$ and $\pi^{\pm}$ in 200-GeV Cu + Cu and Au + Au collisions.  These $R_{AA}$ measurements are the scaled ratios of the integrated particle yields at high $p_{T}$.  The measured values of $R_{\mathrm{AuAu}}^{NPE}$ are consistent with the values of $R_{\mathrm{AuAu}}^{\pi}$, except for the most peripheral centrality class (with the lowest value of $\langle N_{part}\rangle$), where non-photonic $e^{\pm}$ may be more suppressed than pions.  In Cu + Cu collisions, it appears that non-photonic $e^{\pm}$ may be more suppressed than pions.  This is particularly visible for the peripheral 20-60\% and combined 0-60\% centrality classes ($\langle N_{part}\rangle=33$ and 51, respectively).  For the 0-20\% centrality class, the measured value of $R_{\mathrm{CuCu}}^{NPE}$ is less than $R_{\mathrm{CuCu}}^{\pi}$, though the measurements are consistent with uncertainties.

A reason for measuring $R_{\mathrm{CuCu}}^{NPE}$ was to determine if the shape of the QGP, not just its size, plays a role in the suppression of non-photonic $e^{\pm}$ and heavy quarks.  A Cu + Cu collision in the 0-20\% centrality class and a Au + Au collision in the 40-60\% centrality class have similar values of $\langle N_{part}\rangle$ and should therefore produce similar volumes of quark-gluon plasma.  A Cu + Cu collision in the 0-20\% centrality class should have a roughly circular cross section (in the $xy$-plane), while a Au + Au collision in the 40-60\% centrality class would have a more elliptical cross-section.  On average, a heavy quark produced near the center of the medium would have a greater path length through a quark-gluon plasma produced in the Cu + Cu collision than in the Au + Au collision, which might be expected\footnote{assuming that the two collision systems have the same energy density} to result in more heavy-quark energy loss for Cu + Cu collisions, and more suppression of high-$p_{T}$ non-photonic $e^{\pm}$.  As shown in Figure~\ref{fig:results:raa_npart}, the measurements of the nuclear modification factor for non-photonic $e^{\pm}$ in Cu + Cu collisions are consistent within uncertainties with measurements of the same quantity in Au + Au collisions at similar values of $\langle N_{part}\rangle$.  It appears that the central value of $R_{\mathrm{CuCu,0-20\%}}^{NPE}$ (at $\langle N_{part}\rangle=86$) may lie somewhat below the expected value based on the measurements in Au + Au (if $R_{\mathrm{AuAu}}^{NPE}$ varied linearly with $\langle N_{part}\rangle$ between $\langle N_{part}\rangle=60$ and $\langle N_{part}\rangle=140$).  However, the large uncertainties on the $R_{AA}^{NPE}$ measurements prevent a strong statement from being made regarding this issue.

\begin{figure}[htbp]
\begin{center}
\includegraphics[width=0.85\linewidth]{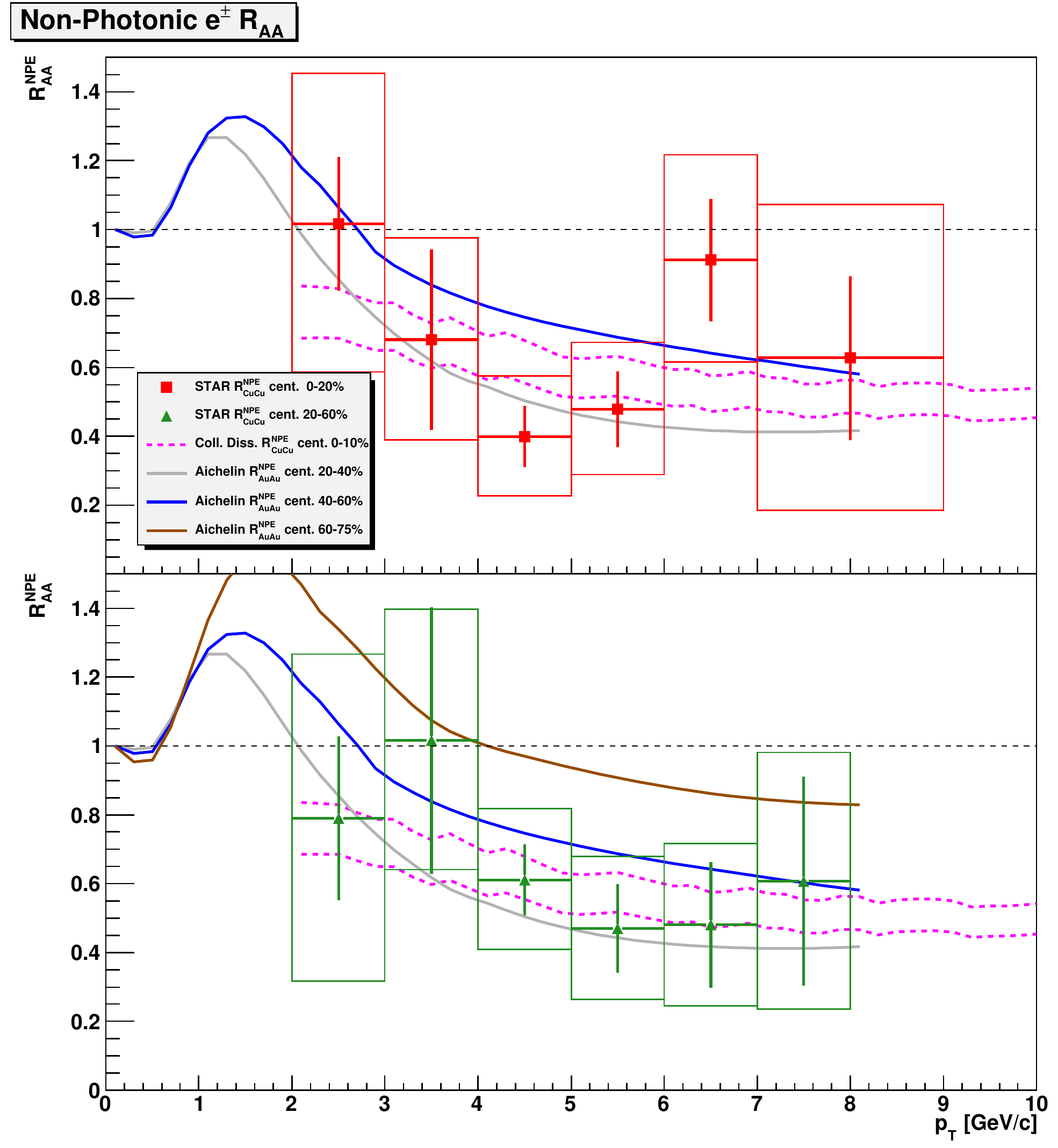}
\caption[The measured nuclear modification factor for non-photonic $e^{\pm}$ in Cu + Cu collisions compared with theoretical predictions.]{The measured nuclear modification factor for non-photonic $e^{\pm}$ in Cu + Cu collisions compared with theoretical predictions.  The range of $R_{\mathrm{CuCu},0-10\%}^{NPE}$ predicted by the Collisional Dissociation model\protect\cite{Adil2007139,PhysRevC.80.054902} is shown as the dashed curves.  Predictions of $R_{\mathrm{AuAu}}^{NPE}$ by the Gossiaux/Aichelin model\protect\cite{PhysRevC.78.014904,PhysRevC.79.044906} for three Au + Au centrality classes are shown as solid curves.}
\label{fig:results:raa_theory}
\end{center}
\end{figure}

Figure~\ref{fig:results:raa_theory} shows the measured values of $R_{\mathrm{CuCu}}^{NPE}$ compared to predictions by two theoretical models\footnote{These models are among those discussed in Section~\ref{sec:theory:heavy_interactions}.} for the nuclear modification factor of non-photonic $e^{\pm}$.  The measured values of $R_{\mathrm{CuCu,0-20\%}}^{NPE}$ are consistent with the prediction of the Collisional Dissociation model~\cite{Adil2007139,PhysRevC.80.054902} for $R_{\mathrm{CuCu,0-10\%}}^{NPE}$.  The measured values of $R_{\mathrm{CuCu}}^{NPE}$ for the 20-60\% centrality class are generally consistent with the prediction of the Collisional Dissociation model for the 0-10\% most central Cu + Cu collisions.  The values of $R_{\mathrm{AuAu}}^{NPE}$ measured~\cite{PHENIX_NPE2010} by the PHENIX collaboration are well described for all but  the most peripheral centrality class by the predictions of the Gossiaux/Aichelin model~\cite{PhysRevC.78.014904,PhysRevC.79.044906}.\footnote{This model predicts that the values of $R_{\mathrm{AuAu}}^{NPE}$ will increase as collisions become more peripheral.  This trend is supported by the PHENIX measurements, except for the measurement in the most peripheral (60-92\%) centrality class.}  The measured values of $R_{\mathrm{CuCu},0-20\%}^{NPE}$ (upper panel) are closer to the Gossiaux/Aichelin prediction of $R_{\mathrm{AuAu},20-40\%}^{NPE}$ than $R_{\mathrm{AuAu},40-60\%}^{NPE}$, even though the value of $\langle N_{part}\rangle$ for the 20\% most central Cu + Cu collisions is closer to the value of $\langle N_{part}\rangle$ for the 40-60\% centrality class of Au + Au collisions.  The measured values of $R_{\mathrm{CuCu},20-60\%}^{NPE}$ (lower panel) are consistent with the Gossiaux/Aichelin prediction for $R_{\mathrm{AuAu}}^{NPE},20-40\%$ despite the factor of $\approx 4$ difference in $\langle N_{part}\rangle$ between the two centrality classes.

\clearpage

In summary, this dissertation has presented the first measurement of the yields of non-photonic $e^{\pm}$ in 200-GeV Cu + Cu collisions.  The limited number of events analyzed, the large photonic $e^{\pm}$ background, and limited knowledge of the high-$p_{T}$ $J/\psi$ spectrum in Cu + Cu collisions contribute to the large uncertainties in this measurement.  Non-photonic $e^{\pm}$ are suppressed at high transverse momentum in Cu + Cu collisions relative to $p+p$ collisions.  For Cu + Cu collisions, non-photonic $e^{\pm}$ appear to be more suppressed at high $p_{T}$ than charged pions, with the difference in suppression becoming larger for more peripheral collisions.  In contrast, measurements in Au + Au collisions indicate that the suppression of non-photonic $e^{\pm}$ is similar in magnitude to the suppression of pions, although non-photonic $e^{\pm}$ may be more suppressed than pions in collisions with low values of $\langle N_{part}\rangle$.  Although the measured values of $R_{\mathrm{CuCu}}^{NPE}$ have large uncertainties, they are generally consistent with measurements of $R_{\mathrm{AuAu}}^{NPE}$ for collision systems with similar values of $\langle N_{part}\rangle$.  The suppression of non-photonic $e^{\pm}$ measured in this dissertation is generally greater than the predictions of the Gossiaux/Aichelin model for similar values of $\langle N_{part}\rangle$.  The Collisional Dissociation model prediction is consistent with the observed suppression of non-photonic $e^{\pm}$ for central Cu + Cu collisions.  However, that model is also consistent with the measured suppression of non-photonic $e^{\pm}$ for the 20-60\% centrality class, despite a factor of $\approx 3$ difference in $\langle N_{part}\rangle$.

\section{Looking Forward}
\label{sec:results:outlook}

The production of heavy quarks and their interactions with the quark-gluon plasma are active subjects of study at both RHIC and the Large Hadron Collider (LHC).  It is intended that the non-photonic $e^{\pm}$ spectra found in this dissertation will be combined with the $D^{0}$ yield measured in~\cite{BaumgartThesis} to provide an updated measurement of the total charm cross-section in 200-GeV Cu + Cu collisions.  The PHENIX collaboration has submitted for publication new measurements~\cite{PHENIX_NPE2010} of $R_{\mathrm{AuAu}}^{NPE}$ and the elliptic-flow parameter $v_{2}$ for non-photonic $e^{\pm}$.  The STAR collaboration will soon submit a new measurement~\cite{STAR_ppNPE2011} of the non-photonic $e^{\pm}$ yield in $p+p$ collisions; a new measurement of the non-photonic $e^{\pm}$ yield in Au + Au collisions is in progress.  Efforts are also underway in the STAR collaboration to measure the flow of heavy quarks by measuring $v_{2}$ for $D^{0}$ in Au + Au collisions.  Measurements of the $J/\psi$ yield in $p+p$ and $d$ + Au collisions and of the $\Upsilon$ yield in $d$ + Au collisions are also in progress.

The LHC has recently begun operations, opening a new energy regime of heavy-ion collisions for study.  Pb + Pb collisions at $\sqrt{s_{NN}}=2.76$ TeV have been recorded~\cite{PhysRevLett.105.252301}, along with $p+p$ collisions at energies including 900 GeV, 2.36 TeV, and 7 TeV~\cite{springerlink:10.1140/epjc/s10052-009-1227-4,springerlink:10.1140/epjc/s10052-010-1339-x,springerlink:10.1140/epjc/s10052-010-1350-2}.  The total charm (bottom) cross-section is predicted~\cite{VogtHeavyCrossSections2009} to increase by a factor of $\approx 10$ $(\approx 100)$ in collisions at $\sqrt{s_{NN}}=5.5$ TeV (the maximum energy for Pb + Pb collisions) in comparison to collisions at $\sqrt{s_{NN}}=200\GeV$, allowing heavy flavor to be measured at higher momenta and with increased statistical significance relative to RHIC.  It may be possible to study the path-length dependence of heavy-flavor suppression by measuring heavy-flavor yields in the reaction plane and perpendicular to it (\textit{cf.} the measurement of jet quenching in and out of the reaction plane shown in Figure~\ref{fig:intro:jet_quenching}).  The ATLAS collaboration~\cite{ATLAS_overview} has measured~\cite{ATLAS_Jpsi} the suppression of $J/\psi$ in Pb + Pb collisions.  The ALICE experiment~\cite{ALICE_overview} is beginning a broad program of heavy-flavor measurements.  The yields of $D^{0}(\bar{D}^{0})$ and $D^{\pm}$ in heavy-ion collisions have been measured~\cite{Grelli_WWND2011} by reconstructing their hadronic decays.  Simulations~\cite{Harris_INTWorkshop2010} indicate that for ten million 5.5-TeV Pb + Pb collisions in the 0-5\% centrality class, the yield of $D^{0}$ can be measured at momenta as high as $20\GeV/c$ with a relative statistical uncertainty of $<20\%$. The yields of heavy quarkonia may be measured by reconstructing their decays to $e^{-}e^{+}$ pairs at mid-rapidity or $\mu^{-}\mu^{+}$ pairs at forward rapidity $(2.5<y<4)$.  Measurements of single muons and non-photonic $e^{\pm}$ are also underway.  In the ALICE experiment, $e^{\pm}$ are identified through measurements of energy loss in the Time Projection Chamber, particle speed in the Time-of-Flight detector, the radiation yield in the Transition Radiation Detector, and/or energy deposition in the Electromagnetic Calorimeter.  It is expected~\cite{Harris_INTWorkshop2010} that for ten million 5.5-TeV Pb + Pb collisions in the 0-5\% centrality class, the yield of non-photonic $e^{\pm}$ from $B$-meson decays may be measured at momenta as high as $20\GeV/c$ with a relative statistical uncertainty of $<10\%$.

\clearpage

\appendix

\chapter{Conversion of $\boldsymbol{\langle dE/dx\rangle}$ to $\boldsymbol{n\sigma_{e}}$}
\label{sec:dedx2nsigma}

This appendix describes the method used to estimate the most probable value of the normalized TPC energy loss $(n\sigma_{e})$ for a given value of $\langle dE/dx\rangle$, the 70\% truncated mean TPC energy loss.  See Section~\ref{sec:experiment:tpc:dedx} and Chapter~\ref{sec:dedx} for more information on $\langle dE/dx\rangle$ and $n\sigma_{e}$.  The method described in this appendix is used to generate the curves shown in several figures, including Figure~\ref{fig:anintro:nsigma_p_e_mb} (page~\pageref{fig:anintro:nsigma_p_e_mb}).

\begin{figure}[htbp]
\begin{center}
\includegraphics[width=0.8\linewidth]{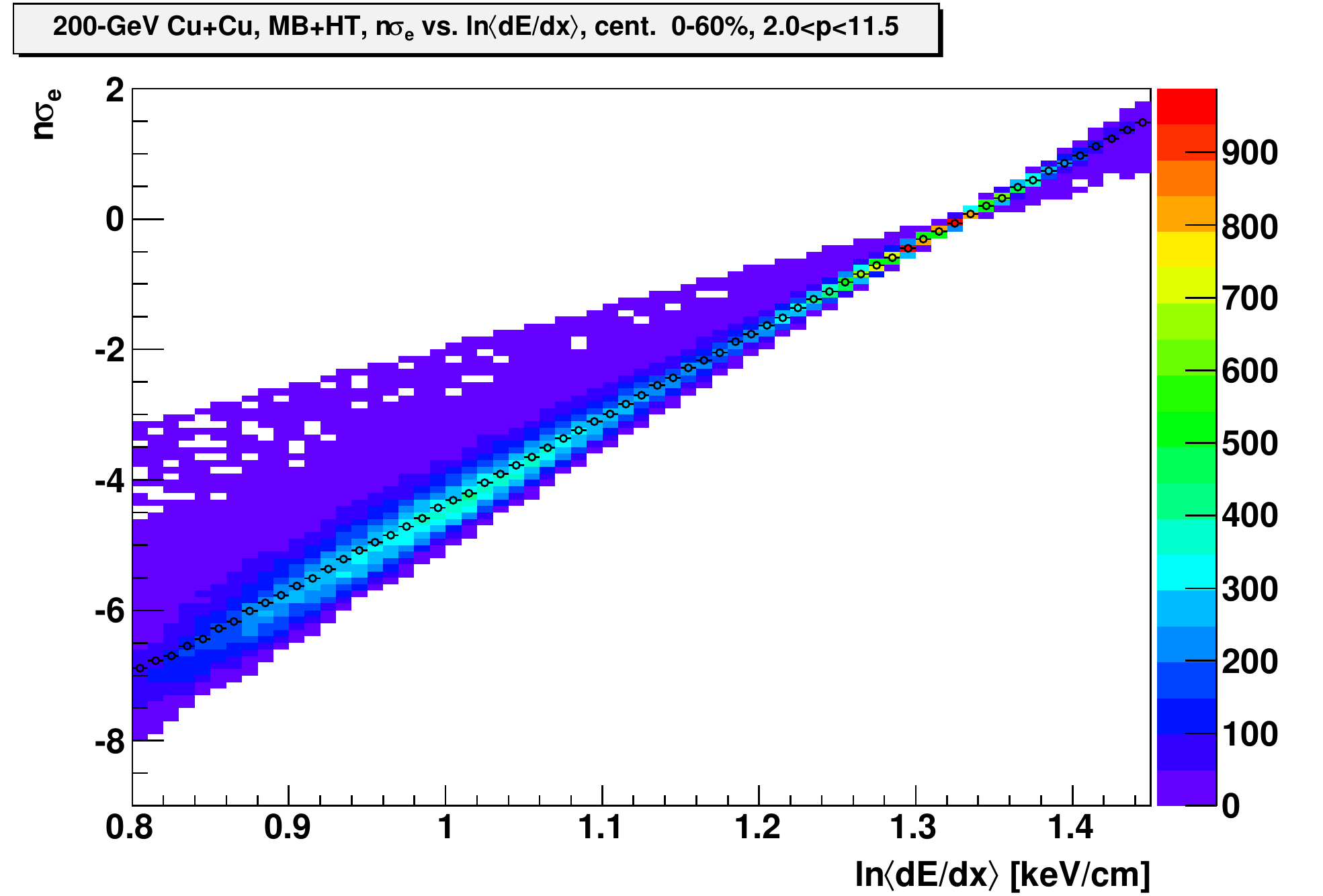}
\caption[The normalized TPC energy loss $n\sigma_{e}$ vs. $\ln\langle dE/dx\rangle$.]{The normalized TPC energy loss $n\sigma_{e}$ vs. $\ln\langle dE/dx\rangle$ for both minimum-bias and high-tower triggered events.  The black circles indicate an estimate of the most probable value of $n\sigma_{e}$ (see the text for details).}
\label{fig:dedx2nsigma:dedx_nsigma}
\includegraphics[width=0.8\linewidth]{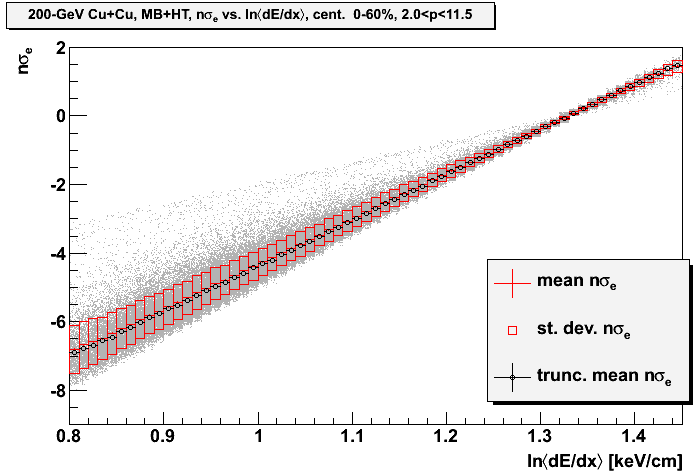}
\caption[The mean and most-probable values of $n\sigma_{e}$ as a function of $\ln\langle dE/dx\rangle$.]{The mean (red, central values) value of $n\sigma_{e}$ as a function of $\ln\langle dE/dx\rangle$, as well as the standard deviation (red boxes).  Also shown (in black) is the estimated most probable (truncated mean) value of $n\sigma_{e}$.}
\label{fig:dedx2nsigma:dedx_nsigma_mean}
\end{center}
\end{figure}

Figure~\ref{fig:dedx2nsigma:dedx_nsigma} shows the relationship between $\ln\langle dE/dx\rangle$ and $n\sigma_{e}$ for particles that pass all other $e^{\pm}$ identification cuts.  The relationship is not one-to-one because $B_{e}$ and $\sigma_{e}$ depend upon the length of the track in the TPC.  When fitting the inclusive $e^{\pm}$ distributions with multiple Gaussians, it can be useful to know the expected locations of the means of the hadron peaks.  Bichsel's parameterization~\cite{Bichsel2006154} can be used to generate an expected value of $\langle dE/dx\rangle$ for hadron tracks with typical characteristics; this section presents a method to convert those values to an approximate most probable value of $n\sigma_{e}$.

The black data points shown in Figure~\ref{fig:dedx2nsigma:dedx_nsigma} give an approximate most probable value of $n\sigma_{e}$ for each bin in $\ln\langle dE/dx\rangle$.  This value is found as follows.  For each bin in $\ln\langle dE/dx\rangle$, the mean value of $n\sigma_{e}$ and standard deviation are calculated.  Then, the mean value of $n\sigma_{e}$ is recalculated including only those values of $n\sigma_{e}$ within one standard deviation of the original mean.  This truncated mean (not related to the 70\% truncated mean used to calculate $\langle dE/dx\rangle$) is shown in Figure~\ref{fig:dedx2nsigma:dedx_nsigma}.  The truncated mean seems to be better than the mean at reproducing the maximum of the $n\sigma_{e}$ distribution in each $\ln\langle dE/dx\rangle$ bin.  Figure~\ref{fig:dedx2nsigma:dedx_nsigma_mean} shows the mean $n\sigma_{e}$, the standard deviation, and the truncated mean $n\sigma_{e}$ as a function of $\ln\langle dE/dx\rangle$.

The truncated mean $n\sigma_{e}$ as a function of $\ln\langle dE/dx\rangle$ can be fit with a linear function of the form

\begin{equation}
n\sigma_{e}=\left[\ln\langle dE/dx\rangle-\ln B_{e0}\right]\frac{1}{\sigma_{e0}}.
\end{equation}

\begin{figure}[htbp]
\begin{center}
\includegraphics[width=0.85\linewidth]{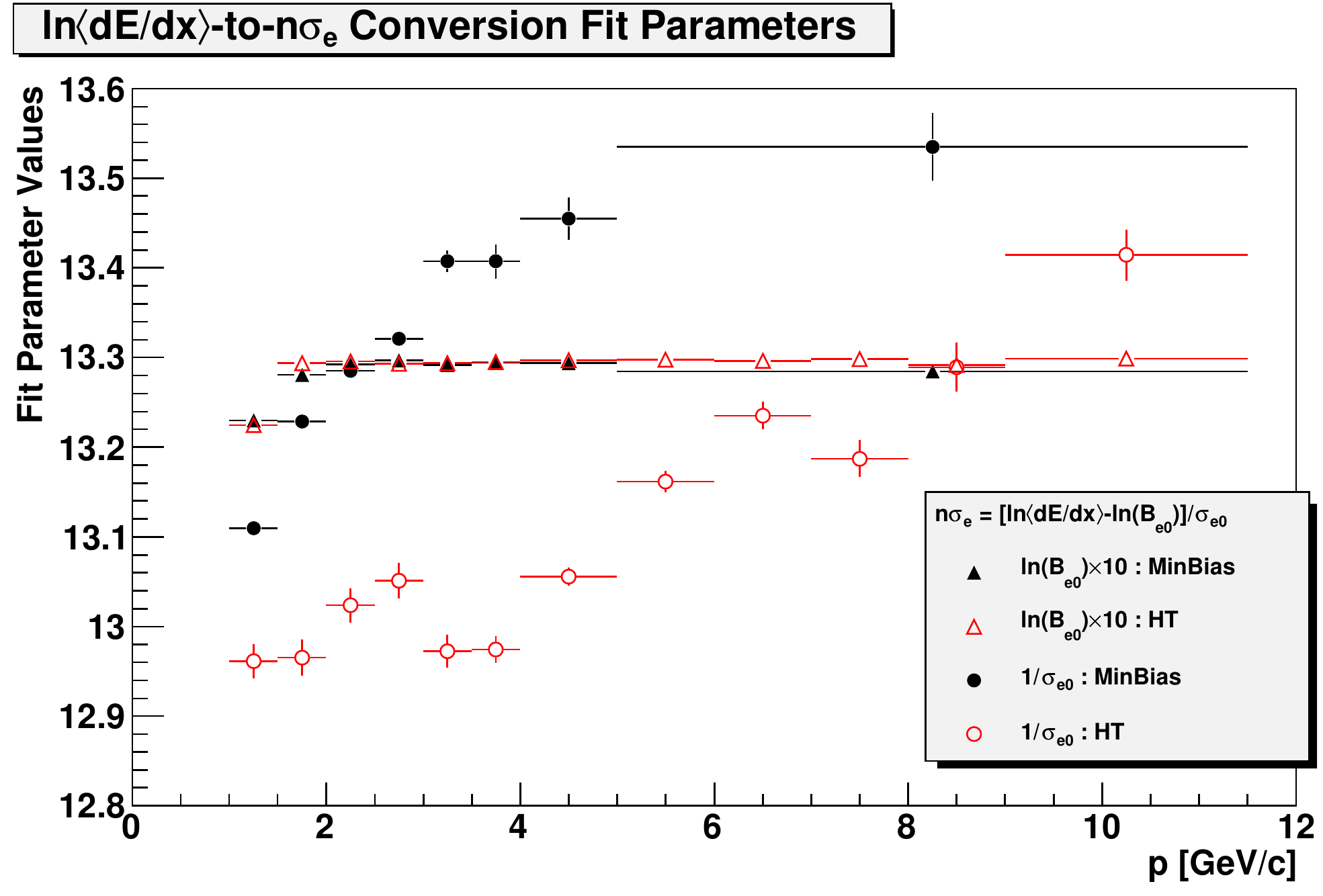}
\caption[Parameters of linear $n\sigma_{e}$ vs. $\ln\langle dE/dx\rangle$ fits.]{The values of the intercept $(\ln B_{e0})$ and slope $(1/\sigma_{e0})$ of linear fits to the most probable value of $n\sigma_{e}$ vs. $\ln\langle dE/dx\rangle$.  These values are used to convert a value of $\langle dE/dx\rangle$ given by Bichsel's parametrization\protect\cite{Bichsel2006154} to an approximate most probable value of $n\sigma_{e}$.}
\label{fig:dedx2nsigma:dedx_nsigma_fit_pars}
\end{center}
\end{figure}

\noindent The values of the fit parameters $\ln B_{e0}$ and $1/\sigma_{e0}$ are shown in Figure~\ref{fig:dedx2nsigma:dedx_nsigma_fit_pars}.  The intercept $\ln B_{e0}$ does not appear to depend on $p$ for the momentum range of interest for this analysis; the slope $1/\sigma_{e0}$ appears to increase slightly with increasing momentum.  These fits are used to calculate an approximate most probable value of $n\sigma_{e}$ for a given input value of $\ln\langle dE/dx\rangle$ obtained from Bichsel's parameterization.  The curves shown in Figures~\ref{fig:dedx:nsigma_p_ht},~\ref{fig:dedx:inc_fit_pars_mb}, and~\ref{fig:dedx:inc_fit_pars_ht} are generated using a single linear conversion function calculated for one large momentum bin.

\clearpage

\clearpage

\chapter{Treatment of Uncertainties}
\label{sec:uncertainties}

This Appendix describes the calculation of uncertainties for three cases.  Section~\ref{sec:uncertainties:weff} describes the estimation of the statistical uncertainties for weighted efficiency calculations.  In Section~\ref{sec:uncertainties:spectra}, the expression for the statistical uncertainty of the efficiency-corrected non-photonic $e^{\pm}$ yield (Equation~\ref{eq:results:stat_err_nec}) is derived.  Section~\ref{sec:uncertainties:spectra_sys} describes the calculation of the systematic uncertainties in the efficiency-corrected $e^{\pm}$ yields.

\section[Statistical Uncertainties of Weighted Efficiencies]{Statistical Uncertainties of \\ Weighted Efficiencies}
\label{sec:uncertainties:weff}

\begin{figure}[htbp]
\begin{center}
\includegraphics[width=0.85\linewidth]{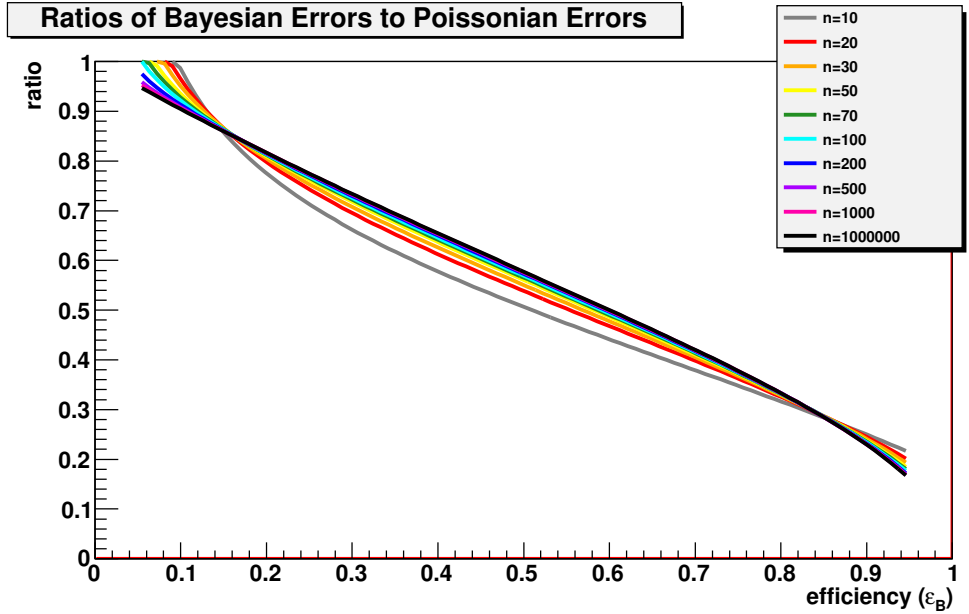}
\caption[Ratios of Bayesian to Poissonian uncertainties as a function of the efficiency $\varepsilon_{B}=k/n$ for various values of $n$.]{Ratios of Bayesian to Poissonian uncertainties as a function of the efficiency $\varepsilon_{B}=k/n$ for various values of $n$.  This ratio (with $n=100$) is used to convert a calculated Poissonian uncertainty in the efficiency calculation into an estimated Bayesian uncertainty.}
\label{fig:bre:error_ratios}
\end{center}
\end{figure}

In Chapters~\ref{sec:rec} and~\ref{sec:bre} the $e^{\pm}$ reconstruction efficiency $(\varepsilon_{R})$ and the background rejection efficiency $(\varepsilon_{B})$ are calculated by finding the ratio (as a function of $p_{T}$) of two simulated $e^{\pm}$ spectra.  Those efficiencies are properly calculated as the ratio of two weighted spectra, although the unweighted ratios are sometimes shown for illustration purposes.  The unweighted efficiencies are calculated using the ``Bayes error formula" from~\cite{UllrichXuErrors}, which gives the variance of an efficiency when the numerator $k$ is a subset of the denominator $n$ and the two are highly correlated.  The uncertainty in the efficiency $\varepsilon=k/n$ is

\begin{equation}
\label{eq:uncertainties:bayes}
\sigma\varepsilon=\sqrt{\frac{k+1}{n+2}\left(\frac{k+2}{n+3}-\frac{k+1}{n+2}\right)}
\end{equation}

For the weighted efficiencies, the statistical uncertainties in the efficiency $\varepsilon$ are estimated as follows.  The uncertainty in $\varepsilon$ is calculated with the assumption that the numerator and the denominator both independently follow Poisson statistics; the resulting uncertainty is then multiplied by a conversion factor to generate an estimated error based on the Bayes error formula.  This calculation is explained in greater detail below.  It should be noted that the systematic uncertainties assigned to $\varepsilon_{R}$ and $\varepsilon_{B}$ are large enough to account for the statistical fluctuations in those efficiency calculations.

For the sake of clarity, the calculation of the uncertainty of the background rejection efficiency $(\varepsilon_{B})$ will be described.  Only a single $e^{\pm}$ transverse-momentum bin is considered and each $e^{\pm}$ is weighted according to the $p_{T}$ of its parent photon.  The numerator and denominator of the efficiency calculation are called $k$ and $n$, respectively.  The number of rejected $e^{\pm}$ that come from a photon with transverse momentum $p_{T}^{\gamma}$ is $N_{rej.}(p_{T}^{\gamma})$ , and the photon weighting function is $W(p_{T}^{\gamma})$.  The numerator $k$ is a weighted sum:

\begin{equation}
k=\sum_{p_{T}^{\gamma}} \left[W(p_{T}^{\gamma})\times N_{rej.}(p_{T}^{\gamma})\right]
\end{equation}

\noindent If each measurement of $N_{rej.}$ follows Poisson statistics, the uncertainty of $k$ would be

\begin{equation}
\sigma_{k}^2=\sum_{p_{T}^{\gamma}} \left[W(p_{T}^{\gamma}) \sqrt{N_{rej.}(p_{T}^{\gamma})}\right]^2=\sum_{p_{T}^{\gamma}} \left[W^2(p_{T}^{\gamma})\times N_{rej.}(p_{T}^{\gamma})\right].
\end{equation}

\noindent Similarly, if $N_{rec.}$ is the number of reconstructed $e^{\pm}$,

\begin{equation}
n=\sum_{p_{T}^{\gamma}} \left[W(p_{T}^{\gamma})\times N_{rec.}(p_{T}^{\gamma})\right]
\end{equation}

\noindent and

\begin{equation}
\sigma_{n}^2=\sum_{p_{T}^{\gamma}} \left[W(p_{T}^{\gamma}) \sqrt{N_{rec.}(p_{T}^{\gamma})}\right]^2=\sum_{p_{T}^{\gamma}} \left[W^2(p_{T}^{\gamma})\times N_{rec.}(p_{T}^{\gamma})\right].
\end{equation}

\noindent If $k$ and $n$ were statistically independent, the uncertainty of the ratio $k/n$ would be

\begin{equation}
\sigma(\varepsilon_{B})=\varepsilon_{B}\sqrt{\left(\frac{\sigma_{k}}{k}\right)^2+\left(\frac{\sigma_{n}}{n}\right)^2}.
\end{equation}

\noindent Of course, $k$ and $n$ are not independent.  In the unweighted case, when $k$ and $n$ are integers, the ratio of the uncertainty calculated using Equation~\ref{eq:uncertainties:bayes} to the uncertainty calculated using the Poisson formula is

\begin{equation}
\begin{split}
R& =\sqrt{\frac{k+1}{n+2}\left(\frac{k+2}{n+3}-\frac{k+1}{n+2}\right)}\frac{n}{k\sqrt{1/k+1/n}}\\
 & =\sqrt{\frac{\varepsilon_{B}n+1}{n+2}\left(\frac{\varepsilon_{B}n+2}{n+3}-\frac{\varepsilon_{B}n+1}{n+2}\right)}\frac{1}{\varepsilon_{B}\sqrt{1/(\varepsilon_{B}n )+1/n}}.
 \end{split}
 \label{eq:uncertainties:bayes_poisson}
\end{equation}

Figure~\ref{fig:bre:error_ratios} shows this ratio as a function of the efficiency $\varepsilon_{B}=k/n$ for different values of $n$.  For the range of efficiencies of interest for this analysis ($0.4\leq\varepsilon_{B}\leq 0.8$) this ratio exhibits less than a 10\% change when $n$ varies from 10 to $10^6$.  For the weighted efficiency calculations shown in Chapter~\ref{sec:bre}, equation ~\ref{eq:uncertainties:bayes_poisson} (with $n=100$) is used to convert the calculated Poissonian uncertainty to an estimated Bayesian uncertainty.  The same arguments apply for the calculation of the uncertainties of the $e^{\pm}$ reconstruction efficiency $(\varepsilon_{R})$.  The numerator is the number of reconstructed $e^{\pm}$, each weighted according its simulated transverse momentum; the denominator is the number of simulated $e^{\pm}$.

\clearpage

\section[Statistical Uncertainties of the Non-Photonic $e^{\pm}$ Yield]{Statistical Uncertainties of \\ the Non-Photonic $\boldsymbol{e^{\pm}}$ Yield}
\label{sec:uncertainties:spectra}

In this section, Equation~\ref{eq:results:stat_err_nec} for the statistical uncertainty of $N_{EC}$, the efficiency-corrected non-photonic $e^{\pm}$ yield is derived, following~\cite{STAR_ppNPE2011}.  The ratio $(r)$ of the efficiency-corrected photonic $e^{\pm}$ yield to the inclusive $e^{\pm}$ yield is

\begin{equation}
r=\frac{P}{K_{inc.}I\varepsilon_{B}},
\end{equation}

\noindent where $P$ and $I$ are the uncorrected photonic and inclusive $e^{\pm}$ yields, respectively.  The uncertainties in $K_{inc.}$ and $\varepsilon_{B}$ are accounted for in the systematic uncertainties of the spectra.  It is assumed that the statistical uncertainties of $I$ and $r$ are uncorrelated.  Therefore,

\begin{equation}
\left(\frac{\sigma P}{P}\right)^{2}=\left(\frac{\sigma I}{I}\right)^{2}+\left(\frac{\sigma r}{r}\right)^{2}\;\;\Rightarrow\;\;\sigma r=r\sqrt{\left(\frac{\sigma P}{P}\right)^{2}-\left(\frac{\sigma I}{I}\right)^{2}},
\end{equation}

\noindent where $\sigma X$ denotes the statistical uncertainty of quantity $X$.  The efficiency-corrected non-photonic $e^{\pm}$ yield is

\begin{equation}
\label{eq:uncertainties:nec}
N_{EC}=\frac{1}{Q}\left(K_{inc.}I-\frac{1}{\varepsilon_{B}}P\right)=\frac{K_{inc.}I}{Q}(1-r),
\end{equation}

\noindent where $Q$ is shorthand for a combination of several correction factors, which do not influence the \textit{statistical} uncertainty of $N_{EC}$.  The statistical uncertainty of $N_{EC}$ is

\begin{equation}
\label{eq:uncertainties:stat_err_nec}
\sigma N_{EC}=N_{EC}\sqrt{\left(\frac{\sigma I}{I}\right)^{2}+\left(\frac{\sigma[1-r]}{1-r}\right)^{2}}.
\end{equation}

\noindent Since $\sigma[1-r]=\sigma r$,

\begin{equation}
\label{eq:uncertainties:stat_err_nec_final}
\begin{split}
\sigma N_{EC}& =N_{EC}\sqrt{\left(\frac{\sigma I}{I}\right)^{2}+\frac{r^{2}}{(1-r)^{2}}\left[\left(\frac{\sigma P}{P}\right)^{2}-\left(\frac{\sigma I}{I}\right)^{2}\right]}\\
 & =N_{EC}\sqrt{\frac{1}{I}+\frac{r^{2}}{(1-r)^{2}}\left[\left(\frac{\sigma P}{P}\right)^{2}-\frac{1}{I}\right]}.
\end{split}
\end{equation}

\clearpage

\section[Systematic Uncertainties of $e^{\pm}$ Spectra]{Systematic Uncertainties of $\boldsymbol{e^{\pm}}$ Spectra}
\label{sec:uncertainties:spectra_sys}

The systematic uncertainties of the inclusive, photonic, and non-photonic $e^{\pm}$ spectra are due to the uncertainties in the various correction factors.  In the calculation of the systematic uncertainties of the efficiency-corrected $e^{\pm}$ yields ($I_{EC}$, $P_{EC}$, and $N_{EC}$) it is assumed that the uncertainties in $\langle A_{BEMC}\rangle$, $\varepsilon_{R}$, $\varepsilon_{T}$, $\varepsilon_{dE/dx}$, $K_{inc.}$, and $\varepsilon_{B}$ are uncorrelated.  In this analysis, the lower and upper systematic uncertainties are not required to be the same.  In the following discussion, $\lambda[X]$ and $\upsilon[X]$ denote the lower and upper systematic uncertainties of quantity $X$, respectively.  It is convenient to define the variable $Q$ as the product of the four correction factors that are common to the calculations of $I_{EC}$, $P_{EC}$, and $N_{EC}$.

\begin{equation}
Q=\langle A_{BEMC}\rangle\varepsilon_{R}\cdot\varepsilon_{T}\cdot\varepsilon_{dE/dx}.
\end{equation}

\noindent Therefore,

\begin{equation}
\lambda[Q]=Q\sqrt{\left(\frac{\lambda[\langle A_{BEMC}\rangle]}{\langle A_{BEMC}\rangle}\right)^{2}+\left(\frac{\lambda[\varepsilon_{R}]}{\varepsilon_{R}}\right)^{2}+\left(\frac{\lambda[\varepsilon_{T}]}{\varepsilon_{T}}\right)^{2}+\left(\frac{\lambda[\varepsilon_{dE/dx}]}{\varepsilon_{dE/dx}}\right)^{2}}
\end{equation}

\noindent and

\begin{equation}
\upsilon[Q]=Q\sqrt{\left(\frac{\upsilon[\langle A_{BEMC}\rangle]}{\langle A_{BEMC}\rangle}\right)^{2}+\left(\frac{\upsilon[\varepsilon_{R}]}{\varepsilon_{R}}\right)^{2}+\left(\frac{\upsilon[\varepsilon_{T}]}{\varepsilon_{T}}\right)^{2}+\left(\frac{\upsilon[\varepsilon_{dE/dx}]}{\varepsilon_{dE/dx}}\right)^{2}}.
\end{equation}

\noindent The systematic uncertainties of $I_{EC}$ are

\begin{equation}
\lambda[I_{EC}]=I_{EC}\sqrt{\left(\frac{\upsilon[Q]}{Q}\right)^{2}+\left(\frac{\lambda[K_{inc.}]}{K_{inc.}}\right)^{2}}
\end{equation}

\noindent and

\begin{equation}
\upsilon[I_{EC}]=I_{EC}\sqrt{\left(\frac{\lambda[Q]}{Q}\right)^{2}+\left(\frac{\upsilon[K_{inc.}]}{K_{inc.}}\right)^{2}}.
\end{equation}

\noindent The systematic uncertainties of $P_{EC}$ are

\begin{equation}
\lambda[P_{EC}]=P_{EC}\sqrt{\left(\frac{\upsilon[Q]}{Q}\right)^{2}+\left(\frac{\upsilon[\varepsilon_{B}]}{\varepsilon_{B}}\right)^{2}}
\end{equation}

\noindent and

\begin{equation}
\upsilon[P_{EC}]=P_{EC}\sqrt{\left(\frac{\lambda[Q]}{Q}\right)^{2}+\left(\frac{\lambda[\varepsilon_{B}]}{\varepsilon_{B}}\right)^{2}}.
\end{equation}

\noindent The systematic uncertainties of $N_{EC}$ are

\begin{equation}
\lambda[N_{EC}]=\sqrt{\left(\frac{\upsilon[Q]}{Q}N_{EC}\right)^{2}+\left(\frac{I}{Q}\lambda[K_{inc.}]\right)^{2}+\left(\frac{P}{\varepsilon_{B}^{2}Q}\lambda[\varepsilon_{B}]\right)^{2}}
\end{equation}

\noindent and

\begin{equation}
\upsilon[N_{EC}]=\sqrt{\left(\frac{\lambda[Q]}{Q}N_{EC}\right)^{2}+\left(\frac{I}{Q}\upsilon[K_{inc.}]\right)^{2}+\left(\frac{P}{\varepsilon_{B}^{2}Q}\upsilon[\varepsilon_{B}]\right)^{2}}.
\end{equation}

\clearpage

\chapter{Additional Material}
This appendix contains supplementary figures which were excluded from the main body of the dissertation for the sake of brevity.  Additional figures related to the calculation of the $e^{\pm}$ reconstruction efficiency ($\varepsilon_{R}$; see Chapter~\ref{sec:rec}) are presented in Section~\ref{sec:add:rec}.  Additional figures and tables related to the inclusive $e^{\pm}$ purity ($K_{inc.}$; see Section~\ref{sec:dedx:purity}) are presented in Section~\ref{sec:add:purity}.

\section[$e^{\pm}$ Reconstruction Efficiency]{$\boldsymbol{e^{\pm}}$ Reconstruction Efficiency}
\label{sec:add:rec}

This section contains additional figures relating to the calculation of the $e^{\pm}$ reconstruction efficiency $(\varepsilon_{R})$, which is described in detail in Chapter~\ref{sec:rec}.

\subsection{Calculated Efficiencies from Embedding}
\label{sec:add:rec:embedding}

This section contains additional plots of $\varepsilon_{R}$, calculated from embedding (simulation data set $N_{1}$), with the values of various $e^{\pm}$ identification cuts changed from their standard values.  These plots are an extension of the series presented in Section~\ref{sec:rec:eff_from_emb}.  In these figures, the black circles indicate the calculation of $\varepsilon_{R}$ with all cuts having their standard values.

\begin{figure}[htbp]
\begin{center}
\includegraphics[width=0.85\linewidth]{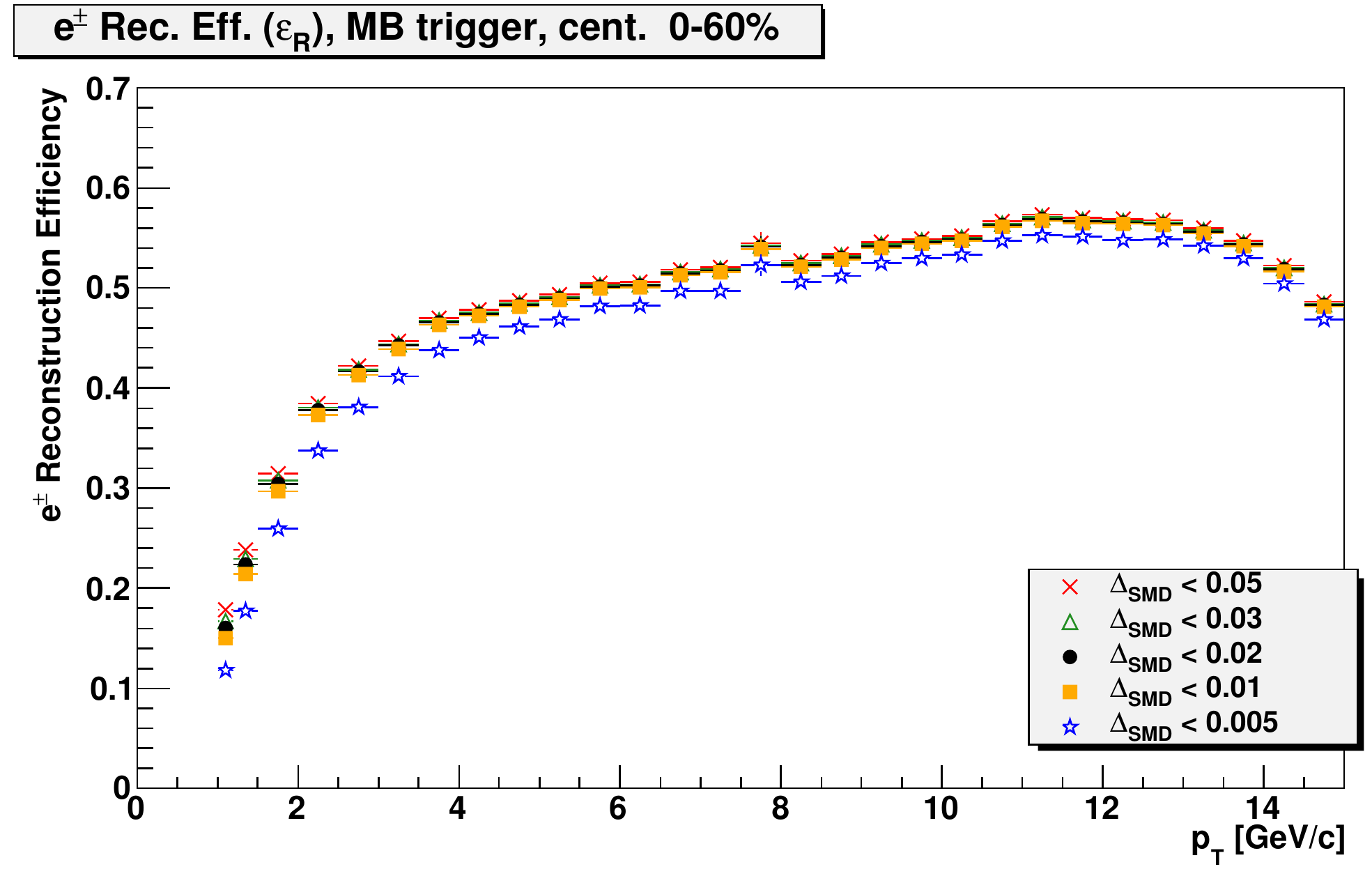}
\caption{The $e^{\pm}$ reconstruction efficiency as a function of $p_{T}$ for various values of the cut on $\Delta_{SMD}$.}
\label{fig:add:rec:eff_dsmd}
\end{center}
\end{figure}

Figures~\ref{fig:add:rec:eff_dsmd} shows $\varepsilon_{R}$ calculated with various values for the cut on the BSMD cluster displacement $\Delta_{SMD}=\sqrt{(\Delta\eta_{SMD})^{2}+(\Delta\phi_{SMD})^{2}}$.  Changing the value of the $\Delta_{SMD}$ cut above 0.01 has negligible effect on $\varepsilon_{R}$.

\begin{figure}[htbp]
\begin{center}
\includegraphics[width=0.85\linewidth]{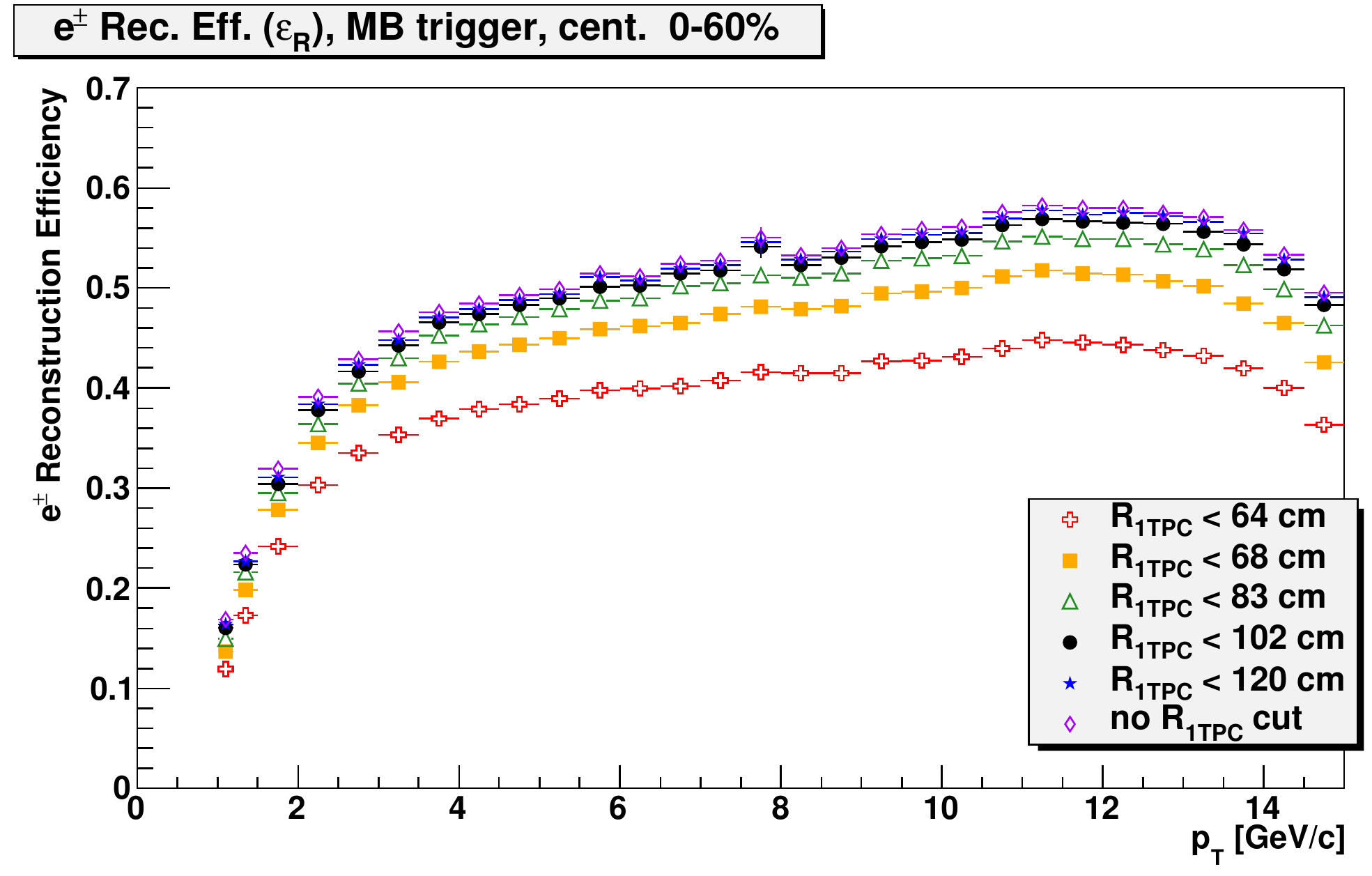}
\caption{The $e^{\pm}$ reconstruction efficiency as a function of $p_{T}$ for various values of the cut on $R_{1TPC}$.}
\label{fig:add:rec:eff_r1tpc}
\includegraphics[width=0.85\linewidth]{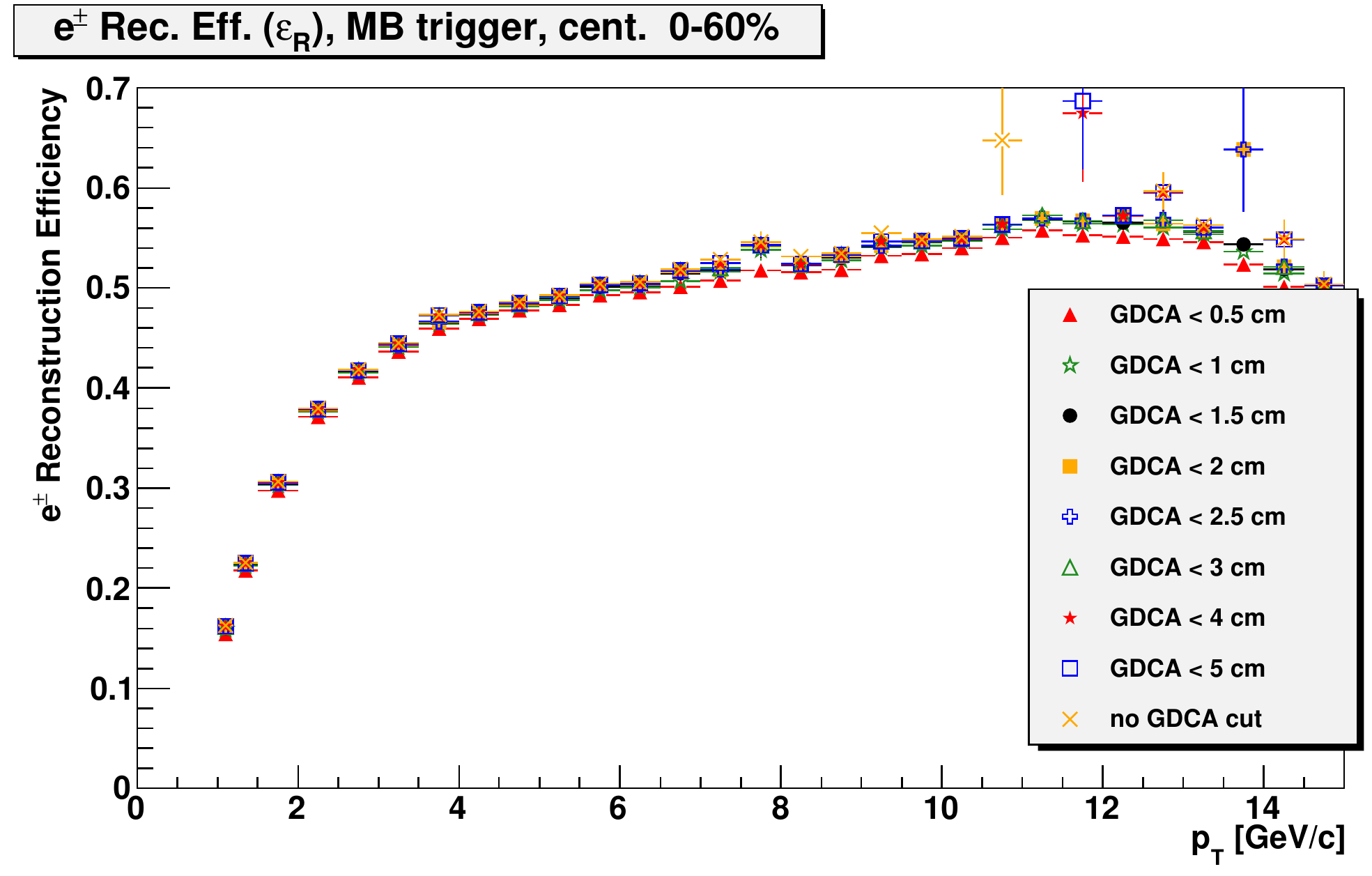}
\caption[The $e^{\pm}$ reconstruction efficiency as a function of $p_{T}$ for various values of the $GDCA$ cut.]{The $e^{\pm}$ reconstruction efficiency as a function of $p_{T}$ for various values of the $GDCA$ cut.  No large change in $\varepsilon_{R}$ is observed as the value of the $GDCA$ cut is varied above 0.5 cm.}
\label{fig:add:rec:eff_gdca}
\end{center}
\end{figure}

Figure~\ref{fig:add:rec:eff_r1tpc} shows the $e^{\pm}$ reconstruction efficiency calculated with various values for the cut on $R_{1TPC}$.

Figure~\ref{fig:add:rec:eff_gdca} shows $\varepsilon_{R}$ calculated with various values for the global DCA cut.  No strong dependence of the efficiency on the $GDCA$ cut is observed.

\begin{figure}[htbp]
\begin{center}
\includegraphics[width=0.85\linewidth]{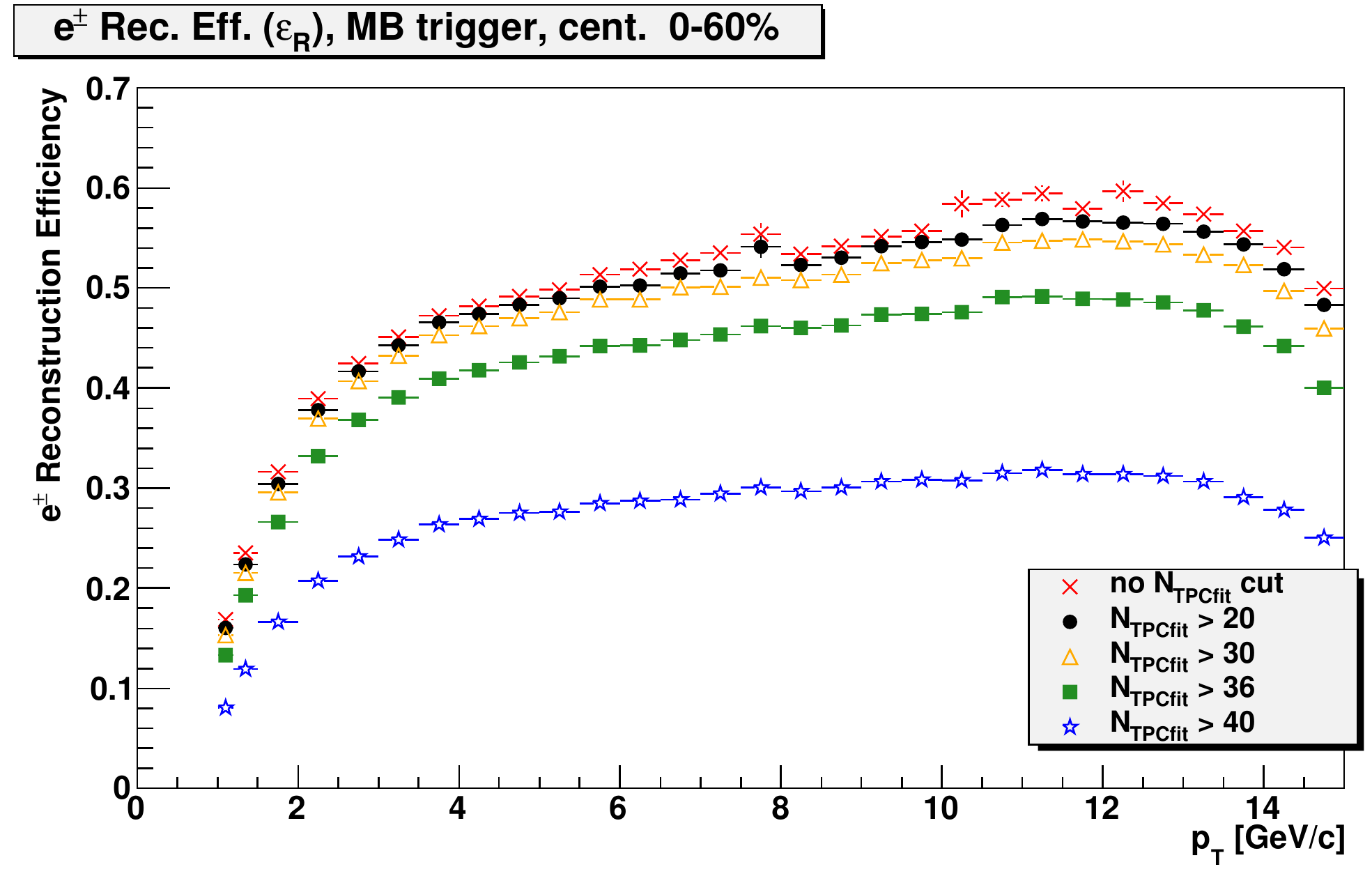}
\caption{The $e^{\pm}$ reconstruction efficiency as a function of $p_{T}$ for various values of the cut on $N_{TPCfit}$.}
\label{fig:add:rec:eff_tfp}
\includegraphics[width=0.85\linewidth]{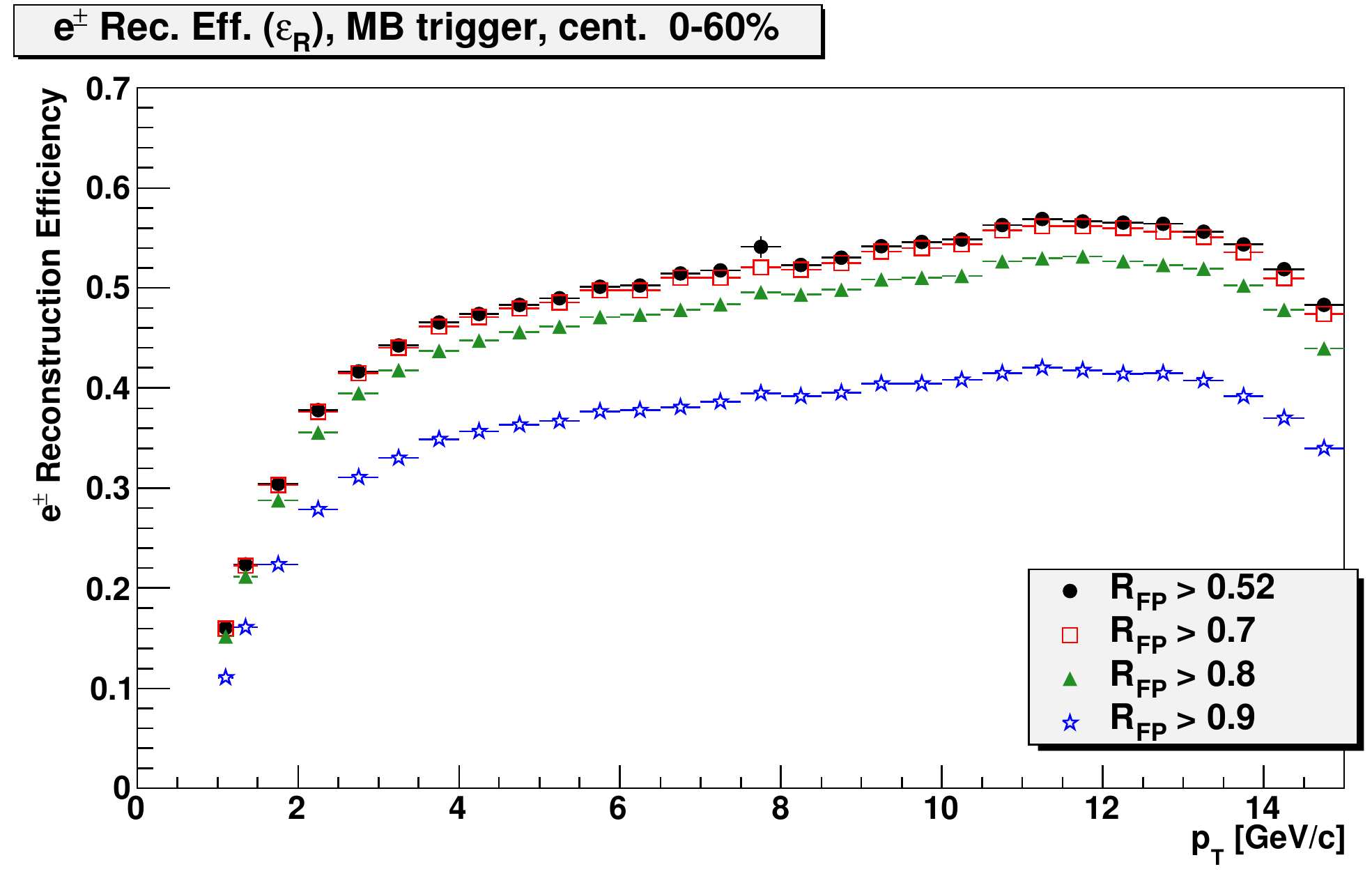}
\caption{The $e^{\pm}$ reconstruction efficiency as a function of $p_{T}$ for various values of the cut on $R_{FP}$.}
\label{fig:add:rec:eff_fpr}
\end{center}
\end{figure}

Figures~\ref{fig:add:rec:eff_tfp} and~\ref{fig:add:rec:eff_fpr} show $\varepsilon_{R}$ calculated with various values for the cuts on $N_{TPCfit}$ and $R_{FP}$, respectively.

\clearpage

\subsection{Comparison of Embedding data to Real Data}
\label{sec:add:rec:partial}

The plots in this section are an extension of the series presented in Section~\ref{sec:rec:compare2real}.

\begin{figure}[htbp]
\begin{center}
\includegraphics[width=0.85\linewidth]{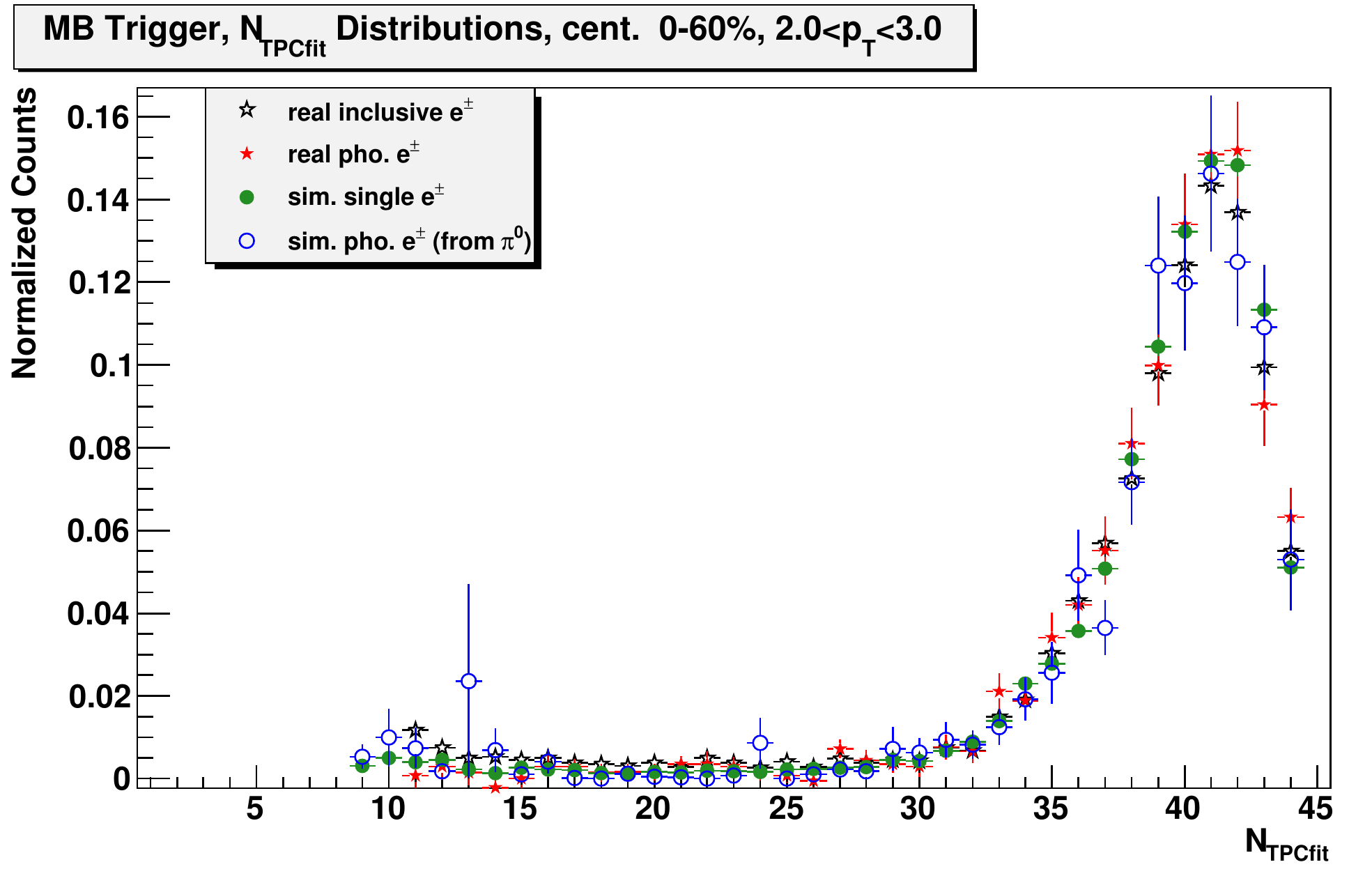}
\caption{Distributions of $N_{TPCfit}$ for real and simulated $e^{\pm}$.}
\label{fig:add:rec:dist_tfp}
\includegraphics[width=0.85\linewidth]{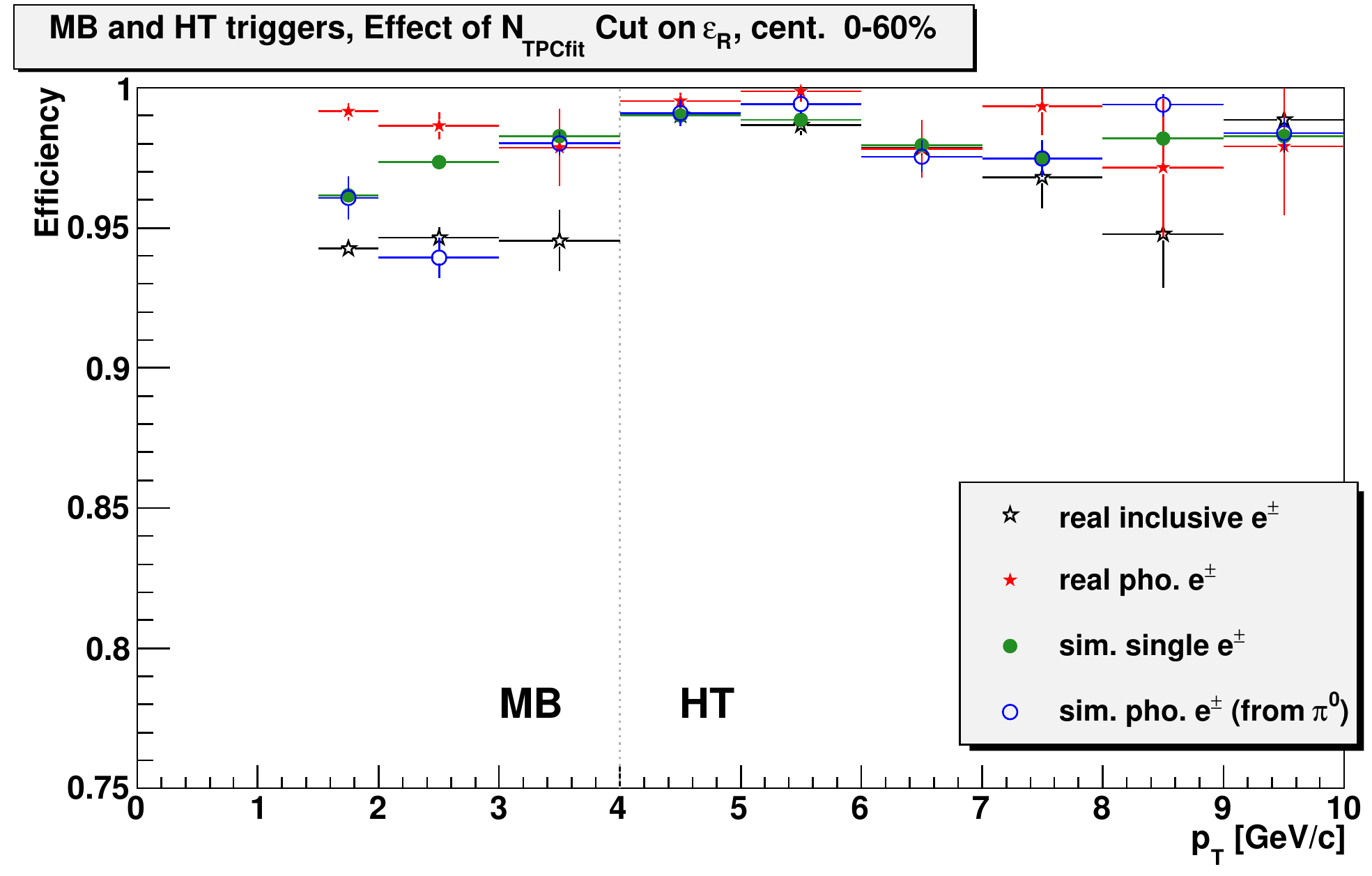}
\caption[Partial efficiencies of the cut $N_{TPCfit}>20$.]{Partial efficiencies of the cut $N_{TPCfit}>20$.  For $p_{T}<4\GeV/c$, the results are for minimum-bias events; for $p_{T}>4\GeV/c$, the results are for high-tower triggered events.}
\label{fig:add:rec:par_eff_tfp}
\end{center}
\end{figure}

Figure~\ref{fig:add:rec:dist_tfp} shows the distributions of $N_{TPCfit}$ and Figure~\ref{fig:add:rec:par_eff_tfp} shows the partial efficiencies of the $N_{TPCfit}>20$ cut.

\begin{figure}[htbp]
\begin{center}
\includegraphics[width=0.85\linewidth]{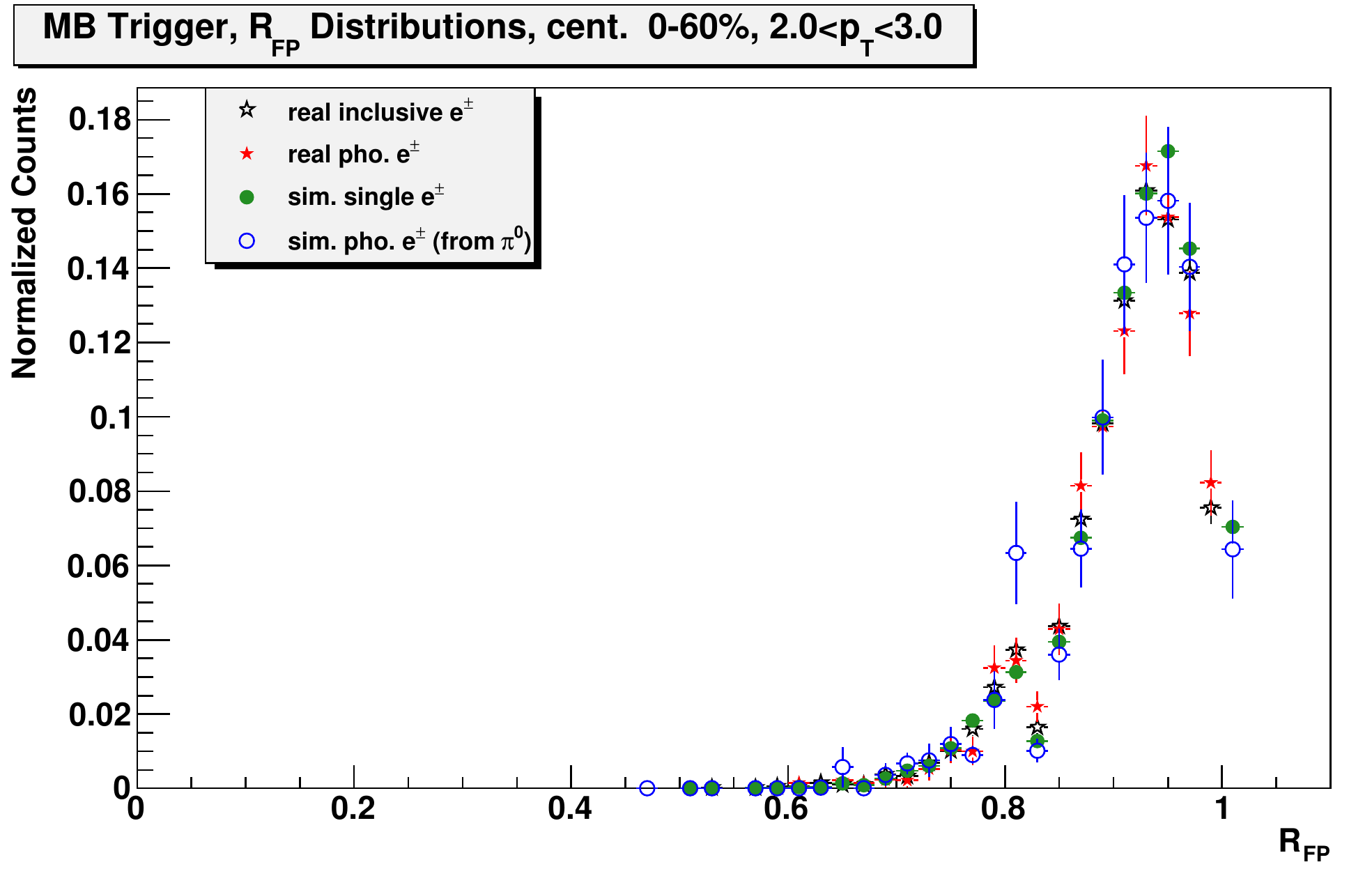}
\caption[Distributions of the TPC-fit-points ratio $(R_{FP})$ for real and simulated $e^{\pm}$.]{Distributions of the TPC-fit-points ratio $(R_{FP})$ for real and simulated $e^{\pm}$.  Nearly all $e^{\pm}$ pass the $R_{FP}>0.52$ cut and the partial efficiency is very close to 1.}
\label{fig:add:rec:dist_fpr}
\end{center}
\end{figure}

Figure~\ref{fig:add:rec:dist_fpr} shows the distributions of $R_{FP}$; the partial efficiencies of the $R_{FP}>0.52$ cut are all very close to unity and are not shown.

\begin{figure}[htbp]
\begin{center}
\includegraphics[width=0.85\linewidth]{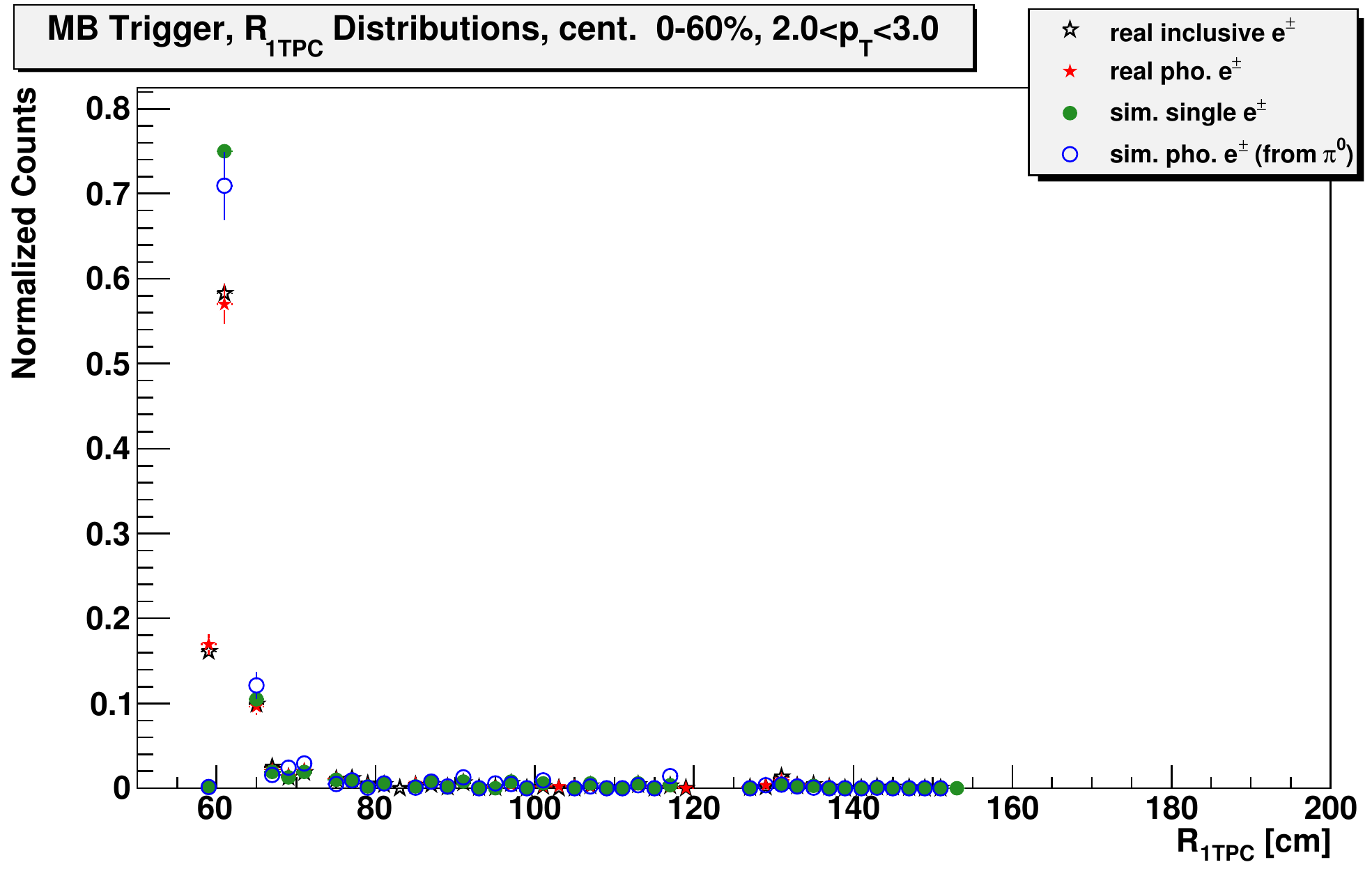}
\caption{Distributions of $R_{1TPC}$ for real and simulated $e^{\pm}$.}
\label{fig:add:rec:dist_r1tpc}
\includegraphics[width=0.85\linewidth]{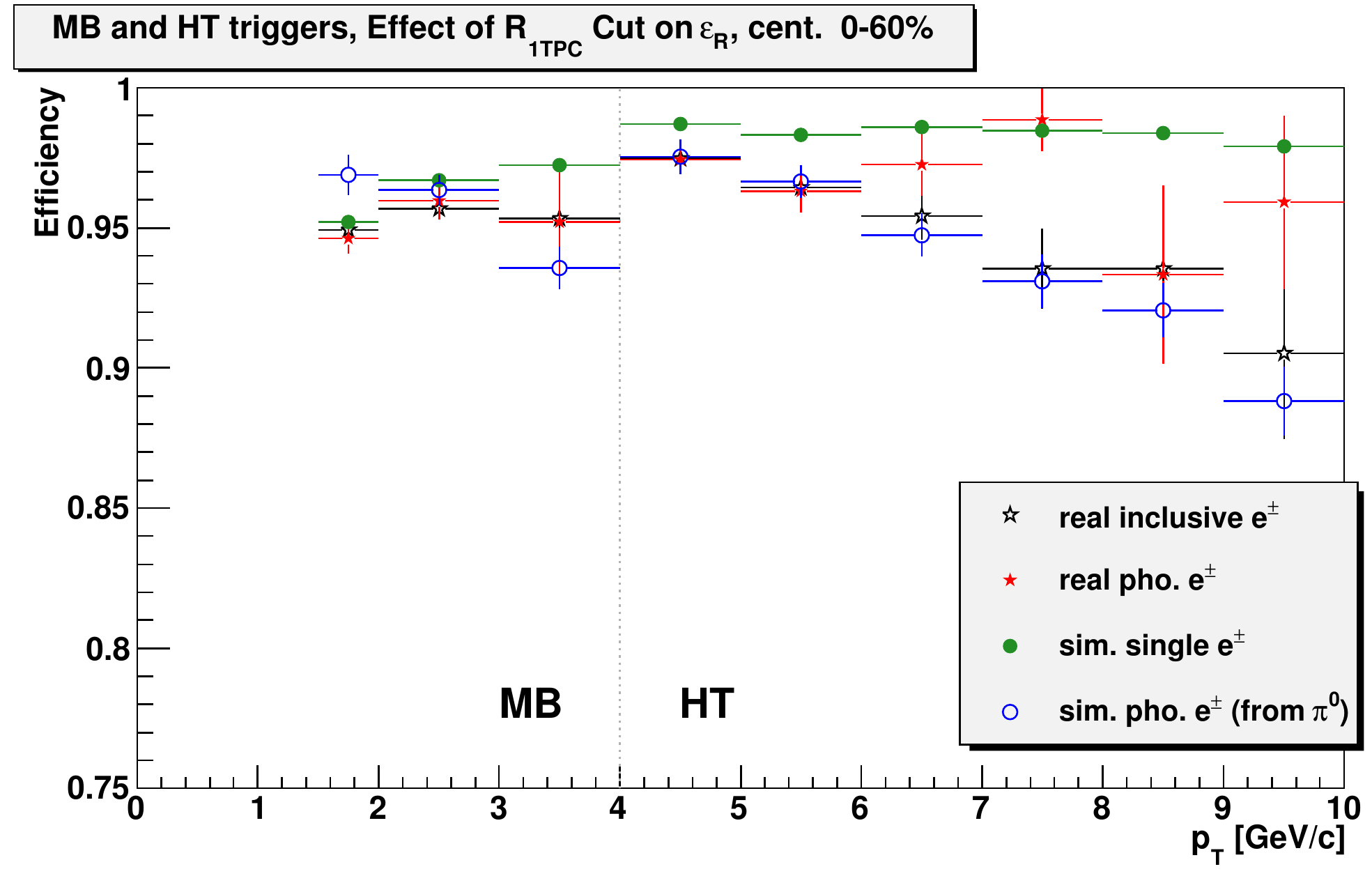}
\caption{Partial efficiencies of the cut $R_{1TPC}<102$ cm.}
\label{fig:add:rec:par_eff_r1tpc}
\end{center}
\end{figure}

Figure~\ref{fig:add:rec:dist_r1tpc} shows the distributions of $R_{1TPC}$ and Figure~\ref{fig:add:rec:par_eff_r1tpc} shows the partial efficiencies of the $R_{1TPC}<102$ cm cut.

\clearpage

\section[Inclusive $e^{\pm}$ Purity]{Inclusive $\boldsymbol{e^{\pm}}$ Purity}
\label{sec:add:purity}

\begin{figure}[htbp]
\begin{center}
\includegraphics[width=0.85\linewidth]{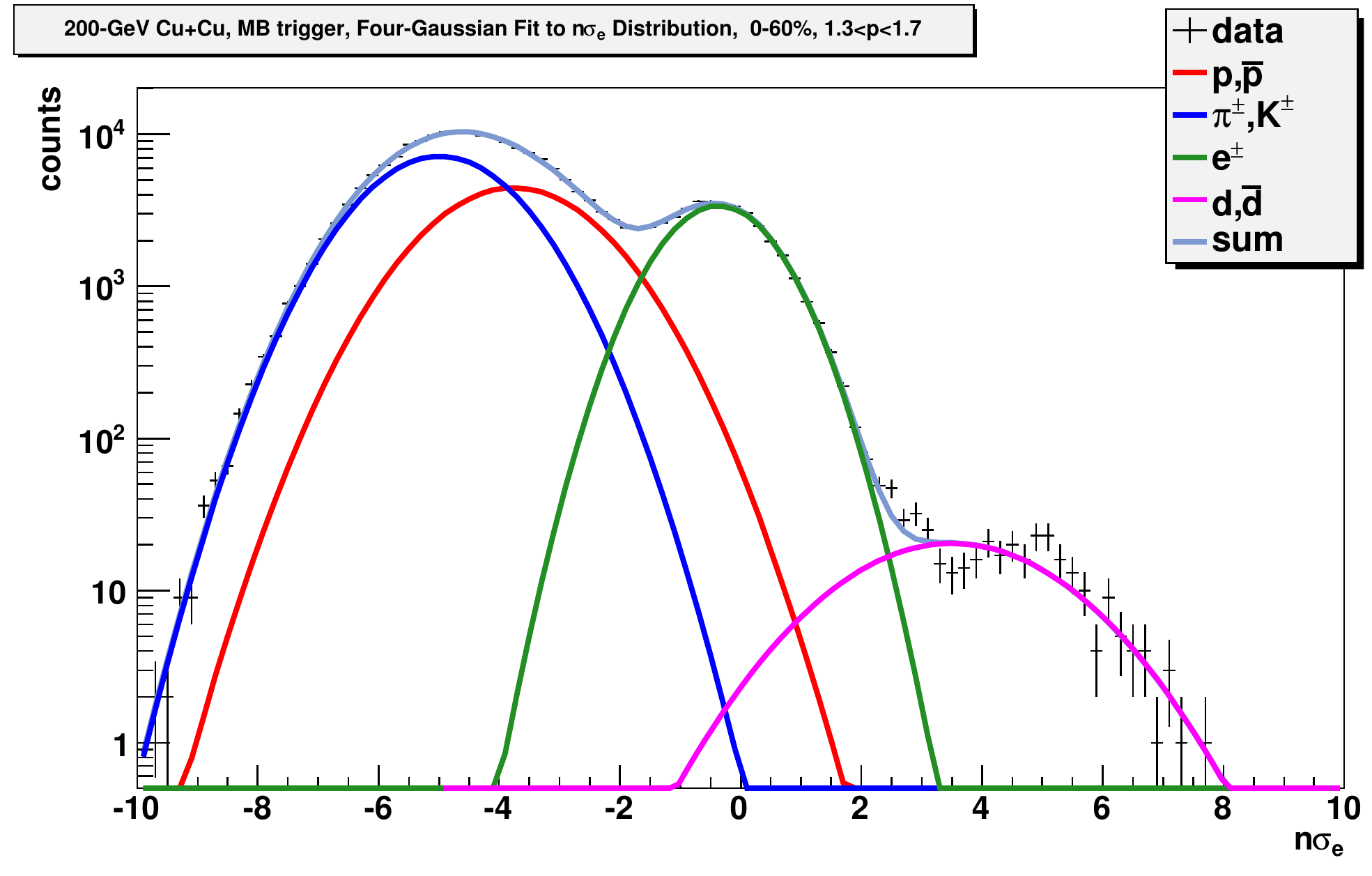}
\caption[Four-Gaussian fit of $n\sigma_{e}$ distribution: 0-60\% centrality class of minimum-bias triggered events; $1.3\GeV/c<p<1.7\GeV/c$.]{Four-Gaussian fit of $n\sigma_{e}$ distribution: 0-60\% centrality class of minimum-bias triggered events; $1.3\GeV/c<p<1.7\GeV/c$.  The magenta peak accounts for the (anti)deuteron contribution.  (identical to Figure~\ref{fig:dedx:4g_fit}, page~\pageref{fig:dedx:4g_fit})}
\label{fig:add:purity:fit_nsigma_mb_c03_p1}
\end{center}
\end{figure}

\begin{figure}[htbp]
\begin{center}
\includegraphics[width=0.85\linewidth]{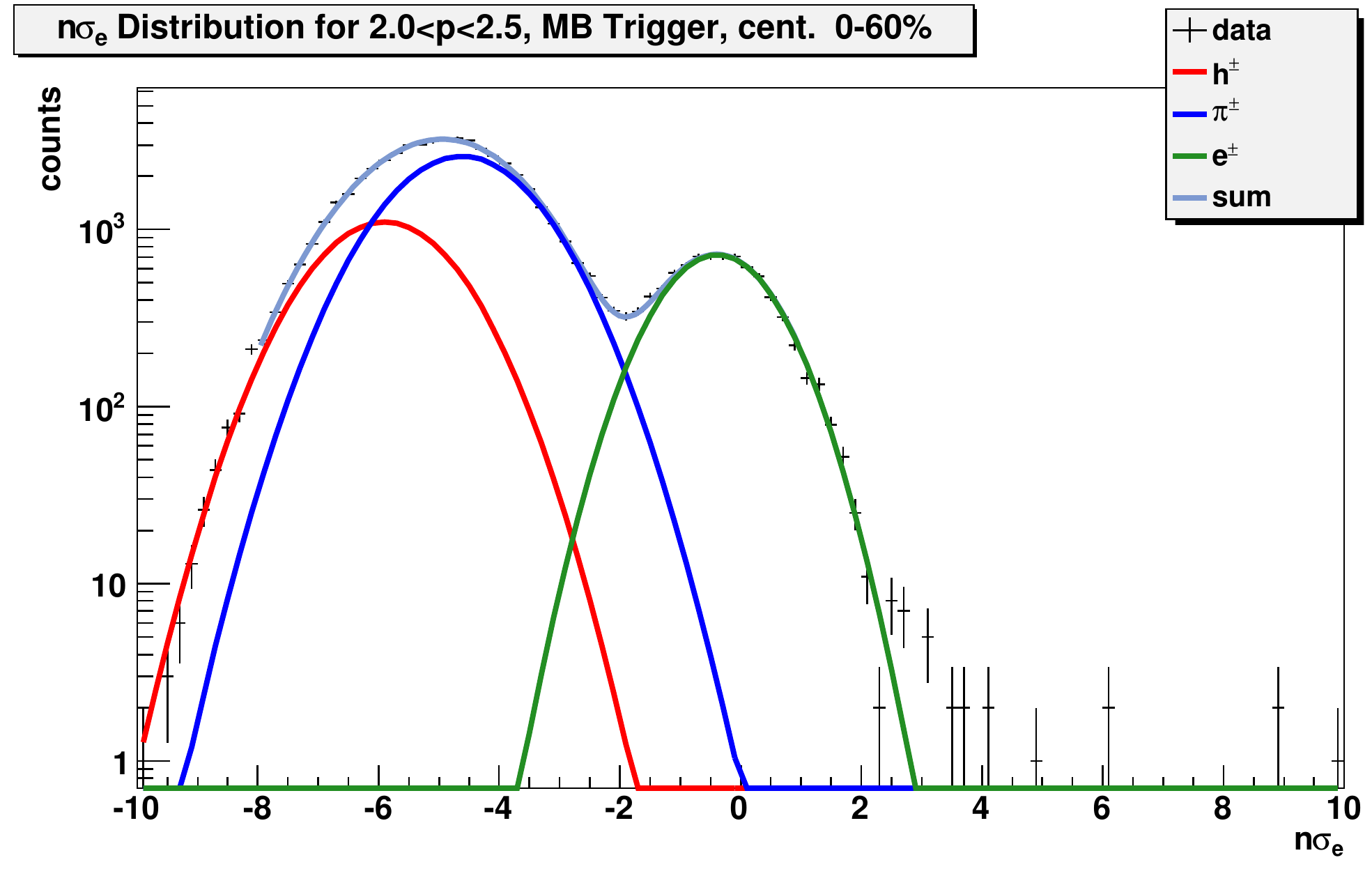}
\caption[Three-Gaussian fit of $n\sigma_{e}$ distribution: 0-60\% centrality class of minimum-bias triggered events; $2\GeV/c<p<2.5\GeV/c$.]{Three-Gaussian fit of $n\sigma_{e}$ distribution: 0-60\% centrality class of minimum-bias triggered events; $2\GeV/c<p<2.5\GeV/c$. (identical to Figure~\ref{fig:dedx:3g_fit}, page~\pageref{fig:dedx:3g_fit})}
\label{fig:add:purity:fit_nsigma_mb_c03_p3}
\includegraphics[width=0.85\linewidth]{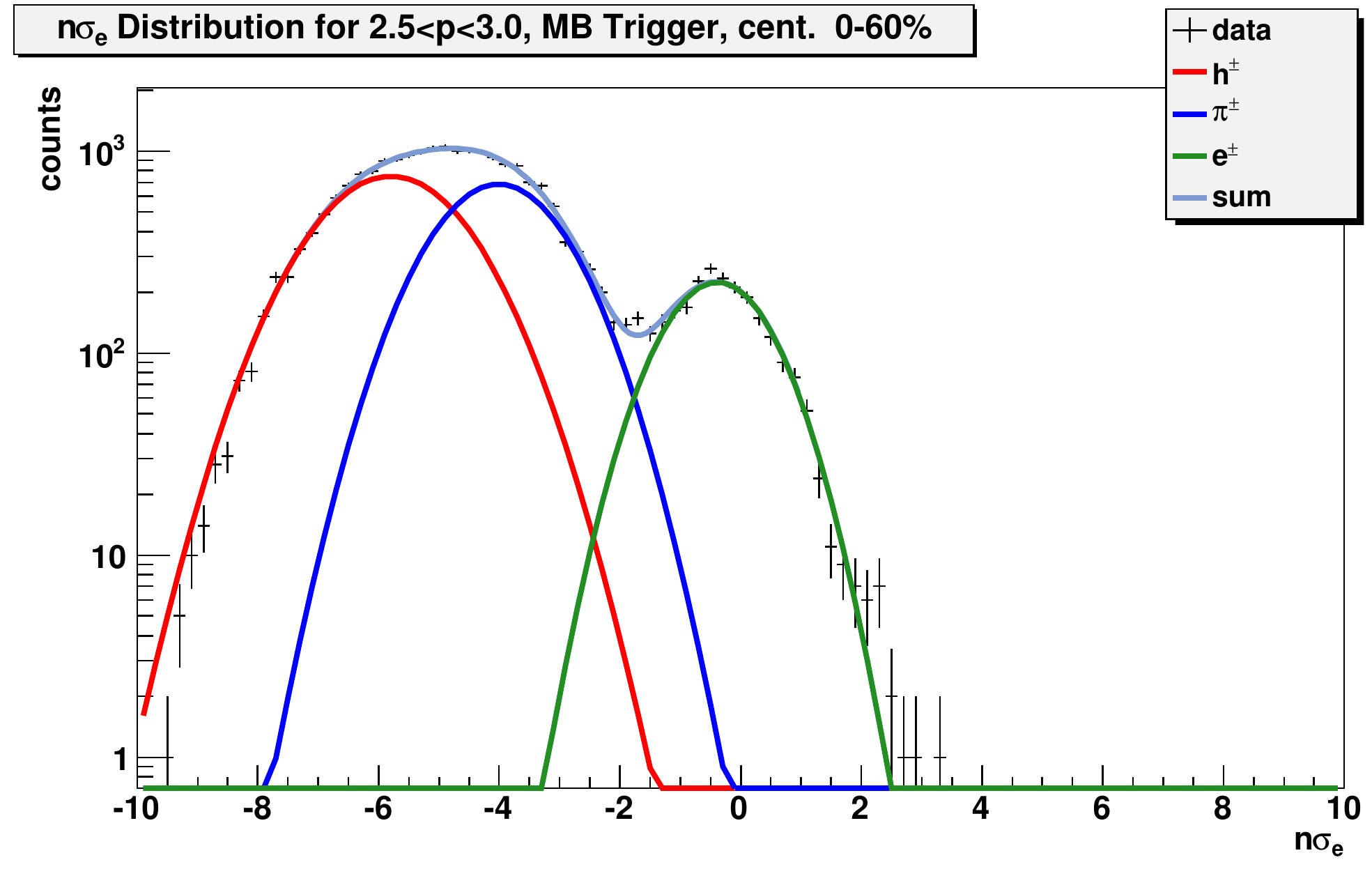}
\caption{Three-Gaussian fit of $n\sigma_{e}$ distribution: 0-60\% centrality class of minimum-bias triggered events; $2.5\GeV/c<p<3\GeV/c$.}
\label{fig:add:purity:fit_nsigma_mb_c03_p4}
\end{center}
\end{figure}

\begin{figure}[htbp]
\begin{center}
\includegraphics[width=0.85\linewidth]{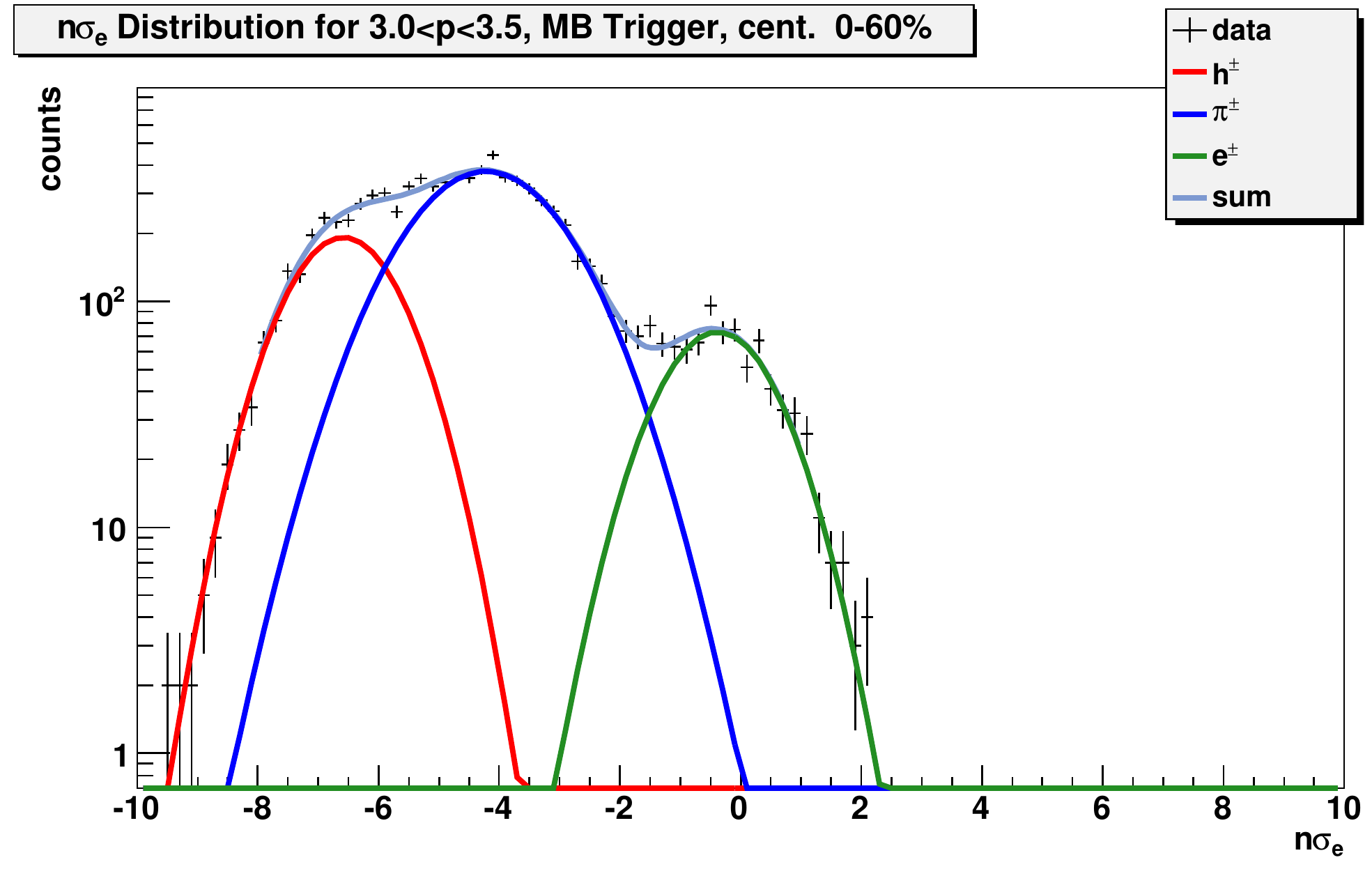}
\caption{Three-Gaussian fit of $n\sigma_{e}$ distribution: 0-60\% centrality class of minimum-bias triggered events; $3\GeV/c<p<3.5\GeV/c$.}
\label{fig:add:purity:fit_nsigma_mb_c03_p5}
\includegraphics[width=0.85\linewidth]{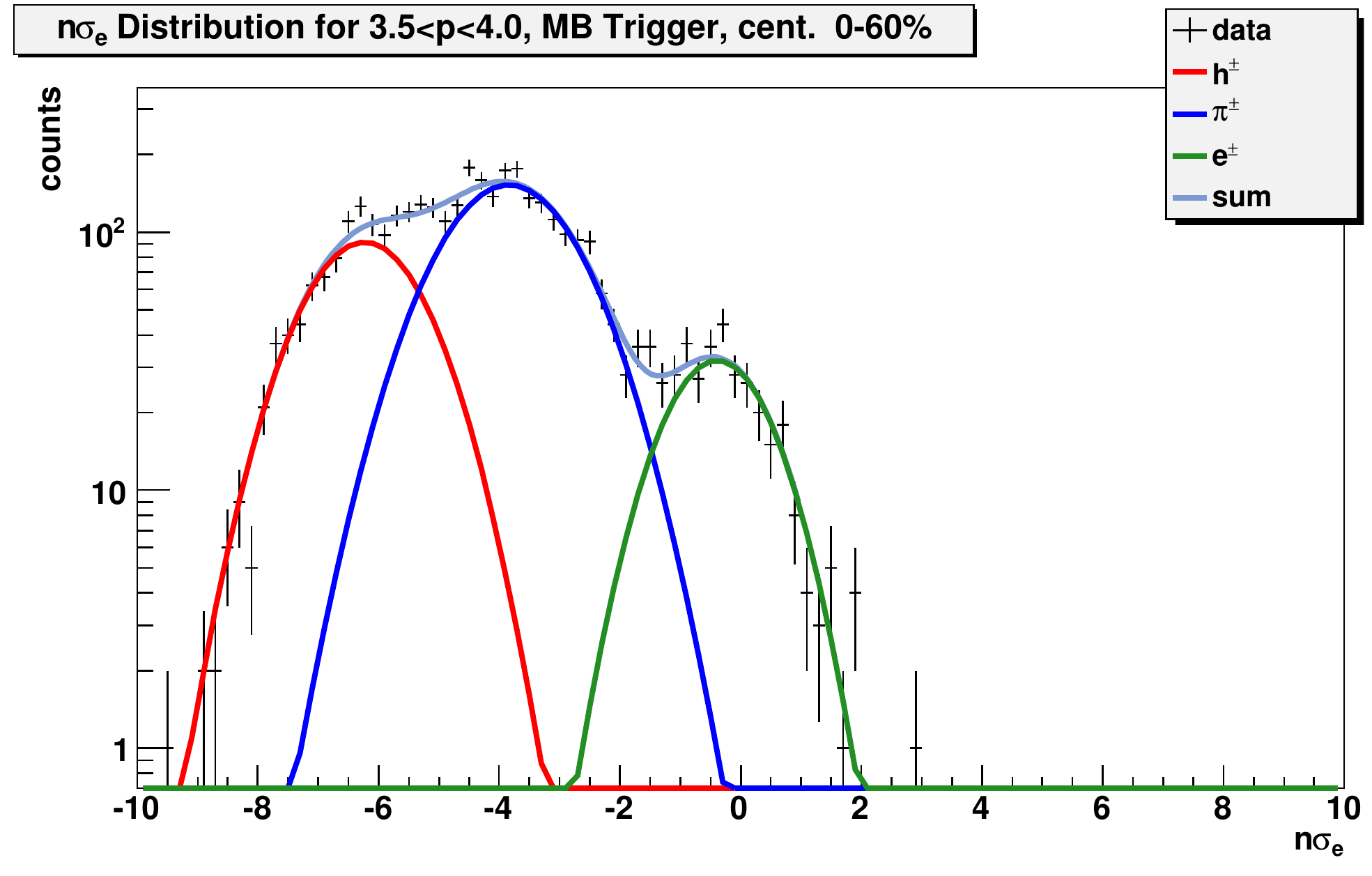}
\caption{Three-Gaussian fit of $n\sigma_{e}$ distribution: 0-60\% centrality class of minimum-bias triggered events; $3.5\GeV/c<p<4\GeV/c$.}
\label{fig:add:purity:fit_nsigma_mb_c03_p6}
\end{center}
\end{figure}

\begin{figure}[htbp]
\begin{center}
\includegraphics[width=0.85\linewidth]{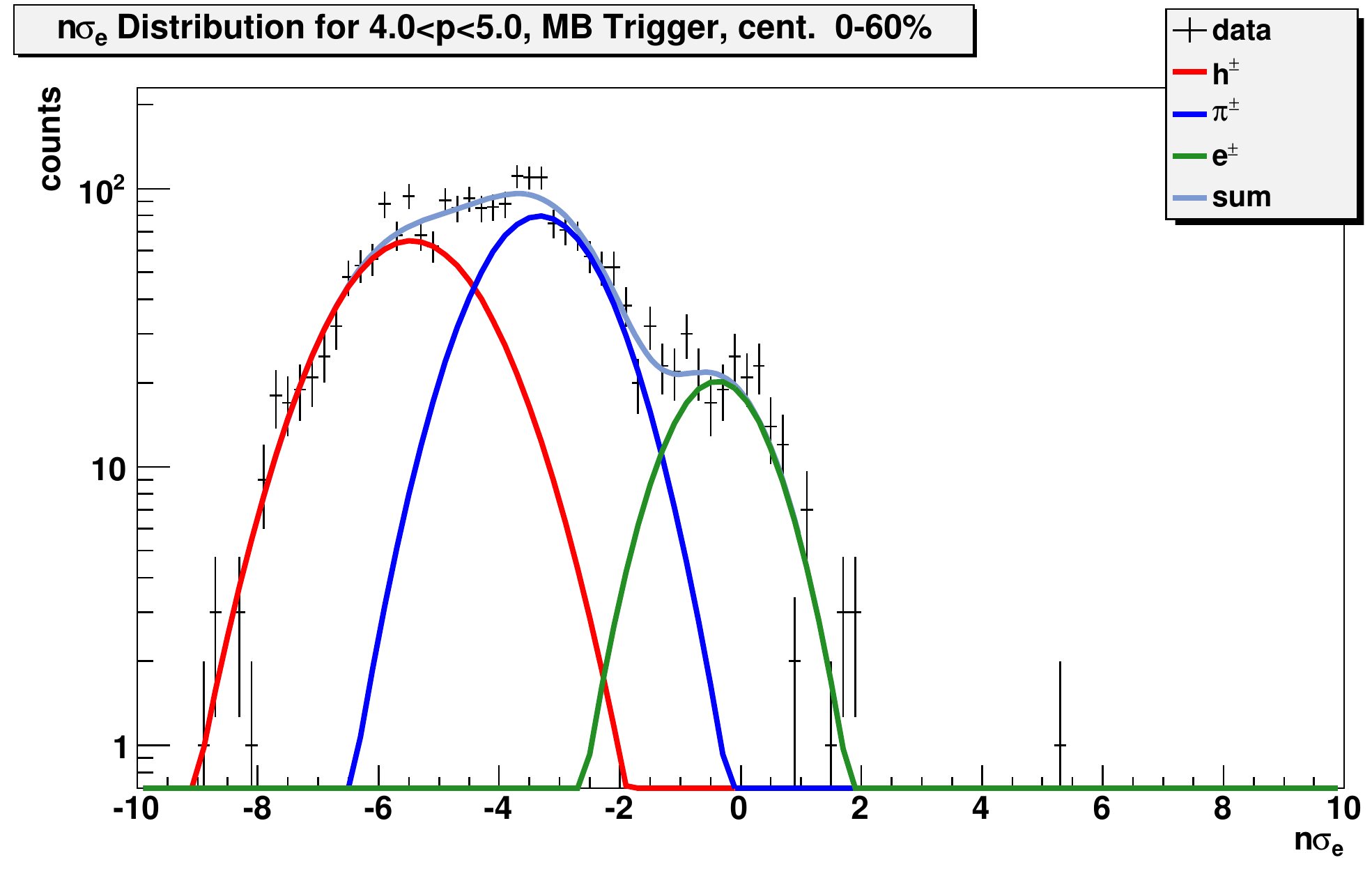}
\caption{Three-Gaussian fit of $n\sigma_{e}$ distribution: 0-60\% centrality class of minimum-bias triggered events; $4\GeV/c<p<5\GeV/c$.}
\label{fig:add:purity:fit_nsigma_mb_c03_p7}
\includegraphics[width=0.85\linewidth]{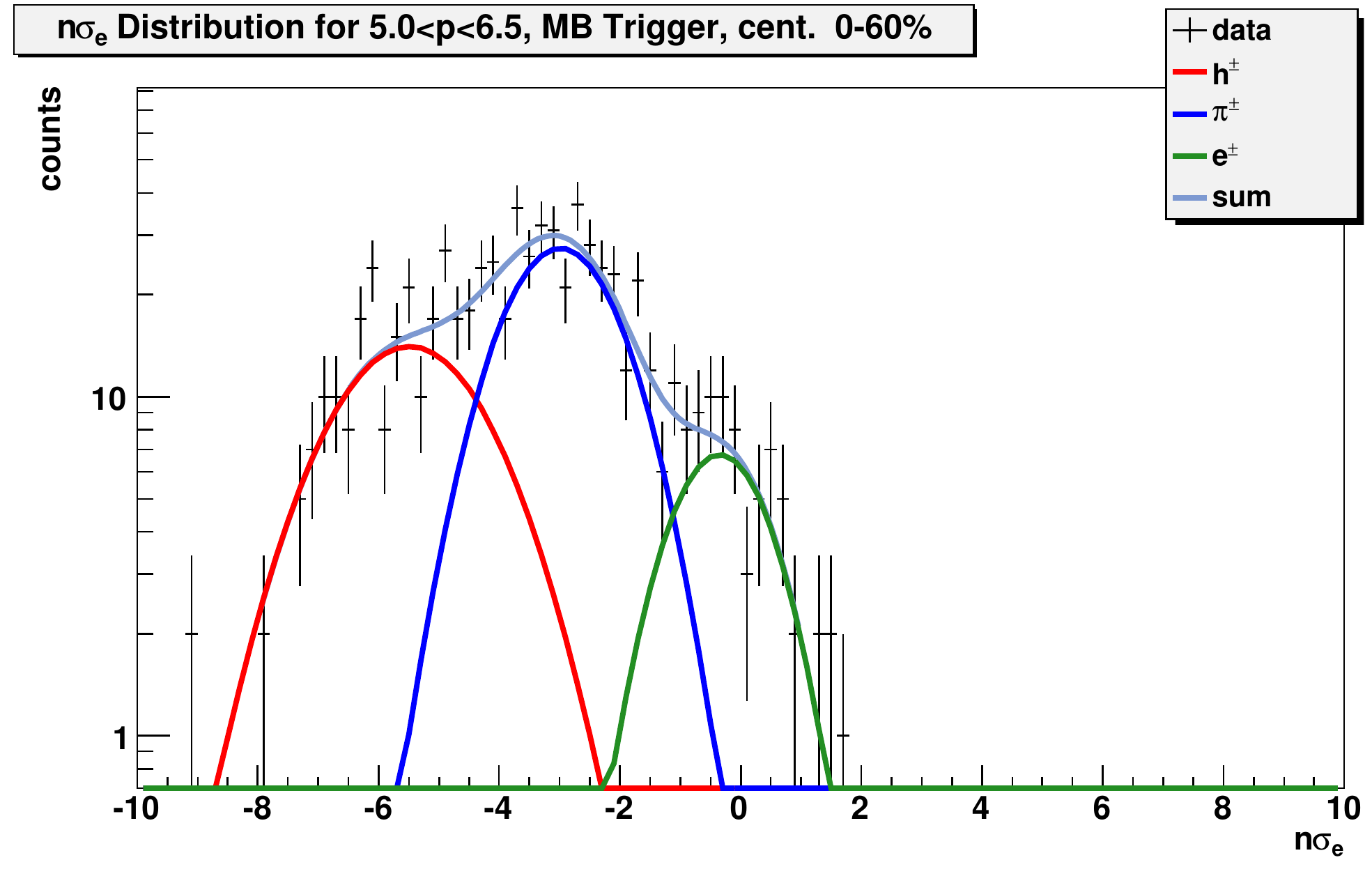}
\caption{Three-Gaussian fit of $n\sigma_{e}$ distribution: 0-60\% centrality class of minimum-bias triggered events; $5\GeV/c<p<6.5\GeV/c$.}
\label{fig:add:purity:fit_nsigma_mb_c03_p8}
\end{center}
\end{figure}

\begin{figure}[htbp]
\begin{center}
\includegraphics[width=0.85\linewidth]{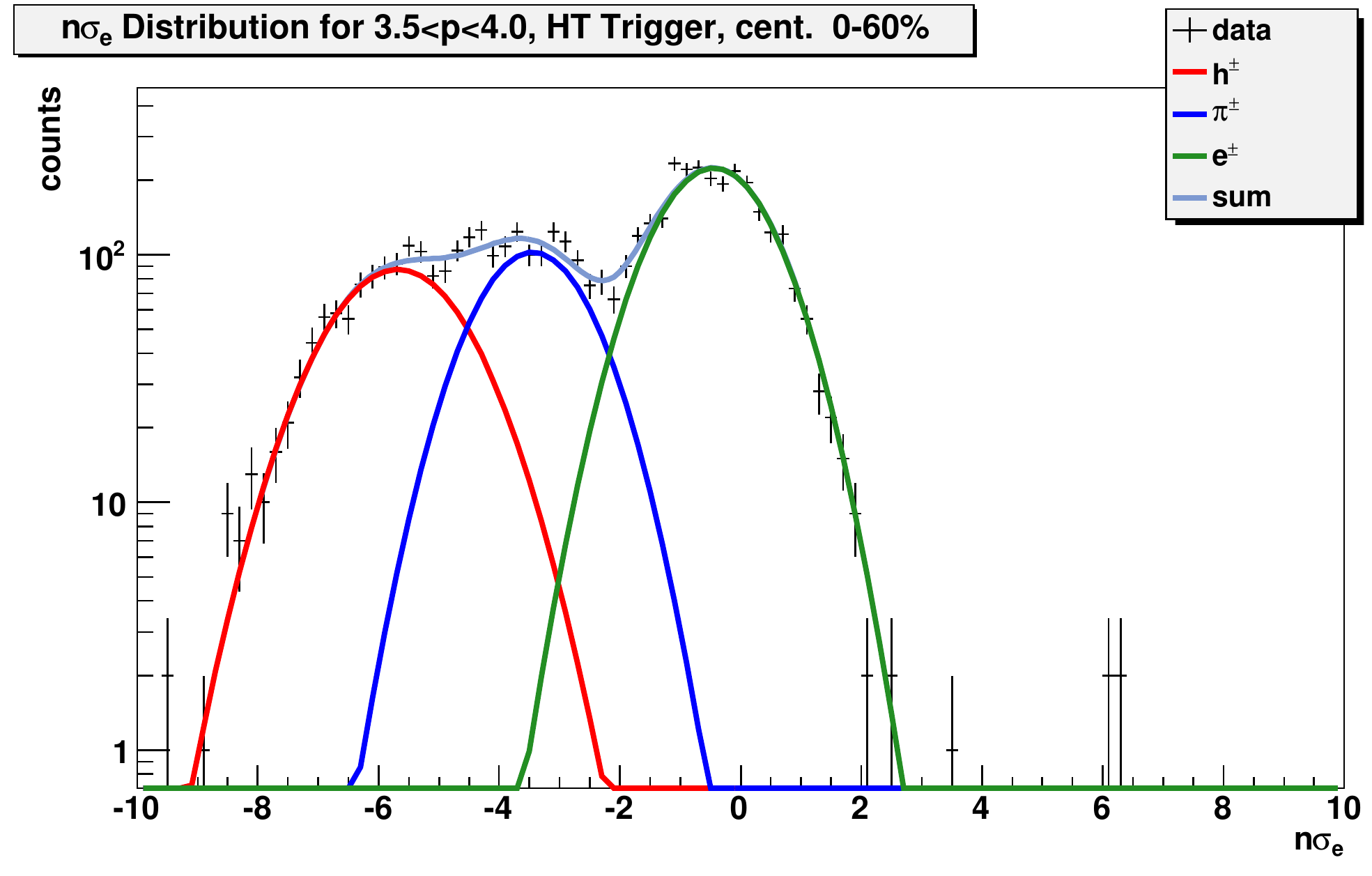}
\caption{Three-Gaussian fit of $n\sigma_{e}$ distribution: 0-60\% centrality class of high-tower triggered events; $3.5\GeV/c<p<4\GeV/c$.}
\label{fig:add:purity:fit_nsigma_ht_c03_p6}
\includegraphics[width=0.85\linewidth]{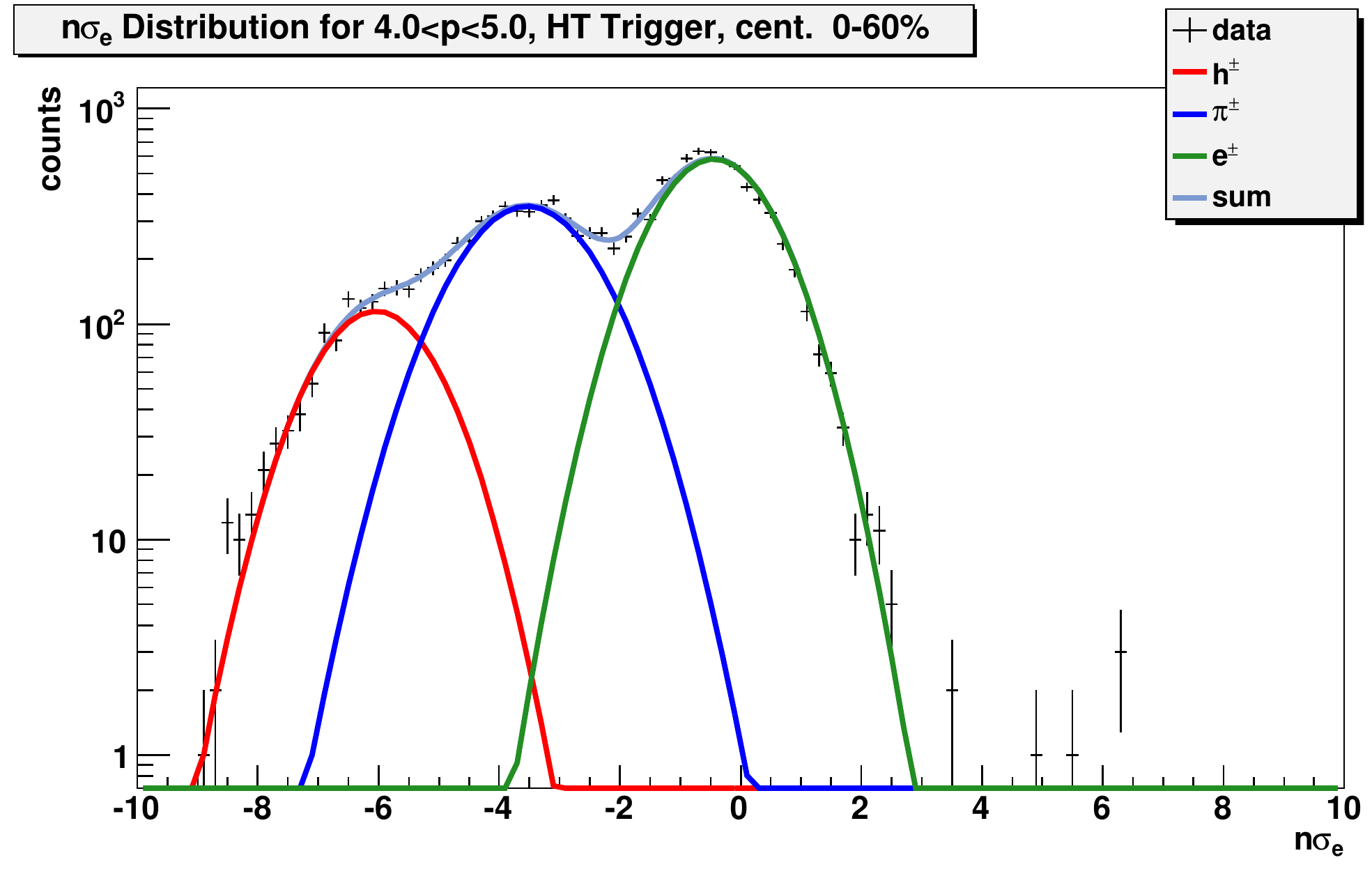}
\caption{Three-Gaussian fit of $n\sigma_{e}$ distribution: 0-60\% centrality class of high-tower triggered events; $4\GeV/c<p<5\GeV/c$.}
\label{fig:add:purity:fit_nsigma_ht_c03_p7}
\end{center}
\end{figure}

\begin{figure}[htbp]
\begin{center}
\includegraphics[width=0.85\linewidth]{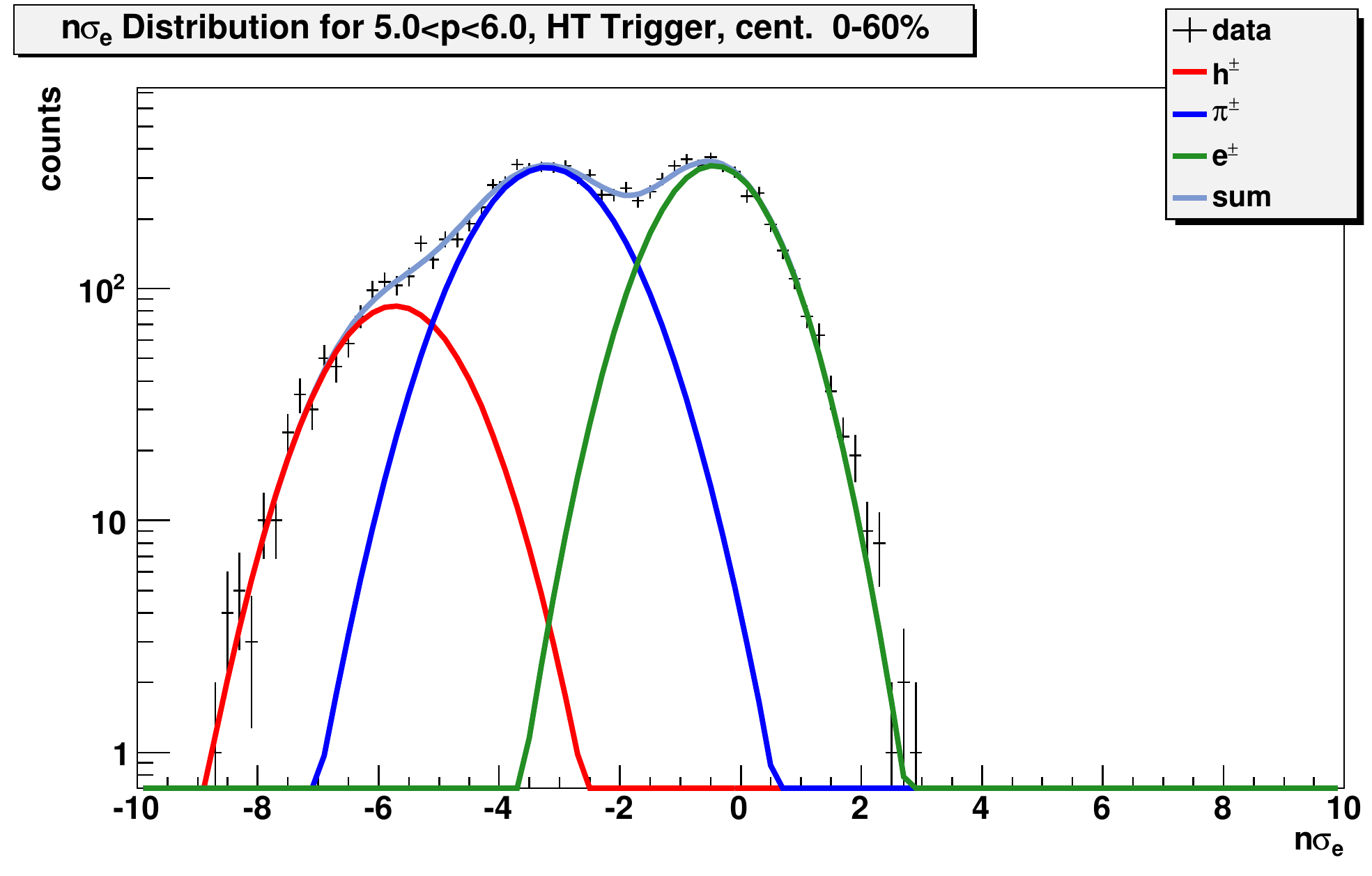}
\caption{Three-Gaussian fit of $n\sigma_{e}$ distribution: 0-60\% centrality class of high-tower triggered events; $5\GeV/c<p<6\GeV/c$.}
\label{fig:add:purity:fit_nsigma_ht_c03_p8}
\includegraphics[width=0.85\linewidth]{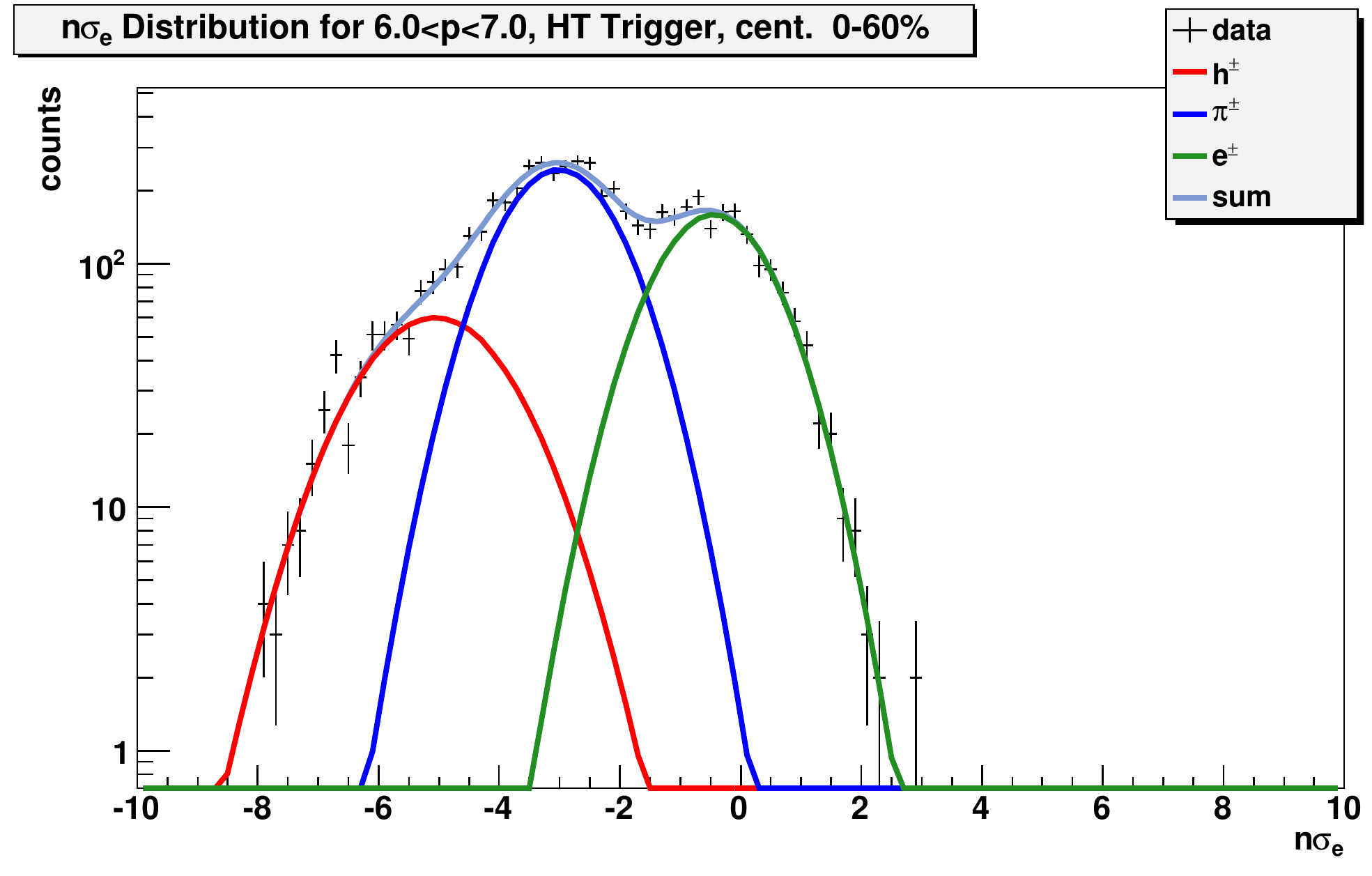}
\caption{Three-Gaussian fit of $n\sigma_{e}$ distribution: 0-60\% centrality class of high-tower triggered events; $6\GeV/c<p<7\GeV/c$.}
\label{fig:add:purity:fit_nsigma_ht_c03_p9}
\end{center}
\end{figure}

\begin{figure}[htbp]
\begin{center}
\includegraphics[width=0.85\linewidth]{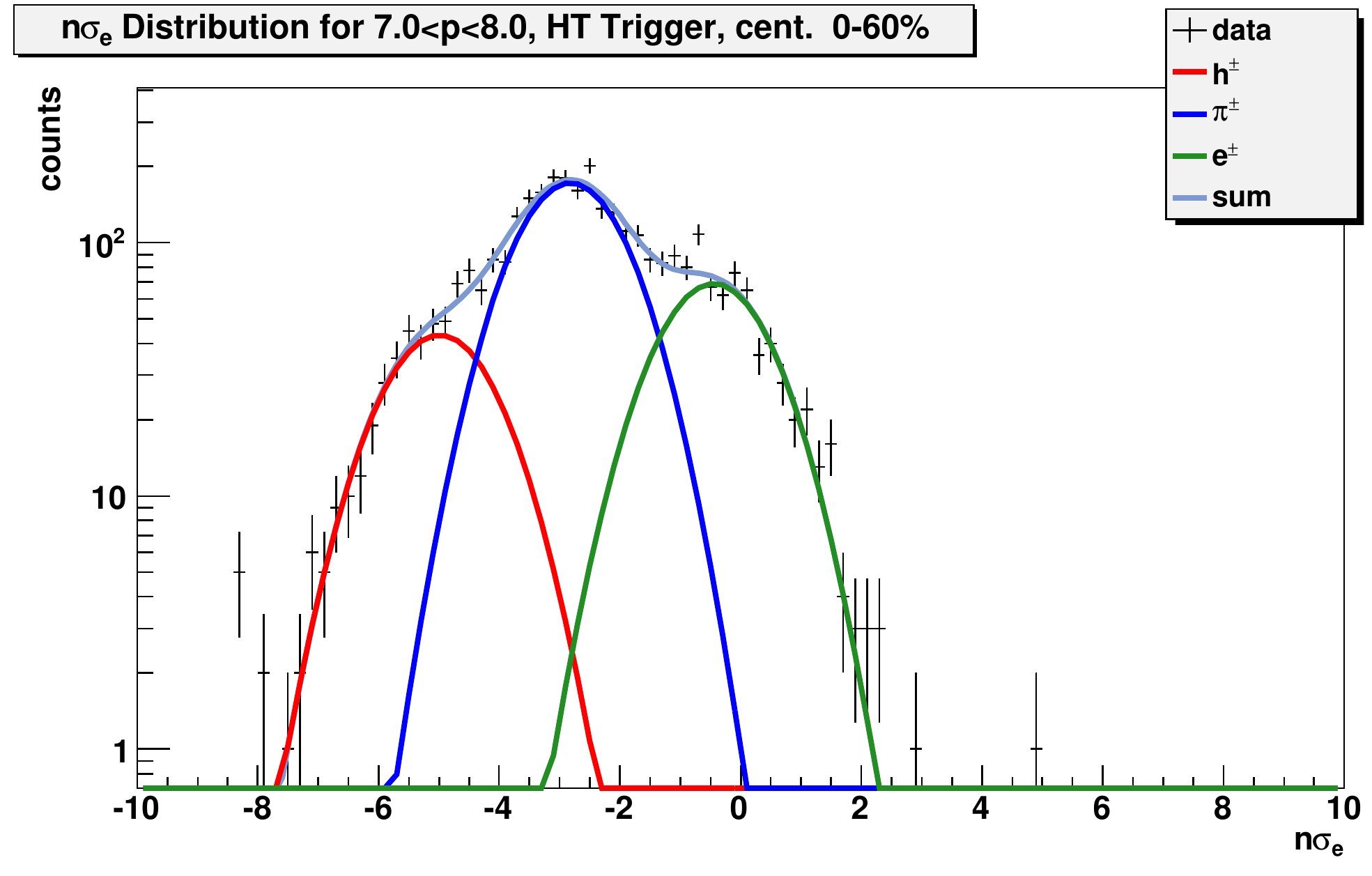}
\caption{Three-Gaussian fit of $n\sigma_{e}$ distribution: 0-60\% centrality class of high-tower triggered events; $7\GeV/c<p<8\GeV/c$.}
\label{fig:add:purity:fit_nsigma_ht_c03_p10}
\includegraphics[width=0.85\linewidth]{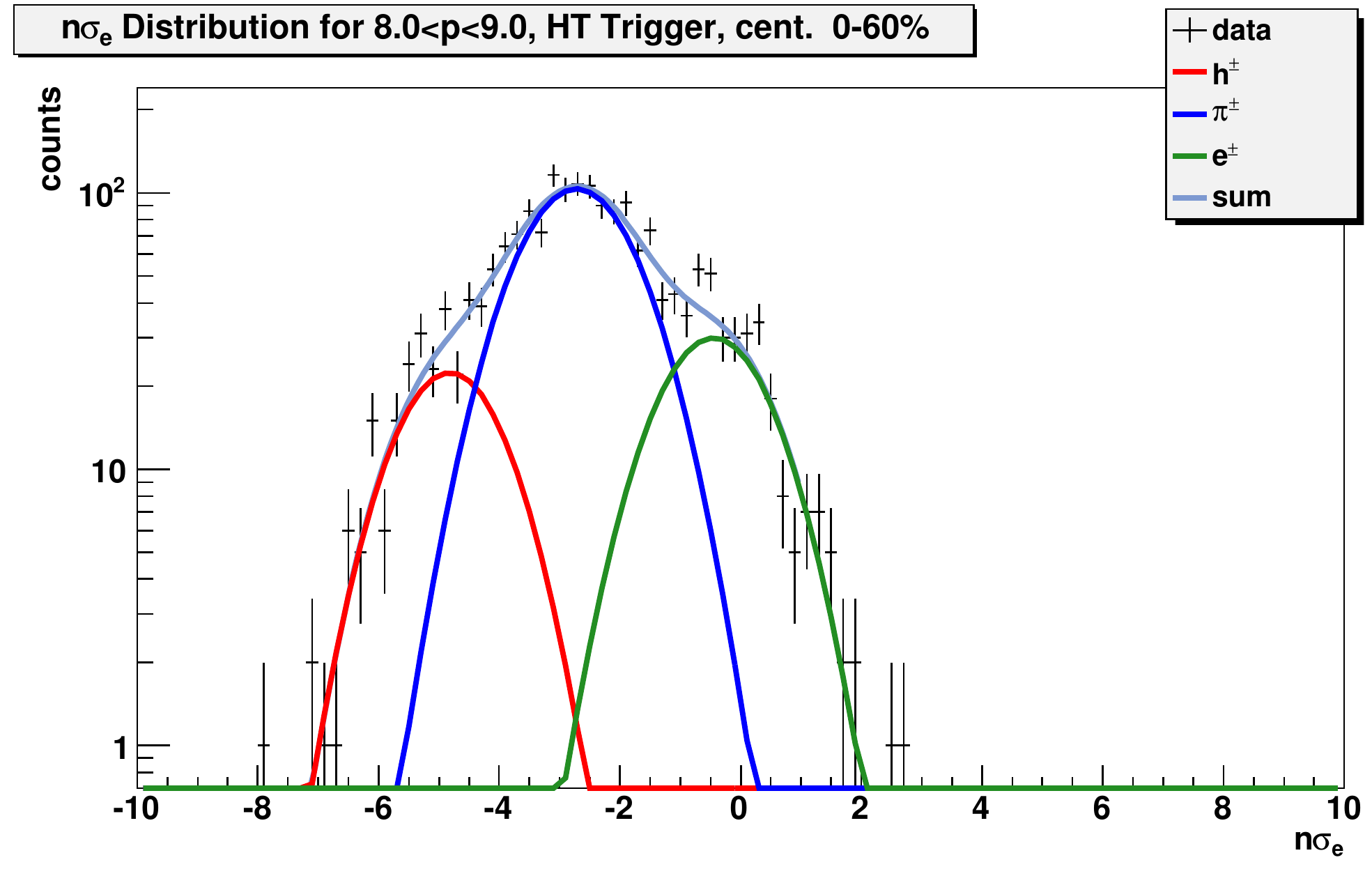}
\caption{Three-Gaussian fit of $n\sigma_{e}$ distribution: 0-60\% centrality class of high-tower triggered events; $8\GeV/c<p<9\GeV/c$.}
\label{fig:add:purity:fit_nsigma_ht_c03_p11}
\end{center}
\end{figure}

\begin{figure}[htbp]
\begin{center}
\includegraphics[width=0.85\linewidth]{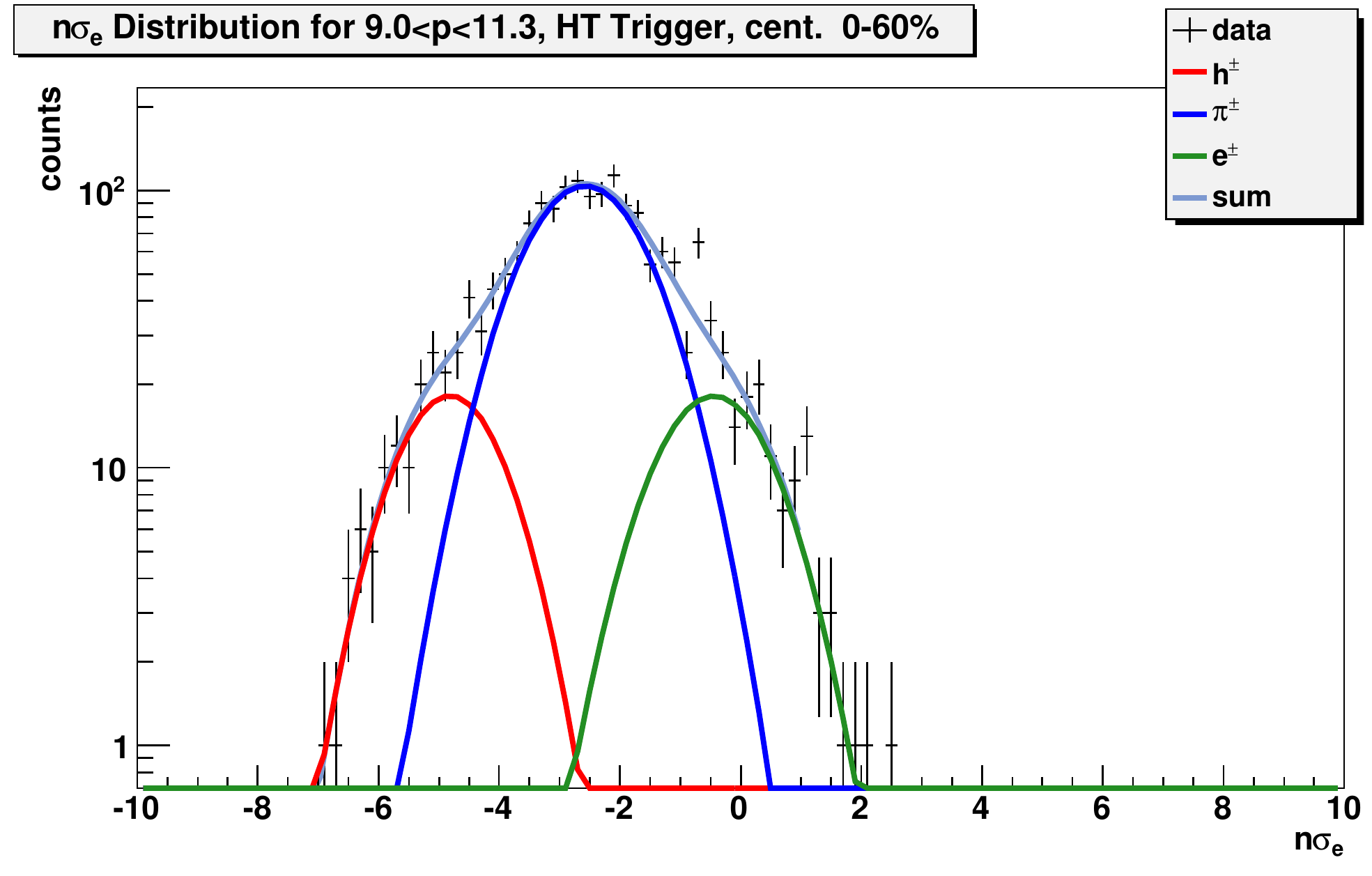}
\caption{Three-Gaussian fit of $n\sigma_{e}$ distribution: 0-60\% centrality class of high-tower triggered events; $9\GeV/c<p<11.3\GeV/c$.}
\label{fig:add:purity:fit_nsigma_ht_c03_p12}
\end{center}
\end{figure}

This section contains additional figures and tables relating to the calculation of the inclusive $e^{\pm}$ purity $(K_{inc.})$, which is described in detail in Section~\ref{sec:dedx:purity}.  Figures~\ref{fig:add:purity:fit_nsigma_mb_c03_p1} through~\ref{fig:add:purity:fit_nsigma_ht_c03_p12} are plots of the the multi-Gaussian fits to the $n\sigma_{e}$ distributions for the 0-60\% centrality class.  In each figure, the green peak accounts for the $e^{\pm}$ contribution, while the blue and red peaks account for hadron contributions (with the blue peak intended to account for charged pions).  Tables~\ref{table:add:purity:mb} and~\ref{table:add:purity:ht} contain the calculated values of $K_{inc.}$ as a function of $p_{T}$ for two event trigger types and three centrality classes.



\clearpage

\clearpage

\clearpage

\begin{table}
\caption{Inclusive $e^{\pm}$ purity $(K_{inc.})$ as a function of $p_{T}$ for the minimum-bias triggered data set.}
\label{table:add:purity:mb}
\begin{tabular}{| c | c || c | c | c |}
\hline
$p_{T}^{min}$ & $p_{T}^{max}$ & & lower & upper\\
$[\GeV/c]$ & $[\GeV/c]$ & purity & uncertainty & uncertainty\\\hline\hline
\multicolumn{5}{| c |}{MB trigger, centrality 0-20\%}\\\hline
2.0 & 2.5 & 0.97009 & 0.01152 & 0.01092\\\hline
2.5 & 3.0 & 0.93983 & 0.01851 & 0.01632\\\hline
3.0 & 3.5 & 0.88753 & 0.04563 & 0.04009\\\hline
3.5 & 4.0 & 0.84976 & 0.10105 & 0.08891\\\hline\hline
\multicolumn{5}{| c |}{MB trigger, centrality 20-60\%}\\\hline
2.0 & 2.5 & 0.97647 & 0.00977 & 0.00945\\\hline
2.5 & 3.0 & 0.96144 & 0.01309 & 0.01115\\\hline
3.0 & 3.5 & 0.93288 & 0.03145 & 0.02550\\\hline
3.5 & 4.0 & 0.87917 & 0.06763 & 0.05778\\\hline
4.0 & 4.5 & 0.83746 & 0.10990 & 0.08943\\\hline\hline
\multicolumn{5}{| c |}{MB trigger, centrality 0-60\%}\\\hline
2.0 & 2.5 & 0.97192 & 0.01088 & 0.01042\\\hline
2.5 & 3.0 & 0.95003 & 0.01716 & 0.01476\\\hline
3.0 & 3.5 & 0.90798 & 0.04234 & 0.03586\\\hline
3.5 & 4.0 & 0.87411 & 0.08126 & 0.06815\\\hline
4.0 & 4.5 & 0.82238 & 0.13420 & 0.11626\\\hline\hline
\end{tabular}
\end{table}

\begin{table}
\caption{Inclusive $e^{\pm}$ purity $(K_{inc.})$ as a function of $p_{T}$ for the high-tower triggered data set.}
\label{table:add:purity:ht}
\begin{tabular}{| c | c || c | c | c |}
\hline
$p_{T}^{min}$ & $p_{T}^{max}$ & & lower & upper\\
$[\GeV/c]$ & $[\GeV/c]$ & purity & uncertainty & uncertainty\\\hline\hline
\multicolumn{5}{| c |}{HT trigger, centrality 0-20\%}\\\hline
3.5 & 4.0 & 0.98305 & 0.01097 & 0.00777\\\hline
4.0 & 4.5 & 0.97306 & 0.00807 & 0.00715\\\hline
4.5 & 5.0 & 0.95449 & 0.01319 & 0.01195\\\hline
5.0 & 5.5 & 0.92818 & 0.02272 & 0.02054\\\hline
5.5 & 6.0 & 0.91508 & 0.02787 & 0.02516\\\hline
6.0 & 6.5 & 0.88956 & 0.03436 & 0.03138\\\hline
6.5 & 7.0 & 0.84440 & 0.03850 & 0.03681\\\hline
7.0 & 7.5 & 0.80547 & 0.04324 & 0.04287\\\hline
7.5 & 8.0 & 0.75479 & 0.06429 & 0.06755\\\hline
8.0 & 8.5 & 0.73205 & 0.06838 & 0.07179\\\hline
8.5 & 9.0 & 0.66771 & 0.10601 & 0.11702\\\hline\hline
\multicolumn{5}{| c |}{HT trigger, centrality 20-60\%}\\\hline
3.5 & 4.0 & 0.98731 & 0.00734 & 0.00485\\\hline
4.0 & 4.5 & 0.97583 & 0.00754 & 0.00658\\\hline
4.5 & 5.0 & 0.96111 & 0.01094 & 0.00982\\\hline
5.0 & 5.5 & 0.92848 & 0.01859 & 0.01725\\\hline
5.5 & 6.0 & 0.90131 & 0.02533 & 0.02391\\\hline
6.0 & 6.5 & 0.86248 & 0.03510 & 0.03356\\\hline
6.5 & 7.0 & 0.82938 & 0.04072 & 0.03925\\\hline
7.0 & 7.5 & 0.78388 & 0.04833 & 0.04759\\\hline
7.5 & 8.0 & 0.67611 & 0.06183 & 0.06770\\\hline\hline
\multicolumn{5}{| c |}{HT trigger, centrality 0-60\%}\\\hline
3.5 & 4.0 & 0.98477 & 0.00896 & 0.00637\\\hline
4.0 & 4.5 & 0.97136 & 0.00837 & 0.00746\\\hline
4.5 & 5.0 & 0.95180 & 0.01261 & 0.01155\\\hline
5.0 & 5.5 & 0.92555 & 0.02076 & 0.01912\\\hline
5.5 & 6.0 & 0.91411 & 0.02615 & 0.02403\\\hline
6.0 & 6.5 & 0.89065 & 0.03350 & 0.03105\\\hline
6.5 & 7.0 & 0.84732 & 0.03840 & 0.03677\\\hline
7.0 & 7.5 & 0.80261 & 0.04532 & 0.04410\\\hline
7.5 & 8.0 & 0.71752 & 0.07533 & 0.07068\\\hline
8.0 & 8.5 & 0.67326 & 0.07989 & 0.07604\\\hline
8.5 & 9.0 & 0.56928 & 0.13911 & 0.12581\\\hline\hline
\end{tabular}
\end{table}

\clearpage

\chapter{Catalog of Functions}
\label{sec:functions}

This Appendix is a catalog of functions used to describe the spectra of pions, $\eta$ mesons, direct photons, (anti)protons, $J/\psi$, and $e^{\pm}$.  Most of the functions take the form of a power-law term and/or an exponential term.  Unless otherwise noted, the parameters $A$ and $C$ have units of $\mathrm{mb}\cdot\GeV\,^{-2}c^{3}$ and the parameters $B$, $D$, and $T$ have units of $\GeV/c$ or $(\GeV/c)^{2}$.

\section{Pions}
\label{sec:functions:pi}
The following is a list of functions which are used to describe pion spectra.  They are used as weighting functions in calculating the background rejection efficiency $\varepsilon_{B}$ (see Chapter~\ref{sec:bre}).  The function $F[\pi^{0},2]$ is used to construct (via $m_{T}$ scaling) weighting functions for other meson species in calculating (see Chapter~\ref{sec:res_back}) the yields of residual photonic $e^{\pm}$ and non-photonic $e^{\pm}$ from light-flavor sources $(Y_{LFP})$ and $(Y_{LFN})$.  These functions are plotted in Figure~\ref{fig:bre:rhic_pion} (page~\pageref{fig:bre:rhic_pion}).

\begin{itemize}

\item $F[\pi^{0},1]$
\begin{description}
\item[Formula] $A(1+p_{T}/B)^{n}$
\item[Parameters] 393, 1.212, -9.97
\item[Data] PHENIX $\pi^{0}$ in $p+p$ (2003)~\cite{PhysRevLett.91.241803}
\end{description}

\item $F[\pi^{0},2]$
\begin{description}
\item[Formula] $A(1+p_{T}^{2}/B)^{n}+Ce^{-p_{T}/D}$
\item[Parameters] 6.79322, 1.4428, -4.21776, 274.899, 0.141229
\item[Data] PHENIX $\pi^{0}$ in $p+p$ (2007)~\cite{PhysRevD.76.051106}
\end{description}

\item $F[\pi^{0},3]$
\begin{description}
\item[Formula] $A(1+p_{T}/B)^{n}$
\item[Parameters] 1690, 0.723, -8.61
\item[Data] STAR $\pi^{0}$ in $p+p$~\cite{PhysRevC.81.064904}
\end{description}

\item $F[\pi^{0},4]$
\begin{description}
\item[Formula] $A(1+p_{T}/B)^{n}$
\item[Parameters] 13131.6, 0.558664, -8.65379
\item[Data] PHENIX $\pi^{0}$ in Cu + Cu 0-10\%~\cite{PhysRevLett.101.162301}
\item[Note] The data were scaled by $1/(\langle T_{AA}\rangle R_{AA})=42\,\mathrm{mb}/(182.7\times 0.515)$ before fitting.
\end{description}

\item $F[\pi^{\pm},1]$
\begin{description}
\item[Formula] $A(1+p_{T}^{2}/B)^{n}+Ce^{-p_{T}/D}$
\item[Parameters] 2.07378, 2.12146, -4.32962, 128.454, 0.188331
\item[Data] STAR $\pi^{0}$ in Cu + Cu 0-60\%~\cite{PhysRevC.81.054907}
\item[Note] The data were scaled by $1/(\langle T_{AA}\rangle R_{AA})=42\,\mathrm{mb}/(80.41\times 0.7)$ before fitting.
\end{description}

\end{itemize}

\section[$\eta$ Mesons]{$\boldsymbol{\eta}$ Mesons}
\label{sec:functions:eta}
The following is a list of functions which are used to describe $\eta$-meson spectra.  They are used as weighting functions in calculating the background rejection efficiency $\varepsilon_{B}$ (see Chapter~\ref{sec:res_back}).  These functions are plotted in Figure~\ref{fig:bre:rhic_eta} (page~\pageref{fig:bre:rhic_eta}).

\begin{itemize}

\item $F[\eta,1]$
\begin{description}
\item[Formula] $A(1+p_{T}/B)^{n}$
\item[Parameters] 94.2433, 1.23777, -9.63473
\item[Data] PHENIX $\eta$ mesons in $p+p$~\cite{PhysRevC.75.024909}
\end{description}

\item $F[\eta,2]$
\begin{description}
\item[Formula] $A(1+q^{2}/B)^{n}+Ce^{-q/D}$, $q=\sqrt{p_{T}^{2}+m(\eta)^2-m(\pi^{0})^2}$
\item[Parameters] $0.48\times 6.79322$, 1.4428, -4.21776, $0.48\times 274.899$, 0.141229
\item[Note] This is not a fit, but rather an expected spectrum derived from $F[\pi^{0},2]$ (see Section~\ref{sec:bre:weighting}).
\item[Data] derived from PHENIX $\pi^{0}$ in $p+p$~\cite{PhysRevD.76.051106}
\end{description}

\item $F[\eta,3]$
\begin{description}
\item[Formula] $A(1+p_{T}/B)^{n}$
\item[Parameters] 70, 1.33, -9.83
\item[Data] STAR $\eta$ mesons in $p+p$~\cite{PhysRevC.81.064904}
\end{description}

\end{itemize}

\section{Photons}
\label{sec:functions:photon}
The following is a list of functions which are used to describe photon spectra.  They are used as weighting functions in calculating the background rejection efficiency $\varepsilon_{B}$ (see Chapter~\ref{sec:res_back}).  In the function names, $\gamma_{\eta}$ indicates photons from $\eta$-meson decays, $\gamma_{D}$ indicates the non-thermal component of a direct photon spectrum, and $\gamma_{T}$ indicates thermal direct photons.

\begin{itemize}

\item $F[\gamma_{\eta},1]$
\begin{description}
\item[Formula] $A(1+p_{T}/B)^{n}$
\item[Parameters] 47.426, 1.04317, -9.69906
\item[Note] This is a fit to a simulated photon spectrum, for which $F[\eta,1]$ was used as a weighting function (see Section~\ref{sec:bre:weighting}).
\item[Data] derived from PHENIX $\eta$ mesons in $p+p$~\cite{PhysRevC.75.024909}
\end{description}

\item $F[\gamma_{\eta},2]$
\begin{description}
\item[Formula] $A(1+p_{T}/B)^{n}$
\item[Parameters] 418.887, 0.675643, -9.07911
\item[Note] This is a fit to a simulated photon spectrum, for which $F[\eta,2]$ was used as a weighting function (see Section~\ref{sec:bre:weighting}).
\item[Data] derived from PHENIX $\pi^{0}$ in $p+p$~\cite{PhysRevD.76.051106}
\end{description}

\item $F[\gamma_{\eta},3]$
\begin{description}
\item[Formula] $A(1+p_{T}/B)^{n}$
\item[Parameters] 35.4076, 1.11683, -9.88181
\item[Note] This is a fit to a simulated photon spectrum, for which $F[\eta,3]$ was used as a weighting function (see Section~\ref{sec:bre:weighting}).
\item[Data] derived from STAR $\eta$ mesons in $p+p$~\cite{PhysRevC.81.064904}
\end{description}

\item $F[\gamma_{D},1]$
\begin{description}
\item[Formula] $A(1+p_{T}/B)^{n}$
\item[Parameters] 163.226, 0.279184, -6.3762
\item[Data] PHENIX direct photons in $p+p$~\cite{PhysRevLett.104.132301,PhysRevLett.98.012002}
\item[Note] $F[\gamma_{D},1]$ and $F[\gamma_{D},2]$ are fits to the same direct-photon spectrum.  $F[\gamma_{D},1]$ was generated by the author of this dissertation.
\end{description}

\item $F[\gamma_{D},2]$
\begin{description}
\item[Formula] $A(1+p_{T}^{2}/B)^{n}$
\item[Parameters] 0.00919245, 1.798, -3.28487
\item[Data] PHENIX direct photons in $p+p$~\cite{PhysRevLett.104.132301,PhysRevLett.98.012002}
\item[Note] $F[\gamma_{D},1]$ and $F[\gamma_{D},2]$ are fits to the same direct-photon spectrum.  $F[\gamma_{D},2]$ is the published fit generated by the PHENIX Collaboration.
\end{description}

\item $F[\gamma_{D},3]$
\begin{description}
\item[Formula] $A(1+p_{T}/B)^{n}$
\item[Parameters] 0.090482, 1.36984, -7.35996
\item[Data] PHENIX direct photons in Au + Au 0-92\%~\cite{PhysRevLett.104.132301,PhysRevLett.94.232301}
\item[Note] The data were fit with $\langle T_{AA}\rangle\times F[\gamma_{D},3]$, with $\langle T_{AA}\rangle=6.14\,\mathrm{mb}^{-1}$.
\end{description}

\item $F[\gamma_{D},4]$
\begin{description}
\item[Formula] $A(1+p_{T}/B)^{n}$
\item[Parameters] 0.248478, 1.2668, -7.60368
\item[Data] PHENIX direct photons in Au + Au 0-20\%~\cite{PhysRevLett.104.132301,PhysRevLett.94.232301}
\item[Note] The data were fit with $\langle T_{AA}\rangle\times F[\gamma_{D},4]$, with $\langle T_{AA}\rangle=18.6\,\mathrm{mb}^{-1}$.
\end{description}

\item $F[\gamma_{D},5]$
\begin{description}
\item[Formula] $A(1+p_{T}/B)^{n}$
\item[Parameters] $2.14531\cdot 10^{14}$, 0.00217448, -6.05396
\item[Data] PHENIX direct photons in Au + Au 20-40\%~\cite{PhysRevLett.104.132301,PhysRevLett.94.232301}
\item[Note] The data were fit with $\langle T_{AA}\rangle\times F[\gamma_{D},5]$, with $\langle T_{AA}\rangle=7.07\,\mathrm{mb}^{-1}$.
\end{description}

\item $F[\gamma_{T},1]$
\begin{description}
\item[Formula] $Ce^{-p_{T}/T}$
\item[Parameters] 3.24934, 0.233
\item[Data] PHENIX direct photons in Au + Au 0-92\%~\cite{PhysRevLett.104.132301,PhysRevLett.94.232301}
\item[Note] The data were fit with $\langle T_{AA}\rangle\times(F[\gamma_{D},2]+F[\gamma_{T},1])$, with $\langle T_{AA}\rangle=6.14\,\mathrm{mb}^{-1}$.
\end{description}

\item $F[\gamma_{T},2]$
\begin{description}
\item[Formula] $Ce^{-p_{T}/T}$
\item[Parameters] 4.38509, 0.221
\item[Data] PHENIX direct photons in Au + Au 0-20\%~\cite{PhysRevLett.104.132301,PhysRevLett.94.232301}
\item[Note] The data were fit with $\langle T_{AA}\rangle\times(F[\gamma_{D},2]+F[\gamma_{T},1])$, with $\langle T_{AA}\rangle=18.6\,\mathrm{mb}^{-1}$.
\end{description}

\item $F[\gamma_{T},3]$
\begin{description}
\item[Formula] $Ce^{-p_{T}/T}$
\item[Parameters] 5.62872, 0.217
\item[Data] PHENIX direct photons in Au + Au 20-40\%~\cite{PhysRevLett.104.132301,PhysRevLett.94.232301}
\item[Note] The data were fit with $\langle T_{AA}\rangle\times(F[\gamma_{D},2]+F[\gamma_{T},1])$, with $\langle T_{AA}\rangle=7.07\,\mathrm{mb}^{-1}$.
\end{description}

\end{itemize}

\section{Protons}
\label{sec:functions:proton}
The following is a list of functions which are used to describe proton spectra.  The function $F[p,4]$ is used as a weighting function in calculating the yield of non-photonic $e^{\pm}$ from baryon decays (which is found to be negligible, see Section~\ref{sec:res_back:lfn}).  These functions are plotted in Figure~\ref{fig:res_back:star_proton} (page~\pageref{fig:res_back:star_proton}).

\begin{itemize}

\item $F[p,1]$
\begin{description}
\item[Formula] $A(1+p_{T}^{2}/B)^{n}+Ce^{-p_{T}/D}$
\item[Parameters] 0.953737, 3.27513, -4.43296, 44.4136, 0.343277
\item[Data] STAR (anti)protons in Au + Au 0-12\%~\cite{PhysRevLett.97.152301}
\item[Note] Parameters $A$ and $C$ are measured in $\GeV\,^{-2}c^{3}$.
\end{description}

\item $F[p,2]$
\begin{description}
\item[Formula] $A(1+p_{T}^{2}/B)^{n}+Ce^{-p_{T}/D}$
\item[Parameters] 0.703249, 3.34638, -4.42059, 32.2110, 0.344818
\item[Data] STAR (anti)protons in Au + Au 10-20\%~\cite{PhysRevLett.97.152301}
\item[Note] Parameters $A$ and $C$ are measured in $\GeV\,^{-2}c^{3}$.
\end{description}

\item $F[p,3]$
\begin{description}
\item[Formula] $A(1+p_{T}^{2}/B)^{n}+Ce^{-p_{T}/D}$
\item[Parameters] 0.122992, 3.62065, -4.07714, 18.3999, 0.347876
\item[Data] STAR (anti)protons in Au + Au 20-40\%~\cite{PhysRevLett.97.152301}
\item[Note] Parameters $A$ and $C$ are measured in $\GeV\,^{-2}c^{3}$.
\end{description}

\item $F[p,4]$
\begin{description}
\item[Formula] $A(1+p_{T}^{2}/B)^{n}+Ce^{-p_{T}/D}$
\item[Parameters] 0.307271, 3.68843, -4.74017, 9.59262, 0.310598
\item[Data] STAR (anti)protons in Au + Au 40-60\%~\cite{PhysRevLett.97.152301}
\item[Note] Parameters $A$ and $C$ are measured in $\GeV\,^{-2}c^{3}$.
\end{description}

\item $F[p,5]$
\begin{description}
\item[Formula] $A(1+p_{T}^{2}/B)^{n}+Ce^{-p_{T}/D}$
\item[Parameters] 1.26394, 1.49505, -4.29711, 0.205822, 0.338809
\item[Data] STAR (anti)protons in Au + Au 60-80\%~\cite{PhysRevLett.97.152301}
\item[Note] Parameters $A$ and $C$ are measured in $\GeV\,^{-2}c^{3}$.
\end{description}

\end{itemize}

\section[$J/\psi$]{$\boldsymbol{J/\psi}$}
\label{sec:functions:jpsi}
These $J/\psi$ weighting functions are used calculating the yield of photonic $e^{\pm}$ from $J/\psi$ decays (see Section~\ref{sec:res_back:jpsi}).  The fit functions for Cu + Cu collisions are plotted in Figure~\ref{fig:res_back:jpsi_spectra}.

\begin{itemize}
\item $F[J/\psi,pp]$, $F[J/\psi,pp-]$, $F[J/\psi,pp+]$
\begin{description}
\item[Formula] $A[\exp(-ap_{T}-bp_{T}^2)+p_{T}/p_{0}]^{-n}$
\item[Parameters] (5.23757, 0.323357, 0.0573737, 2.59389, 8.44212), (3.74232, 0.354190, 0.0320993, 2.74964, 8.71965), (6.86912, 0.297896, 0.0749802, 2.50527, 8.27808)
\item[Data] PHENIX~\cite{PhysRevD.82.012001} and STAR~\cite{PhysRevC.80.041902} $J/\psi$ in $p+p$
\item[Note] Three fits were performed to obtain a ``central" fit $(F[J/\psi,pp])$, and lower $(F[J/\psi,pp-])$ and upper $(F[J/\psi,pp+])$ limits.  Units of parameters: $A$ in $\mathrm{nb\, GeV\,^{-2}}c^{3}$; $a$ in $(\GeV/c)^{-1}$; $b$ in $(\GeV/c)^{-2}$; $p_{0}$ in $\GeV/c$; $n$ unit-less
\end{description}

\item $F[J/\psi,0-20\%]$, $F[J/\psi,0-20\%-]$, $F[J/\psi,0-20\%+]$
\begin{description}
\item[Formula] $A[\exp(-ap_{T}-bp_{T}^2)+p_{T}/p_{0}]^{-n}$
\item[Parameters] (9.10922, 0.221385, -0.0100289 ,4.06224, 10.0477), (5.35754, 0.449842, -0.0632743, 2.87785, 5.89605), (11.5138, 0.278625, 0.00748965, 3.07351, 8.44259)
\item[Data] PHENIX~\cite{PhysRevLett.101.122301} and STAR~\cite{PhysRevC.80.041902} $J/\psi$ in Cu + Cu 0-20\%
\item[Note] Three fits were performed to obtain a ``central" fit $(F[J/\psi,0-20\%])$, and lower $(F[J/\psi,0-20\%-])$ and upper $(F[J/\psi,0-20\%+])$ limits.  Units of parameters: $A$ in $\mathrm{nb\, GeV\,^{-2}}c^{3}$; $a$ in $(\GeV/c)^{-1}$; $b$ in $(\GeV/c)^{-2}$; $p_{0}$ in $\GeV/c$; $n$ unit-less
\end{description}

\item $F[J/\psi,20-60\%]$ and $F[J/\psi,20-60\%-]$
\begin{description}
\item[Formula] $A[\exp(-ap_{T}-bp_{T}^2)+p_{T}/p_{0}]^{-n}$
\item[Parameters] (3.71419, 0.113357, -0.0073416, 7.67173, 18.2388), (3.236, 0.111221, -0.0090455, 7.87763, 17.3856)
\item[Data] PHENIX~\cite{PhysRevLett.101.122301} $J/\psi$ in Cu + Cu 20-60\%
\item[Note] Two fits were performed to obtain a ``central" fit $(F[J/\psi,20-60\%])$ and a lower limit $(F[J/\psi,20-60\%-])$.  See Section~\ref{sec:res_back:jpsi} for a description of the calculation of the upper limit $(F[J/\psi,20-60\%+])$.  Units of parameters: $A$ in $\mathrm{nb\, GeV\,^{-2}}c^{3}$; $a$ in $(\GeV/c)^{-1}$; $b$ in $(\GeV/c)^{-2}$; $p_{0}$ in $\GeV/c$; $n$ unit-less
\end{description}
\end{itemize}

\section[Non-Photonic $e^{\pm}$]{Non-Photonic $\boldsymbol{e^{\pm}}$}
\label{sec:functions:npe}
The following is a list of functions which are used to describe non-photonic $e^{\pm}$ spectra.

\begin{itemize}

\item $F[e_{N},1]$, $F[e_{N},1-]$, and $F[e_{N},1+]$
\begin{description}
\item[Formula] $A(1+p_{T}/B)^{n}$
\item[Parameters] $(A,B,n)$=(0.0757745, 1.31713, -9.20756), (0.0691622, 1.27749, -9.05677), (0.167812, 1.02855, -8.54378)
\item[Data] PHENIX NPE in $p+p$~\cite{PhysRevLett.97.252002}
\item[Note] Three fits were performed to obtain a ``central" fit $(F[e_{N},1])$, and lower $(F[e_{N},1-])$ and upper $(F[e_{N},1+])$ limits.
\end{description}

\item $F[e_{N},2]$, $F[e_{N},2-]$, and $F[e_{N},2+]$
\begin{description}
\item[Formula] $A(1+p_{T}/B)^{n}$
\item[Parameters] $(A,B,n)$=(0.00228122, 3.27957, -11.8724), ($1.85453\cdot 10^{-4}$, 15.6626, -31.0754), (0.0499159, 1.23543, -8.62345)
\item[Data] STAR NPE in $p+p$~\cite{STAR_ppNPE2011}
\item[Note] Three fits were performed to obtain a ``central" fit $(F[e_{N},2])$, and lower $(F[e_{N},2-])$ and upper $(F[e_{N},2+])$ limits.
\end{description}

\item $F[e_{N},I_{0}]$
\begin{description}
\item[Formula] $A(1+p_{T}/B)^{n}$
\item[Parameters] 3488.44, 0.208018, -7.80847
\item[Data] STAR NPE in Cu + Cu [this dissertation]
\item[Note] fit to iteration 0 of the efficiency-corrected non-photonic $e^{\pm}$ yield $(N_{EC})$; parameter $A$ in $\GeV\,^{-2}c^{3}$
\end{description}

\item $F[e_{N},I_{1}]$
\begin{description}
\item[Formula] $A(1+p_{T}/B)^{n}$
\item[Parameters] 3866.20, 0.207808, -7.83930
\item[Data] STAR NPE in Cu + Cu [this dissertation]
\item[Note] fit to iteration 1 of the efficiency-corrected non-photonic $e^{\pm}$ yield $(N_{EC})$; parameter $A$ in $\GeV\,^{-2}c^{3}$
\end{description}

\end{itemize}

\clearpage

\bibliographystyle{h-physrev}
\addcontentsline{toc}{chapter}{Bibliography} 
\bibliography{bibliography/thesis_misc,bibliography/thesis_star,bibliography/thesis_phenix,bibliography/thesis_lhc}

\begin{thebibliography}{100}

\bibitem{FinnegansWake}
J.~Joyce,
\newblock {\em Finnegans Wake} (Farber and Farber Limited, London, 1939).

\bibitem{Restaurant}
D.~Adams,
\newblock {\em The Restaurant at the End of the Universe} (Harmony Books, New
  York, 1980).

\bibitem{GriffithsParticles}
D.~Griffiths,
\newblock {\em Introduction to Elementary Particles} (Wiley-VCH, Weinheim,
  Germany, 1987).

\bibitem{HalzenMartin}
F.~Halzen and A.~Martin,
\newblock {\em Quarks and Leptons} (John Wiley \& Sons, New York, New York,
  1987).

\bibitem{PeskinSchroeder}
M.~Peskin and D.~Schroeder,
\newblock {\em An Introduction to Quantum Field Theory} (Perseus Books,
  Cambridge, Massachusetts, 1995).

\bibitem{Bethke2007351}
S.~Bethke, {\em Experimental tests of asymptotic freedom},
\newblock Progress in Particle and Nuclear Physics {\bf 58}, 351  (2007).

\bibitem{STAR_white_paper2005}
J.~Adams {\em et~al.}, {\em Experimental and theoretical challenges in the
  search for the quark-gluon plasma: The STAR Collaboration's critical
  assessment of the evidence from RHIC collisions},
\newblock Nuclear Physics A {\bf 757}, 102  (2005),
\newblock First Three Years of Operation of RHIC.

\bibitem{KarschLattice2002}
F.~Karsch, {\em Lattice results on QCD thermodynamics},
\newblock Nuclear Physics A {\bf 698}, 199  (2002).

\bibitem{StephanovQCDphase}
M.~Stephanov, {\em QCD Phase Diagram: An Overview},
\newblock PoSLAT2006:024, arXiv:hep-lat/0701002  (2006).

\bibitem{STAR_CriticalPoint2010}
M.~M. Aggarwal {\em et~al.} ({STAR} Collaboration), {\em An Experimental
  Exploration of the {QCD} Phase Diagram: The Search for the Critical Point and
  the Onset of De-confinement},
\newblock {arXiv}:1007.2613  (2010).

\bibitem{RHIC_ScienceMag1999}
D.~Voss, {\em Making the Stuff of the Big Bang},
\newblock Science {\bf 285}, 1194 (1999).

\bibitem{PhysRevC.70.054907}
J.~Adams {\em et~al.} ({STAR} Collaboration), {\em Measurements of transverse
  energy distributions in $Au+Au$ collisions at $\sqrt{s_{NN}}=200$ {GeV}},
\newblock Phys. Rev. C {\bf 70}, 054907 (2004).

\bibitem{UllrichRHICOverview2007}
T.~Ullrich,
\newblock {RHIC} experimental overview - what we have learned,
\newblock in {\em {Colliders to Cosmic Rays 2007} conference}, 2007.

\bibitem{BassQM2001}
S.~Bass,
\newblock Collision dynamics and cascades,
\newblock in {\em {Quark Matter 2001}}, 2001.

\bibitem{Glauber1959}
R.~Glauber,
\newblock {\em Lectures in Theoretical Physics} (Interscience, New York, 1959).

\bibitem{AnnualReviewGlauber}
M.~L. Miller, K.~Reygers, S.~J. Sanders, and P.~Steinberg, {\em Glauber
  Modeling in High-Energy Nuclear Collisions},
\newblock Annual Review of Nuclear and Particle Science {\bf 57}, 205 (2007).

\bibitem{PDG_review}
K.~Nakamura {\em et~al.} (Particle Data Group Collaboration), {\em Review of
  Particle Physics},
\newblock Journal of Physics G: Nuclear and Particle Physics {\bf 37}, 075021
  (2010).

\bibitem{PhysRevLett.70.525}
M.~Honda {\em et~al.}, {\em Inelastic cross section for p-air collisions from
  air shower experiments and total cross section for p-p collisions up to
  $\sqrt{s}$ =24 TeV},
\newblock Phys. Rev. Lett. {\bf 70}, 525 (1993).

\bibitem{MillerThesis}
M.~Miller,
\newblock {\em Measurement of Jets and Jet Quenching at {RHIC}},
\newblock PhD thesis, Yale University, 2003.

\bibitem{PhysRevLett.90.082302}
C.~Adler {\em et~al.} ({STAR} Collaboration), {\em Disappearance of
  Back-To-Back High-$p_{T}$ Hadron Correlations in Central $Au+Au$ Collisions
  at $\sqrt{s_{NN}}=200\,\,GeV$},
\newblock Phys. Rev. Lett. {\bf 90}, 082302 (2003).

\bibitem{PhysRevLett.91.072304}
J.~Adams {\em et~al.} ({STAR} Collaboration), {\em Evidence from $d+Au$
  Measurements for Final-State Suppression of High-$p_{T}$ Hadrons in $Au+Au$
  Collisions at RHIC},
\newblock Phys. Rev. Lett. {\bf 91}, 072304 (2003).

\bibitem{PhysRevLett.93.252301}
J.~Adams {\em et~al.} ({STAR} Collaboration), {\em Azimuthal Anisotropy and
  Correlations at Large Transverse Momenta in $p+p$ and $Au+Au$ Collisions at
  $\sqrt{s_{NN}}=200\,\,GeV$},
\newblock Phys. Rev. Lett. {\bf 93}, 252301 (2004).

\bibitem{KolbHeinzHydro2003}
P.~Kolb and U.~Heinz, {\em Hydrodynamic description of ultrarelativistic
  heavy-ion collisions},
\newblock nucl-th/0305084  (2003).

\bibitem{PhysRevLett.87.182301}
C.~Adler {\em et~al.} ({STAR} Collaboration), {\em Identified Particle Elliptic
  Flow in $Au+Au$ Collisions at $\sqrt{s_{NN}}=130\;GeV$},
\newblock Phys. Rev. Lett. {\bf 87}, 182301 (2001).

\bibitem{Huovinen200158}
P.~Huovinen, P.~Kolb, U.~Heinz, P.~Ruuskanen, and S.~Voloshin, {\em Radial and
  elliptic flow at RHIC: further predictions},
\newblock Physics Letters B {\bf 503}, 58  (2001).

\bibitem{PhysRevLett.92.052302}
J.~Adams {\em et~al.} ({STAR} Collaboration), {\em Particle-Type Dependence of
  Azimuthal Anisotropy and Nuclear Modification of Particle Production in
  $Au+Au$ Collisions at $\sqrt{s_{NN}}=200\,\,GeV$},
\newblock Phys. Rev. Lett. {\bf 92}, 052302 (2004).

\bibitem{PhysRevLett.95.122301}
J.~Adams {\em et~al.} ({STAR} Collaboration), {\em Multistrange Baryon Elliptic
  Flow in $Au+Au$ Collisions at $\sqrt{s_{NN}}=200\,\,GeV$},
\newblock Phys. Rev. Lett. {\bf 95}, 122301 (2005).

\bibitem{PhysRevLett.98.162301}
A.~Adare {\em et~al.} ({PHENIX} Collaboration), {\em Scaling Properties of
  Azimuthal Anisotropy in $Au+Au$ and $Cu+Cu$ Collisions at
  $\sqrt{s_{NN}}=200\,\,GeV$},
\newblock Phys. Rev. Lett. {\bf 98}, 162301 (2007).

\bibitem{VoloshinZhang1996}
S.~Voloshin and Y.~Zhang, {\em Flow study in relativistic nuclear collisions by
  Fourier expansion of azimuthal particle distributions},
\newblock Zeitschrift f\"ur Physik C {\bf 70}, 665 (1996).

\bibitem{PhysRevLett.91.072301}
S.~S. Adler {\em et~al.} ({PHENIX} Collaboration), {\em Suppressed $\pi{}^{0}$
  Production at Large Transverse Momentum in Central $Au+Au$ Collisions at
  $\sqrt{s_{NN}}=200\,\,GeV$},
\newblock Phys. Rev. Lett. {\bf 91}, 072301 (2003).

\bibitem{PhysRevC.76.034904}
S.~S. Adler {\em et~al.} ({PHENIX} Collaboration), {\em Detailed study of
  high-$p_{T}$ neutral pion suppression and azimuthal anisotropy in $Au+Au$
  collisions at $\sqrt{s_{NN}}=200$ GeV},
\newblock Phys. Rev. C {\bf 76}, 034904 (2007).

\bibitem{PhysRevC.75.024909}
S.~S. Adler {\em et~al.} ({PHENIX} Collaboration), {\em High transverse
  momentum $\eta$ meson production in $p+p$, $d+Au$, and $Au+Au$ collisions at
  $\sqrt{s_{NN}}=200$ GeV},
\newblock Phys. Rev. C {\bf 75}, 024909 (2007).

\bibitem{PhysRevLett.94.232301}
S.~S. Adler {\em et~al.} ({PHENIX} Collaboration), {\em Centrality Dependence
  of Direct Photon Production in $\sqrt{s_{NN}}=200\,\,GeV$ $Au+Au$
  Collisions},
\newblock Phys. Rev. Lett. {\bf 94}, 232301 (2005).

\bibitem{PhysRevLett.89.252301}
I.~Vitev and M.~Gyulassy, {\em High-$p_{T}$ Tomography of $d+Au$ and $Au+Au$ at
  SPS, RHIC, and LHC},
\newblock Phys. Rev. Lett. {\bf 89}, 252301 (2002).

\bibitem{VitevJetTomographyQM204}
I.~Vitev, {\em Jet Tomography},
\newblock Journal of Physics G {\bf 30}, S791 (2004).

\bibitem{PhysRevC.72.014905}
M.~G. Mustafa, {\em Energy loss of charm quarks in the quark-gluon plasma:
  Collisional vs radiative losses},
\newblock Phys. Rev. C {\bf 72}, 014905 (2005).

\bibitem{PhysRevC.74.064907}
M.~Djordjevic, {\em Collisional energy loss in a finite size QCD matter},
\newblock Phys. Rev. C {\bf 74}, 064907 (2006).

\bibitem{BraunMunzingerThermal2003}
P.~Braun-Munzinger, K.~Redlich, and J.~Stachel, {\em Particle Production in
  Heavy Ion Collisions},
\newblock arXiv:nucl-th/0304013v1  (2003).

\bibitem{Cleymans1986217}
J.~Cleymans, R.~V. Gavai, and E.~Suhonen, {\em Quarks and gluons at high
  temperatures and densities},
\newblock Physics Reports {\bf 130}, 217  (1986).

\bibitem{PhysRevLett.94.062301}
J.~Adams {\em et~al.} ({STAR} Collaboration), {\em Open Charm Yields in $d+Au$
  Collisions at $\sqrt{s_{NN}}=200\,\hskip0.056em\hskip0.056emGeV$},
\newblock Phys. Rev. Lett. {\bf 94}, 062301 (2005).

\bibitem{Zhang2006701}
H.~Zhang, {\em Open Charm Production in $Au+Au$ Collisions at STAR},
\newblock Nuclear Physics A {\bf 774}, 701  (2006),
\newblock QUARK MATTER 2005 - Proceedings of the 18th International Conference
  on Ultra-Relativistic Nucleus--Nucleus Collisions.

\bibitem{BaumgartThesis}
S.~Baumgart,
\newblock {\em A Study of Open Charm Production in Heavy Ion Collisions of
  Center-of-Mass Energy 200 GeV per Nucleon},
\newblock PhD thesis, Yale University, 2009.

\bibitem{PhysRevLett.98.232301}
A.~Adare {\em et~al.} ({PHENIX} Collaboration), {\em $J/\psi{}$ Production
  versus Centrality, Transverse Momentum, and Rapidity in $Au+Au$ Collisions at
  $\sqrt{s_{NN}}=200\,\,GeV$},
\newblock Phys. Rev. Lett. {\bf 98}, 232301 (2007).

\bibitem{PhysRevLett.96.012304}
S.~S. Adler {\em et~al.} ({PHENIX} Collaboration), {\em $J/\psi{}$ Production
  and Nuclear Effects for $d+Au$ and $p+p$ Collisions at
  $\sqrt{s_{NN}}=200\,\,GeV$},
\newblock Phys. Rev. Lett. {\bf 96}, 012304 (2006).

\bibitem{AverbeckQM2006}
R.~Averbeck, {\em Heavy-quark and electromagnetic probes},
\newblock Journal of Physics G: Nuclear and Particle Physics {\bf 34}, S567
  (2007).

\bibitem{PhysRevD.82.012001}
A.~Adare {\em et~al.} ({PHENIX} Collaboration), {\em Transverse momentum
  dependence of $J/\psi{}$ polarization at midrapidity in $p+p$ collisions at
  $\sqrt{s}=200\,\,GeV$},
\newblock Phys. Rev. D {\bf 82}, 012001 (2010).

\bibitem{PhysRevC.80.041902}
B.~I. Abelev {\em et~al.} ({STAR} Collaboration), {\em $J/\psi{}$ production at
  high transverse momenta in $p+p$ and $Cu+Cu$ collisions at
  $\sqrt{s_{NN}}=200$ GeV},
\newblock Phys. Rev. C {\bf 80}, 041902 (2009).

\bibitem{PhysRevD.82.012004}
B.~I. Abelev {\em et~al.} ({STAR} Collaboration), {\em $\Upsilon{}$ cross
  section in $p+p$ collisions at $\sqrt{s}=200\,\,GeV$},
\newblock Phys. Rev. D {\bf 82}, 012004 (2010).

\bibitem{PhysRevD.76.092002}
S.~S. Adler {\em et~al.} ({PHENIX} Collaboration), {\em Measurement of single
  muons at forward rapidity in $p+p$ collisions at $\sqrt{s}=200\,\,GeV$ and
  implications for charm production},
\newblock Phys. Rev. D {\bf 76}, 092002 (2007).

\bibitem{ZhongQM2006}
C.~Zhong (for {STAR} Collaboration), {\em Scaling of the charm cross-section
  and modification of charm $p_{T}$ spectra at {RHIC}},
\newblock Journal of Physics G: Nuclear and Particle Physics {\bf 34}, S741
  (2007).

\bibitem{PhysRevLett.97.252002}
A.~Adare {\em et~al.} ({PHENIX} Collaboration), {\em Measurement of
  High-$p_{T}$ Single Electrons from Heavy-Flavor Decays in $p+p$ Collisions at
  $\sqrt{s}=200\,\,GeV$},
\newblock Phys. Rev. Lett. {\bf 97}, 252002 (2006).

\bibitem{PhysRevLett.98.172301}
A.~Adare {\em et~al.} ({PHENIX} Collaboration), {\em Energy Loss and Flow of
  Heavy Quarks in $Au+Au$ Collisions at $\sqrt{s_{NN}}=200\,\,GeV$},
\newblock Phys. Rev. Lett. {\bf 98}, 172301 (2007).

\bibitem{VogtHeavyCrossSections2009}
R.~Vogt, {\em Determining the uncertainty on the total heavy-flavor cross
  section},
\newblock The European Physical Journal C - Particles and Fields {\bf 61}, 793
  (2009).

\bibitem{PhysRevLett.95.122001}
M.~Cacciari, P.~Nason, and R.~Vogt, {\em QCD Predictions for Charm and Bottom
  Quark Production at RHIC},
\newblock Phys. Rev. Lett. {\bf 95}, 122001 (2005).

\bibitem{PHENIX_NPE2010}
A.~Adare {\em et~al.} ({PHENIX} Collaboration), {\em Heavy Quark Production in
  $p+p$ and Energy Loss and Flow of Heavy Quarks in $Au+Au$ Collisions at
  $\sqrt{s_{NN}}$=200 {GeV}},
\newblock arXiv:1005.1627v2  (2010).

\bibitem{PhysRevLett.94.112301}
M.~Djordjevic, M.~Gyulassy, and S.~Wicks, {\em Open Charm and Beauty at
  Ultrarelativistic Heavy Ion Colliders},
\newblock Phys. Rev. Lett. {\bf 94}, 112301 (2005).

\bibitem{Djordjevic200681}
M.~Djordjevic, M.~Gyulassy, R.~Vogt, and S.~Wicks, {\em Influence of bottom
  quark jet quenching on single electron tomography of Au + Au},
\newblock Physics Letters B {\bf 632}, 81  (2006).

\bibitem{Djordjevic2004265}
M.~Djordjevic and M.~Gyulassy, {\em Heavy quark radiative energy loss in QCD
  matter},
\newblock Nuclear Physics A {\bf 733}, 265  (2004).

\bibitem{Wicks2007426}
S.~Wicks, W.~Horowitz, M.~Djordjevic, and M.~Gyulassy, {\em Elastic, inelastic,
  and path length fluctuations in jet tomography},
\newblock Nuclear Physics A {\bf 784}, 426  (2007).

\bibitem{Armesto2006362}
N.~Armesto, M.~Cacciari, A.~Dainese, C.~A. Salgado, and U.~A. Wiedemann, {\em
  How sensitive are high-pT electron spectra at RHIC to heavy quark energy
  loss?},
\newblock Physics Letters B {\bf 637}, 362  (2006).

\bibitem{Baier1997291}
R.~Baier, Y.~L. Dokshitzer, A.~H. Mueller, S.~Peigné, and D.~Schiff, {\em
  Radiative energy loss of high energy quarks and gluons in a finite-volume
  quark-gluon plasma},
\newblock Nuclear Physics B {\bf 483}, 291  (1997).

\bibitem{PhysRevC.71.064904}
G.~D. Moore and D.~Teaney, {\em How much do heavy quarks thermalize in a heavy
  ion collision?},
\newblock Phys. Rev. C {\bf 71}, 064904 (2005).

\bibitem{PhysRevC.73.034913}
H.~van Hees, V.~Greco, and R.~Rapp, {\em Heavy-quark probes of the quark-gluon
  plasma and interpretation of recent data taken at the BNL Relativistic Heavy
  Ion Collider},
\newblock Phys. Rev. C {\bf 73}, 034913 (2006).

\bibitem{PhysRevC.71.034907}
H.~van Hees and R.~Rapp, {\em Thermalization of heavy quarks in the quark-gluon
  plasma},
\newblock Phys. Rev. C {\bf 71}, 034907 (2005).

\bibitem{PhysRevC.78.014904}
P.~B. Gossiaux and J.~Aichelin, {\em Toward an understanding of the single
  electron data measured at the BNL Relativistic Heavy Ion Collider (RHIC)},
\newblock Phys. Rev. C {\bf 78}, 014904 (2008).

\bibitem{PhysRevC.79.044906}
P.~B. Gossiaux, R.~Bierkandt, and J.~Aichelin, {\em Tomography of quark gluon
  plasma at energies available at the BNL Relativistic Heavy Ion Collider
  (RHIC) and the CERN Large Hadron Collider (LHC)},
\newblock Phys. Rev. C {\bf 79}, 044906 (2009).

\bibitem{Adil2007139}
A.~Adil and I.~Vitev, {\em Collisional dissociation of heavy mesons in dense
  QCD matter},
\newblock Physics Letters B {\bf 649}, 139  (2007).

\bibitem{PhysRevC.80.054902}
R.~Sharma, I.~Vitev, and B.-W. Zhang, {\em Light-cone wave function approach to
  open heavy flavor dynamics in QCD matter},
\newblock Phys. Rev. C {\bf 80}, 054902 (2009).

\bibitem{Dokshitzer2001199}
Y.~L. Dokshitzer and D.~E. Kharzeev, {\em Heavy-quark colorimetry of QCD
  matter},
\newblock Physics Letters B {\bf 519}, 199  (2001).

\bibitem{PhysRevLett.96.032301}
S.~S. Adler {\em et~al.} ({PHENIX} Collaboration), {\em Nuclear Modification of
  Electron Spectra and Implications for Heavy Quark Energy Loss in $Au+Au$
  Collisions at $\sqrt{s_{NN}}=200\,\,GeV$},
\newblock Phys. Rev. Lett. {\bf 96}, 032301 (2006).

\bibitem{Wong1994}
C.-Y. Wong,
\newblock {\em Introduction to High-Energy Heavy-Ion Collisions} (World
  Scientific, Singapore, 1994).

\bibitem{PhysRevD.77.014015}
S.~Peign\'e and A.~Peshier, {\em Collisional energy loss of a fast muon in a
  hot QED plasma},
\newblock Phys. Rev. D {\bf 77}, 014015 (2008).

\bibitem{XioyanLinThesis}
X.~Lin,
\newblock {\em Non-Photonic Electron Angular Correlations with Charged Hadrons
  from the {STAR} Experiment\: First Measurement of Bottom Contribution to
  Non-Photonic Electrons at {RHIC}},
\newblock PhD thesis, Central China Normal University, 2007.

\bibitem{Biritz2009849c}
B.~Biritz, {\em Non-photonic electron-hadron azimuthal correlation for {AuAu},
  {CuCu} and pp collisions at},
\newblock Nuclear Physics A {\bf 830}, 849c  (2009),
\newblock Quark Matter 2009 - The 21st International Conference on
  Ultrarelativistic Nucleus-Nucleus Collisions.

\bibitem{PYTHIA6.4}
T.~Sj\"ostrand, S.~Mrenna, and P.~Skands, {\em {PYTHIA} 6.4 physics and
  manual},
\newblock Journal of High Energy Physics {\bf 2006}, 026 (2006).

\bibitem{Dunlop2009419c}
J.~Dunlop, {\em Open Heavy Flavor Production in Heavy Ion Collisions},
\newblock Nuclear Physics A {\bf 830}, 419c  (2009),
\newblock Quark Matter 2009 - The 21st International Conference on
  Ultrarelativistic Nucleus-Nucleus Collisions.

\bibitem{PhysRevLett.98.192301}
B.~I. Abelev {\em et~al.} ({STAR} Collaboration), {\em Transverse Momentum and
  Centrality Dependence of High-$p_{T}$ Nonphotonic Electron Suppression in
  $Au+Au$ Collisions at $\sqrt{s_{NN}}=200\,\,GeV$},
\newblock Phys. Rev. Lett. {\bf 98}, 192301 (2007).

\bibitem{PhysRevLett.98.192301Erratum}
B.~I. Abelev {\em et~al.} ({STAR} Collaboration),
\newblock {\em Erratum: Transverse Momentum and Centrality Dependence of
  High-$p_{T}$ Nonphotonic Electron Suppression in $Au+Au$ Collisions at
  $\sqrt{s_{NN}}=200\,\,GeV$ [{Phys. Rev. Lett.} 98, 192301 (2007)]},
\newblock in preparation.

\bibitem{Mischke2009361}
A.~Mischke, {\em A new correlation method to identify and separate charm and
  bottom production processes at RHIC},
\newblock Physics Letters B {\bf 671}, 361  (2009).

\bibitem{STAR_ppNPE2011}
H.~Agakishiev {\em et~al.} ({STAR} Collaboration), {\em High $p_{T}$
  non-photonic electron production in $p+p$ collisions at $\sqrt{s}$ = 200
  GeV},
\newblock {arXiv:1102.2611v1} [nucl-ex]  (2011).

\bibitem{Harrison2003235}
M.~Harrison, T.~Ludlam, and S.~Ozaki, {\em {RHIC} project overview},
\newblock Nuclear Instruments and Methods in Physics Research Section A:
  Accelerators, Spectrometers, Detectors and Associated Equipment {\bf 499},
  235  (2003),
\newblock The {Relativistic Heavy Ion Collider} Project: {RHIC} and its
  Detectors.

\bibitem{Hahn2003245}
H.~Hahn {\em et~al.}, {\em The {RHIC} design overview},
\newblock Nuclear Instruments and Methods in Physics Research Section A:
  Accelerators, Spectrometers, Detectors and Associated Equipment {\bf 499},
  245  (2003),
\newblock The {Relativistic Heavy Ion Collider} Project: {RHIC} and its
  Detectors.

\bibitem{RHIC_Design}
N.~Samios,
\newblock {\em Conceptual Design of the {Relativistic Heavy Ion Collider}}
  (National Technical Information Service, Springfield, Virginia, 1989).

\bibitem{Anerella2003280}
M.~Anerella {\em et~al.}, {\em The {RHIC} magnet system},
\newblock Nuclear Instruments and Methods in Physics Research Section A:
  Accelerators, Spectrometers, Detectors and Associated Equipment {\bf 499},
  280  (2003),
\newblock The {Relativistic Heavy Ion Collider} Project: {RHIC} and its
  Detectors.

\bibitem{Back2003603}
B.~B. Back {\em et~al.}, {\em The {PHOBOS} detector at {RHIC}},
\newblock Nuclear Instruments and Methods in Physics Research Section A:
  Accelerators, Spectrometers, Detectors and Associated Equipment {\bf 499},
  603  (2003),
\newblock The {Relativistic Heavy Ion Collider} Project: {RHIC} and its
  Detectors.

\bibitem{Adamczyk2003437}
M.~Adamczyk {\em et~al.}, {\em The {BRAHMS} experiment at {RHIC}},
\newblock Nuclear Instruments and Methods in Physics Research Section A:
  Accelerators, Spectrometers, Detectors and Associated Equipment {\bf 499},
  437  (2003),
\newblock The {Relativistic Heavy Ion Collider} Project: {RHIC} and its
  Detectors.

\bibitem{Adcox2003469}
K.~Adcox {\em et~al.}, {\em {PHENIX} detector overview},
\newblock Nuclear Instruments and Methods in Physics Research Section A:
  Accelerators, Spectrometers, Detectors and Associated Equipment {\bf 499},
  469  (2003),
\newblock The {Relativistic Heavy Ion Collider} Project: {RHIC} and its
  Detectors.

\bibitem{Adcox2003489}
K.~Adcox {\em et~al.}, {\em {PHENIX} central arm tracking detectors},
\newblock Nuclear Instruments and Methods in Physics Research Section A:
  Accelerators, Spectrometers, Detectors and Associated Equipment {\bf 499},
  489  (2003),
\newblock The {Relativistic Heavy Ion Collider} Project: {RHIC} and its
  Detectors.

\bibitem{Aizawa2003508}
M.~Aizawa {\em et~al.}, {\em {PHENIX} central arm particle {ID} detectors},
\newblock Nuclear Instruments and Methods in Physics Research Section A:
  Accelerators, Spectrometers, Detectors and Associated Equipment {\bf 499},
  508  (2003),
\newblock The {Relativistic Heavy Ion Collider} Project: {RHIC} and its
  Detectors.

\bibitem{Aphecetche2003521}
L.~Aphecetche {\em et~al.}, {\em {PHENIX} calorimeter},
\newblock Nuclear Instruments and Methods in Physics Research Section A:
  Accelerators, Spectrometers, Detectors and Associated Equipment {\bf 499},
  521  (2003),
\newblock The {Relativistic Heavy Ion Collider} Project: {RHIC} and its
  Detectors.

\bibitem{Aronson2003480}
S.~H. Aronson {\em et~al.}, {\em {PHENIX} magnet system},
\newblock Nuclear Instruments and Methods in Physics Research Section A:
  Accelerators, Spectrometers, Detectors and Associated Equipment {\bf 499},
  480  (2003),
\newblock The {Relativistic Heavy Ion Collider} Project: {RHIC} and its
  Detectors.

\bibitem{Akikawa2003537}
H.~Akikawa {\em et~al.}, {\em {PHENIX} Muon Arms},
\newblock Nuclear Instruments and Methods in Physics Research Section A:
  Accelerators, Spectrometers, Detectors and Associated Equipment {\bf 499},
  537  (2003),
\newblock The {Relativistic Heavy Ion Collider} Project: {RHIC} and its
  Detectors.

\bibitem{Allen2003549}
M.~Allen {\em et~al.}, {\em {PHENIX} inner detectors},
\newblock Nuclear Instruments and Methods in Physics Research Section A:
  Accelerators, Spectrometers, Detectors and Associated Equipment {\bf 499},
  549  (2003),
\newblock The {Relativistic Heavy Ion Collider} Project: {RHIC} and its
  Detectors.

\bibitem{Adler2003433}
C.~Adler {\em et~al.}, {\em The {RHIC} zero-degree calorimeters},
\newblock Nuclear Instruments and Methods in Physics Research Section A:
  Accelerators, Spectrometers, Detectors and Associated Equipment {\bf 499},
  433  (2003),
\newblock The {Relativistic Heavy Ion Collider} Project: {RHIC} and its
  Detectors.

\bibitem{Bieser2003766}
F.~S. Bieser {\em et~al.}, {\em The {STAR} trigger},
\newblock Nuclear Instruments and Methods in Physics Research Section A:
  Accelerators, Spectrometers, Detectors and Associated Equipment {\bf 499},
  766  (2003),
\newblock The {Relativistic Heavy Ion Collider} Project: {RHIC} and its
  Detectors.

\bibitem{Ackermann2003624}
K.~H. Ackermann {\em et~al.}, {\em {STAR} detector overview},
\newblock Nuclear Instruments and Methods in Physics Research Section A:
  Accelerators, Spectrometers, Detectors and Associated Equipment {\bf 499},
  624  (2003),
\newblock The {Relativistic Heavy Ion Collider} Project: {RHIC} and its
  Detectors.

\bibitem{Anderson2003659}
M.~Anderson {\em et~al.}, {\em The {STAR} time projection chamber: a unique
  tool for studying high multiplicity events at {RHIC}},
\newblock Nuclear Instruments and Methods in Physics Research Section A:
  Accelerators, Spectrometers, Detectors and Associated Equipment {\bf 499},
  659  (2003),
\newblock The {Relativistic Heavy Ion Collider} Project: {RHIC} and its
  Detectors.

\bibitem{Beddo2003725}
M.~Beddo {\em et~al.}, {\em The {STAR Barrel Electromagnetic Calorimeter}},
\newblock Nuclear Instruments and Methods in Physics Research Section A:
  Accelerators, Spectrometers, Detectors and Associated Equipment {\bf 499},
  725  (2003),
\newblock The {Relativistic Heavy Ion Collider} Project: {RHIC} and its
  Detectors.

\bibitem{Bergsma2003633}
F.~Bergsma {\em et~al.}, {\em The {STAR} detector magnet subsystem},
\newblock Nuclear Instruments and Methods in Physics Research Section A:
  Accelerators, Spectrometers, Detectors and Associated Equipment {\bf 499},
  633  (2003),
\newblock The {Relativistic Heavy Ion Collider} Project: {RHIC} and its
  Detectors.

\bibitem{Bellwied2003640}
R.~Bellwied {\em et~al.}, {\em The {STAR Silicon Vertex Tracker}: A large area
  Silicon Drift Detector},
\newblock Nuclear Instruments and Methods in Physics Research Section A:
  Accelerators, Spectrometers, Detectors and Associated Equipment {\bf 499},
  640  (2003),
\newblock The {Relativistic Heavy Ion Collider} Project: {RHIC} and its
  Detectors.

\bibitem{Arnold2003652}
L.~Arnold {\em et~al.}, {\em The {STAR} silicon strip detector ({SSD})},
\newblock Nuclear Instruments and Methods in Physics Research Section A:
  Accelerators, Spectrometers, Detectors and Associated Equipment {\bf 499},
  652  (2003),
\newblock The {Relativistic Heavy Ion Collider} Project: {RHIC} and its
  Detectors.

\bibitem{Ackermann2003713}
K.~H. Ackermann {\em et~al.}, {\em The forward time projection chamber in
  {STAR}},
\newblock Nuclear Instruments and Methods in Physics Research Section A:
  Accelerators, Spectrometers, Detectors and Associated Equipment {\bf 499},
  713  (2003),
\newblock The {Relativistic Heavy Ion Collider} Project: {RHIC} and its
  Detectors.

\bibitem{Allgower2003740}
C.~E. Allgower {\em et~al.}, {\em The {STAR} endcap electromagnetic
  calorimeter},
\newblock Nuclear Instruments and Methods in Physics Research Section A:
  Accelerators, Spectrometers, Detectors and Associated Equipment {\bf 499},
  740  (2003),
\newblock The {Relativistic Heavy Ion Collider} Project: {RHIC} and its
  Detectors.

\bibitem{Anderson2003679}
M.~Anderson {\em et~al.}, {\em A readout system for the {STAR} time projection
  chamber},
\newblock Nuclear Instruments and Methods in Physics Research Section A:
  Accelerators, Spectrometers, Detectors and Associated Equipment {\bf 499},
  679  (2003),
\newblock The {Relativistic Heavy Ion Collider} Project: {RHIC} and its
  Detectors.

\bibitem{STAR_Note190}
J.~Mitchell and I.~Sakrejda, {\em Tracking for the {STAR TPC}: Documentation
  and User's Guide},
\newblock {STAR} Note {\bf 190} (1994).

\bibitem{STAR_Note89}
S.~Margetis and D.~Cebra, {\em Main Vertex reconstruction in STAR},
\newblock {STAR} Note {\bf 89} (1992).

\bibitem{Fernow}
R.~Fernow,
\newblock {\em Introduction to Experimental Particle Physics} (Cambridge
  University Press, Cambridge, UK, 1986).

\bibitem{Bichsel2006154}
H.~Bichsel, {\em A method to improve tracking and particle identification in
  {TPCs} and silicon detectors},
\newblock Nuclear Instruments and Methods in Physics Research Section A:
  Accelerators, Spectrometers, Detectors and Associated Equipment {\bf 562},
  154  (2006).

\bibitem{Xu201028}
Y.~Xu {\em et~al.}, {\em Improving the {dE/dx} calibration of the {STAR TPC}
  for the high-{pT} hadron identification},
\newblock Nuclear Instruments and Methods in Physics Research Section A:
  Accelerators, Spectrometers, Detectors and Associated Equipment {\bf 614}, 28
   (2010).

\bibitem{FabjanCalorimetry}
C.~Fabjan,
\newblock Calorimetry in high-energy physics,
\newblock in {\em Experimental Techniques in High-Energy Nuclear and Particle
  Physic}, edited by T.~Ferbel, pp. 257--324, World Scientific, Singapore,
  1991.

\bibitem{PhysRevC.81.054907}
B.~I. Abelev {\em et~al.} ({STAR} Collaboration), {\em Spectra of identified
  high-$p_{T}$ $\pi^{0}$ and $p(\bar{p})$ in $\mathrm{Cu+Cu}$ collisions at
  $\sqrt{s_{NN}}=200\;\mathrm{GeV}$},
\newblock Phys. Rev. C {\bf 81}, 054907 (2010).

\bibitem{Caines_Private_2011}
H.~Caines,
\newblock private communication, 2011.

\bibitem{Akimenko199592}
S.~A. Akimenko {\em et~al.}, {\em Study of strip fiber prototype shower maximum
  detector for the STAR experiment at RHIC},
\newblock Nuclear Instruments and Methods in Physics Research Section A:
  Accelerators, Spectrometers, Detectors and Associated Equipment {\bf 365}, 92
   (1995).

\bibitem{Vogelsang_Private_2010}
W.~Vogelsang,
\newblock private communication, 2010.

\bibitem{Agostinelli2003250}
S.~Agostinelli {\em et~al.}, {\em {GEANT4}--a simulation toolkit},
\newblock Nuclear Instruments and Methods in Physics Research Section A:
  Accelerators, Spectrometers, Detectors and Associated Equipment {\bf 506},
  250  (2003).

\bibitem{PhysRev.98.1355}
N.~M. Kroll and W.~Wada, {\em Internal Pair Production Associated with the
  Emission of High-Energy Gamma Rays},
\newblock Phys. Rev. {\bf 98}, 1355 (1955).

\bibitem{PhysRevC.81.064904}
B.~I. Abelev {\em et~al.} ({STAR} Collaboration), {\em Inclusive $\pi{}^{0}$,
  $\eta{}$, and direct photon production at high transverse momentum in $p+p$
  and $d+Au$ collisions at $\sqrt{s_{NN}}=200\;GeV$},
\newblock Phys. Rev. C {\bf 81}, 064904 (2010).

\bibitem{STAR_CuCu_freezeout}
M.~M. Aggarwal {\em et~al.} ({STAR} Collaboration), {\em Scaling properties at
  freeze-out in relativistic heavy ion collisions},
\newblock {arXiv}:1008.3133  (2010).

\bibitem{PhysRevLett.91.241803}
S.~S. Adler {\em et~al.} ({PHENIX} Collaboration), {\em Midrapidity
  Neutral-Pion Production in Proton-Proton Collisions at $\sqrt{s}=200\;GeV$},
\newblock Phys. Rev. Lett. {\bf 91}, 241803 (2003).

\bibitem{PhysRevD.76.051106}
A.~Adare {\em et~al.} ({PHENIX} Collaboration), {\em Inclusive cross section
  and double helicity asymmetry for $\pi{}^{0}$ production in $p+p$ collisions
  at $\sqrt{s}=200\;GeV$: Implications for the polarized gluon distribution in
  the proton},
\newblock Phys. Rev. D {\bf 76}, 051106 (2007).

\bibitem{PhysRevLett.101.162301}
A.~Adare {\em et~al.} ({PHENIX} Collaboration), {\em Onset of $\pi{}^{0}$
  Suppression Studied in $Cu+Cu$ Collisions at $\sqrt{s_{NN}}=22.4$, 62.4, and
  200 {GeV}},
\newblock Phys. Rev. Lett. {\bf 101}, 162301 (2008).

\bibitem{PhysRevLett.104.132301}
A.~Adare {\em et~al.} ({PHENIX} Collaboration), {\em Enhanced Production of
  Direct Photons in $Au+Au$ Collisions at $\sqrt{s_{NN}}=200\;GeV$ and
  Implications for the Initial Temperature},
\newblock Phys. Rev. Lett. {\bf 104}, 132301 (2010).

\bibitem{PhysRevLett.98.012002}
S.~S. Adler {\em et~al.} ({PHENIX} Collaboration), {\em Measurement of Direct
  Photon Production in $p+p$ Collisions at $\sqrt{s}=200\;GeV$},
\newblock Phys. Rev. Lett. {\bf 98}, 012002 (2007).

\bibitem{PhysRevC.69.034909}
S.~S. Adler {\em et~al.} ({PHENIX} Collaboration), {\em Identified charged
  particle spectra and yields in $Au+Au$ collisions at
  $\sqrt{s_{NN}}=200\;GeV$},
\newblock Phys. Rev. C {\bf 69}, 034909 (2004).

\bibitem{PhysRevLett.97.152301}
B.~I. Abelev {\em et~al.} ({STAR} Collaboration), {\em Identified Baryon and
  Meson Distributions at Large Transverse Momenta from $Au+Au$ Collisions at
  $\sqrt{s_{NN}}=200\,\,GeV$},
\newblock Phys. Rev. Lett. {\bf 97}, 152301 (2006).

\bibitem{PhysRevLett.98.062301}
J.~Adams {\em et~al.} ({STAR} Collaboration), {\em Scaling Properties of
  Hyperon Production in $Au+Au$ Collisions at $\sqrt{s_{NN}}=200\,\,GeV$},
\newblock Phys. Rev. Lett. {\bf 98}, 062301 (2007).

\bibitem{PhysRevC.79.034909}
B.~I. Abelev {\em et~al.} ({STAR} Collaboration), {\em Systematic measurements
  of identified particle spectra in $pp$, $d+Au$, and $Au+Au$ collisions at the
  STAR detector},
\newblock Phys. Rev. C {\bf 79}, 034909 (2009).

\bibitem{PhysRevLett.101.122301}
A.~Adare {\em et~al.} (PHENIX Collaboration Collaboration), {\em $J/\psi{}$
  Production in $\sqrt{s_{NN}}=200\,\,GeV$ $Cu+Cu$ Collisions},
\newblock Phys. Rev. Lett. {\bf 101}, 122301 (2008).

\bibitem{Frawley2008125}
A.~Frawley, T.~Ullrich, and R.~Vogt, {\em Heavy flavor in heavy-ion collisions
  at RHIC and RHIC II},
\newblock Physics Reports {\bf 462}, 125  (2008).

\bibitem{Barger1980253}
V.~Barger, W.~Y. Keung, and R.~J.~N. Phillips, {\em On $\psi$ and $\Upsilon$
  production via gluons},
\newblock Physics Letters B {\bf 91}, 253  (1980).

\bibitem{Vogt_RHICII2005}
R.~Vogt,
\newblock {\em Baseline Predictions for Open and Hidden Heavy Flavor Production
  at RHIC II},
\newblock presented at {RHIC II Workshop}, 2005.

\bibitem{Gavai_IJMPA1995}
R.~Gavai {\em et~al.}, {\em Quarkonium Production in Hadronic Collisions},
\newblock International Journal of Modern Physics A {\bf 10}, 3043 (1995).

\bibitem{Lafferty1995541}
G.~D. Lafferty and T.~R. Wyatt, {\em Where to stick your data points: The
  treatment of measurements within wide bins},
\newblock Nuclear Instruments and Methods in Physics Research Section A:
  Accelerators, Spectrometers, Detectors and Associated Equipment {\bf 355},
  541  (1995).

\bibitem{PhysRevLett.105.252301}
K.~Aamodt {\em et~al.} ({ALICE} Collaboration), {\em Charged-Particle
  Multiplicity Density at Midrapidity in Central {Pb-Pb} Collisions at
  $\sqrt{s_{NN}}=2.76$ {TeV}},
\newblock Phys. Rev. Lett. {\bf 105}, 252301 (2010).

\bibitem{springerlink:10.1140/epjc/s10052-009-1227-4}
K.~Aamodt {\em et~al.} ({ALICE} Collaboration), {\em First proton-proton
  collisions at the {LHC} as observed with the {ALICE} detector: measurement of
  the charged-particle pseudorapidity density at $\sqrt{s}=900$ {GeV}},
\newblock The European Physical Journal C - Particles and Fields {\bf 65}, 111
  (2010).

\bibitem{springerlink:10.1140/epjc/s10052-010-1339-x}
K.~Aamodt {\em et~al.} ({ALICE} Collaboration), {\em Charged-particle
  multiplicity measurement in proton-proton collisions at $\sqrt{s}=0.9$ and
  2.36 {TeV} with {ALICE at LHC}},
\newblock The European Physical Journal C - Particles and Fields {\bf 68}, 89
  (2010).

\bibitem{springerlink:10.1140/epjc/s10052-010-1350-2}
K.~Aamodt {\em et~al.} ({ALICE} Collaboration), {\em Charged-particle
  multiplicity measurement in proton-proton collisions at $\sqrt{s}=7$ {TeV}
  with {ALICE at LHC}},
\newblock The European Physical Journal C - Particles and Fields {\bf 68}, 345
  (2010).

\bibitem{ATLAS_overview}
G.~Aad {\em et~al.} ({ATLAS} Collaboration), {\em The {ATLAS} Experiment at the
  {CERN Large Hadron Collider}},
\newblock Journal of Instrumentation , S08003 (2008).

\bibitem{ATLAS_Jpsi}
G.~Aad {\em et~al.} ({ATLAS} Collaboration), {\em Measurement of the centrality
  dependence of $\mathit{J/\psi}$ yields and observation of $\mathit{Z}$
  production in lead-lead collisions with the {ATLAS} detector at the {LHC}},
\newblock ar{X}iv:1012.5419  (2010).

\bibitem{ALICE_overview}
K.~Aamodt {\em et~al.} ({ALICE} Collaboration), {\em The {ALICE} experiment at
  the {CERN LHC}},
\newblock Journal of Instrumentation , S08002 (2008).

\bibitem{Grelli_WWND2011}
A.~Grelli ({ALICE} Collaboration),
\newblock {\em Heavy Flavour Physics with the {ALICE} detector at {LHC}},
\newblock presented at {Winter Workshop on Nuclear Dynamics}, 2011.

\bibitem{Harris_INTWorkshop2010}
J.~Harris ({ALICE} Collaboration),
\newblock {\em Prospects for First Measurements with {ALICE} in $Pb+Pb$
  Collisions at the {LHC}},
\newblock presented at {INT Workshop, Seattle, WA}, 2010.

\bibitem{UllrichXuErrors}
T.~Ullrich and Z.~Xu, {\em Treatment of Errors in Efficiency Calculations},
\newblock {arXiv}:physics/0701199v1  (2007).

\end{thebibliography}
\printindex

\end{document}